\documentstyle[12pt]{article}
\renewcommand{\theequation}{\arabic{section}.\arabic{equation}}
\renewcommand{\thefootnote}{\star}

\catcode`\@=11
\def\fleqnarray{\stepcounter{equation}\let\@currentlabel=\theequation
\global\@eqnswtrue
\global\@eqcnt\z@\tabskip\mathindent\let\\=\@eqncr
\abovedisplayskip\topsep\ifvmode\advance\abovedisplayskip\partopsep\fi
\belowdisplayskip\abovedisplayskip
\belowdisplayshortskip\abovedisplayskip
\abovedisplayshortskip\abovedisplayskip
$$\halign to
\linewidth\bgroup\@eqnsel\hskip\@centering$\displaystyle\tabskip\z@
 {##}$&\global\@eqcnt\@ne \hskip 2\arraycolsep \hfil${##}$\hfil
 &\global\@eqcnt\tw@ \hskip 2\arraycolsep $\displaystyle{##}$\hfil
 \tabskip\@centering&\llap{##}\tabskip\z@\cr}
\def\endfleqnarray{\@@eqncr\egroup
 \global\advance\c@equation\m@ne$$\global\@ignoretrue }
\newdimen\mathindent
\mathindent = \parindent
\catcode`\@=\active

\begin{document}
\hsize=6.5 in
\vsize=9.0 in
\voffset=-23mm
\hoffset=-14mm
\textheight=9.0 in
\textwidth=6.5 in
\linewidth=6.5 in

\begin{center}
{\large\bf Covariant Perturbation Theory\\ 
 (IV). Third Order in the Curvature}\footnote{\normalsize This paper appeared in February 1993 as the University of Manitoba report, SPIRES-HEP:
PRINT-93-0274 (MANITOBA). The purpose of
the present publication is to make it more
accessible. As compared to the original text,
a minor error in the Appendix is corrected. {\em Mathematica} files with the results of this paper are included in the source file of the present submission.}\\
 \hphantom{jjj}
\vspace{5mm}

  {\sc A. O. Barvinsky,
  Yu. V. Gusev,
  V. V. Zhytnikov}\\
  {\it Nuclear Safety Institute, Bolshaya Tulskaya 52,}\\
  {\it Moscow 113191, Russia}\\ \vspace{\baselineskip}
   {\rm and}\\ \vspace{\baselineskip}
  {\sc G. A. Vilkovisky}\\
  {\it Lebedev Physics Institute, Leninsky Prospect 53,}\\
  {\it Moscow 117924, Russia}
\end{center}
\vspace{5mm}
\baselineskip=14 pt
\centerline{
\bf Abstract
}
\vspace{4pt}
{
\par
The trace of the heat kernel and the one-loop effective
action for the generic differential operator
are calculated to third order in the background
curvatures: the Riemann curvature, the commutator
curvature and the potential. In the case of
effective action, this is equivalent to a
calculation (in the covariant form) of the one-loop
vertices in all models of gravitating fields.
The basis of nonlocal invariants of third order
in the curvature is built, and constraints
arising between these invariants in
low-dimensional
manifolds are obtained.
All third-order form factors in the heat kernel
and effective action are calculated, and
several integral representations for them are
obtained. In the case of effective action, this
includes a specially generalized spectral
representation used in applications to the
expectation-value equations. The results
for the heat kernel are checked  by deriving
all the known coefficients of the
Schwinger-DeWitt expansion including
$a_3$ and the cubic terms of $a_4$.
The results for the effective action are
checked by deriving the trace anomaly in two and
four dimensions. In four dimensions, this
derivation is carried out by several
different techniques elucidating the mechanism
by which the local anomaly emerges from
the nonlocal action. In two dimensions, it is shown
by a direct calculation that the series for
the effective action terminates  at second order
in the curvature. The asymptotic behaviours of the
form factors are calculated including the
late-time behaviour in the heat kernel
and the small-$\Box$ behaviour in the
effective action. In quantum gravity, the latter
behaviour contains the effects of vacuum
radiation including the Hawking effect.}\\

\thispagestyle{empty}
\pagebreak
\tableofcontents
\thispagestyle{empty}
\pagebreak
\setcounter{page}{1}
\hfuzz=20pt
\renewcommand{\thefootnote}{\star}

\section{Introduction}
\setcounter{equation}{0}

\hspace{\parindent}
The present paper is a sequel of the series of papers
where covariant perturbation theory was proposed [1] and
used [2,3] for the calculation of the heat kernel,
effective action, and expectation-value equations in
field theory. We refer to ref. [1] as paper I; papers
II and III are refs. [2] and [3] respectively.

In the present paper, the trace of the heat kernel and
the one-loop effective action for the general background
are calculated to third order in the field strengths
(curvatures). The calculation of  the heat kernel is
carried out for an arbitrary space-time dimension, and
the effective action is computed in dimensions two and four.
For the form factors in the heat kernel and
four-dimensional effective action, several integral
representations are obtained
including the spectral representation used in the
expectation-value equations [1,4].
Also, tables of various asymptotic behaviours
of the form factors are presented. The results are
checked by comparison with the Schwinger-DeWitt
expansion for the heat kernel, and by derivation of the
trace anomaly in four dimensions. In two dimensions,
it is explicitly checked that the expansion of the
effective action terminates at second order in
the curvature; terms of third order vanish. The
mechanism of this vanishing is revealed
and shown to be the same by which, in four dimensions,
the local
trace anomaly emerges from the nonlocal effective
action. The derivation of the trace anomaly
from the effective action is carried out in
several different ways including the technique of
spectral representation.

For the present work one needs a classification of
nonlocal curvature invariants, including
constrains arising between these invariants in
low-di\-men\-sional manifolds. Such a classification is
given to third order in the curvature. The paper is supplied
with an appendix which contains a systematic analysis of
identities for nonlocal cubic invariants in four
dimensions. As a by-product of this analysis, a
mechanism is discovered by which the Gauss-Bonnet
invariant becomes topological in four dimensions
(eq. (A.39) of Appendix).

The present paper, like the preceding ones, deals only
with the version of covariant perturbation theory appropriate
for noncompact asymptotically flat manifolds. The effective
action is computed for euclidean signature of the metric,
which, for certain quantum states, is sufficient for
obtaining the expectation-value equations of lorentzian
field theory [1].

For the motivation of the present study, discussion of the
method, and relevant physical problems we refer to the
preceding papers and the recent report [5] where covariant
perturbation theory along with some of its applications is
reviewed. Here we only remind of the notation.

The subject of calculation is the heat kernel
\begin{equation} K(s)=\exp sH \end{equation}
where $H$ is the generic second-order operator
\begin{equation}
H=g^{\mu\nu}\nabla_\mu\nabla_\nu\hat{1}+(\hat{P}-\frac16R\hat{1}),\hspace{7mm}
g^{\mu\nu}\nabla_\mu\nabla_\nu\equiv\Box
\end{equation}
acting on small disturbances of an arbitrary set of fields
$\varphi^A(x)$. Here $A$ stands for any set of discrete
indices, and the hat indicates that the quantity is a matrix
acting on the vector of small disturbances
$\delta\varphi^A$ :
\begin{equation}\hat{1}=\delta^A{}_B,\hspace{7mm} \hat{P}=P^A{}_B,\ {\rm etc.} \end{equation}
The matrix trace will be denoted by ${\rm tr}$ :
\begin{equation}{\rm tr}\hat{1}=\delta^A{}_A,\hspace{7mm} {\rm tr}\hat{P}=P^A{}_A,\ {\rm etc.} \end{equation}
In (1.2), $g_{\mu\nu}$ is a positive-definite metric
characterized by its Riemann curvature
\footnote{\normalsize We use the conventions
$R^\mu_{\,\cdot\, \alpha\nu\beta}=\partial_\nu
\Gamma^\mu_{\alpha\beta}-\cdots,\  R_{\alpha\beta}=
R^\mu_{\,\cdot\, \alpha\mu\beta},\
R=g^{\alpha\beta}R_{\alpha\beta}$.}
$R^{\alpha\beta\mu\nu}$,
$\nabla_\mu$ is a covariant derivative (with respect to an
arbitrary connection) characterized by its commutator
curvature
\begin{equation}(\nabla_\mu\nabla_\nu-\nabla_\nu\nabla_\mu)\delta\varphi^A =
{\cal R}^A{}_{B\mu\nu}\delta\varphi^B,\hspace{7mm}
{\cal R}^A{}_{B\mu\nu}\equiv\hat{\cal R}_{\mu\nu},
\end{equation}
and $\hat{P}$ is an arbitrary matrix. The redefinition of the
potential in (1.2) by inclusion of the term in the Ricci
scalar $R$ is a matter of convenience.

For the set of the field strengths (curvatures)
\begin{equation} R^{\alpha\beta\mu\nu},\hspace{7mm} \hat{\cal R}_{\mu\nu},\hspace{7mm} \hat{P}\end{equation}
characterizing the background we use the collective
notation $\Re$. The calculations in covariant perturbation
theory are carried out with accuracy ${\rm O}[\Re^n]$, i.e.
up to terms of $n$th and higher power in the curvatures (1.6).
It is worth noting that, since the calculations are covariant,
any term in $g_{\mu\nu}$ is in fact of infinite power in the
curvature, and ${\rm O}[\Re^n]$ means terms containing $n$ or
more curvatures {\em explicitly}.

The heat kernel (1.1) governs the covariant diagrammatic
technique [5--8] to all loop orders. At one-loop order,
the effective action is given by the trace of the heat
kernel
\begin{equation}-W=\frac12\int^\infty_0 \frac{ds}s{\rm Tr} K(s)+
\int d^{2\omega}x\delta^{(2\omega)}(x,x)(\dots)\end{equation}
where ${\rm Tr}$, as distinct from ${\rm tr}$ in (1.4),
denotes the functional trace
\begin{equation}{\rm Tr} K(s)=\int d^{2\omega}x\ {\rm tr}[K(s)
\delta^{(2\omega)}(x,y)]\Big|_{y=x}.\end{equation}
Here and below, $2\omega$ is the space-time dimension,
and the term with ellipses $(\dots)$ stands for the
contribution of the local functional measure [9],
proportional to the delta-function at  coincident
points. As shown in [10,11], this contribution always
cancels the volume divergences of the loop
\footnote{\normalsize Loops with the heat kernels are
finite, and the ultraviolet divergences appear as
divergences of the integrals over $s$ at the lower limits.}
which otherwise would appear in (1.7) in the form of
a divergent cosmological term. {\em In the case of a massless
operator} (1.2), the result of this cancellation can be
written down as
\begin{equation}
-W=\frac12\int^\infty_0 \frac{ds}s\Big({\rm Tr} K(s)-{\rm Tr} K(s)\Big|_{\Re=0}\Big)
\end{equation}
where subtracted is the zeroth-order term of the
covariant expansion in powers of the curvature. The
masslessness of the operator (1.2) means that, like the
Riemann and commutator curvatures, the potential $\hat{P}$
falls off at infinity of the manifold. For the
precise conditions of this fall off see [2].

The paper is organized so that the first twelve sections
contain only the presentation of the final results, and
the remaining eight sections contain their derivations.
The final result for the trace of the heat kernel is given
in sec. 2. The late-time and early-time asymptotic
behaviours of ${\rm Tr} K(s)$ are considered in sects. 3 and 4
where also a comparison with the Schwinger-DeWitt
expansion is carried out. Sect. 5 contains the calculation
of the effective action in two dimensions. The final result
for the effective action in four dimensions is given in
sect. 6. Integral representations of the form factors
in the effective action are given in sects. 7 -- 9 :
the $\alpha$-representation (sect. 7),
the Laplace representation (sect. 8), and the spectral
representation (sect. 9). Sects. 10 and 11 contain the
tables of asymptotic behaviours of the form factors. The
trace anomaly is derived in sect. 12.
Sects. 13--16 contain the derivation of the
result for the trace of the heat kernel, and in
sect. 15 an alternative representation of the form
factors in ${\rm Tr} K(s)$ is given. Sects. 17--20 contain
the derivation of the results for the effective action
in four dimensions. Appendix contains the analysis of
identities for local and nonlocal cubic invariants in
low-dimensional manifolds.

The authors dispose of the results of the present paper
in the format of the computer algebra program
{\it Mathematica}. These files are available from 
the source file submitted at {\tt http://arxiv.org}.

\section{Final result for the trace of the heat
kernel to third order in the curvature}
\setcounter{equation}{0}

\hspace{\parindent}
The result is
\begin{eqnarray}
{\rm Tr} K(s) &=& \frac1{(4\pi s)^\omega}\int\! dx\, g^{1/2}\, \,{\rm tr}\Big\{\hat{1}+s\hat{P}\nonumber\\&&\mbox{}
+s^2\sum^{5}_{i=1}f_{i}(-s\Box_2)\Re_1\Re_2({i})\nonumber\\&&\mbox{}
+s^3\sum^{11}_{i=1}F_{i}(-s\Box_1,-s\Box_2,-s\Box_3) \Re_1\Re_2\Re_3({i})\nonumber\\&&\mbox{}
+s^4\sum^{25}_{i=12}F_{i}(-s\Box_1,-s\Box_2,-s\Box_3) \Re_1\Re_2\Re_3({i})\nonumber\\&&\mbox{}
+s^5\sum^{28}_{i=26}F_{i}(-s\Box_1,-s\Box_2,-s\Box_3) \Re_1\Re_2\Re_3({i})\nonumber\\&&\mbox{}
+s^6F_{29}(-s\Box_1,-s\Box_2,-s\Box_3) \Re_1\Re_2\Re_3({29})\nonumber\\&&\ \ \ \ \ \ \ \ \mbox{}
+{\rm O}[\Re^4]\big\}.
\end{eqnarray}
Here terms of zeroth, first and second order in
the curvature reproduce the results of paper II
\footnote{
\normalsize To second order in the curvature, the result for
${\rm Tr} K(s)$ was first published in
A.O.Barvinsky and G.A.Vilkovisky,
Proceedings of the Fourth Seminar on
Quantum Gravity,
May 25-29, 1987, Moscow,
eds. M.A.Markov, V.A.Berezin and V.P.Frolov
(World Scientific, Singapore, 1988) p.217.
}.
There are five quadratic structures
\begin{eqnarray}
\Re_1\Re_2({1})&=&R_{1\,\mu\nu} R_2^{\mu\nu}\hat{1},\\[\baselineskip]
\Re_1\Re_2({2})&=&R_1 R_2\hat{1},\\[\baselineskip]
\Re_1\Re_2({3})&=&\hat{P}_1 R_2,\\[\baselineskip]
\Re_1\Re_2({4})&=&\hat{P}_1\hat{P}_2,\\[\baselineskip]
\Re_1\Re_2({5})&=&\hat{\cal R}_{1\mu\nu}\hat{\cal R}_2^{\mu\nu}
\end{eqnarray}
whose contributions are of the form
\begin{equation}\int\! dx\, g^{1/2}\,  f(-s\Box_2)\Re_1\Re_2=
\int\! dx\, g^{1/2}\, \Re f(-s\Box)\Re,\end{equation}
and the notation on the left-hand side of (2.7)
assumes that $\Box_2$ acts on $\Re_2$. The form factors
$f_{i}$, $i=1$ to $5$, are functions of the operator
\begin{equation}-s\Box=\xi\end{equation}
(with $\Box$ defined in (1.2)) and are expressed
through the basic second-order form factor
\begin{equation} f(\xi)=\int_{\alpha\geq0}\!d^2\alpha\,
\delta(1-\alpha_1-\alpha_2)\exp(-\alpha_1\alpha_2\xi)
=\int^1_0\!d\alpha\,{\rm e}^{-\alpha(1-\alpha)\xi}\end{equation}
as follows
\begin{eqnarray}
f_1(\xi) &=& \frac{(f(\xi)-1+\frac16\xi)}{\xi^2},\\[\baselineskip]
f_2(\xi) &=& \frac18\left[
  \frac1{36}f(\xi)+\frac13\frac{(f(\xi)-1)}{\xi}-\frac{(f(\xi)-1+\frac16\xi)}{\xi^2}\right],\\[\baselineskip]
f_3(\xi) &=&
  \frac1{12}f(\xi)+\frac12\frac{(f(\xi)-1)}{\xi},\\[\baselineskip]
f_4(\xi) &=& \frac12f(\xi),\\[\baselineskip]
f_5(\xi) &=& -\frac12\frac{(f(\xi)-1)}{\xi}.
\end{eqnarray}

Terms of third order in the curvature in (2.1) are
given by a sum of contributions of twenty nine
cubic structures. Eleven of them contain no derivatives
\begin{eqnarray}
\Re_1\Re_2\Re_3({1})&=&\hat{P}_1\hat{P}_2\hat{P}_3,\\[\baselineskip]
\Re_1\Re_2\Re_3({2})&=&\hat{\cal R}^{\ \mu}_{1\ \alpha}\hat{\cal R}^{\ \alpha}_{2\ \beta}\hat{\cal R}^{\ \beta}_{3\ \mu},\\[\baselineskip]
\Re_1\Re_2\Re_3({3})&=&\hat{\cal R}^{\mu\nu}_1\hat{\cal R}_{2\,\mu\nu}\hat{P}_3,\\[\baselineskip]
\Re_1\Re_2\Re_3({4})&=&R_1 R_2 \hat{P}_3,\\[\baselineskip]
\Re_1\Re_2\Re_3({5})&=&R_1^{\mu\nu}R_{2\,\mu\nu}\hat{P}_3,\\[\baselineskip]
\Re_1\Re_2\Re_3({6})&=&\hat{P}_1\hat{P}_2 R_3,\\[\baselineskip]
\Re_1\Re_2\Re_3({7})&=&R_1\hat{\cal R}^{\mu\nu}_2\hat{\cal R}_{3\,\mu\nu},\\[\baselineskip]
\Re_1\Re_2\Re_3({8})&=&R_1^{\alpha\beta}\hat{\cal R}_{2\,\alpha}^{\ \ \ \,\mu}\hat{\cal R}_{3\,\beta\mu},\\[\baselineskip]
\Re_1\Re_2\Re_3({9})&=&R_1 R_2 R_3\hat{1},\\[\baselineskip]
\Re_1\Re_2\Re_3({10})&=&R_{1\,\alpha}^\mu R_{2\,\beta}^{\alpha} R_{3\,\mu}^\beta\hat{1},\\[\baselineskip]
\Re_1\Re_2\Re_3({11})&=&R_1^{\mu\nu}R_{2\,\mu\nu}R_3\hat{1},
\end{eqnarray}
fourteen contain two derivatives
\begin{eqnarray}
\Re_1\Re_2\Re_3({12})&=&\hat{\cal R}_1^{\alpha\beta}\nabla^\mu\hat{\cal R}_{2\mu\alpha}\nabla^\nu\hat{\cal R}_{3\nu\beta},\\[\baselineskip]
\Re_1\Re_2\Re_3({13})&=&\hat{\cal R}_1^{\mu\nu} \nabla_\mu\hat{P}_2\nabla_\nu\hat{P}_3,\\[\baselineskip]
\Re_1\Re_2\Re_3({14})&=&\nabla_\mu \hat{\cal R}_1^{\mu\alpha}\nabla^\nu \hat{\cal R}_{2\,\nu\alpha}\hat{P}_3,\\[\baselineskip]
\Re_1\Re_2\Re_3({15})&=&R_1^{\mu\nu}\nabla_\mu R_2\nabla_\nu \hat{P}_3,\\[\baselineskip]
\Re_1\Re_2\Re_3({16})&=&\nabla^\mu R_1^{\nu\alpha}\nabla_\nu R_{2\,\mu\alpha}\hat{P}_3,\\[\baselineskip]
\Re_1\Re_2\Re_3({17})&=&R_1^{\mu\nu}\nabla_\mu\nabla_\nu\hat{P}_2\hat{P}_3,\\[\baselineskip]
\Re_1\Re_2\Re_3({18})&=&R_{1\,\alpha\beta}\nabla_\mu\hat{\cal R}_2^{\mu\alpha}\nabla_\nu\hat{\cal R}_3^{\nu\beta},\\[\baselineskip]
\Re_1\Re_2\Re_3({19})&=&R_1^{\alpha\beta}\nabla_\alpha\hat{\cal R}_2^{\mu\nu}\nabla_\beta\hat{\cal R}_{3\,\mu\nu},\\[\baselineskip]
\Re_1\Re_2\Re_3({20})&=&R_1\nabla_\alpha\hat{\cal R}_2^{\alpha\mu}\nabla^\beta\hat{\cal R}_{3\,\beta\mu},\\[\baselineskip]
\Re_1\Re_2\Re_3({21})&=&R_1^{\mu\nu}\nabla_\mu\nabla_\lambda\hat{\cal R}_2^{\lambda\alpha}\hat{\cal R}_{3\,\alpha\nu},\\[\baselineskip]
\Re_1\Re_2\Re_3({22})&=&R_1^{\alpha\beta}\nabla_\alpha R_2 \nabla_\beta R_3\hat{1},\\[\baselineskip]
\Re_1\Re_2\Re_3({23})&=&\nabla^\mu R_1^{\nu\alpha}\nabla_\nu R_{2\,\mu\alpha}R_3\hat{1},\\[\baselineskip]
\Re_1\Re_2\Re_3({24})&=&R_1^{\mu\nu}\nabla_\mu R_2^{\alpha\beta}\nabla_\nu R_{3\,\alpha\beta}\hat{1},\\[\baselineskip]
\Re_1\Re_2\Re_3({25})&=&R_1^{\mu\nu}\nabla_\alpha R_{2\,\beta\mu}\nabla^\beta R_{3\,\nu}^\alpha\hat{1},
\end{eqnarray}
three contain four derivatives
\begin{eqnarray}
\Re_1\Re_2\Re_3({26})&=&\nabla_\alpha\nabla_\beta R_1^{\mu\nu}\nabla_\mu\nabla_\nu R_2^{\alpha\beta}\hat{P}_3,\\[\baselineskip]
\Re_1\Re_2\Re_3({27})&=&\nabla_\alpha\nabla_\beta R_1^{\mu\nu}\nabla_\mu\nabla_\nu R_2^{\alpha\beta} R_3\hat{1},\\[\baselineskip]
\Re_1\Re_2\Re_3({28})&=&\nabla_\mu R_1^{\alpha\lambda} \nabla_\nu R_{2\,\lambda}^\beta\nabla_\alpha\nabla_\beta R_3^{\mu\nu}\hat{1},
\end{eqnarray}
and one contains six derivatives
\begin{eqnarray}
\Re_1\Re_2\Re_3({29})&=&\nabla_\lambda\nabla_\sigma R_1^{\alpha\beta}\nabla_\alpha\nabla_\beta R_2^{\mu\nu}\nabla_\mu\nabla_\nu R_3^{\lambda\sigma}\hat{1}.
\end{eqnarray}
These twenty nine structures form a complete basis
of nonlocal invariants of third order in the
curvature. (Ten of them are purely gravitational,
and with gravity switched off there are six.)

The third-order form factors $F_{i}$, $i=1$ to $29$,
in the expressions
\begin{equation} F(-s\Box_1,-s\Box_2,-s\Box_3)\Re_1\Re_2\Re_3\end{equation}
are functions of three operator arguments
\begin{equation}
-s\Box_1=\xi_1,\hspace{7mm}
-s\Box_2=\xi_2,\hspace{7mm}
-s\Box_3=\xi_3,
\end{equation}
and it is assumed that
$\Box_1$ acts on the curvature with the label 1,
$\Box_2$ acts on the curvature with the label 2,
$\Box_3$ acts on the curvature with the label 3.
There is no question about a commutativity of
$\Box_{m}$ with $\nabla{}$'s acting on $\Re_m$ in
(2.26)--(2.43) since a contribution of any such
commutator is already ${\rm O}[\Re^4]$.

In expression (2.1), the form factors $F_{i}$
automatically acquire symmetries (if any) of their
respective curvature structures ${\rm tr}\Re_1\Re_2\Re_3({i})$.
These symmetries (under permutations of the
labels 1,2,3) follow from the table (2.15)--(2.43)
and imply the following symmetrization of the
form factors:
\begin{eqnarray}
F^{\rm sym}_{1}(\xi_1,\xi_2,\xi_3) &=&
  \frac13\Big( F_1(\xi_1,\xi_2,\xi_3)
          +F_1(\xi_3,\xi_1,\xi_2)\nonumber\\&&\ \ \ \ \mbox{}
          +F_1(\xi_2,\xi_3,\xi_1)\Big),\\[\baselineskip]
F^{\rm sym}_{2}(\xi_1,\xi_2,\xi_3) &=&
  \frac13\Big( F_2(\xi_1,\xi_2,\xi_3)
          +F_2(\xi_3,\xi_1,\xi_2)\nonumber\\&&\ \ \ \ \mbox{}
          +F_2(\xi_2,\xi_3,\xi_1)\Big),\\[\baselineskip]
F^{\rm sym}_{3}(\xi_1,\xi_2,\xi_3) &=&F_{3}(\xi_1,\xi_2,\xi_3),\\[\baselineskip]
F^{\rm sym}_{4}(\xi_1,\xi_2,\xi_3) &=&
  \frac12\Big(F_4(\xi_1,\xi_2,\xi_3)
        +F_4(\xi_2,\xi_1,\xi_3)\Big),\\[\baselineskip]
F^{\rm sym}_{5}(\xi_1,\xi_2,\xi_3) &=&
  \frac12\Big(F_5(\xi_1,\xi_2,\xi_3)
        +F_5(\xi_2,\xi_1,\xi_3)\Big),\\[\baselineskip]
F^{\rm sym}_{6}(\xi_1,\xi_2,\xi_3) &=&
  \frac12\Big(F_6(\xi_1,\xi_2,\xi_3)
        +F_6(\xi_2,\xi_1,\xi_3)\Big),\\[\baselineskip]
F^{\rm sym}_{7}(\xi_1,\xi_2,\xi_3) &=&
  \frac12\Big(F_7(\xi_1,\xi_2,\xi_3)
        +F_7(\xi_1,\xi_3,\xi_2)\Big),\\[\baselineskip]
F^{\rm sym}_{8}(\xi_1,\xi_2,\xi_3) &=&
  \frac12\Big(F_8(\xi_1,\xi_2,\xi_3)
        +F_8(\xi_1,\xi_3,\xi_2)\Big),\\[\baselineskip]
F^{\rm sym}_{9}(\xi_1,\xi_2,\xi_3) &=&
  \frac16\Big(F_9(\xi_1,\xi_2,\xi_3)
          +F_9(\xi_3,\xi_1,\xi_2)\nonumber\\&&\ \ \ \ \mbox{}
          +F_9(\xi_2,\xi_3,\xi_1)
          +F_9(\xi_2,\xi_1,\xi_3)\nonumber\\&&\ \ \ \ \mbox{}
          +F_9(\xi_3,\xi_2,\xi_1)
          +F_9(\xi_1,\xi_3,\xi_2)\Big),\\[\baselineskip]
F^{\rm sym}_{10}(\xi_1,\xi_2,\xi_3) &=&
  \frac16\Big(F_{10}(\xi_1,\xi_2,\xi_3)
          +F_{10}(\xi_3,\xi_1,\xi_2)\nonumber\\&&\ \ \ \ \mbox{}
          +F_{10}(\xi_2,\xi_3,\xi_1)
          +F_{10}(\xi_2,\xi_1,\xi_3)\nonumber\\&&\ \ \ \ \mbox{}
          +F_{10}(\xi_3,\xi_2,\xi_1)
          +F_{10}(\xi_1,\xi_3,\xi_2)\Big),\\[\baselineskip]
F^{\rm sym}_{11}(\xi_1,\xi_2,\xi_3) &=&
  \frac12\Big(F_{11}(\xi_1,\xi_2,\xi_3)
+F_{11}(\xi_2,\xi_1,\xi_3)\Big),\\[\baselineskip]
F^{\rm sym}_{12}(\xi_1,\xi_2,\xi_3)&=&F_{12}(\xi_1,\xi_2,\xi_3),\\[\baselineskip]
F^{\rm sym}_{13}(\xi_1,\xi_2,\xi_3)&=&F_{13}(\xi_1,\xi_2,\xi_3),\\[\baselineskip]
F^{\rm sym}_{14}(\xi_1,\xi_2,\xi_3)&=&F_{14}(\xi_1,\xi_2,\xi_3),\\[\baselineskip]
F^{\rm sym}_{15}(\xi_1,\xi_2,\xi_3)&=&F_{15}(\xi_1,\xi_2,\xi_3),\\[\baselineskip]
F^{\rm sym}_{16}(\xi_1,\xi_2,\xi_3) &=&
  \frac12\Big(F_{16}(\xi_1,\xi_2,\xi_3)
   +F_{16}(\xi_2,\xi_1,\xi_3)\Big),\\[\baselineskip]
F^{\rm sym}_{17}(\xi_1,\xi_2,\xi_3)&=&F_{17}(\xi_1,\xi_2,\xi_3),\\[\baselineskip]
F^{\rm sym}_{18}(\xi_1,\xi_2,\xi_3) &=&
  \frac12\Big(F_{18}(\xi_1,\xi_2,\xi_3)
  +F_{18}(\xi_1,\xi_3,\xi_2)\Big),\\[\baselineskip]
F^{\rm sym}_{19}(\xi_1,\xi_2,\xi_3) &=&
  \frac12\Big(F_{19}(\xi_1,\xi_2,\xi_3)
  +F_{19}(\xi_1,\xi_3,\xi_2)\Big),\\[\baselineskip]
F^{\rm sym}_{20}(\xi_1,\xi_2,\xi_3) &=&
  \frac12\Big(F_{20}(\xi_1,\xi_2,\xi_3)
  +F_{20}(\xi_1,\xi_3,\xi_2)\Big),\\[\baselineskip]
F^{\rm sym}_{21}(\xi_1,\xi_2,\xi_3)&=&F_{21}(\xi_1,\xi_2,\xi_3),\\[\baselineskip]
F^{\rm sym}_{22}(\xi_1,\xi_2,\xi_3) &=&
  \frac12\Big(F_{22}(\xi_1,\xi_2,\xi_3)
  +F_{22}(\xi_1,\xi_3,\xi_2)\Big),\\[\baselineskip]
F^{\rm sym}_{23}(\xi_1,\xi_2,\xi_3) &=&
  \frac12\Big(F_{23}(\xi_1,\xi_2,\xi_3)
  +F_{23}(\xi_2,\xi_1,\xi_3)\Big),\\[\baselineskip]
F^{\rm sym}_{24}(\xi_1,\xi_2,\xi_3) &=&
  \frac12\Big(F_{24}(\xi_1,\xi_2,\xi_3)
  +F_{24}(\xi_1,\xi_3,\xi_2)\Big),\\[\baselineskip]
F^{\rm sym}_{25}(\xi_1,\xi_2,\xi_3) &=&
  \frac12\Big(F_{25}(\xi_1,\xi_2,\xi_3)
  +F_{25}(\xi_1,\xi_3,\xi_2)\Big),\\[\baselineskip]
F^{\rm sym}_{26}(\xi_1,\xi_2,\xi_3) &=&
  \frac12\Big(F_{26}(\xi_1,\xi_2,\xi_3)
  +F_{26}(\xi_2,\xi_1,\xi_3)\Big),\\[\baselineskip]
F^{\rm sym}_{27}(\xi_1,\xi_2,\xi_3) &=&
  \frac12\Big(F_{27}(\xi_1,\xi_2,\xi_3)
  +F_{27}(\xi_2,\xi_1,\xi_3)\Big),\\[\baselineskip]
F^{\rm sym}_{28}(\xi_1,\xi_2,\xi_3) &=&
  \frac12\Big(F_{28}(\xi_1,\xi_2,\xi_3)
  +F_{28}(\xi_2,\xi_1,\xi_3)\Big),\\[\baselineskip]
F^{\rm sym}_{29}(\xi_1,\xi_2,\xi_3) &=&
  \frac13( F_{29}(\xi_1,\xi_2,\xi_3)
          +F_{29}(\xi_3,\xi_1,\xi_2)\nonumber\\&&\ \ \ \ \mbox{}
          +F_{29}(\xi_2,\xi_3,\xi_1)).
\end{eqnarray}
When taken separately from their curvature structures,
the functions $F_i$ make sense only being explicitly
symmetrized as above.

All the twenty nine functions $F_{i}(\xi_1,\xi_2,\xi_3)$ are expressed
through the basic third-order form factor
\begin{eqnarray}
F(\xi_1,\xi_2,\xi_3)&=&\int_{\alpha\geq0}d^3\alpha\,
\delta(1-\alpha_1-\alpha_2-\alpha_3)\nonumber\\&&\ \ \ \ \ \ \ \ \mbox{}
\times\exp(-\alpha_1\alpha_2\xi_3-\alpha_2\alpha_3\xi_1-\alpha_1\alpha_3\xi_2)
\end{eqnarray}
(which is completely symmetric in $\xi_1,\xi_2,\xi_3$)
and the basic second order form factor $f(\xi)$
introduced in (2.9). The coefficients of these
expressions are rational functions with a universal
denominator
\begin{equation}\Delta={\xi_1}^2+{\xi_2}^2+{\xi_3}^2
-2\xi_1\xi_2-2\xi_1\xi_3-2\xi_2\xi_3\end{equation}
raised to a certain power. In terms of (2.9), (2.75) and
(2.76), the explicit expressions for the (not symmetrized)
third-order form factors are as follows:
\arraycolsep=0pt
\mathindent=0pt
\begin{fleqnarray}&&
F_{1}(\xi_1,\xi_2,\xi_3)={1\over 3}F(\xi_1,\xi_2,\xi_3),
\end{fleqnarray}
\begin{fleqnarray}&&
F_{2}(\xi_1,\xi_2,\xi_3)=F(\xi_1,\xi_2,\xi_3)\Big[{{4 \xi_1 \xi_2 \xi_3}\over {3 {{\Delta}^3}}}
 ( -3 {{\xi_1}^2} \xi_2 - 3 {{\xi_1}^2} \xi_3 + 2 \xi_1 \xi_2 \xi_3 +
    3 {{\xi_3}^3} )\nonumber\\&&\ \ \ \ \mbox{}
+\frac{4}{{\Delta}^{2}}
( -{{\xi_1}^2} \xi_2 - {{\xi_1}^2} \xi_3+ 2 \xi_1 \xi_2 \xi_3 +
    {{\xi_3}^3} )\Big]\nonumber\\&&\ \ \ \ \mbox{}
+f(\xi_1)\frac{8 \xi_1 \xi_2 \xi_3}{{\Delta}^{3}}
 ( {{\xi_1}^2} - {{\xi_2}^2} + 2 \xi_2 \xi_3 - {{\xi_3}^2}
)\nonumber\\&&\ \ \ \ \mbox{}
+\left(\frac{f(\xi_1)-1}{\xi_1}\right)\frac{4\xi_1}{{\Delta}^{2}}
 ( 3 {{\xi_1}^2} - 2 \xi_1 \xi_2 - {{\xi_2}^2} - 2 \xi_1 \xi_3 + 2 \xi_2 \xi_3 -
    {{\xi_3}^2} )\nonumber\\&&\ \ \ \ \mbox{}
-2\frac1{\xi_1-\xi_2}\left(\frac{f(\xi_1)-1}{\xi_1}-\frac{f(\xi_2)-1}{\xi_2}\right),
\end{fleqnarray}
\begin{fleqnarray}&&
F_{3}(\xi_1,\xi_2,\xi_3)=F(\xi_1,\xi_2,\xi_3)\Big[
\frac{2 \xi_1 \xi_2}{{\Delta}^{2}}
 ( \xi_1 - \xi_2 - \xi_3 )( -\xi_1 + \xi_2 - \xi_3 )\nonumber\\&&\ \ \ \ \mbox{}
-{2\over {\Delta}}
( \xi_1 + \xi_2 - \xi_3 )\Big]
+f(\xi_1)\frac{4 \xi_1 \xi_2}{{\Delta}^{2}}( -\xi_1 + \xi_2 - \xi_3 )\nonumber\\&&\ \ \ \ \mbox{}
+f(\xi_2)\frac{4 \xi_1 \xi_2}{{\Delta}^{2}}
 ( \xi_1 - \xi_2 - \xi_3 )\nonumber\\&&\ \ \ \ \mbox{}
+f(\xi_3)
\frac{1}{{\Delta}^{2}}(
{{\xi_1}^3} - {{\xi_1}^2} \xi_2
 - \xi_1 {{\xi_2}^2} + {{\xi_2}^3} - 3 {{\xi_1}^2} \xi_3 \nonumber\\&&\ \ \ \ \mbox{}+
  6 \xi_1 \xi_2 \xi_3 - 3 {{\xi_2}^2} \xi_3 + 3 \xi_1 {{\xi_3}^2}
+ 3 \xi_2 {{\xi_3}^2} -
  {{\xi_3}^3}),
\end{fleqnarray}\begin{fleqnarray}&&
F_{4}(\xi_1,\xi_2,\xi_3)=F(\xi_1,\xi_2,\xi_3)\Big[{1\over {36 {{\Delta}^4}}}
(-4 {{\xi_1}^8} - 4 {{\xi_1}^7} \xi_2 + 32 {{\xi_1}^6} {{\xi_2}^2} -
  28 {{\xi_1}^5} {{\xi_2}^3} \nonumber\\&&\ \ \ \ \mbox{}
+ 4 {{\xi_1}^4} {{\xi_2}^4} + 2 {{\xi_1}^7} \xi_3 +
  26 {{\xi_1}^6} \xi_2 \xi_3 - 90 {{\xi_1}^5} {{\xi_2}^2} \xi_3 +
  62 {{\xi_1}^4} {{\xi_2}^3} \xi_3 \nonumber\\&&\ \ \ \ \mbox{}+ 38 {{\xi_1}^6} {{\xi_3}^2} -
  60 {{\xi_1}^5} \xi_2 {{\xi_3}^2} + 42 {{\xi_1}^4} {{\xi_2}^2} {{\xi_3}^2} -
  20 {{\xi_1}^3} {{\xi_2}^3} {{\xi_3}^2} - 82 {{\xi_1}^5} {{\xi_3}^3} \nonumber\\&&\ \ \ \ \mbox{}+
  62 {{\xi_1}^4} \xi_2 {{\xi_3}^3} + 20 {{\xi_1}^3} {{\xi_2}^2} {{\xi_3}^3} +
  50 {{\xi_1}^4} {{\xi_3}^4} - 28 {{\xi_1}^3} \xi_2 {{\xi_3}^4} +
  6 {{\xi_1}^2} {{\xi_2}^2} {{\xi_3}^4} \nonumber\\&&\ \ \ \ \mbox{}+ 14 {{\xi_1}^3} {{\xi_3}^5} +
  6 {{\xi_1}^2} \xi_2 {{\xi_3}^5}
- 22 {{\xi_1}^2} {{\xi_3}^6} - 2 \xi_1 \xi_2 {{\xi_3}^6} +
  2 \xi_1 {{\xi_3}^7} + {{\xi_3}^8})\nonumber\\&&\ \ \ \ \mbox{}
+{4\over {3 {{\Delta}^3}}}
( -3 {{\xi_1}^5} + 3 {{\xi_1}^3} {{\xi_2}^2} + 5 {{\xi_1}^4} \xi_3 -
    2 {{\xi_1}^3} \xi_2 \xi_3
- 3 {{\xi_1}^2} {{\xi_2}^2} \xi_3 \nonumber\\&&\ \ \ \ \mbox{}+ {{\xi_1}^3} {{\xi_3}^2} +
    2 {{\xi_1}^2} \xi_2 {{\xi_3}^2}
- 3 {{\xi_1}^2} {{\xi_3}^3} + \xi_1 \xi_2 {{\xi_3}^3} -
    2 \xi_1 {{\xi_3}^4} + {{\xi_3}^5} )\Big]\nonumber\\&&\ \ \ \ \mbox{}
-\left(F(\xi_1,\xi_2,\xi_3)-\frac12\right)\frac{2}{{\Delta}^{2}}
( 4 {{\xi_1}^2} + 2 \xi_1 \xi_2 - 2 \xi_1 \xi_3- {{\xi_3}^2} )\nonumber\\&&\ \ \ \ \mbox{}
-f(\xi_1){1\over {24 {{\Delta}^4} \xi_2}}
( {{\xi_1}^8} - 2 {{\xi_1}^7} \xi_2+ 34 {{\xi_1}^6} {{\xi_2}^2} -
    74 {{\xi_1}^5} {{\xi_2}^3} + 52 {{\xi_1}^4} {{\xi_2}^4} \nonumber\\&&\ \ \ \ \mbox{}-
    38 {{\xi_1}^3} {{\xi_2}^5}
+ 46 {{\xi_1}^2} {{\xi_2}^6} - 14 \xi_1 {{\xi_2}^7} -
    5 {{\xi_2}^8} - 8 {{\xi_1}^7} \xi_3 + 38 {{\xi_1}^6} \xi_2 \xi_3 \nonumber\\&&\ \ \ \ \mbox{}-
    76 {{\xi_1}^5} {{\xi_2}^2} \xi_3 + 90 {{\xi_1}^4} {{\xi_2}^3} \xi_3 +
    16 {{\xi_1}^3} {{\xi_2}^4} \xi_3 - 134 {{\xi_1}^2} {{\xi_2}^5} \xi_3 +
    68 \xi_1 {{\xi_2}^6} \xi_3 \nonumber\\&&\ \ \ \ \mbox{}
+ 6 {{\xi_2}^7} \xi_3 + 28 {{\xi_1}^6} {{\xi_3}^2} -
    106 {{\xi_1}^5} \xi_2 {{\xi_3}^2}
+ 30 {{\xi_1}^4} {{\xi_2}^2} {{\xi_3}^2} +
    28 {{\xi_1}^3} {{\xi_2}^3} {{\xi_3}^2} \nonumber\\&&\ \ \ \ \mbox{}
+ 88 {{\xi_1}^2} {{\xi_2}^4} {{\xi_3}^2} -
    114 \xi_1 {{\xi_2}^5} {{\xi_3}^2} + 46 {{\xi_2}^6} {{\xi_3}^2} -
    56 {{\xi_1}^5} {{\xi_3}^3} + 78 {{\xi_1}^4} \xi_2 {{\xi_3}^3} \nonumber\\&&\ \ \ \ \mbox{}-
    8 {{\xi_1}^3} {{\xi_2}^2} {{\xi_3}^3}
- 12 {{\xi_1}^2} {{\xi_2}^3} {{\xi_3}^3} +
    48 \xi_1 {{\xi_2}^4} {{\xi_3}^3} - 146 {{\xi_2}^5} {{\xi_3}^3} +
    70 {{\xi_1}^4} {{\xi_3}^4}\nonumber\\&&\ \ \ \ \mbox{} + 58 {{\xi_1}^3} \xi_2 {{\xi_3}^4} +
    94 {{\xi_1}^2} {{\xi_2}^2} {{\xi_3}^4}
+ 78 \xi_1 {{\xi_2}^3} {{\xi_3}^4} +
    180 {{\xi_2}^4} {{\xi_3}^4} - 56 {{\xi_1}^3} {{\xi_3}^5} \nonumber\\&&\ \ \ \ \mbox{}-
    110 {{\xi_1}^2} \xi_2 {{\xi_3}^5} - 108 \xi_1 {{\xi_2}^2} {{\xi_3}^5} -
    110 {{\xi_2}^3} {{\xi_3}^5} + 28 {{\xi_1}^2} {{\xi_3}^6}
+ 50 \xi_1 \xi_2 {{\xi_3}^6} \nonumber\\&&\ \ \ \ \mbox{}+
    34 {{\xi_2}^2} {{\xi_3}^6} - 8 \xi_1 {{\xi_3}^7} - 6 \xi_2 {{\xi_3}^7}
+ {{\xi_3}^8} )\nonumber\\&&\ \ \ \ \mbox{}
-f(\xi_3){1\over {24 {{\Delta}^4} \xi_2}}
( -{{\xi_1}^8} + {{\xi_1}^7} \xi_2 - 14 {{\xi_1}^6} {{\xi_2}^2} +
    46 {{\xi_1}^5} {{\xi_2}^3}\nonumber\\&&\ \ \ \ \mbox{}
- 32 {{\xi_1}^4} {{\xi_2}^4} + 8 {{\xi_1}^7} \xi_3 -
    30 {{\xi_1}^6} \xi_2 \xi_3 + 44 {{\xi_1}^5} {{\xi_2}^2} \xi_3 -
    34 {{\xi_1}^4} {{\xi_2}^3} \xi_3 \nonumber\\&&\ \ \ \ \mbox{}+ 12 {{\xi_1}^3} {{\xi_2}^4} \xi_3 -
    28 {{\xi_1}^6} {{\xi_3}^2} + 78 {{\xi_1}^5} \xi_2 {{\xi_3}^2} -
    18 {{\xi_1}^4} {{\xi_2}^2} {{\xi_3}^2}
- 32 {{\xi_1}^3} {{\xi_2}^3} {{\xi_3}^2} \nonumber\\&&\ \ \ \ \mbox{}+
    56 {{\xi_1}^5} {{\xi_3}^3} - 22 {{\xi_1}^4} \xi_2 {{\xi_3}^3} -
    16 {{\xi_1}^3} {{\xi_2}^2} {{\xi_3}^3}
- 18 {{\xi_1}^2} {{\xi_2}^3} {{\xi_3}^3} -
    70 {{\xi_1}^4} {{\xi_3}^4} \nonumber\\&&\ \ \ \ \mbox{}- 128 {{\xi_1}^3} \xi_2 {{\xi_3}^4} -
    82 {{\xi_1}^2} {{\xi_2}^2} {{\xi_3}^4} + 56 {{\xi_1}^3} {{\xi_3}^5} +
    166 {{\xi_1}^2} \xi_2 {{\xi_3}^5} + 78 \xi_1 {{\xi_2}^2} {{\xi_3}^5} \nonumber\\&&\ \ \ \ \mbox{}-
    28 {{\xi_1}^2} {{\xi_3}^6}
- 78 \xi_1 \xi_2 {{\xi_3}^6} + 8 \xi_1 {{\xi_3}^7} +
    7 \xi_2 {{\xi_3}^7} - {{\xi_3}^8} )\nonumber\\&&\ \ \ \ \mbox{}
-\left(\frac{f(\xi_1)-1}{\xi_1}\right){1\over {4 {{\Delta}^3} \xi_2}}
( {{\xi_1}^6} + 24 {{\xi_1}^5} \xi_2+ 41 {{\xi_1}^4} {{\xi_2}^2} -
    24 {{\xi_1}^3} {{\xi_2}^3}
- 13 {{\xi_1}^2} {{\xi_2}^4} \nonumber\\&&\ \ \ \ \mbox{}- 32 \xi_1 {{\xi_2}^5} +
    3 {{\xi_2}^6} - 6 {{\xi_1}^5} \xi_3 - 8 {{\xi_1}^4} \xi_2 \xi_3 +
    12 {{\xi_1}^3} {{\xi_2}^2} \xi_3 + 40 {{\xi_1}^2} {{\xi_2}^3} \xi_3 \nonumber\\&&\ \ \ \ \mbox{}+
    74 \xi_1 {{\xi_2}^4} \xi_3
- 16 {{\xi_2}^5} \xi_3 + 15 {{\xi_1}^4} {{\xi_3}^2} -
    32 {{\xi_1}^3} \xi_2 {{\xi_3}^2}
- 26 {{\xi_1}^2} {{\xi_2}^2} {{\xi_3}^2} \nonumber\\&&\ \ \ \ \mbox{}-
    24 \xi_1 {{\xi_2}^3} {{\xi_3}^2} + 35 {{\xi_2}^4} {{\xi_3}^2} -
    20 {{\xi_1}^3} {{\xi_3}^3} - 16 {{\xi_1}^2} \xi_2 {{\xi_3}^3} -
    52 \xi_1 {{\xi_2}^2} {{\xi_3}^3} \nonumber\\&&\ \ \ \ \mbox{}- 40 {{\xi_2}^3} {{\xi_3}^3} +
    15 {{\xi_1}^2} {{\xi_3}^4}
+ 40 \xi_1 \xi_2 {{\xi_3}^4} + 25 {{\xi_2}^2} {{\xi_3}^4} -
    6 \xi_1 {{\xi_3}^5} - 8 \xi_2 {{\xi_3}^5} + {{\xi_3}^6} )\nonumber\\&&\ \ \ \ \mbox{}
-\left(\frac{f(\xi_3)-1}{\xi_3}\right){1\over {4 {{\Delta}^3} \xi_2}}
( -{{\xi_1}^6} + 7 {{\xi_1}^5} \xi_2 - 15 {{\xi_1}^4} {{\xi_2}^2} +
    9 {{\xi_1}^3} {{\xi_2}^3}
+ 6 {{\xi_1}^5} \xi_3 \nonumber\\&&\ \ \ \ \mbox{}- 42 {{\xi_1}^4} \xi_2 \xi_3 +
    18 {{\xi_1}^3} {{\xi_2}^2} \xi_3 + 18 {{\xi_1}^2} {{\xi_2}^3} \xi_3 -
    15 {{\xi_1}^4} {{\xi_3}^2} + 17 {{\xi_1}^3} \xi_2 {{\xi_3}^2} \nonumber\\&&\ \ \ \ \mbox{}-
    2 {{\xi_1}^2} {{\xi_2}^2} {{\xi_3}^2} + 20 {{\xi_1}^3} {{\xi_3}^3} +
    52 {{\xi_1}^2} \xi_2 {{\xi_3}^3} + 8 \xi_1 {{\xi_2}^2} {{\xi_3}^3} -
    15 {{\xi_1}^2} {{\xi_3}^4} \nonumber\\&&\ \ \ \ \mbox{}
- 23 \xi_1 \xi_2 {{\xi_3}^4} + 6 \xi_1 {{\xi_3}^5} -
    5 \xi_2 {{\xi_3}^5} - {{\xi_3}^6} )\nonumber\\&&\ \ \ \ \mbox{}
-\frac1{\xi_2-\xi_3}\Big(f(\xi_2)-f(\xi_3)\Big)\frac{(\xi_3-\xi_2-\xi_1)}{24\xi_1}\nonumber\\&&\ \ \ \ \mbox{}
-\frac1{\xi_2-\xi_3}\left(\frac{f(\xi_2)-1}{\xi_2}-\frac{f(\xi_3)-1}{\xi_3}\right)\frac{(\xi_3-\xi_2-\xi_1)}{4\xi_1},
\end{fleqnarray}
\begin{fleqnarray}&&
F_{5}(\xi_1,\xi_2,\xi_3)=\left(F(\xi_1,\xi_2,\xi_3)-\frac12\right){2\over {\xi_1 \xi_2}}
-f(\xi_3){{(-2 \xi_2 + \xi_3)}\over {8 \xi_1 \xi_2}}\nonumber\\&&\ \ \ \ \mbox{}
-\left(\frac{f(\xi_3)-1}{\xi_3}\right){{(-2 \xi_2 + 5 \xi_3)}\over {4 \xi_1 \xi_2}},
\end{fleqnarray}\begin{fleqnarray}&&
F_{6}(\xi_1,\xi_2,\xi_3)=F(\xi_1,\xi_2,\xi_3)\Big[-{1\over {6 {{\Delta}^2}}}
( 2 {{\xi_1}^4} + 4 {{\xi_1}^3} \xi_2 - 6 {{\xi_1}^2} {{\xi_2}^2} \nonumber\\&&\ \ \ \ \mbox{}-
    2 {{\xi_1}^3} \xi_3
+ 2 {{\xi_1}^2} \xi_2 \xi_3 - 2 \xi_1 \xi_2 {{\xi_3}^2} -
    2\xi_1 {{\xi_3}^3} + {{\xi_3}^4} )
-{{2 \xi_1}\over {\Delta}}\Big]\nonumber\\&&\ \ \ \ \mbox{}
-f(\xi_1){1\over {2 {{\Delta}^2}}}
( {{\xi_1}^3} + 5 {{\xi_1}^2} \xi_2 - 5 \xi_1 {{\xi_2}^2}
- {{\xi_2}^3} \nonumber\\&&\ \ \ \ \mbox{}+
    {{\xi_1}^2} \xi_3
+ 6 \xi_1 \xi_2 \xi_3 + {{\xi_2}^2} \xi_3 - \xi_1 {{\xi_3}^2} +
    \xi_2 {{\xi_3}^2} - {{\xi_3}^3} )\nonumber\\&&\ \ \ \ \mbox{}
-f(\xi_3){1\over {2 {{\Delta}^2}}}
( -2 {{\xi_1}^3} + 2 {{\xi_1}^2} \xi_2
+ 2 {{\xi_1}^2} \xi_3 - 2 \xi_1 \xi_2 \xi_3 -
 2 \xi_1 {{\xi_3}^2} + {{\xi_3}^3} ),
\end{fleqnarray}\begin{fleqnarray}&&
F_{7}(\xi_1,\xi_2,\xi_3)=F(\xi_1,\xi_2,\xi_3)\Big[{{2 \xi_2 \xi_3}\over {3 {{\Delta}^4}}}
 ( {{\xi_1}^6} - 5 {{\xi_1}^5} \xi_3 + 6 {{\xi_1}^4} \xi_2 \xi_3 +
    {{\xi_1}^4} {{\xi_3}^2} \nonumber\\&&\ \ \ \ \mbox{}- 6 {{\xi_1}^3} \xi_2 {{\xi_3}^2} +
    8 {{\xi_1}^2} {{\xi_2}^2} {{\xi_3}^2} + 6 {{\xi_1}^3} {{\xi_3}^3} -
    4 {{\xi_1}^2} \xi_2 {{\xi_3}^3} - 2 \xi_1 {{\xi_2}^2} {{\xi_3}^3} \nonumber\\&&\ \ \ \ \mbox{}+
    8 {{\xi_2}^3} {{\xi_3}^3}
- 4 {{\xi_1}^2} {{\xi_3}^4} + 3 \xi_1 \xi_2 {{\xi_3}^4} -
    9 {{\xi_2}^2} {{\xi_3}^4} - \xi_1 {{\xi_3}^5} + {{\xi_3}^6} )\nonumber\\&&\ \ \ \ \mbox{}
+{2\over {3 {{\Delta}^3}}}( {{\xi_1}^5}
- 7 {{\xi_1}^4} \xi_3 + 24 {{\xi_1}^3} \xi_2 \xi_3 +
    8 {{\xi_1}^3} {{\xi_3}^2} - 42 {{\xi_1}^2} \xi_2 {{\xi_3}^2} +
    34 \xi_1 {{\xi_2}^2} {{\xi_3}^2} \nonumber\\&&\ \ \ \ \mbox{}- 2 {{\xi_1}^2} {{\xi_3}^3} -
    32 \xi_1 \xi_2 {{\xi_3}^3}
- 34 {{\xi_2}^2} {{\xi_3}^3} - 2 \xi_1 {{\xi_3}^4} +
    33 \xi_2 {{\xi_3}^4} + {{\xi_3}^5} )\Big]\nonumber\\&&\ \ \ \ \mbox{}
+\left(F(\xi_1,\xi_2,\xi_3)-\frac12\right)\frac{2}{{\Delta}^{2}}
( {{\xi_1}^2} - 4 \xi_1 \xi_3 + 10 \xi_2 \xi_3 + 2 {{\xi_3}^2} )\nonumber\\&&\ \ \ \ \mbox{}
+f(\xi_1){1\over {12 {{\Delta}^4}}}
(-{{\xi_1}^7} + 14 {{\xi_1}^6} \xi_3 - 10 {{\xi_1}^5} \xi_2 \xi_3 -
  42 {{\xi_1}^5} {{\xi_3}^2} + 2 {{\xi_1}^4} \xi_2 {{\xi_3}^2} \nonumber\\&&\ \ \ \ \mbox{}+
  62 {{\xi_1}^3} {{\xi_2}^2} {{\xi_3}^2} + 70 {{\xi_1}^4} {{\xi_3}^3} +
  8 {{\xi_1}^3} \xi_2 {{\xi_3}^3} + 4 {{\xi_1}^2} {{\xi_2}^2} {{\xi_3}^3} +
  116 \xi_1 {{\xi_2}^3} {{\xi_3}^3} \nonumber\\&&\ \ \ \ \mbox{}- 70 {{\xi_1}^3} {{\xi_3}^4} -
  46 {{\xi_1}^2} \xi_2 {{\xi_3}^4} - 178 \xi_1 {{\xi_2}^2} {{\xi_3}^4} -
  58 {{\xi_2}^3} {{\xi_3}^4}
+ 42 {{\xi_1}^2} {{\xi_3}^5} \nonumber\\&&\ \ \ \ \mbox{}+ 76 \xi_1 \xi_2 {{\xi_3}^5} +
  90 {{\xi_2}^2} {{\xi_3}^5}
- 14 \xi_1 {{\xi_3}^6} - 34 \xi_2 {{\xi_3}^6} + 2 {{\xi_3}^7})\nonumber\\&&\ \ \ \ \mbox{}
+f(\xi_3){{4 \xi_2 \xi_3}\over {3 {{\Delta}^4}}}
 ( -2 {{\xi_1}^5} + 7 {{\xi_1}^4} \xi_2 - 8 {{\xi_1}^3} {{\xi_2}^2} +
    2 {{\xi_1}^2} {{\xi_2}^3} + 2 \xi_1 {{\xi_2}^4}
- {{\xi_2}^5} \nonumber\\&&\ \ \ \ \mbox{}+ 3 {{\xi_1}^4} \xi_3 -
    8 {{\xi_1}^3} \xi_2 \xi_3
+ 6 {{\xi_1}^2} {{\xi_2}^2} \xi_3 - {{\xi_2}^4} \xi_3 +
    2 {{\xi_1}^3} {{\xi_3}^2} - 4 {{\xi_1}^2} \xi_2 {{\xi_3}^2} \nonumber\\&&\ \ \ \ \mbox{}-
    6 \xi_1 {{\xi_2}^2} {{\xi_3}^2} + 8 {{\xi_2}^3} {{\xi_3}^2} -
    4 {{\xi_1}^2} {{\xi_3}^3}
+ 4 \xi_1 \xi_2 {{\xi_3}^3} - 8 {{\xi_2}^2} {{\xi_3}^3} +
    \xi_2 {{\xi_3}^4} + {{\xi_3}^5} )\nonumber\\&&\ \ \ \ \mbox{}
-\left(\frac{f(\xi_1)-1}{\xi_1}\right){1\over {2 {{\Delta}^3}}}
( -{{\xi_1}^5} - 2 {{\xi_1}^4} \xi_3 - 36 {{\xi_1}^3} \xi_2 \xi_3 +
    20 {{\xi_1}^3} {{\xi_3}^2} \nonumber\\&&\ \ \ \ \mbox{}+ 28 {{\xi_1}^2} \xi_2 {{\xi_3}^2} -
    54 \xi_1 {{\xi_2}^2} {{\xi_3}^2} - 28 {{\xi_1}^2} {{\xi_3}^3} +
    40 \xi_1 \xi_2 {{\xi_3}^3} \nonumber\\&&\ \ \ \ \mbox{}
- 4 {{\xi_2}^2} {{\xi_3}^3} + 14 \xi_1 {{\xi_3}^4} +
    6 \xi_2 {{\xi_3}^4} - 2 {{\xi_3}^5} )\nonumber\\&&\ \ \ \ \mbox{}
+\left(\frac{f(\xi_3)-1}{\xi_3}\right){1\over {2 {{\Delta}^3} \xi_1}}
({{\xi_1}^6} - 6 {{\xi_1}^5} \xi_2 + 15 {{\xi_1}^4} {{\xi_2}^2} -
  20 {{\xi_1}^3} {{\xi_2}^3}
+ 15 {{\xi_1}^2} {{\xi_2}^4}\nonumber\\&&\ \ \ \ \mbox{} - 6 \xi_1 {{\xi_2}^5} +
  {{\xi_2}^6} - 8 {{\xi_1}^5} \xi_3 + 28 {{\xi_1}^4} \xi_2 \xi_3 -
  32 {{\xi_1}^3} {{\xi_2}^2} \xi_3 + 8 {{\xi_1}^2} {{\xi_2}^3} \xi_3 \nonumber\\&&\ \ \ \ \mbox{}+
  8 \xi_1 {{\xi_2}^4} \xi_3
- 4 {{\xi_2}^5} \xi_3 + 13 {{\xi_1}^4} {{\xi_3}^2} -
  76 {{\xi_1}^3} \xi_2 {{\xi_3}^2}
+ 118 {{\xi_1}^2} {{\xi_2}^2} {{\xi_3}^2} \nonumber\\&&\ \ \ \ \mbox{}-
  60 \xi_1 {{\xi_2}^3} {{\xi_3}^2} + 5 {{\xi_2}^4} {{\xi_3}^2} -
  32 {{\xi_1}^2} \xi_2 {{\xi_3}^3} - 32 \xi_1 {{\xi_2}^2} {{\xi_3}^3} -
  13 {{\xi_1}^2} {{\xi_3}^4} \nonumber\\&&\ \ \ \ \mbox{}
+ 82 \xi_1 \xi_2 {{\xi_3}^4} - 5 {{\xi_2}^2} {{\xi_3}^4} +
  8 \xi_1 {{\xi_3}^5} + 4 \xi_2 {{\xi_3}^5} - {{\xi_3}^6})\nonumber\\&&\ \ \ \ \mbox{}
+\frac1{\xi_2-\xi_3}\left(\frac{f(\xi_2)-1}{\xi_2}-\frac{f(\xi_3)-1}{\xi_3}\right)\frac{\xi_2}{2\xi_1},
\end{fleqnarray}\begin{fleqnarray}&&
F_{8}(\xi_1,\xi_2,\xi_3)=F(\xi_1,\xi_2,\xi_3)\Big[\frac{4 \xi_1 \xi_2 \xi_3}{{\Delta}^{4}}
 ( -{{\xi_1}^5} + 6 {{\xi_1}^4} \xi_3 - 8 {{\xi_1}^3} \xi_2 \xi_3 -
    4 {{\xi_1}^3} {{\xi_3}^2} \nonumber\\&&\ \ \ \ \mbox{}+ 12 {{\xi_1}^2} \xi_2 {{\xi_3}^2} -
    6 \xi_1 {{\xi_2}^2} {{\xi_3}^2} - 4 {{\xi_1}^2} {{\xi_3}^3} +
    4 {{\xi_2}^2} {{\xi_3}^3} + 6 \xi_1 {{\xi_3}^4} - 2 \xi_2 {{\xi_3}^4} -
 2 {{\xi_3}^5} )\nonumber\\&&\ \ \ \ \mbox{}
+\frac{8 \xi_2 \xi_3}{{\Delta}^{3}}
 ( -7 {{\xi_1}^3} + 18 {{\xi_1}^2} \xi_3 - 14 \xi_1 \xi_2 \xi_3 +
    6 \xi_1 {{\xi_3}^2} + 10 \xi_2 {{\xi_3}^2} - 10 {{\xi_3}^3} )
\Big]\nonumber\\&&\ \ \ \ \mbox{}
-\left(F(\xi_1,\xi_2,\xi_3)-\frac12\right){8\over {{{\Delta}^2} \xi_1}}
( -{{\xi_1}^3} + 6 {{\xi_1}^2} \xi_3
+ 10 \xi_1 \xi_2 \xi_3 - 6 \xi_1 {{\xi_3}^2}\nonumber\\&&\ \ \ \ \mbox{} -
 2 \xi_2 {{\xi_3}^2} + 2 {{\xi_3}^3} )\nonumber\\&&\ \ \ \ \mbox{}
-f(\xi_1)\frac{8 \xi_1 \xi_2 \xi_3}{{\Delta}^{4}}
 ( {{\xi_1}^4} - 4 {{\xi_1}^3} \xi_3 + 4 {{\xi_1}^2} \xi_2 \xi_3 -
    4 \xi_1 \xi_2 {{\xi_3}^2}
+ 2 {{\xi_2}^2} {{\xi_3}^2} \nonumber\\&&\ \ \ \ \mbox{}+ 4 \xi_1 {{\xi_3}^3} -
    2 {{\xi_3}^4} )\nonumber\\&&\ \ \ \ \mbox{}
+f(\xi_2)\frac{16 \xi_1 \xi_2 \xi_3}{{\Delta}^{4}}
 ( {{\xi_1}^4} - 2 {{\xi_1}^3} \xi_2
+ 2 \xi_1 {{\xi_2}^3} - {{\xi_2}^4} - 4 {{\xi_1}^3} \xi_3 \nonumber\\&&\ \ \ \ \mbox{}
+ 6 {{\xi_1}^2} \xi_2 \xi_3 - 2 {{\xi_2}^3} \xi_3 +
    6 {{\xi_1}^2} {{\xi_3}^2}
- 6 \xi_1 \xi_2 {{\xi_3}^2} - 4 \xi_1 {{\xi_3}^3} +
    2 \xi_2 {{\xi_3}^3} + {{\xi_3}^4} )\nonumber\\&&\ \ \ \ \mbox{}
-\left(\frac{f(\xi_1)-1}{\xi_1}\right)
\frac{32 \xi_1 \xi_2 \xi_3 }{{\Delta}^{3}}
( 3 {{\xi_1}^2} - 2 \xi_1 \xi_3 +
 4 \xi_2 \xi_3 - 4 {{\xi_3}^2} )\nonumber\\&&\ \ \ \ \mbox{}
-\left(\frac{f(\xi_2)-1}{\xi_2}\right){2\over {{{\Delta}^3} \xi_1}}
( {{\xi_1}^6} - 2 {{\xi_1}^5} \xi_2 - 5 {{\xi_1}^4} {{\xi_2}^2} +
    20 {{\xi_1}^3} {{\xi_2}^3}
- 25 {{\xi_1}^2} {{\xi_2}^4}\nonumber\\&&\ \ \ \ \mbox{} + 14 \xi_1 {{\xi_2}^5} -
    3 {{\xi_2}^6} - 6 {{\xi_1}^5} \xi_3 + 2 {{\xi_1}^4} \xi_2 \xi_3 -
    44 {{\xi_1}^3} {{\xi_2}^2} \xi_3 - 44 {{\xi_1}^2} {{\xi_2}^3} \xi_3 \nonumber\\&&\ \ \ \ \mbox{}+
    82 \xi_1 {{\xi_2}^4} \xi_3
+ 10 {{\xi_2}^5} \xi_3 + 15 {{\xi_1}^4} {{\xi_3}^2} +
    12 {{\xi_1}^3} \xi_2 {{\xi_3}^2}
+ 114 {{\xi_1}^2} {{\xi_2}^2} {{\xi_3}^2} \nonumber\\&&\ \ \ \ \mbox{}-
    36 \xi_1 {{\xi_2}^3} {{\xi_3}^2} - 9 {{\xi_2}^4} {{\xi_3}^2} -
    20 {{\xi_1}^3} {{\xi_3}^3} - 28 {{\xi_1}^2} \xi_2 {{\xi_3}^3} -
    76 \xi_1 {{\xi_2}^2} {{\xi_3}^3} - 4 {{\xi_2}^3} {{\xi_3}^3}\nonumber\\&&\ \ \ \ \mbox{} +
    15 {{\xi_1}^2} {{\xi_3}^4}
+ 22 \xi_1 \xi_2 {{\xi_3}^4} + 11 {{\xi_2}^2} {{\xi_3}^4} -
    6 \xi_1 {{\xi_3}^5} - 6 \xi_2 {{\xi_3}^5} + {{\xi_3}^6} ),
\end{fleqnarray}
\begin{fleqnarray}&&
F_{9}(\xi_1,\xi_2,\xi_3)=F(\xi_1,\xi_2,\xi_3)\Big[
{1\over {108 {{\Delta}^6}}}(
6 {{\xi_1}^{11}} \xi_2
- 24 {{\xi_1}^{10}} {{\xi_2}^2} - 26 {{\xi_1}^9} {{\xi_2}^3} +
  126 {{\xi_1}^8} {{\xi_2}^4} \nonumber\\&&\ \ \ \ \mbox{}- 108 {{\xi_1}^7} {{\xi_2}^5} +
  24 {{\xi_1}^6} {{\xi_2}^6} - 54 {{\xi_1}^9} {{\xi_2}^2} \xi_3 +
  150 {{\xi_1}^8} {{\xi_2}^3} \xi_3 - 156 {{\xi_1}^7} {{\xi_2}^4} \xi_3 \nonumber\\&&\ \ \ \ \mbox{}+
  60 {{\xi_1}^6} {{\xi_2}^5} \xi_3
- 456 {{\xi_1}^7} {{\xi_2}^3} {{\xi_3}^2} -
  60 {{\xi_1}^6} {{\xi_2}^4} {{\xi_3}^2}
+ 222 {{\xi_1}^5} {{\xi_2}^5} {{\xi_3}^2} -
  396 {{\xi_1}^5} {{\xi_2}^4} {{\xi_3}^3} \nonumber\\&&\ \ \ \ \mbox{}
+ 162 {{\xi_1}^4} {{\xi_2}^4} {{\xi_3}^4} +
  364 {{\xi_1}^3} {{\xi_2}^3} {{\xi_3}^6}
+ 186 {{\xi_1}^2} {{\xi_2}^2} {{\xi_3}^8} -
  3 \xi_1 \xi_2 {{\xi_3}^{10}} + {{\xi_3}^{12}})\nonumber\\&&\ \ \ \ \mbox{}
-{1\over {36 {{\Delta}^5}}}
( 130 {{\xi_1}^8} \xi_2 + 400 {{\xi_1}^7} {{\xi_2}^2} -
    560 {{\xi_1}^6} {{\xi_2}^3} + 172 {{\xi_1}^5} {{\xi_2}^4}\nonumber\\&&\ \ \ \ \mbox{} -
    8 {{\xi_1}^6} {{\xi_2}^2} \xi_3 + 760 {{\xi_1}^5} {{\xi_2}^3} \xi_3 -
    570 {{\xi_1}^4} {{\xi_2}^4} \xi_3
+ 400 {{\xi_1}^4} {{\xi_2}^3} {{\xi_3}^2} \nonumber\\&&\ \ \ \ \mbox{}-
    200 {{\xi_1}^3} {{\xi_2}^3} {{\xi_3}^3}
- 396 {{\xi_1}^2} {{\xi_2}^2} {{\xi_3}^5} -
    156 \xi_1 \xi_2 {{\xi_3}^7} - 71 {{\xi_3}^9} )\Big]\nonumber\\&&\ \ \ \ \mbox{}-
\left(F(\xi_1,\xi_2,\xi_3)-\frac12\right){4\over {{{\Delta}^4} \xi_3}}
 ( -{{\xi_1}^7} + 5 {{\xi_1}^6} \xi_2 - 9 {{\xi_1}^5} {{\xi_2}^2} +
    5 {{\xi_1}^4} {{\xi_2}^3} + 12 {{\xi_1}^5} \xi_2 \xi_3 \nonumber\\&&\ \ \ \ \mbox{}+
    4 {{\xi_1}^4} {{\xi_2}^2} \xi_3 - {{\xi_1}^3} {{\xi_2}^3} \xi_3 -
    {{\xi_1}^3} {{\xi_2}^2} {{\xi_3}^2}
+ 7 {{\xi_1}^2} {{\xi_2}^2} {{\xi_3}^3} -
    8 \xi_1 \xi_2 {{\xi_3}^5} - 7 {{\xi_3}^7} )\nonumber\\&&\ \ \ \ \mbox{}
+\left(F(\xi_1,\xi_2,\xi_3)-\frac12+\frac{\xi_1+\xi_2+\xi_3}{24}\right){2\over {{{\Delta}^3} \xi_1 \xi_2 \xi_3}}
( -2 {{\xi_1}^5} \xi_2
-10{{\xi_1}^4} {{\xi_2}^2} \nonumber\\&&\ \ \ \ \mbox{}+ 10 {{\xi_1}^3} {{\xi_2}^3} +
    18 {{\xi_1}^3} {{\xi_2}^2} \xi_3
- 20 {{\xi_1}^2} {{\xi_2}^2} {{\xi_3}^2} +
22 \xi_1 \xi_2 {{\xi_3}^4} + {{\xi_3}^6} )\nonumber\\&&\ \ \ \ \mbox{}+
f(\xi_1){1\over {288 {{\Delta}^6}}}(
5 {{\xi_1}^{11}} + 50 {{\xi_1}^{10}} \xi_3 - 2 {{\xi_1}^9} \xi_2 \xi_3 -
  122 {{\xi_1}^9} {{\xi_3}^2} \nonumber\\&&\ \ \ \ \mbox{}- 134 {{\xi_1}^8} \xi_2 {{\xi_3}^2} +
  508 {{\xi_1}^7} {{\xi_2}^2} {{\xi_3}^2} - 114 {{\xi_1}^8} {{\xi_3}^3} -
  48 {{\xi_1}^7} \xi_2 {{\xi_3}^3} \nonumber\\&&\ \ \ \ \mbox{}
- 664 {{\xi_1}^6} {{\xi_2}^2} {{\xi_3}^3} +
  1144 {{\xi_1}^5} {{\xi_2}^3} {{\xi_3}^3}
+ 228 {{\xi_1}^7} {{\xi_3}^4} +
  68 {{\xi_1}^6} \xi_2 {{\xi_3}^4} \nonumber\\&&\ \ \ \ \mbox{}
- 1196 {{\xi_1}^5} {{\xi_2}^2} {{\xi_3}^4} -
  1100 {{\xi_1}^4} {{\xi_2}^3} {{\xi_3}^4}
+ 1462 {{\xi_1}^3} {{\xi_2}^4} {{\xi_3}^4}+
  180 {{\xi_1}^6} {{\xi_3}^5}\nonumber\\&&\ \ \ \ \mbox{} + 232 {{\xi_1}^5} \xi_2 {{\xi_3}^5} +
  1292 {{\xi_1}^4} {{\xi_2}^2} {{\xi_3}^5}
- 2000 {{\xi_1}^3} {{\xi_2}^3} {{\xi_3}^5} -
  1268 {{\xi_1}^2} {{\xi_2}^4} {{\xi_3}^5}\nonumber\\&&\ \ \ \ \mbox{}
- 140 \xi_1 {{\xi_2}^5} {{\xi_3}^5} -
  180 {{\xi_1}^5} {{\xi_3}^6} + 36 {{\xi_1}^4} \xi_2 {{\xi_3}^6} +
  728 {{\xi_1}^3} {{\xi_2}^2} {{\xi_3}^6}\nonumber\\&&\ \ \ \ \mbox{}
+ 2344 {{\xi_1}^2} {{\xi_2}^3} {{\xi_3}^6} +
  476 \xi_1 {{\xi_2}^4} {{\xi_3}^6} + 180 {{\xi_2}^5} {{\xi_3}^6} -
  228 {{\xi_1}^4} {{\xi_3}^7} \nonumber\\&&\ \ \ \ \mbox{}- 304 {{\xi_1}^3} \xi_2 {{\xi_3}^7} -
  1208 {{\xi_1}^2} {{\xi_2}^2} {{\xi_3}^7}
- 496 \xi_1 {{\xi_2}^3} {{\xi_3}^7} -
  132 {{\xi_2}^4} {{\xi_3}^7} \nonumber\\&&\ \ \ \ \mbox{}+ 114 {{\xi_1}^3} {{\xi_3}^8} +
  10 {{\xi_1}^2} \xi_2 {{\xi_3}^8} + 86 \xi_1 {{\xi_2}^2} {{\xi_3}^8} -
  210 {{\xi_2}^3} {{\xi_3}^8} + 122 {{\xi_1}^2} {{\xi_3}^9} \nonumber\\&&\ \ \ \ \mbox{}
  + 124 \xi_1 \xi_2 {{\xi_3}^9} + 202 {{\xi_2}^2} {{\xi_3}^9}
  - 50 \xi_1 {{\xi_3}^{10}}
- 30 \xi_2 {{\xi_3}^{10}} - 10 {{\xi_3}^{11}})\nonumber\\&&\ \ \ \ \mbox{}-
\left(\frac{f(\xi_1)-1}{\xi_1}\right){1\over {12 {{\Delta}^5}}}
( -47 {{\xi_1}^9} - {{\xi_1}^8} \xi_3 - 103 {{\xi_1}^7} \xi_2 \xi_3 \nonumber\\&&\ \ \ \ \mbox{}+
    256 {{\xi_1}^7} {{\xi_3}^2} + 42 {{\xi_1}^6} \xi_2 {{\xi_3}^2} -
    174 {{\xi_1}^5} {{\xi_2}^2} {{\xi_3}^2}
- 124 {{\xi_1}^6} {{\xi_3}^3} \nonumber\\&&\ \ \ \ \mbox{}+
    418 {{\xi_1}^5} \xi_2 {{\xi_3}^3}
+ 294 {{\xi_1}^4} {{\xi_2}^2} {{\xi_3}^3} -
    30 {{\xi_1}^3} {{\xi_2}^3} {{\xi_3}^3}
- 12 {{\xi_1}^5} {{\xi_3}^4} \nonumber\\&&\ \ \ \ \mbox{}-
    416 {{\xi_1}^4} \xi_2 {{\xi_3}^4}
- 8 {{\xi_1}^3} {{\xi_2}^2} {{\xi_3}^4} +
    128 {{\xi_1}^2} {{\xi_2}^3} {{\xi_3}^4}
- 190 \xi_1 {{\xi_2}^4} {{\xi_3}^4}\nonumber\\&&\ \ \ \ \mbox{} +
    122 {{\xi_1}^4} {{\xi_3}^5}
+ 278 {{\xi_1}^3} \xi_2 {{\xi_3}^5} -
    186 {{\xi_1}^2} {{\xi_2}^2} {{\xi_3}^5}
+ 330 \xi_1 {{\xi_2}^3} {{\xi_3}^5} \nonumber\\&&\ \ \ \ \mbox{}+
    16 {{\xi_2}^4} {{\xi_3}^5} - 240 {{\xi_1}^3} {{\xi_3}^6} +
    54 {{\xi_1}^2} \xi_2 {{\xi_3}^6} - 60 \xi_1 {{\xi_2}^2} {{\xi_3}^6} -
    26 {{\xi_2}^3} {{\xi_3}^6} \nonumber\\&&\ \ \ \ \mbox{}
+ 4{{\xi_1}^2}{{\xi_3}^7}-170 \xi_1 \xi_2 {{\xi_3}^7} +
    10 {{\xi_2}^2} {{\xi_3}^7} +90 \xi_1 {{\xi_3}^8} + \xi_2 {{\xi_3}^8}
- {{\xi_3}^9} )\nonumber\\&&\ \ \ \ \mbox{}+
\left(\frac{f(\xi_1)-1+\frac16\xi_1}{\xi_1^2}\right){1\over {8 {{\Delta}^4} \xi_2 \xi_3}}(
-4 {{\xi_1}^9} + 116 {{\xi_1}^8} \xi_3 + 369 {{\xi_1}^7} \xi_2 \xi_3 \nonumber\\&&\ \ \ \ \mbox{}-
  400 {{\xi_1}^7} {{\xi_3}^2} - 146 {{\xi_1}^6} \xi_2 {{\xi_3}^2} +
  174 {{\xi_1}^5} {{\xi_2}^2} {{\xi_3}^2} + 536 {{\xi_1}^6} {{\xi_3}^3} \nonumber\\&&\ \ \ \ \mbox{}+
  362 {{\xi_1}^5} \xi_2 {{\xi_3}^3} - 46 {{\xi_1}^4} {{\xi_2}^2} {{\xi_3}^3} -
  1070 {{\xi_1}^3} {{\xi_2}^3} {{\xi_3}^3} - 176 {{\xi_1}^5} {{\xi_3}^4} \nonumber\\&&\ \ \ \ \mbox{}-
  354 {{\xi_1}^4} \xi_2 {{\xi_3}^4} + 696 {{\xi_1}^3} {{\xi_2}^2} {{\xi_3}^4} -
  740 {{\xi_1}^2} {{\xi_2}^3} {{\xi_3}^4}
- 180 \xi_1 {{\xi_2}^4} {{\xi_3}^4} \nonumber\\&&\ \ \ \ \mbox{}-
  272 {{\xi_1}^4} {{\xi_3}^5} + 70 {{\xi_1}^3} \xi_2 {{\xi_3}^5} +
  1650 {{\xi_1}^2} {{\xi_2}^2} {{\xi_3}^5}
+ 306 \xi_1 {{\xi_2}^3} {{\xi_3}^5} \nonumber\\&&\ \ \ \ \mbox{}-
  26 {{\xi_2}^4} {{\xi_3}^5} + 304 {{\xi_1}^3} {{\xi_3}^6} -
  790 {{\xi_1}^2} \xi_2 {{\xi_3}^6} - 132 \xi_1 {{\xi_2}^2} {{\xi_3}^6} +
  58 {{\xi_2}^3} {{\xi_3}^6} \nonumber\\&&\ \ \ \ \mbox{}
- 120 {{\xi_1}^2} {{\xi_3}^7}- 18 \xi_1 \xi_2 {{\xi_3}^7}-
  50 {{\xi_2}^2} {{\xi_3}^7} + 24 \xi_1 {{\xi_3}^8} + 22 \xi_2 {{\xi_3}^8}
- 4 {{\xi_3}^9})\nonumber\\&&\ \ \ \ \mbox{}
+\frac{1}{576}\frac1{\xi_1-\xi_2}\Big(f(\xi_1)-f(\xi_2)\Big)+
\frac{1}{48}\frac1{\xi_1-\xi_2}\left(\frac{f(\xi_1)-1}{\xi_1}-\frac{f(\xi_2)-1}{\xi_2}\right)\nonumber\\&&\ \ \ \ \mbox{}
-\frac1{\xi_1-\xi_2}\left(\frac{f(\xi_1)-1+\frac16\xi_1}{{\xi_1}^2}-\frac{f(\xi_2)-1+\frac16\xi_2}{{\xi_2}^2}\right)\left(\frac{\xi_1}{2\xi_3}-\frac{3}{16}\right),
\end{fleqnarray}\begin{fleqnarray}&&
F_{10}(\xi_1,\xi_2,\xi_3)=\left(F(\xi_1,\xi_2,\xi_3)-\frac12+\frac{\xi_1+\xi_2+\xi_3}{24}\right) {8\over {3 \xi_1 \xi_2 \xi_3}}\nonumber\\&&\ \ \ \ \mbox{}-
\left(\frac{f(\xi_1)-1+\frac16\xi_1}{\xi_1^2}\right){{2\xi_1}\over {\xi_2 \xi_3}},
\end{fleqnarray}\begin{fleqnarray}&&
F_{11}(\xi_1,\xi_2,\xi_3)=-\left(F(\xi_1,\xi_2,\xi_3)-\frac12\right) {2\over {3 {{\Delta}^2} \xi_1 \xi_2}}
( {{\xi_1}^4} + 2 {{\xi_1}^3} \xi_2 - 3 {{\xi_1}^2} {{\xi_2}^2}\nonumber\\&&\ \ \ \ \mbox{} -
    {{\xi_1}^3} \xi_3 + {{\xi_1}^2} \xi_2 \xi_3 - 3 {{\xi_1}^2} {{\xi_3}^2} -
    4 \xi_1 \xi_2 {{\xi_3}^2} + 5 \xi_1 {{\xi_3}^3} - {{\xi_3}^4} )\nonumber\\&&\ \ \ \ \mbox{}+
\left(F(\xi_1,\xi_2,\xi_3)-\frac12+\frac{\xi_1+\xi_2+\xi_3}{24}\right){2\over {\Delta \xi_1 \xi_2}}
( -2 \xi_1 + \xi_3 )\nonumber\\&&\ \ \ \ \mbox{}
-f(\xi_3){1\over {96 \xi_1 \xi_2}}( -2 \xi_1 + \xi_3 )
-\left(\frac{f(\xi_3)-1}{\xi_3}\right){1\over {6 \xi_1 \xi_2}}( -\xi_1 + \xi_3 )\nonumber\\&&\ \ \ \ \mbox{}
-\left(\frac{f(\xi_1)-1+\frac16\xi_1}{\xi_1^2}\right){1\over {{{\Delta}^2} \xi_2}}
( {{\xi_1}^4} + 6 {{\xi_1}^3} \xi_2 - 8 {{\xi_1}^2} {{\xi_2}^2} +
    2 \xi_1 {{\xi_2}^3} \nonumber\\&&\ \ \ \ \mbox{}
- {{\xi_2}^4} + 2 {{\xi_1}^3} \xi_3 + 2 \xi_1 {{\xi_2}^2} \xi_3 +
    4 {{\xi_2}^3} \xi_3
- 8 {{\xi_1}^2} {{\xi_3}^2} \nonumber\\&&\ \ \ \ \mbox{}- 10 \xi_1 \xi_2 {{\xi_3}^2} -
    6 {{\xi_2}^2} {{\xi_3}^2} + 6 \xi_1 {{\xi_3}^3} + 4 \xi_2 {{\xi_3}^3}
- {{\xi_3}^4})\nonumber\\&&\ \ \ \ \mbox{}+
\left(\frac{f(\xi_3)-1+\frac16\xi_3}{\xi_3^2}\right){1\over {8 {{\Delta}^2} \xi_1 \xi_2}}
(-2 {{\xi_1}^5}
+ 6 {{\xi_1}^4} \xi_2 - 4 {{\xi_1}^3} {{\xi_2}^2} \nonumber\\&&\ \ \ \ \mbox{}+ 2 {{\xi_1}^4} \xi_3 +
  24 {{\xi_1}^3} \xi_2 \xi_3 - 26 {{\xi_1}^2} {{\xi_2}^2} \xi_3 +
  28 {{\xi_1}^3} {{\xi_3}^2} \nonumber\\&&\ \ \ \ \mbox{}
- 60 {{\xi_1}^2} \xi_2 {{\xi_3}^2}
- 44 {{\xi_1}^2} {{\xi_3}^3} + 12 \xi_1 \xi_2 {{\xi_3}^3}
+ 6 \xi_1 {{\xi_3}^4} +  5 {{\xi_3}^5})\nonumber\\&&\ \ \ \ \mbox{}+
 \frac{1}{2}\frac1{\xi_1-\xi_2}\left(\frac{f(\xi_1)-1+\frac16\xi_1}{{\xi_1}^2}-\frac{f(\xi_2)-1+\frac16\xi_2}{{\xi_2}^2}\right),
\end{fleqnarray}
\begin{fleqnarray}&&
F_{12}(\xi_1,\xi_2,\xi_3)=F(\xi_1,\xi_2,\xi_3)\Big[-\frac{2 \xi_1 }{{\Delta}^{3}}
{{( \xi_1 - \xi_2 - \xi_3 ) }^2}
( -\xi_1 - \xi_2 + \xi_3 )\times\nonumber\\&&\ \ \ \ \mbox{}\times( \xi_1 - \xi_2 + \xi_3 )
-\frac{8}{{\Delta}^{2}}( -2 {{\xi_1}^2} + \xi_1 \xi_2
+ {{\xi_2}^2} + \xi_1 \xi_3 - 2 \xi_2 \xi_3 +{{\xi_3}^2} )\Big]\nonumber\\&&\ \ \ \ \mbox{}
-f(\xi_1)\frac{4 \xi_1 }{{\Delta}^{3}}
( \xi_1 - \xi_2 - \xi_3 )( -\xi_1 - \xi_2 + \xi_3 )
  ( \xi_1 - \xi_2 + \xi_3 )\nonumber\\&&\ \ \ \ \mbox{}
+f(\xi_2)\frac{4 \xi_1}{{\Delta}^{3}}
 {{( \xi_1 - \xi_2 - \xi_3 ) }^2} ( -\xi_1 - \xi_2 + \xi_3 )\nonumber\\&&\ \ \ \ \mbox{}
-f(\xi_3)
\frac{4 \xi_1}{{\Delta}^{3}}( \xi_1 - \xi_2 - \xi_3 )
  ( {{\xi_1}^2} - 2 \xi_1 \xi_2 + {{\xi_2}^2} - {{\xi_3}^2} )\nonumber\\&&\ \ \ \ \mbox{}
+\left(\frac{f(\xi_1)-1}{\xi_1}\right){2\over {{{\Delta}^2} \xi_2 \xi_3}}
 ( -{{\xi_1}^3} \xi_2  + 3 {{\xi_1}^2} {{\xi_2}^2} -
    3 \xi_1 {{\xi_2}^3} + {{\xi_2}^4} \nonumber\\&&\ \ \ \ \mbox{}
- {{\xi_1}^3} \xi_3 + 18 {{\xi_1}^2} \xi_2 \xi_3 +
    3 \xi_1 {{\xi_2}^2} \xi_3
- 4 {{\xi_2}^3} \xi_3 + 3 {{\xi_1}^2} {{\xi_3}^2} \nonumber\\&&\ \ \ \ \mbox{}+
    3 \xi_1 \xi_2 {{\xi_3}^2}
+ 6 {{\xi_2}^2} {{\xi_3}^2} - 3 \xi_1 {{\xi_3}^3} -
    4 \xi_2 {{\xi_3}^3} + {{\xi_3}^4} )\nonumber\\&&\ \ \ \ \mbox{}
+\left(\frac{f(\xi_2)-1}{\xi_2}\right){2\over {{{\Delta}^2} \xi_3}}
( {{\xi_1}^3} - 3 {{\xi_1}^2} \xi_2 + 3 \xi_1 {{\xi_2}^2} - {{\xi_2}^3} \nonumber\\&&\ \ \ \ \mbox{}-
 3 {{\xi_1}^2} \xi_3 - 6 \xi_1 \xi_2 \xi_3
- 7 {{\xi_2}^2} \xi_3 + 3 \xi_1 {{\xi_3}^2}+
    9 \xi_2 {{\xi_3}^2} - {{\xi_3}^3} )\nonumber\\&&\ \ \ \ \mbox{}
+\left(\frac{f(\xi_3)-1}{\xi_3}\right){2\over {{{\Delta}^2} \xi_2}}
 ( {{\xi_1}^3} - 3 {{\xi_1}^2} \xi_2
+ 3 \xi_1 {{\xi_2}^2} - {{\xi_2}^3} \nonumber\\&&\ \ \ \ \mbox{}-
    3 {{\xi_1}^2} \xi_3 - 6 \xi_1 \xi_2 \xi_3
+ 9 {{\xi_2}^2} \xi_3 + 3 \xi_1 {{\xi_3}^2} -
    7 \xi_2 {{\xi_3}^2} - {{\xi_3}^3} )\nonumber\\&&\ \ \ \ \mbox{}
+\frac1{\xi_1-\xi_2}\left(\frac{f(\xi_1)-1}{\xi_1}-\frac{f(\xi_2)-1}{\xi_2}\right)\frac{2}{\xi_3}\nonumber\\&&\ \ \ \ \mbox{}
+\frac1{\xi_1-\xi_3}\left(\frac{f(\xi_1)-1}{\xi_1}-\frac{f(\xi_3)-1}{\xi_3}\right)\frac{2}{\xi_2},
\end{fleqnarray}\begin{fleqnarray}&&
F_{13}(\xi_1,\xi_2,\xi_3)=F(\xi_1,\xi_2,\xi_3)
{2\over {\Delta}}( -\xi_1 + \xi_2 + \xi_3 )
-f(\xi_1){4\over {\Delta}}\nonumber\\&&\ \ \ \ \mbox{}
-f(\xi_2){2\over {\Delta \xi_1}}
( -\xi_1 - \xi_2 + \xi_3 )
-f(\xi_3){2\over {\Delta \xi_1}}
( -\xi_1 + \xi_2 - \xi_3 )\nonumber\\&&\ \ \ \ \mbox{}
-\frac1{\xi_2-\xi_3}\Big(f(\xi_2)-f(\xi_3)\Big)\frac{2}{\xi_1},
\end{fleqnarray}\begin{fleqnarray}&&
F_{14}(\xi_1,\xi_2,\xi_3)=F(\xi_1,\xi_2,\xi_3)\Big[
-\frac{2}{{\Delta}^{2}}( -\xi_1 + \xi_2 - \xi_3 )\times\nonumber\\&&\ \ \ \ \mbox{}\times
  ( -\xi_1 - \xi_2 + \xi_3 )
  ( -\xi_1 + \xi_2 + \xi_3 )+{8\over {\Delta}}\Big]\nonumber\\&&\ \ \ \ \mbox{}
+f(\xi_1)\frac{4}{{\Delta}^{2}}
( -\xi_1
+\xi_2 - \xi_3 )  ( -\xi_1 - \xi_2 + \xi_3 )\nonumber\\&&\ \ \ \ \mbox{}
-f(\xi_2)\frac{4}{{\Delta}^{2}}
( -\xi_1 - \xi_2 + \xi_3 )
 ( -\xi_1 + \xi_2 + \xi_3 )\nonumber\\&&\ \ \ \ \mbox{}
-f(\xi_3)\frac{4}{{\Delta}^{2}}
( {{\xi_1}^2}- 2 \xi_1 \xi_2 + {{\xi_2}^2} - {{\xi_3}^2} ),
\end{fleqnarray}\begin{fleqnarray}&&
F_{15}(\xi_1,\xi_2,\xi_3)=F(\xi_1,\xi_2,\xi_3)\Big[-{{2 \xi_1}\over {3 {{\Delta}^4}}}
 {{( -\xi_1 + \xi_2 + \xi_3 ) }^2}
  ( {{\xi_1}^4} + 2 {{\xi_1}^3} \xi_2 \nonumber\\&&\ \ \ \ \mbox{}- 6 {{\xi_1}^2} {{\xi_2}^2} +
    2 \xi_1 {{\xi_2}^3} + {{\xi_2}^4}
+ 2 {{\xi_1}^3} \xi_3 + 4 {{\xi_1}^2} \xi_2 \xi_3 -
    2 \xi_1 {{\xi_2}^2} \xi_3
- 4 {{\xi_2}^3} \xi_3\nonumber\\&&\ \ \ \ \mbox{} - 6 {{\xi_1}^2} {{\xi_3}^2} -
    2 \xi_1 \xi_2 {{\xi_3}^2}
+ 6 {{\xi_2}^2} {{\xi_3}^2} + 2 \xi_1 {{\xi_3}^3} -
    4 \xi_2 {{\xi_3}^3} + {{\xi_3}^4} )\nonumber\\&&\ \ \ \ \mbox{}
-{{4}\over {3 {{\Delta}^3}}}
( 19 {{\xi_1}^4} - 22 {{\xi_1}^3} \xi_2 - 12 {{\xi_1}^2} {{\xi_2}^2} +
    14 \xi_1 {{\xi_2}^3} + {{\xi_2}^4} \nonumber\\&&\ \ \ \ \mbox{}
- 22 {{\xi_1}^3} \xi_3 + 40 {{\xi_1}^2} \xi_2 \xi_3 -
    14 \xi_1 {{\xi_2}^2} \xi_3 - 4 {{\xi_2}^3} \xi_3
- 12 {{\xi_1}^2} {{\xi_3}^2} \nonumber\\&&\ \ \ \ \mbox{}-
    14 \xi_1 \xi_2 {{\xi_3}^2} + 6 {{\xi_2}^2} {{\xi_3}^2}
+ 14 \xi_1 {{\xi_3}^3} -
    4 \xi_2 {{\xi_3}^3} + {{\xi_3}^4} )\Big]\nonumber\\&&\ \ \ \ \mbox{}
-\left(F(\xi_1,\xi_2,\xi_3)-\frac12\right)\frac{48 \xi_1}{{\Delta}^{2}}\nonumber\\&&\ \ \ \ \mbox{}
+f(\xi_1){{4 \xi_1}\over {3 {{\Delta}^4}}}
 ( -\xi_1 + \xi_2 + \xi_3 )
  ( {{\xi_1}^4} + 2 {{\xi_1}^3} \xi_2 - 6 {{\xi_1}^2} {{\xi_2}^2} \nonumber\\&&\ \ \ \ \mbox{}+
    2 \xi_1 {{\xi_2}^3} + {{\xi_2}^4}
+ 2 {{\xi_1}^3} \xi_3 + 4 {{\xi_1}^2} \xi_2 \xi_3 -
    2 \xi_1 {{\xi_2}^2} \xi_3
- 4 {{\xi_2}^3} \xi_3 \nonumber\\&&\ \ \ \ \mbox{}- 6 {{\xi_1}^2} {{\xi_3}^2} -
    2 \xi_1 \xi_2 {{\xi_3}^2}
+ 6 {{\xi_2}^2} {{\xi_3}^2} + 2 \xi_1 {{\xi_3}^3} -
    4 \xi_2 {{\xi_3}^3} + {{\xi_3}^4} )\nonumber\\&&\ \ \ \ \mbox{}
-f(\xi_2){1\over {6 {{\Delta}^4} \xi_1}}( -\xi_1 - \xi_2 + \xi_3 )
      ( 9 {{\xi_1}^6} - 8 {{\xi_1}^5} \xi_2-35 {{\xi_1}^4} {{\xi_2}^2}\nonumber\\&&\ \ \ \ \mbox{}+
      64 {{\xi_1}^3} {{\xi_2}^3}
- 37 {{\xi_1}^2} {{\xi_2}^4} + 8 \xi_1 {{\xi_2}^5} -
      {{\xi_2}^6} - 8 {{\xi_1}^5} \xi_3 \nonumber\\&&\ \ \ \ \mbox{}+ 6 {{\xi_1}^4} \xi_2 \xi_3 +
      20 {{\xi_1}^2} {{\xi_2}^3} \xi_3
- 24 \xi_1 {{\xi_2}^4} \xi_3 + 6 {{\xi_2}^5} \xi_3 \nonumber\\&&\ \ \ \ \mbox{}-
      35 {{\xi_1}^4} {{\xi_3}^2} + 34 {{\xi_1}^2} {{\xi_2}^2} {{\xi_3}^2} +
      16 \xi_1 {{\xi_2}^3} {{\xi_3}^2} - 15 {{\xi_2}^4} {{\xi_3}^2} \nonumber\\&&\ \ \ \ \mbox{}+
      64 {{\xi_1}^3} {{\xi_3}^3} + 20 {{\xi_1}^2} \xi_2 {{\xi_3}^3} +
      16 \xi_1 {{\xi_2}^2} {{\xi_3}^3} + 20 {{\xi_2}^3} {{\xi_3}^3}\nonumber\\&&\ \ \ \ \mbox{} -
      37 {{\xi_1}^2} {{\xi_3}^4} - 24 \xi_1 \xi_2 {{\xi_3}^4} -
      15 {{\xi_2}^2} {{\xi_3}^4} + 8 \xi_1 {{\xi_3}^5}
+ 6 \xi_2 {{\xi_3}^5} - {{\xi_3}^6} ) \nonumber\\&&\ \ \ \ \mbox{}
-f(\xi_3){1\over {6 {{\Delta}^4} \xi_1}}
( -9 {{\xi_1}^7} + 17 {{\xi_1}^6} \xi_2
+27{{\xi_1}^5} {{\xi_2}^2}
-99 {{\xi_1}^4} {{\xi_2}^3} \nonumber\\&&\ \ \ \ \mbox{}+ 101 {{\xi_1}^3} {{\xi_2}^4} -
    45 {{\xi_1}^2} {{\xi_2}^5}
+ 9 \xi_1 {{\xi_2}^6} - {{\xi_2}^7} - {{\xi_1}^6} \xi_3 -
    6 {{\xi_1}^5} \xi_2 \xi_3 \nonumber\\&&\ \ \ \ \mbox{}+  41 {{\xi_1}^4} {{\xi_2}^2} \xi_3 -
    84 {{\xi_1}^3} {{\xi_2}^3} \xi_3 + 81 {{\xi_1}^2} {{\xi_2}^4} \xi_3 -
    38 \xi_1 {{\xi_2}^5} \xi_3\nonumber\\&&\ \ \ \ \mbox{}
+ 7 {{\xi_2}^6} \xi_3 + 43 {{\xi_1}^5} {{\xi_3}^2} -
    41 {{\xi_1}^4} \xi_2 {{\xi_3}^2}
- 34 {{\xi_1}^3} {{\xi_2}^2} {{\xi_3}^2} \nonumber\\&&\ \ \ \ \mbox{}-
    2 {{\xi_1}^2} {{\xi_2}^3} {{\xi_3}^2} + 55 \xi_1 {{\xi_2}^4} {{\xi_3}^2} -
    21 {{\xi_2}^5} {{\xi_3}^2} - 29 {{\xi_1}^4} {{\xi_3}^3} \nonumber\\&&\ \ \ \ \mbox{}+
    44 {{\xi_1}^3} \xi_2 {{\xi_3}^3} - 30 {{\xi_1}^2} {{\xi_2}^2} {{\xi_3}^3} -
    20 \xi_1 {{\xi_2}^3} {{\xi_3}^3} + 35 {{\xi_2}^4} {{\xi_3}^3} \nonumber\\&&\ \ \ \ \mbox{}-
    27 {{\xi_1}^3} {{\xi_3}^4} - 33 {{\xi_1}^2} \xi_2 {{\xi_3}^4} -
    25 \xi_1 {{\xi_2}^2} {{\xi_3}^4} - 35 {{\xi_2}^3} {{\xi_3}^4} \nonumber\\&&\ \ \ \ \mbox{}+
    29 {{\xi_1}^2} {{\xi_3}^5} + 26 \xi_1 \xi_2 {{\xi_3}^5}
+ 21 {{\xi_2}^2} {{\xi_3}^5} -
    7 \xi_1 {{\xi_3}^6} - 7 \xi_2 {{\xi_3}^6} + {{\xi_3}^7} )\nonumber\\&&\ \ \ \ \mbox{}
-\left(\frac{f(\xi_1)-1}{\xi_1}\right)\frac{16{{\xi_1}^2}}{{\Delta}^{3}}
 ( 3 {{\xi_1}^2} - \xi_1 \xi_2 - 2 {{\xi_2}^2} - \xi_1 \xi_3 +
    4 \xi_2 \xi_3 - 2 {{\xi_3}^2} )\nonumber\\&&\ \ \ \ \mbox{}
+\left(\frac{f(\xi_2)-1}{\xi_2}\right){1\over {{{\Delta}^3} \xi_1}}( -\xi_1 - \xi_2 + \xi_3 )
( 3 {{\xi_1}^4} - 46 {{\xi_1}^3} \xi_2 + 44 {{\xi_1}^2} {{\xi_2}^2}
\nonumber\\&&\ \ \ \ \mbox{}-   2 \xi_1 {{\xi_2}^3} + {{\xi_2}^4}
- 10 {{\xi_1}^3} \xi_3 + 40 {{\xi_1}^2} \xi_2 \xi_3 -
    2 \xi_1 {{\xi_2}^2} \xi_3
- 4 {{\xi_2}^3} \xi_3\nonumber\\&&\ \ \ \ \mbox{} + 12 {{\xi_1}^2} {{\xi_3}^2} +
    10 \xi_1 \xi_2 {{\xi_3}^2}
+ 6 {{\xi_2}^2} {{\xi_3}^2} - 6 \xi_1 {{\xi_3}^3} -
    4 \xi_2 {{\xi_3}^3} + {{\xi_3}^4} )\nonumber\\&&\ \ \ \ \mbox{}
+\left(\frac{f(\xi_3)-1}{\xi_3}\right){1\over {{{\Delta}^3} \xi_1}}(
-3 {{\xi_1}^5} + 13 {{\xi_1}^4} \xi_2 - 22 {{\xi_1}^3} {{\xi_2}^2} +
  18 {{\xi_1}^2} {{\xi_2}^3}\nonumber\\&&\ \ \ \ \mbox{}
- 7\xi_1 {{\xi_2}^4}+ {{\xi_2}^5} + 43 {{\xi_1}^4} \xi_3 -
76 {{\xi_1}^3} \xi_2 \xi_3
+ 18 {{\xi_1}^2} {{\xi_2}^2} \xi_3 + 20 \xi_1 {{\xi_2}^3} \xi_3\nonumber\\&&\ \ \ \ \mbox{} -
  5 {{\xi_2}^4} \xi_3
+ 2 {{\xi_1}^3} {{\xi_3}^2} + 6 {{\xi_1}^2} \xi_2 {{\xi_3}^2} -
  18 \xi_1 {{\xi_2}^2} {{\xi_3}^2} + 10 {{\xi_2}^3} {{\xi_3}^2} -
  42 {{\xi_1}^2} {{\xi_3}^3}\nonumber\\&&\ \ \ \ \mbox{}
+ 4 \xi_1 \xi_2 {{\xi_3}^3} - 10 {{\xi_2}^2} {{\xi_3}^3} +
  \xi_1 {{\xi_3}^4} + 5 \xi_2 {{\xi_3}^4} - {{\xi_3}^5})\nonumber\\&&\ \ \ \ \mbox{}
+\frac1{\xi_2-\xi_3}\Big(f(\xi_2)-f(\xi_3)\Big)\frac{1}{6\xi_1}\nonumber\\&&\ \ \ \ \mbox{}
+\frac1{\xi_2-\xi_3}\left(\frac{f(\xi_2)-1}{\xi_2}-\frac{f(\xi_3)-1}{\xi_3}\right)\frac{1}{\xi_1},
\end{fleqnarray}
\begin{fleqnarray}&&
F_{16}(\xi_1,\xi_2,\xi_3)=F(\xi_1,\xi_2,\xi_3)\frac{8}{{\Delta}^{2}}
( -2 {{\xi_1}^2} + 2 \xi_1 \xi_2 + {{\xi_3}^2} )\nonumber\\&&\ \ \ \ \mbox{}
-\left(F(\xi_1,\xi_2,\xi_3)-\frac12\right){8\over {\Delta \xi_1 \xi_2}}( 2 \xi_1 - \xi_3 )\nonumber\\&&\ \ \ \ \mbox{}
-f(\xi_3){1\over {2 \xi_1 \xi_2}}+\left(\frac{f(\xi_1)-1}{\xi_1}\right)\frac{32 \xi_1}{{\Delta}^{2}}
 ( -\xi_1 + \xi_2 - \xi_3 )\nonumber\\&&\ \ \ \ \mbox{}
-\left(\frac{f(\xi_3)-1}{\xi_3}\right){1\over {{{\Delta}^2} \xi_1 \xi_2}}
( 2 {{\xi_1}^4} - 8 {{\xi_1}^3} \xi_2 + 6 {{\xi_1}^2} {{\xi_2}^2} -
    16 {{\xi_1}^3} \xi_3 \nonumber\\&&\ \ \ \ \mbox{}
+ 16 {{\xi_1}^2} \xi_2 \xi_3 + 36 {{\xi_1}^2} {{\xi_3}^2} -
 20 \xi_1 \xi_2 {{\xi_3}^2} - 32 \xi_1 {{\xi_3}^3} + 5 {{\xi_3}^4} ),
\end{fleqnarray}\begin{fleqnarray}&&
F_{17}(\xi_1,\xi_2,\xi_3)=F(\xi_1,\xi_2,\xi_3)\Big[-\frac{2 \xi_1}{{\Delta}^{2}}
 {{( -\xi_1 + \xi_2 + \xi_3 ) }^2}
-{4\over {\Delta}}\Big]\nonumber\\&&\ \ \ \ \mbox{}
+f(\xi_1)\frac{4 \xi_1}{{\Delta}^{2}}
 ( -\xi_1 + \xi_2 + \xi_3 )
-f(\xi_2){1\over {{{\Delta}^2} \xi_1}}
 ( -\xi_1 - \xi_2 + \xi_3 )\times\nonumber\\&&\ \ \ \ \mbox{}\times
( 3 {{\xi_1}^2} - 4 \xi_1 \xi_2 + {{\xi_2}^2}
- 4 \xi_1 \xi_3 - 2 \xi_2 \xi_3 + {{\xi_3}^2} )\nonumber\\&&\ \ \ \ \mbox{}
-f(\xi_3){1\over {{{\Delta}^2} \xi_1}}
 ( -\xi_1 + \xi_2 - \xi_3 )
  ( 3 {{\xi_1}^2} - 4 \xi_1 \xi_2 + {{\xi_2}^2}
- 4 \xi_1 \xi_3 - 2 \xi_2 \xi_3 + {{\xi_3}^2} )
\nonumber\\&&\ \ \ \ \mbox{}
-\frac1{\xi_2-\xi_3}\Big(f(\xi_2)-f(\xi_3)\Big)\frac{1}{\xi_1},
\end{fleqnarray}\begin{fleqnarray}&&
F_{18}(\xi_1,\xi_2,\xi_3)=F(\xi_1,\xi_2,\xi_3)\Big[-\frac{2 \xi_1 }{{\Delta}^{4}}
( {{\xi_1}^6} - 8 {{\xi_1}^5} \xi_3 + 14 {{\xi_1}^4} \xi_2 \xi_3 +
    10 {{\xi_1}^4} {{\xi_3}^2} \nonumber\\&&\ \ \ \ \mbox{}- 32 {{\xi_1}^3} \xi_2 {{\xi_3}^2} +
    18 {{\xi_1}^2} {{\xi_2}^2} {{\xi_3}^2} + 8 {{\xi_1}^2} \xi_2 {{\xi_3}^3} -
    16 \xi_1 {{\xi_2}^2} {{\xi_3}^3} + 4 {{\xi_2}^3} {{\xi_3}^3}\nonumber\\&&\ \ \ \ \mbox{} -
    10 {{\xi_1}^2} {{\xi_3}^4}
+ 8 \xi_1 \xi_2 {{\xi_3}^4} + 2 {{\xi_2}^2} {{\xi_3}^4} +
    8 \xi_1 {{\xi_3}^5} - 4 \xi_2 {{\xi_3}^5} - 2 {{\xi_3}^6} )\nonumber\\&&\ \ \ \ \mbox{}
-\frac{4}{{\Delta}^{3}}
( 7 {{\xi_1}^4} - 32 {{\xi_1}^3} \xi_3 + 32 {{\xi_1}^2} \xi_2 \xi_3 +
 12 {{\xi_1}^2} {{\xi_3}^2} \nonumber\\&&\ \ \ \ \mbox{}
- 32 \xi_1 \xi_2 {{\xi_3}^2} + 10 {{\xi_2}^2} {{\xi_3}^2} +
    16 \xi_1 {{\xi_3}^3} - 10 {{\xi_3}^4} )\Big]\nonumber\\&&\ \ \ \ \mbox{}
-\left(F(\xi_1,\xi_2,\xi_3)-\frac12\right){{32}\over {{{\Delta}^2} \xi_1}}
( {{\xi_1}^2} - \xi_1 \xi_3 + \xi_2 \xi_3 - {{\xi_3}^2} )\nonumber\\&&\ \ \ \ \mbox{}
+f(\xi_1)\frac{4 \xi_1}{{\Delta}^{4}}
 ( -{{\xi_1}^5} + 6 {{\xi_1}^4} \xi_3 - 8 {{\xi_1}^3} \xi_2 \xi_3 -
    4 {{\xi_1}^3} {{\xi_3}^2} + 12 {{\xi_1}^2} \xi_2 {{\xi_3}^2} \nonumber\\&&\ \ \ \ \mbox{}-
    6 \xi_1 {{\xi_2}^2} {{\xi_3}^2} - 4 {{\xi_1}^2} {{\xi_3}^3} +
    4 {{\xi_2}^2} {{\xi_3}^3}
+ 6 \xi_1 {{\xi_3}^4} - 2 \xi_2 {{\xi_3}^4} - 2 {{\xi_3}^5} )\nonumber\\&&\ \ \ \ \mbox{}
-f(\xi_2)\frac{8 \xi_1}{{\Delta}^{4}} ( -\xi_1 + \xi_2 + \xi_3 )
  ( {{\xi_1}^4} - 2 {{\xi_1}^3} \xi_2 + 2 \xi_1 {{\xi_2}^3}
- {{\xi_2}^4} - 4 {{\xi_1}^3} \xi_3 \nonumber\\&&\ \ \ \ \mbox{}
+ 6 {{\xi_1}^2} \xi_2 \xi_3 - 2 {{\xi_2}^3} \xi_3 +
    6 {{\xi_1}^2} {{\xi_3}^2}
- 6 \xi_1 \xi_2 {{\xi_3}^2} - 4 \xi_1 {{\xi_3}^3} +
    2 \xi_2 {{\xi_3}^3} + {{\xi_3}^4} )\nonumber\\&&\ \ \ \ \mbox{}
+\left(\frac{f(\xi_1)-1}{\xi_1}\right)\frac{16 \xi_1}{{\Delta}^{3}}
 ( -3 {{\xi_1}^3} + 8 {{\xi_1}^2} \xi_3 - 6 \xi_1 \xi_2 \xi_3 +
    2 \xi_1 {{\xi_3}^2} + 4 \xi_2 {{\xi_3}^2} - 4 {{\xi_3}^3} )\nonumber\\&&\ \ \ \ \mbox{}
+\left(\frac{f(\xi_2)-1}{\xi_2}\right){4\over {{{\Delta}^3} \xi_1}}
( -{{\xi_1}^5} + 17 {{\xi_1}^4} \xi_2 - 2 {{\xi_1}^3} {{\xi_2}^2} -
    46 {{\xi_1}^2} {{\xi_2}^3} + 35 \xi_1 {{\xi_2}^4} \nonumber\\&&\ \ \ \ \mbox{}- 3 {{\xi_2}^5} +
    5 {{\xi_1}^4} \xi_3
- 44 {{\xi_1}^3} \xi_2 \xi_3 + 22 {{\xi_1}^2} {{\xi_2}^2} \xi_3 +
    4 \xi_1 {{\xi_2}^3} \xi_3
+ 13 {{\xi_2}^4} \xi_3\nonumber\\&&\ \ \ \ \mbox{} - 10 {{\xi_1}^3} {{\xi_3}^2} +
    30 {{\xi_1}^2} \xi_2 {{\xi_3}^2} - 38 \xi_1 {{\xi_2}^2} {{\xi_3}^2} -
    22 {{\xi_2}^3} {{\xi_3}^2}
+ 10 {{\xi_1}^2} {{\xi_3}^3}\nonumber\\&&\ \ \ \ \mbox{} + 4 \xi_1 \xi_2 {{\xi_3}^3}+
    18 {{\xi_2}^2} {{\xi_3}^3} - 5 \xi_1 {{\xi_3}^4} - 7 \xi_2 {{\xi_3}^4}
+ {{\xi_3}^5} ),
\end{fleqnarray}\begin{fleqnarray}&&
F_{19}(\xi_1,\xi_2,\xi_3)=F(\xi_1,\xi_2,\xi_3)\Big[\frac{4 \xi_1 \xi_2 \xi_3 }{{\Delta}^{4}}
( {{\xi_1}^4} - 4 {{\xi_1}^3} \xi_3 + 4 {{\xi_1}^2} \xi_2 \xi_3 -
    4 \xi_1 \xi_2 {{\xi_3}^2} \nonumber\\&&\ \ \ \ \mbox{}
+ 2 {{\xi_2}^2} {{\xi_3}^2} + 4 \xi_1 {{\xi_3}^3} -
    2 {{\xi_3}^4} )
+\frac{4}{{\Delta}^{3}}
 ( {{\xi_1}^4} - 2 {{\xi_1}^3} \xi_3 + 12 {{\xi_1}^2} \xi_2 \xi_3 -
    6 {{\xi_1}^2} {{\xi_3}^2} \nonumber\\&&\ \ \ \ \mbox{}
- 18 \xi_1 \xi_2 {{\xi_3}^2} + 8 {{\xi_2}^2} {{\xi_3}^2} +
10 \xi_1 {{\xi_3}^3} - 4 \xi_2 {{\xi_3}^3}- 4 {{\xi_3}^4} )\Big]\nonumber\\&&\ \ \ \ \mbox{}
+\left(F(\xi_1,\xi_2,\xi_3)-\frac12\right){{16}\over {{{\Delta}^2} \xi_1}}
( {{\xi_1}^2} - \xi_1 \xi_3 + \xi_2 \xi_3 - {{\xi_3}^2} )\nonumber\\&&\ \ \ \ \mbox{}
-f(\xi_1)\frac{8 \xi_1 \xi_2 \xi_3}{{\Delta}^{4}}
 ( -{{\xi_1}^3} + 2 {{\xi_1}^2} \xi_3 - 2 \xi_1 \xi_2 \xi_3 +
    2 \xi_1 {{\xi_3}^2} + 2 \xi_2 {{\xi_3}^2} - 2 {{\xi_3}^3} )\nonumber\\&&\ \ \ \ \mbox{}
+f(\xi_3)\frac{16 \xi_1 \xi_2 \xi_3}{{\Delta}^{4}}
 ( -{{\xi_1}^3} + 3 {{\xi_1}^2} \xi_2 - 3 \xi_1 {{\xi_2}^2} +
    {{\xi_2}^3} + {{\xi_1}^2} \xi_3\nonumber\\&&\ \ \ \ \mbox{} - 2 \xi_1 \xi_2 \xi_3 + {{\xi_2}^2} \xi_3 +
    \xi_1 {{\xi_3}^2} - \xi_2 {{\xi_3}^2} - {{\xi_3}^3} )\nonumber\\&&\ \ \ \ \mbox{}
-\left(\frac{f(\xi_1)-1}{\xi_1}\right)\frac{8 \xi_1}{{\Delta}^{3}}
 ( -{{\xi_1}^3} - 10 \xi_1 \xi_2 \xi_3 + 6 \xi_1 {{\xi_3}^2} +
    4 \xi_2 {{\xi_3}^2} - 4 {{\xi_3}^3} )\nonumber\\&&\ \ \ \ \mbox{}
-\left(\frac{f(\xi_3)-1}{\xi_3}\right){{4 \xi_3}\over {{{\Delta}^3} \xi_1}}
 ( 3 {{\xi_1}^4} - 8 {{\xi_1}^3} \xi_2 + 6 {{\xi_1}^2} {{\xi_2}^2} -
    {{\xi_2}^4} \nonumber\\&&\ \ \ \ \mbox{}+ 4 {{\xi_1}^3} \xi_3 + 20 {{\xi_1}^2} \xi_2 \xi_3 -
    28 \xi_1 {{\xi_2}^2} \xi_3
+ 4 {{\xi_2}^3} \xi_3 - 18 {{\xi_1}^2} {{\xi_3}^2} \nonumber\\&&\ \ \ \ \mbox{}+
    16 \xi_1 \xi_2 {{\xi_3}^2}
- 6 {{\xi_2}^2} {{\xi_3}^2} + 12 \xi_1 {{\xi_3}^3} +
     \xi_2 {{\xi_3}^3} - {{\xi_3}^4} )\nonumber\\&&\ \ \ \ \mbox{}
+\frac1{\xi_2-\xi_3}\left(\frac{f(\xi_2)-1}{\xi_2}-\frac{f(\xi_3)-1}{\xi_3}\right)\frac{1}{\xi_1},
\end{fleqnarray}
\begin{fleqnarray}&&
F_{20}(\xi_1,\xi_2,\xi_3)=F(\xi_1,\xi_2,\xi_3)\Big[{2\over {3 {{\Delta}^4}}}
 ( {{\xi_1}^7} - 7 {{\xi_1}^6} \xi_3 + 11 {{\xi_1}^5} \xi_2 \xi_3 +
    6 {{\xi_1}^5} {{\xi_3}^2} \nonumber\\&&\ \ \ \ \mbox{} - 19 {{\xi_1}^4} \xi_2 {{\xi_3}^2} +
    14 {{\xi_1}^3} {{\xi_2}^2} {{\xi_3}^2} + 5 {{\xi_1}^4} {{\xi_3}^3} -
    4 {{\xi_1}^3} \xi_2 {{\xi_3}^3}
- 14 {{\xi_1}^2} {{\xi_2}^2} {{\xi_3}^3} \nonumber\\&&\ \ \ \ \mbox{}+
    10 \xi_1 {{\xi_2}^3} {{\xi_3}^3} - 10 {{\xi_1}^3} {{\xi_3}^4} +
    11 {{\xi_1}^2} \xi_2 {{\xi_3}^4} - 10 \xi_1 {{\xi_2}^2} {{\xi_3}^4} -
    7 {{\xi_2}^3} {{\xi_3}^4}\nonumber\\&&\ \ \ \ \mbox{}
+ 3 {{\xi_1}^2} {{\xi_3}^5} - 2 \xi_1 \xi_2 {{\xi_3}^5} +
    9 {{\xi_2}^2} {{\xi_3}^5} + 2 \xi_1 {{\xi_3}^6} - \xi_2 {{\xi_3}^6}
- {{\xi_3}^7} )\nonumber\\&&\ \ \ \ \mbox{}
-{2\over {3 {{\Delta}^3}}}
( -19 {{\xi_1}^4} + 80 {{\xi_1}^3} \xi_3 - 76 {{\xi_1}^2} \xi_2 \xi_3 -
    12 {{\xi_1}^2} {{\xi_3}^2} \nonumber\\&&\ \ \ \ \mbox{}+ 64 \xi_1 \xi_2 {{\xi_3}^2}
 - 42 {{\xi_2}^2} {{\xi_3}^2} -
 64 \xi_1 {{\xi_3}^3} + 8 \xi_2 {{\xi_3}^3} + 34 {{\xi_3}^4} )\Big]\nonumber\\&&\ \ \ \ \mbox{}
+\left(F(\xi_1,\xi_2,\xi_3)-\frac12\right)\frac{24}{{\Delta}^{2}}( \xi_1 - 2 \xi_3 )\nonumber\\&&\ \ \ \ \mbox{}
+f(\xi_1){4\over {3 {{\Delta}^4}}}
( {{\xi_1}^6} - 5 {{\xi_1}^5} \xi_3 + 6 {{\xi_1}^4} \xi_2 \xi_3 +
    {{\xi_1}^4} {{\xi_3}^2} \nonumber\\&&\ \ \ \ \mbox{}- 6 {{\xi_1}^3} \xi_2 {{\xi_3}^2} +
    8 {{\xi_1}^2} {{\xi_2}^2} {{\xi_3}^2} + 6 {{\xi_1}^3} {{\xi_3}^3} -
    4 {{\xi_1}^2} \xi_2 {{\xi_3}^3} - 2 \xi_1 {{\xi_2}^2} {{\xi_3}^3} \nonumber\\&&\ \ \ \ \mbox{}+
  8 {{\xi_2}^3} {{\xi_3}^3}
- 4 {{\xi_1}^2} {{\xi_3}^4} + 3 \xi_1 \xi_2 {{\xi_3}^4} -
 9 {{\xi_2}^2} {{\xi_3}^4} - \xi_1 {{\xi_3}^5} + {{\xi_3}^6} )\nonumber\\&&\ \ \ \ \mbox{}
-f(\xi_3){4\over {3 {{\Delta}^4}}}
( 2 {{\xi_1}^6} - 9 {{\xi_1}^5} \xi_2 + 15 {{\xi_1}^4} {{\xi_2}^2} -
    10 {{\xi_1}^3} {{\xi_2}^3} + 3 \xi_1 {{\xi_2}^5} \nonumber\\&&\ \ \ \ \mbox{}
- {{\xi_2}^6} - 5 {{\xi_1}^5} \xi_3 +
    18 {{\xi_1}^4} \xi_2 \xi_3 - 22 {{\xi_1}^3} {{\xi_2}^2} \xi_3 +
    8 {{\xi_1}^2} {{\xi_2}^3} \xi_3\nonumber\\&&\ \ \ \ \mbox{}
+ 3 \xi_1 {{\xi_2}^4} \xi_3 - 2 {{\xi_2}^5} \xi_3 +
    {{\xi_1}^4} {{\xi_3}^2} - 2 {{\xi_1}^3} \xi_2 {{\xi_3}^2} +
    8 {{\xi_1}^2} {{\xi_2}^2} {{\xi_3}^2} \nonumber\\&&\ \ \ \ \mbox{}
- 14 \xi_1 {{\xi_2}^3} {{\xi_3}^2} +
    7 {{\xi_2}^4} {{\xi_3}^2} + 6 {{\xi_1}^3} {{\xi_3}^3} -
    12 {{\xi_1}^2} \xi_2 {{\xi_3}^3} + 6 \xi_1 {{\xi_2}^2} {{\xi_3}^3} \nonumber\\&&\ \ \ \ \mbox{}-
    4 {{\xi_1}^2} {{\xi_3}^4}
+ 3 \xi_1 \xi_2 {{\xi_3}^4} - 7 {{\xi_2}^2} {{\xi_3}^4} -
    \xi_1 {{\xi_3}^5} + 2 \xi_2 {{\xi_3}^5} + {{\xi_3}^6} )\nonumber\\&&\ \ \ \ \mbox{}
-\left(\frac{f(\xi_1)-1}{\xi_1}\right)\frac{2}{{\Delta}^{3}}
( -11 {{\xi_1}^4} + 24 {{\xi_1}^3} \xi_3 - 20 {{\xi_1}^2} \xi_2 \xi_3 +
 20 {{\xi_1}^2} {{\xi_3}^2} \nonumber\\&&\ \ \ \ \mbox{}
+ 24 \xi_1 \xi_2 {{\xi_3}^2} + 6 {{\xi_2}^2} {{\xi_3}^2} -
 24 \xi_1 {{\xi_3}^3} - 8 \xi_2 {{\xi_3}^3} + 2 {{\xi_3}^4} )\nonumber\\&&\ \ \ \ \mbox{}
-\left(\frac{f(\xi_3)-1}{\xi_3}\right){2\over {{{\Delta}^3} \xi_1}}
( -{{\xi_1}^5} + 5 {{\xi_1}^4} \xi_2
 - 10 {{\xi_1}^3} {{\xi_2}^2} +
    10 {{\xi_1}^2} {{\xi_2}^3} \nonumber\\&&\ \ \ \ \mbox{}- 5 \xi_1 {{\xi_2}^4}
 + {{\xi_2}^5} + 19 {{\xi_1}^4} \xi_3 -
    52 {{\xi_1}^3} \xi_2 \xi_3
+ 42 {{\xi_1}^2} {{\xi_2}^2} \xi_3 \nonumber\\&&\ \ \ \ \mbox{}- 4 \xi_1 {{\xi_2}^3} \xi_3 -
    5 {{\xi_2}^4} \xi_3
- 10 {{\xi_1}^3} {{\xi_3}^2} + 30 {{\xi_1}^2} \xi_2 {{\xi_3}^2}\nonumber\\&&\ \ \ \ \mbox{}-
    30 \xi_1 {{\xi_2}^2} {{\xi_3}^2} + 10 {{\xi_2}^3} {{\xi_3}^2} -
    34 {{\xi_1}^2} {{\xi_3}^3}
 + 12 \xi_1 \xi_2 {{\xi_3}^3}\nonumber\\&&\ \ \ \ \mbox{} - 10 {{\xi_2}^2} {{\xi_3}^3} +
    27 \xi_1 {{\xi_3}^4} + 5 \xi_2 {{\xi_3}^4} - {{\xi_3}^5} )\nonumber\\&&\ \ \ \ \mbox{}
-\frac1{\xi_2-\xi_3}\left(\frac{f(\xi_2)-1}{\xi_2}-\frac{f(\xi_3)-1}{\xi_3}\right)\frac{1}{\xi_1},
\end{fleqnarray}\begin{fleqnarray}&&
F_{21}(\xi_1,\xi_2,\xi_3)=F(\xi_1,\xi_2,\xi_3)\Big[
-\frac{8 \xi_1\xi_3}{{\Delta}^{4}} ( -\xi_1 + \xi_2 - \xi_3 )
( -\xi_1 - \xi_2 + \xi_3 )\times\nonumber\\&&\ \ \ \ \mbox{}\times
  {{( -\xi_1 + \xi_2 + \xi_3 ) }^3}
+\frac{16}{{\Delta}^{3}}( -\xi_1 + \xi_2 + \xi_3 )\times\nonumber\\&&\ \ \ \ \mbox{}\times
  ( -{{\xi_1}^3} + 3 {{\xi_1}^2} \xi_2 - 3 \xi_1 {{\xi_2}^2}
+ {{\xi_2}^3} -    6 {{\xi_1}^2} \xi_3 \nonumber\\&&\ \ \ \ \mbox{}
+ 4 \xi_1 \xi_2 \xi_3 + 2 {{\xi_2}^2} \xi_3 + 3 \xi_1 {{\xi_3}^2} -
    7 \xi_2 {{\xi_3}^2} + 4 {{\xi_3}^3} )\Big]\nonumber\\&&\ \ \ \ \mbox{}
+\left(F(\xi_1,\xi_2,\xi_3)-\frac12\right){{32}\over {{{\Delta}^2} \xi_1}}
( {{\xi_1}^2} - 2 \xi_1 \xi_2
+ {{\xi_2}^2} + 4 \xi_1 \xi_3 - 2 \xi_2 \xi_3 +{{\xi_3}^2} )\nonumber\\&&\ \ \ \ \mbox{}
+f(\xi_1)\frac{16 \xi_3\xi_1}{{\Delta}^{4}}
 ( -\xi_1 + \xi_2 - \xi_3 )( -\xi_1 - \xi_2 + \xi_3 )
  {{( -\xi_1 + \xi_2 + \xi_3 ) }^2}\nonumber\\&&\ \ \ \ \mbox{}
-f(\xi_2)\frac{16 \xi_1 \xi_3}{{\Delta}^{4}}
 ( -\xi_1 - \xi_2 + \xi_3 )
  {{( -\xi_1 + \xi_2 + \xi_3 ) }^3}\nonumber\\&&\ \ \ \ \mbox{}
-f(\xi_3)\frac{16\xi_1 \xi_3}{{\Delta}^{4}}
 ( -\xi_1 + \xi_2 + \xi_3 )
  ( -{{\xi_1}^3} + 3 {{\xi_1}^2} \xi_2
- 3 \xi_1 {{\xi_2}^2} \nonumber\\&&\ \ \ \ \mbox{} + {{\xi_2}^3} +
    {{\xi_1}^2} \xi_3
- 2 \xi_1 \xi_2 \xi_3 + {{\xi_2}^2} \xi_3 + \xi_1 {{\xi_3}^2} -
    \xi_2 {{\xi_3}^2} - {{\xi_3}^3} )\nonumber\\&&\ \ \ \ \mbox{}
-\left(\frac{f(\xi_1)-1}{\xi_1}\right)\frac{32 \xi_1}{{\Delta}^{3}}
 ( -{{\xi_1}^3} + 3 {{\xi_1}^2} \xi_2
- 3 \xi_1 {{\xi_2}^2} + {{\xi_2}^3} \nonumber\\&&\ \ \ \ \mbox{} -  5 {{\xi_1}^2} \xi_3
 + 4 \xi_1 \xi_2 \xi_3 + {{\xi_2}^2} \xi_3 + 3 \xi_1 {{\xi_3}^2} -
    5 \xi_2 {{\xi_3}^2} + 3 {{\xi_3}^3} )\nonumber\\&&\ \ \ \ \mbox{}
-\left(\frac{f(\xi_2)-1}{\xi_2}\right){8\over {{{\Delta}^3} \xi_1}}
( -{{\xi_1}^5} + 9 {{\xi_1}^4} \xi_2
 - 22 {{\xi_1}^3} {{\xi_2}^2}
+ 22 {{\xi_1}^2} {{\xi_2}^3}\nonumber\\&&\ \ \ \ \mbox{} - 9 \xi_1 {{\xi_2}^4}
+ {{\xi_2}^5} + 4 {{\xi_1}^4} \xi_3 -  4 {{\xi_1}^3} \xi_2 \xi_3
+ 16 {{\xi_1}^2} {{\xi_2}^2} \xi_3 - 12 \xi_1 {{\xi_2}^3} \xi_3 \nonumber\\&&\ \ \ \ \mbox{}-
    4 {{\xi_2}^4} \xi_3
- 6 {{\xi_1}^3} {{\xi_3}^2} - 18 {{\xi_1}^2} \xi_2 {{\xi_3}^2} +
    10 \xi_1 {{\xi_2}^2} {{\xi_3}^2} + 6 {{\xi_2}^3} {{\xi_3}^2} \nonumber\\&&\ \ \ \ \mbox{}+
    4 {{\xi_1}^2} {{\xi_3}^3}
+ 12 \xi_1 \xi_2 {{\xi_3}^3} - 4 {{\xi_2}^2} {{\xi_3}^3} -
    \xi_1 {{\xi_3}^4} + \xi_2 {{\xi_3}^4} )\nonumber\\&&\ \ \ \ \mbox{}
-\left(\frac{f(\xi_3)-1}{\xi_3}\right){{8 \xi_3}\over {{{\Delta}^3} \xi_1}}
 ( {{\xi_1}^4} - 4 {{\xi_1}^3} \xi_2 + 6 {{\xi_1}^2} {{\xi_2}^2} -
    4 \xi_1 {{\xi_2}^3} \nonumber\\&&\ \ \ \ \mbox{}+ {{\xi_2}^4}
 + 20 {{\xi_1}^3} \xi_3 - 44 {{\xi_1}^2} \xi_2 \xi_3 +
    28 \xi_1 {{\xi_2}^2} \xi_3
- 4 {{\xi_2}^3} \xi_3 - 2 {{\xi_1}^2} {{\xi_3}^2} \nonumber\\&&\ \ \ \ \mbox{}-
    4 \xi_1 \xi_2 {{\xi_3}^2}
+ 6 {{\xi_2}^2} {{\xi_3}^2} - 20 \xi_1 {{\xi_3}^3} -
    4 \xi_2 {{\xi_3}^3} + {{\xi_3}^4} )\nonumber\\&&\ \ \ \ \mbox{}
-\frac1{\xi_2-\xi_3}\left(\frac{f(\xi_2)-1}{\xi_2}-\frac{f(\xi_3)-1}{\xi_3}\right)\frac{4}{\xi_1},
\end{fleqnarray}
\begin{fleqnarray}&&
F_{22}(\xi_1,\xi_2,\xi_3)=F(\xi_1,\xi_2,\xi_3)\Big[{{\xi_1}\over {18 {{\Delta}^6}}}
 ( {{\xi_1}^{10}} + 4 {{\xi_1}^9} \xi_3+ 2 {{\xi_1}^8} \xi_2 \xi_3 -
    30 {{\xi_1}^8} {{\xi_3}^2} \nonumber\\&&\ \ \ \ \mbox{}- 32 {{\xi_1}^7} \xi_2 {{\xi_3}^2} +
    148 {{\xi_1}^6} {{\xi_2}^2} {{\xi_3}^2} + 16 {{\xi_1}^6} \xi_2 {{\xi_3}^3} -
    240 {{\xi_1}^5} {{\xi_2}^2} {{\xi_3}^3}
+ 248 {{\xi_1}^4} {{\xi_2}^3} {{\xi_3}^3}\nonumber\\&&\ \ \ \ \mbox{} +
    156 {{\xi_1}^6} {{\xi_3}^4} - 8 {{\xi_1}^5} \xi_2 {{\xi_3}^4} -
    284 {{\xi_1}^4} {{\xi_2}^2} {{\xi_3}^4}
- 192 {{\xi_1}^3} {{\xi_2}^3} {{\xi_3}^4} \nonumber\\&&\ \ \ \ \mbox{}+
    230 {{\xi_1}^2} {{\xi_2}^4} {{\xi_3}^4}
 - 264 {{\xi_1}^5} {{\xi_3}^5} +
    136 {{\xi_1}^4} \xi_2 {{\xi_3}^5}
+ 416 {{\xi_1}^3} {{\xi_2}^2} {{\xi_3}^5} \nonumber\\&&\ \ \ \ \mbox{}-
    240 {{\xi_1}^2} {{\xi_2}^3} {{\xi_3}^5}
- 104 \xi_1 {{\xi_2}^4} {{\xi_3}^5} +
    28 {{\xi_2}^5} {{\xi_3}^5}+ 156 {{\xi_1}^4} {{\xi_3}^6} \nonumber\\&&\ \ \ \ \mbox{}-
    224 {{\xi_1}^3} \xi_2 {{\xi_3}^6}
- 72 {{\xi_1}^2} {{\xi_2}^2} {{\xi_3}^6} +
    176 \xi_1 {{\xi_2}^3} {{\xi_3}^6} - 28 {{\xi_2}^4} {{\xi_3}^6} \nonumber\\&&\ \ \ \ \mbox{}+
    112 {{\xi_1}^2} \xi_2 {{\xi_3}^7} - 80 \xi_1 {{\xi_2}^2} {{\xi_3}^7} -
    16 {{\xi_2}^3} {{\xi_3}^7} - 30 {{\xi_1}^2} {{\xi_3}^8}\nonumber\\&&\ \ \ \ \mbox{}
+ 4 \xi_1 \xi_2 {{\xi_3}^8} +
    26 {{\xi_2}^2} {{\xi_3}^8} + 4 \xi_1 {{\xi_3}^9} - 12 \xi_2 {{\xi_3}^9} +
    2 {{\xi_3}^{10}} )\nonumber\\&&\ \ \ \ \mbox{}+
{1\over {9 {{\Delta}^5}}}(
73 {{\xi_1}^8} - 52 {{\xi_1}^7} \xi_3 + 148 {{\xi_1}^6} \xi_2 \xi_3 -
  844 {{\xi_1}^6} {{\xi_3}^2} \nonumber\\&&\ \ \ \ \mbox{}+ 180 {{\xi_1}^5} \xi_2 {{\xi_3}^2} +
  1008 {{\xi_1}^4} {{\xi_2}^2} {{\xi_3}^2} + 1004 {{\xi_1}^5} {{\xi_3}^3} -
  1648 {{\xi_1}^4} \xi_2 {{\xi_3}^3} \nonumber\\&&\ \ \ \ \mbox{}
- 760 {{\xi_1}^3} {{\xi_2}^2} {{\xi_3}^3} +
  952 {{\xi_1}^2} {{\xi_2}^3} {{\xi_3}^3} + 320 {{\xi_1}^4} {{\xi_3}^4} +
  1556 {{\xi_1}^3} \xi_2 {{\xi_3}^4} \nonumber\\&&\ \ \ \ \mbox{}
- 1116 {{\xi_1}^2} {{\xi_2}^2} {{\xi_3}^4} -
  244 \xi_1 {{\xi_2}^3} {{\xi_3}^4} + 250 {{\xi_2}^4} {{\xi_3}^4} -
  796 {{\xi_1}^3} {{\xi_3}^5} \nonumber\\&&\ \ \ \ \mbox{}+ 72 {{\xi_1}^2} \xi_2 {{\xi_3}^5} +
  612 \xi_1 {{\xi_2}^2} {{\xi_3}^5} - 256 {{\xi_2}^3} {{\xi_3}^5} +
  92 {{\xi_1}^2} {{\xi_3}^6} \nonumber\\&&\ \ \ \ \mbox{}- 532 \xi_1 \xi_2 {{\xi_3}^6}
- 88 {{\xi_2}^2} {{\xi_3}^6} +  164 \xi_1 {{\xi_3}^7}
+ 128 \xi_2 {{\xi_3}^7} - 34 {{\xi_3}^8})\Big]\nonumber\\&&\ \ \ \ \mbox{}
-\left(F(\xi_1,\xi_2,\xi_3)-\frac12\right){2\over {{{\Delta}^4} \xi_1 \xi_2 \xi_3}}
( {{\xi_1}^8} - 16 {{\xi_1}^7} \xi_3 - 44 {{\xi_1}^6} \xi_2 \xi_3 \nonumber\\&&\ \ \ \ \mbox{}+
    44 {{\xi_1}^6} {{\xi_3}^2} + 36 {{\xi_1}^5} \xi_2 {{\xi_3}^2} -
    116 {{\xi_1}^4} {{\xi_2}^2} {{\xi_3}^2} - 52 {{\xi_1}^5} {{\xi_3}^3} \nonumber\\&&\ \ \ \ \mbox{}+
    208 {{\xi_1}^4} \xi_2 {{\xi_3}^3} + 64 {{\xi_1}^3} {{\xi_2}^2} {{\xi_3}^3} -
    64 {{\xi_1}^2} {{\xi_2}^3} {{\xi_3}^3} + 20 {{\xi_1}^4} {{\xi_3}^4} \nonumber\\&&\ \ \ \ \mbox{}-
    72 {{\xi_1}^3} \xi_2 {{\xi_3}^4} + 164 {{\xi_1}^2} {{\xi_2}^2} {{\xi_3}^4} +
    52 \xi_1 {{\xi_2}^3} {{\xi_3}^4} - 10 {{\xi_2}^4} {{\xi_3}^4} \nonumber\\&&\ \ \ \ \mbox{}+
    8 {{\xi_1}^3} {{\xi_3}^5} - 96 {{\xi_1}^2} \xi_2 {{\xi_3}^5} -
    84 \xi_1 {{\xi_2}^2} {{\xi_3}^5} + 8 {{\xi_2}^3} {{\xi_3}^5} -
    4 {{\xi_1}^2} {{\xi_3}^6} \nonumber\\&&\ \ \ \ \mbox{}+ 36 \xi_1 \xi_2 {{\xi_3}^6}
+ 8 {{\xi_2}^2} {{\xi_3}^6} -  4 \xi_1 {{\xi_3}^7}
- 8 \xi_2 {{\xi_3}^7} + 2 {{\xi_3}^8} )\nonumber\\&&\ \ \ \ \mbox{}
+\left(F(\xi_1,\xi_2,\xi_3)-\frac12+\frac{\xi_1+\xi_2+\xi_3}{24}\right){16\over {{{\Delta}^3} \xi_1 \xi_2}}
( 2 {{\xi_1}^4} + 10 {{\xi_1}^3} \xi_2 - 5 {{\xi_1}^3} \xi_3 \nonumber\\&&\ \ \ \ \mbox{}+
    5 {{\xi_1}^2} \xi_2 \xi_3 - \xi_1 {{\xi_2}^2} \xi_3
+ 3 {{\xi_1}^2} {{\xi_3}^2} -
    2 {{\xi_2}^2} {{\xi_3}^2}
+ \xi_1 {{\xi_3}^3} + 3 \xi_2 {{\xi_3}^3} - {{\xi_3}^4} )\nonumber\\&&\ \ \ \ \mbox{}
-f(\xi_1){\xi_1\over {9 {{\Delta}^6}}}
(-{{\xi_1}^9} - 6 {{\xi_1}^8} \xi_3 - 8 {{\xi_1}^7} \xi_2 \xi_3 +
      24 {{\xi_1}^7} {{\xi_3}^2} \nonumber\\&&\ \ \ \ \mbox{}+ 40 {{\xi_1}^6} \xi_2 {{\xi_3}^2} -
      108 {{\xi_1}^5} {{\xi_2}^2} {{\xi_3}^2} + 24 {{\xi_1}^6} {{\xi_3}^3} +
      48 {{\xi_1}^5} \xi_2 {{\xi_3}^3}\nonumber\\&&\ \ \ \ \mbox{}
+ 72 {{\xi_1}^4} {{\xi_2}^2} {{\xi_3}^3} -
      176 {{\xi_1}^3} {{\xi_2}^3} {{\xi_3}^3} - 132 {{\xi_1}^5} {{\xi_3}^4} -
      76 {{\xi_1}^4} \xi_2 {{\xi_3}^4}\nonumber\\&&\ \ \ \ \mbox{}
+ 280 {{\xi_1}^3} {{\xi_2}^2} {{\xi_3}^4} +
      120 {{\xi_1}^2} {{\xi_2}^3} {{\xi_3}^4}
- 110 \xi_1 {{\xi_2}^4} {{\xi_3}^4} +
      132 {{\xi_1}^4} {{\xi_3}^5} \nonumber\\&&\ \ \ \ \mbox{}- 80 {{\xi_1}^3} \xi_2 {{\xi_3}^5} -
      216 {{\xi_1}^2} {{\xi_2}^2} {{\xi_3}^5}
+ 144 \xi_1 {{\xi_2}^3} {{\xi_3}^5} +
      28 {{\xi_2}^4} {{\xi_3}^5} \nonumber\\&&\ \ \ \ \mbox{} - 24 {{\xi_1}^3} {{\xi_3}^6} +
      120 {{\xi_1}^2} \xi_2 {{\xi_3}^6} - 24 \xi_1 {{\xi_2}^2} {{\xi_3}^6} -
      56 {{\xi_2}^3} {{\xi_3}^6} - 24 {{\xi_1}^2} {{\xi_3}^7}\nonumber\\&&\ \ \ \ \mbox{} -
      16 \xi_1 \xi_2 {{\xi_3}^7}
+ 40 {{\xi_2}^2} {{\xi_3}^7} + 6 \xi_1 {{\xi_3}^8} -
      14 \xi_2 {{\xi_3}^8} + 2 {{\xi_3}^9} ) \nonumber\\&&\ \ \ \ \mbox{}
+f(\xi_3){1\over {72 {{\Delta}^6} \xi_1}}
(-15 {{\xi_1}^{11}} - 29 {{\xi_1}^{10}} \xi_2
 + 335 {{\xi_1}^9} {{\xi_2}^2} -
  555 {{\xi_1}^8} {{\xi_2}^3} \nonumber\\&&\ \ \ \ \mbox{}- 198 {{\xi_1}^7} {{\xi_2}^4} +
  1518 {{\xi_1}^6} {{\xi_2}^5} - 1842 {{\xi_1}^5} {{\xi_2}^6} +
  1050 {{\xi_1}^4} {{\xi_2}^7} \nonumber\\&&\ \ \ \ \mbox{}
- 315 {{\xi_1}^3} {{\xi_2}^8} +
  63 {{\xi_1}^2} {{\xi_2}^9} - 13 \xi_1 {{\xi_2}^{10}}
+ {{\xi_2}^{11}} \nonumber\\&&\ \ \ \ \mbox{}-
  59 {{\xi_1}^{10}} \xi_3 + 106 {{\xi_1}^9} \xi_2 \xi_3
+ 65 {{\xi_1}^8} {{\xi_2}^2} \xi_3 +
  152 {{\xi_1}^7} {{\xi_2}^3} \xi_3 \nonumber\\&&\ \ \ \ \mbox{}- 1366 {{\xi_1}^6} {{\xi_2}^4} \xi_3 +
  2300 {{\xi_1}^5} {{\xi_2}^5} \xi_3 - 1846 {{\xi_1}^4} {{\xi_2}^6} \xi_3 +
  920 {{\xi_1}^3} {{\xi_2}^7} \xi_3 \nonumber\\&&\ \ \ \ \mbox{}- 367 {{\xi_1}^2} {{\xi_2}^8} \xi_3 +
  106 \xi_1 {{\xi_2}^9} \xi_3 - 11 {{\xi_2}^{10}} \xi_3
+ 247 {{\xi_1}^9} {{\xi_3}^2}\nonumber\\&&\ \ \ \ \mbox{} -
  305 {{\xi_1}^8} \xi_2 {{\xi_3}^2}
- 948 {{\xi_1}^7} {{\xi_2}^2} {{\xi_3}^2} +
  1820 {{\xi_1}^6} {{\xi_2}^3} {{\xi_3}^2}
- 494 {{\xi_1}^5} {{\xi_2}^4} {{\xi_3}^2} \nonumber\\&&\ \ \ \ \mbox{}-
  366 {{\xi_1}^4} {{\xi_2}^5} {{\xi_3}^2}
- 484 {{\xi_1}^3} {{\xi_2}^6} {{\xi_3}^2} +
  844 {{\xi_1}^2} {{\xi_2}^7} {{\xi_3}^2}
- 369 \xi_1 {{\xi_2}^8} {{\xi_3}^2} \nonumber\\&&\ \ \ \ \mbox{}+
  55 {{\xi_2}^9} {{\xi_3}^2} + 27 {{\xi_1}^8} {{\xi_3}^3} +
  440 {{\xi_1}^7} \xi_2 {{\xi_3}^3}
- 668 {{\xi_1}^6} {{\xi_2}^2} {{\xi_3}^3} \nonumber\\&&\ \ \ \ \mbox{}-
  1080 {{\xi_1}^5} {{\xi_2}^3} {{\xi_3}^3}
+2242{{\xi_1}^4} {{\xi_2}^4} {{\xi_3}^3}-
  568 {{\xi_1}^3} {{\xi_2}^5} {{\xi_3}^3}
- 924 {{\xi_1}^2} {{\xi_2}^6} {{\xi_3}^3} \nonumber\\&&\ \ \ \ \mbox{}+
  696 \xi_1 {{\xi_2}^7} {{\xi_3}^3} - 165 {{\xi_2}^8} {{\xi_3}^3} -
  726 {{\xi_1}^7} {{\xi_3}^4} + 150 {{\xi_1}^6} \xi_2 {{\xi_3}^4} \nonumber\\&&\ \ \ \ \mbox{}+
  1874 {{\xi_1}^5} {{\xi_2}^2} {{\xi_3}^4}
- 1298 {{\xi_1}^4} {{\xi_2}^3} {{\xi_3}^4} -
  50 {{\xi_1}^3} {{\xi_2}^4} {{\xi_3}^4}
+ 434 {{\xi_1}^2} {{\xi_2}^5} {{\xi_3}^4} \nonumber\\&&\ \ \ \ \mbox{}-
  714 \xi_1 {{\xi_2}^6} {{\xi_3}^4} + 330 {{\xi_2}^7} {{\xi_3}^4} +
  594 {{\xi_1}^6} {{\xi_3}^5} - 1028 {{\xi_1}^5} \xi_2 {{\xi_3}^5} \nonumber\\&&\ \ \ \ \mbox{}-
  258 {{\xi_1}^4} {{\xi_2}^2} {{\xi_3}^5}
+ 1000 {{\xi_1}^3} {{\xi_2}^3} {{\xi_3}^5}-
  98 {{\xi_1}^2} {{\xi_2}^4} {{\xi_3}^5}
+ 252 \xi_1 {{\xi_2}^5} {{\xi_3}^5} \nonumber\\&&\ \ \ \ \mbox{}-
  462 {{\xi_2}^6} {{\xi_3}^5} + 270 {{\xi_1}^5} {{\xi_3}^6} +
  998 {{\xi_1}^4} \xi_2 {{\xi_3}^6}
- 260 {{\xi_1}^3} {{\xi_2}^2} {{\xi_3}^6}\nonumber\\&&\ \ \ \ \mbox{} +
  252 {{\xi_1}^2} {{\xi_2}^3} {{\xi_3}^6}
+ 294 \xi_1 {{\xi_2}^4} {{\xi_3}^6} +
  462 {{\xi_2}^5} {{\xi_3}^6} - 522 {{\xi_1}^4} {{\xi_3}^7} \nonumber\\&&\ \ \ \ \mbox{}-
  456 {{\xi_1}^3} \xi_2 {{\xi_3}^7}
- 364 {{\xi_1}^2} {{\xi_2}^2} {{\xi_3}^7} -
  456 \xi_1 {{\xi_2}^3} {{\xi_3}^7} - 330 {{\xi_2}^4} {{\xi_3}^7} \nonumber\\&&\ \ \ \ \mbox{}+
  213 {{\xi_1}^3} {{\xi_3}^8} + 199 {{\xi_1}^2} \xi_2 {{\xi_3}^8} +
  279 \xi_1 {{\xi_2}^2} {{\xi_3}^8} + 165 {{\xi_2}^3} {{\xi_3}^8}\nonumber\\&&\ \ \ \ \mbox{} -
  39 {{\xi_1}^2} {{\xi_3}^9}
- 86 \xi_1 \xi_2 {{\xi_3}^9} - 55 {{\xi_2}^2} {{\xi_3}^9}+
  11 \xi_1 {{\xi_3}^{10}} + 11 \xi_2 {{\xi_3}^{10}} - {{\xi_3}^{11}})\nonumber\\&&\ \ \ \ \mbox{}
-\left(\frac{f(\xi_1)-1}{\xi_1}\right){8\xi_1\over {3 {{\Delta}^5}}}
(-6 {{\xi_1}^7} - 7 {{\xi_1}^6} \xi_3 - 18 {{\xi_1}^5} \xi_2 \xi_3 \nonumber\\&&\ \ \ \ \mbox{}+
    62 {{\xi_1}^5} {{\xi_3}^2} + 9 {{\xi_1}^4} \xi_2 {{\xi_3}^2} -
    68 {{\xi_1}^3} {{\xi_2}^2} {{\xi_3}^2} - 25 {{\xi_1}^4} {{\xi_3}^3} \nonumber\\&&\ \ \ \ \mbox{}+
    112 {{\xi_1}^3} \xi_2 {{\xi_3}^3}
+ 38 {{\xi_1}^2} {{\xi_2}^2} {{\xi_3}^3} -
    28 \xi_1 {{\xi_2}^3} {{\xi_3}^3} - 44 {{\xi_1}^3} {{\xi_3}^4} \nonumber\\&&\ \ \ \ \mbox{}-
    57 {{\xi_1}^2} \xi_2 {{\xi_3}^4} + 54 \xi_1 {{\xi_2}^2} {{\xi_3}^4} +
    15 {{\xi_2}^3} {{\xi_3}^4} + 19 {{\xi_1}^2} {{\xi_3}^5} \nonumber\\&&\ \ \ \ \mbox{}
- 36 \xi_1 \xi_2 {{\xi_3}^5} -  27 {{\xi_2}^2} {{\xi_3}^5}
+ 10 \xi_1 {{\xi_3}^6} + 15 \xi_2 {{\xi_3}^6} - 3 {{\xi_3}^7})\nonumber\\&&\ \ \ \ \mbox{}
-\left(\frac{f(\xi_3)-1}{\xi_3}\right){1\over {6 {{\Delta}^5} \xi_1}}
( -3 {{\xi_1}^9} + 5 {{\xi_1}^8} \xi_2 + 38 {{\xi_1}^7} {{\xi_2}^2} -
    166 {{\xi_1}^6} {{\xi_2}^3} \nonumber\\&&\ \ \ \ \mbox{}+ 296 {{\xi_1}^5} {{\xi_2}^4} -
    292 {{\xi_1}^4} {{\xi_2}^5} + 170 {{\xi_1}^3} {{\xi_2}^6} -
    58 {{\xi_1}^2} {{\xi_2}^7} + 11 \xi_1 {{\xi_2}^8}\nonumber\\&&\ \ \ \ \mbox{} - {{\xi_2}^9} +
    185 {{\xi_1}^8} \xi_3 - 258 {{\xi_1}^7} \xi_2 \xi_3 -
696 {{\xi_1}^6} {{\xi_2}^2} \xi_3 +
    1586 {{\xi_1}^5} {{\xi_2}^3} \xi_3 \nonumber\\&&\ \ \ \ \mbox{}- 690 {{\xi_1}^4} {{\xi_2}^4} \xi_3 -
    510 {{\xi_1}^3} {{\xi_2}^5} \xi_3 + 488 {{\xi_1}^2} {{\xi_2}^6} \xi_3 -
    114 \xi_1 {{\xi_2}^7} \xi_3 \nonumber\\&&\ \ \ \ \mbox{}
+ 9 {{\xi_2}^8} \xi_3 + 116 {{\xi_1}^7} {{\xi_3}^2} +
    90 {{\xi_1}^6} \xi_2 {{\xi_3}^2}
- 790 {{\xi_1}^5} {{\xi_2}^2} {{\xi_3}^2} \nonumber\\&&\ \ \ \ \mbox{}+
    480 {{\xi_1}^4} {{\xi_2}^3} {{\xi_3}^2}
+ 696 {{\xi_1}^3} {{\xi_2}^4} {{\xi_3}^2} -
    854 {{\xi_1}^2} {{\xi_2}^5} {{\xi_3}^2}
+ 298 \xi_1 {{\xi_2}^6} {{\xi_3}^2} \nonumber\\&&\ \ \ \ \mbox{}-
    36 {{\xi_2}^7} {{\xi_3}^2} - 892 {{\xi_1}^6} {{\xi_3}^3} +
    1014 {{\xi_1}^5} \xi_2 {{\xi_3}^3}
+ 820 {{\xi_1}^4} {{\xi_2}^2} {{\xi_3}^3} \nonumber\\&&\ \ \ \ \mbox{}-
    1148 {{\xi_1}^3} {{\xi_2}^3} {{\xi_3}^3}
+ 324 {{\xi_1}^2}{{\xi_2}^4}{{\xi_3}^3}-
    202 \xi_1 {{\xi_2}^5} {{\xi_3}^3} + 84 {{\xi_2}^6} {{\xi_3}^3} \nonumber\\&&\ \ \ \ \mbox{}+
    198 {{\xi_1}^5} {{\xi_3}^4} - 1164 {{\xi_1}^4} \xi_2 {{\xi_3}^4} +
    1050 {{\xi_1}^3} {{\xi_2}^2} {{\xi_3}^4}
+ 282 {{\xi_1}^2} {{\xi_2}^3} {{\xi_3}^4} \nonumber\\&&\ \ \ \ \mbox{}-
    240 \xi_1 {{\xi_2}^4} {{\xi_3}^4} - 126 {{\xi_2}^5} {{\xi_3}^4} +
    846 {{\xi_1}^4} {{\xi_3}^5} + 90 {{\xi_1}^3} \xi_2 {{\xi_3}^5} \nonumber\\&&\ \ \ \ \mbox{}-
    352 {{\xi_1}^2} {{\xi_2}^2} {{\xi_3}^5}
+ 394 \xi_1 {{\xi_2}^3} {{\xi_3}^5} +
    126 {{\xi_2}^4} {{\xi_3}^5} - 348 {{\xi_1}^3} {{\xi_3}^6} \nonumber\\&&\ \ \ \ \mbox{}+
    310 {{\xi_1}^2} \xi_2 {{\xi_3}^6} - 106 \xi_1 {{\xi_2}^2} {{\xi_3}^6} -
    84 {{\xi_2}^3} {{\xi_3}^6}
- 140 {{\xi_1}^2} {{\xi_3}^7} \nonumber\\&&\ \ \ \ \mbox{}- 78 \xi_1 \xi_2 {{\xi_3}^7} +
    36 {{\xi_2}^2} {{\xi_3}^7} + 37 \xi_1 {{\xi_3}^8}
- 9 \xi_2 {{\xi_3}^8} + {{\xi_3}^9} )\nonumber\\&&\ \ \ \ \mbox{}
-\left(\frac{f(\xi_1)-1+\frac16\xi_1}{\xi_1^2}\right){2\over {{{\Delta}^4} \xi_2 \xi_3}}
( 3 {{\xi_1}^8}
- 42 {{\xi_1}^7} \xi_3 - 72 {{\xi_1}^6} \xi_2 \xi_3 \nonumber\\&&\ \ \ \ \mbox{}+
    104 {{\xi_1}^6} {{\xi_3}^2} - 114 {{\xi_1}^5} \xi_2 {{\xi_3}^2} -
    128 {{\xi_1}^4} {{\xi_2}^2} {{\xi_3}^2} - 126 {{\xi_1}^5} {{\xi_3}^3} \nonumber\\&&\ \ \ \ \mbox{}+
    288 {{\xi_1}^4} \xi_2 {{\xi_3}^3} - 12 {{\xi_1}^3} {{\xi_2}^2} {{\xi_3}^3} -
    48 {{\xi_1}^2} {{\xi_2}^3} {{\xi_3}^3} + 96 {{\xi_1}^4} {{\xi_3}^4} \nonumber\\&&\ \ \ \ \mbox{}+
    82 {{\xi_1}^3} \xi_2 {{\xi_3}^4} + 144 {{\xi_1}^2} {{\xi_2}^2} {{\xi_3}^4} +
    90 \xi_1 {{\xi_2}^3} {{\xi_3}^4} + 70 {{\xi_2}^4} {{\xi_3}^4}\nonumber\\&&\ \ \ \ \mbox{} -
    70 {{\xi_1}^3} {{\xi_3}^5} - 144 {{\xi_1}^2} \xi_2 {{\xi_3}^5} -
    162 \xi_1 {{\xi_2}^2} {{\xi_3}^5} - 112 {{\xi_2}^3} {{\xi_3}^5} \nonumber\\&&\ \ \ \ \mbox{}+
    48 {{\xi_1}^2} {{\xi_3}^6} + 90 \xi_1 \xi_2 {{\xi_3}^6}
+ 56 {{\xi_2}^2} {{\xi_3}^6} - 18 \xi_1 {{\xi_3}^7}
 - 16 \xi_2 {{\xi_3}^7} + 2 {{\xi_3}^8} )\nonumber\\&&\ \ \ \ \mbox{}
-\left(\frac{f(\xi_3)-1+\frac16\xi_3}{\xi_3^2}\right){1\over {2 {{\Delta}^4} \xi_1 \xi_2}}
( 8 {{\xi_1}^8} - 55 {{\xi_1}^7} \xi_2 + 167 {{\xi_1}^6} {{\xi_2}^2} \nonumber\\&&\ \ \ \ \mbox{} -
    295 {{\xi_1}^5} {{\xi_2}^3} + 335 {{\xi_1}^4} {{\xi_2}^4} -
    253 {{\xi_1}^3} {{\xi_2}^5} + 125 {{\xi_1}^2} {{\xi_2}^6}
- 37 \xi_1 {{\xi_2}^7} +
    5 {{\xi_2}^8} \nonumber\\&&\ \ \ \ \mbox{}- 72 {{\xi_1}^7} \xi_3 + 285 {{\xi_1}^6} \xi_2 \xi_3 -
    414 {{\xi_1}^5} {{\xi_2}^2} \xi_3 + 195 {{\xi_1}^4} {{\xi_2}^3} \xi_3 +
    180 {{\xi_1}^3} {{\xi_2}^4} \xi_3 \nonumber\\&&\ \ \ \ \mbox{}- 333 {{\xi_1}^2} {{\xi_2}^5} \xi_3 +
    210 \xi_1 {{\xi_2}^6} \xi_3 - 51 {{\xi_2}^7} \xi_3
+ 280 {{\xi_1}^6} {{\xi_3}^2} +
    189 {{\xi_1}^5} \xi_2 {{\xi_3}^2} \nonumber\\&&\ \ \ \ \mbox{}
- 859 {{\xi_1}^4} {{\xi_2}^2} {{\xi_3}^2} -
    70 {{\xi_1}^3} {{\xi_2}^3} {{\xi_3}^2}
+ 714 {{\xi_1}^2} {{\xi_2}^4} {{\xi_3}^2} -
    407 \xi_1 {{\xi_2}^5} {{\xi_3}^2} \nonumber\\&&\ \ \ \ \mbox{}+ 153 {{\xi_2}^6} {{\xi_3}^2} -
    520 {{\xi_1}^5} {{\xi_3}^3} + 641 {{\xi_1}^4} \xi_2 {{\xi_3}^3} +
 204 {{\xi_1}^3} {{\xi_2}^2} {{\xi_3}^3}\nonumber\\&&\ \ \ \ \mbox{}
- 250 {{\xi_1}^2} {{\xi_2}^3} {{\xi_3}^3}+
    116 \xi_1 {{\xi_2}^4} {{\xi_3}^3} - 191 {{\xi_2}^5} {{\xi_3}^3} +
    456 {{\xi_1}^4} {{\xi_3}^4} \nonumber\\&&\ \ \ \ \mbox{}- 421 {{\xi_1}^3} \xi_2 {{\xi_3}^4} +
    609 {{\xi_1}^2} {{\xi_2}^2} {{\xi_3}^4}
+ 573 \xi_1 {{\xi_2}^3} {{\xi_3}^4} +
    95 {{\xi_2}^4} {{\xi_3}^4}\nonumber\\&&\ \ \ \ \mbox{} - 152 {{\xi_1}^3} {{\xi_3}^5} -
    873 {{\xi_1}^2} \xi_2 {{\xi_3}^5} - 718 \xi_1 {{\xi_2}^2} {{\xi_3}^5} -
    25 {{\xi_2}^3} {{\xi_3}^5}
+ 8 {{\xi_1}^2} {{\xi_3}^6}\nonumber\\&&\ \ \ \ \mbox{} + 287 \xi_1 \xi_2 {{\xi_3}^6} +
    51 {{\xi_2}^2} {{\xi_3}^6} - 24 \xi_1 {{\xi_3}^7} - 53 \xi_2 {{\xi_3}^7} +
    16 {{\xi_3}^8} )\nonumber\\&&\ \ \ \ \mbox{}
+\frac1{\xi_2-\xi_3}\Big(f(\xi_2)-f(\xi_3)\Big)\frac{1}{144\xi_1}
+\frac1{\xi_2-\xi_3}\left(\frac{f(\xi_2)-1}{\xi_2}-\frac{f(\xi_3)-1}{\xi_3}\right)\frac{1}{12\xi_1}\nonumber\\&&\ \ \ \ \mbox{}
-\frac1{\xi_1-\xi_3}\left(\frac{f(\xi_1)-1+\frac16\xi_1}{{\xi_1}^2}-\frac{f(\xi_3)-1+\frac16\xi_3}{{\xi_3}^2}\right)\frac{2}{\xi_2}\nonumber\\&&\ \ \ \ \mbox{}
-\frac1{\xi_2-\xi_3}\left(\frac{f(\xi_2)-1+\frac16\xi_2}{{\xi_2}^2}-\frac{f(\xi_3)-1+\frac16\xi_3}{{\xi_3}^2}\right)\frac{1}{4\xi_1},
\end{fleqnarray}
\begin{fleqnarray}&&
F_{23}(\xi_1,\xi_2,\xi_3)=F(\xi_1,\xi_2,\xi_3){8\over {3 {{\Delta}^4}}}
({{\xi_1}^6} - 9 {{\xi_1}^4} {{\xi_2}^2} + 8 {{\xi_1}^3} {{\xi_2}^3} -
    {{\xi_1}^5} \xi_3 \nonumber\\&&\ \ \ \ \mbox{}
+ 3 {{\xi_1}^4} \xi_2 \xi_3 - 2 {{\xi_1}^3} {{\xi_2}^2} \xi_3 -
    4 {{\xi_1}^4} {{\xi_3}^2} - 4 {{\xi_1}^3} \xi_2 {{\xi_3}^2} +
    8 {{\xi_1}^2} {{\xi_2}^2} {{\xi_3}^2} \nonumber\\&&\ \ \ \ \mbox{}+ 6 {{\xi_1}^3} {{\xi_3}^3} -
    6 {{\xi_1}^2} \xi_2 {{\xi_3}^3}
+ {{\xi_1}^2} {{\xi_3}^4} + 6 \xi_1 \xi_2 {{\xi_3}^4} -
    5 \xi_1 {{\xi_3}^5} + {{\xi_3}^6} )\nonumber\\&&\ \ \ \ \mbox{}
-\left(F(\xi_1,\xi_2,\xi_3)-\frac12\right){8\over {3 {{\Delta}^3} \xi_1 \xi_2}}
( -{{\xi_1}^5} - 33 {{\xi_1}^4} \xi_2 + 34 {{\xi_1}^3} {{\xi_2}^2} \nonumber\\&&\ \ \ \ \mbox{} +
    2 {{\xi_1}^4} \xi_3 + 32 {{\xi_1}^3} \xi_2 \xi_3
- 34 {{\xi_1}^2} {{\xi_2}^2} \xi_3 +
    2 {{\xi_1}^3} {{\xi_3}^2} + 42 {{\xi_1}^2} \xi_2 {{\xi_3}^2} \nonumber\\&&\ \ \ \ \mbox{}-
    8 {{\xi_1}^2} {{\xi_3}^3} - 24 \xi_1 \xi_2 {{\xi_3}^3} + 7 \xi_1 {{\xi_3}^4}
- {{\xi_3}^5})\nonumber\\&&\ \ \ \ \mbox{}
+\left(F(\xi_1,\xi_2,\xi_3)-\frac12+\frac{\xi_1+\xi_2+\xi_3}{24}\right){8\over {{{\Delta}^2} \xi_1 \xi_2}}
( 2 {{\xi_1}^2} + 10 \xi_1 \xi_2 - 4 \xi_1 \xi_3
+ {{\xi_3}^2} )\nonumber\\&&\ \ \ \ \mbox{}
-f(\xi_3){1\over {24 \xi_1 \xi_2}}
-\left(\frac{f(\xi_1)-1}{\xi_1}\right){{16\xi_1}\over {3 {{\Delta}^4}}}
( -\xi_1 + \xi_2 - \xi_3 )
  ( {{\xi_1}^4} + 2 {{\xi_1}^3} \xi_2 \nonumber\\&&\ \ \ \ \mbox{}- 6 {{\xi_1}^2} {{\xi_2}^2}+
    2 \xi_1 {{\xi_2}^3}
+ {{\xi_2}^4} - {{\xi_1}^3} \xi_3 + {{\xi_1}^2} \xi_2 \xi_3 +
    \xi_1 {{\xi_2}^2} \xi_3 - {{\xi_2}^3} \xi_3 \nonumber\\&&\ \ \ \ \mbox{}- 3 {{\xi_1}^2} {{\xi_3}^2} -
    8 \xi_1 \xi_2 {{\xi_3}^2} - 3 {{\xi_2}^2} {{\xi_3}^2} + 5 \xi_1 {{\xi_3}^3} +
    5 \xi_2 {{\xi_3}^3} - 2 {{\xi_3}^4} )\nonumber\\&&\ \ \ \ \mbox{}
-\left(\frac{f(\xi_3)-1}{\xi_3}\right){2\over {3 {{\Delta}^4} \xi_1 \xi_2}}
( {{\xi_1}^8} - 8 {{\xi_1}^7} \xi_2 + 28 {{\xi_1}^6} {{\xi_2}^2} -
    56 {{\xi_1}^5} {{\xi_2}^3} \nonumber\\&&\ \ \ \ \mbox{}
+ 35 {{\xi_1}^4} {{\xi_2}^4} - 9 {{\xi_1}^7} \xi_3 +
    57 {{\xi_1}^6} \xi_2 \xi_3 - 117 {{\xi_1}^5} {{\xi_2}^2} \xi_3 \nonumber\\&&\ \ \ \ \mbox{}+
    69 {{\xi_1}^4} {{\xi_2}^3} \xi_3 + 35 {{\xi_1}^6} {{\xi_3}^2} -
    110 {{\xi_1}^5} \xi_2 {{\xi_3}^2}
+ 125 {{\xi_1}^4} {{\xi_2}^2} {{\xi_3}^2} \nonumber\\&&\ \ \ \ \mbox{}-
    50 {{\xi_1}^3} {{\xi_2}^3} {{\xi_3}^2} - 77 {{\xi_1}^5} {{\xi_3}^3} +
    63 {{\xi_1}^4} \xi_2 {{\xi_3}^3}
+ 14 {{\xi_1}^3} {{\xi_2}^2} {{\xi_3}^3} \nonumber\\&&\ \ \ \ \mbox{}+
    105 {{\xi_1}^4} {{\xi_3}^4} + 36 {{\xi_1}^3} \xi_2 {{\xi_3}^4} -
    13 {{\xi_1}^2} {{\xi_2}^2} {{\xi_3}^4} - 91 {{\xi_1}^3} {{\xi_3}^5} \nonumber\\&&\ \ \ \ \mbox{}-
    73 {{\xi_1}^2} \xi_2 {{\xi_3}^5} + 49 {{\xi_1}^2} {{\xi_3}^6} +
    25 \xi_1 \xi_2 {{\xi_3}^6} - 15 \xi_1 {{\xi_3}^7} + {{\xi_3}^8} )\nonumber\\&&\ \ \ \ \mbox{}
+\left(\frac{f(\xi_1)-1+\frac16\xi_1}{\xi_1^2}\right){4\over {{{\Delta}^3} \xi_2 \xi_3}}
 ( -{{\xi_1}^5} \xi_2  + 5 {{\xi_1}^4} {{\xi_2}^2} \nonumber\\&&\ \ \ \ \mbox{}-
    10 {{\xi_1}^3} {{\xi_2}^3}
+ 10 {{\xi_1}^2} {{\xi_2}^4} - 5 \xi_1 {{\xi_2}^5} +
    {{\xi_2}^6} + {{\xi_1}^5} \xi_3 \nonumber\\&&\ \ \ \ \mbox{}+ 50 {{\xi_1}^4} \xi_2 \xi_3 -
    22 {{\xi_1}^3} {{\xi_2}^2} \xi_3 - 36 {{\xi_1}^2} {{\xi_2}^3} \xi_3 +
    13 \xi_1 {{\xi_2}^4} \xi_3
- 6 {{\xi_2}^5} \xi_3 \nonumber\\&&\ \ \ \ \mbox{}+ {{\xi_1}^4} {{\xi_3}^2} -
    22 {{\xi_1}^3} \xi_2 {{\xi_3}^2}
+ 56 {{\xi_1}^2} {{\xi_2}^2} {{\xi_3}^2} -
    2 \xi_1 {{\xi_2}^3} {{\xi_3}^2} + 15 {{\xi_2}^4} {{\xi_3}^2} \nonumber\\&&\ \ \ \ \mbox{}-
    10 {{\xi_1}^3} {{\xi_3}^3} - 44 {{\xi_1}^2} \xi_2 {{\xi_3}^3} -
    22 \xi_1 {{\xi_2}^2} {{\xi_3}^3} - 20 {{\xi_2}^3} {{\xi_3}^3} +
    14 {{\xi_1}^2} {{\xi_3}^4} \nonumber\\&&\ \ \ \ \mbox{}
+ 23 \xi_1 \xi_2 {{\xi_3}^4} + 15 {{\xi_2}^2} {{\xi_3}^4} -
    7 \xi_1 {{\xi_3}^5} - 6 \xi_2 {{\xi_3}^5} + {{\xi_3}^6} )\nonumber\\&&\ \ \ \ \mbox{}
+\left(\frac{f(\xi_3)-1+\frac16\xi_3}{\xi_3^2}\right){1\over {2 {{\Delta}^3} \xi_1 \xi_2}}(
2 {{\xi_1}^6} - 12 {{\xi_1}^5} \xi_2 + 30 {{\xi_1}^4} {{\xi_2}^2} \nonumber\\&&\ \ \ \ \mbox{}-
  20 {{\xi_1}^3} {{\xi_2}^3} - 4 {{\xi_1}^5} \xi_3 + 12 {{\xi_1}^4} \xi_2 \xi_3 -
  8 {{\xi_1}^3} {{\xi_2}^2} \xi_3 - 26 {{\xi_1}^4} {{\xi_3}^2} \nonumber\\&&\ \ \ \ \mbox{}-
  184 {{\xi_1}^3} \xi_2 {{\xi_3}^2} + 210 {{\xi_1}^2} {{\xi_2}^2} {{\xi_3}^2} +
  72 {{\xi_1}^3} {{\xi_3}^3} - 136 {{\xi_1}^2} \xi_2 {{\xi_3}^3} \nonumber\\&&\ \ \ \ \mbox{}-
  50 {{\xi_1}^2} {{\xi_3}^4}
+ 162 \xi_1 \xi_2 {{\xi_3}^4} - 4 \xi_1 {{\xi_3}^5} +
  5 {{\xi_3}^6})\nonumber\\&&\ \ \ \ \mbox{}
+\frac1{\xi_1-\xi_2}\left(\frac{f(\xi_1)-1+\frac16\xi_1}{{\xi_1}^2}-\frac{f(\xi_2)-1+\frac16\xi_2}{{\xi_2}^2}\right)\frac{2}{\xi_3},
\end{fleqnarray}\begin{fleqnarray}&&
F_{24}(\xi_1,\xi_2,\xi_3)=\left(F(\xi_1,\xi_2,\xi_3)-\frac12\right){{4\xi_1}\over {{{\Delta}^2} \xi_2 \xi_3}}
( {{\xi_1}^2} - 4 \xi_1 \xi_3
 + 2 \xi_2 \xi_3 + 2 {{\xi_3}^2} )\nonumber\\&&\ \ \ \ \mbox{}+
\left(F(\xi_1,\xi_2,\xi_3)-\frac12+\frac{\xi_1+\xi_2+\xi_3}{24}\right){8\over {\Delta \xi_2 \xi_3}}\nonumber\\&&\ \ \ \ \mbox{}
+\left(\frac{f(\xi_1)-1+\frac16\xi_1}{\xi_1^2}\right){2\over {{{\Delta}^2} \xi_2 \xi_3}}
 (5 {{\xi_1}^4} - 16 {{\xi_1}^3} \xi_3
+ 4 {{\xi_1}^2} \xi_2 \xi_3 \nonumber\\&&\ \ \ \ \mbox{}+
    12 {{\xi_1}^2} {{\xi_3}^2} + 8 \xi_1 \xi_2 {{\xi_3}^2}
+ 6 {{\xi_2}^2} {{\xi_3}^2} -
    8 \xi_1 {{\xi_3}^3} - 8 \xi_2 {{\xi_3}^3} + 2 {{\xi_3}^4} )\nonumber\\&&\ \ \ \ \mbox{}
+\left(\frac{f(\xi_3)-1+\frac16\xi_3}{\xi_3^2}\right){4\over {{{\Delta}^2} \xi_1 \xi_2}}
( {{\xi_1}^4}
- 4 {{\xi_1}^3} \xi_2 + 6 {{\xi_1}^2} {{\xi_2}^2} \nonumber\\&&\ \ \ \ \mbox{}- 4 \xi_1 {{\xi_2}^3}
+ {{\xi_2}^4} - 7 {{\xi_1}^3} \xi_3 + 11 {{\xi_1}^2} \xi_2 \xi_3
-  \xi_1 {{\xi_2}^2} \xi_3 - 3 {{\xi_2}^3} \xi_3 \nonumber\\&&\ \ \ \ \mbox{}+ 7 {{\xi_1}^2} {{\xi_3}^2} +
    6 \xi_1 \xi_2 {{\xi_3}^2} + 3 {{\xi_2}^2} {{\xi_3}^2} - \xi_1 {{\xi_3}^3} -
    \xi_2 {{\xi_3}^3} )\nonumber\\&&\ \ \ \ \mbox{}
+\frac1{\xi_2-\xi_3}\left(\frac{f(\xi_2)-1+\frac16\xi_2}{{\xi_2}^2}-\frac{f(\xi_3)-1+\frac16\xi_3}{{\xi_3}^2}\right)\frac{2}{\xi_1},
\end{fleqnarray}\begin{fleqnarray}&&
F_{25}(\xi_1,\xi_2,\xi_3)=-\left(F(\xi_1,\xi_2,\xi_3)-\frac12\right){{16}\over {{{\Delta}^2} \xi_1}}
( -{{\xi_1}^2} - 2 \xi_2 \xi_3 + 2 {{\xi_3}^2} )\nonumber\\&&\ \ \ \ \mbox{}
+\left(F(\xi_1,\xi_2,\xi_3)-\frac12+\frac{\xi_1+\xi_2+\xi_3}{24}\right){{16}\over {\Delta \xi_1 \xi_2 \xi_3}}
( \xi_1 - 2 \xi_3 )\nonumber\\&&\ \ \ \ \mbox{}+
\left(\frac{f(\xi_1)-1+\frac16\xi_1}{\xi_1^2}\right){4\over {{{\Delta}^2} \xi_2 \xi_3}}
( -{{\xi_1}^4} + 4 {{\xi_1}^3} \xi_3 + 16 {{\xi_1}^2} \xi_2 \xi_3 \nonumber\\&&\ \ \ \ \mbox{}+
    4 \xi_1 \xi_2 {{\xi_3}^2} + 6 {{\xi_2}^2} {{\xi_3}^2}
 - 4 \xi_1 {{\xi_3}^3} - 8 \xi_2 {{\xi_3}^3} + 2 {{\xi_3}^4} )\nonumber\\&&\ \ \ \ \mbox{}+
\left(\frac{f(\xi_3)-1+\frac16\xi_3}{\xi_3^2}\right){8\over {{{\Delta}^2} \xi_1 \xi_2}}
( {{\xi_1}^4} - 4 {{\xi_1}^3} \xi_2
+ 6 {{\xi_1}^2} {{\xi_2}^2} \nonumber\\&&\ \ \ \ \mbox{}-
    4 \xi_1 {{\xi_2}^3} + {{\xi_2}^4} - 4 {{\xi_1}^3} \xi_3
+ 4 {{\xi_1}^2} \xi_2 \xi_3 + 4 \xi_1 {{\xi_2}^2} \xi_3 - 4 {{\xi_2}^3} \xi_3 \nonumber\\&&\ \ \ \ \mbox{}
+ 6 {{\xi_1}^2} {{\xi_3}^2}
- 4 \xi_1 \xi_2 {{\xi_3}^2} + 14 {{\xi_2}^2} {{\xi_3}^2}
- 4 \xi_1 {{\xi_3}^3} - 12 \xi_2 {{\xi_3}^3} + {{\xi_3}^4} )\nonumber\\&&\ \ \ \ \mbox{}
+\frac1{\xi_1-\xi_2}\left(\frac{f(\xi_1)-1+\frac16\xi_1}{{\xi_1}^2}-\frac{f(\xi_2)-1+\frac16\xi_2}{{\xi_2}^2}\right)\frac{8}{\xi_3},
\end{fleqnarray}
\begin{fleqnarray}&&
F_{26}(\xi_1,\xi_2,\xi_3)=F(\xi_1,\xi_2,\xi_3)\Big[
\frac{4 \xi_1 \xi_2}{{\Delta}^{4}}
 ( 2 {{\xi_1}^4} - 8 {{\xi_1}^3} \xi_2 + 6 {{\xi_1}^2} {{\xi_2}^2} -
    4 {{\xi_1}^2} {{\xi_3}^2} \nonumber\\&&\ \ \ \ \mbox{}+ 4 \xi_1 \xi_2 {{\xi_3}^2} + {{\xi_3}^4} )
-\frac{16}{{\Delta}^{3}}
( -3 {{\xi_1}^3} + 3 {{\xi_1}^2} \xi_2 + 4 {{\xi_1}^2} \xi_3\nonumber\\&&\ \ \ \ \mbox{}
    - 4 \xi_1 \xi_2 \xi_3 +  \xi_1 {{\xi_3}^2} - {{\xi_3}^3} )\Big]\nonumber\\&&\ \ \ \ \mbox{}
+\left(F(\xi_1,\xi_2,\xi_3)-\frac12\right){8\over {{{\Delta}^2} \xi_1 \xi_2}}
( 2 {{\xi_1}^2} + 4 \xi_1 \xi_2
- 4 \xi_1 \xi_3 + {{\xi_3}^2} )\nonumber\\&&\ \ \ \ \mbox{}
-f(\xi_1)\frac{16 \xi_1 \xi_2}{{\Delta}^{4}}
 ( -\xi_1 + \xi_2 - \xi_3 )^2
  ( -\xi_1 + \xi_2 + \xi_3 )\nonumber\\&&\ \ \ \ \mbox{}
-f(\xi_3){1\over {2 {{\Delta}^4} \xi_1 \xi_2}}
( -2 {{\xi_1}^7} + 18 {{\xi_1}^6} \xi_2 - 42 {{\xi_1}^5} {{\xi_2}^2} +
    26 {{\xi_1}^4} {{\xi_2}^3} \nonumber\\&&\ \ \ \ \mbox{}
+ 14 {{\xi_1}^6} \xi_3 - 76 {{\xi_1}^5} \xi_2 \xi_3 +
    178 {{\xi_1}^4} {{\xi_2}^2} \xi_3 - 116 {{\xi_1}^3} {{\xi_2}^3} \xi_3 \nonumber\\&&\ \ \ \ \mbox{}-
    42 {{\xi_1}^5} {{\xi_3}^2} + 110 {{\xi_1}^4} \xi_2 {{\xi_3}^2} -
    68 {{\xi_1}^3} {{\xi_2}^2} {{\xi_3}^2} + 70 {{\xi_1}^4} {{\xi_3}^3} \nonumber\\&&\ \ \ \ \mbox{}-
    40 {{\xi_1}^3} \xi_2 {{\xi_3}^3} - 30 {{\xi_1}^2} {{\xi_2}^2} {{\xi_3}^3} -
    70 {{\xi_1}^3} {{\xi_3}^4} - 50 {{\xi_1}^2} \xi_2 {{\xi_3}^4}\nonumber\\&&\ \ \ \ \mbox{} +
    42 {{\xi_1}^2} {{\xi_3}^5}
+ 26 \xi_1 \xi_2 {{\xi_3}^5} - 14 \xi_1 {{\xi_3}^6} +
    {{\xi_3}^7} )\nonumber\\&&\ \ \ \ \mbox{}
+\left(\frac{f(\xi_1)-1}{\xi_1}\right)\frac{32 \xi_1}{{\Delta}^{3}}
 ( 3 {{\xi_1}^2} - \xi_1 \xi_2
- 2 {{\xi_2}^2} - \xi_1 \xi_3 + 4 \xi_2 \xi_3 - 2 {{\xi_3}^2} )\nonumber\\&&\ \ \ \ \mbox{}
-\left(\frac{f(\xi_3)-1}{\xi_3}\right){1\over {{{\Delta}^3} \xi_1 \xi_2}}
( -2 {{\xi_1}^5} + 6 {{\xi_1}^4} \xi_2 - 4 {{\xi_1}^3} {{\xi_2}^2} +
    18 {{\xi_1}^4} \xi_3 \nonumber\\&&\ \ \ \ \mbox{}
+ 24 {{\xi_1}^3} \xi_2 \xi_3 - 42 {{\xi_1}^2} {{\xi_2}^2} \xi_3 -
    52 {{\xi_1}^3} {{\xi_3}^2} + 52 {{\xi_1}^2} \xi_2 {{\xi_3}^2} \nonumber\\&&\ \ \ \ \mbox{}+
    68 {{\xi_1}^2} {{\xi_3}^3}
- 20 \xi_1 \xi_2 {{\xi_3}^3} - 42 \xi_1 {{\xi_3}^4} +
     5{{\xi_3}^5} ),
\end{fleqnarray}\begin{fleqnarray}&&
F_{27}(\xi_1,\xi_2,\xi_3)=F(\xi_1,\xi_2,\xi_3)\Big[-{{2\xi_1\xi_2}\over {3 {{\Delta}^6}}}
( 2 {{\xi_1}^8} - 4 {{\xi_1}^7} \xi_2 - 16 {{\xi_1}^6} {{\xi_2}^2} +
    68 {{\xi_1}^5} {{\xi_2}^3}\nonumber\\&&\ \ \ \ \mbox{}
     - 50 {{\xi_1}^4} {{\xi_2}^4} - 2 {{\xi_1}^7} \xi_3 +
    10 {{\xi_1}^6} \xi_2 \xi_3 - 18 {{\xi_1}^5} {{\xi_2}^2} \xi_3 +
    10 {{\xi_1}^4} {{\xi_2}^3} \xi_3 - 4 {{\xi_1}^6} {{\xi_3}^2}\nonumber\\&&\ \ \ \ \mbox{} -
    4 {{\xi_1}^5} \xi_2 {{\xi_3}^2} + 52 {{\xi_1}^4} {{\xi_2}^2} {{\xi_3}^2} -
    44 {{\xi_1}^3} {{\xi_2}^3} {{\xi_3}^2} + 2 {{\xi_1}^5} {{\xi_3}^3} -
    6 {{\xi_1}^4} \xi_2 {{\xi_3}^3} \nonumber\\&&\ \ \ \ \mbox{}
+ 4 {{\xi_1}^3} {{\xi_2}^2} {{\xi_3}^3} +
    4 {{\xi_1}^4} {{\xi_3}^4} + 4 {{\xi_1}^3} \xi_2 {{\xi_3}^4} -
    8 {{\xi_1}^2} {{\xi_2}^2} {{\xi_3}^4} + 2 {{\xi_1}^3} {{\xi_3}^5} \nonumber\\&&\ \ \ \ \mbox{}-
    2 {{\xi_1}^2} \xi_2 {{\xi_3}^5} - 4 {{\xi_1}^2} {{\xi_3}^6}
+ 2 \xi_1 \xi_2 {{\xi_3}^6} - 2 \xi_1 {{\xi_3}^7} + {{\xi_3}^8} )\nonumber\\&&\ \ \ \ \mbox{}
+{8\over {3 {{\Delta}^5}}}
 ( -3 {{\xi_1}^7} - 30 {{\xi_1}^6} \xi_2 + 108 {{\xi_1}^5} {{\xi_2}^2} -
    75 {{\xi_1}^4} {{\xi_2}^3}
+ {{\xi_1}^6} \xi_3 \nonumber\\&&\ \ \ \ \mbox{}+ 30 {{\xi_1}^5} \xi_2 \xi_3 -
    129 {{\xi_1}^4} {{\xi_2}^2} \xi_3 + 98 {{\xi_1}^3} {{\xi_2}^3} \xi_3 +
    15 {{\xi_1}^5} {{\xi_3}^2} \nonumber\\&&\ \ \ \ \mbox{}+ 21 {{\xi_1}^4} \xi_2 {{\xi_3}^2} -
    36 {{\xi_1}^3} {{\xi_2}^2} {{\xi_3}^2} - 12 {{\xi_1}^4} {{\xi_3}^3} -
    12 {{\xi_1}^3} \xi_2 {{\xi_3}^3} \nonumber\\&&\ \ \ \ \mbox{}
+ 24 {{\xi_1}^2} {{\xi_2}^2} {{\xi_3}^3} -
    17 {{\xi_1}^3} {{\xi_3}^4} - 12 {{\xi_1}^2} \xi_2 {{\xi_3}^4} +
    21 {{\xi_1}^2} {{\xi_3}^5} \nonumber\\&&\ \ \ \ \mbox{}+ 3 \xi_1 \xi_2 {{\xi_3}^5}
 - 3 \xi_1 {{\xi_3}^6} - {{\xi_3}^7})\Big]\nonumber\\&&\ \ \ \ \mbox{}
-\left(F(\xi_1,\xi_2,\xi_3)-\frac12\right){4\over {3 {{\Delta}^4} \xi_1 \xi_2 \xi_3}}
 ( -6 {{\xi_1}^7}
+ 30 {{\xi_1}^6} \xi_2 - 54 {{\xi_1}^5} {{\xi_2}^2} \nonumber\\&&\ \ \ \ \mbox{}+
    30 {{\xi_1}^4} {{\xi_2}^3}
+ 38 {{\xi_1}^6} \xi_3 + 120 {{\xi_1}^5} \xi_2 \xi_3 +
    258 {{\xi_1}^4} {{\xi_2}^2} \xi_3 \nonumber\\&&\ \ \ \ \mbox{}
- 416 {{\xi_1}^3} {{\xi_2}^3} \xi_3 -
    66 {{\xi_1}^5} {{\xi_3}^2} - 222 {{\xi_1}^4} \xi_2 {{\xi_3}^2} +
    288 {{\xi_1}^3} {{\xi_2}^2} {{\xi_3}^2} \nonumber\\&&\ \ \ \ \mbox{}+ 6 {{\xi_1}^4} {{\xi_3}^3} -
    12 {{\xi_1}^3} \xi_2 {{\xi_3}^3}
- 174 {{\xi_1}^2} {{\xi_2}^2} {{\xi_3}^3} +
    86 {{\xi_1}^3} {{\xi_3}^4} \nonumber\\&&\ \ \ \ \mbox{}+ 78 {{\xi_1}^2} \xi_2 {{\xi_3}^4} -
    78 {{\xi_1}^2} {{\xi_3}^5}
- 6 \xi_1 \xi_2 {{\xi_3}^5} + 18 \xi_1 {{\xi_3}^6} +
    {{\xi_3}^7} )\nonumber\\&&\ \ \ \ \mbox{}
+\left(F(\xi_1,\xi_2,\xi_3)-\frac12+\frac{\xi_1+\xi_2+\xi_3}{24}\right){16\over {{{\Delta}^3} \xi_1 \xi_2 \xi_3}}
( {{\xi_1}^4} - 4 {{\xi_1}^3} \xi_2 + 3 {{\xi_1}^2} {{\xi_2}^2} \nonumber\\&&\ \ \ \ \mbox{}-
    7 {{\xi_1}^3} \xi_3 - 23 {{\xi_1}^2} \xi_2 \xi_3 + 9 {{\xi_1}^2} {{\xi_3}^2} +
    11 \xi_1 \xi_2 {{\xi_3}^2} - \xi_1 {{\xi_3}^3} - {{\xi_3}^4} )\nonumber\\&&\ \ \ \ \mbox{}
+f(\xi_1){{8\xi_1\xi_2}\over {3 {{\Delta}^6}}}
{{( -\xi_1 + \xi_2 - \xi_3 ) }^2}
  ( -\xi_1 + \xi_2 + \xi_3 )\times\nonumber\\&&\ \ \ \ \mbox{}\times
  ( {{\xi_1}^4} + 2 {{\xi_1}^3} \xi_2 - 6 {{\xi_1}^2} {{\xi_2}^2} +
    2 \xi_1 {{\xi_2}^3}
+ {{\xi_2}^4} - {{\xi_1}^3} \xi_3 + {{\xi_1}^2} \xi_2 \xi_3 \nonumber\\&&\ \ \ \ \mbox{}+
    \xi_1 {{\xi_2}^2} \xi_3
- {{\xi_2}^3} \xi_3 - 2 \xi_1 \xi_2 {{\xi_3}^2} - \xi_1 {{\xi_3}^3} -
    \xi_2 {{\xi_3}^3} + {{\xi_3}^4} )\nonumber\\&&\ \ \ \ \mbox{}
-f(\xi_3){1\over {24 {{\Delta}^6} \xi_1 \xi_2}}
( -2 {{\xi_1}^{11}} + 26 {{\xi_1}^{10}} \xi_2
- 222 {{\xi_1}^9} {{\xi_2}^2} + 790 {{\xi_1}^8} {{\xi_2}^3}\nonumber\\&&\ \ \ \ \mbox{}
- 1268 {{\xi_1}^7} {{\xi_2}^4} +
    676 {{\xi_1}^6} {{\xi_2}^5}
+ 22 {{\xi_1}^{10}} \xi_3 - 212 {{\xi_1}^9} \xi_2 \xi_3\nonumber\\&&\ \ \ \ \mbox{}+
    894 {{\xi_1}^8} {{\xi_2}^2} \xi_3 - 2224 {{\xi_1}^7} {{\xi_2}^3} \xi_3 +
    3692 {{\xi_1}^6} {{\xi_2}^4} \xi_3
- 2172 {{\xi_1}^5} {{\xi_2}^5} \xi_3 \nonumber\\&&\ \ \ \ \mbox{}-
    110 {{\xi_1}^9} {{\xi_3}^2} + 738 {{\xi_1}^8} \xi_2 {{\xi_3}^2} -
 1816 {{\xi_1}^7} {{\xi_2}^2} {{\xi_3}^2}
+ 2120 {{\xi_1}^6} {{\xi_2}^3} {{\xi_3}^2}\nonumber\\&&\ \ \ \ \mbox{} -
    932 {{\xi_1}^5} {{\xi_2}^4} {{\xi_3}^2} + 330 {{\xi_1}^8} {{\xi_3}^3} -
    1392 {{\xi_1}^7} \xi_2 {{\xi_3}^3}
+ 1912 {{\xi_1}^6} {{\xi_2}^2} {{\xi_3}^3} \nonumber\\&&\ \ \ \ \mbox{}-
    400 {{\xi_1}^5} {{\xi_2}^3} {{\xi_3}^3}
- 450 {{\xi_1}^4} {{\xi_2}^4} {{\xi_3}^3} -
    660 {{\xi_1}^7} {{\xi_3}^4} + 1428 {{\xi_1}^6} \xi_2 {{\xi_3}^4} \nonumber\\&&\ \ \ \ \mbox{}-
    484 {{\xi_1}^5} {{\xi_2}^2} {{\xi_3}^4}
- 284 {{\xi_1}^4} {{\xi_2}^3} {{\xi_3}^4}+
    924 {{\xi_1}^6} {{\xi_3}^5} - 504 {{\xi_1}^5} \xi_2 {{\xi_3}^5} \nonumber\\&&\ \ \ \ \mbox{}-
    380 {{\xi_1}^4} {{\xi_2}^2} {{\xi_3}^5}
- 40 {{\xi_1}^3} {{\xi_2}^3} {{\xi_3}^5} -
    924 {{\xi_1}^5} {{\xi_3}^6} - 588 {{\xi_1}^4} \xi_2 {{\xi_3}^6}\nonumber\\&&\ \ \ \ \mbox{} -
    504 {{\xi_1}^3} {{\xi_2}^2} {{\xi_3}^6} + 660 {{\xi_1}^4} {{\xi_3}^7} +
    912 {{\xi_1}^3} \xi_2 {{\xi_3}^7}
+ 524 {{\xi_1}^2} {{\xi_2}^2} {{\xi_3}^7} \nonumber\\&&\ \ \ \ \mbox{}-
    330 {{\xi_1}^3} {{\xi_3}^8} - 558 {{\xi_1}^2} \xi_2 {{\xi_3}^8} +
    110 {{\xi_1}^2} {{\xi_3}^9} \nonumber\\&&\ \ \ \ \mbox{}
+ 86 \xi_1 \xi_2 {{\xi_3}^9} - 22 \xi_1 {{\xi_3}^{10}} +
    {{\xi_3}^{11}} )\nonumber\\&&\ \ \ \ \mbox{}
-\left(\frac{f(\xi_1)-1}{\xi_1}\right){{16\xi_1}\over {3 {{\Delta}^5}}}
( 3 {{\xi_1}^6} + 32 {{\xi_1}^5} \xi_2 - 76 {{\xi_1}^4} {{\xi_2}^2} +
    8 {{\xi_1}^3} {{\xi_2}^3}
+ 67 {{\xi_1}^2} {{\xi_2}^4} \nonumber\\&&\ \ \ \ \mbox{}- 32 \xi_1 {{\xi_2}^5} -
    2 {{\xi_2}^6} + 2 {{\xi_1}^5} \xi_3 + 3 {{\xi_1}^4} \xi_2 \xi_3 +
    51 {{\xi_1}^3} {{\xi_2}^2} \xi_3 - 119 {{\xi_1}^2} {{\xi_2}^3} \xi_3 \nonumber\\&&\ \ \ \ \mbox{}+
    63 \xi_1 {{\xi_2}^4} \xi_3 - 13 {{\xi_1}^4} {{\xi_3}^2} -
    30 {{\xi_1}^3} \xi_2 {{\xi_3}^2} + 53 {{\xi_1}^2} {{\xi_2}^2} {{\xi_3}^2} -
    28 \xi_1 {{\xi_2}^3} {{\xi_3}^2} \nonumber\\&&\ \ \ \ \mbox{}+ 18 {{\xi_2}^4} {{\xi_3}^2} -
    {{\xi_1}^3} {{\xi_3}^3} - 17 {{\xi_1}^2} \xi_2 {{\xi_3}^3} -
    10 \xi_1 {{\xi_2}^2} {{\xi_3}^3} - 32 {{\xi_2}^3} {{\xi_3}^3} \nonumber\\&&\ \ \ \ \mbox{}+
    16 {{\xi_1}^2} {{\xi_3}^4}
+ 12 \xi_1 \xi_2 {{\xi_3}^4} + 18 {{\xi_2}^2} {{\xi_3}^4} -
    5 \xi_1 {{\xi_3}^5} - 2 {{\xi_3}^6} )\nonumber\\&&\ \ \ \ \mbox{}
-\left(\frac{f(\xi_3)-1}{\xi_3}\right){1\over {3 {{\Delta}^5} \xi_1 \xi_2}}
( -2 {{\xi_1}^9} + 20 {{\xi_1}^8} \xi_2 - 70 {{\xi_1}^7} {{\xi_2}^2} \nonumber\\&&\ \ \ \ \mbox{}+
    110 {{\xi_1}^6} {{\xi_2}^3}
- 58 {{\xi_1}^5} {{\xi_2}^4} + 20 {{\xi_1}^8} \xi_3 -
    194 {{\xi_1}^7} \xi_2 \xi_3 + 44 {{\xi_1}^6} {{\xi_2}^2} \xi_3 \nonumber\\&&\ \ \ \ \mbox{}+
    1250 {{\xi_1}^5} {{\xi_2}^3} \xi_3 - 1120 {{\xi_1}^4} {{\xi_2}^4} \xi_3 -
    88 {{\xi_1}^7} {{\xi_3}^2} + 390 {{\xi_1}^6} \xi_2 {{\xi_3}^2}\nonumber\\&&\ \ \ \ \mbox{} -
    642 {{\xi_1}^5} {{\xi_2}^2} {{\xi_3}^2}
+ 340 {{\xi_1}^4} {{\xi_2}^3} {{\xi_3}^2} +
    224 {{\xi_1}^6} {{\xi_3}^3} - 378 {{\xi_1}^5} \xi_2 {{\xi_3}^3}\nonumber\\&&\ \ \ \ \mbox{} +
    552 {{\xi_1}^4} {{\xi_2}^2} {{\xi_3}^3}
- 398 {{\xi_1}^3} {{\xi_2}^3} {{\xi_3}^3} -
    364 {{\xi_1}^5} {{\xi_3}^4} + 326 {{\xi_1}^4} \xi_2 {{\xi_3}^4}\nonumber\\&&\ \ \ \ \mbox{} +
    38 {{\xi_1}^3} {{\xi_2}^2} {{\xi_3}^4} + 392 {{\xi_1}^4} {{\xi_3}^5} -
    86 {{\xi_1}^3} \xi_2 {{\xi_3}^5}
+ 102 {{\xi_1}^2} {{\xi_2}^2} {{\xi_3}^5} \nonumber\\&&\ \ \ \ \mbox{}-
    280 {{\xi_1}^3} {{\xi_3}^6} - 254 {{\xi_1}^2} \xi_2 {{\xi_3}^6} +
    128 {{\xi_1}^2} {{\xi_3}^7}
+ 105 \xi_1 \xi_2 {{\xi_3}^7} - 34 \xi_1 {{\xi_3}^8} +
    2 {{\xi_3}^9} )\nonumber\\&&\ \ \ \ \mbox{}
+\left(\frac{f(\xi_1)-1+\frac16\xi_1}{\xi_1^2}\right){{4\xi_1}\over {{{\Delta}^4} \xi_2 \xi_3}}
( 5 {{\xi_1}^6} - 22 {{\xi_1}^5} \xi_2 + 35 {{\xi_1}^4} {{\xi_2}^2} -
    20 {{\xi_1}^3} {{\xi_2}^3}\nonumber\\&&\ \ \ \ \mbox{}
- 5 {{\xi_1}^2} {{\xi_2}^4} + 10 \xi_1 {{\xi_2}^5} -
    3 {{\xi_2}^6} - 27 {{\xi_1}^5} \xi_3 - 97 {{\xi_1}^4} \xi_2 \xi_3 \nonumber\\&&\ \ \ \ \mbox{}-
    166 {{\xi_1}^3} {{\xi_2}^2} \xi_3 + 222 {{\xi_1}^2} {{\xi_2}^3} \xi_3 +
    49 \xi_1 {{\xi_2}^4} \xi_3
+ 19 {{\xi_2}^5} \xi_3 + 36 {{\xi_1}^4} {{\xi_3}^2} \nonumber\\&&\ \ \ \ \mbox{}+
    80 {{\xi_1}^3} \xi_2 {{\xi_3}^2}
- 192 {{\xi_1}^2} {{\xi_2}^2} {{\xi_3}^2} -
    120 \xi_1 {{\xi_2}^3} {{\xi_3}^2} - 44 {{\xi_2}^4} {{\xi_3}^2}\nonumber\\&&\ \ \ \ \mbox{} +
    2 {{\xi_1}^3} {{\xi_3}^3} + 50 {{\xi_1}^2} \xi_2 {{\xi_3}^3} +
    62 \xi_1 {{\xi_2}^2} {{\xi_3}^3} + 46 {{\xi_2}^3} {{\xi_3}^3} \nonumber\\&&\ \ \ \ \mbox{}-
    27 {{\xi_1}^2} {{\xi_3}^4}
- 10 \xi_1 \xi_2 {{\xi_3}^4} - 19 {{\xi_2}^2} {{\xi_3}^4}+
   9 \xi_1 {{\xi_3}^5} - \xi_2 {{\xi_3}^5} + 2 {{\xi_3}^6} )\nonumber\\&&\ \ \ \ \mbox{}
-\left(\frac{f(\xi_3)-1+\frac16\xi_3}{\xi_3^2}\right){1\over {2 {{\Delta}^4} \xi_1 \xi_2}}
( -6 {{\xi_1}^7} + 30 {{\xi_1}^6} \xi_2 - 54 {{\xi_1}^5} {{\xi_2}^2} \nonumber\\&&\ \ \ \ \mbox{}+
    30 {{\xi_1}^4} {{\xi_2}^3}
+ 74 {{\xi_1}^6} \xi_3 - 204 {{\xi_1}^5} \xi_2 \xi_3 +
    150 {{\xi_1}^4} {{\xi_2}^2} \xi_3 - 20 {{\xi_1}^3} {{\xi_2}^3} \xi_3 \nonumber\\&&\ \ \ \ \mbox{}-
    294 {{\xi_1}^5} {{\xi_3}^2} - 558 {{\xi_1}^4} \xi_2 {{\xi_3}^2} +
    852 {{\xi_1}^3} {{\xi_2}^2} {{\xi_3}^2} + 378 {{\xi_1}^4} {{\xi_3}^3} \nonumber\\&&\ \ \ \ \mbox{}+
    728 {{\xi_1}^3} \xi_2 {{\xi_3}^3}
- 1106 {{\xi_1}^2} {{\xi_2}^2} {{\xi_3}^3} -
    34 {{\xi_1}^3} {{\xi_3}^4} - 382 {{\xi_1}^2} \xi_2 {{\xi_3}^4} \nonumber\\&&\ \ \ \ \mbox{}-
    210 {{\xi_1}^2} {{\xi_3}^5}
+ 154 \xi_1 \xi_2 {{\xi_3}^5} + 78 \xi_1 {{\xi_3}^6} +
    7 {{\xi_3}^7} ),
\end{fleqnarray}
\begin{fleqnarray}&&
F_{28}(\xi_1,\xi_2,\xi_3)=F(\xi_1,\xi_2,\xi_3){\frac{16\xi_3}{{\Delta}^{4}}}
 ( -2 {{\xi_1}^4} + 2 {{\xi_1}^2} {{\xi_2}^2} \nonumber\\&&\ \ \ \ \mbox{}+ 4 {{\xi_1}^3} \xi_3 -
    4 {{\xi_1}^2} \xi_2 \xi_3 + 4 \xi_1 \xi_2 {{\xi_3}^2} - 4 \xi_1 {{\xi_3}^3}
+ {{\xi_3}^4} )\nonumber\\&&\ \ \ \ \mbox{}
+\left(F(\xi_1,\xi_2,\xi_3)-\frac12\right){{16}\over {{{\Delta}^3} \xi_1 \xi_2}}
(-4 {{\xi_1}^4} - 4 {{\xi_1}^3} \xi_2 + 8 {{\xi_1}^2} {{\xi_2}^2} +
    10 {{\xi_1}^3} \xi_3 \nonumber\\&&\ \ \ \ \mbox{}-
   18 {{\xi_1}^2} \xi_2 \xi_3 - 6 {{\xi_1}^2} {{\xi_3}^2} +
    12 \xi_1 \xi_2 {{\xi_3}^2} - 2 \xi_1 {{\xi_3}^3} + {{\xi_3}^4})\nonumber\\&&\ \ \ \ \mbox{}
+\left(F(\xi_1,\xi_2,\xi_3)-\frac12+\frac{\xi_1+\xi_2+\xi_3}{24}\right){{64}\over {{{\Delta}^2} \xi_1 \xi_2 \xi_3}}
( -{{\xi_1}^2}+ \xi_1 \xi_2 - \xi_1 \xi_3 + {{\xi_3}^2} )\nonumber\\&&\ \ \ \ \mbox{}
+\left(\frac{f(\xi_1)-1}{\xi_1}\right)\frac{64\xi_1\xi_3}{{\Delta}^{4}}
( -\xi_1 + \xi_2 - \xi_3 )
  ( {{\xi_1}^2} + 2 \xi_1 \xi_2 + {{\xi_2}^2}
- 2 \xi_1 \xi_3 \nonumber\\&&\ \ \ \ \mbox{}- 2 \xi_2 \xi_3 + {{\xi_3}^2})\nonumber\\&&\ \ \ \ \mbox{}
-\left(\frac{f(\xi_3)-1}{\xi_3}\right)
\frac{32{{\xi_3}^2}}{{\Delta}^{4}}
( -2 {{\xi_1}^3} + 2 {{\xi_1}^2} \xi_2
+ 2 {{\xi_1}^2} \xi_3 -2 \xi_1 \xi_2 \xi_3 + 2 \xi_1 {{\xi_3}^2} -{{\xi_3}^3})\nonumber\\&&\ \ \ \ \mbox{}
+\left(\frac{f(\xi_1)-1+\frac16\xi_1}{\xi_1^2}\right){{16\xi_1}\over {{{\Delta}^3} \xi_2 \xi_3}}
( {{\xi_1}^4}
- 4 {{\xi_1}^3} \xi_2 + 6 {{\xi_1}^2} {{\xi_2}^2} -
    4 \xi_1 {{\xi_2}^3} \nonumber\\&&\ \ \ \ \mbox{}
+ {{\xi_2}^4}- 12 {{\xi_1}^3} \xi_3 - 16 {{\xi_1}^2} \xi_2 \xi_3 +
    28 \xi_1 {{\xi_2}^2} \xi_3
+ 18 {{\xi_1}^2} {{\xi_3}^2} - 20 \xi_1 \xi_2 {{\xi_3}^2}\nonumber\\&&\ \ \ \ \mbox{} -
    6 {{\xi_2}^2} {{\xi_3}^2}
- 4 \xi_1 {{\xi_3}^3} + 8 \xi_2 {{\xi_3}^3} - 3 {{\xi_3}^4}
)\nonumber\\&&\ \ \ \ \mbox{}
-\left(\frac{f(\xi_3)-1+\frac16\xi_3}{\xi_3^2}\right){{32{{\xi_3}^2}}\over {{{\Delta}^3} \xi_1 \xi_2}}
  ( -4 {{\xi_1}^3} + 4 {{\xi_1}^2} \xi_2 + 6 {{\xi_1}^2} \xi_3
\nonumber\\&&\ \ \ \ \mbox{}
-  10 \xi_1 \xi_2 \xi_3 - {{\xi_3}^3} ),
\end{fleqnarray}\begin{fleqnarray}&&
F_{29}(\xi_1,\xi_2,\xi_3)= F(\xi_1,\xi_2,\xi_3)\Big[{{8 \xi_1 \xi_2 \xi_3}\over {3 {{\Delta}^6}}}
 ( 6 {{\xi_1}^5} \xi_2 + 3 {{\xi_1}^4} {{\xi_2}^2} -
    12 {{\xi_1}^3} {{\xi_2}^3} + 6 {{\xi_1}^5} \xi_3 \nonumber\\&&\ \ \ \ \mbox{}+
    12 {{\xi_1}^3} {{\xi_2}^2} \xi_3 +
    3 {{\xi_1}^4} {{\xi_3}^2} + 12 {{\xi_1}^3} \xi_2 {{\xi_3}^2} -
    10 {{\xi_1}^2} {{\xi_2}^2} {{\xi_3}^2}
- 18 \xi_1 \xi_2 {{\xi_3}^4} - 3 {{\xi_3}^6}
)\nonumber\\&&\ \ \ \ \mbox{}
+\frac{16}{{\Delta}^{5}}
( 3 {{\xi_1}^5} \xi_2 + 6 {{\xi_1}^4} {{\xi_2}^2}
- 14 {{\xi_1}^3} {{\xi_2}^3} +
    3 {{\xi_1}^5} \xi_3
+ 18 {{\xi_1}^3} {{\xi_2}^2} \xi_3 \nonumber\\&&\ \ \ \ \mbox{}+ 6 {{\xi_1}^4} {{\xi_3}^2} +
    18 {{\xi_1}^3} \xi_2 {{\xi_3}^2}
- 16 {{\xi_1}^2} {{\xi_2}^2} {{\xi_3}^2} -
    21 \xi_1 \xi_2 {{\xi_3}^4} - 2 {{\xi_3}^6} )\Big]\nonumber\\&&\ \ \ \ \mbox{}+
\left(F(\xi_1,\xi_2,\xi_3)-\frac12\right){{16}\over {{{\Delta}^4} \xi_1 \xi_2 \xi_3}}
( 9 {{\xi_1}^4} {{\xi_2}^2} - 16 {{\xi_1}^3} {{\xi_2}^3} +
    30 {{\xi_1}^3} {{\xi_2}^2} \xi_3 \nonumber\\&&\ \ \ \ \mbox{}+ 9 {{\xi_1}^4} {{\xi_3}^2} +
    30 {{\xi_1}^3} \xi_2 {{\xi_3}^2} - 34 {{\xi_1}^2} {{\xi_2}^2} {{\xi_3}^2} -
    30 \xi_1 \xi_2 {{\xi_3}^4} - {{\xi_3}^6} )\nonumber\\&&\ \ \ \ \mbox{}
-\left(F(\xi_1,\xi_2,\xi_3)-\frac12+\frac{\xi_1+\xi_2+\xi_3}{24}\right){{32}\over {{{\Delta}^3} \xi_1 \xi_2 \xi_3}}
(-3 {{\xi_1}^2} \xi_2 - 3 {{\xi_1}^2} \xi_3\nonumber\\&&\ \ \ \ \mbox{}
+ 4 \xi_1 \xi_2 \xi_3 + 3 {{\xi_3}^3} )\nonumber\\&&\ \ \ \ \mbox{}
+f(\xi_1)\frac{16 \xi_1 \xi_2 \xi_3 }{{\Delta}^{6}}
(-{{\xi_1}^5} + {{\xi_1}^4} \xi_2 + 2 {{\xi_1}^3} {{\xi_2}^2} -
    2 {{\xi_1}^2} {{\xi_2}^3}
- \xi_1 {{\xi_2}^4} \nonumber\\&&\ \ \ \ \mbox{}+ {{\xi_2}^5} + {{\xi_1}^4} \xi_3 -
    4 {{\xi_1}^3} \xi_2 \xi_3
+ 2 {{\xi_1}^2} {{\xi_2}^2} \xi_3 + 4 \xi_1 {{\xi_2}^3} \xi_3 -
    3 {{\xi_2}^4} \xi_3 \nonumber\\&&\ \ \ \ \mbox{}
+ 2 {{\xi_1}^3} {{\xi_3}^2} + 2 {{\xi_1}^2} \xi_2 {{\xi_3}^2} -
    6 \xi_1 {{\xi_2}^2} {{\xi_3}^2} + 2 {{\xi_2}^3} {{\xi_3}^2} \nonumber\\&&\ \ \ \ \mbox{}-
    2 {{\xi_1}^2} {{\xi_3}^3}
+ 4 \xi_1 \xi_2 {{\xi_3}^3} + 2 {{\xi_2}^2} {{\xi_3}^3} -
    \xi_1 {{\xi_3}^4} - 3 \xi_2 {{\xi_3}^4} + {{\xi_3}^5} )\nonumber\\&&\ \ \ \ \mbox{}
+\left(\frac{f(\xi_1)-1}{\xi_1}\right)\frac{32 \xi_1}{{\Delta}^{5}}
 ( -2 {{\xi_1}^5} + {{\xi_1}^4} \xi_2 + 7 {{\xi_1}^3} {{\xi_2}^2} -
    7 {{\xi_1}^2} {{\xi_2}^3} \nonumber\\&&\ \ \ \ \mbox{}
- \xi_1 {{\xi_2}^4} + 2 {{\xi_2}^5} + {{\xi_1}^4} \xi_3-
    18 {{\xi_1}^3} \xi_2 \xi_3
+ 7 {{\xi_1}^2} {{\xi_2}^2} \xi_3 + 16 \xi_1 {{\xi_2}^3}\xi_3\nonumber\\&&\ \ \ \ \mbox{} -
    6 {{\xi_2}^4} \xi_3
+ 7 {{\xi_1}^3} {{\xi_3}^2} + 7 {{\xi_1}^2} \xi_2 {{\xi_3}^2} -
    30 \xi_1 {{\xi_2}^2} {{\xi_3}^2} + 4 {{\xi_2}^3} {{\xi_3}^2} \nonumber\\&&\ \ \ \ \mbox{}-
    7 {{\xi_1}^2} {{\xi_3}^3}
+ 16 \xi_1 \xi_2 {{\xi_3}^3} + 4 {{\xi_2}^2} {{\xi_3}^3} -
    \xi_1 {{\xi_3}^4} - 6 \xi_2 {{\xi_3}^4} + 2 {{\xi_3}^5} )\nonumber\\&&\ \ \ \ \mbox{}
+\left(\frac{f(\xi_1)-1+\frac16\xi_1}{\xi_1^2}\right){{24 \xi_1}\over {{{\Delta}^4} \xi_2 \xi_3}}
 ( -{{\xi_1}^5} - 3 {{\xi_1}^4} \xi_2 + 14 {{\xi_1}^3} {{\xi_2}^2} \nonumber\\&&\ \ \ \ \mbox{}-
    14 {{\xi_1}^2} {{\xi_2}^3} + 3 \xi_1 {{\xi_2}^4} + {{\xi_2}^5}
- 3 {{\xi_1}^4} \xi_3 -
    36 {{\xi_1}^3} \xi_2 \xi_3 + 22 {{\xi_1}^2} {{\xi_2}^2} \xi_3 \nonumber\\&&\ \ \ \ \mbox{}+
    20 \xi_1 {{\xi_2}^3} \xi_3 - 3 {{\xi_2}^4} \xi_3 + 14 {{\xi_1}^3} {{\xi_3}^2} +
    22 {{\xi_1}^2} \xi_2 {{\xi_3}^2} - 46 \xi_1 {{\xi_2}^2} {{\xi_3}^2} \nonumber\\&&\ \ \ \ \mbox{}+
    2 {{\xi_2}^3} {{\xi_3}^2}
- 14 {{\xi_1}^2} {{\xi_3}^3} + 20 \xi_1 \xi_2 {{\xi_3}^3} +
    2 {{\xi_2}^2} {{\xi_3}^3}\nonumber\\&&\ \ \ \ \mbox{} + 3 \xi_1 {{\xi_3}^4} - 3 \xi_2 {{\xi_3}^4}
+ {{\xi_3}^5} ).
\end{fleqnarray}
\mathindent = \parindent
\arraycolsep = 5pt

For the derivation of these results see sects. 13--16.
The differential equations for the basic form factors,
and comments on the form of the expressions above will be
found in sect. 16. Another representation for the
third-order form factors is given in sect. 15.

\section{The late-time behaviour of the
trace of the heat kernel}
\setcounter{equation}{0}

\hspace{\parindent}Derivation of the late-time
behaviour of the form factors in the heat kernel was
given in paper II to all orders in the curvature
(see also sect. 15 below). For the basic form factors
(2.9) and (2.75) this behaviour is
\begin{equation} f(-s\Box) = -\frac1s\frac2{\Box}
+{\rm O}\left(\frac1{s^2}\right),\hspace{7mm} s\rightarrow\infty\end{equation}
\begin{equation}
F(-s\Box_1,-s\Box_2,-s\Box_3) =
\frac1{s^2}\Big(\frac1{\Box_1\Box_2}+\frac1{\Box_1\Box_3}+\frac1{\Box_2\Box_3}\Big)
+{\rm O}\left(\frac1{s^3}\right),\hspace{7mm} s\rightarrow\infty.\end{equation}
The late-time behaviour of all second-order and
third-order form factors follows then from the explicit
expressions above
\footnote{\normalsize Another way is to use the
$\alpha$-representation of the form factors in sect. 15,
and eqs. (15.54), (15.55).
}.
With the symmetries (2.46)--(2.74) taken into account,
one obtains
\begin{eqnarray}
f_1(-s\Box) &=&-\frac1s\,\frac1{6\Box} +
{\rm O}\left(\frac1{s^2}\right),
\\[\baselineskip]
f_2(-s\Box) &=&\frac1s\,\frac1{18\Box} +
{\rm O}\left(\frac1{s^2}\right),
\\[\baselineskip]
f_3(-s\Box) &=&\frac1s\,\frac1{3\Box} +
{\rm O}\left(\frac1{s^2}\right),
\\[\baselineskip]
f_4(-s\Box) &=&-\frac1s\,\frac1{\Box} +
{\rm O}\left(\frac1{s^2}\right),
\\[\baselineskip]
f_5(-s\Box) &=&-\frac1s\,\frac1{2\Box} +
{\rm O}\left(\frac1{s^2}\right),
\end{eqnarray}
and
\arraycolsep=0pt
\mathindent=0pt
\begin{fleqnarray}&&
F^{\rm sym}_{1}(-s\Box_1,-s\Box_2,-s\Box_3)=\frac1{s^2}\frac13\left(\frac1{\Box_1\Box_2}+\frac1{\Box_1\Box_3}+\frac1{\Box_2\Box_3}\right)+{\rm O}\left(\frac1{s^3}\right),\\[\baselineskip] &&
F^{\rm sym}_{2}(-s\Box_1,-s\Box_2,-s\Box_3)=-\frac1{s^2}\frac23\left(\frac1{\Box_1\Box_2}+\frac1{\Box_1\Box_3}+\frac1{\Box_2\Box_3}\right)\!+\!{\rm O}\left(\frac1{s^3}\right)\!,\\[\baselineskip] &&
F^{\rm sym}_{3}(-s\Box_1,-s\Box_2,-s\Box_3)={\rm O}\left(\frac1{s^3}\right),\\[\baselineskip] &&
F^{\rm sym}_{4}(-s\Box_1,-s\Box_2,-s\Box_3)=\frac1{s^2}\frac1{36}\left(\frac1{\Box_1\Box_2}+\frac1{\Box_1\Box_3}+\frac1{\Box_2\Box_3}\right)\!+\!{\rm O}\left(\frac1{s^3}\right)\!\!,\\[\baselineskip] &&
F^{\rm sym}_{5}(-s\Box_1,-s\Box_2,-s\Box_3)={\rm O}\left(\frac1{s^3}\right),\\[\baselineskip] &&
F^{\rm sym}_{6}(-s\Box_1,-s\Box_2,-s\Box_3)=-\frac1{s^2}\frac16\left(\frac1{\Box_1\Box_2}+\frac1{\Box_1\Box_3}+\frac1{\Box_2\Box_3}\right)+{\rm O}\left(\frac1{s^3}\right) ,\nonumber\\&&\mbox{}
\\[\baselineskip] &&
F^{\rm sym}_{7}(-s\Box_1,-s\Box_2,-s\Box_3)={\rm O}\left(\frac1{s^3}\right),\\[\baselineskip] &&
F^{\rm sym}_{8}(-s\Box_1,-s\Box_2,-s\Box_3)=\frac1{s^2}\big(\frac1{\Box_1\Box_2}
  +\frac1{\Box_1\Box_3}\big)+{\rm O}\left(\frac1{s^3}\right),\\[\baselineskip] &&
F^{\rm sym}_{9}(-s\Box_1,-s\Box_2,-s\Box_3)=-\frac1{s^2}\frac1{648}\left(\frac1{\Box_1\Box_2}+\frac1{\Box_1\Box_3}+\frac1{\Box_2\Box_3}\right)
\nonumber\\&&\mbox{}
+{\rm O}\left(\frac1{s^3}\right),
\\[\baselineskip] &&
F^{\rm sym}_{10}(-s\Box_1,-s\Box_2,-s\Box_3)={\rm O}\left(\frac1{s^3}\right),\\[\baselineskip] &&
F^{\rm sym}_{11}(-s\Box_1,-s\Box_2,-s\Box_3)=\frac1{s^2}\frac1{12}\big(
  \frac1{\Box_1\Box_2}-\frac1{\Box_1\Box_3}-\frac1{\Box_2\Box_3}\big)
  +{\rm O}\left(\frac1{s^3}\right),\\[\baselineskip] &&
 sF^{\rm sym}_{12}(-s\Box_1,-s\Box_2,-s\Box_3)=-\frac1{s^2}2\frac1{\Box_1\Box_2\Box_3}+{\rm O}\left(\frac1{s^3}\right),\\[\baselineskip] &&
 sF^{\rm sym}_{13}(-s\Box_1,-s\Box_2,-s\Box_3)=-\frac1{s^2}2\frac1{\Box_1\Box_2\Box_3}+{\rm O}\left(\frac1{s^3}\right),\\[\baselineskip] &&
 sF^{\rm sym}_{14}(-s\Box_1,-s\Box_2,-s\Box_3)=-\frac1{s^2}2\frac1{\Box_1\Box_2\Box_3}+{\rm O}\left(\frac1{s^3}\right),\\[\baselineskip] &&
 sF^{\rm sym}_{15}(-s\Box_1,-s\Box_2,-s\Box_3)={\rm O}\left(\frac1{s^3}\right),\\[\baselineskip] &&
 sF^{\rm sym}_{16}(-s\Box_1,-s\Box_2,-s\Box_3)={\rm O}\left(\frac1{s^3}\right),\\[\baselineskip] &&
 sF^{\rm sym}_{17}(-s\Box_1,-s\Box_2,-s\Box_3)={\rm O}\left(\frac1{s^3}\right),\\[\baselineskip] &&
 sF^{\rm sym}_{18}(-s\Box_1,-s\Box_2,-s\Box_3)=\frac1{s^2}2\frac1{\Box_1\Box_2\Box_3}+{\rm O}\left(\frac1{s^3}\right),\\[\baselineskip] &&
 sF^{\rm sym}_{19}(-s\Box_1,-s\Box_2,-s\Box_3)=-\frac1{s^2}\frac1{\Box_1\Box_2\Box_3}+{\rm O}\left(\frac1{s^3}\right),\\[\baselineskip] &&
 sF^{\rm sym}_{20}(-s\Box_1,-s\Box_2,-s\Box_3)=\frac1{s^2}\frac13\frac1{\Box_1\Box_2\Box_3}+{\rm O}\left(\frac1{s^3}\right),\\[\baselineskip] &&
 sF^{\rm sym}_{21}(-s\Box_1,-s\Box_2,-s\Box_3)=\frac1{s^2}4\frac1{\Box_1\Box_2\Box_3}+{\rm O}\left(\frac1{s^3}\right),\\[\baselineskip] &&
 sF^{\rm sym}_{22}(-s\Box_1,-s\Box_2,-s\Box_3)=\frac1{s^2}\frac16\frac1{\Box_1\Box_2\Box_3}+{\rm O}\left(\frac1{s^3}\right),\\[\baselineskip] &&
 sF^{\rm sym}_{23}(-s\Box_1,-s\Box_2,-s\Box_3)=-\frac1{s^2}\frac13\frac1{\Box_1\Box_2\Box_3}+{\rm O}\left(\frac1{s^3}\right),\\[\baselineskip] &&
 sF^{\rm sym}_{24}(-s\Box_1,-s\Box_2,-s\Box_3)=-\frac1{s^2}\frac13\frac1{\Box_1\Box_2\Box_3}+{\rm O}\left(\frac1{s^3}\right),\\[\baselineskip] &&
 sF^{\rm sym}_{25}(-s\Box_1,-s\Box_2,-s\Box_3)={\rm O}\left(\frac1{s^3}\right),\\[\baselineskip] &&
 s^2F^{\rm sym}_{26}(-s\Box_1,-s\Box_2,-s\Box_3)={\rm O}\left(\frac1{s^3}\right),\\[\baselineskip] &&
 s^2F^{\rm sym}_{27}(-s\Box_1,-s\Box_2,-s\Box_3)={\rm O}\left(\frac1{s^3}\right),\\[\baselineskip] &&
 s^2F^{\rm sym}_{28}(-s\Box_1,-s\Box_2,-s\Box_3)={\rm O}\left(\frac1{s^3}\right),\\[\baselineskip] &&
 s^3F^{\rm sym}_{29}(-s\Box_1,-s\Box_2,-s\Box_3)={\rm O}\left(\frac1{s^3}\right).
\end{fleqnarray}
\arraycolsep=5pt
\mathindent=\parindent

The result is that the behaviour of the trace of the
heat kernel at large $s$ is $s^{-\omega+1}$, and the
coefficient of this asymptotic behaviour is obtained
to third order in the curvature
\footnote{\normalsize
As shown in paper II, this behaviour holds at all orders in
the curvature except zeroth.
This power asymptotic behaviour is characteristic of
a non-compact manifold.  For a compact manifold it will
be replaced by the exponential behaviour
\[{\rm Tr} K(s) \propto \exp(-\lambda_{\rm min}s),\hspace{7mm}
s\rightarrow\infty\]
where $\lambda_{\rm min}$ is the minimum eigenvalue of the
operator $(-H)$ in (1.2). By applying the modification of
covariant perturbation theory, appropriate for
compact manifolds, one should be able to obtain this minimum
eigenvalue as a nonlocal expansion in powers of the
curvature (or a deviation of the curvature from the
reference one).
}.
As seen from the expressions above, not all basis structures
contribute to the leading asymptotic behaviour. The
asymptotic form of ${\rm Tr} K(s)$ is as follows:
\begin{eqnarray}
{\rm Tr} K(s) &=&
{\frac{s}{(4\pi s)^\omega}\int\! dx\, g^{1/2}\,{\rm tr}\ }
\Big\{\hat{P}
\nonumber\\&& \hspace{7mm}\mbox{}
-\hat{P}\frac1{\Box}\hat{P}
	-\frac12\hat{\cal R}_{\mu\nu}\frac1{\Box}\hat{\cal R}^{\mu\nu}
	+\frac13\hat{P}\frac1{\Box} R
\nonumber\\&& \hspace{7mm}\mbox{}
	-\frac16R_{\mu\nu}\frac1{\Box} R^{\mu\nu}\hat{1}
	+\frac1{18}R\frac1{\Box} R\hat{1}
\nonumber\\&& \hspace{7mm}\mbox{}
       +{\hat{P}\frac1{\Box}\hat{P}\frac1{\Box}\hat{P}}
       -2{\hat{\cal R}^{\mu}_{\ \alpha}\frac1{\Box}\hat{\cal R}^{\alpha}_{\ \beta}
	\frac1{\Box}\hat{\cal R}^{\beta}_{\ \mu}}
\nonumber\\&& \hspace{7mm}\mbox{}
       +\frac1{36}{\frac1{\Box} R\frac1{\Box} R \hat{P}}
       +\frac1{18}{R\frac1{\Box} R\frac1{\Box} \hat{P}}
\nonumber\\&& \hspace{7mm}\mbox{}
       -\frac16{\frac1{\Box}\hat{P}\frac1{\Box} \hat{P} R}
       -\frac13{\hat{P}\frac1{\Box} \hat{P}\frac1{\Box} R}
\nonumber\\&& \hspace{7mm}\mbox{}
       +2{\frac1{\Box} R^{\alpha\beta}\frac1{\Box}\hat{\cal R}_{\alpha}^{\ \mu}
	\hat{\cal R}_{\beta\mu}}
       -\frac1{216}{R\frac1{\Box} R\frac1{\Box} R\hat{1}}
\nonumber\\&& \hspace{7mm}\mbox{}
       +\frac1{12}{\frac1{\Box} R^{\mu\nu}\frac1{\Box} R_{\mu\nu}R\hat{1}}
       -\frac16{R^{\mu\nu}\frac1{\Box} R_{\mu\nu}\frac1{\Box} R\hat{1}}
\nonumber\\&& \hspace{7mm}\mbox{}
       -2{\frac1{\Box}\hat{\cal R}^{\alpha\beta}\nabla^\mu\frac1{\Box}
	\hat{\cal R}_{\mu\alpha}\nabla^\nu\frac1{\Box}\hat{\cal R}_{\nu\beta}}
       -2{\frac1{\Box}\hat{\cal R}^{\mu\nu} \nabla_\mu\frac1{\Box}\hat{P}\nabla_\nu\frac1{\Box}\hat{P}}
\nonumber\\&& \hspace{7mm}\mbox{}
       -2{\nabla_\mu\frac1{\Box}\hat{\cal R}^{\mu\alpha}
	\nabla^\nu\frac1{\Box}\hat{\cal R}_{\nu\alpha}\frac1{\Box}\hat{P}}
       +2{\frac1{\Box} R_{\alpha\beta}\nabla_\mu
	\frac1{\Box}\hat{\cal R}^{\mu\alpha}\nabla_\nu\frac1{\Box}\hat{\cal R}^{\nu\beta}}
\nonumber\\&& \hspace{7mm}\mbox{}
       -{\frac1{\Box} R^{\alpha\beta}\nabla_\alpha
	\frac1{\Box}\hat{\cal R}^{\mu\nu}\nabla_\beta\frac1{\Box}\hat{\cal R}_{\mu\nu}}
       +\frac13{\frac1{\Box} R\nabla_\alpha\frac1{\Box}\hat{\cal R}^{\alpha\mu}
	\nabla^\beta\frac1{\Box}\hat{\cal R}_{\beta\mu}}
\nonumber\\&& \hspace{7mm}\mbox{}
       +4{\frac1{\Box} R^{\mu\nu}\nabla_\mu\nabla_\lambda
	\frac1{\Box}\hat{\cal R}^{\lambda\alpha}\frac1{\Box}\hat{\cal R}_{\alpha\nu}}
       +\frac16{\frac1{\Box} R^{\alpha\beta}
	\nabla_\alpha\frac1{\Box} R \nabla_\beta\frac1{\Box} R\hat{1}}
\nonumber\\&& \hspace{7mm}\mbox{}
       -\frac13{\nabla^\mu\frac1{\Box} R^{\nu\alpha}
	\nabla_\nu\frac1{\Box} R_{\mu\alpha}\frac1{\Box} R\hat{1}}
       -\frac13{\frac1{\Box} R^{\mu\nu}\nabla_\mu\frac1{\Box}
	R^{\alpha\beta}\nabla_\nu\frac1{\Box} R_{\alpha\beta}\hat{1}}
\nonumber\\&& \hspace{15mm}\mbox{}
 +{\rm O}[\Re^4]\Big\} +
{\rm O}\left(\frac1{s^\omega}\right),\hspace{7mm}
	s\rightarrow\infty.
\end{eqnarray}

\section{The early-time behaviour of the trace
of the heat kernel, and comparison with the Schwinger-DeWitt
expansion}
\setcounter{equation}{0}

\hspace{\parindent}
Derivation of the early-time behaviour of the form factors in
the heat kernel presents no problem. One may use either the
explicit expressions in sect. 2 or the $\alpha$-representation in
sect. 15 and eqs. (15.51), (15.53). For the basic form factors (2.9)
and (2.75) one obtains
\begin{equation}
f(-s\Box)=1 + \frac16 s\Box +\frac{1}{60}s^2 \Box^2
+{\rm O}\left(s^3\right), \hspace{7mm} s\rightarrow 0
\end{equation}
\begin{equation}
F(-s\Box_1,-s\Box_2,-s\Box_3) = \frac12 +\frac{1}{24}s({\Box_1}+{\Box_2}+\Box_3) +{\rm O}\left(s^2\right),
\hspace{7mm} s\rightarrow 0
\end{equation}
and the full table of asymptotic behaviours for the second-order and
third-order form factors is as follows:
\mathindent=0pt
\arraycolsep=0pt
\begin{fleqnarray}&& f_1(-s\Box)= \frac{1}{60} +\frac{1}{840}s\Box
		  +\frac{1}{15120}s^2\Box^2
				      +{\rm O}\left(s^3\right),\end{fleqnarray} \begin{fleqnarray}&&
f_2(-s\Box)=
                      -\frac{1}{180}
                        -\frac{1}{3780}s\Box
				      +{\rm O}\left(s^3\right),\end{fleqnarray} \begin{fleqnarray}&&
f_3(-s\Box)=
                        \frac{1}{180}s\Box
                       +\frac{1}{1260}s^2\Box^2
				      +{\rm O}\left(s^3\right),\end{fleqnarray} \begin{fleqnarray}&&
f_4(-s\Box)=
                         \frac{1}{2}
                   +\frac{1}{12}s\Box
                  +\frac{1}{120}s^2\Box^2
				      +{\rm O}\left(s^3\right),\end{fleqnarray} \begin{fleqnarray}&&
f_5(-s\Box)=
                          \frac{1}{12}
                        +\frac{1}{120}s\Box
                         +\frac{1}{1680}s^2\Box^2
				      +{\rm O}\left(s^3\right),
\end{fleqnarray}
\begin{fleqnarray}&&
F^{\rm sym}_{1}(-s\Box_1,-s\Box_2,-s\Box_3)=
\frac16
                                    +s\Big(
{ { \frac{{\Box_3}}{72}}+ { \frac{{\Box_2}}{72}}+ { \frac{{\Box_1}}{72}}}
                                     \Big)
				      +{\rm O}\left(s^2\right),\end{fleqnarray} \begin{fleqnarray}&&
F^{\rm sym}_{2}(-s\Box_1,-s\Box_2,-s\Box_3)=
-\frac1{45}
                                    +s\Big(
{ {- \frac{{\Box_3}}{1890}} {- \frac{{\Box_2}}{1890}}
{- \frac{{\Box_1}}{1890}}}
                                     \Big)
				      +{\rm O}\left(s^2\right),\end{fleqnarray} \begin{fleqnarray}&&
F^{\rm sym}_{3}(-s\Box_1,-s\Box_2,-s\Box_3)=
\frac1{12}
                                    +s\Big(
{ { \frac{{\Box_1}}{180}}+ { \frac{{\Box_2}}{180}}+ { \frac{{\Box_3}}{90}}}
                                     \Big)
				      +{\rm O}\left(s^2\right),\end{fleqnarray} \begin{fleqnarray}&&
F^{\rm sym}_{4}(-s\Box_1,-s\Box_2,-s\Box_3)=
                                    s\Big(
{ { \frac{{\Box_2}}{15120}} {- \frac{{\Box_3}}{15120}}+
{ \frac{{\Box_1}}{15120}}}
                                     \Big)
				      +{\rm O}\left(s^2\right),\end{fleqnarray} \begin{fleqnarray}&&
F^{\rm sym}_{5}(-s\Box_1,-s\Box_2,-s\Box_3)=
                                     \Big(
 \frac{1}{180}
{ -  \frac{{\Box_3}}{{ 90 {\Box_1}}}}
+  { \frac{{ {{\Box_3}}^{2}}}{{ 180{\Box_1} {\Box_2}}}}
+  { \frac{{\Box_1}}{{ 180 {\Box_2}}}}
+  { \frac{{\Box_2}}{{ 180 {\Box_1}}}}
\nonumber\\&&\ \ \ \ \mbox{}
{ - \frac{{\Box_3}}{{ 90 {\Box_2}}}}
                                     \Big)
                                    +s\Big(
 { \frac{{ {\Box_3}^{3}}}{{ 1120 {\Box_1} {\Box_2}}}}
+  { \frac{{ {\Box_1}^{2}}}{{ 3360 {\Box_2}}}}
+  { \frac{{ {\Box_2}^{2}}}{{ 3360 {\Box_1}}}}
+ { \frac{{\Box_2}}{3360}}
+ { \frac{{\Box_1}}{3360}}
\nonumber\\&&\ \ \ \ \mbox{}
{ -  \frac{{ {\Box_3}^{2}}}{{ 672 {\Box_1}}}}
+  { \frac{{ {\Box_2} {\Box_3}}}{{ 3360{\Box_1}}}}
+  { \frac{{ {\Box_1} {\Box_3}}}{{ 3360 {\Box_2}}}}
{ -  \frac{{ {\Box_3}^{2}}}{{ 672 {\Box_2}}}}
+ { \frac{{\Box_3}}{2520}}
                                     \Big)
				      +{\rm O}\left(s^2\right),\end{fleqnarray} \begin{fleqnarray}&&
F^{\rm sym}_{6}(-s\Box_1,-s\Box_2,-s\Box_3)=
                                    s{ \frac{{\Box_3}}{720}}
				      +{\rm O}\left(s^2\right),\end{fleqnarray} \begin{fleqnarray}&&
F^{\rm sym}_{7}(-s\Box_1,-s\Box_2,-s\Box_3)=
                                    s\Big(
{ { \frac{{ 13 {\Box_1}}}{30240}} {- \frac{{\Box_2}}{30240}}
{- \frac{{\Box_3}}{30240}}}
                                     \Big)
				      +{\rm O}\left(s^2\right),\end{fleqnarray} \begin{fleqnarray}&&
F^{\rm sym}_{8}(-s\Box_1,-s\Box_2,-s\Box_3)=
                                     \Big(
{ \frac1{45} +  { \frac{{\Box_3}}{{ 36 {\Box_1}}}} +
{ \frac{{\Box_2}}{{ 36 {\Box_1}}}}}
                                     \Big)
                                    +s\Big(
 { \frac{{\Box_1}}{840}}+ { \frac{{\Box_2}}{420}}+
{ \frac{{\Box_3}}{420}} \nonumber\\&&\ \ \ \ \mbox{}
+  { \frac{{ {\Box_2} {\Box_3}}}{{ 210 {\Box_1}}}}
+  { \frac{{ {\Box_2}^{2}}}{{ 420 {\Box_1}}}} +  { \frac{{ {\Box_3}
^{2}}}{{ 420 {\Box_1}}}}
                                     \Big)
				      +{\rm O}\left(s^2\right),\end{fleqnarray} \begin{fleqnarray}&&
F^{\rm sym}_{9}(-s\Box_1,-s\Box_2,-s\Box_3)=
                                 \frac1{s}\Big(
{ \frac{{\Box_1}}{1080 {\Box_2} {\Box_3}}}
{+ \frac{{\Box_2}}{1080 {\Box_1} {\Box_3}}}
{+ \frac{{\Box_3}}{1080 {\Box_1} {\Box_2}}}
                                     \Big)
\nonumber\\&&\ \ \ \ \mbox{}
                                     +\Big(
 { \frac{{\Box_3}}{{ 15120 {\Box_1}}}}
{-\frac{1}{3240}}
+  { \frac{{\Box_2}}{{ 15120{\Box_1}}}}
+  { \frac{{\Box_1}}{{ 15120 {\Box_3}}}}
+{ \frac{{ {\Box_2}^{2}}}{{ 10080 {\Box_1} {\Box_3}}}}
\nonumber\\&&\ \ \ \ \mbox{}
+  { \frac{{ {\Box_3}^{2}}}{{ 10080 {\Box_1} {\Box_2}}}}
+  { \frac{{\Box_1}}{{ 15120 {\Box_2}}}}
+  { \frac{{\Box_2}}{{ 15120 {\Box_3}}}}
+  { \frac{{\Box_3}}{{ 15120 {\Box_2}}}}
+  { \frac{{ {\Box_1}^{2}}}{{ 10080 {\Box_2} {\Box_3}}}}
                                     \Big)
\nonumber\\&&\ \ \ \ \mbox{}
                                    +s\Big(
 { \frac{{ {\Box_2}^{2}}}{{ 151200 {\Box_1}}}}
+  { \frac{{ {\Box_1} {\Box_2}}}{{ 151200 {\Box_3}}}}
+  { \frac{{ {\Box_3}^{2}}}{{ 151200 {\Box_2}}}}
+  { \frac{{ {\Box_1}^{2}}}{{ 151200{\Box_2}}}}
+  { \frac{{ {\Box_1} {\Box_3}}}{{ 151200 {\Box_2}}}}
\nonumber\\&&\ \ \ \ \mbox{}
+  { \frac{{ {\Box_2} {\Box_3}}}{{151200 {\Box_1}}}}
+  { \frac{{ {\Box_2}^{2}}}{{ 151200 {\Box_3}}}}
+  { \frac{{ {\Box_2}^{3}}}{{ 151200 {\Box_1} {\Box_3}}}}
+  { \frac{{ {\Box_1}^{3}}}{{ 151200 {\Box_2} {\Box_3}}}}
\nonumber\\&&\ \ \ \ \mbox{}
+  { \frac{{ {\Box_3}^{3}}}{{ 151200 {\Box_1} {\Box_2}}}}
+  { \frac{{ {\Box_1}^{2}}}{{ 151200 {\Box_3}}}}
+  { \frac{{ {\Box_3}^{2}}}{{ 151200 {\Box_1}}}}
{- \frac{{\Box_3}}{75600}}
{- \frac{{\Box_2}}{75600}}
\nonumber\\&&\ \ \ \ \mbox{}
{- \frac{{\Box_1}}{75600}}
                                     \Big)
				      +{\rm O}\left(s^2\right),\end{fleqnarray} \begin{fleqnarray}&&
F^{\rm sym}_{10}(-s\Box_1,-s\Box_2,-s\Box_3)=
                                 \frac1{s}\Big(
 { -  \frac{1}{{ 135 {\Box_1}}}}
{ -  \frac{1}{{ 135 {\Box_3}}}}
{ -  \frac{1}{{135 {\Box_2}}}}
+  { \frac{{\Box_3}}{{ 270 {\Box_1} {\Box_2}}}}
\nonumber\\&&\ \ \ \ \mbox{}
+  { \frac{{\Box_1}}{{ 270 {\Box_2} {\Box_3}}}}
 +  { \frac{{\Box_2}}{{ 270 {\Box_1} {\Box_3}}}}
                                     \Big)
                                     +\Big(
 { \frac{{ {\Box_1}^{2}}}{{ 2520 {\Box_2} {\Box_3}}}}
{ -  \frac{{\Box_3}}{{ 2520 {\Box_2}}}}
{ - \frac{{\Box_2}}{{ 2520 {\Box_1}}}}
\nonumber\\&&\ \ \ \ \mbox{}
+  { \frac{{ {\Box_2}^{2}}}{{ 2520 {\Box_1} {\Box_3}}}}
{ - \frac{{\Box_1}}{{ 2520 {\Box_3}}}}
+  { \frac{{ {\Box_3}^{2}}}{{ 2520 {\Box_1} {\Box_2}}}}
{ - \frac{{\Box_2}}{{ 2520 {\Box_3}}}}
{-\frac{1}{1890}}
{ -  \frac{{\Box_3}}{{ 2520 {\Box_1}}}}
\nonumber\\&&\ \ \ \ \mbox{}
{ -  \frac{{\Box_1}}{{ 2520 {\Box_2}}}}
                                     \Big)
                                    +s\Big(
 { \frac{{ {\Box_1}^{3}}}{{ 37800 {\Box_2} {\Box_3}}}}
+  { \frac{{ {\Box_3}^{3}}}{{ 37800 {\Box_1} {\Box_2}}}}
+  { \frac{{ {\Box_2}^{3}}}{{ 37800 {\Box_1} {\Box_3}}}}
\nonumber\\&&\ \ \ \ \mbox{}
{ -  \frac{{ {\Box_2} {\Box_3}}}{{56700 {\Box_1}}}}
{ -  \frac{{ {\Box_1}^{2}}}{{ 56700 {\Box_2}}}}
{ -  \frac{{ {\Box_2}^{2}}}{{ 56700 {\Box_1}}}}
{ -  \frac{{ {\Box_3}^{2}}}{{ 56700 {\Box_2}}}}
{ -  \frac{{ {\Box_2}^{2}}}{{ 56700 {\Box_3}}}}
\nonumber\\&&\ \ \ \ \mbox{}
{ -  \frac{{ {\Box_1} {\Box_3}}}{{ 56700 {\Box_2}}}}
{ -  \frac{{ {\Box_1}^{2}}}{{ 56700 {\Box_3}}}}
{ -  \frac{{ {\Box_1} {\Box_2}}}{{ 56700 {\Box_3}}}}
{ -  \frac{{ {\Box_3}^{2}}}{{ 56700 {\Box_1}}}}
{- \frac{{\Box_3}}{37800}}
{- \frac{{\Box_2}}{37800}}
\nonumber\\&&\ \ \ \ \mbox{}
{- \frac{{\Box_1}}{37800}}
                                     \Big)
				      +{\rm O}\left(s^2\right),\end{fleqnarray} \begin{fleqnarray}&&
F^{\rm sym}_{11}(-s\Box_1,-s\Box_2,-s\Box_3)=
                                 \frac1{s}\Big(
{ { \frac{1}{{ 180 {\Box_1}}}}
{ -  \frac{{\Box_3}}{{ 180 {\Box_1} {\Box_2}}}}
+  { \frac{1}{{ 180 {\Box_2}}}}}
                                     \Big)
\nonumber\\&&\ \ \ \ \mbox{}
                                     +\Big(
 \frac{1}{15120}
+  { \frac{{\Box_3}}{{ 3780 {\Box_2}}}}
{ -  \frac{{\Box_2}}{{ 30240{\Box_1}}}}
{ -  \frac{{\Box_1}}{{ 30240 {\Box_2}}}}
+  { \frac{{\Box_3}}{{ 3780 {\Box_1}}}}
{ - \frac{{ {\Box_3}^{2}}}{{ 4320 {\Box_1} {\Box_2}}}}
                                     \Big)
\nonumber\\&&\ \ \ \ \mbox{}
                                    +s\Big(
 { \frac{{\Box_3}}{25200}}
{ -  \frac{{ 13 { {\Box_3}^{2}}}}{{ 302400 {\Box_1}}}}
+  { \frac{{ {\Box_2} {\Box_3}}}{{ 37800 {\Box_1}}}}
+ { \frac{{\Box_1}}{302400}}
+ { \frac{{\Box_2}}{302400}}
+  { \frac{{ {\Box_3}^{3}}}{{ 50400 {\Box_1} {\Box_2}}}}
\nonumber\\&&\ \ \ \ \mbox{}
{ -  \frac{{ 13 { {\Box_3}^{2}}}}{{ 302400 {\Box_2}}}}
{ -  \frac{{ {\Box_1}^{2}}}{{ 302400 {\Box_2}}}}
+  { \frac{{ {\Box_1} {\Box_3}}}{{ 37800 {\Box_2}}}}
{ -  \frac{{ {\Box_2}^{2}}}{{ 302400 {\Box_1}}}}
                                     \Big)
				      +{\rm O}\left(s^2\right),\end{fleqnarray} \begin{fleqnarray}&&
sF^{\rm sym}_{12}(-s\Box_1,-s\Box_2,-s\Box_3)=
s\frac{1}{252}
				      +{\rm O}\left(s^2\right),\end{fleqnarray} \begin{fleqnarray}&&
sF^{\rm sym}_{13}(-s\Box_1,-s\Box_2,-s\Box_3)=
s\frac{1}{60}
				      +{\rm O}\left(s^2\right),\end{fleqnarray} \begin{fleqnarray}&&
sF^{\rm sym}_{14}(-s\Box_1,-s\Box_2,-s\Box_3)=
                                    s\frac{1}{180}
				      +{\rm O}\left(s^2\right),\end{fleqnarray} \begin{fleqnarray}&&
sF^{\rm sym}_{15}(-s\Box_1,-s\Box_2,-s\Box_3)=
                                   - s{\frac{1}{1890}}
				      +{\rm O}\left(s^2\right),\end{fleqnarray} \begin{fleqnarray}&&
sF^{\rm sym}_{16}(-s\Box_1,-s\Box_2,-s\Box_3)=
                                     \Big(
{ { \frac{1}{{ 45 {\Box_1}}}}
{ -  \frac{{\Box_3}}{{ 45 {\Box_1} {\Box_2}}}}
+ { \frac{1}{{ 45{\Box_2}}}}}
                                     \Big)
\nonumber\\&&\ \ \ \ \mbox{}
                                    +s\Big(
 { \frac{{\Box_3}}{{ 420 {\Box_1}}}}
 { -  \frac{{ {\Box_3}^{2}}}{{ 280 {\Box_1} {\Box_2}}}}
+  { \frac{{\Box_3}}{{ 420 {\Box_2}}}}
+  { \frac{{\Box_1}}{{ 840 {\Box_2}}}}
+ \frac{1}{630}
+  { \frac{{\Box_2}}{{ 840 {\Box_1}}}}
                                     \Big)
				      +{\rm O}\left(s^2\right),
\nonumber\\
\end{fleqnarray} \begin{fleqnarray}&&
sF^{\rm sym}_{17}(-s\Box_1,-s\Box_2,-s\Box_3)=
                                    s\frac{1}{180}
				      +{\rm O}\left(s^2\right),\end{fleqnarray} \begin{fleqnarray}&&
sF^{\rm sym}_{18}(-s\Box_1,-s\Box_2,-s\Box_3)=
{ \frac{1}{{ 18 {\Box_1}}}}
                                    +s\Big(
{ \frac{1}{1260}
+  { \frac{{\Box_2}}{{ 210 {\Box_1}}}}
+  { \frac{{\Box_3}}{{ 210 {\Box_1}}}}}
                                     \Big)
				      +{\rm O}\left(s^2\right),
\nonumber\\
\end{fleqnarray} \begin{fleqnarray}&&
sF^{\rm sym}_{19}(-s\Box_1,-s\Box_2,-s\Box_3)=
{- \frac{1}{{ 36 {\Box_1}}}}
                                    +s\Big(
{ {-\frac{1}{1260}}
{ -  \frac{{\Box_2}}{{ 420 {\Box_1}}}}
{ -  \frac{{\Box_3}}{{ 420 {\Box_1}}}}}
                                     \Big)
				      \!+\!{\rm O}\left(s^2\right)\!,
\nonumber\\
\end{fleqnarray} \begin{fleqnarray}&&
sF^{\rm sym}_{20}(-s\Box_1,-s\Box_2,-s\Box_3)=
                                    -s{\frac{1}{7560}}
				      +{\rm O}\left(s^2\right),\end{fleqnarray} \begin{fleqnarray}&&
sF^{\rm sym}_{21}(-s\Box_1,-s\Box_2,-s\Box_3)=
{ \frac{1}{{ 9 {\Box_1}}}}
                                    +s\Big(
{ { \frac{{\Box_3}}{{ 105 {\Box_1}}}}
+ \frac{1}{630}
+  { \frac{{\Box_2}}{{ 105 {\Box_1}}}}}
                                     \Big)
				      +{\rm O}\left(s^2\right),
\nonumber\\
\end{fleqnarray} \begin{fleqnarray}&&
sF^{\rm sym}_{22}(-s\Box_1,-s\Box_2,-s\Box_3)=
                                 \frac1{s}\Big(
{ { -  \frac{1}{{ 60 {\Box_2} {\Box_3}}}}
{ -  \frac{1}{{ 180 {\Box_1} {\Box_3}}}}
{ - \frac{1}{{ 180 {\Box_1} {\Box_2}}}}}
                                     \Big)\nonumber\\&&\ \ \ \ \mbox{}
                                     +\Big(
{ { -  \frac{{\Box_2}}{{ 2520 {\Box_1} {\Box_3}}}}
{ -  \frac{{\Box_1}}{{ 630 {\Box_2} {\Box_3}}}}
{ - \frac{1}{{ 2520 {\Box_3}}}}
{ -  \frac{1}{{ 2520 {\Box_2}}}}
+  { \frac{1}{{ 2520 {\Box_1}}}}
{ -  \frac{{\Box_3}}{{ 2520 {\Box_1} {\Box_2}}}}}
                                     \Big)
\nonumber\\&&\ \ \ \ \mbox{}
                                    +s\Big(
 \frac{1}{226800}
{ -  \frac{{\Box_1}}{{ 16800 {\Box_2}}}}
{ -  \frac{{\Box_3}}{{ 50400 {\Box_2}}}}
{ -  \frac{{ {\Box_1}^{2}}}{{ 10080 {\Box_2} {\Box_3}}}}
{ -  \frac{{\Box_2}}{{ 50400{\Box_3}}}}
+  { \frac{{\Box_3}}{{ 50400 {\Box_1}}}}
\nonumber\\&&\ \ \ \ \mbox{}
{ -  \frac{{\Box_1}}{{ 16800 {\Box_3}}}}
+  { \frac{{\Box_2}}{{ 50400 {\Box_1}}}}
{ -  \frac{{ {\Box_3}^{2}}}{{ 50400 {\Box_1} {\Box_2}}}}
{ - \frac{{ {\Box_2}^{2}}}{{ 50400 {\Box_1} {\Box_3}}}}
                                     \Big)
				      +{\rm O}\left(s^2\right),\end{fleqnarray} \begin{fleqnarray}&&
sF^{\rm sym}_{23}(-s\Box_1,-s\Box_2,-s\Box_3)=
                                 \frac1{s}
{ \frac{1}{{ 45 {\Box_1} {\Box_2}}}}
                                     +\Big(
{ { -  \frac{1}{{ 7560 {\Box_1}}}}
{ -  \frac{1}{{ 7560 {\Box_2}}}}
+  { \frac{{\Box_3}}{{ 1080 {\Box_1} {\Box_2}}}}}
                                     \Big)\nonumber\\&&\mbox{}\ \
                                    +s\Big(
{ { -  \frac{{\Box_2}}{{ 75600 {\Box_1}}}}
{ -  \frac{{\Box_1}}{{ 75600 {\Box_2}}}}
+  { \frac{{\Box_3}}{{ 10800 {\Box_2}}}}
+  { \frac{{\Box_3}}{{ 10800 {\Box_1}}}}
{ -  \frac{{ {\Box_3}^{2}}}{{ 12600 {\Box_1} {\Box_2}}}}}
                                     \Big)
				      \!+{\rm O}\left(s^2\right)\!,
\nonumber\\
\end{fleqnarray} \begin{fleqnarray}&&
sF^{\rm sym}_{24}(-s\Box_1,-s\Box_2,-s\Box_3)=
                                 \frac1{s}
{ \frac{1}{{ 45 {\Box_2} {\Box_3}}}}
                                     +\Big(
{ { -  \frac{1}{{ 840 {\Box_3}}}}
{ -  \frac{1}{{ 840 {\Box_2}}}}
+  { \frac{{\Box_1}}{{ 504 {\Box_2} {\Box_3}}}}}
                                     \Big)
\nonumber\\&&\ \ \ \ \mbox{}
                                    +s\Big(
{ {-\frac{1}{9450}}
{ -  \frac{{\Box_3}}{{ 12600 {\Box_2}}}}
{ -  \frac{{\Box_1}}{{25200 {\Box_2}}}}
+  { \frac{{ {\Box_1}^{2}}}{{ 8400 {\Box_2} {\Box_3}}}}
{ -  \frac{{\Box_1}}{{25200 {\Box_3}}}}
{ -  \frac{{\Box_2}}{{ 12600 {\Box_3}}}}}
                                     \Big)
\nonumber\\&&\ \ \ \ \mbox{}
				      +{\rm O}\left(s^2\right),\end{fleqnarray} \begin{fleqnarray}&&
sF^{\rm sym}_{25}(-s\Box_1,-s\Box_2,-s\Box_3)=
                                 \frac1{s}\Big(
{ { \frac{1}{{ 45 {\Box_1} {\Box_2}}}}
{ -  \frac{1}{{ 45 {\Box_2} {\Box_3}}}}
+  { \frac{1}{{ 45 {\Box_1} {\Box_3}}}}}
                                     \Big)
\nonumber\\&&\ \ \ \ \mbox{}
                                     +\Big(
{ { -  \frac{{\Box_1}}{{ 420 {\Box_2} {\Box_3}}}}
{ -  \frac{1}{{ 315 {\Box_1}}}}
+  { \frac{{\Box_3}}{{ 420 {\Box_1} {\Box_2}}}}
+  { \frac{{\Box_2}}{{ 420 {\Box_1} {\Box_3}}}}}
                                     \Big)
\nonumber\\&&\ \ \ \ \mbox{}
                                    +s\Big(
 { \frac{{\Box_3}}{{ 18900 {\Box_2}}}}
{ -  \frac{{\Box_1}}{{ 18900 {\Box_2}}}}
{ -  \frac{{\Box_3}}{{ 6300 {\Box_1}}}}
+  { \frac{{\Box_2}}{{ 18900 {\Box_3}}}}
{ -  \frac{{\Box_1}}{{ 18900 {\Box_3}}}}
{-\frac{1}{3150}}
\nonumber\\&&\ \ \ \ \mbox{}
{ -  \frac{{ {\Box_1}^{2}}}{{ 6300 {\Box_2} {\Box_3}}}}
{ -  \frac{{\Box_2}}{{ 6300 {\Box_1}}}}
+  { \frac{{ {\Box_3}^{2}}}{{ 6300 {\Box_1} {\Box_2}}}}
+  { \frac{{ {\Box_2}^{2}}}{{ 6300 {\Box_1} {\Box_3}}}}
                                     \Big)
				      +{\rm O}\left(s^2\right),\end{fleqnarray} \begin{fleqnarray}&&
s^2 F^{\rm sym}_{26}(-s\Box_1,-s\Box_2,-s\Box_3)=
{ \frac{1}{{ 45 {\Box_1} {\Box_2}}}}
                                    +s\Big(
{ { \frac{1}{{ 840 {\Box_2}}}}
+  { \frac{{\Box_3}}{{ 280 {\Box_1} {\Box_2}}}}
+  { \frac{1}{{ 840 {\Box_1}}}}}
                                     \Big)
\nonumber\\&&\ \ \ \ \mbox{}
				      +{\rm O}\left(s^2\right),\end{fleqnarray} \begin{fleqnarray}&&
s^2 F^{\rm sym}_{27}(-s\Box_1,-s\Box_2,-s\Box_3)=
                                - \frac1{s}
{   \frac{1}{{ 45 {\Box_1} {\Box_2} {\Box_3}}}}
                                     +\Big(
 { -  \frac{1}{{ 504 {\Box_1} {\Box_3}}}}
{ -  \frac{1}{{ 504 {\Box_2} {\Box_3}}}}
\nonumber\\&&\ \ \ \ \mbox{}
{ - \frac{1}{{ 756 {\Box_1} {\Box_2}}}}
                                     \Big)
                                    +s\Big(
 { -  \frac{1}{{ 6300 {\Box_3}}}}
{ -  \frac{{\Box_1}}{{ 8400 {\Box_2} {\Box_3}}}}
{ - \frac{1}{{ 7560 {\Box_2}}}}
{ -  \frac{1}{{ 7560 {\Box_1}}}}
\nonumber\\&&\ \ \ \ \mbox{}
{ -  \frac{{\Box_2}}{{8400 {\Box_1} {\Box_3}}}}
+  { \frac{{\Box_3}}{{ 25200 {\Box_1} {\Box_2}}}}
                                     \Big)
				      +{\rm O}\left(s^2\right),\end{fleqnarray} \begin{fleqnarray}&&
s^2 F^{\rm sym}_{28}(-s\Box_1,-s\Box_2,-s\Box_3)=
                                 -\frac1{s}
{  \frac{2}{{ 45 {\Box_1} {\Box_2} {\Box_3}}}}
                                     +\Big(
 { -  \frac{1}{{ 210 {\Box_2} {\Box_3}}}}
{ -  \frac{1}{{ 315 {\Box_1} {\Box_2}}}}
\nonumber\\&&\ \ \ \ \mbox{}
{ - \frac{1}{{ 210 {\Box_1} {\Box_3}}}}
                                     \Big)
                                    +s\Big(
 { -  \frac{1}{{ 3150 {\Box_1}}}}
{ -  \frac{{\Box_3}}{{ 6300 {\Box_1} {\Box_2}}}}
{ - \frac{1}{{ 3150 {\Box_2}}}}
{ -  \frac{2}{{ 4725 {\Box_3}}}}
\nonumber\\&&\ \ \ \ \mbox{}
{ -  \frac{{\Box_1}}{{3150 {\Box_2} {\Box_3}}}}
{ -  \frac{{\Box_2}}{{ 3150 {\Box_1} {\Box_3}}}}
                                     \Big)
				      +{\rm O}\left(s^2\right),\end{fleqnarray} \begin{fleqnarray}&&
 s^3 F^{\rm sym}_{29}(-s\Box_1,-s\Box_2,-s\Box_3)=
 \frac{1}{{ 1890 {\Box_1} {\Box_2} {\Box_3}}}}
                                    +s\Big(
{ { \frac{1}{{ 18900 {\Box_2} {\Box_3}}}}
+  { \frac{1}{{ 18900 {\Box_1} {\Box_2}}}}
\nonumber\\&&\ \ \ \ \mbox{}
+  { \frac{1}{{ 18900 {\Box_1} {\Box_3}}}}
                                     \Big)
				      +{\rm O}\left(s^2\right).
\end{fleqnarray}
\mathindent = \parindent
\arraycolsep=5pt

It is striking that, in the early-time expansion, the third-order
form factors are still nonlocal and,
for some of them, the expansion
starts with a negative power of $s$.
By making a comparison with the
table of tensor structures in (2.15)--(2.43),
one can see that such a
behaviour is inherent only in the gravitational form factors, and,
moreover, the nonlocal operators $1/\Box$ in the asymptotic
expressions above act only on the gravitational curvatures. As
discussed in paper II, these features will persist at all higher
orders in $\Re$, and the cause is that the
basis set of curvatures for
the heat kernel does not contain the Riemann tensor. In covariant
perturbation theory, the Riemann tensor gets automatically excluded
via the Bianchi identities [2]. Below we show that restoring the
Riemann tensor restores the locality of the early-time expansion.

The early-time expansion of the heat kernel is known as the
Schwinger--DeWitt expansion. For ${\rm Tr} K(s)$ it is of the form [6]
\begin{equation}
{\rm Tr} K(s)=\frac1{(4\pi s)^\omega} \sum_{n=0}^{\infty} s^n \int\! dx\, g^{1/2}\,  {\rm tr}\ \hat{a}_n (x,x)
\end{equation}
where $\hat{a}_n (x,x)$ are the DeWitt coefficients with coincident
arguments. All $\hat{a}_n (x,x)$ are {\em local} functions of the
background fields entering the operator (1.2). There exist
independent methods for obtaining these coefficients, and, for
$n=0,1,2,3,4$, the $\hat{a}_n (x,x)$ have been calculated explicitly
[6,7,12--15]. A comparison with these known expressions, carried out
below, provides a powerful check of the present results.

By inserting the expansions (4.3)--(4.36) in (2.1), one arrives at
eq. (4.37) with the following results for the (integrated) DeWitt
coefficients $a_0$ to $a_4$:
\mathindent = 0pt
\arraycolsep=0pt
\begin{fleqnarray}&&
\int\! dx\, g^{1/2}\,  {\rm tr}\, \hat{a}_0 (x,x)=\int\! dx\, g^{1/2}\,  {\rm tr}\, \hat{1},
\end{fleqnarray}
\begin{fleqnarray}&&
\int\! dx\, g^{1/2}\,  {\rm tr}\, \hat{a}_1 (x,x)=\int\! dx\, g^{1/2}\,  {\rm tr}\, \hat{P},
\end{fleqnarray}

\begin{fleqnarray}&&
\int\! dx\, g^{1/2}\,  {\rm tr}\,\hat{a_2}(x,x)=
\int\! dx\, g^{1/2}\,  {\rm tr}\left\{ \frac12\hat{P}_1\hat{P}_2+
\frac1{12}\hat{\cal R}_{1\mu\nu}\hat{\cal R}_2^{\mu\nu}
\right.\nonumber\\&&\mbox{}
+\frac1{60}R_{1\,\mu\nu} R_2^{\mu\nu}\hat{1}
-\frac1{180}R_1 R_2\hat{1}
+{{\Box_3}\over {360 \Box_1 {\Box_2}}}
R_1 R_2 R_3\hat{1}
\nonumber\\&&\mbox{}
+\Big(-{{1}\over {45 \Box_3}} + {{\Box_3}\over {90 \Box_1 \Box_2}}\Big)
R_{1\,\alpha}^\mu R_{2\,\beta}^{\alpha} R_{3\,\mu}^\beta\hat{1}
+\Big({1\over {90 \Box_2}} - {{\Box_3}\over {180 \Box_1 \Box_2}}\Big)
R_1^{\mu\nu}R_{2\,\mu\nu}R_3\hat{1}
\nonumber\\&&\mbox{}
+\Big(-{{1}\over {90 \Box_1 \Box_2}} - {1\over {60 \Box_2 \Box_3}}\Big)
R_1^{\alpha\beta}\nabla_\alpha R_2 \nabla_\beta R_3\hat{1}
+{1\over {45 \Box_1 \Box_2}}
\nabla^\mu R_1^{\nu\alpha}\nabla_\nu R_{2\,\mu\alpha}R_3\hat{1}
\nonumber\\&&\mbox{}
+{1\over {45 \Box_2 \Box_3}}
R_1^{\mu\nu}\nabla_\mu R_2^{\alpha\beta}\nabla_\nu R_{3\,\alpha\beta}\hat{1}
+\Big({2\over {45 \Box_1 \Box_2}} - {1\over {45 \Box_2 \Box_3}}\Big)
R_1^{\mu\nu}\nabla_\alpha R_{2\,\beta\mu}\nabla^\beta R_{3\,\nu}^\alpha\hat{1}
\nonumber\\&&\mbox{}
-{{1}\over {45 \Box_1 \Box_2 \Box_3}}
\nabla_\alpha\nabla_\beta R_1^{\mu\nu}\nabla_\mu\nabla_\nu R_2^{\alpha\beta} R_3\hat{1}
\nonumber\\&&\mbox{}\left.
-{{2}\over {45 \Box_1 \Box_2 \Box_3}}
\nabla_\mu R_1^{\alpha\lambda} \nabla_\nu R_{2\,\lambda}^\beta\nabla_\alpha\nabla_\beta R_3^{\mu\nu}\hat{1} \right\}
+{\rm O}[\Re^4],
\end{fleqnarray}

\begin{fleqnarray}&&
\int\! dx\, g^{1/2}\,  {\rm tr}\,\hat{a_3}(x,x)=
\int\! dx\, g^{1/2}\, {\rm tr} \left\{\frac{\Box_2}{12}\hat{P}_1\hat{P}_2
+\frac{\Box_2}{120}\hat{\cal R}_{1\mu\nu}\hat{\cal R}_2^{\mu\nu}+\frac{\Box_2}{180}\hat{P}_1 R_2
\right.\nonumber\\&&\mbox{}
+\frac{\Box_2}{840}R_{1\,\mu\nu} R_2^{\mu\nu}\hat{1}-\frac{\Box_2}{3780}R_1 R_2\hat{1}
+{1\over 6}
\hat{P}_1\hat{P}_2\hat{P}_3
-{1\over {45}}
\hat{\cal R}^{\ \mu}_{1\ \alpha}\hat{\cal R}^{\ \alpha}_{2\ \beta}\hat{\cal R}^{\ \beta}_{3\ \mu}
\nonumber\\&&\mbox{}
+{1\over {12}}
\hat{\cal R}^{\mu\nu}_1\hat{\cal R}_{2\,\mu\nu}\hat{P}_3
+\Big({1\over {180}} + {{\Box_1}\over {90 \Box_2}}
- {{\Box_3}\over {45 \Box_2}} +
  {{{{\Box_3}^2}}\over {180 \Box_1 \Box_2}}\Big)
  R_1^{\mu\nu}R_{2\,\mu\nu}\hat{P}_3
\nonumber\\&&\mbox{}
+\Big({1\over {45}} + {{\Box_3}\over {18 \Box_1}}\Big)
R_1^{\alpha\beta}\hat{\cal R}_{2\,\alpha}^{\ \ \ \,\mu}\hat{\cal R}_{3\,\beta\mu}
+\Big(-{1\over {3240}} + {{\Box_1}\over {2520 \Box_3}} +
  {{{{\Box_3}^2}}\over {3360 \Box_1 \Box_2}}\Big)
R_1 R_2 R_3\hat{1}
\nonumber\\&&\mbox{}
+\Big(-{1\over {1890}} - {{\Box_1}\over {420 \Box_3}} +
  {{{{\Box_3}^2}}\over {840 \Box_1 \Box_2}}\Big)
R_{1\,\alpha}^\mu R_{2\,\beta}^{\alpha} R_{3\,\mu}^\beta\hat{1}
\nonumber\\&&\mbox{}
+\Big({1\over {15120}} - {{\Box_1}\over {15120 \Box_2}}
+ {{\Box_3}\over {1890 \Box_2}} -
  {{{{\Box_3}^2}}\over {4320 \Box_1 \Box_2}}\Big)
R_1^{\mu\nu}R_{2\,\mu\nu}R_3\hat{1}
\nonumber\\&&\mbox{}
+\Big({2\over {45 \Box_2}} - {{\Box_3}\over {45 \Box_1 \Box_2}}\Big)
\nabla^\mu R_1^{\nu\alpha}\nabla_\nu R_{2\,\mu\alpha}\hat{P}_3
+{1\over {18 \Box_1}}
R_{1\,\alpha\beta}\nabla_\mu\hat{\cal R}_2^{\mu\alpha}\nabla_\nu\hat{\cal R}_3^{\nu\beta}
\nonumber\\&&\mbox{}
-{{1}\over {36 \Box_1}}
R_1^{\alpha\beta}\nabla_\alpha\hat{\cal R}_2^{\mu\nu}\nabla_\beta\hat{\cal R}_{3\,\mu\nu}
+{1\over {9 \Box_1}}
R_1^{\mu\nu}\nabla_\mu\nabla_\lambda\hat{\cal R}_2^{\lambda\alpha}\hat{\cal R}_{3\,\alpha\nu}
\nonumber\\&&\mbox{}
+\Big({1\over {2520 \Box_1}} - {1\over {1260 \Box_2}}
- {{\Box_1}\over {630 \Box_2 \Box_3}} -
  {{\Box_3}\over {1260 \Box_1 \Box_2}}\Big)
R_1^{\alpha\beta}\nabla_\alpha R_2 \nabla_\beta R_3\hat{1}
\nonumber\\&&\mbox{}
+\Big(-{{1}\over {3780 \Box_2}} + {{\Box_3}\over {1080 \Box_1 \Box_2}}\Big)
\nabla^\mu R_1^{\nu\alpha}\nabla_\nu R_{2\,\mu\alpha}R_3\hat{1}
\nonumber\\&&\mbox{}
+\Big(-{{1}\over {420 \Box_2}} + {{\Box_1}\over {504 \Box_2 \Box_3}}\Big)
R_1^{\mu\nu}\nabla_\mu R_2^{\alpha\beta}\nabla_\nu R_{3\,\alpha\beta}\hat{1}
\nonumber\\&&\mbox{}
+\Big(-{{1}\over {315 \Box_1}} - {{\Box_1}\over {420 \Box_2 \Box_3}} +
  {{\Box_3}\over {210 \Box_1 \Box_2}}\Big)
R_1^{\mu\nu}\nabla_\alpha R_{2\,\beta\mu}\nabla^\beta R_{3\,\nu}^\alpha\hat{1}
\nonumber\\&&\mbox{}
+{1\over {45 \Box_1 \Box_2}}
\nabla_\alpha\nabla_\beta R_1^{\mu\nu}\nabla_\mu\nabla_\nu R_2^{\alpha\beta}\hat{P}_3
\nonumber\\&&\mbox{}
+\Big(-{{1}\over {756 \Box_1 \Box_2}} - {1\over {252 \Box_2 \Box_3}}\Big)
\nabla_\alpha\nabla_\beta R_1^{\mu\nu}\nabla_\mu\nabla_\nu R_2^{\alpha\beta} R_3\hat{1}
\nonumber\\&&\mbox{}
+\Big(-{{1}\over {315 \Box_1 \Box_2}} - {1\over {105 \Box_2 \Box_3}}\Big)
\nabla_\mu R_1^{\alpha\lambda} \nabla_\nu R_{2\,\lambda}^\beta\nabla_\alpha\nabla_\beta R_3^{\mu\nu}\hat{1}
\nonumber\\&&\mbox{}\left.
+{1\over {1890 \Box_1 \Box_2 \Box_3}}
 \nabla_\lambda\nabla_\sigma R_1^{\alpha\beta}\nabla_\alpha\nabla_\beta R_2^{\mu\nu}\nabla_\mu\nabla_\nu R_3^{\lambda\sigma}\hat{1} \right\}
+{\rm O}[\Re^4],
\end{fleqnarray}

\begin{fleqnarray}&&
\int\! dx\, g^{1/2}\,  {\rm tr}\,\hat{a_4}(x,x)=
\int\! dx\, g^{1/2}\, {\rm tr}\left.\{\frac{{\Box_2}^2}{120}\hat{P}_1\hat{P}_2
+\frac{{\Box_2}^2}{1260}\hat{P}_1 R_2+\frac{{\Box_2}^2}{1680}\hat{\cal R}_{1\mu\nu}\hat{\cal R}_2^{\mu\nu}
\right.\nonumber\\&&\mbox{}
+\frac{{\Box_2}^2}{15120}R_{1\,\mu\nu} R_2^{\mu\nu}\hat{1}
+{{\Box_3}\over {24}}
\hat{P}_1\hat{P}_2\hat{P}_3
-{{\Box_3}\over {630}}
\hat{\cal R}^{\ \mu}_{1\ \alpha}\hat{\cal R}^{\ \alpha}_{2\ \beta}\hat{\cal R}^{\ \beta}_{3\ \mu}
\nonumber\\&&\mbox{}
+\Big({{\Box_1}\over {180}} + {{\Box_2}\over {180}}
+ {{\Box_3}\over {90}}\Big)
\hat{\cal R}^{\mu\nu}_1\hat{\cal R}_{2\,\mu\nu}\hat{P}_3
+\Big({{\Box_1}\over {7560}} - {{\Box_3}\over {15120}}\Big)
R_1 R_2 \hat{P}_3
\nonumber\\&&\mbox{}
+\Big({{\Box_1}\over {1680}} + {{{{\Box_1}^2}}\over {1680 \Box_2}} +
  {{\Box_3}\over {2520}} + {{\Box_1 \Box_3}\over {1680 \Box_2}} -
  {{{{\Box_3}^2}}\over {336 \Box_2}}
+ {{{{\Box_3}^3}}\over {1120 \Box_1 \Box_2}}\Big)
R_1^{\mu\nu}R_{2\,\mu\nu}\hat{P}_3
\nonumber\\&&\mbox{}
+{{\Box_3}\over {720}}
\hat{P}_1\hat{P}_2 R_3
+\Big({{13 \Box_1}\over {30240}} - {{\Box_3}\over {15120}}\Big)
R_1\hat{\cal R}^{\mu\nu}_2\hat{\cal R}_{3\,\mu\nu}
\nonumber\\&&\mbox{}
+\Big({{\Box_1}\over {840}} + {{\Box_3}\over {210}}
+ {{\Box_2 \Box_3}\over {210 \Box_1}} +
  {{{{\Box_3}^2}}\over {210 \Box_1}}\Big)
R_1^{\alpha\beta}\hat{\cal R}_{2\,\alpha}^{\ \ \ \,\mu}\hat{\cal R}_{3\,\beta\mu}
\nonumber\\&&\mbox{}
+\Big({{{{\Box_1}^2}}\over {25200 \Box_3}}
+ {{\Box_1 \Box_2}\over {50400 \Box_3}} -
  {{\Box_3}\over {25200}}
+ {{{{\Box_3}^3}}\over {50400 \Box_1 \Box_2}}\Big)
R_1 R_2 R_3\hat{1}
\nonumber\\&&\mbox{}
+\Big(-{{{{\Box_1}^2}}\over {9450 \Box_3}}
- {{\Box_1 \Box_2}\over {18900 \Box_3}} -
  {{\Box_3}\over {12600}}
+ {{{{\Box_3}^3}}\over {12600 \Box_1 \Box_2}}\Big)
R_{1\,\alpha}^\mu R_{2\,\beta}^{\alpha} R_{3\,\mu}^\beta\hat{1}
\nonumber\\&&\mbox{}
+\Big({{\Box_1}\over {151200}}
- {{{{\Box_1}^2}}\over {151200 \Box_2}} +
  {{\Box_3}\over {25200}} + {{\Box_1 \Box_3}\over {18900 \Box_2}} -
  {{13 {{\Box_3}^2}}\over {151200 \Box_2}}
\nonumber\\&&\mbox{}
+ {{{{\Box_3}^3}}\over {50400 \Box_1 \Box_2}}\Big)
R_1^{\mu\nu}R_{2\,\mu\nu}R_3\hat{1}
+{1\over {252}}
\hat{\cal R}_1^{\alpha\beta}\nabla^\mu\hat{\cal R}_{2\mu\alpha}\nabla^\nu\hat{\cal R}_{3\nu\beta}
\nonumber\\&&\mbox{}
+{1\over {60}}
 \hat{\cal R}_1^{\mu\nu} \nabla_\mu\hat{P}_2\nabla_\nu\hat{P}_3
+{1\over {180}}
 \nabla_\mu \hat{\cal R}_1^{\mu\alpha}\nabla^\nu \hat{\cal R}_{2\,\nu\alpha}\hat{P}_3
-{1\over {1890}}
 R_1^{\mu\nu}\nabla_\mu R_2\nabla_\nu \hat{P}_3
\nonumber\\&&\mbox{}
+\Big({1\over {630}} + {{\Box_1}\over {420 \Box_2}}
+ {{\Box_3}\over {210 \Box_2}} -
  {{{{\Box_3}^2}}\over {280 \Box_1 \Box_2}}\Big)
\nabla^\mu R_1^{\nu\alpha}\nabla_\nu R_{2\,\mu\alpha}\hat{P}_3
\nonumber\\&&\mbox{}
+{1\over {180}}
R_1^{\mu\nu}\nabla_\mu\nabla_\nu\hat{P}_2\hat{P}_3
+\Big({1\over {1260}} + {{\Box_3}\over {105 \Box_1}}\Big)
R_{1\,\alpha\beta}\nabla_\mu\hat{\cal R}_2^{\mu\alpha}\nabla_\nu\hat{\cal R}_3^{\nu\beta}
\nonumber\\&&\mbox{}
+ \Big(-{1\over {1260}} - {{\Box_3}\over {210 \Box_1}}\Big)
R_1^{\alpha\beta}\nabla_\alpha\hat{\cal R}_2^{\mu\nu}\nabla_\beta\hat{\cal R}_{3\,\mu\nu}
-{1\over {7560}}
R_1\nabla_\alpha\hat{\cal R}_2^{\alpha\mu}\nabla^\beta\hat{\cal R}_{3\,\beta\mu}
\nonumber\\&&\mbox{}
+\Big({1\over {630}} + {\Box_2\over {105 \Box_1}}
+ {\Box_3\over {105 \Box_1}}\Big)
R_1^{\mu\nu}\nabla_\mu\nabla_\lambda\hat{\cal R}_2^{\lambda\alpha}\hat{\cal R}_{3\,\alpha\nu}
+ \Big({1\over {226800}} - {{\Box_1}\over {8400 \Box_2}}
\nonumber\\&&\mbox{}
-  {{{{\Box_1}^2}}\over {10080 \Box_2 \Box_3}}
+ {{\Box_3}\over {25200 \Box_1}} -
  {{\Box_3}\over {25200 \Box_2}}
- {{{{\Box_3}^2}}\over {25200 \Box_1 \Box_2}}\Big)
R_1^{\alpha\beta}\nabla_\alpha R_2 \nabla_\beta R_3\hat{1}
\nonumber\\&&\mbox{}
+\Big(-{{\Box_1}\over {37800 \Box_2}} + {{\Box_3}\over {5400 \Box_2}} -
  {{{{\Box_3}^2}}\over {12600 \Box_1 \Box_2}}\Big)
\nabla^\mu R_1^{\nu\alpha}\nabla_\nu R_{2\,\mu\alpha}R_3\hat{1}
\nonumber\\&&\mbox{}
+\Big(-{1\over {9450}} - {{\Box_1}\over {12600 \Box_2}} +
  {{{{\Box_1}^2}}\over {8400 \Box_2 \Box_3}}
- {{\Box_3}\over {6300 \Box_2}}\Big)
R_1^{\mu\nu}\nabla_\mu R_2^{\alpha\beta}\nabla_\nu R_{3\,\alpha\beta}\hat{1}
\nonumber\\&&\mbox{}
+\Big(-{1\over {3150}} - {{\Box_1}\over {9450 \Box_2}} -
  {{{{\Box_1}^2}}\over {6300 \Box_2 \Box_3}}
- {{\Box_3}\over {3150 \Box_1}} +
  {{\Box_3}\over {9450 \Box_2}}
+ {{{{\Box_3}^2}}\over {3150 \Box_1 \Box_2}}\Big)
\nonumber\\&&\mbox{}\times
R_1^{\mu\nu}\nabla_\alpha R_{2\,\beta\mu}\nabla^\beta R_{3\,\nu}^\alpha\hat{1}
+\Big({1\over {420 \Box_2}} + {{\Box_3}\over {280 \Box_1 \Box_2}}\Big)
\nabla_\alpha\nabla_\beta R_1^{\mu\nu}\nabla_\mu\nabla_\nu R_2^{\alpha\beta}\hat{P}_3
\nonumber\\&&\mbox{}
+\Big(-{{1}\over {3780 \Box_2}} - {1\over {6300 \Box_3}} -
  {{\Box_1}\over {4200 \Box_2 \Box_3}}
+ {{\Box_3}\over {25200 \Box_1 \Box_2}}\Big)
\nabla_\alpha\nabla_\beta R_1^{\mu\nu}\nabla_\mu\nabla_\nu R_2^{\alpha\beta} R_3\hat{1}
\nonumber\\&&\mbox{}
+\Big(-{{1}\over {1575 \Box_2}} - {2\over {4725 \Box_3}}
-  {{\Box_1}\over {1575 \Box_2 \Box_3}}
- {{\Box_3}\over {6300 \Box_1 \Box_2}}\Big)
\nabla_\mu R_1^{\alpha\lambda} \nabla_\nu R_{2\,\lambda}^\beta\nabla_\alpha\nabla_\beta R_3^{\mu\nu}\hat{1}
\nonumber\\&&\mbox{}\left.
+{1\over {6300 \Box_1 \Box_2}}
 \nabla_\lambda\nabla_\sigma R_1^{\alpha\beta}\nabla_\alpha\nabla_\beta R_2^{\mu\nu}\nabla_\mu\nabla_\nu R_3^{\lambda\sigma}\hat{1} \right\}
+{\rm O}[\Re^4].
\end{fleqnarray}
\mathindent = \parindent
\arraycolsep=5pt

The task is now to bring expressions (4.40)--(4.42) to a local form
by restoring the Riemann tensor. The expression for the Riemann
tensor solving the Bianchi identities to second order in the Ricci
tensor is given in Appendix (eq. (A.32)). The procedure that we use
is as follows. For each $a_n$, we first consider
a linear combination
of all possible local invariants of the appropriate dimension with
unknown coefficients. Next, in this combination, we exclude the
Riemann tensor, and equate the result to the nonlocal expression
above. This gives a set of equations for the coefficients,
which, in
each case, has a {\em unique} solution. In the case of $a_2$,
there is
only one local invariant with explicit participation of the Riemann
tensor:
\begin{equation}
\int\! dx\, g^{1/2}\,
R_{\alpha\beta\mu\nu}
R^{\alpha\beta\mu\nu}.
\end{equation}
In the case of $a_3$, there are seven (the integral over space-time
is assumed):
\begin{equation}
\begin{array}{ll}
{\rm tr} {\hat{P} R^{\mu\nu\alpha\beta}
          R_{\mu\nu\alpha\beta}},
& \hspace{7mm}
  {\rm tr}  {\hat{\cal R}^{\alpha\beta}\hat{\cal R}^{\mu\nu}
               R_{\alpha\beta\mu\nu}} ,
\\[3mm]
               {R^{\alpha\beta}_{\ \ \mu\nu}
                     R^{\mu\nu}_{\ \ \sigma\rho}
                         R^{\sigma\rho}_{\ \ \alpha\beta}},
& \hspace{7mm}
		{R^{\alpha\ \beta}_{\ \mu \ \nu}
                     R^{\mu \ \nu}_{\ \sigma \ \rho}
                         R^{\sigma \ \rho}_{\ \alpha \ \beta}},
\\[3mm]
                {R^{\alpha}_{\beta}
                     R_{\alpha\mu\nu\sigma}
                         R^{\beta\mu\nu\sigma}},
& \hspace{7mm}
  {R R^{\mu\nu\alpha\beta}
          R_{\mu\nu\alpha\beta} },

\\[3mm]
                {R^{\alpha\mu}
                    R^{\beta\nu}
                         R_{\alpha\beta\mu\nu}},
\end{array}
\end{equation}
and the coefficient of the sixth turns out to be zero. In the case of
$a_4$, there are ten (counting only cubic):

\begin{equation}
\begin{array}{ll}
{\rm tr}    {\Box\hat{P}
                 R^{\alpha\beta\mu\nu}
                            R_{\alpha\beta\mu\nu}},
& \hspace{7mm}
{\rm tr}    {\hat{P}
                 \nabla_\mu\nabla_\alpha R_{\nu\beta}
                         R^{\mu\nu\alpha\beta}},
\\[3mm]
{\rm tr}    { \hat{\cal R}^{\alpha\beta}
                   \Box
                    \hat{\cal R}^{\mu\nu}
                           R_{\alpha\beta\mu\nu}},
& \hspace{7mm}
               {\Box R^{\alpha}_{\beta}
                         R_{\alpha\mu\nu\sigma}
                                 R^{\beta\mu\nu\sigma}},
\\[3mm]
    {\Box R
        R^{\mu\nu\alpha\beta}
          R_{\mu\nu\alpha\beta}},
& \hspace{7mm}
               {R_{\mu\nu}
                       \nabla^\mu R_{\alpha\beta\sigma\rho}
                              \nabla^\nu R^{\alpha\beta\sigma\rho}},
\\[3mm]
    {R
	\nabla_\mu\nabla_\alpha R_{\nu\beta}
                     R^{\mu\nu\alpha\beta}},
& \hspace{7mm}
    {\nabla_\mu R_{\nu\alpha}
		      \nabla^\alpha R_{\rho\sigma}
                            R^{\mu\rho\nu\sigma}},
\\[3mm]
    { \nabla_\alpha R_{\beta\lambda}
		      \nabla_\mu R_{\nu}^{\lambda}
                                  R^{\alpha\beta\mu\nu}},
& \hspace{7mm}
    { R^{\alpha\mu}
	    \Box
		 R^{\beta\nu}
                      R_{\alpha\beta\mu\nu}},
\end{array}
\end{equation}
and the last one proves to be absent. The number of invariants with
the Riemann tensor does not grow fast owing to the Bianchi identities
and, particularly, their corollary
\begin{eqnarray}&&
\Box
  R^{\alpha\beta\mu\nu}\equiv
       \frac12\Big(\nabla^\mu \nabla^\alpha R^{\nu\beta}
+\nabla^\alpha \nabla^\mu R^{\nu\beta}
-\nabla^\nu \nabla^\alpha R^{\mu\beta}
-\nabla^\alpha \nabla^\nu R^{\mu\beta}
\nonumber\\&& \hspace{20mm}\mbox{}
-\nabla^\mu \nabla^\beta R^{\nu\alpha}
-\nabla^\beta \nabla^\mu R^{\nu\alpha}
+\nabla^\nu \nabla^\beta R^{\mu\alpha}
+\nabla^\beta \nabla^\nu R^{\mu\alpha}
\Big)
\nonumber\\&& \hspace{20mm}\mbox{}
+ R^{[\mu}_{\lambda}
  R^{\nu] \lambda\beta\alpha}
- R^{[\alpha}_{\lambda}
  R^{\beta] \lambda\mu\nu}
\nonumber\\&& \hspace{20mm}\mbox{}
-4  R^{\alpha\ [\mu}_{\  \sigma \ \ \lambda}
  R^{\nu] \lambda\beta\sigma}
-  R^{\alpha\beta}_{\ \ \ \sigma\lambda}
  R^{\mu\nu\sigma\lambda}
\end{eqnarray}
which excludes
$\Box  R_{\alpha\beta\mu\nu}$ in a local way
\footnote{ \normalsize
On the bases of local and nonlocal invariants see Appendix.}.

The final results are as follows. The expressions (4.38) and (4.39)
are already in the local form. The expression (4.40) is brought to a
local form  by using eq.(A.38) of Appendix:
\begin{fleqnarray}&&
\int\! dx\, g^{1/2}\,  {\rm tr}\,\hat{a_2}(x,x)=\int\! dx\, g^{1/2}\,  {\rm tr}\left\{ \frac12 \hat{P}\hat{P}+
\frac1{12}\hat{\cal R}_{\mu\nu}\hat{\cal R}^{\mu\nu}\right.\nonumber\\&&\ \ \ \ \ \ \ \ \mbox{}\left.
+\left[\frac1{180} R_{\alpha\beta\mu\nu}R^{\alpha\beta\mu\nu}
-\frac1{180} R_{\mu\nu}R^{\mu\nu}\right]\hat{1}\right\}
+{\rm O}[\Re^4],
\end{fleqnarray}
and the expressions (4.41), (4.42) rewritten in terms of invariants
(4.44) and (4.45) take the form
\begin{fleqnarray}&&
\int\! dx\, g^{1/2}\,  {\rm tr}\,\hat{a_3}(x,x)=\int\! dx\, g^{1/2}\, {\rm tr} \left\{\frac{1}{12}\hat{P}\Box\hat{P}
+\frac{1}{120}\hat{\cal R}_{\mu\nu}\Box\hat{\cal R}^{\mu\nu}\right.
\nonumber\\&&\ \ \ \ \ \ \ \ \mbox{}
+\frac{1}{180}\hat{P}\Box R
+\left[\frac{1}{840}R_{\mu\nu}\Box R^{\mu\nu}
-\frac{1}{3780}R\Box R\right]\hat{1}
\nonumber\\&&\ \ \ \ \ \ \ \ \mbox{}
+{1\over 6}\hat{P}\hat{P}\hat{P}
-{1\over {45}}
  	{\hat{\cal R}^{\mu}_{\ \alpha}
  		\hat{\cal R}^{\alpha}_{\ \beta}\hat{\cal R}^{\beta}_{\ \mu}}
\nonumber\\&&\ \ \ \ \ \ \ \ \mbox{}
+{1\over {12}}
       {\hat{P}\hat{\cal R}^{\alpha\beta}\hat{\cal R}_{\alpha\beta}}
+{1\over {72}}
    {R^{\mu\nu}_{\ \ \alpha\beta}
          \hat{\cal R}^{\alpha\beta}\hat{\cal R}_{\mu\nu}}
\nonumber\\&&\ \ \ \ \ \ \ \ \mbox{}
-{1\over {180}}
      {R^{\mu\nu}\hat{\cal R}^{\alpha}_{\ \mu}\hat{\cal R}_{\alpha\nu}}
+{1\over {180}}
    {R^{\mu\nu\alpha\beta}
          R_{\mu\nu\alpha\beta}\hat{P}}
\nonumber\\&&\ \ \ \ \ \ \ \ \mbox{}
-{1\over {180}}
    {R^{\alpha\beta}
          R_{\alpha\beta}\hat{P}}
+\left[-\frac{1}{1620}
		{R^{\alpha\ \beta}_{\ \mu \ \nu}
                     R^{\mu \ \nu}_{\ \sigma \ \rho}
                         R^{\sigma \ \rho}_{\ \alpha \ \beta}}
\right.
\nonumber\\&&\ \ \ \ \ \ \ \ \mbox{}
+\frac{17}{45360}
               {R^{\alpha\beta}_{\ \ \mu\nu}
                     R^{\mu\nu}_{\ \ \sigma\rho}
                         R^{\sigma\rho}_{\ \ \alpha\beta}}
+\frac{1}{7560}
               {R_{\alpha\beta}
                     R_{\ \mu\nu\lambda}^{\alpha}
                         R^{\beta\mu\nu\lambda}}
\nonumber\\&&\ \ \ \ \ \ \ \ \mbox{}\left.\left.
+\frac{1}{945}
               {R_{\alpha\beta}
                     R^{\mu\nu}
                         R^{\alpha\ \beta}_{\ \mu \ \nu}}
-\frac{4}{2835}
               {R^{\alpha}_{\beta}
                     R^{\beta}_{\mu}
                         R^{\mu}_{\alpha}}
\right]\hat{1}\right\}
+{\rm O}[\Re^4],
\end{fleqnarray}
\begin{fleqnarray}&&
\int\! dx\, g^{1/2}\,  {\rm tr}\,\hat{a_4}(x,x)=\int\! dx\, g^{1/2}\, {\rm tr}\left\{\frac1{120}\hat{P}{\Box^2}\hat{P}
+\frac1{1260}\hat{P}{\Box^2}R
\right.
\nonumber\\&&\ \ \ \ \ \ \ \ \mbox{}
+\frac1{1680}\hat{\cal R}^{\mu\nu}{\Box^2}\hat{\cal R}_{\mu\nu}
+\frac1{15120}R^{\mu\nu}{\Box^2}R_{\mu\nu}\hat{1}
\nonumber\\&&\ \ \ \ \ \ \ \ \mbox{}
+\frac1{24}\Box\hat{P}\hat{P}\hat{P}
-\frac1{630}\Box
  	{\hat{\cal R}^{\mu}_{\ \alpha}
  		\hat{\cal R}^{\alpha}_{\ \beta}\hat{\cal R}^{\beta}_{\ \mu}}
\nonumber\\&&\ \ \ \ \ \ \ \ \mbox{}
+\frac1{252}
	{\hat{\cal R}^{\alpha\beta}
  		\nabla^\mu\hat{\cal R}_{\mu\alpha}\nabla^\nu\hat{\cal R}_{\nu\beta}}
+\frac1{180}
	{\Box\hat{\cal R}^{\mu\nu}\hat{\cal R}_{\mu\nu}\hat{P}}
\nonumber\\&&\ \ \ \ \ \ \ \ \mbox{}
+\frac1{180}
	{\hat{\cal R}^{\mu\nu}\Box\hat{\cal R}_{\mu\nu}\hat{P}}
+\frac1{90}
	{\hat{\cal R}^{\mu\nu}\hat{\cal R}_{\mu\nu}\Box\hat{P}}
\nonumber\\&&\ \ \ \ \ \ \ \ \mbox{}
+\frac1{180}
  	{\nabla_\mu \hat{\cal R}^{\mu\alpha}\nabla^\nu \hat{\cal R}_{\nu\alpha}\hat{P}}
+\frac1{720}
       \hat{P}\hat{P}\Box R
\nonumber\\&&\ \ \ \ \ \ \ \ \mbox{}
+\frac1{180}
	{R^{\mu\nu}\nabla_\mu\nabla_\nu\hat{P}\hat{P}}
+\frac1{60}
 	{\hat{\cal R}^{\mu\nu} \nabla_\mu\hat{P}\nabla_\nu\hat{P}}
\nonumber\\&&\ \ \ \ \ \ \ \ \mbox{}
-\frac1{1890}
	{R^{\mu\nu}\nabla_\mu R\nabla_\nu \hat{P}}
-\frac1{15120}
      {\Box\hat{P} R R}
\nonumber\\&&\ \ \ \ \ \ \ \ \mbox{}
+\frac1{7560}
      {\hat{P} R\Box R}
-\frac1{1260}
 	{\nabla^\mu R^{\nu\alpha}\nabla_\nu R_{\mu\alpha}\hat{P}}
\nonumber\\&&\ \ \ \ \ \ \ \ \mbox{}
-\frac1{840}
	{R^{\mu\nu}\Box R_{\mu\nu}\hat{P}}
-\frac1{5040}
	{R^{\mu\nu} R_{\mu\nu}\Box \hat{P}}
\nonumber\\&&\ \ \ \ \ \ \ \ \mbox{}
+\frac1{1120}
    {R^{\mu\nu\alpha\beta}
          R_{\mu\nu\alpha\beta}\Box\hat{P}}
+\frac1{420}
    {R^{\mu\nu\alpha\beta}
		 \nabla_\mu\nabla_\alpha R_{\nu\beta}
				\hat{P}}
\nonumber\\&&\ \ \ \ \ \ \ \ \mbox{}
+\frac{13}{30240}
	{\Box R\hat{\cal R}^{\mu\nu}\hat{\cal R}_{\mu\nu}}
-\frac{1}{15120}
	{R\hat{\cal R}^{\mu\nu}\Box\hat{\cal R}_{\mu\nu}}
\nonumber\\&&\ \ \ \ \ \ \ \ \mbox{}
-\frac{1}{7560}
	{R\nabla_\alpha\hat{\cal R}^{\alpha\mu}\nabla^\beta\hat{\cal R}_{\beta\mu}}
+\frac{1}{840}
	{\Box R^{\alpha\beta}\hat{\cal R}_{\alpha}^{\ \mu}
        \hat{\cal R}_{\beta\mu}}
\nonumber\\&&\ \ \ \ \ \ \ \ \mbox{}
-\frac{1}{1260}
	{R^{\alpha\beta}\nabla_\alpha\hat{\cal R}^{\mu\nu}\nabla_\beta
        \hat{\cal R}_{\mu\nu}}
+\frac{1}{630}
	{R^{\mu\nu}\nabla_\mu\nabla_\lambda\hat{\cal R}^{\lambda\alpha}
        \hat{\cal R}_{\alpha\nu}}
\nonumber\\&&\ \ \ \ \ \ \ \ \mbox{}
+\frac{1}{1260}
	{R_{\alpha\beta}\nabla_\mu\hat{\cal R}^{\mu\alpha}\nabla_\nu
        \hat{\cal R}^{\nu\beta}}
+\frac{1}{420}
    {R_{\mu\nu\alpha\beta}
          \hat{\cal R}^{\alpha\beta}\Box\hat{\cal R}^{\mu\nu}}
\nonumber\\&&\ \ \ \ \ \ \ \ \mbox{}
+\left[
\frac{1}{50400}
    {R^{\mu\nu\alpha\beta}
          R_{\mu\nu\alpha\beta}\Box R}
+\frac{1}{6300}
               {\Box R_{\alpha\beta}
                         R_{\ \mu\nu\lambda}^{\alpha}
                                 R^{\beta\mu\nu\lambda}}
\right.
\nonumber\\&&\ \ \ \ \ \ \ \ \mbox{}
-\frac{1}{25200}
               {R_{\lambda\sigma}
                       \nabla^\lambda R^{\mu\nu\alpha\beta}
                              \nabla^\sigma R_{\mu\nu\alpha\beta}}
-\frac1{37800}
    {R^{\mu\nu\alpha\beta}
		 \nabla_\mu\nabla_\alpha R_{\nu\beta}
				R}
\nonumber\\&&\ \ \ \ \ \ \ \ \mbox{}
-\frac1{6300}
    {R^{\mu\alpha\nu\beta}
		 \nabla_\mu R_{\nu\lambda}
		      \nabla^\lambda R_{\alpha\beta}}
-\frac2{4725}
    {R^{\alpha\beta\mu\nu}
		 \nabla_\alpha R_{\beta\lambda}
		      \nabla_\mu R_{\nu}^{\lambda}}
\nonumber\\&&\ \ \ \ \ \ \ \ \mbox{}
+\frac1{37800}
	{R^{\mu\nu}
               \nabla_\alpha R_{\beta\mu}
                       \nabla^\beta R_{\nu}^\alpha}
-\frac1{9450}
	{R^{\mu\nu}
                \nabla_\mu R^{\alpha\beta}
                          \nabla_\nu R_{\alpha\beta}}
\nonumber\\&&\ \ \ \ \ \ \ \ \mbox{}
-\frac1{18900}
	{\nabla^\mu R^{\nu\alpha}\nabla_\nu R_{\mu\alpha}R}
+\frac{29}{453600}
	{R^{\alpha\beta}\nabla_\alpha R \nabla_\beta R}
\nonumber\\&&\ \ \ \ \ \ \ \ \mbox{}
+\frac{1}{37800}
	{R R^{\mu\nu}\Box R_{\mu\nu}}
-\frac{1}{75600}
	{\Box R R^{\mu\nu}R_{\mu\nu}}
\nonumber\\&&\ \ \ \ \ \ \ \ \mbox{}
\left.\left.
-\frac{1}{7560}
  	{\Box R^{\mu}_{\alpha}
  		R^{\alpha}_{\beta} R^{\beta}_{\mu}}
-\frac{1}{100800}
	       {\Box R R R}
\right]\hat{1}
\right\}
+{\rm O}[\Re^4].
\end{fleqnarray}

The expressions (4.38), (4.39) and (4.47) for $a_0, a_1$ and $a_2$
coincide with the results obtained by other methods [6,7,12--14].
It is easy to compare expression (4.48) for $a_3$
with the result in [14,15] since they differ only by the substitution
\footnote{\normalsize
Which in eq. (8.21) of paper II
figures with the cubic terms omitted.}
\begin{eqnarray}&&
\int\! dx\, g^{1/2}\,  {\rm tr} \nabla_\alpha \hat{\cal R}^{\alpha\mu} \nabla^\beta \hat{\cal R}_{\beta\mu}
=\int\! dx\, g^{1/2}\,  {\rm tr}\Big(
-\frac12 \hat{\cal R}_{\mu\nu}\Box \hat{\cal R}^{\mu\nu}
+2
  	{\hat{\cal R}^{\mu}_{\ \alpha}
  		\hat{\cal R}^{\alpha}_{\ \beta}\hat{\cal R}^{\beta}_{\ \mu}}
\nonumber\\&& \hspace{50mm}\mbox{}
 +    {R^{\mu\nu}\hat{\cal R}^{\alpha}_{\ \mu}\hat{\cal R}_{\alpha\nu}}
- \frac12
    {R^{\mu\nu}_{\ \ \alpha\beta}
          \hat{\cal R}^{\alpha\beta}\hat{\cal R}_{\mu\nu}}
\Big),
\end{eqnarray}
and it is very difficult to compare expression (4.49) for $a_4$ with
the result in [15]. Some identities used for this purpose will be
found in sect. 14. The coincidence {\em does} take place with
accuracy ${\rm O}[\Re^4]$ but expression (4.49) is a result of such
drastic simplifications  that it should be considered as {\em new}.
It goes without saying that, although all the equations
(4.47)--(4.49) are presently obtained with accuracy ${\rm O}[\Re^4]$,
the results for $a_2$ and $a_3$ are exact.

\section{The effective action in two dimensions}
\setcounter{equation}{0}

\hspace{\parindent}
As discussed in paper II, the effective action (1.9)
\begin{eqnarray} -W=\frac12  \int^{\infty}_{0}\frac{ds}{s}
\left(\left.{\rm Tr} K(s)-{\rm Tr} K(s)\right|_{\Re=0}\right)
\end{eqnarray}
in two dimensions is generally nonanalytic in the curvature.
An exceptional case is the conformal invariant scalar
quantum field for which
\begin{eqnarray}
&&  {\rm tr} \hat{1}=1,\hspace{7mm}  \hat{\cal R}_{\mu\nu}=0,\hspace{7mm} \hat{P}=\frac16 R\hat{1},\hspace{7mm}
R_{\mu\nu}=\frac12 g_{\mu\nu}R,\nonumber\\&&\mbox{}
g^{1/2}R={\rm \,a\, total\, derivative},\hspace{7mm} \omega=1.
\end{eqnarray}
In this case, the effective action  is expandable in
powers of the
curvature because the integral (5.1) converges
at the upper limit
at each order of this
expansion owing to specific cancellations in the
asymptotic behaviours
of the form factors at large $s$. Furthermore, in
the case (5.2),
the expansion of $W$ in powers of the curvature should
terminate at the
second power thereby yielding an exact result; the terms of
third and
higher powers in the curvature should vanish order by order.
Our present
purpose  is to  check explicitly the vanishing of the
third-order terms.

By using the conditions (5.2) in (2.1), we
obtain ${\rm Tr} K(s)$
 as an expansion in powers  of the  Ricci scalar only:
\begin{fleqnarray}&&
{\rm Tr} K(s)=
\frac1{4\pi s}\int\! dx\, g^{1/2}\, \left\{1+s^2\sum^{5}_{i=1}
c_{i} f_{i}(-s\Box_2) R_1 R_2
\right.\nonumber\\&&\mbox{} \
 \left.+s^3\sum^{29}_{i=1} C_{i}  F_{i}(-s\Box_1,-s\Box_2,-s\Box_3)
R_1 R_2 R_3 +{\rm O}[R^4]
 \right\}, \hspace{7mm} \omega=1
\end{fleqnarray}
 where
\mathindent=0pt
\arraycolsep=0pt
\[ \hspace{-15mm}
c_{1}=\frac12,\ \ \ \  c_{2}=1,\ \ \ \  c_{3}=\frac16,\ \ \ \
c_{4}=\frac{1}{36},\ \ \ \  c_{5}=0,
\]
and
\begin{fleqnarray}&&
C_{1}=\frac{1}{216},\ \
C_{4}=\frac{1}{6},\ \
C_{5}=\frac{1}{12},\ \
C_{6}=\frac{1}{36},\ \
C_{9}=1,\ \
C_{10}=\frac{1}{4},\ \
C_{11}=\frac{1}{2},
\nonumber\\[1.5mm]&&
C_{15}=\frac{s}{24}(\Box_1-\Box_2-\Box_3),\ \
C_{16}=\frac{s}{48}(\Box_3-\Box_2-\Box_1),\ \
C_{17}=\frac{s}{72}\Box_2,
\nonumber\\[1.5mm]&&
C_{22}=\frac{s}{4}(\Box_1-\Box_2-\Box_3),\ \
C_{23}=\frac{s}{8}(\Box_3-\Box_2-\Box_1),\ \
C_{24}=\frac{s}{8}(\Box_1-\Box_2-\Box_3),
\nonumber\\[1.5mm]&&
C_{25}=\frac{s}{16}(\Box_1-\Box_2-\Box_3),\ \
C_{26}=\frac{s^2}{24}\Box_1\Box_2,\ \
 C_{27}=\frac{s^2}{4}\Box_1\Box_2,
\nonumber\\[1.5mm]&&
C_{28}=\frac{s^2}{16}\Box_3(\Box_3-\Box_2-\Box_1),\ \
C_{29}=\frac{s^3}{8}\Box_1\Box_2\Box_3,
\nonumber\\[1.5mm]&&
C_{2}=C_{3}=C_{7}=C_{8}=C_{12}=C_{13}=
C_{14}=C_{18}=C_{19}=C_{20}=C_{21}=0.
\nonumber\\
\end{fleqnarray}
\mathindent=\parindent
\arraycolsep=5pt
After insertion in (5.3)  of the
coefficients (5.4)
and the expressions for the form factors $f_{i}$
and $F_{i}$ given in sect. 2,
 ${\rm Tr} K(s)$ divided by $s$ takes the form
\begin{eqnarray}
\frac1s{\rm Tr} K(s) &=&
\frac1{4\pi}\int\! dx\, g^{1/2}\,  \left\{
\frac{1}{s^2}+\left[ \frac1{32}f(-s\Box_2)-\frac18
\left(\frac{f(-s\Box_2)-1}{s\Box_2}\right)
\right.\right.
\nonumber\\&&\mbox{}
\left.
+\frac38\left(\frac{f(-s\Box_2)-1-\frac16s\Box_2}{(s\Box_2)^2}\right)
\right] R_1 R_2
\nonumber\\&&\mbox{}
+\left[-s F(-s\Box_1,-s\Box_2,-s\Box_3)
{{ {{\Box_1}^2} {{\Box_2}^2} {{\Box_3}^2}}\over{3 {{D}^3}}}
\right.
\nonumber\\&&\mbox{}
+f(-s\Box_1)\frac{1}{32 {{D}^3} \Box_2}
( {{\Box_1}^6} - 4 {{\Box_1}^5} \Box_2 - 4 {{\Box_1}^5} \Box_3
     + 3 {{\Box_1}^4} \Box_2
     \Box_3 \nonumber\\&&\mbox{}+ 24 {{\Box_1}^3} {{\Box_2}^2} \Box_3
     + 5 {{\Box_1}^4} {{\Box_3}^2} + 24
     {{\Box_1}^3} \Box_2 {{\Box_3}^2}
      - 2 {{\Box_1}^2} {{\Box_2}^2} {{\Box_3}^2} \nonumber\\&&\mbox{}+
     32 \Box_1 {{\Box_2}^3} {{\Box_3}^2}
     - 25 {{\Box_1}^2} \Box_2 {{\Box_3}^3} - 36 \Box_1
     {{\Box_2}^2} {{\Box_3}^3} + 5 {{\Box_2}^3} {{\Box_3}^3} \nonumber\\&&\mbox{}
     - 5 {{\Box_1}^2}
     {{\Box_3}^4} - 9 {{\Box_2}^2} {{\Box_3}^4}
     + 4 \Box_1 {{\Box_3}^5} + 5 \Box_2
     {{\Box_3}^5} - {{\Box_3}^6} )\nonumber\\&&\mbox{}
-\left(\frac{f(-s\Box_1)-1}{s\Box_1}\right)
\frac{1}{8 {{D}^2}\Box_2}
  ({{\Box_1}^4} - 2 {{\Box_1}^3} \Box_3
 - 12 {{\Box_1}^2}  \Box_2 \Box_3 \nonumber\\&&\mbox{}- 10 \Box_1 {{\Box_2}^2} \Box_3 +
   8 \Box_1 \Box_2 {{\Box_3}^2} - 2 {{\Box_2}^2} {{\Box_3}^2}
   + 2 \Box_1 {{\Box_3}^3}
\nonumber\\&&\mbox{}
 +   3 \Box_2 {{\Box_3}^3} - {{\Box_3}^4})
\nonumber\\&&\mbox{}
+\!\left(\frac{f(-s\Box_1)-1-\frac16s\Box_1}{(s\Box_1)^2}\right)\!
\frac{3}{8 D \Box_2}
( {{\Box_1}^2} + 4 \Box_1 \Box_2 + \Box_2 \Box_3 - {{\Box_3}^2} )
\nonumber\\&&\mbox{}
-\frac1{\Box_2-\Box_3}\frac{\Box_2}{32\Box_1}
\Big(f(-s\Box_2)-f(-s\Box_3)\Big)
\nonumber\\&&\mbox{}
+\frac{1}{\Box_2-\Box_3}\frac{\Box_2}{8\Box_1}
\left(\frac{f(-s\Box_2)-1}{s\Box_2}
-\frac{f(-s\Box_3)-1}{s\Box_3}\right)\nonumber\\&&\mbox{}
-\frac{1}{\Box_2-\Box_3}\frac{3\Box_2}{8\Box_1}
\left(\frac{f(-s\Box_2)-1-\frac16s\Box_2}{(s\Box_2)^2}\right.
\nonumber\\&&\mbox{}
\left.\left.\left.
-\frac{f(-s\Box_3)-1-\frac16s\Box_3}
{(s\Box_3)^2}\right)\right]
R_1 R_2 R_3 +{\rm O}[R^4]\right\}, \hspace{7mm} \omega=1
\end{eqnarray}
in terms of the basic form factors (2.9) and (2.75), and
\[
D={\Box_1}^2+{\Box_2}^2+{\Box_3}^2-2\Box_1\Box_2-2\Box_1\Box_3-2\Box_2\Box_3.
\]

By using the asymptotic behaviours (3.1)
and (3.2), one can now check
that, at $s\rightarrow\infty$, the
leading terms $1/s$ in (5.5)
cancel at both second order and third
order in the curvature so that
\begin{eqnarray}
\frac1s{\rm Tr} K(s)&=&{\rm O}\left(\frac1{s^2}\right),\hspace{7mm}  s\rightarrow\infty.
\end{eqnarray}
As a result, the integral (5.1)
converges at the upper limit.
The convergence at the lower limit
in the curvature-dependent terms holds trivially.
Only the term of zeroth order in the curvature
 is ultraviolet divergent
but, in the
effective action (5.1), this
term gets subtracted [10,11].

For the calculation of the integral
(5.1), one may use the
differential equations for the basic
 form factors (eqs. (16.39), (16.40)
and (16.42) of sect.~16) to make the
 following substitutions in (5.5):
\begin{fleqnarray}&&
-s\frac{\Box_1\Box_2\Box_3}{D}F(-s\Box_1,-s\Box_2,-s\Box_3)=
\frac{d}{ds}\Big(sF(-s\Box_1,-s\Box_2,-s\Box_3)\Big)\nonumber\\&&\ \ \ \ \ \ \ \ \mbox{}
+\frac{\Box_1(\Box_3+\Box_2-\Box_1)}{2D}f(-s\Box_1)
+\frac{\Box_2(\Box_1+\Box_3-\Box_2)}{2D}f(-s\Box_2)\nonumber\\&&\ \ \ \ \ \ \ \ \mbox{}
+\frac{\Box_3(\Box_1+\Box_2-\Box_3)}{2D}f(-s\Box_3),
\end{fleqnarray}
\begin{fleqnarray}&&
\frac{f(-s\Box)-1}{s\Box}=\frac{d}{ds}
\left(-\frac{2}{\Box}f(-s\Box)\right)
+\frac12f(-s\Box),
\end{fleqnarray}
\begin{fleqnarray}&&
\frac{f(-s\Box)-1-\frac16s\Box}{(s\Box)^2}=
\frac{d}{ds}\left(-\frac{2}{3\Box}
\frac{f(-s\Box)-1}{s\Box}
-\frac{1}{3\Box}f(-s\Box)\right)\nonumber\\&&\ \ \ \ \ \ \ \ \mbox{}
+\frac{1}{12}f(-s\Box).
\end{fleqnarray}
The result of these substitutions
is that the expression (5.5)
becomes a total derivative in $s$:
\begin{eqnarray}
\frac1s{\rm Tr} K(s) &=&
\frac1{4\pi}\int\! dx\, g^{1/2}\,  \frac{d}{ds}\left\{
-\frac1s+l(s,\Box_2) R_1 R_2
\right.
\nonumber\\&&\mbox{}
\left.
+h(s,\Box_1,\Box_2,\Box_3) R_1 R_2 R_3 +{\rm O}[R^4]\right\},
\hspace{7mm} \omega=1
\end{eqnarray}
where
\begin{eqnarray}
l(s,\Box)&=&\frac{1}{\Box}\left(\frac18f(-s\Box)-\frac14
\frac{f(-s\Box)-1}{s\Box}\right),
\end{eqnarray}
\begin{eqnarray}
h(s,\Box_1,\Box_2,\Box_3)&=&h_1^{{\rm sym}}
+h_2^{{\rm sym}}+h_3^{{\rm sym}},
\end{eqnarray}
and $h_1^{{\rm sym}},h_2^{{\rm sym}},h_3^{{\rm sym}}$
are the completely
symmetrized in $\Box_1,\Box_2,\Box_3$ functions
\begin{eqnarray}
h_1&=&
s F(-s\Box_1,-s\Box_2,-s\Box_3)
{{ {\Box_1} {\Box_2} {\Box_3}}\over{3 {{D}^2}}},
\\[\baselineskip]
h_2&=&
f(-s\Box_1)\frac{1}{8 {{D}^2}{\Box_1} {\Box_2}}
  ({{\Box_1}^4} - 2 {{\Box_1}^3} {\Box_3}+ 2 {\Box_1} {{\Box_3}^3}
- {{\Box_3}^4}- 2 {{\Box_1}^3} {{\Box_2}}
\nonumber\\&&\mbox{}
+3{\Box_2}{{\Box_3}^3}
- 8 {{\Box_1}^2}  {\Box_2} {\Box_3}
+8 {\Box_1} {\Box_2} {{\Box_3}^2}
- 10 {\Box_1} {{\Box_2}^2} {\Box_3} -2 {{\Box_2}^2} {{\Box_3}^2})
\nonumber\\&&\mbox{}
-\left(\frac{f(-s\Box_1)-1}{s{\Box_1}}\right)
\frac{1}{4 D{\Box_1} {\Box_2}}
( {{\Box_1}^2} + 4 {\Box_1} {\Box_2}
+ {\Box_2} {\Box_3} - {{\Box_3}^2} ),
\\[\baselineskip]
h_3&=&\frac{1}{{\Box_2}-{\Box_3}}\frac{\Box_2}{\Box_1}\left[
-\frac18\left(\frac1{\Box_2}f(-s\Box_2)
-\frac1{\Box_3}f(-s\Box_3)\right)
\right.
\nonumber\\&&\mbox{}
\left.
+\frac14
\left(\frac1{\Box_2}\frac{f(-s\Box_2)-1}{s{\Box_2}}
-\frac1{\Box_3}\frac{f(-s\Box_3)-1}{s{\Box_3}}\right)\right].
\end{eqnarray}

Insertion of (5.10) in (5.1)
gives for the effective action:
\begin{eqnarray}
W &=&
\frac1{8\pi}\int\! dx\, g^{1/2}\,
\Big(l(0,\Box_2)R_1 R_2
\nonumber\\&&\mbox{}
+h(0,\Box_1,\Box_2,\Box_3) R_1 R_2 R_3
+{\rm O}[R^4]\Big),
\hspace{7mm} \omega=1
\end{eqnarray}
where use is made of the fact that the functions
$l$ and $h$ vanish at
\hbox{$s\rightarrow\infty$}. With the asymptotic behaviours
 (4.1) and (4.2),
and the explicit expressions above for the
 functions $l$ and $h$, we obtain
\begin{eqnarray}
\left.h_1^{\rm sym}\right|_{s=0}&=&0,\nonumber\\
\left.h_2^{\rm sym}\right|_{s=0}&=&
\left.-h_3^{\rm sym}\right|_{s=0}=
-\frac{1}{36}\frac{\Box_1+\Box_2+\Box_3}{\Box_1\Box_2\Box_3}.
\end{eqnarray}
The result is
\begin{eqnarray}
l(0,\Box)=\frac{1}{12}\frac{1}{\Box}, &&
h(0,\Box_1,\Box_2,\Box_3)=0,
\end{eqnarray}
and
\begin{equation}
W=\frac1{96\pi}\int\! dx\, g^{1/2}\,  R\frac{1}{\Box}R+{\rm O}[R^4],\hspace{7mm}  \omega=1.
\end{equation}
Here the term of second order in the
curvature reproduces the result of paper
II (and the results of refs. [16,17]
obtained by integrating the trace
anomaly).

Thus the third--order contribution in $W$
really vanishes, and the mechanism
of this vanishing is that, under special
conditions like (5.2),
the third-order contribution in $s^{-1}{\rm Tr} K(s)$
becomes a total derivative
of a function vanishing at both $s=0$ and $s=\infty$.
This mechanism underlies
all "miraculous" cancellations of nonlocal
terms including the trace anomaly
in four dimensions.

\section{Final result for the effective action in four
dimensions. Explicit representation of the form factors}
\setcounter{equation}{0}

\hspace{\parindent}
The result for the one-loop effective action (1.9)
to third order in the curvature is of the form ($\omega=2$)
\begin{eqnarray}
-W&=&
{\frac1{2(4\pi)^2}\int\! dx\, g^{1/2}\, \,{\rm tr}\,}
\left\{
\sum^{5}_{i=1}\gamma_i(-\Box_2)\Re_1\Re_2({i})
\right.
\nonumber\\&&\mbox{}
\left.
+\sum^{29}_{i=1}\Gamma_{i}(-\Box_1,-\Box_2,-\Box_3)\Re_1\Re_2\Re_3({i})+{\rm O}[\Re^4]\right\}.
\end{eqnarray}
Here terms of zeroth and first order in the curvature
are omitted
\footnote{\normalsize
Since these terms are local and at most quadratic in
derivatives, they must be removed by renormalization.
The zeroth-order term violates the boundary condition
of asymptotic flatness but, in the present case of massless
quantum fields, it is cancelled by the contribution
of the functional measure [10,11].
}. Terms of second order in the curvature are given
by five quadratic structures (2.2)--(2.6), and terms of
third order by twenty nine cubic structures (2.15)--(2.43).
The second-order form factors
$\gamma_i(-\Box),\ i=1$ to $5$, are of the form
\begin{eqnarray}
\gamma_1(-\Box)&=&\frac1{60}
  \left(-\ln\Big(-\frac{\Box}{\mu^2}\Big)+\frac{16}{15}\right),\\[\baselineskip]
\gamma_2(-\Box)&=&\frac1{180}
  \left(\ln\Big(-\frac{\Box}{\mu^2}\Big)-\frac{37}{30}\right),\\[\baselineskip]
\gamma_3(-\Box)&=&-\frac1{18},\\[\baselineskip]
\gamma_4(-\Box)&=&-\frac12\ln\Big(-\frac{\Box}{\mu^2}\Big),\\[\baselineskip]
\gamma_5(-\Box)&=&\frac1{12}
  \left(-\ln\Big(-\frac{\Box}{\mu^2}\Big)+\frac23\right)
\end{eqnarray}
where the parameter $\mu^2>0$ accounts for the ultraviolet
arbitrariness. The form factor $\gamma_3(-\Box)$ is local
and independent of this arbitrariness. To second order in the
curvature, the expressions above reproduce the results
of the paper II.

The third-order form factors
\begin{equation}
\Gamma_{i}(-\Box_1,-\Box_2,-\Box_3),\hspace{7mm} i=1\ {\rm to}\ 29,
\end{equation}
{\it contain no arbitrary parameters} and are expressed through
the basic third-order form factor
\begin{eqnarray}
\Gamma(-\Box_1,-\Box_2,-\Box_3)&=&\int_{\alpha\geq 0}
d^3\,\alpha\,\delta(1-\alpha_1-\alpha_2-\alpha_3)
\nonumber\\&&\ \ \ \ \mbox{}\times(
-\alpha_1\alpha_2\Box_3
-\alpha_1\alpha_3\Box_2
-\alpha_2\alpha_3\Box_1)^{-1}
\end{eqnarray}
and the second-order form factors
\begin{equation}
{\rm a)\ }\ln({\Box_{n}}/{\Box_{m}}),\hspace{7mm}
{\rm b)\ }\frac{\ln(\Box_{n}/\Box_{m})}{\Box_{n}-\Box_{m}},\hspace{7mm}
n,m=1,2,3.
\end{equation}
The coefficients of these expressions are rational
functions of the following general form:
\begin{equation}
\frac{P(\Box)}{D^6\Box_1\Box_2\Box_3}
\end{equation}
where $P(\Box)$ is a polynomial, and
\begin{equation}
D={\Box_1}^2+{\Box_2}^2+{\Box_3}^2
-2\Box_1\Box_2-2\Box_1\Box_3-2\Box_2\Box_3.
\end{equation}
There are also purely rational contributions of the form (6.10).
In this representation, the explicit expressions for $\Gamma_i$
are given below. In sects. 7--9 we present also several integral
representations of the form factors $\Gamma_i$. The derivations
of these results are given in sects. 17--20.

When taken separately from their curvature structures, the
form factors (6.7) should be explicitly symmetrized
\begin{equation}
\Gamma_i\rightarrow\Gamma_i^{\rm sym}
\end{equation}
according to the same laws (2.46)--(2.74) as the form
factors in the heat kernel. In a not symmetrized form,
the explicit expressions for the third-order form factors
are as follows:

\arraycolsep 0pt
\begin{fleqnarray}&&\Gamma_{1}(-\Box_1,-\Box_2,-\Box_3) =
\frac13\Gamma(-\Box_1,-\Box_2,-\Box_3)
,\end{fleqnarray}

\begin{fleqnarray}&&\Gamma_{2}(-\Box_1,-\Box_2,-\Box_3) =
\Gamma(-\Box_1,-\Box_2,-\Box_3)\frac1{3 {D^3}}\Big(
   8 {{\Box_1}^4} \Box_2 \Box_3   \nonumber\\&&\ \ \ \ \mbox{}
  -8 {{\Box_1}^3} {{\Box_2}^2} \Box_3
  -8 {{\Box_1}^2} {{\Box_2}^3} \Box_3
  +8 \Box_1 {{\Box_2}^4} \Box_3
  -8 {{\Box_1}^3} \Box_2 {{\Box_3}^2}  \nonumber\\&&\ \ \ \ \mbox{}
  +32 {{\Box_1}^2} {{\Box_2}^2} {{\Box_3}^2}
  -8 \Box_1 {{\Box_2}^3} {{\Box_3}^2}
  -8 {{\Box_1}^2} \Box_2 {{\Box_3}^3} \nonumber\\&&\ \ \ \ \mbox{}
  -8 \Box_1 {{\Box_2}^2} {{\Box_3}^3}
  +8 \Box_1 \Box_2 {{\Box_3}^4}
\Big) \nonumber\\&&\ \ \ \ \mbox{}
+\frac{\ln(\Box_1/\Box_2)}{9 {D^3}}\Big(
   8 {{\Box_1}^4} \Box_2
  -24 {{\Box_1}^3} {{\Box_2}^2}
  +24 {{\Box_1}^2} {{\Box_2}^3}  \nonumber\\&&\ \ \ \ \mbox{}
  -8 \Box_1 {{\Box_2}^4}
  +4 {{\Box_1}^4} \Box_3
  +96 {{\Box_1}^3} \Box_2 \Box_3
  -96 \Box_1 {{\Box_2}^3} \Box_3   \nonumber\\&&\ \ \ \ \mbox{}
  -4 {{\Box_2}^4} \Box_3
  -12 {{\Box_1}^3} {{\Box_3}^2}
  -108 {{\Box_1}^2} \Box_2 {{\Box_3}^2}
  +108 \Box_1 {{\Box_2}^2} {{\Box_3}^2} \nonumber\\&&\ \ \ \ \mbox{}
  +12 {{\Box_2}^3} {{\Box_3}^2}
  +12 {{\Box_1}^2} {{\Box_3}^3}
  -12 {{\Box_2}^2} {{\Box_3}^3}
  -4 \Box_1 {{\Box_3}^4}
  +4 \Box_2 {{\Box_3}^4}
\Big) \nonumber\\&&\ \ \ \ \mbox{}
+\frac13\frac{\ln(\Box_1/\Box_2)}{(\Box_1-\Box_2)} \nonumber\\&&\ \ \ \ \mbox{}
+\frac1{9 {D^2}}\Big(
  -2 {{\Box_1}^3}
  +2 {{\Box_1}^2} \Box_2
  +2 \Box_1 {{\Box_2}^2}
  -2 {{\Box_2}^3}
  +2 {{\Box_1}^2} \Box_3 \nonumber\\&&\ \ \ \ \mbox{}
  -28 \Box_1 \Box_2 \Box_3
  +2 {{\Box_2}^2} \Box_3
  +2 \Box_1 {{\Box_3}^2}
  +2 \Box_2 {{\Box_3}^2}
  -2 {{\Box_3}^3}
\Big)
,\end{fleqnarray}

\begin{fleqnarray}&&\Gamma_{3}(-\Box_1,-\Box_2,-\Box_3) =
\Gamma(-\Box_1,-\Box_2,-\Box_3)\frac1{{D^2}}\Big(
  -2 {{\Box_1}^3} \Box_2     \nonumber\\&&\ \ \ \ \mbox{}
  +4 {{\Box_1}^2} {{\Box_2}^2}
  -2 \Box_1 {{\Box_2}^3}
  -2 {{\Box_1}^2} \Box_2 \Box_3
  -2 \Box_1 {{\Box_2}^2} \Box_3
  +4 \Box_1 \Box_2 {{\Box_3}^2}
\Big) \nonumber\\&&\ \ \ \ \mbox{}
+\frac{\ln(\Box_1/\Box_2)}{3 {D^2}}\Big(
  -{{\Box_1}^3}
  -9 {{\Box_1}^2} \Box_2
  +9 \Box_1 {{\Box_2}^2}
  +{{\Box_2}^3}           \nonumber\\&&\ \ \ \ \mbox{}
  +2 {{\Box_1}^2} \Box_3
  -2 {{\Box_2}^2} \Box_3
  -\Box_1 {{\Box_3}^2}
  +\Box_2 {{\Box_3}^2}
\Big) \nonumber\\&&\ \ \ \ \mbox{}
+\frac{\ln(\Box_1/\Box_3)}{3 {D^2}}\Big(
  -2 {{\Box_1}^3}
  -3 {{\Box_1}^2} \Box_2
  +6 \Box_1 {{\Box_2}^2}
  -{{\Box_2}^3}
  +4 {{\Box_1}^2} \Box_3   \nonumber\\&&\ \ \ \ \mbox{}
  -12 \Box_1 \Box_2 \Box_3
  +2 {{\Box_2}^2} \Box_3
  -2 \Box_1 {{\Box_3}^2}
  -\Box_2 {{\Box_3}^2}
\Big) \nonumber\\&&\ \ \ \ \mbox{}
+\frac{\ln(\Box_2/\Box_3)}{3 {D^2}}\Big(
  -{{\Box_1}^3}
  +6 {{\Box_1}^2} \Box_2
  -3 \Box_1 {{\Box_2}^2}
  -2 {{\Box_2}^3}         \nonumber\\&&\ \ \ \ \mbox{}
  +2 {{\Box_1}^2} \Box_3
  -12 \Box_1 \Box_2 \Box_3
  +4 {{\Box_2}^2} \Box_3
  -\Box_1 {{\Box_3}^2}
  -2 \Box_2 {{\Box_3}^2}
\Big) \nonumber\\&&\ \ \ \ \mbox{}
+\frac1{D}\Big(
   \Box_1
  +\Box_2
  -\Box_3
\Big)
,\end{fleqnarray}

\begin{fleqnarray}&&\Gamma_{4}(-\Box_1,-\Box_2,-\Box_3) =
\Gamma(-\Box_1,-\Box_2,-\Box_3)\frac1{36 {D^4}}\Big(
  -4 {{\Box_1}^8}             \nonumber\\&&\ \ \ \ \mbox{}
  -4 {{\Box_1}^7} \Box_2
  +32 {{\Box_1}^6} {{\Box_2}^2}
  -28 {{\Box_1}^5} {{\Box_2}^3}
  +4 {{\Box_1}^4} {{\Box_2}^4}  \nonumber\\&&\ \ \ \ \mbox{}
  +2 {{\Box_1}^7} \Box_3
  -118 {{\Box_1}^6} \Box_2 \Box_3
  -90 {{\Box_1}^5} {{\Box_2}^2} \Box_3
  +206 {{\Box_1}^4} {{\Box_2}^3} \Box_3 \nonumber\\&&\ \ \ \ \mbox{}
  +38 {{\Box_1}^6} {{\Box_3}^2}
  +180 {{\Box_1}^5} \Box_2 {{\Box_3}^2}
  -198 {{\Box_1}^4} {{\Box_2}^2} {{\Box_3}^2}
  -236 {{\Box_1}^3} {{\Box_2}^3} {{\Box_3}^2} \nonumber\\&&\ \ \ \ \mbox{}
  -82 {{\Box_1}^5} {{\Box_3}^3}
  +110 {{\Box_1}^4} \Box_2 {{\Box_3}^3}
  +188 {{\Box_1}^3} {{\Box_2}^2} {{\Box_3}^3}
  +50 {{\Box_1}^4} {{\Box_3}^4}             \nonumber\\&&\ \ \ \ \mbox{}
  -172 {{\Box_1}^3} \Box_2 {{\Box_3}^4}
  +90 {{\Box_1}^2} {{\Box_2}^2} {{\Box_3}^4}
  +14 {{\Box_1}^3} {{\Box_3}^5}
  -90 {{\Box_1}^2} \Box_2 {{\Box_3}^5}        \nonumber\\&&\ \ \ \ \mbox{}
  -22 {{\Box_1}^2} {{\Box_3}^6}
  +46 \Box_1 \Box_2 {{\Box_3}^6}
  +2 \Box_1 {{\Box_3}^7}
  +{{\Box_3}^8}
\Big) \nonumber\\&&\ \ \ \ \mbox{}
+\frac{\ln(\Box_1/\Box_2)}{18 {D^4}}\Big(
  -6 {{\Box_1}^7}
  -8 {{\Box_1}^6} \Box_2
  +24 {{\Box_1}^5} {{\Box_2}^2}    \nonumber\\&&\ \ \ \ \mbox{}
  +10 {{\Box_1}^4} {{\Box_2}^3}
  +11 {{\Box_1}^6} \Box_3
  -64 {{\Box_1}^5} \Box_2 \Box_3
  -125 {{\Box_1}^4} {{\Box_2}^2} \Box_3 \nonumber\\&&\ \ \ \ \mbox{}
  +10 {{\Box_1}^5} {{\Box_3}^2}
  +112 {{\Box_1}^4} \Box_2 {{\Box_3}^2}
  +54 {{\Box_1}^3} {{\Box_2}^2} {{\Box_3}^2}
  -30 {{\Box_1}^4} {{\Box_3}^3}       \nonumber\\&&\ \ \ \ \mbox{}
  +12 {{\Box_1}^3} \Box_2 {{\Box_3}^3}
  +10 {{\Box_1}^3} {{\Box_3}^4}
  -58 {{\Box_1}^2} \Box_2 {{\Box_3}^4}
  +11 {{\Box_1}^2} {{\Box_3}^5}
  -6 \Box_1 {{\Box_3}^6}
\Big) \nonumber\\&&\ \ \ \ \mbox{}
+\frac{\ln(\Box_1/\Box_3)}{18 {D^4}}\Big(
  -6 {{\Box_1}^7}
  +2 {{\Box_1}^6} \Box_2
  +12 {{\Box_1}^5} {{\Box_2}^2}       \nonumber\\&&\ \ \ \ \mbox{}
  +2 {{\Box_1}^4} {{\Box_2}^3}
  -8 {{\Box_1}^3} {{\Box_2}^4}
  -12 {{\Box_1}^2} {{\Box_2}^5}
  +10 \Box_1 {{\Box_2}^6}             \nonumber\\&&\ \ \ \ \mbox{}
  +{{\Box_1}^6} \Box_3
  -80 {{\Box_1}^5} \Box_2 \Box_3
  -46 {{\Box_1}^4} {{\Box_2}^2} \Box_3
  +72 {{\Box_1}^3} {{\Box_2}^3} \Box_3  \nonumber\\&&\ \ \ \ \mbox{}
  +79 {{\Box_1}^2} {{\Box_2}^4} \Box_3
  -16 \Box_1 {{\Box_2}^5} \Box_3
  -10 {{\Box_2}^6} \Box_3
  +38 {{\Box_1}^5} {{\Box_3}^2}       \nonumber\\&&\ \ \ \ \mbox{}
  +56 {{\Box_1}^4} \Box_2 {{\Box_3}^2}
  -60 {{\Box_1}^3} {{\Box_2}^2} {{\Box_3}^2}
  -114 {{\Box_1}^2} {{\Box_2}^3} {{\Box_3}^2}
  -56 \Box_1 {{\Box_2}^4} {{\Box_3}^2}  \nonumber\\&&\ \ \ \ \mbox{}
  +28 {{\Box_2}^5} {{\Box_3}^2}
  -45 {{\Box_1}^4} {{\Box_3}^3}
  +72 {{\Box_1}^3} \Box_2 {{\Box_3}^3}
  +18 {{\Box_1}^2} {{\Box_2}^2} {{\Box_3}^3} \nonumber\\&&\ \ \ \ \mbox{}
  +60 \Box_1 {{\Box_2}^3} {{\Box_3}^3}
  -15 {{\Box_2}^4} {{\Box_3}^3}
  -10 {{\Box_1}^3} {{\Box_3}^4}
  -2 {{\Box_1}^2} \Box_2 {{\Box_3}^4}        \nonumber\\&&\ \ \ \ \mbox{}
  +56 \Box_1 {{\Box_2}^2} {{\Box_3}^4}
  -20 {{\Box_2}^3} {{\Box_3}^4}
  +31 {{\Box_1}^2} {{\Box_3}^5}
  -48 \Box_1 \Box_2 {{\Box_3}^5}             \nonumber\\&&\ \ \ \ \mbox{}
  +20 {{\Box_2}^2} {{\Box_3}^5}
  -6 \Box_1 {{\Box_3}^6}
  -3 {{\Box_3}^7}
\Big) \nonumber\\&&\ \ \ \ \mbox{}
+\frac1{24 {D^3}}\Big(
   16 {{\Box_1}^5}
  +4 {{\Box_1}^4} \Box_2
  -20 {{\Box_1}^3} {{\Box_2}^2}
  -22 {{\Box_1}^4} \Box_3             \nonumber\\&&\ \ \ \ \mbox{}
  +44 {{\Box_1}^3} \Box_2 \Box_3
  +50 {{\Box_1}^2} {{\Box_2}^2} \Box_3
  -20 {{\Box_1}^3} {{\Box_3}^2}
  -24 {{\Box_1}^2} \Box_2 {{\Box_3}^2}  \nonumber\\&&\ \ \ \ \mbox{}
  +32 {{\Box_1}^2} {{\Box_3}^3}
  -14 \Box_1 \Box_2 {{\Box_3}^3}
  +4 \Box_1 {{\Box_3}^4}
  -5 {{\Box_3}^5}
\Big)
,\end{fleqnarray}

\begin{fleqnarray}&&\Gamma_{5}(-\Box_1,-\Box_2,-\Box_3) =
\Gamma(-\Box_1,-\Box_2,-\Box_3)\frac{\Box_1 \Box_2 {{\Box_3}^2}}{D^2}
\nonumber\\&&\ \ \ \ \mbox{}
+\frac{\ln(\Box_1/\Box_2)}{18 {D^2} \Box_2}\Big(
  -{{\Box_1}^4}
  +3 {{\Box_1}^3} \Box_2
  -4 {{\Box_1}^2} {{\Box_2}^2}
  +3 {{\Box_1}^3} \Box_3        \nonumber\\&&\ \ \ \ \mbox{}
  +4 {{\Box_1}^2} \Box_2 \Box_3
  -3 {{\Box_1}^2} {{\Box_3}^2}
  -7 \Box_1 \Box_2 {{\Box_3}^2}
  +\Box_1 {{\Box_3}^3}
\Big) \nonumber\\&&\ \ \ \ \mbox{}
+\frac{\ln(\Box_1/\Box_3)}{18 {D^2} \Box_1 \Box_2}\Big(
  -2 {{\Box_1}^5}
  +6 {{\Box_1}^4} \Box_2
  -5 {{\Box_1}^3} {{\Box_2}^2}
  -{{\Box_1}^2} {{\Box_2}^3}    \nonumber\\&&\ \ \ \ \mbox{}
  +3 \Box_1 {{\Box_2}^4}
  -{{\Box_2}^5}
  +6 {{\Box_1}^4} \Box_3
  +8 {{\Box_1}^3} \Box_2 \Box_3
  -21 {{\Box_1}^2} {{\Box_2}^2} \Box_3 \nonumber\\&&\ \ \ \ \mbox{}
  +4 \Box_1 {{\Box_2}^3} \Box_3
  +3 {{\Box_2}^4} \Box_3
  -6 {{\Box_1}^3} {{\Box_3}^2}
  -14 {{\Box_1}^2} \Box_2 {{\Box_3}^2} \nonumber\\&&\ \ \ \ \mbox{}
  -7 \Box_1 {{\Box_2}^2} {{\Box_3}^2}
  -3 {{\Box_2}^3} {{\Box_3}^2}
  +2 {{\Box_1}^2} {{\Box_3}^3}
  +{{\Box_2}^2} {{\Box_3}^3}
\Big) \nonumber\\&&\ \ \ \ \mbox{}
+\frac1{24 D \Box_1 \Box_2}\Big(
   6 {{\Box_1}^3}
  -6 {{\Box_1}^2} \Box_2
  -14 {{\Box_1}^2} \Box_3       \nonumber\\&&\ \ \ \ \mbox{}
  -14 \Box_1 \Box_2 \Box_3
  +10 \Box_1 {{\Box_3}^2}
  -{{\Box_3}^3}
\Big)
,\end{fleqnarray}

\begin{fleqnarray}&&\Gamma_{6}(-\Box_1,-\Box_2,-\Box_3) =
\Gamma(-\Box_1,-\Box_2,-\Box_3)\frac1{6 {D^2}}\Big(
  -2 {{\Box_1}^4}             \nonumber\\&&\ \ \ \ \mbox{}
  -4 {{\Box_1}^3} \Box_2
  +6 {{\Box_1}^2} {{\Box_2}^2}
  +2 {{\Box_1}^3} \Box_3
  -14 {{\Box_1}^2} \Box_2 \Box_3  \nonumber\\&&\ \ \ \ \mbox{}
  +2 \Box_1 \Box_2 {{\Box_3}^2}
  +2 \Box_1 {{\Box_3}^3}
  -{{\Box_3}^4}
\Big) \nonumber\\&&\ \ \ \ \mbox{}
+\frac{\ln(\Box_1/\Box_2)}{3 {D^2}}\Big(
  -2 {{\Box_1}^3}
  -6 {{\Box_1}^2} \Box_2
  +{{\Box_1}^2} \Box_3
  +\Box_1 {{\Box_3}^2}
\Big) \nonumber\\&&\ \ \ \ \mbox{}
+\frac{\ln(\Box_1/\Box_3)}{6 {D^2}}\Big(
  -5 {{\Box_1}^3}
  -3 {{\Box_1}^2} \Box_2
  +9 \Box_1 {{\Box_2}^2}
  -{{\Box_2}^3}
  +{{\Box_1}^2} \Box_3          \nonumber\\&&\ \ \ \ \mbox{}
  -12 \Box_1 \Box_2 \Box_3
  -{{\Box_2}^2} \Box_3
  +\Box_1 {{\Box_3}^2}
  -\Box_2 {{\Box_3}^2}
  +3 {{\Box_3}^3}
\Big) \nonumber\\&&\ \ \ \ \mbox{}
+\frac{\Box_1}{D}
,\end{fleqnarray}

\begin{fleqnarray}&&\Gamma_{7}(-\Box_1,-\Box_2,-\Box_3) =
\Gamma(-\Box_1,-\Box_2,-\Box_3)\frac1{3 {D^4}}\Big(
   4 {{\Box_1}^6} \Box_2 \Box_3
  -24 {{\Box_1}^5} {{\Box_2}^2} \Box_3 \nonumber\\&&\ \ \ \ \mbox{}
  +18 {{\Box_1}^4} {{\Box_2}^3} \Box_3
  +8 {{\Box_1}^3} {{\Box_2}^4} \Box_3
  -12 {{\Box_1}^2} {{\Box_2}^5} \Box_3
  +2 {{\Box_2}^7} \Box_3
  +63 {{\Box_1}^4} {{\Box_2}^2} {{\Box_3}^2}  \nonumber\\&&\ \ \ \ \mbox{}
  -108 {{\Box_1}^3} {{\Box_2}^3} {{\Box_3}^2}
  -66 {{\Box_1}^2} {{\Box_2}^4} {{\Box_3}^2}
  +72 \Box_1 {{\Box_2}^5} {{\Box_3}^2}
  +114 {{\Box_1}^2} {{\Box_2}^3} {{\Box_3}^3}  \nonumber\\&&\ \ \ \ \mbox{}
  -72 \Box_1 {{\Box_2}^4} {{\Box_3}^3}
  -18 {{\Box_2}^5} {{\Box_3}^3}
  +16 {{\Box_2}^4} {{\Box_3}^4}
\Big) \nonumber\\&&\ \ \ \ \mbox{}
+\frac{\ln(\Box_1/\Box_2)}{36 {D^4} \Box_1}\Big(
   3 {{\Box_1}^8}
  -12 {{\Box_1}^7} \Box_2
  +12 {{\Box_1}^6} {{\Box_2}^2}
  +16 {{\Box_1}^5} {{\Box_2}^3}
  -50 {{\Box_1}^4} {{\Box_2}^4}            \nonumber\\&&\ \ \ \ \mbox{}
  +52 {{\Box_1}^3} {{\Box_2}^5}
  -28 {{\Box_1}^2} {{\Box_2}^6}
  +8 \Box_1 {{\Box_2}^7}
  -{{\Box_2}^8}
  -18 {{\Box_1}^7} \Box_3                  \nonumber\\&&\ \ \ \ \mbox{}
  +198 {{\Box_1}^6} \Box_2 \Box_3
  -194 {{\Box_1}^5} {{\Box_2}^2} \Box_3
  -350 {{\Box_1}^4} {{\Box_2}^3} \Box_3
  +498 {{\Box_1}^3} {{\Box_2}^4} \Box_3      \nonumber\\&&\ \ \ \ \mbox{}
  -46 {{\Box_1}^2} {{\Box_2}^5} \Box_3
  -94 \Box_1 {{\Box_2}^6} \Box_3
  +6 {{\Box_2}^7} \Box_3
  +48 {{\Box_1}^6} {{\Box_3}^2}
  -430 {{\Box_1}^5} \Box_2 {{\Box_3}^2}      \nonumber\\&&\ \ \ \ \mbox{}
  +1230 {{\Box_1}^4} {{\Box_2}^2} {{\Box_3}^2}
  +40 {{\Box_1}^3} {{\Box_2}^3} {{\Box_3}^2}
  -960 {{\Box_1}^2} {{\Box_2}^4} {{\Box_3}^2}
  +86 \Box_1 {{\Box_2}^5} {{\Box_3}^2}       \nonumber\\&&\ \ \ \ \mbox{}
  -14 {{\Box_2}^6} {{\Box_3}^2}
  -76 {{\Box_1}^5} {{\Box_3}^3}
  +218 {{\Box_1}^4} \Box_2 {{\Box_3}^3}
  -1096 {{\Box_1}^3} {{\Box_2}^2} {{\Box_3}^3} \nonumber\\&&\ \ \ \ \mbox{}
  +1224 {{\Box_1}^2} {{\Box_2}^3} {{\Box_3}^3}
  +228 \Box_1 {{\Box_2}^4} {{\Box_3}^3}
  +14 {{\Box_2}^5} {{\Box_3}^3}
  +80 {{\Box_1}^4} {{\Box_3}^4}              \nonumber\\&&\ \ \ \ \mbox{}
  +132 {{\Box_1}^3} \Box_2 {{\Box_3}^4}
  -96 {{\Box_1}^2} {{\Box_2}^2} {{\Box_3}^4}
  -372 \Box_1 {{\Box_2}^3} {{\Box_3}^4}
  -58 {{\Box_1}^3} {{\Box_3}^5}              \nonumber\\&&\ \ \ \ \mbox{}
  -122 {{\Box_1}^2} \Box_2 {{\Box_3}^5}
  +130 \Box_1 {{\Box_2}^2} {{\Box_3}^5}
  -14 {{\Box_2}^3} {{\Box_3}^5}
  +28 {{\Box_1}^2} {{\Box_3}^6}              \nonumber\\&&\ \ \ \ \mbox{}
  +22 \Box_1 \Box_2 {{\Box_3}^6}
  +14 {{\Box_2}^2} {{\Box_3}^6}
  -8 \Box_1 {{\Box_3}^7}
  -6 \Box_2 {{\Box_3}^7}
  +{{\Box_3}^8}
\Big) \nonumber\\&&\ \ \ \ \mbox{}
+\frac{\ln(\Box_2/\Box_3)}{18 {D^4} \Box_1}\Big(
  -3 {{\Box_1}^7} \Box_2
  +18 {{\Box_1}^6} {{\Box_2}^2}
  -46 {{\Box_1}^5} {{\Box_2}^3}
  +65 {{\Box_1}^4} {{\Box_2}^4}              \nonumber\\&&\ \ \ \ \mbox{}
  -55 {{\Box_1}^3} {{\Box_2}^5}
  +28 {{\Box_1}^2} {{\Box_2}^6}
  -8 \Box_1 {{\Box_2}^7}
  +{{\Box_2}^8}
  -118 {{\Box_1}^5} {{\Box_2}^2} \Box_3        \nonumber\\&&\ \ \ \ \mbox{}
  +284 {{\Box_1}^4} {{\Box_2}^3} \Box_3
  -183 {{\Box_1}^3} {{\Box_2}^4} \Box_3
  -38 {{\Box_1}^2} {{\Box_2}^5} \Box_3
  +58 \Box_1 {{\Box_2}^6} \Box_3               \nonumber\\&&\ \ \ \ \mbox{}
  -6 {{\Box_2}^7} \Box_3
  -568 {{\Box_1}^3} {{\Box_2}^3} {{\Box_3}^2}
  +432 {{\Box_1}^2} {{\Box_2}^4} {{\Box_3}^2}
  +22 \Box_1 {{\Box_2}^5} {{\Box_3}^2}         \nonumber\\&&\ \ \ \ \mbox{}
  +14 {{\Box_2}^6} {{\Box_3}^2}
  -300 \Box_1 {{\Box_2}^4} {{\Box_3}^3}
  -14 {{\Box_2}^5} {{\Box_3}^3}
\Big) \nonumber\\&&\ \ \ \ \mbox{}
+\frac{\ln(\Box_2/\Box_3)}{(\Box_2-\Box_3)}\Big({{-\Box_2}\over {12 \Box_1}}\Big) \nonumber\\&&\ \ \ \ \mbox{}
+\frac1{24 {D^3}}\Big(
  -3 {{\Box_1}^5}
  +26 {{\Box_1}^4} \Box_2
  -44 {{\Box_1}^3} {{\Box_2}^2}
  +36 {{\Box_1}^2} {{\Box_2}^3}              \nonumber\\&&\ \ \ \ \mbox{}
  -14 \Box_1 {{\Box_2}^4}
  +2 {{\Box_2}^5}
  -92 {{\Box_1}^3} \Box_2 \Box_3
  +180 {{\Box_1}^2} {{\Box_2}^2} \Box_3        \nonumber\\&&\ \ \ \ \mbox{}
  +88 \Box_1 {{\Box_2}^3} \Box_3
  -110 {{\Box_2}^4} \Box_3
  -218 \Box_1 {{\Box_2}^2} {{\Box_3}^2}
  +108 {{\Box_2}^3} {{\Box_3}^2}
\Big)
,\end{fleqnarray}
\begin{fleqnarray}&&\Gamma_{8}(-\Box_1,-\Box_2,-\Box_3) =
\Gamma(-\Box_1,-\Box_2,-\Box_3)\frac1{{D^4}}\Big(
  -4 {{\Box_1}^6} \Box_2 \Box_3                \nonumber\\&&\ \ \ \ \mbox{}
  +24 {{\Box_1}^5} {{\Box_2}^2} \Box_3
  -16 {{\Box_1}^4} {{\Box_2}^3} \Box_3
  -16 {{\Box_1}^3} {{\Box_2}^4} \Box_3
  +24 {{\Box_1}^2} {{\Box_2}^5} \Box_3         \nonumber\\&&\ \ \ \ \mbox{}
  -8 \Box_1 {{\Box_2}^6} \Box_3
  -84 {{\Box_1}^4} {{\Box_2}^2} {{\Box_3}^2}
  +168 {{\Box_1}^3} {{\Box_2}^3} {{\Box_3}^2}
  +72 {{\Box_1}^2} {{\Box_2}^4} {{\Box_3}^2}   \nonumber\\&&\ \ \ \ \mbox{}
  -96 \Box_1 {{\Box_2}^5} {{\Box_3}^2}
  -176 {{\Box_1}^2} {{\Box_2}^3} {{\Box_3}^3}
  +104 \Box_1 {{\Box_2}^4} {{\Box_3}^3}
\Big) \nonumber\\&&\ \ \ \ \mbox{}
+\frac{\ln(\Box_1/\Box_2)}{9 {D^4} \Box_1}\Big(
  -3 {{\Box_1}^8}
  +16 {{\Box_1}^7} \Box_2
  -32 {{\Box_1}^6} {{\Box_2}^2}
  +24 {{\Box_1}^5} {{\Box_2}^3}
  +10 {{\Box_1}^4} {{\Box_2}^4}              \nonumber\\&&\ \ \ \ \mbox{}
  -32 {{\Box_1}^3} {{\Box_2}^5}
  +24 {{\Box_1}^2} {{\Box_2}^6}
  -8 \Box_1 {{\Box_2}^7}
  +{{\Box_2}^8}
  +20 {{\Box_1}^7} \Box_3
  -192 {{\Box_1}^6} \Box_2 \Box_3              \nonumber\\&&\ \ \ \ \mbox{}
  +180 {{\Box_1}^5} {{\Box_2}^2} \Box_3
  +392 {{\Box_1}^4} {{\Box_2}^3} \Box_3
  -628 {{\Box_1}^3} {{\Box_2}^4} \Box_3
  +192 {{\Box_1}^2} {{\Box_2}^5} \Box_3        \nonumber\\&&\ \ \ \ \mbox{}
  +44 \Box_1 {{\Box_2}^6} \Box_3
  -8 {{\Box_2}^7} \Box_3
  -58 {{\Box_1}^6} {{\Box_3}^2}
  +396 {{\Box_1}^5} \Box_2 {{\Box_3}^2}        \nonumber\\&&\ \ \ \ \mbox{}
  -1374 {{\Box_1}^4} {{\Box_2}^2} {{\Box_3}^2}
  +80 {{\Box_1}^3} {{\Box_2}^3} {{\Box_3}^2}
  +930 {{\Box_1}^2} {{\Box_2}^4} {{\Box_3}^2}
  +4 \Box_1 {{\Box_2}^5} {{\Box_3}^2}          \nonumber\\&&\ \ \ \ \mbox{}
  +22 {{\Box_2}^6} {{\Box_3}^2}
  +96 {{\Box_1}^5} {{\Box_3}^3}
  -176 {{\Box_1}^4} \Box_2 {{\Box_3}^3}
  +1408 {{\Box_1}^3} {{\Box_2}^2} {{\Box_3}^3} \nonumber\\&&\ \ \ \ \mbox{}
  -1128 {{\Box_1}^2} {{\Box_2}^3} {{\Box_3}^3}
  -176 \Box_1 {{\Box_2}^4} {{\Box_3}^3}
  -24 {{\Box_2}^5} {{\Box_3}^3}
  -100 {{\Box_1}^4} {{\Box_3}^4}             \nonumber\\&&\ \ \ \ \mbox{}
  -176 {{\Box_1}^3} \Box_2 {{\Box_3}^4}
  -156 {{\Box_1}^2} {{\Box_2}^2} {{\Box_3}^4}
  +176 \Box_1 {{\Box_2}^3} {{\Box_3}^4}
  +68 {{\Box_1}^3} {{\Box_3}^5}              \nonumber\\&&\ \ \ \ \mbox{}
  +168 {{\Box_1}^2} \Box_2 {{\Box_3}^5}
  -4 \Box_1 {{\Box_2}^2} {{\Box_3}^5}
  +24 {{\Box_2}^3} {{\Box_3}^5}
  -30 {{\Box_1}^2} {{\Box_3}^6}              \nonumber\\&&\ \ \ \ \mbox{}
  -44 \Box_1 \Box_2 {{\Box_3}^6}
  -22 {{\Box_2}^2} {{\Box_3}^6}
  +8 \Box_1 {{\Box_3}^7}
  +8 \Box_2 {{\Box_3}^7}
  -{{\Box_3}^8}
\Big) \nonumber\\&&\ \ \ \ \mbox{}
+\frac{\ln(\Box_2/\Box_3)}{9 {D^4} \Box_1}\Big(
   4 {{\Box_1}^7} \Box_2
  -26 {{\Box_1}^6} {{\Box_2}^2}
  +72 {{\Box_1}^5} {{\Box_2}^3}
  -110 {{\Box_1}^4} {{\Box_2}^4}             \nonumber\\&&\ \ \ \ \mbox{}
  +100 {{\Box_1}^3} {{\Box_2}^5}
  -54 {{\Box_1}^2} {{\Box_2}^6}
  +16 \Box_1 {{\Box_2}^7}
  -2 {{\Box_2}^8}
  +216 {{\Box_1}^5} {{\Box_2}^2} \Box_3        \nonumber\\&&\ \ \ \ \mbox{}
  -568 {{\Box_1}^4} {{\Box_2}^3} \Box_3
  +452 {{\Box_1}^3} {{\Box_2}^4} \Box_3
  -24 {{\Box_1}^2} {{\Box_2}^5} \Box_3
  -88 \Box_1 {{\Box_2}^6} \Box_3               \nonumber\\&&\ \ \ \ \mbox{}
  +16 {{\Box_2}^7} \Box_3
  +1328 {{\Box_1}^3} {{\Box_2}^3} {{\Box_3}^2}
  -1086 {{\Box_1}^2} {{\Box_2}^4} {{\Box_3}^2}
  -8 \Box_1 {{\Box_2}^5} {{\Box_3}^2}          \nonumber\\&&\ \ \ \ \mbox{}
  -44 {{\Box_2}^6} {{\Box_3}^2}
  +352 \Box_1 {{\Box_2}^4} {{\Box_3}^3}
  +48 {{\Box_2}^5} {{\Box_3}^3}
\Big) \nonumber\\&&\ \ \ \ \mbox{}
+\frac1{6 {D^3} \Box_1}\Big(
   {{\Box_1}^6}
  -8 {{\Box_1}^5} \Box_2
  +10 {{\Box_1}^4} {{\Box_2}^2}
  -10 {{\Box_1}^2} {{\Box_2}^4}
  +8 \Box_1 {{\Box_2}^5}
  -2 {{\Box_2}^6}                          \nonumber\\&&\ \ \ \ \mbox{}
  +74 {{\Box_1}^4} \Box_2 \Box_3
  -192 {{\Box_1}^3} {{\Box_2}^2} \Box_3
  -24 {{\Box_1}^2} {{\Box_2}^3} \Box_3
  +72 \Box_1 {{\Box_2}^4} \Box_3
  +4 {{\Box_2}^5} \Box_3                     \nonumber\\&&\ \ \ \ \mbox{}
  +274 {{\Box_1}^2} {{\Box_2}^2} {{\Box_3}^2}
  -80 \Box_1 {{\Box_2}^3} {{\Box_3}^2}
  +2 {{\Box_2}^4} {{\Box_3}^2}
  -4 {{\Box_2}^3} {{\Box_3}^3}
\Big)
,\end{fleqnarray}

\begin{fleqnarray}&&\Gamma_{9}(-\Box_1,-\Box_2,-\Box_3) =
\Gamma(-\Box_1,-\Box_2,-\Box_3)\frac1{324 {D^6}}\Big(
   3 {{\Box_1}^{12}}
  +18 {{\Box_1}^{11}} \Box_2                 \nonumber\\&&\ \ \ \ \mbox{}
  -72 {{\Box_1}^{10}} {{\Box_2}^2}
  -78 {{\Box_1}^9} {{\Box_2}^3}
  +378 {{\Box_1}^8} {{\Box_2}^4}
  -324 {{\Box_1}^7} {{\Box_2}^5}             \nonumber\\&&\ \ \ \ \mbox{}
  +72 {{\Box_1}^6} {{\Box_2}^6}
  +792 {{\Box_1}^{10}} \Box_2 \Box_3
  -2304 {{\Box_1}^9} {{\Box_2}^2} \Box_3
  -3150 {{\Box_1}^8} {{\Box_2}^3} \Box_3       \nonumber\\&&\ \ \ \ \mbox{}
  +7164 {{\Box_1}^7} {{\Box_2}^4} \Box_3
  -3312 {{\Box_1}^6} {{\Box_2}^5} \Box_3
  +9090 {{\Box_1}^8} {{\Box_2}^2} {{\Box_3}^2} \nonumber\\&&\ \ \ \ \mbox{}
  -16200 {{\Box_1}^7} {{\Box_2}^3} {{\Box_3}^2}
  -1836 {{\Box_1}^6} {{\Box_2}^4} {{\Box_3}^2}
  +2232 {{\Box_1}^5} {{\Box_2}^5} {{\Box_3}^2}     \nonumber\\&&\ \ \ \ \mbox{}
  +16320 {{\Box_1}^6} {{\Box_2}^3} {{\Box_3}^3}
  -6732 {{\Box_1}^5} {{\Box_2}^4} {{\Box_3}^3}
  -2466 {{\Box_1}^4} {{\Box_2}^4} {{\Box_3}^4}
\Big) \nonumber\\&&\ \ \ \ \mbox{}
+\frac{\ln(\Box_1/\Box_2)}{1080 {D^6} \Box_2 \Box_3}\Big(
   6 {{\Box_1}^{12}} \Box_2
  -72 {{\Box_1}^{11}} {{\Box_2}^2}
  +360 {{\Box_1}^{10}} {{\Box_2}^3}
  -984 {{\Box_1}^9} {{\Box_2}^4}                 \nonumber\\&&\ \ \ \ \mbox{}
  +1566 {{\Box_1}^8} {{\Box_2}^5}
  -1296 {{\Box_1}^7} {{\Box_2}^6}
  +3 {{\Box_1}^{12}} \Box_3
  -33 {{\Box_1}^{11}} \Box_2 \Box_3                \nonumber\\&&\ \ \ \ \mbox{}
  +1216 {{\Box_1}^{10}} {{\Box_2}^2} \Box_3
  -3944 {{\Box_1}^9} {{\Box_2}^3} \Box_3
  +3827 {{\Box_1}^8} {{\Box_2}^4} \Box_3           \nonumber\\&&\ \ \ \ \mbox{}
  +1143 {{\Box_1}^7} {{\Box_2}^5} \Box_3
  -5576 {{\Box_1}^6} {{\Box_2}^6} \Box_3
  -36 {{\Box_1}^{11}} {{\Box_3}^2}
  +824 {{\Box_1}^{10}} \Box_2 {{\Box_3}^2}         \nonumber\\&&\ \ \ \ \mbox{}
  +6216 {{\Box_1}^9} {{\Box_2}^2} {{\Box_3}^2}
  -1176 {{\Box_1}^8} {{\Box_2}^3} {{\Box_3}^2}
  -20260 {{\Box_1}^7} {{\Box_2}^4} {{\Box_3}^2}    \nonumber\\&&\ \ \ \ \mbox{}
  +15280 {{\Box_1}^6} {{\Box_2}^5} {{\Box_3}^2}
  +180 {{\Box_1}^{10}} {{\Box_3}^3}
  -2860 {{\Box_1}^9} \Box_2 {{\Box_3}^3}           \nonumber\\&&\ \ \ \ \mbox{}
  -13704 {{\Box_1}^8} {{\Box_2}^2} {{\Box_3}^3}
  +60960 {{\Box_1}^7} {{\Box_2}^3} {{\Box_3}^3}
  -11220 {{\Box_1}^6} {{\Box_2}^4} {{\Box_3}^3}    \nonumber\\&&\ \ \ \ \mbox{}
  -18396 {{\Box_1}^5} {{\Box_2}^5} {{\Box_3}^3}
  -492 {{\Box_1}^9} {{\Box_3}^4}
  +3376 {{\Box_1}^8} \Box_2 {{\Box_3}^4}           \nonumber\\&&\ \ \ \ \mbox{}
  +4276 {{\Box_1}^7} {{\Box_2}^2} {{\Box_3}^4}
  -61416 {{\Box_1}^6} {{\Box_2}^3} {{\Box_3}^4}
  +14112 {{\Box_1}^5} {{\Box_2}^4} {{\Box_3}^4}    \nonumber\\&&\ \ \ \ \mbox{}
  +783 {{\Box_1}^8} {{\Box_3}^5}
  -495 {{\Box_1}^7} \Box_2 {{\Box_3}^5}
  +176 {{\Box_1}^6} {{\Box_2}^2} {{\Box_3}^5}      \nonumber\\&&\ \ \ \ \mbox{}
  +12696 {{\Box_1}^5} {{\Box_2}^3} {{\Box_3}^5}
  +10506 {{\Box_1}^4} {{\Box_2}^4} {{\Box_3}^5}
  -648 {{\Box_1}^7} {{\Box_3}^6}                 \nonumber\\&&\ \ \ \ \mbox{}
  -2500 {{\Box_1}^6} \Box_2 {{\Box_3}^6}
  +4632 {{\Box_1}^5} {{\Box_2}^2} {{\Box_3}^6}
  +31092 {{\Box_1}^4} {{\Box_2}^3} {{\Box_3}^6}    \nonumber\\&&\ \ \ \ \mbox{}
  +3076 {{\Box_1}^5} \Box_2 {{\Box_3}^7}
  -15104 {{\Box_1}^4} {{\Box_2}^2} {{\Box_3}^7}
  -50196 {{\Box_1}^3} {{\Box_2}^3} {{\Box_3}^7}    \nonumber\\&&\ \ \ \ \mbox{}
  +648 {{\Box_1}^5} {{\Box_3}^8}
  -1638 {{\Box_1}^4} \Box_2 {{\Box_3}^8}
  +24536 {{\Box_1}^3} {{\Box_2}^2} {{\Box_3}^8}
  -783 {{\Box_1}^4} {{\Box_3}^9}                 \nonumber\\&&\ \ \ \ \mbox{}
  -451 {{\Box_1}^3} \Box_2 {{\Box_3}^9}
  -12528 {{\Box_1}^2} {{\Box_2}^2} {{\Box_3}^9}
  +492 {{\Box_1}^3} {{\Box_3}^{10}}
  +1084 {{\Box_1}^2} \Box_2 {{\Box_3}^{10}}        \nonumber\\&&\ \ \ \ \mbox{}
  -180 {{\Box_1}^2} {{\Box_3}^{11}}
  -392 \Box_1 \Box_2 {{\Box_3}^{11}}
  +36 \Box_1 {{\Box_3}^{12}}
  -3 {{\Box_3}^{13}}
\Big) \nonumber\\&&\ \ \ \ \mbox{}
+\frac{\ln(\Box_1/\Box_2)}{(\Box_1-\Box_2)}\Big({{-6 \Box_1 + \Box_3}\over {720 \Box_3}}\Big) \nonumber\\&&\ \ \ \ \mbox{}
+\frac1{2160 {D^5} \Box_2 \Box_3}\Big(
  -{{\Box_1}^{11}}
  +32 {{\Box_1}^{10}} \Box_2
  -140 {{\Box_1}^9} {{\Box_2}^2}
  +224 {{\Box_1}^8} {{\Box_2}^3}                 \nonumber\\&&\ \ \ \ \mbox{}
  +66 {{\Box_1}^7} {{\Box_2}^4}
  -768 {{\Box_1}^6} {{\Box_2}^5}
  +588 {{\Box_1}^5} {{\Box_2}^6}
  -418 {{\Box_1}^9} \Box_2 \Box_3                  \nonumber\\&&\ \ \ \ \mbox{}
  +120 {{\Box_1}^8} {{\Box_2}^2} \Box_3
  +4556 {{\Box_1}^7} {{\Box_2}^3} \Box_3
  -6508 {{\Box_1}^6} {{\Box_2}^4} \Box_3           \nonumber\\&&\ \ \ \ \mbox{}
  +2636 {{\Box_1}^5} {{\Box_2}^5} \Box_3
  -10326 {{\Box_1}^7} {{\Box_2}^2} {{\Box_3}^2}
  +20320 {{\Box_1}^6} {{\Box_2}^3} {{\Box_3}^2}    \nonumber\\&&\ \ \ \ \mbox{}
  +760 {{\Box_1}^5} {{\Box_2}^4} {{\Box_3}^2}
  -408 {{\Box_1}^4} {{\Box_2}^5} {{\Box_3}^2}
  -27064 {{\Box_1}^5} {{\Box_2}^3} {{\Box_3}^3}    \nonumber\\&&\ \ \ \ \mbox{}
  +7428 {{\Box_1}^4} {{\Box_2}^4} {{\Box_3}^3}
  +5978 {{\Box_1}^3} {{\Box_2}^4} {{\Box_3}^4}
\Big)
,\end{fleqnarray}

\begin{fleqnarray}&&\Gamma_{10}(-\Box_1,-\Box_2,-\Box_3) =
\Gamma(-\Box_1,-\Box_2,-\Box_3)\frac1{3 {D^3}}\Big(
   6 {{\Box_1}^4} \Box_2 \Box_3                    \nonumber\\&&\ \ \ \ \mbox{}
  -12 {{\Box_1}^3} {{\Box_2}^2} \Box_3
  +8 {{\Box_1}^2} {{\Box_2}^2} {{\Box_3}^2}
\Big) \nonumber\\&&\ \ \ \ \mbox{}
+\frac{\ln(\Box_1/\Box_2)}{9 {D^3}}\Big(
   4 {{\Box_1}^4} \Box_2
  -12 {{\Box_1}^3} {{\Box_2}^2}
  +2 {{\Box_1}^4} \Box_3
  +48 {{\Box_1}^3} \Box_2 \Box_3                   \nonumber\\&&\ \ \ \ \mbox{}
  -6 {{\Box_1}^3} {{\Box_3}^2}
  -54 {{\Box_1}^2} \Box_2 {{\Box_3}^2}
  +6 {{\Box_1}^2} {{\Box_3}^3}
  -2 \Box_1 {{\Box_3}^4}
\Big) \nonumber\\&&\ \ \ \ \mbox{}
+\frac1{540 {D^2} \Box_2 \Box_3}\Big(
  -{{\Box_1}^5}
  +12 {{\Box_1}^4} \Box_2
  -30 {{\Box_1}^3} {{\Box_2}^2}
  +20 {{\Box_1}^2} {{\Box_2}^3}     \nonumber\\&&\ \ \ \ \mbox{}
  -108 {{\Box_1}^3} \Box_2 \Box_3
  +204 {{\Box_1}^2} {{\Box_2}^2} \Box_3
  -418 \Box_1 {{\Box_2}^2} {{\Box_3}^2}
\Big)
,\end{fleqnarray}

\begin{fleqnarray}&&\Gamma_{11}(-\Box_1,-\Box_2,-\Box_3) =
\Gamma(-\Box_1,-\Box_2,-\Box_3)\frac1{3 {D^4}}\Big(
   5 {{\Box_1}^5} \Box_2 {{\Box_3}^2}              \nonumber\\&&\ \ \ \ \mbox{}
  -2 {{\Box_1}^4} {{\Box_2}^2} {{\Box_3}^2}
  -3 {{\Box_1}^3} {{\Box_2}^3} {{\Box_3}^2}
  -11 {{\Box_1}^4} \Box_2 {{\Box_3}^3}
  +17 {{\Box_1}^3} {{\Box_2}^2} {{\Box_3}^3}       \nonumber\\&&\ \ \ \ \mbox{}
  +3 {{\Box_1}^3} \Box_2 {{\Box_3}^4}
  -11 {{\Box_1}^2} {{\Box_2}^2} {{\Box_3}^4}
  +7 {{\Box_1}^2} \Box_2 {{\Box_3}^5}
  -2 \Box_1 \Box_2 {{\Box_3}^6}
\Big) \nonumber\\&&\ \ \ \ \mbox{}
+\frac{\ln(\Box_1/\Box_2)}{540 {D^4} \Box_2}\Big(
   2 {{\Box_1}^8}
  -20 {{\Box_1}^7} \Box_2
  +74 {{\Box_1}^6} {{\Box_2}^2}
  -150 {{\Box_1}^5} {{\Box_2}^3}                 \nonumber\\&&\ \ \ \ \mbox{}
  +202 {{\Box_1}^4} {{\Box_2}^4}
  -20 {{\Box_1}^7} \Box_3
  +144 {{\Box_1}^6} \Box_2 \Box_3
  -208 {{\Box_1}^5} {{\Box_2}^2} \Box_3            \nonumber\\&&\ \ \ \ \mbox{}
  +64 {{\Box_1}^4} {{\Box_2}^3} \Box_3
  +81 {{\Box_1}^6} {{\Box_3}^2}
  -321 {{\Box_1}^5} \Box_2 {{\Box_3}^2}
  +1251 {{\Box_1}^4} {{\Box_2}^2} {{\Box_3}^2}     \nonumber\\&&\ \ \ \ \mbox{}
  +1173 {{\Box_1}^3} {{\Box_2}^3} {{\Box_3}^2}
  -178 {{\Box_1}^5} {{\Box_3}^3}
  +232 {{\Box_1}^4} \Box_2 {{\Box_3}^3}
  -2266 {{\Box_1}^3} {{\Box_2}^2} {{\Box_3}^3}     \nonumber\\&&\ \ \ \ \mbox{}
  +235 {{\Box_1}^4} {{\Box_3}^4}
  +99 {{\Box_1}^3} \Box_2 {{\Box_3}^4}
  +1208 {{\Box_1}^2} {{\Box_2}^2} {{\Box_3}^4}
  -192 {{\Box_1}^3} {{\Box_3}^5}                 \nonumber\\&&\ \ \ \ \mbox{}
  -216 {{\Box_1}^2} \Box_2 {{\Box_3}^5}
  +95 {{\Box_1}^2} {{\Box_3}^6}
  +85 \Box_1 \Box_2 {{\Box_3}^6}
  -26 \Box_1 {{\Box_3}^7}
  +3 {{\Box_3}^8}
\Big) \nonumber\\&&\ \ \ \ \mbox{}
+\frac{\ln(\Box_1/\Box_3)}{540 {D^4} \Box_1 \Box_2}\Big(
   4 {{\Box_1}^9}
  -31 {{\Box_1}^8} \Box_2
  +97 {{\Box_1}^7} {{\Box_2}^2}
  -159 {{\Box_1}^6} {{\Box_2}^3}                 \nonumber\\&&\ \ \ \ \mbox{}
  +143 {{\Box_1}^5} {{\Box_2}^4}
  -59 {{\Box_1}^4} {{\Box_2}^5}
  -9 {{\Box_1}^3} {{\Box_2}^6}
  +23 {{\Box_1}^2} {{\Box_2}^7}
  -11 \Box_1 {{\Box_2}^8}
  +2 {{\Box_2}^9}                              \nonumber\\&&\ \ \ \ \mbox{}
  -40 {{\Box_1}^8} \Box_3
  +216 {{\Box_1}^7} \Box_2 \Box_3
  -368 {{\Box_1}^6} {{\Box_2}^2} \Box_3
  +272 {{\Box_1}^5} {{\Box_2}^3} \Box_3            \nonumber\\&&\ \ \ \ \mbox{}
  -180 {{\Box_1}^4} {{\Box_2}^4} \Box_3
  +208 {{\Box_1}^3} {{\Box_2}^5} \Box_3
  -160 {{\Box_1}^2} {{\Box_2}^6} \Box_3
  +72 \Box_1 {{\Box_2}^7} \Box_3                   \nonumber\\&&\ \ \ \ \mbox{}
  -20 {{\Box_2}^8} \Box_3
  +162 {{\Box_1}^7} {{\Box_3}^2}
  -390 {{\Box_1}^6} \Box_2 {{\Box_3}^2}
  +1395 {{\Box_1}^5} {{\Box_2}^2} {{\Box_3}^2}     \nonumber\\&&\ \ \ \ \mbox{}
  -75 {{\Box_1}^4} {{\Box_2}^3} {{\Box_3}^2}
  -1248 {{\Box_1}^3} {{\Box_2}^4} {{\Box_3}^2}
  +144 {{\Box_1}^2} {{\Box_2}^5} {{\Box_3}^2}
  -69 \Box_1 {{\Box_2}^6} {{\Box_3}^2}             \nonumber\\&&\ \ \ \ \mbox{}
  +81 {{\Box_2}^7} {{\Box_3}^2}
  -356 {{\Box_1}^6} {{\Box_3}^3}
  -40 {{\Box_1}^5} \Box_2 {{\Box_3}^3}
  -1262 {{\Box_1}^4} {{\Box_2}^2} {{\Box_3}^3}     \nonumber\\&&\ \ \ \ \mbox{}
  +2184 {{\Box_1}^3} {{\Box_2}^3} {{\Box_3}^3}
  +1004 {{\Box_1}^2} {{\Box_2}^4} {{\Box_3}^3}
  -272 \Box_1 {{\Box_2}^5} {{\Box_3}^3}
  -178 {{\Box_2}^6} {{\Box_3}^3}                 \nonumber\\&&\ \ \ \ \mbox{}
  +470 {{\Box_1}^5} {{\Box_3}^4}
  +828 {{\Box_1}^4} \Box_2 {{\Box_3}^4}
  -389 {{\Box_1}^3} {{\Box_2}^2} {{\Box_3}^4}
  -1597 {{\Box_1}^2} {{\Box_2}^3} {{\Box_3}^4}     \nonumber\\&&\ \ \ \ \mbox{}
  +729 \Box_1 {{\Box_2}^4} {{\Box_3}^4}
  +235 {{\Box_2}^5} {{\Box_3}^4}
  -384 {{\Box_1}^4} {{\Box_3}^5}
  -936 {{\Box_1}^3} \Box_2 {{\Box_3}^5}            \nonumber\\&&\ \ \ \ \mbox{}
  +216 {{\Box_1}^2} {{\Box_2}^2} {{\Box_3}^5}
  -720 \Box_1 {{\Box_2}^3} {{\Box_3}^5}
  -192 {{\Box_2}^4} {{\Box_3}^5}
  +190 {{\Box_1}^3} {{\Box_3}^6}                 \nonumber\\&&\ \ \ \ \mbox{}
  +422 {{\Box_1}^2} \Box_2 {{\Box_3}^6}
  +337 \Box_1 {{\Box_2}^2} {{\Box_3}^6}
  +95 {{\Box_2}^3} {{\Box_3}^6}
  -52 {{\Box_1}^2} {{\Box_3}^7}                  \nonumber\\&&\ \ \ \ \mbox{}
  -72 \Box_1 \Box_2 {{\Box_3}^7}
  -26 {{\Box_2}^2} {{\Box_3}^7}
  +6 \Box_1 {{\Box_3}^8}
  +3 \Box_2 {{\Box_3}^8}
\Big) \nonumber\\&&\ \ \ \ \mbox{}
+\frac1{120}\frac{\ln(\Box_1/\Box_2)}{(\Box_1-\Box_2)} \nonumber\\&&\ \ \ \ \mbox{}
+\frac1{1080 {D^3} \Box_1 \Box_2}\Big(
   19 {{\Box_1}^7}
  -77 {{\Box_1}^6} \Box_2
  +117 {{\Box_1}^5} {{\Box_2}^2}
  -59 {{\Box_1}^4} {{\Box_2}^3}                  \nonumber\\&&\ \ \ \ \mbox{}
  -100 {{\Box_1}^6} \Box_3
  -48 {{\Box_1}^5} \Box_2 \Box_3
  +228 {{\Box_1}^4} {{\Box_2}^2} \Box_3
  -80 {{\Box_1}^3} {{\Box_2}^3} \Box_3             \nonumber\\&&\ \ \ \ \mbox{}
  +219 {{\Box_1}^5} {{\Box_3}^2}
  +555 {{\Box_1}^4} \Box_2 {{\Box_3}^2}
  -1854 {{\Box_1}^3} {{\Box_2}^2} {{\Box_3}^2}
  -260 {{\Box_1}^4} {{\Box_3}^3}                 \nonumber\\&&\ \ \ \ \mbox{}
  -632 {{\Box_1}^3} \Box_2 {{\Box_3}^3}
  +696 {{\Box_1}^2} {{\Box_2}^2} {{\Box_3}^3}
  +185 {{\Box_1}^3} {{\Box_3}^4}
  +201 {{\Box_1}^2} \Box_2 {{\Box_3}^4}            \nonumber\\&&\ \ \ \ \mbox{}
  -84 {{\Box_1}^2} {{\Box_3}^5}
  -12 \Box_1 \Box_2 {{\Box_3}^5}
  +25 \Box_1 {{\Box_3}^6}
  -2 {{\Box_3}^7}
\Big)
,\end{fleqnarray}

\begin{fleqnarray}&&\Gamma_{12}(-\Box_1,-\Box_2,-\Box_3) =
\Gamma(-\Box_1,-\Box_2,-\Box_3)\frac1{{D^3}}\Big(
  -2 {{\Box_1}^5}                              \nonumber\\&&\ \ \ \ \mbox{}
  +4 {{\Box_1}^4} \Box_2
  -4 {{\Box_1}^2} {{\Box_2}^3}
  +2 \Box_1 {{\Box_2}^4}
  +4 {{\Box_1}^4} \Box_3                         \nonumber\\&&\ \ \ \ \mbox{}
  -24 {{\Box_1}^3} \Box_2 \Box_3
  +12 {{\Box_1}^2} {{\Box_2}^2} \Box_3
  +8 \Box_1 {{\Box_2}^3} \Box_3
  +12 {{\Box_1}^2} \Box_2 {{\Box_3}^2}             \nonumber\\&&\ \ \ \ \mbox{}
  -20 \Box_1 {{\Box_2}^2} {{\Box_3}^2}
  -4 {{\Box_1}^2} {{\Box_3}^3}
  +8 \Box_1 \Box_2 {{\Box_3}^3}
  +2 \Box_1 {{\Box_3}^4}
\Big) \nonumber\\&&\ \ \ \ \mbox{}
+\frac{\ln(\Box_1/\Box_2)}{9 {D^3} \Box_2 \Box_3}\Big(
  -2 {{\Box_1}^5} \Box_2
  +10 {{\Box_1}^4} {{\Box_2}^2}
  -20 {{\Box_1}^3} {{\Box_2}^3}                  \nonumber\\&&\ \ \ \ \mbox{}
  +20 {{\Box_1}^2} {{\Box_2}^4}
  -10 \Box_1 {{\Box_2}^5}
  +2 {{\Box_2}^6}
  -{{\Box_1}^5} \Box_3
  -21 {{\Box_1}^4} \Box_2 \Box_3                   \nonumber\\&&\ \ \ \ \mbox{}
  -6 {{\Box_1}^3} {{\Box_2}^2} \Box_3
  +66 {{\Box_1}^2} {{\Box_2}^3} \Box_3
  -25 \Box_1 {{\Box_2}^4} \Box_3
  -13 {{\Box_2}^5} \Box_3                        \nonumber\\&&\ \ \ \ \mbox{}
  +5 {{\Box_1}^4} {{\Box_3}^2}
  +36 {{\Box_1}^3} \Box_2 {{\Box_3}^2}
  -162 {{\Box_1}^2} {{\Box_2}^2} {{\Box_3}^2}
  -36 \Box_1 {{\Box_2}^3} {{\Box_3}^2}             \nonumber\\&&\ \ \ \ \mbox{}
  +29 {{\Box_2}^4} {{\Box_3}^2}
  -10 {{\Box_1}^3} {{\Box_3}^3}
  -6 {{\Box_1}^2} \Box_2 {{\Box_3}^3}
  +78 \Box_1 {{\Box_2}^2} {{\Box_3}^3}             \nonumber\\&&\ \ \ \ \mbox{}
  -30 {{\Box_2}^3} {{\Box_3}^3}
  +10 {{\Box_1}^2} {{\Box_3}^4}
  -2 \Box_1 \Box_2 {{\Box_3}^4}
  +16 {{\Box_2}^2} {{\Box_3}^4}                  \nonumber\\&&\ \ \ \ \mbox{}
  -5 \Box_1 {{\Box_3}^5}
  -5 \Box_2 {{\Box_3}^5}
  +{{\Box_3}^6}
\Big) \nonumber\\&&\ \ \ \ \mbox{}
+\frac{\ln(\Box_1/\Box_3)}{9 {D^3} \Box_2 \Box_3}\Big(
  -{{\Box_1}^5} \Box_2
  +5 {{\Box_1}^4} {{\Box_2}^2}
  -10 {{\Box_1}^3} {{\Box_2}^3}                  \nonumber\\&&\ \ \ \ \mbox{}
  +10 {{\Box_1}^2} {{\Box_2}^4}
  -5 \Box_1 {{\Box_2}^5}
  +{{\Box_2}^6}
  -2 {{\Box_1}^5} \Box_3                         \nonumber\\&&\ \ \ \ \mbox{}
  -21 {{\Box_1}^4} \Box_2 \Box_3
  +36 {{\Box_1}^3} {{\Box_2}^2} \Box_3
  -6 {{\Box_1}^2} {{\Box_2}^3} \Box_3
  -2 \Box_1 {{\Box_2}^4} \Box_3                    \nonumber\\&&\ \ \ \ \mbox{}
  -5 {{\Box_2}^5} \Box_3
  +10 {{\Box_1}^4} {{\Box_3}^2}
  -6 {{\Box_1}^3} \Box_2 {{\Box_3}^2}
  -162 {{\Box_1}^2} {{\Box_2}^2} {{\Box_3}^2}      \nonumber\\&&\ \ \ \ \mbox{}
  +78 \Box_1 {{\Box_2}^3} {{\Box_3}^2}
  +16 {{\Box_2}^4} {{\Box_3}^2}
  -20 {{\Box_1}^3} {{\Box_3}^3}
  +66 {{\Box_1}^2} \Box_2 {{\Box_3}^3}             \nonumber\\&&\ \ \ \ \mbox{}
  -36 \Box_1 {{\Box_2}^2} {{\Box_3}^3}
  -30 {{\Box_2}^3} {{\Box_3}^3}
  +20 {{\Box_1}^2} {{\Box_3}^4}
  -25 \Box_1 \Box_2 {{\Box_3}^4}                   \nonumber\\&&\ \ \ \ \mbox{}
  +29 {{\Box_2}^2} {{\Box_3}^4}
  -10 \Box_1 {{\Box_3}^5}
  -13 \Box_2 {{\Box_3}^5}
  +2 {{\Box_3}^6}
\Big) \nonumber\\&&\ \ \ \ \mbox{}
+\frac{\ln(\Box_2/\Box_3)}{9 {D^3} \Box_2 \Box_3}\Big(
   {{\Box_1}^5} \Box_2
  -5 {{\Box_1}^4} {{\Box_2}^2}
  +10 {{\Box_1}^3} {{\Box_2}^3}                  \nonumber\\&&\ \ \ \ \mbox{}
  -10 {{\Box_1}^2} {{\Box_2}^4}
  +5 \Box_1 {{\Box_2}^5}
  -{{\Box_2}^6}
  -{{\Box_1}^5} \Box_3                           \nonumber\\&&\ \ \ \ \mbox{}
  +42 {{\Box_1}^3} {{\Box_2}^2} \Box_3
  -72 {{\Box_1}^2} {{\Box_2}^3} \Box_3
  +23 \Box_1 {{\Box_2}^4} \Box_3
  +8 {{\Box_2}^5} \Box_3                         \nonumber\\&&\ \ \ \ \mbox{}
  +5 {{\Box_1}^4} {{\Box_3}^2}
  -42 {{\Box_1}^3} \Box_2 {{\Box_3}^2}
  +114 \Box_1 {{\Box_2}^3} {{\Box_3}^2}
  -13 {{\Box_2}^4} {{\Box_3}^2}                  \nonumber\\&&\ \ \ \ \mbox{}
  -10 {{\Box_1}^3} {{\Box_3}^3}
  +72 {{\Box_1}^2} \Box_2 {{\Box_3}^3}
  -114 \Box_1 {{\Box_2}^2} {{\Box_3}^3}
  +10 {{\Box_1}^2} {{\Box_3}^4}                  \nonumber\\&&\ \ \ \ \mbox{}
  -23 \Box_1 \Box_2 {{\Box_3}^4}
  +13 {{\Box_2}^2} {{\Box_3}^4}
  -5 \Box_1 {{\Box_3}^5}
  -8 \Box_2 {{\Box_3}^5}
  +{{\Box_3}^6}
\Big) \nonumber\\&&\ \ \ \ \mbox{}
+\frac{\ln(\Box_1/\Box_2)}{(\Box_1-\Box_2)}{1\over {3 \Box_3}} \nonumber\\&&\ \ \ \ \mbox{}
+\frac{\ln(\Box_1/\Box_3)}{(\Box_1-\Box_3)}{1\over {3 \Box_2}} \nonumber\\&&\ \ \ \ \mbox{}
+\frac1{3 {D^2}}\Big(
   16 {{\Box_1}^2}
  -12 \Box_1 \Box_2
  -4 {{\Box_2}^2}
  -12 \Box_1 \Box_3
  +8 \Box_2 \Box_3
  -4 {{\Box_3}^2}
\Big)
,\end{fleqnarray}

\begin{fleqnarray}&&\Gamma_{13}(-\Box_1,-\Box_2,-\Box_3) =
\Gamma(-\Box_1,-\Box_2,-\Box_3)\frac1{D}\Big(
   2 \Box_1
  -2 \Box_2
  -2 \Box_3
\Big) \nonumber\\&&\ \ \ \ \mbox{}
+\frac{\ln(\Box_1/\Box_2)}{3 D \Box_1}\Big(
   6 \Box_1
  +2 \Box_2
  -2 \Box_3
\Big) \nonumber\\&&\ \ \ \ \mbox{}
+\frac{\ln(\Box_1/\Box_3)}{3 D \Box_1}\Big(
   6 \Box_1
  -2 \Box_2
  +2 \Box_3
\Big) \nonumber\\&&\ \ \ \ \mbox{}
+\frac{\ln(\Box_2/\Box_3)}{3 D \Box_1}\Big(
  -4 \Box_2
  +4 \Box_3
\Big) \nonumber\\&&\ \ \ \ \mbox{}
+\frac{\ln(\Box_2/\Box_3)}{(\Box_2-\Box_3)}{2\over {\Box_1}}
,\end{fleqnarray}

\begin{fleqnarray}&&\Gamma_{14}(-\Box_1,-\Box_2,-\Box_3) =
\Gamma(-\Box_1,-\Box_2,-\Box_3)\frac1{{D^2}}\Big(
  -2 {{\Box_1}^3}                              \nonumber\\&&\ \ \ \ \mbox{}
  +2 {{\Box_1}^2} \Box_2
  +2 \Box_1 {{\Box_2}^2}
  -2 {{\Box_2}^3}
  +2 {{\Box_1}^2} \Box_3
  -12 \Box_1 \Box_2 \Box_3                         \nonumber\\&&\ \ \ \ \mbox{}
  +2 {{\Box_2}^2} \Box_3
  +2 \Box_1 {{\Box_3}^2}
  +2 \Box_2 {{\Box_3}^2}
  -2 {{\Box_3}^3}
\Big) \nonumber\\&&\ \ \ \ \mbox{}
+\frac{\ln(\Box_1/\Box_2)}{{D^2}}\Big(
  -4 {{\Box_1}^2}
  +4 {{\Box_2}^2}
  +4 \Box_1 \Box_3
  -4 \Box_2 \Box_3
\Big) \nonumber\\&&\ \ \ \ \mbox{}
+\frac{\ln(\Box_1/\Box_3)}{{D^2}}\Big(
  -4 {{\Box_1}^2}
  +4 \Box_1 \Box_2
  -4 \Box_2 \Box_3
  +4 {{\Box_3}^2}
\Big) \nonumber\\&&\ \ \ \ \mbox{}
+\frac{\ln(\Box_2/\Box_3)}{{D^2}}\Big(
   4 \Box_1 \Box_2
  -4 {{\Box_2}^2}
  -4 \Box_1 \Box_3
  +4 {{\Box_3}^2}
\Big) \nonumber\\&&\ \ \ \ \mbox{}
+4\frac1{D}
,\end{fleqnarray}

\begin{fleqnarray}&&\Gamma_{15}(-\Box_1,-\Box_2,-\Box_3) =
\Gamma(-\Box_1,-\Box_2,-\Box_3)\frac1{3 {D^4}}\Big(
   2 {{\Box_1}^7}
  -18 {{\Box_1}^5} {{\Box_2}^2}                  \nonumber\\&&\ \ \ \ \mbox{}
  +32 {{\Box_1}^4} {{\Box_2}^3}
  -18 {{\Box_1}^3} {{\Box_2}^4}
  +2 \Box_1 {{\Box_2}^6}
  +72 {{\Box_1}^5} \Box_2 \Box_3
  -72 {{\Box_1}^4} {{\Box_2}^2} \Box_3             \nonumber\\&&\ \ \ \ \mbox{}
  -72 {{\Box_1}^3} {{\Box_2}^3} \Box_3
  +72 {{\Box_1}^2} {{\Box_2}^4} \Box_3
  -18 {{\Box_1}^5} {{\Box_3}^2}
  -72 {{\Box_1}^4} \Box_2 {{\Box_3}^2}             \nonumber\\&&\ \ \ \ \mbox{}
  +252 {{\Box_1}^3} {{\Box_2}^2} {{\Box_3}^2}
  -72 {{\Box_1}^2} {{\Box_2}^3} {{\Box_3}^2}
  -18 \Box_1 {{\Box_2}^4} {{\Box_3}^2}
  +32 {{\Box_1}^4} {{\Box_3}^3}                  \nonumber\\&&\ \ \ \ \mbox{}
  -72 {{\Box_1}^3} \Box_2 {{\Box_3}^3}
  -72 {{\Box_1}^2} {{\Box_2}^2} {{\Box_3}^3}
  +32 \Box_1 {{\Box_2}^3} {{\Box_3}^3}
  -18 {{\Box_1}^3} {{\Box_3}^4}                  \nonumber\\&&\ \ \ \ \mbox{}
  +72 {{\Box_1}^2} \Box_2 {{\Box_3}^4}
  -18 \Box_1 {{\Box_2}^2} {{\Box_3}^4}
  +2 \Box_1 {{\Box_3}^6}
\Big) \nonumber\\&&\ \ \ \ \mbox{}
+\frac{\ln(\Box_1/\Box_2)}{9 {D^4}}\Big(
   18 {{\Box_1}^6}
  +10 {{\Box_1}^5} \Box_2
  -120 {{\Box_1}^4} {{\Box_2}^2}                 \nonumber\\&&\ \ \ \ \mbox{}
  +120 {{\Box_1}^3} {{\Box_2}^3}
  -10 {{\Box_1}^2} {{\Box_2}^4}
  -18 \Box_1 {{\Box_2}^5}
  -22 {{\Box_1}^5} \Box_3                        \nonumber\\&&\ \ \ \ \mbox{}
  +264 {{\Box_1}^4} \Box_2 \Box_3
  -264 {{\Box_1}^2} {{\Box_2}^3} \Box_3
  +22 \Box_1 {{\Box_2}^4} \Box_3
  -42 {{\Box_1}^4} {{\Box_3}^2}                  \nonumber\\&&\ \ \ \ \mbox{}
  -306 {{\Box_1}^3} \Box_2 {{\Box_3}^2}
  +306 {{\Box_1}^2} {{\Box_2}^2} {{\Box_3}^2}
  +42 \Box_1 {{\Box_2}^3} {{\Box_3}^2}
  +78 {{\Box_1}^3} {{\Box_3}^3}                  \nonumber\\&&\ \ \ \ \mbox{}
  -78 \Box_1 {{\Box_2}^2} {{\Box_3}^3}
  -32 {{\Box_1}^2} {{\Box_3}^4}
  +32 \Box_1 \Box_2 {{\Box_3}^4}
\Big) \nonumber\\&&\ \ \ \ \mbox{}
+\frac{\ln(\Box_1/\Box_3)}{9 {D^4}}\Big(
   18 {{\Box_1}^6}
  -22 {{\Box_1}^5} \Box_2
  -42 {{\Box_1}^4} {{\Box_2}^2}                  \nonumber\\&&\ \ \ \ \mbox{}
  +78 {{\Box_1}^3} {{\Box_2}^3}
  -32 {{\Box_1}^2} {{\Box_2}^4}
  +10 {{\Box_1}^5} \Box_3
  +264 {{\Box_1}^4} \Box_2 \Box_3                  \nonumber\\&&\ \ \ \ \mbox{}
  -306 {{\Box_1}^3} {{\Box_2}^2} \Box_3
  +32 \Box_1 {{\Box_2}^4} \Box_3
  -120 {{\Box_1}^4} {{\Box_3}^2}
  +306 {{\Box_1}^2} {{\Box_2}^2} {{\Box_3}^2}      \nonumber\\&&\ \ \ \ \mbox{}
  -78 \Box_1 {{\Box_2}^3} {{\Box_3}^2}
  +120 {{\Box_1}^3} {{\Box_3}^3}
  -264 {{\Box_1}^2} \Box_2 {{\Box_3}^3}
  +42 \Box_1 {{\Box_2}^2} {{\Box_3}^3}             \nonumber\\&&\ \ \ \ \mbox{}
  -10 {{\Box_1}^2} {{\Box_3}^4}
  +22 \Box_1 \Box_2 {{\Box_3}^4}
  -18 \Box_1 {{\Box_3}^5}
\Big) \nonumber\\&&\ \ \ \ \mbox{}
+\frac{\ln(\Box_2/\Box_3)}{9 {D^4}}\Big(
  -32 {{\Box_1}^5} \Box_2
  +78 {{\Box_1}^4} {{\Box_2}^2}
  -42 {{\Box_1}^3} {{\Box_2}^3}                  \nonumber\\&&\ \ \ \ \mbox{}
  -22 {{\Box_1}^2} {{\Box_2}^4}
  +18 \Box_1 {{\Box_2}^5}
  +32 {{\Box_1}^5} \Box_3
  -306 {{\Box_1}^3} {{\Box_2}^2} \Box_3            \nonumber\\&&\ \ \ \ \mbox{}
  +264 {{\Box_1}^2} {{\Box_2}^3} \Box_3
  +10 \Box_1 {{\Box_2}^4} \Box_3
  -78 {{\Box_1}^4} {{\Box_3}^2}
  +306 {{\Box_1}^3} \Box_2 {{\Box_3}^2}            \nonumber\\&&\ \ \ \ \mbox{}
  -120 \Box_1 {{\Box_2}^3} {{\Box_3}^2}
  +42 {{\Box_1}^3} {{\Box_3}^3}
  -264 {{\Box_1}^2} \Box_2 {{\Box_3}^3}
  +120 \Box_1 {{\Box_2}^2} {{\Box_3}^3}            \nonumber\\&&\ \ \ \ \mbox{}
  +22 {{\Box_1}^2} {{\Box_3}^4}
  -10 \Box_1 \Box_2 {{\Box_3}^4}
  -18 \Box_1 {{\Box_3}^5}
\Big) \nonumber\\&&\ \ \ \ \mbox{}
+\frac1{3 {D^3}}\Big(
  -13 {{\Box_1}^4}
  +13 {{\Box_1}^3} \Box_2
  +13 {{\Box_1}^2} {{\Box_2}^2}                  \nonumber\\&&\ \ \ \ \mbox{}
  -13 \Box_1 {{\Box_2}^3}
  +13 {{\Box_1}^3} \Box_3
  -62 {{\Box_1}^2} \Box_2 \Box_3
  +13 \Box_1 {{\Box_2}^2} \Box_3                   \nonumber\\&&\ \ \ \ \mbox{}
  +13 {{\Box_1}^2} {{\Box_3}^2}
  +13 \Box_1 \Box_2 {{\Box_3}^2}
  -13 \Box_1 {{\Box_3}^3}
\Big)
,\end{fleqnarray}

\begin{fleqnarray}&&\Gamma_{16}(-\Box_1,-\Box_2,-\Box_3) =
\Gamma(-\Box_1,-\Box_2,-\Box_3)\frac1{{D^3}}\Big(
   16 {{\Box_1}^3} \Box_2 \Box_3                   \nonumber\\&&\ \ \ \ \mbox{}
  -16 {{\Box_1}^2} {{\Box_2}^2} \Box_3
  +8 {{\Box_1}^2} \Box_2 {{\Box_3}^2}
  -12 \Box_1 \Box_2 {{\Box_3}^3}
\Big) \nonumber\\&&\ \ \ \ \mbox{}
+\frac{\ln(\Box_1/\Box_2)}{9 {D^3} \Box_2}\Big(
  -2 {{\Box_1}^5}
  +12 {{\Box_1}^4} \Box_2
  -18 {{\Box_1}^3} {{\Box_2}^2}                  \nonumber\\&&\ \ \ \ \mbox{}
  +8 {{\Box_1}^4} \Box_3
  +168 {{\Box_1}^2} {{\Box_2}^2} \Box_3
  -12 {{\Box_1}^3} {{\Box_3}^2}
  -36 {{\Box_1}^2} \Box_2 {{\Box_3}^2}             \nonumber\\&&\ \ \ \ \mbox{}
  +8 {{\Box_1}^2} {{\Box_3}^3}
  +24 \Box_1 \Box_2 {{\Box_3}^3}
  -2 \Box_1 {{\Box_3}^4}
\Big) \nonumber\\&&\ \ \ \ \mbox{}
+\frac{\ln(\Box_1/\Box_3)}{9 {D^3} \Box_1 \Box_2}\Big(
  -4 {{\Box_1}^6}
  +24 {{\Box_1}^5} \Box_2
  -54 {{\Box_1}^4} {{\Box_2}^2}                  \nonumber\\&&\ \ \ \ \mbox{}
  +60 {{\Box_1}^3} {{\Box_2}^3}
  -36 {{\Box_1}^2} {{\Box_2}^4}
  +12 \Box_1 {{\Box_2}^5}
  -2 {{\Box_2}^6}                              \nonumber\\&&\ \ \ \ \mbox{}
  +16 {{\Box_1}^5} \Box_3
  +72 {{\Box_1}^3} {{\Box_2}^2} \Box_3
  -96 {{\Box_1}^2} {{\Box_2}^3} \Box_3
  +8 {{\Box_2}^5} \Box_3                         \nonumber\\&&\ \ \ \ \mbox{}
  -24 {{\Box_1}^4} {{\Box_3}^2}
  -72 {{\Box_1}^3} \Box_2 {{\Box_3}^2}
  +216 {{\Box_1}^2} {{\Box_2}^2} {{\Box_3}^2}
  -36 \Box_1 {{\Box_2}^3} {{\Box_3}^2}             \nonumber\\&&\ \ \ \ \mbox{}
  -12 {{\Box_2}^4} {{\Box_3}^2}
  +16 {{\Box_1}^3} {{\Box_3}^3}
  +48 {{\Box_1}^2} \Box_2 {{\Box_3}^3}
  +24 \Box_1 {{\Box_2}^2} {{\Box_3}^3}             \nonumber\\&&\ \ \ \ \mbox{}
  +8 {{\Box_2}^3} {{\Box_3}^3}
  -4 {{\Box_1}^2} {{\Box_3}^4}
  -2 {{\Box_2}^2} {{\Box_3}^4}
\Big) \nonumber\\&&\ \ \ \ \mbox{}
+\frac1{6 {D^2} \Box_1 \Box_2}\Big(
   6 {{\Box_1}^4}
  -32 {{\Box_1}^3} \Box_2
  +26 {{\Box_1}^2} {{\Box_2}^2}
  -20 {{\Box_1}^3} \Box_3
  -4 {{\Box_1}^2} \Box_2 \Box_3                    \nonumber\\&&\ \ \ \ \mbox{}
  +24 {{\Box_1}^2} {{\Box_3}^2}
  +24 \Box_1 \Box_2 {{\Box_3}^2}
  -12 \Box_1 {{\Box_3}^3}
  +{{\Box_3}^4}
\Big)
,\end{fleqnarray}

\begin{fleqnarray}&&\Gamma_{17}(-\Box_1,-\Box_2,-\Box_3) =
\Gamma(-\Box_1,-\Box_2,-\Box_3)\frac1{{D^2}}\Big(
   2 {{\Box_1}^3}                              \nonumber\\&&\ \ \ \ \mbox{}
  -4 {{\Box_1}^2} \Box_2
  +2 \Box_1 {{\Box_2}^2}
  -4 {{\Box_1}^2} \Box_3
  +8 \Box_1 \Box_2 \Box_3
  +2 \Box_1 {{\Box_3}^2}
\Big) \nonumber\\&&\ \ \ \ \mbox{}
+\frac{\ln(\Box_1/\Box_2)}{3 {D^2} \Box_1}\Big(
   9 {{\Box_1}^3}
  -5 {{\Box_1}^2} \Box_2
  -5 \Box_1 {{\Box_2}^2}
  +{{\Box_2}^3}                                \nonumber\\&&\ \ \ \ \mbox{}
  -13 {{\Box_1}^2} \Box_3
  -3 {{\Box_2}^2} \Box_3
  +5 \Box_1 {{\Box_3}^2}
  +3 \Box_2 {{\Box_3}^2}
  -{{\Box_3}^3}
\Big) \nonumber\\&&\ \ \ \ \mbox{}
+\frac{\ln(\Box_1/\Box_3)}{3 {D^2} \Box_1}\Big(
   9 {{\Box_1}^3}
  -13 {{\Box_1}^2} \Box_2
  +5 \Box_1 {{\Box_2}^2}
  -{{\Box_2}^3}
  -5 {{\Box_1}^2} \Box_3                          \nonumber\\&&\ \ \ \ \mbox{}
  +3 {{\Box_2}^2} \Box_3
  -5 \Box_1 {{\Box_3}^2}
  -3 \Box_2 {{\Box_3}^2}
  +{{\Box_3}^3}
\Big) \nonumber\\&&\ \ \ \ \mbox{}
+\frac{\ln(\Box_2/\Box_3)}{3 {D^2} \Box_1}\Big(
  -8 {{\Box_1}^2} \Box_2
  +10 \Box_1 {{\Box_2}^2}
  -2 {{\Box_2}^3}
  +8 {{\Box_1}^2} \Box_3                         \nonumber\\&&\ \ \ \ \mbox{}
  +6 {{\Box_2}^2} \Box_3
  -10 \Box_1 {{\Box_3}^2}
  -6 \Box_2 {{\Box_3}^2}
  +2 {{\Box_3}^3}
\Big) \nonumber\\&&\ \ \ \ \mbox{}
+\frac{\ln(\Box_2/\Box_3)}{(\Box_2-\Box_3)}{1\over {\Box_1}} \nonumber\\&&\ \ \ \ \mbox{}
+(-2)\frac1{D}
,\end{fleqnarray}

\begin{fleqnarray}&&\Gamma_{18}(-\Box_1,-\Box_2,-\Box_3) =
\Gamma(-\Box_1,-\Box_2,-\Box_3)\frac1{{D^4}}\Big(
   2 {{\Box_1}^7}                              \nonumber\\&&\ \ \ \ \mbox{}
  -16 {{\Box_1}^6} \Box_2
  +20 {{\Box_1}^5} {{\Box_2}^2}
  -20 {{\Box_1}^3} {{\Box_2}^4}
  +16 {{\Box_1}^2} {{\Box_2}^5}                  \nonumber\\&&\ \ \ \ \mbox{}
  -4 \Box_1 {{\Box_2}^6}
  +56 {{\Box_1}^5} \Box_2 \Box_3
  -192 {{\Box_1}^4} {{\Box_2}^2} \Box_3            \nonumber\\&&\ \ \ \ \mbox{}
  +64 {{\Box_1}^3} {{\Box_2}^3} \Box_3
  +80 {{\Box_1}^2} {{\Box_2}^4} \Box_3
  -48 \Box_1 {{\Box_2}^5} \Box_3
  +180 {{\Box_1}^3} {{\Box_2}^2} {{\Box_3}^2}      \nonumber\\&&\ \ \ \ \mbox{}
  -176 {{\Box_1}^2} {{\Box_2}^3} {{\Box_3}^2}
  -12 \Box_1 {{\Box_2}^4} {{\Box_3}^2}
  +64 \Box_1 {{\Box_2}^3} {{\Box_3}^3}
\Big) \nonumber\\&&\ \ \ \ \mbox{}
+\frac{\ln(\Box_1/\Box_2)}{9 {D^4} \Box_1}\Big(
   78 {{\Box_1}^7}
  -186 {{\Box_1}^6} \Box_2
  -42 {{\Box_1}^5} {{\Box_2}^2}                  \nonumber\\&&\ \ \ \ \mbox{}
  +470 {{\Box_1}^4} {{\Box_2}^3}
  -470 {{\Box_1}^3} {{\Box_2}^4}
  +162 {{\Box_1}^2} {{\Box_2}^5}
  -14 \Box_1 {{\Box_2}^6}                        \nonumber\\&&\ \ \ \ \mbox{}
  +2 {{\Box_2}^7}
  -294 {{\Box_1}^6} \Box_3
  +1056 {{\Box_1}^5} \Box_2 \Box_3
  -582 {{\Box_1}^4} {{\Box_2}^2} \Box_3            \nonumber\\&&\ \ \ \ \mbox{}
  -752 {{\Box_1}^3} {{\Box_2}^3} \Box_3
  +478 {{\Box_1}^2} {{\Box_2}^4} \Box_3
  +112 \Box_1 {{\Box_2}^5} \Box_3
  -18 {{\Box_2}^6} \Box_3                        \nonumber\\&&\ \ \ \ \mbox{}
  +402 {{\Box_1}^5} {{\Box_3}^2}
  -1554 {{\Box_1}^4} \Box_2 {{\Box_3}^2}
  +1284 {{\Box_1}^3} {{\Box_2}^2} {{\Box_3}^2}
  -12 {{\Box_1}^2} {{\Box_2}^3} {{\Box_3}^2}       \nonumber\\&&\ \ \ \ \mbox{}
  -182 \Box_1 {{\Box_2}^4} {{\Box_3}^2}
  +62 {{\Box_2}^5} {{\Box_3}^2}
  -230 {{\Box_1}^4} {{\Box_3}^3}
  +608 {{\Box_1}^3} \Box_2 {{\Box_3}^3}            \nonumber\\&&\ \ \ \ \mbox{}
  -780 {{\Box_1}^2} {{\Box_2}^2} {{\Box_3}^3}
  -110 {{\Box_2}^4} {{\Box_3}^3}
  +50 {{\Box_1}^3} {{\Box_3}^4}
  +170 {{\Box_1}^2} \Box_2 {{\Box_3}^4}            \nonumber\\&&\ \ \ \ \mbox{}
  +182 \Box_1 {{\Box_2}^2} {{\Box_3}^4}
  +110 {{\Box_2}^3} {{\Box_3}^4}
  -18 {{\Box_1}^2} {{\Box_3}^5}
  -112 \Box_1 \Box_2 {{\Box_3}^5}                  \nonumber\\&&\ \ \ \ \mbox{}
  -62 {{\Box_2}^2} {{\Box_3}^5}
  +14 \Box_1 {{\Box_3}^6}
  +18 \Box_2 {{\Box_3}^6}
  -2 {{\Box_3}^7}
\Big) \nonumber\\&&\ \ \ \ \mbox{}
+\frac{\ln(\Box_2/\Box_3)}{9 {D^4} \Box_1}\Big(
  -108 {{\Box_1}^6} \Box_2
  +444 {{\Box_1}^5} {{\Box_2}^2}
  -700 {{\Box_1}^4} {{\Box_2}^3}                 \nonumber\\&&\ \ \ \ \mbox{}
  +520 {{\Box_1}^3} {{\Box_2}^4}
  -180 {{\Box_1}^2} {{\Box_2}^5}
  +28 \Box_1 {{\Box_2}^6}
  -4 {{\Box_2}^7}                              \nonumber\\&&\ \ \ \ \mbox{}
  -972 {{\Box_1}^4} {{\Box_2}^2} \Box_3
  +1360 {{\Box_1}^3} {{\Box_2}^3} \Box_3
  -308 {{\Box_1}^2} {{\Box_2}^4} \Box_3            \nonumber\\&&\ \ \ \ \mbox{}
  -224 \Box_1 {{\Box_2}^5} \Box_3
  +36 {{\Box_2}^6} \Box_3
  -768 {{\Box_1}^2} {{\Box_2}^3} {{\Box_3}^2}
  +364 \Box_1 {{\Box_2}^4} {{\Box_3}^2}            \nonumber\\&&\ \ \ \ \mbox{}
  -124 {{\Box_2}^5} {{\Box_3}^2}
  +220 {{\Box_2}^4} {{\Box_3}^3}
\Big) \nonumber\\&&\ \ \ \ \mbox{}
+\frac1{3 {D^3} \Box_1}\Big(
  -20 {{\Box_1}^5}
  +106 {{\Box_1}^4} \Box_2
  -76 {{\Box_1}^3} {{\Box_2}^2}
  -8 {{\Box_1}^2} {{\Box_2}^3}                   \nonumber\\&&\ \ \ \ \mbox{}
  +20 \Box_1 {{\Box_2}^4}
  -2 {{\Box_2}^5}
  -128 {{\Box_1}^3} \Box_2 \Box_3
  +128 {{\Box_1}^2} {{\Box_2}^2} \Box_3            \nonumber\\&&\ \ \ \ \mbox{}
  +16 \Box_1 {{\Box_2}^3} \Box_3
  +6 {{\Box_2}^4} \Box_3
  -36 \Box_1 {{\Box_2}^2} {{\Box_3}^2}
  -4 {{\Box_2}^3} {{\Box_3}^2}
\Big)
,\end{fleqnarray}

\begin{fleqnarray}&&\Gamma_{19}(-\Box_1,-\Box_2,-\Box_3) =
\Gamma(-\Box_1,-\Box_2,-\Box_3)\frac1{{D^4}}\Big(
  -8 {{\Box_1}^5} \Box_2 \Box_3                    \nonumber\\&&\ \ \ \ \mbox{}
  +24 {{\Box_1}^4} {{\Box_2}^2} \Box_3
  +24 {{\Box_1}^3} {{\Box_2}^3} \Box_3
  -56 {{\Box_1}^2} {{\Box_2}^4} \Box_3
  +24 \Box_1 {{\Box_2}^5} \Box_3                   \nonumber\\&&\ \ \ \ \mbox{}
  -72 {{\Box_1}^3} {{\Box_2}^2} {{\Box_3}^2}
  +96 {{\Box_1}^2} {{\Box_2}^3} {{\Box_3}^2}
  +24 \Box_1 {{\Box_2}^4} {{\Box_3}^2}
  -48 \Box_1 {{\Box_2}^3} {{\Box_3}^3}
\Big) \nonumber\\&&\ \ \ \ \mbox{}
+\frac{\ln(\Box_1/\Box_2)}{9 {D^4} \Box_1}\Big(
  -12 {{\Box_1}^6} \Box_2
  +48 {{\Box_1}^5} {{\Box_2}^2}
  -72 {{\Box_1}^4} {{\Box_2}^3}                  \nonumber\\&&\ \ \ \ \mbox{}
  +48 {{\Box_1}^3} {{\Box_2}^4}
  -12 {{\Box_1}^2} {{\Box_2}^5}
  -6 {{\Box_1}^6} \Box_3
  -228 {{\Box_1}^5} \Box_2 \Box_3                  \nonumber\\&&\ \ \ \ \mbox{}
  +78 {{\Box_1}^4} {{\Box_2}^2} \Box_3
  +528 {{\Box_1}^3} {{\Box_2}^3} \Box_3
  -346 {{\Box_1}^2} {{\Box_2}^4} \Box_3
  -28 \Box_1 {{\Box_2}^5} \Box_3                   \nonumber\\&&\ \ \ \ \mbox{}
  +2 {{\Box_2}^6} \Box_3
  +24 {{\Box_1}^5} {{\Box_3}^2}
  +402 {{\Box_1}^4} \Box_2 {{\Box_3}^2}
  -912 {{\Box_1}^3} {{\Box_2}^2} {{\Box_3}^2}      \nonumber\\&&\ \ \ \ \mbox{}
  -72 {{\Box_1}^2} {{\Box_2}^3} {{\Box_3}^2}
  +56 \Box_1 {{\Box_2}^4} {{\Box_3}^2}
  -10 {{\Box_2}^5} {{\Box_3}^2}
  -36 {{\Box_1}^4} {{\Box_3}^3}                  \nonumber\\&&\ \ \ \ \mbox{}
  -48 {{\Box_1}^3} \Box_2 {{\Box_3}^3}
  +576 {{\Box_1}^2} {{\Box_2}^2} {{\Box_3}^3}
  +20 {{\Box_2}^4} {{\Box_3}^3}
  +24 {{\Box_1}^3} {{\Box_3}^4}                  \nonumber\\&&\ \ \ \ \mbox{}
  -140 {{\Box_1}^2} \Box_2 {{\Box_3}^4}
  -56 \Box_1 {{\Box_2}^2} {{\Box_3}^4}
  -20 {{\Box_2}^3} {{\Box_3}^4}
  -6 {{\Box_1}^2} {{\Box_3}^5}                   \nonumber\\&&\ \ \ \ \mbox{}
  +28 \Box_1 \Box_2 {{\Box_3}^5}
  +10 {{\Box_2}^2} {{\Box_3}^5}
  -2 \Box_2 {{\Box_3}^6}
\Big) \nonumber\\&&\ \ \ \ \mbox{}
+\frac{\ln(\Box_2/\Box_3)}{9 {D^4} \Box_1}\Big(
   6 {{\Box_1}^6} \Box_2
  -24 {{\Box_1}^5} {{\Box_2}^2}
  +36 {{\Box_1}^4} {{\Box_2}^3}                  \nonumber\\&&\ \ \ \ \mbox{}
  -24 {{\Box_1}^3} {{\Box_2}^4}
  +6 {{\Box_1}^2} {{\Box_2}^5}
  +324 {{\Box_1}^4} {{\Box_2}^2} \Box_3
  -576 {{\Box_1}^3} {{\Box_2}^3} \Box_3            \nonumber\\&&\ \ \ \ \mbox{}
  +206 {{\Box_1}^2} {{\Box_2}^4} \Box_3
  +56 \Box_1 {{\Box_2}^5} \Box_3
  -4 {{\Box_2}^6} \Box_3
  +648 {{\Box_1}^2} {{\Box_2}^3} {{\Box_3}^2}      \nonumber\\&&\ \ \ \ \mbox{}
  -112 \Box_1 {{\Box_2}^4} {{\Box_3}^2}
  +20 {{\Box_2}^5} {{\Box_3}^2}
  -40 {{\Box_2}^4} {{\Box_3}^3}
\Big) \nonumber\\&&\ \ \ \ \mbox{}
+\frac{\ln(\Box_2/\Box_3)}{(\Box_2-\Box_3)}{1\over {6 \Box_1}} \nonumber\\&&\ \ \ \ \mbox{}
+\frac1{3 {D^3} \Box_1}\Big(
   {{\Box_1}^5}
  -{{\Box_1}^4} \Box_2
  -10 {{\Box_1}^3} {{\Box_2}^2}
  +16 {{\Box_1}^2} {{\Box_2}^3}                  \nonumber\\&&\ \ \ \ \mbox{}
  -8 \Box_1 {{\Box_2}^4}
  +{{\Box_2}^5}
  +48 {{\Box_1}^3} \Box_2 \Box_3
  -76 {{\Box_1}^2} {{\Box_2}^2} \Box_3             \nonumber\\&&\ \ \ \ \mbox{}
  -16 \Box_1 {{\Box_2}^3} \Box_3
  -3 {{\Box_2}^4} \Box_3
  +24 \Box_1 {{\Box_2}^2} {{\Box_3}^2}
  +2 {{\Box_2}^3} {{\Box_3}^2}
\Big)
,\end{fleqnarray}

\begin{fleqnarray}&&\Gamma_{20}(-\Box_1,-\Box_2,-\Box_3) =
\Gamma(-\Box_1,-\Box_2,-\Box_3)\frac1{3 {D^4}}\Big(
  -2 {{\Box_1}^7}                              \nonumber\\&&\ \ \ \ \mbox{}
  +14 {{\Box_1}^6} \Box_2
  -12 {{\Box_1}^5} {{\Box_2}^2}
  -10 {{\Box_1}^4} {{\Box_2}^3}
  +20 {{\Box_1}^3} {{\Box_2}^4}                  \nonumber\\&&\ \ \ \ \mbox{}
  -6 {{\Box_1}^2} {{\Box_2}^5}
  -4 \Box_1 {{\Box_2}^6}
  +2 {{\Box_2}^7}
  -60 {{\Box_1}^5} \Box_2 \Box_3                   \nonumber\\&&\ \ \ \ \mbox{}
  +198 {{\Box_1}^4} {{\Box_2}^2} \Box_3
  -16 {{\Box_1}^3} {{\Box_2}^3} \Box_3
  -150 {{\Box_1}^2} {{\Box_2}^4} \Box_3
  +72 \Box_1 {{\Box_2}^5} \Box_3                   \nonumber\\&&\ \ \ \ \mbox{}
  +2 {{\Box_2}^6} \Box_3
  -216 {{\Box_1}^3} {{\Box_2}^2} {{\Box_3}^2}
  +228 {{\Box_1}^2} {{\Box_2}^3} {{\Box_3}^2}
  +36 \Box_1 {{\Box_2}^4} {{\Box_3}^2}             \nonumber\\&&\ \ \ \ \mbox{}
  -18 {{\Box_2}^5} {{\Box_3}^2}
  -104 \Box_1 {{\Box_2}^3} {{\Box_3}^3}
  +14 {{\Box_2}^4} {{\Box_3}^3}
\Big) \nonumber\\&&\ \ \ \ \mbox{}
+\frac{\ln(\Box_1/\Box_2)}{9 {D^4} \Box_1}\Big(
  -27 {{\Box_1}^7}
  +49 {{\Box_1}^6} \Box_2
  +61 {{\Box_1}^5} {{\Box_2}^2}                  \nonumber\\&&\ \ \ \ \mbox{}
  -195 {{\Box_1}^4} {{\Box_2}^3}
  +135 {{\Box_1}^3} {{\Box_2}^4}
  -13 {{\Box_1}^2} {{\Box_2}^5}
  -9 \Box_1 {{\Box_2}^6}                         \nonumber\\&&\ \ \ \ \mbox{}
  -{{\Box_2}^7}
  +89 {{\Box_1}^6} \Box_3
  -384 {{\Box_1}^5} \Box_2 \Box_3
  +189 {{\Box_1}^4} {{\Box_2}^2} \Box_3            \nonumber\\&&\ \ \ \ \mbox{}
  +400 {{\Box_1}^3} {{\Box_2}^3} \Box_3
  -269 {{\Box_1}^2} {{\Box_2}^4} \Box_3
  -32 \Box_1 {{\Box_2}^5} \Box_3
  +7 {{\Box_2}^6} \Box_3                         \nonumber\\&&\ \ \ \ \mbox{}
  -91 {{\Box_1}^5} {{\Box_3}^2}
  +591 {{\Box_1}^4} \Box_2 {{\Box_3}^2}
  -594 {{\Box_1}^3} {{\Box_2}^2} {{\Box_3}^2}
  +6 {{\Box_1}^2} {{\Box_2}^3} {{\Box_3}^2}        \nonumber\\&&\ \ \ \ \mbox{}
  +109 \Box_1 {{\Box_2}^4} {{\Box_3}^2}
  -21 {{\Box_2}^5} {{\Box_3}^2}
  +15 {{\Box_1}^4} {{\Box_3}^3}
  -172 {{\Box_1}^3} \Box_2 {{\Box_3}^3}            \nonumber\\&&\ \ \ \ \mbox{}
  +414 {{\Box_1}^2} {{\Box_2}^2} {{\Box_3}^3}
  -36 \Box_1 {{\Box_2}^3} {{\Box_3}^3}
  +35 {{\Box_2}^4} {{\Box_3}^3}
  +15 {{\Box_1}^3} {{\Box_3}^4}                  \nonumber\\&&\ \ \ \ \mbox{}
  -145 {{\Box_1}^2} \Box_2 {{\Box_3}^4}
  -91 \Box_1 {{\Box_2}^2} {{\Box_3}^4}
  -35 {{\Box_2}^3} {{\Box_3}^4}
  +7 {{\Box_1}^2} {{\Box_3}^5}                   \nonumber\\&&\ \ \ \ \mbox{}
  +68 \Box_1 \Box_2 {{\Box_3}^5}
  +21 {{\Box_2}^2} {{\Box_3}^5}
  -9 \Box_1 {{\Box_3}^6}
  -7 \Box_2 {{\Box_3}^6}
  +{{\Box_3}^7}
\Big) \nonumber\\&&\ \ \ \ \mbox{}
+\frac{\ln(\Box_2/\Box_3)}{9 {D^4} \Box_1}\Big(
   40 {{\Box_1}^6} \Box_2
  -152 {{\Box_1}^5} {{\Box_2}^2}
  +210 {{\Box_1}^4} {{\Box_2}^3}                 \nonumber\\&&\ \ \ \ \mbox{}
  -120 {{\Box_1}^3} {{\Box_2}^4}
  +20 {{\Box_1}^2} {{\Box_2}^5}
  +2 {{\Box_2}^7}
  +402 {{\Box_1}^4} {{\Box_2}^2} \Box_3            \nonumber\\&&\ \ \ \ \mbox{}
  -572 {{\Box_1}^3} {{\Box_2}^3} \Box_3
  +124 {{\Box_1}^2} {{\Box_2}^4} \Box_3
  +100 \Box_1 {{\Box_2}^5} \Box_3
  -14 {{\Box_2}^6} \Box_3                        \nonumber\\&&\ \ \ \ \mbox{}
  +408 {{\Box_1}^2} {{\Box_2}^3} {{\Box_3}^2}
  -200 \Box_1 {{\Box_2}^4} {{\Box_3}^2}
  +42 {{\Box_2}^5} {{\Box_3}^2}
  -70 {{\Box_2}^4} {{\Box_3}^3}
\Big) \nonumber\\&&\ \ \ \ \mbox{}
+\frac{\ln(\Box_2/\Box_3)}{(\Box_2-\Box_3)}\Big({{-1}\over {6 \Box_1}}\Big) \nonumber\\&&\ \ \ \ \mbox{}
+\frac1{6 {D^3}}\Big(
   15 {{\Box_1}^4}
  -68 {{\Box_1}^3} \Box_2
  +24 {{\Box_1}^2} {{\Box_2}^2}
  +36 \Box_1 {{\Box_2}^3}
  -22 {{\Box_2}^4}                             \nonumber\\&&\ \ \ \ \mbox{}
  +96 {{\Box_1}^2} \Box_2 \Box_3
  -108 \Box_1 {{\Box_2}^2} \Box_3
  -16 {{\Box_2}^3} \Box_3
  +38 {{\Box_2}^2} {{\Box_3}^2}
\Big)
,\end{fleqnarray}

\begin{fleqnarray}&&\Gamma_{21}(-\Box_1,-\Box_2,-\Box_3) =
\Gamma(-\Box_1,-\Box_2,-\Box_3)\frac1{{D^4}}\Big(
  -8 {{\Box_1}^6} \Box_3                         \nonumber\\&&\ \ \ \ \mbox{}
  +8 {{\Box_1}^5} \Box_2 \Box_3
  +48 {{\Box_1}^4} {{\Box_2}^2} \Box_3
  -112 {{\Box_1}^3} {{\Box_2}^3} \Box_3
  +88 {{\Box_1}^2} {{\Box_2}^4} \Box_3             \nonumber\\&&\ \ \ \ \mbox{}
  -24 \Box_1 {{\Box_2}^5} \Box_3
  +24 {{\Box_1}^5} {{\Box_3}^2}
  -144 {{\Box_1}^4} \Box_2 {{\Box_3}^2}
  +144 {{\Box_1}^3} {{\Box_2}^2} {{\Box_3}^2}      \nonumber\\&&\ \ \ \ \mbox{}
  +48 {{\Box_1}^2} {{\Box_2}^3} {{\Box_3}^2}
  -72 \Box_1 {{\Box_2}^4} {{\Box_3}^2}
  -16 {{\Box_1}^4} {{\Box_3}^3}
  +192 {{\Box_1}^3} \Box_2 {{\Box_3}^3}            \nonumber\\&&\ \ \ \ \mbox{}
  -336 {{\Box_1}^2} {{\Box_2}^2} {{\Box_3}^3}
  +128 \Box_1 {{\Box_2}^3} {{\Box_3}^3}
  -16 {{\Box_1}^3} {{\Box_3}^4}
  +16 {{\Box_1}^2} \Box_2 {{\Box_3}^4}             \nonumber\\&&\ \ \ \ \mbox{}
  +48 \Box_1 {{\Box_2}^2} {{\Box_3}^4}
  +24 {{\Box_1}^2} {{\Box_3}^5}
  -72 \Box_1 \Box_2 {{\Box_3}^5}
  -8 \Box_1 {{\Box_3}^6}
\Big) \nonumber\\&&\ \ \ \ \mbox{}
+\frac{\ln(\Box_1/\Box_2)}{9 {D^4} \Box_1}\Big(
  -8 {{\Box_1}^7}
  +40 {{\Box_1}^6} \Box_2
  -80 {{\Box_1}^5} {{\Box_2}^2}                  \nonumber\\&&\ \ \ \ \mbox{}
  +80 {{\Box_1}^4} {{\Box_2}^3}
  -40 {{\Box_1}^3} {{\Box_2}^4}
  +8 {{\Box_1}^2} {{\Box_2}^5}
  -116 {{\Box_1}^6} \Box_3                       \nonumber\\&&\ \ \ \ \mbox{}
  -80 {{\Box_1}^5} \Box_2 \Box_3
  +900 {{\Box_1}^4} {{\Box_2}^2} \Box_3
  -1056 {{\Box_1}^3} {{\Box_2}^3} \Box_3
  +308 {{\Box_1}^2} {{\Box_2}^4} \Box_3            \nonumber\\&&\ \ \ \ \mbox{}
  +48 \Box_1 {{\Box_2}^5} \Box_3
  -4 {{\Box_2}^6} \Box_3
  +340 {{\Box_1}^5} {{\Box_3}^2}
  -1044 {{\Box_1}^4} \Box_2 {{\Box_3}^2}           \nonumber\\&&\ \ \ \ \mbox{}
  +96 {{\Box_1}^3} {{\Box_2}^2} {{\Box_3}^2}
  +672 {{\Box_1}^2} {{\Box_2}^3} {{\Box_3}^2}
  -84 \Box_1 {{\Box_2}^4} {{\Box_3}^2}
  +20 {{\Box_2}^5} {{\Box_3}^2}                  \nonumber\\&&\ \ \ \ \mbox{}
  -240 {{\Box_1}^4} {{\Box_3}^3}
  +1248 {{\Box_1}^3} \Box_2 {{\Box_3}^3}
  -936 {{\Box_1}^2} {{\Box_2}^2} {{\Box_3}^3}
  -32 \Box_1 {{\Box_2}^3} {{\Box_3}^3}             \nonumber\\&&\ \ \ \ \mbox{}
  -40 {{\Box_2}^4} {{\Box_3}^3}
  -40 {{\Box_1}^3} {{\Box_3}^4}
  -120 {{\Box_1}^2} \Box_2 {{\Box_3}^4}
  +120 \Box_1 {{\Box_2}^2} {{\Box_3}^4}            \nonumber\\&&\ \ \ \ \mbox{}
  +40 {{\Box_2}^3} {{\Box_3}^4}
  +68 {{\Box_1}^2} {{\Box_3}^5}
  -48 \Box_1 \Box_2 {{\Box_3}^5}
  -20 {{\Box_2}^2} {{\Box_3}^5}                  \nonumber\\&&\ \ \ \ \mbox{}
  -4 \Box_1 {{\Box_3}^6}
  +4 \Box_2 {{\Box_3}^6}
\Big) \nonumber\\&&\ \ \ \ \mbox{}
+\frac{\ln(\Box_1/\Box_3)}{9 {D^4} \Box_1}\Big(
  -4 {{\Box_1}^7}
  +20 {{\Box_1}^6} \Box_2
  -40 {{\Box_1}^5} {{\Box_2}^2}                  \nonumber\\&&\ \ \ \ \mbox{}
  +40 {{\Box_1}^4} {{\Box_2}^3}
  -20 {{\Box_1}^3} {{\Box_2}^4}
  +4 {{\Box_1}^2} {{\Box_2}^5}
  -160 {{\Box_1}^6} \Box_3                       \nonumber\\&&\ \ \ \ \mbox{}
  +320 {{\Box_1}^5} \Box_2 \Box_3
  +36 {{\Box_1}^4} {{\Box_2}^2} \Box_3
  -432 {{\Box_1}^3} {{\Box_2}^3} \Box_3
  +280 {{\Box_1}^2} {{\Box_2}^4} \Box_3            \nonumber\\&&\ \ \ \ \mbox{}
  -48 \Box_1 {{\Box_2}^5} \Box_3
  +4 {{\Box_2}^6} \Box_3
  +284 {{\Box_1}^5} {{\Box_3}^2}
  -1548 {{\Box_1}^4} \Box_2 {{\Box_3}^2}           \nonumber\\&&\ \ \ \ \mbox{}
  +1824 {{\Box_1}^3} {{\Box_2}^2} {{\Box_3}^2}
  -624 {{\Box_1}^2} {{\Box_2}^3} {{\Box_3}^2}
  +84 \Box_1 {{\Box_2}^4} {{\Box_3}^2}
  -20 {{\Box_2}^5} {{\Box_3}^2}                  \nonumber\\&&\ \ \ \ \mbox{}
  +144 {{\Box_1}^4} {{\Box_3}^3}
  +336 {{\Box_1}^3} \Box_2 {{\Box_3}^3}
  -696 {{\Box_1}^2} {{\Box_2}^2} {{\Box_3}^3}
  +32 \Box_1 {{\Box_2}^3} {{\Box_3}^3}             \nonumber\\&&\ \ \ \ \mbox{}
  +40 {{\Box_2}^4} {{\Box_3}^3}
  -476 {{\Box_1}^3} {{\Box_3}^4}
  +828 {{\Box_1}^2} \Box_2 {{\Box_3}^4}
  -120 \Box_1 {{\Box_2}^2} {{\Box_3}^4}            \nonumber\\&&\ \ \ \ \mbox{}
  -40 {{\Box_2}^3} {{\Box_3}^4}
  +208 {{\Box_1}^2} {{\Box_3}^5}
  +48 \Box_1 \Box_2 {{\Box_3}^5}
  +20 {{\Box_2}^2} {{\Box_3}^5}                  \nonumber\\&&\ \ \ \ \mbox{}
  +4 \Box_1 {{\Box_3}^6}
  -4 \Box_2 {{\Box_3}^6}
\Big) \nonumber\\&&\ \ \ \ \mbox{}
+\frac{\ln(\Box_2/\Box_3)}{9 {D^4} \Box_1}\Big(
   4 {{\Box_1}^7}
  -20 {{\Box_1}^6} \Box_2
  +40 {{\Box_1}^5} {{\Box_2}^2}
  -40 {{\Box_1}^4} {{\Box_2}^3}                  \nonumber\\&&\ \ \ \ \mbox{}
  +20 {{\Box_1}^3} {{\Box_2}^4}
  -4 {{\Box_1}^2} {{\Box_2}^5}
  -44 {{\Box_1}^6} \Box_3
  +400 {{\Box_1}^5} \Box_2 \Box_3                  \nonumber\\&&\ \ \ \ \mbox{}
  -864 {{\Box_1}^4} {{\Box_2}^2} \Box_3
  +624 {{\Box_1}^3} {{\Box_2}^3} \Box_3
  -28 {{\Box_1}^2} {{\Box_2}^4} \Box_3
  -96 \Box_1 {{\Box_2}^5} \Box_3                   \nonumber\\&&\ \ \ \ \mbox{}
  +8 {{\Box_2}^6} \Box_3
  -56 {{\Box_1}^5} {{\Box_3}^2}
  -504 {{\Box_1}^4} \Box_2 {{\Box_3}^2}
  +1728 {{\Box_1}^3} {{\Box_2}^2} {{\Box_3}^2}     \nonumber\\&&\ \ \ \ \mbox{}
  -1296 {{\Box_1}^2} {{\Box_2}^3} {{\Box_3}^2}
  +168 \Box_1 {{\Box_2}^4} {{\Box_3}^2}
  -40 {{\Box_2}^5} {{\Box_3}^2}
  +384 {{\Box_1}^4} {{\Box_3}^3}                 \nonumber\\&&\ \ \ \ \mbox{}
  -912 {{\Box_1}^3} \Box_2 {{\Box_3}^3}
  +240 {{\Box_1}^2} {{\Box_2}^2} {{\Box_3}^3}
  +64 \Box_1 {{\Box_2}^3} {{\Box_3}^3}
  +80 {{\Box_2}^4} {{\Box_3}^3}                  \nonumber\\&&\ \ \ \ \mbox{}
  -436 {{\Box_1}^3} {{\Box_3}^4}
  +948 {{\Box_1}^2} \Box_2 {{\Box_3}^4}
  -240 \Box_1 {{\Box_2}^2} {{\Box_3}^4}
  -80 {{\Box_2}^3} {{\Box_3}^4}                  \nonumber\\&&\ \ \ \ \mbox{}
  +140 {{\Box_1}^2} {{\Box_3}^5}
  +96 \Box_1 \Box_2 {{\Box_3}^5}
  +40 {{\Box_2}^2} {{\Box_3}^5}
  +8 \Box_1 {{\Box_3}^6}
  -8 \Box_2 {{\Box_3}^6}
\Big) \nonumber\\&&\ \ \ \ \mbox{}
+\frac{\ln(\Box_2/\Box_3)}{(\Box_2-\Box_3)}\Big({{-2}\over {3 \Box_1}}\Big) \nonumber\\&&\ \ \ \ \mbox{}
+\frac1{3 {D^3} \Box_1}\Big(
   6 {{\Box_1}^5}
  -26 {{\Box_1}^4} \Box_2
  +44 {{\Box_1}^3} {{\Box_2}^2}
  -36 {{\Box_1}^2} {{\Box_2}^3}
  +14 \Box_1 {{\Box_2}^4}                        \nonumber\\&&\ \ \ \ \mbox{}
  -2 {{\Box_2}^5}
  +66 {{\Box_1}^4} \Box_3
  -92 {{\Box_1}^3} \Box_2 \Box_3
  -8 {{\Box_1}^2} {{\Box_2}^2} \Box_3
  +28 \Box_1 {{\Box_2}^3} \Box_3                   \nonumber\\&&\ \ \ \ \mbox{}
  +6 {{\Box_2}^4} \Box_3
  -128 {{\Box_1}^3} {{\Box_3}^2}
  +252 {{\Box_1}^2} \Box_2 {{\Box_3}^2}
  -72 \Box_1 {{\Box_2}^2} {{\Box_3}^2}             \nonumber\\&&\ \ \ \ \mbox{}
  -4 {{\Box_2}^3} {{\Box_3}^2}
  +32 {{\Box_1}^2} {{\Box_3}^3}
  +4 \Box_1 \Box_2 {{\Box_3}^3}
  -4 {{\Box_2}^2} {{\Box_3}^3}
  +26 \Box_1 {{\Box_3}^4}                        \nonumber\\&&\ \ \ \ \mbox{}
  +6 \Box_2 {{\Box_3}^4}
  -2 {{\Box_3}^5}
\Big)
,\end{fleqnarray}

\begin{fleqnarray}&&\Gamma_{22}(-\Box_1,-\Box_2,-\Box_3) =
\Gamma(-\Box_1,-\Box_2,-\Box_3)\frac1{18 {D^6}}\Big(
  -{{\Box_1}^{11}}                             \nonumber\\&&\ \ \ \ \mbox{}
  -4 {{\Box_1}^{10}} \Box_2
  +30 {{\Box_1}^9} {{\Box_2}^2}
  -156 {{\Box_1}^7} {{\Box_2}^4}
  +264 {{\Box_1}^6} {{\Box_2}^5}                 \nonumber\\&&\ \ \ \ \mbox{}
  -156 {{\Box_1}^5} {{\Box_2}^6}
  +30 {{\Box_1}^3} {{\Box_2}^8}
  -4 {{\Box_1}^2} {{\Box_2}^9}
  -2 \Box_1 {{\Box_2}^{10}}                      \nonumber\\&&\ \ \ \ \mbox{}
  -184 {{\Box_1}^9} \Box_2 \Box_3
  +352 {{\Box_1}^8} {{\Box_2}^2} \Box_3
  +1672 {{\Box_1}^7} {{\Box_2}^3} \Box_3           \nonumber\\&&\ \ \ \ \mbox{}
  -2720 {{\Box_1}^6} {{\Box_2}^4} \Box_3
  +304 {{\Box_1}^5} {{\Box_2}^5} \Box_3
  +1168 {{\Box_1}^4} {{\Box_2}^6} \Box_3           \nonumber\\&&\ \ \ \ \mbox{}
  -152 {{\Box_1}^3} {{\Box_2}^7} \Box_3
  -332 {{\Box_1}^2} {{\Box_2}^8} \Box_3
  +80 \Box_1 {{\Box_2}^9} \Box_3                   \nonumber\\&&\ \ \ \ \mbox{}
  -1884 {{\Box_1}^7} {{\Box_2}^2} {{\Box_3}^2}
  +1320 {{\Box_1}^6} {{\Box_2}^3} {{\Box_3}^2}
  +7036 {{\Box_1}^5} {{\Box_2}^4} {{\Box_3}^2}     \nonumber\\&&\ \ \ \ \mbox{}
  -3456 {{\Box_1}^4} {{\Box_2}^5} {{\Box_3}^2}
  -3240 {{\Box_1}^3} {{\Box_2}^6} {{\Box_3}^2}
  +2008 {{\Box_1}^2} {{\Box_2}^7} {{\Box_3}^2}     \nonumber\\&&\ \ \ \ \mbox{}
  -282 \Box_1 {{\Box_2}^8} {{\Box_3}^2}
  -4640 {{\Box_1}^5} {{\Box_2}^3} {{\Box_3}^3}
  +1136 {{\Box_1}^4} {{\Box_2}^4} {{\Box_3}^3}     \nonumber\\&&\ \ \ \ \mbox{}
  +7512 {{\Box_1}^3} {{\Box_2}^5} {{\Box_3}^3}
  -3992 {{\Box_1}^2} {{\Box_2}^6} {{\Box_3}^3}
  +192 \Box_1 {{\Box_2}^7} {{\Box_3}^3}            \nonumber\\&&\ \ \ \ \mbox{}
  -4150 {{\Box_1}^3} {{\Box_2}^4} {{\Box_3}^4}
  +2320 {{\Box_1}^2} {{\Box_2}^5} {{\Box_3}^4}
  +540 \Box_1 {{\Box_2}^6} {{\Box_3}^4}            \nonumber\\&&\ \ \ \ \mbox{}
  -528 \Box_1 {{\Box_2}^5} {{\Box_3}^5}
\Big) \nonumber\\&&\ \ \ \ \mbox{}
+\frac{\ln(\Box_1/\Box_2)}{270 {D^6} \Box_1 \Box_2 \Box_3}\Big(
  -12 {{\Box_1}^{12}} \Box_2
  +144 {{\Box_1}^{11}} {{\Box_2}^2}
  -756 {{\Box_1}^{10}} {{\Box_2}^3}              \nonumber\\&&\ \ \ \ \mbox{}
  +2316 {{\Box_1}^9} {{\Box_2}^4}
  -4632 {{\Box_1}^8} {{\Box_2}^5}
  +6384 {{\Box_1}^7} {{\Box_2}^6}                \nonumber\\&&\ \ \ \ \mbox{}
  -6216 {{\Box_1}^6} {{\Box_2}^7}
  +4296 {{\Box_1}^5} {{\Box_2}^8}
  -2076 {{\Box_1}^4} {{\Box_2}^9}                \nonumber\\&&\ \ \ \ \mbox{}
  +672 {{\Box_1}^3} {{\Box_2}^{10}}
  -132 {{\Box_1}^2} {{\Box_2}^{11}}
  +12 \Box_1 {{\Box_2}^{12}}                     \nonumber\\&&\ \ \ \ \mbox{}
  -6 {{\Box_1}^{12}} \Box_3
  +117 {{\Box_1}^{11}} \Box_2 \Box_3
  -2129 {{\Box_1}^{10}} {{\Box_2}^2} \Box_3        \nonumber\\&&\ \ \ \ \mbox{}
  +7757 {{\Box_1}^9} {{\Box_2}^3} \Box_3
  -10791 {{\Box_1}^8} {{\Box_2}^4} \Box_3
  +4142 {{\Box_1}^7} {{\Box_2}^5} \Box_3           \nonumber\\&&\ \ \ \ \mbox{}
  +5386 {{\Box_1}^6} {{\Box_2}^6} \Box_3
  -8706 {{\Box_1}^5} {{\Box_2}^7} \Box_3
  +7208 {{\Box_1}^4} {{\Box_2}^8} \Box_3           \nonumber\\&&\ \ \ \ \mbox{}
  -4259 {{\Box_1}^3} {{\Box_2}^9} \Box_3
  +1479 {{\Box_1}^2} {{\Box_2}^{10}} \Box_3
  -203 \Box_1 {{\Box_2}^{11}} \Box_3               \nonumber\\&&\ \ \ \ \mbox{}
  +5 {{\Box_2}^{12}} \Box_3
  +72 {{\Box_1}^{11}} {{\Box_3}^2}
  -1567 {{\Box_1}^{10}} \Box_2 {{\Box_3}^2}        \nonumber\\&&\ \ \ \ \mbox{}
  -2598 {{\Box_1}^9} {{\Box_2}^2} {{\Box_3}^2}
  -12647 {{\Box_1}^8} {{\Box_2}^3} {{\Box_3}^2}
  +47868 {{\Box_1}^7} {{\Box_2}^4} {{\Box_3}^2}    \nonumber\\&&\ \ \ \ \mbox{}
  -18226 {{\Box_1}^6} {{\Box_2}^5} {{\Box_3}^2}
  -31616 {{\Box_1}^5} {{\Box_2}^6} {{\Box_3}^2}
  +10014 {{\Box_1}^4} {{\Box_2}^7} {{\Box_3}^2}    \nonumber\\&&\ \ \ \ \mbox{}
  +15452 {{\Box_1}^3} {{\Box_2}^8} {{\Box_3}^2}
  -7919 {{\Box_1}^2} {{\Box_2}^9} {{\Box_3}^2}
  +1222 \Box_1 {{\Box_2}^{10}} {{\Box_3}^2}        \nonumber\\&&\ \ \ \ \mbox{}
  -55 {{\Box_2}^{11}} {{\Box_3}^2}
  -378 {{\Box_1}^{10}} {{\Box_3}^3}
  +6091 {{\Box_1}^9} \Box_2 {{\Box_3}^3}           \nonumber\\&&\ \ \ \ \mbox{}
  +1199 {{\Box_1}^8} {{\Box_2}^2} {{\Box_3}^3}
  -49968 {{\Box_1}^7} {{\Box_2}^3} {{\Box_3}^3}
  -22068 {{\Box_1}^6} {{\Box_2}^4} {{\Box_3}^3}    \nonumber\\&&\ \ \ \ \mbox{}
  +84806 {{\Box_1}^5} {{\Box_2}^5} {{\Box_3}^3}
  +1758 {{\Box_1}^4} {{\Box_2}^6} {{\Box_3}^3}
  -36528 {{\Box_1}^3} {{\Box_2}^7} {{\Box_3}^3}    \nonumber\\&&\ \ \ \ \mbox{}
  +18734 {{\Box_1}^2} {{\Box_2}^8} {{\Box_3}^3}
  -3921 \Box_1 {{\Box_2}^9} {{\Box_3}^3}
  +275 {{\Box_2}^{10}} {{\Box_3}^3}              \nonumber\\&&\ \ \ \ \mbox{}
  +1158 {{\Box_1}^9} {{\Box_3}^4}
  -10413 {{\Box_1}^8} \Box_2 {{\Box_3}^4}
  +27708 {{\Box_1}^7} {{\Box_2}^2} {{\Box_3}^4}    \nonumber\\&&\ \ \ \ \mbox{}
  +43152 {{\Box_1}^6} {{\Box_2}^3} {{\Box_3}^4}
  -83700 {{\Box_1}^5} {{\Box_2}^4} {{\Box_3}^4}
  -26906 {{\Box_1}^4} {{\Box_2}^5} {{\Box_3}^4}    \nonumber\\&&\ \ \ \ \mbox{}
  +59036 {{\Box_1}^3} {{\Box_2}^6} {{\Box_3}^4}
  -16944 {{\Box_1}^2} {{\Box_2}^7} {{\Box_3}^4}
  +7734 \Box_1 {{\Box_2}^8} {{\Box_3}^4}           \nonumber\\&&\ \ \ \ \mbox{}
  -825 {{\Box_2}^9} {{\Box_3}^4}
  -2316 {{\Box_1}^8} {{\Box_3}^5}
  +8794 {{\Box_1}^7} \Box_2 {{\Box_3}^5}           \nonumber\\&&\ \ \ \ \mbox{}
  -30746 {{\Box_1}^6} {{\Box_2}^2} {{\Box_3}^5}
  +55810 {{\Box_1}^5} {{\Box_2}^3} {{\Box_3}^5}
  +42470 {{\Box_1}^4} {{\Box_2}^4} {{\Box_3}^5}    \nonumber\\&&\ \ \ \ \mbox{}
  -59838 {{\Box_1}^3} {{\Box_2}^5} {{\Box_3}^5}
  -5906 {{\Box_1}^2} {{\Box_2}^6} {{\Box_3}^5}
  -9918 \Box_1 {{\Box_2}^7} {{\Box_3}^5}           \nonumber\\&&\ \ \ \ \mbox{}
  +1650 {{\Box_2}^8} {{\Box_3}^5}
  +3192 {{\Box_1}^7} {{\Box_3}^6}
  -3454 {{\Box_1}^6} \Box_2 {{\Box_3}^6}           \nonumber\\&&\ \ \ \ \mbox{}
  -6040 {{\Box_1}^5} {{\Box_2}^2} {{\Box_3}^6}
  -48270 {{\Box_1}^4} {{\Box_2}^3} {{\Box_3}^6}
  +28900 {{\Box_1}^3} {{\Box_2}^4} {{\Box_3}^6}    \nonumber\\&&\ \ \ \ \mbox{}
  +23762 {{\Box_1}^2} {{\Box_2}^5} {{\Box_3}^6}
  +8316 \Box_1 {{\Box_2}^6} {{\Box_3}^6}
  -2310 {{\Box_2}^7} {{\Box_3}^6}                \nonumber\\&&\ \ \ \ \mbox{}
  -3108 {{\Box_1}^6} {{\Box_3}^7}
  +282 {{\Box_1}^5} \Box_2 {{\Box_3}^7}
  +16386 {{\Box_1}^4} {{\Box_2}^2} {{\Box_3}^7}    \nonumber\\&&\ \ \ \ \mbox{}
  +120 {{\Box_1}^3} {{\Box_2}^3} {{\Box_3}^7}
  -17796 {{\Box_1}^2} {{\Box_2}^4} {{\Box_3}^7}
  -4338 \Box_1 {{\Box_2}^5} {{\Box_3}^7}           \nonumber\\&&\ \ \ \ \mbox{}
  +2310 {{\Box_2}^6} {{\Box_3}^7}
  +2148 {{\Box_1}^5} {{\Box_3}^8}
  +454 {{\Box_1}^4} \Box_2 {{\Box_3}^8}            \nonumber\\&&\ \ \ \ \mbox{}
  -3404 {{\Box_1}^3} {{\Box_2}^2} {{\Box_3}^8}
  +5116 {{\Box_1}^2} {{\Box_2}^3} {{\Box_3}^8}
  +1176 \Box_1 {{\Box_2}^4} {{\Box_3}^8}           \nonumber\\&&\ \ \ \ \mbox{}
  -1650 {{\Box_2}^5} {{\Box_3}^8}
  -1038 {{\Box_1}^4} {{\Box_3}^9}
  -487 {{\Box_1}^3} \Box_2 {{\Box_3}^9}            \nonumber\\&&\ \ \ \ \mbox{}
  -541 {{\Box_1}^2} {{\Box_2}^2} {{\Box_3}^9}
  -39 \Box_1 {{\Box_2}^3} {{\Box_3}^9}
  +825 {{\Box_2}^4} {{\Box_3}^9}                 \nonumber\\&&\ \ \ \ \mbox{}
  +336 {{\Box_1}^3} {{\Box_3}^{10}}
  +213 {{\Box_1}^2} \Box_2 {{\Box_3}^{10}}
  -34 \Box_1 {{\Box_2}^2} {{\Box_3}^{10}}          \nonumber\\&&\ \ \ \ \mbox{}
  -275 {{\Box_2}^3} {{\Box_3}^{10}}
  -66 {{\Box_1}^2} {{\Box_3}^{11}}
  -13 \Box_1 \Box_2 {{\Box_3}^{11}}
  +55 {{\Box_2}^2} {{\Box_3}^{11}}               \nonumber\\&&\ \ \ \ \mbox{}
  +6 \Box_1 {{\Box_3}^{12}}
  -5 \Box_2 {{\Box_3}^{12}}
\Big) \nonumber\\&&\ \ \ \ \mbox{}
+\frac{\ln(\Box_2/\Box_3)}{135 {D^6} \Box_1 \Box_3}\Big(
   3 {{\Box_1}^{12}}
  -36 {{\Box_1}^{11}} \Box_2
  +189 {{\Box_1}^{10}} {{\Box_2}^2}              \nonumber\\&&\ \ \ \ \mbox{}
  -579 {{\Box_1}^9} {{\Box_2}^3}
  +1158 {{\Box_1}^8} {{\Box_2}^4}
  -1596 {{\Box_1}^7} {{\Box_2}^5}                \nonumber\\&&\ \ \ \ \mbox{}
  +1554 {{\Box_1}^6} {{\Box_2}^6}
  -1074 {{\Box_1}^5} {{\Box_2}^7}
  +519 {{\Box_1}^4} {{\Box_2}^8}                 \nonumber\\&&\ \ \ \ \mbox{}
  -168 {{\Box_1}^3} {{\Box_2}^9}
  +33 {{\Box_1}^2} {{\Box_2}^{10}}
  -3 \Box_1 {{\Box_2}^{11}}                      \nonumber\\&&\ \ \ \ \mbox{}
  +281 {{\Box_1}^{10}} \Box_2 \Box_3
  -833 {{\Box_1}^9} {{\Box_2}^2} \Box_3
  +189 {{\Box_1}^8} {{\Box_2}^3} \Box_3            \nonumber\\&&\ \ \ \ \mbox{}
  +2326 {{\Box_1}^7} {{\Box_2}^4} \Box_3
  -4420 {{\Box_1}^6} {{\Box_2}^5} \Box_3
  +4494 {{\Box_1}^5} {{\Box_2}^6} \Box_3           \nonumber\\&&\ \ \ \ \mbox{}
  -3377 {{\Box_1}^4} {{\Box_2}^7} \Box_3
  +1886 {{\Box_1}^3} {{\Box_2}^8} \Box_3
  -633 {{\Box_1}^2} {{\Box_2}^9} \Box_3            \nonumber\\&&\ \ \ \ \mbox{}
  +95 \Box_1 {{\Box_2}^{10}} \Box_3
  -5 {{\Box_2}^{11}} \Box_3
  +6923 {{\Box_1}^8} {{\Box_2}^2} {{\Box_3}^2}     \nonumber\\&&\ \ \ \ \mbox{}
  -10080 {{\Box_1}^7} {{\Box_2}^3} {{\Box_3}^2}
  -6260 {{\Box_1}^6} {{\Box_2}^4} {{\Box_3}^2}
  +12788 {{\Box_1}^5} {{\Box_2}^5} {{\Box_3}^2}    \nonumber\\&&\ \ \ \ \mbox{}
  +3186 {{\Box_1}^4} {{\Box_2}^6} {{\Box_3}^2}
  -9428 {{\Box_1}^3} {{\Box_2}^7} {{\Box_3}^2}
  +3689 {{\Box_1}^2} {{\Box_2}^8} {{\Box_3}^2}     \nonumber\\&&\ \ \ \ \mbox{}
  -628 \Box_1 {{\Box_2}^9} {{\Box_3}^2}
  +55 {{\Box_2}^{10}} {{\Box_3}^2}
  +32610 {{\Box_1}^6} {{\Box_2}^3} {{\Box_3}^3}    \nonumber\\&&\ \ \ \ \mbox{}
  -14498 {{\Box_1}^5} {{\Box_2}^4} {{\Box_3}^3}
  -25014 {{\Box_1}^4} {{\Box_2}^5} {{\Box_3}^3}
  +18324 {{\Box_1}^3} {{\Box_2}^6} {{\Box_3}^3}    \nonumber\\&&\ \ \ \ \mbox{}
  -6809 {{\Box_1}^2} {{\Box_2}^7} {{\Box_3}^3}
  +1941 \Box_1 {{\Box_2}^8} {{\Box_3}^3}
  -275 {{\Box_2}^9} {{\Box_3}^3}                 \nonumber\\&&\ \ \ \ \mbox{}
  +34688 {{\Box_1}^4} {{\Box_2}^4} {{\Box_3}^4}
  -15068 {{\Box_1}^3} {{\Box_2}^5} {{\Box_3}^4}
  -426 {{\Box_1}^2} {{\Box_2}^6} {{\Box_3}^4}      \nonumber\\&&\ \ \ \ \mbox{}
  -3279 \Box_1 {{\Box_2}^7} {{\Box_3}^4}
  +825 {{\Box_2}^8} {{\Box_3}^4}
  +14834 {{\Box_1}^2} {{\Box_2}^5} {{\Box_3}^5}    \nonumber\\&&\ \ \ \ \mbox{}
  +2790 \Box_1 {{\Box_2}^6} {{\Box_3}^5}
  -1650 {{\Box_2}^7} {{\Box_3}^5}
  +2310 {{\Box_2}^6} {{\Box_3}^6}
\Big) \nonumber\\&&\ \ \ \ \mbox{}
+\frac{\ln(\Box_1/\Box_2)}{(\Box_1-\Box_2)}{1\over {30 \Box_3}} \nonumber\\&&\ \ \ \ \mbox{}
+\frac{\ln(\Box_2/\Box_3)}{(\Box_2-\Box_3)}{1\over {90 \Box_1}} \nonumber\\&&\ \ \ \ \mbox{}
+\frac1{270 {D^5} \Box_1 \Box_3}\Big(
  -7 {{\Box_1}^{10}}
  +37 {{\Box_1}^9} \Box_2
  -63 {{\Box_1}^8} {{\Box_2}^2}                  \nonumber\\&&\ \ \ \ \mbox{}
  -6 {{\Box_1}^7} {{\Box_2}^3}
  +168 {{\Box_1}^6} {{\Box_2}^4}
  -252 {{\Box_1}^5} {{\Box_2}^5}
  +168 {{\Box_1}^4} {{\Box_2}^6}                 \nonumber\\&&\ \ \ \ \mbox{}
  -42 {{\Box_1}^3} {{\Box_2}^7}
  -9 {{\Box_1}^2} {{\Box_2}^8}
  +7 \Box_1 {{\Box_2}^9}                         \nonumber\\&&\ \ \ \ \mbox{}
  -{{\Box_2}^{10}}
  +205 {{\Box_1}^9} \Box_3
  +179 {{\Box_1}^8} \Box_2 \Box_3
  -3712 {{\Box_1}^7} {{\Box_2}^2} \Box_3           \nonumber\\&&\ \ \ \ \mbox{}
  +5174 {{\Box_1}^6} {{\Box_2}^3} \Box_3
  -1150 {{\Box_1}^5} {{\Box_2}^4} \Box_3
  -1528 {{\Box_1}^4} {{\Box_2}^5} \Box_3           \nonumber\\&&\ \ \ \ \mbox{}
  +428 {{\Box_1}^3} {{\Box_2}^6} \Box_3
  +269 {{\Box_1}^2} {{\Box_2}^7} \Box_3
  -72 \Box_1 {{\Box_2}^8} \Box_3                   \nonumber\\&&\ \ \ \ \mbox{}
  +9 {{\Box_2}^9} \Box_3
  +4988 {{\Box_1}^7} \Box_2 {{\Box_3}^2}
  -3970 {{\Box_1}^6} {{\Box_2}^2} {{\Box_3}^2}     \nonumber\\&&\ \ \ \ \mbox{}
  -20416 {{\Box_1}^5} {{\Box_2}^3} {{\Box_3}^2}
  +8116 {{\Box_1}^4} {{\Box_2}^4} {{\Box_3}^2}
  +9800 {{\Box_1}^3} {{\Box_2}^5} {{\Box_3}^2}     \nonumber\\&&\ \ \ \ \mbox{}
  -3874 {{\Box_1}^2} {{\Box_2}^6} {{\Box_3}^2}
  +187 \Box_1 {{\Box_2}^7} {{\Box_3}^2}
  -35 {{\Box_2}^8} {{\Box_3}^2}                  \nonumber\\&&\ \ \ \ \mbox{}
  +14802 {{\Box_1}^5} {{\Box_2}^2} {{\Box_3}^3}
  +1884 {{\Box_1}^4} {{\Box_2}^3} {{\Box_3}^3}
  -23084 {{\Box_1}^3} {{\Box_2}^4} {{\Box_3}^3}    \nonumber\\&&\ \ \ \ \mbox{}
  +10098 {{\Box_1}^2} {{\Box_2}^5} {{\Box_3}^3}
  -32 \Box_1 {{\Box_2}^6} {{\Box_3}^3}
  +75 {{\Box_2}^7} {{\Box_3}^3}                  \nonumber\\&&\ \ \ \ \mbox{}
  +12898 {{\Box_1}^3} {{\Box_2}^3} {{\Box_3}^4}
  -6484 {{\Box_1}^2} {{\Box_2}^4} {{\Box_3}^4}
  -578 \Box_1 {{\Box_2}^5} {{\Box_3}^4}            \nonumber\\&&\ \ \ \ \mbox{}
  -90 {{\Box_2}^6} {{\Box_3}^4}
  +488 \Box_1 {{\Box_2}^4} {{\Box_3}^5}
  +42 {{\Box_2}^5} {{\Box_3}^5}
\Big)
,\end{fleqnarray}

\begin{fleqnarray}&&\Gamma_{23}(-\Box_1,-\Box_2,-\Box_3) =
\Gamma(-\Box_1,-\Box_2,-\Box_3)\frac1{3 {D^5}}\Big(
  -8 {{\Box_1}^7} \Box_2 \Box_3                    \nonumber\\&&\ \ \ \ \mbox{}
  +72 {{\Box_1}^5} {{\Box_2}^3} \Box_3
  -64 {{\Box_1}^4} {{\Box_2}^4} \Box_3
  +28 {{\Box_1}^6} \Box_2 {{\Box_3}^2}
  -228 {{\Box_1}^5} {{\Box_2}^2} {{\Box_3}^2}      \nonumber\\&&\ \ \ \ \mbox{}
  +200 {{\Box_1}^4} {{\Box_2}^3} {{\Box_3}^2}
  -32 {{\Box_1}^5} \Box_2 {{\Box_3}^3}
  +280 {{\Box_1}^4} {{\Box_2}^2} {{\Box_3}^3}      \nonumber\\&&\ \ \ \ \mbox{}
  -296 {{\Box_1}^3} {{\Box_2}^3} {{\Box_3}^3}
  +8 {{\Box_1}^4} \Box_2 {{\Box_3}^4}
  +120 {{\Box_1}^3} {{\Box_2}^2} {{\Box_3}^4}      \nonumber\\&&\ \ \ \ \mbox{}
  +8 {{\Box_1}^3} \Box_2 {{\Box_3}^5}
  -84 {{\Box_1}^2} {{\Box_2}^2} {{\Box_3}^5}
  -4 {{\Box_1}^2} \Box_2 {{\Box_3}^6}
\Big) \nonumber\\&&\ \ \ \ \mbox{}
+\frac{\ln(\Box_1/\Box_2)}{135 {D^5} \Box_2 \Box_3}\Big(
   6 {{\Box_1}^9} \Box_2
  -54 {{\Box_1}^8} {{\Box_2}^2}
  +216 {{\Box_1}^7} {{\Box_2}^3}                 \nonumber\\&&\ \ \ \ \mbox{}
  -504 {{\Box_1}^6} {{\Box_2}^4}
  +756 {{\Box_1}^5} {{\Box_2}^5}
  +2 {{\Box_1}^9} \Box_3
  -80 {{\Box_1}^8} \Box_2 \Box_3                   \nonumber\\&&\ \ \ \ \mbox{}
  +414 {{\Box_1}^7} {{\Box_2}^2} \Box_3
  -896 {{\Box_1}^6} {{\Box_2}^3} \Box_3
  +860 {{\Box_1}^5} {{\Box_2}^4} \Box_3
  -22 {{\Box_1}^8} {{\Box_3}^2}                  \nonumber\\&&\ \ \ \ \mbox{}
  +424 {{\Box_1}^7} \Box_2 {{\Box_3}^2}
  -1978 {{\Box_1}^6} {{\Box_2}^2} {{\Box_3}^2}
  -42 {{\Box_1}^5} {{\Box_2}^3} {{\Box_3}^2}       \nonumber\\&&\ \ \ \ \mbox{}
  +7246 {{\Box_1}^4} {{\Box_2}^4} {{\Box_3}^2}
  +101 {{\Box_1}^7} {{\Box_3}^3}
  -1110 {{\Box_1}^6} \Box_2 {{\Box_3}^3}           \nonumber\\&&\ \ \ \ \mbox{}
  +3130 {{\Box_1}^5} {{\Box_2}^2} {{\Box_3}^3}
  -8758 {{\Box_1}^4} {{\Box_2}^3} {{\Box_3}^3}
  -259 {{\Box_1}^6} {{\Box_3}^4}                 \nonumber\\&&\ \ \ \ \mbox{}
  +1555 {{\Box_1}^5} \Box_2 {{\Box_3}^4}
  -873 {{\Box_1}^4} {{\Box_2}^2} {{\Box_3}^4}
  +10225 {{\Box_1}^3} {{\Box_2}^3} {{\Box_3}^4}    \nonumber\\&&\ \ \ \ \mbox{}
  +413 {{\Box_1}^5} {{\Box_3}^5}
  -1124 {{\Box_1}^4} \Box_2 {{\Box_3}^5}
  -1435 {{\Box_1}^3} {{\Box_2}^2} {{\Box_3}^5}     \nonumber\\&&\ \ \ \ \mbox{}
  -427 {{\Box_1}^4} {{\Box_3}^6}
  +285 {{\Box_1}^3} \Box_2 {{\Box_3}^6}
  +760 {{\Box_1}^2} {{\Box_2}^2} {{\Box_3}^6}      \nonumber\\&&\ \ \ \ \mbox{}
  +287 {{\Box_1}^3} {{\Box_3}^7}
  +106 {{\Box_1}^2} \Box_2 {{\Box_3}^7}
  -121 {{\Box_1}^2} {{\Box_3}^8}                 \nonumber\\&&\ \ \ \ \mbox{}
  -65 \Box_1 \Box_2 {{\Box_3}^8}
  +29 \Box_1 {{\Box_3}^9}
  -3 {{\Box_3}^{10}}
\Big) \nonumber\\&&\ \ \ \ \mbox{}
+\frac{\ln(\Box_1/\Box_3)}{135 {D^5} \Box_1 \Box_2 \Box_3}\Big(
   3 {{\Box_1}^{10}} \Box_2
  -27 {{\Box_1}^9} {{\Box_2}^2}
  +108 {{\Box_1}^8} {{\Box_2}^3}
  -252 {{\Box_1}^7} {{\Box_2}^4}                 \nonumber\\&&\ \ \ \ \mbox{}
  +378 {{\Box_1}^6} {{\Box_2}^5}
  -378 {{\Box_1}^5} {{\Box_2}^6}
  +252 {{\Box_1}^4} {{\Box_2}^7}
  -108 {{\Box_1}^3} {{\Box_2}^8}                 \nonumber\\&&\ \ \ \ \mbox{}
  +27 {{\Box_1}^2} {{\Box_2}^9}
  -3 \Box_1 {{\Box_2}^{10}}
  +4 {{\Box_1}^{10}} \Box_3
  -70 {{\Box_1}^9} \Box_2 \Box_3                   \nonumber\\&&\ \ \ \ \mbox{}
  +342 {{\Box_1}^8} {{\Box_2}^2} \Box_3
  -808 {{\Box_1}^7} {{\Box_2}^3} \Box_3
  +1060 {{\Box_1}^6} {{\Box_2}^4} \Box_3           \nonumber\\&&\ \ \ \ \mbox{}
  -756 {{\Box_1}^5} {{\Box_2}^5} \Box_3
  +200 {{\Box_1}^4} {{\Box_2}^6} \Box_3
  +88 {{\Box_1}^3} {{\Box_2}^7} \Box_3
  -72 {{\Box_1}^2} {{\Box_2}^8} \Box_3             \nonumber\\&&\ \ \ \ \mbox{}
  +10 \Box_1 {{\Box_2}^9} \Box_3
  +2 {{\Box_2}^{10}} \Box_3
  -44 {{\Box_1}^9} {{\Box_3}^2}
  +443 {{\Box_1}^8} \Box_2 {{\Box_3}^2}            \nonumber\\&&\ \ \ \ \mbox{}
  -2189 {{\Box_1}^7} {{\Box_2}^2} {{\Box_3}^2}
  +2523 {{\Box_1}^6} {{\Box_2}^3} {{\Box_3}^2}
  +2081 {{\Box_1}^5} {{\Box_2}^4} {{\Box_3}^2}     \nonumber\\&&\ \ \ \ \mbox{}
  -5165 {{\Box_1}^4} {{\Box_2}^5} {{\Box_3}^2}
  +2565 {{\Box_1}^3} {{\Box_2}^6} {{\Box_3}^2}
  -211 {{\Box_1}^2} {{\Box_2}^7} {{\Box_3}^2}      \nonumber\\&&\ \ \ \ \mbox{}
  +19 \Box_1 {{\Box_2}^8} {{\Box_3}^2}
  -22 {{\Box_2}^9} {{\Box_3}^2}
  +202 {{\Box_1}^8} {{\Box_3}^3}
  -1140 {{\Box_1}^7} \Box_2 {{\Box_3}^3}           \nonumber\\&&\ \ \ \ \mbox{}
  +2585 {{\Box_1}^6} {{\Box_2}^2} {{\Box_3}^3}
  -12098 {{\Box_1}^5} {{\Box_2}^3} {{\Box_3}^3}
  +14265 {{\Box_1}^4} {{\Box_2}^4} {{\Box_3}^3}    \nonumber\\&&\ \ \ \ \mbox{}
  -3340 {{\Box_1}^3} {{\Box_2}^5} {{\Box_3}^3}
  -545 {{\Box_1}^2} {{\Box_2}^6} {{\Box_3}^3}
  -30 \Box_1 {{\Box_2}^7} {{\Box_3}^3}
  +101 {{\Box_2}^8} {{\Box_3}^3}                 \nonumber\\&&\ \ \ \ \mbox{}
  -518 {{\Box_1}^7} {{\Box_3}^4}
  +1220 {{\Box_1}^6} \Box_2 {{\Box_3}^4}
  +1917 {{\Box_1}^5} {{\Box_2}^2} {{\Box_3}^4}     \nonumber\\&&\ \ \ \ \mbox{}
  +1625 {{\Box_1}^4} {{\Box_2}^3} {{\Box_3}^4}
  -8600 {{\Box_1}^3} {{\Box_2}^4} {{\Box_3}^4}
  +2790 {{\Box_1}^2} {{\Box_2}^5} {{\Box_3}^4}     \nonumber\\&&\ \ \ \ \mbox{}
  -335 \Box_1 {{\Box_2}^6} {{\Box_3}^4}
  -259 {{\Box_2}^7} {{\Box_3}^4}
  +826 {{\Box_1}^6} {{\Box_3}^5}
  +20 {{\Box_1}^5} \Box_2 {{\Box_3}^5}             \nonumber\\&&\ \ \ \ \mbox{}
  -3551 {{\Box_1}^4} {{\Box_2}^2} {{\Box_3}^5}
  +8664 {{\Box_1}^3} {{\Box_2}^3} {{\Box_3}^5}
  -2116 {{\Box_1}^2} {{\Box_2}^4} {{\Box_3}^5}     \nonumber\\&&\ \ \ \ \mbox{}
  +1144 \Box_1 {{\Box_2}^5} {{\Box_3}^5}
  +413 {{\Box_2}^6} {{\Box_3}^5}
  -854 {{\Box_1}^5} {{\Box_3}^6}                 \nonumber\\&&\ \ \ \ \mbox{}
  -1320 {{\Box_1}^4} \Box_2 {{\Box_3}^6}
  -319 {{\Box_1}^3} {{\Box_2}^2} {{\Box_3}^6}
  -1079 {{\Box_1}^2} {{\Box_2}^3} {{\Box_3}^6}     \nonumber\\&&\ \ \ \ \mbox{}
  -1605 \Box_1 {{\Box_2}^4} {{\Box_3}^6}
  -427 {{\Box_2}^5} {{\Box_3}^6}
  +574 {{\Box_1}^4} {{\Box_3}^7}                 \nonumber\\&&\ \ \ \ \mbox{}
  +1292 {{\Box_1}^3} \Box_2 {{\Box_3}^7}
  +1683 {{\Box_1}^2} {{\Box_2}^2} {{\Box_3}^7}
  +1186 \Box_1 {{\Box_2}^3} {{\Box_3}^7}           \nonumber\\&&\ \ \ \ \mbox{}
  +287 {{\Box_2}^4} {{\Box_3}^7}
  -242 {{\Box_1}^3} {{\Box_3}^8}
  -535 {{\Box_1}^2} \Box_2 {{\Box_3}^8}
  -470 \Box_1 {{\Box_2}^2} {{\Box_3}^8}            \nonumber\\&&\ \ \ \ \mbox{}
  -121 {{\Box_2}^3} {{\Box_3}^8}
  +58 {{\Box_1}^2} {{\Box_3}^9}
  +90 \Box_1 \Box_2 {{\Box_3}^9}
  +29 {{\Box_2}^2} {{\Box_3}^9}                  \nonumber\\&&\ \ \ \ \mbox{}
  -6 \Box_1 {{\Box_3}^{10}}
  -3 \Box_2 {{\Box_3}^{10}}
\Big) \nonumber\\&&\ \ \ \ \mbox{}
+\frac{\ln(\Box_1/\Box_2)}{(\Box_1-\Box_2)}\Big({{-1}\over {30 \Box_3}}\Big) \nonumber\\&&\ \ \ \ \mbox{}
+\frac1{270 {D^4} \Box_1 \Box_2}\Big(
   19 {{\Box_1}^8}
  -116 {{\Box_1}^7} \Box_2
  +460 {{\Box_1}^6} {{\Box_2}^2}                 \nonumber\\&&\ \ \ \ \mbox{}
  -1100 {{\Box_1}^5} {{\Box_2}^3}
  +737 {{\Box_1}^4} {{\Box_2}^4}
  -119 {{\Box_1}^7} \Box_3                       \nonumber\\&&\ \ \ \ \mbox{}
  +229 {{\Box_1}^6} \Box_2 \Box_3
  +2979 {{\Box_1}^5} {{\Box_2}^2} \Box_3
  -3089 {{\Box_1}^4} {{\Box_2}^3} \Box_3           \nonumber\\&&\ \ \ \ \mbox{}
  +319 {{\Box_1}^6} {{\Box_3}^2}
  +264 {{\Box_1}^5} \Box_2 {{\Box_3}^2}
  -4983 {{\Box_1}^4} {{\Box_2}^2} {{\Box_3}^2}     \nonumber\\&&\ \ \ \ \mbox{}
  +6560 {{\Box_1}^3} {{\Box_2}^3} {{\Box_3}^2}
  -479 {{\Box_1}^5} {{\Box_3}^3}
  -1055 {{\Box_1}^4} \Box_2 {{\Box_3}^3}           \nonumber\\&&\ \ \ \ \mbox{}
  -1082 {{\Box_1}^3} {{\Box_2}^2} {{\Box_3}^3}
  +445 {{\Box_1}^4} {{\Box_3}^4}
  +1060 {{\Box_1}^3} \Box_2 {{\Box_3}^4}           \nonumber\\&&\ \ \ \ \mbox{}
  +1503 {{\Box_1}^2} {{\Box_2}^2} {{\Box_3}^4}
  -269 {{\Box_1}^3} {{\Box_3}^5}
  -489 {{\Box_1}^2} \Box_2 {{\Box_3}^5}            \nonumber\\&&\ \ \ \ \mbox{}
  +109 {{\Box_1}^2} {{\Box_3}^6}
  +68 \Box_1 \Box_2 {{\Box_3}^6}
  -29 \Box_1 {{\Box_3}^7}
  +2 {{\Box_3}^8}
\Big)
,\end{fleqnarray}

\begin{fleqnarray}&&\Gamma_{24}(-\Box_1,-\Box_2,-\Box_3) =
\Gamma(-\Box_1,-\Box_2,-\Box_3)\frac1{{D^4}}\Big(
   2 {{\Box_1}^5} \Box_2 \Box_3                    \nonumber\\&&\ \ \ \ \mbox{}
  -12 {{\Box_1}^3} {{\Box_2}^3} \Box_3
  +8 {{\Box_1}^2} {{\Box_2}^4} \Box_3
  +8 {{\Box_1}^3} {{\Box_2}^2} {{\Box_3}^2}
  -8 {{\Box_1}^2} {{\Box_2}^3} {{\Box_3}^2}
\Big) \nonumber\\&&\ \ \ \ \mbox{}
+\frac{\ln(\Box_1/\Box_2)}{45 {D^4} \Box_1 \Box_2 \Box_3}\Big(
   2 {{\Box_1}^8} \Box_2
  -16 {{\Box_1}^7} {{\Box_2}^2}
  +52 {{\Box_1}^6} {{\Box_2}^3}                  \nonumber\\&&\ \ \ \ \mbox{}
  -90 {{\Box_1}^5} {{\Box_2}^4}
  +90 {{\Box_1}^4} {{\Box_2}^5}
  -52 {{\Box_1}^3} {{\Box_2}^6}
  +16 {{\Box_1}^2} {{\Box_2}^7}                  \nonumber\\&&\ \ \ \ \mbox{}
  -2 \Box_1 {{\Box_2}^8}
  +{{\Box_1}^8} \Box_3
  -24 {{\Box_1}^7} \Box_2 \Box_3
  +120 {{\Box_1}^6} {{\Box_2}^2} \Box_3            \nonumber\\&&\ \ \ \ \mbox{}
  -184 {{\Box_1}^5} {{\Box_2}^3} \Box_3
  +36 {{\Box_1}^4} {{\Box_2}^4} \Box_3
  +136 {{\Box_1}^3} {{\Box_2}^5} \Box_3
  -108 {{\Box_1}^2} {{\Box_2}^6} \Box_3            \nonumber\\&&\ \ \ \ \mbox{}
  +24 \Box_1 {{\Box_2}^7} \Box_3
  -{{\Box_2}^8} \Box_3
  -8 {{\Box_1}^7} {{\Box_3}^2}
  +102 {{\Box_1}^6} \Box_2 {{\Box_3}^2}            \nonumber\\&&\ \ \ \ \mbox{}
  +141 {{\Box_1}^5} {{\Box_2}^2} {{\Box_3}^2}
  +89 {{\Box_1}^4} {{\Box_2}^3} {{\Box_3}^2}
  -392 {{\Box_1}^3} {{\Box_2}^4} {{\Box_3}^2}
  +146 {{\Box_1}^2} {{\Box_2}^5} {{\Box_3}^2}      \nonumber\\&&\ \ \ \ \mbox{}
  -85 \Box_1 {{\Box_2}^6} {{\Box_3}^2}
  +7 {{\Box_2}^7} {{\Box_3}^2}
  +26 {{\Box_1}^6} {{\Box_3}^3}
  -176 {{\Box_1}^5} \Box_2 {{\Box_3}^3}            \nonumber\\&&\ \ \ \ \mbox{}
  -203 {{\Box_1}^4} {{\Box_2}^2} {{\Box_3}^3}
  +456 {{\Box_1}^3} {{\Box_2}^3} {{\Box_3}^3}
  +38 {{\Box_1}^2} {{\Box_2}^4} {{\Box_3}^3}
  +136 \Box_1 {{\Box_2}^5} {{\Box_3}^3}            \nonumber\\&&\ \ \ \ \mbox{}
  -21 {{\Box_2}^6} {{\Box_3}^3}
  -45 {{\Box_1}^5} {{\Box_3}^4}
  +123 {{\Box_1}^4} \Box_2 {{\Box_3}^4}
  -106 {{\Box_1}^3} {{\Box_2}^2} {{\Box_3}^4}      \nonumber\\&&\ \ \ \ \mbox{}
  -158 {{\Box_1}^2} {{\Box_2}^3} {{\Box_3}^4}
  -105 \Box_1 {{\Box_2}^4} {{\Box_3}^4}
  +35 {{\Box_2}^5} {{\Box_3}^4}
  +45 {{\Box_1}^4} {{\Box_3}^5}                  \nonumber\\&&\ \ \ \ \mbox{}
  -16 {{\Box_1}^3} \Box_2 {{\Box_3}^5}
  +70 {{\Box_1}^2} {{\Box_2}^2} {{\Box_3}^5}
  +32 \Box_1 {{\Box_2}^3} {{\Box_3}^5}
  -35 {{\Box_2}^4} {{\Box_3}^5}                  \nonumber\\&&\ \ \ \ \mbox{}
  -26 {{\Box_1}^3} {{\Box_3}^6}
  -12 {{\Box_1}^2} \Box_2 {{\Box_3}^6}
  +\Box_1 {{\Box_2}^2} {{\Box_3}^6}
  +21 {{\Box_2}^3} {{\Box_3}^6}                  \nonumber\\&&\ \ \ \ \mbox{}
  +8 {{\Box_1}^2} {{\Box_3}^7}
  -7 {{\Box_2}^2} {{\Box_3}^7}
  -\Box_1 {{\Box_3}^8}
  +\Box_2 {{\Box_3}^8}
\Big) \nonumber\\&&\ \ \ \ \mbox{}
+\frac{\ln(\Box_2/\Box_3)}{45 {D^4} \Box_1 \Box_3}\Big(
  -{{\Box_1}^8}
  +8 {{\Box_1}^7} \Box_2
  -26 {{\Box_1}^6} {{\Box_2}^2}
  +45 {{\Box_1}^5} {{\Box_2}^3}                  \nonumber\\&&\ \ \ \ \mbox{}
  -45 {{\Box_1}^4} {{\Box_2}^4}
  +26 {{\Box_1}^3} {{\Box_2}^5}
  -8 {{\Box_1}^2} {{\Box_2}^6}
  +\Box_1 {{\Box_2}^7}                           \nonumber\\&&\ \ \ \ \mbox{}
  -18 {{\Box_1}^6} \Box_2 \Box_3
  +8 {{\Box_1}^5} {{\Box_2}^2} \Box_3
  +87 {{\Box_1}^4} {{\Box_2}^3} \Box_3
  -152 {{\Box_1}^3} {{\Box_2}^4} \Box_3            \nonumber\\&&\ \ \ \ \mbox{}
  +96 {{\Box_1}^2} {{\Box_2}^5} \Box_3
  -24 \Box_1 {{\Box_2}^6} \Box_3
  +2 {{\Box_2}^7} \Box_3
  -292 {{\Box_1}^4} {{\Box_2}^2} {{\Box_3}^2}      \nonumber\\&&\ \ \ \ \mbox{}
  +286 {{\Box_1}^3} {{\Box_2}^3} {{\Box_3}^2}
  -76 {{\Box_1}^2} {{\Box_2}^4} {{\Box_3}^2}
  +86 \Box_1 {{\Box_2}^5} {{\Box_3}^2}
  -14 {{\Box_2}^6} {{\Box_3}^2}                  \nonumber\\&&\ \ \ \ \mbox{}
  -196 {{\Box_1}^2} {{\Box_2}^3} {{\Box_3}^3}
  -104 \Box_1 {{\Box_2}^4} {{\Box_3}^3}
  +42 {{\Box_2}^5} {{\Box_3}^3}
  -70 {{\Box_2}^4} {{\Box_3}^4}
\Big) \nonumber\\&&\ \ \ \ \mbox{}
+\frac{\ln(\Box_2/\Box_3)}{(\Box_2-\Box_3)}\Big({{-1}\over {30 \Box_1}}\Big) \nonumber\\&&\ \ \ \ \mbox{}
+\frac1{540 {D^3} \Box_2 \Box_3}\Big(
   {{\Box_1}^6}
  +6 {{\Box_1}^5} \Box_2
  -24 {{\Box_1}^4} {{\Box_2}^2}
  -4 {{\Box_1}^3} {{\Box_2}^3}                   \nonumber\\&&\ \ \ \ \mbox{}
  +66 {{\Box_1}^2} {{\Box_2}^4}
  -66 \Box_1 {{\Box_2}^5}
  +20 {{\Box_2}^6}
  -90 {{\Box_1}^4} \Box_2 \Box_3                   \nonumber\\&&\ \ \ \ \mbox{}
  -312 {{\Box_1}^3} {{\Box_2}^2} \Box_3
  +1092 {{\Box_1}^2} {{\Box_2}^3} \Box_3
  -558 \Box_1 {{\Box_2}^4} \Box_3
  -48 {{\Box_2}^5} \Box_3                        \nonumber\\&&\ \ \ \ \mbox{}
  -1230 {{\Box_1}^2} {{\Box_2}^2} {{\Box_3}^2}
  +624 \Box_1 {{\Box_2}^3} {{\Box_3}^2}
  +12 {{\Box_2}^4} {{\Box_3}^2}
  +16 {{\Box_2}^3} {{\Box_3}^3}
\Big)
,\end{fleqnarray}

\begin{fleqnarray}&&\Gamma_{25}(-\Box_1,-\Box_2,-\Box_3) =
\Gamma(-\Box_1,-\Box_2,-\Box_3)\frac1{{D^4}}\Big(
  -4 {{\Box_1}^5} \Box_2 \Box_3                    \nonumber\\&&\ \ \ \ \mbox{}
  +16 {{\Box_1}^4} {{\Box_2}^2} \Box_3
  -16 {{\Box_1}^2} {{\Box_2}^4} \Box_3
  +8 \Box_1 {{\Box_2}^5} \Box_3
  -32 {{\Box_1}^3} {{\Box_2}^2} {{\Box_3}^2}       \nonumber\\&&\ \ \ \ \mbox{}
  +32 {{\Box_1}^2} {{\Box_2}^3} {{\Box_3}^2}
  +16 \Box_1 {{\Box_2}^4} {{\Box_3}^2}
  -24 \Box_1 {{\Box_2}^3} {{\Box_3}^3}
\Big) \nonumber\\&&\ \ \ \ \mbox{}
+\frac{\ln(\Box_1/\Box_2)}{45 {D^4} \Box_1 \Box_2 \Box_3}\Big(
   4 {{\Box_1}^8} \Box_2
  -28 {{\Box_1}^7} {{\Box_2}^2}
  +84 {{\Box_1}^6} {{\Box_2}^3}                  \nonumber\\&&\ \ \ \ \mbox{}
  -140 {{\Box_1}^5} {{\Box_2}^4}
  +140 {{\Box_1}^4} {{\Box_2}^5}
  -84 {{\Box_1}^3} {{\Box_2}^6}
  +28 {{\Box_1}^2} {{\Box_2}^7}                  \nonumber\\&&\ \ \ \ \mbox{}
  -4 \Box_1 {{\Box_2}^8}
  +2 {{\Box_1}^8} \Box_3
  -48 {{\Box_1}^7} \Box_2 \Box_3
  +148 {{\Box_1}^6} {{\Box_2}^2} \Box_3            \nonumber\\&&\ \ \ \ \mbox{}
  -112 {{\Box_1}^5} {{\Box_2}^3} \Box_3
  -144 {{\Box_1}^4} {{\Box_2}^4} \Box_3
  +304 {{\Box_1}^3} {{\Box_2}^5} \Box_3
  -196 {{\Box_1}^2} {{\Box_2}^6} \Box_3            \nonumber\\&&\ \ \ \ \mbox{}
  +48 \Box_1 {{\Box_2}^7} \Box_3
  -2 {{\Box_2}^8} \Box_3
  -14 {{\Box_1}^7} {{\Box_3}^2}
  +158 {{\Box_1}^6} \Box_2 {{\Box_3}^2}            \nonumber\\&&\ \ \ \ \mbox{}
  -726 {{\Box_1}^5} {{\Box_2}^2} {{\Box_3}^2}
  +214 {{\Box_1}^4} {{\Box_2}^3} {{\Box_3}^2}
  +998 {{\Box_1}^3} {{\Box_2}^4} {{\Box_3}^2}
  -422 {{\Box_1}^2} {{\Box_2}^5} {{\Box_3}^2}      \nonumber\\&&\ \ \ \ \mbox{}
  -226 \Box_1 {{\Box_2}^6} {{\Box_3}^2}
  +18 {{\Box_2}^7} {{\Box_3}^2}
  +42 {{\Box_1}^6} {{\Box_3}^3}
  -224 {{\Box_1}^5} \Box_2 {{\Box_3}^3}            \nonumber\\&&\ \ \ \ \mbox{}
  +902 {{\Box_1}^4} {{\Box_2}^2} {{\Box_3}^3}
  -1824 {{\Box_1}^3} {{\Box_2}^3} {{\Box_3}^3}
  -242 {{\Box_1}^2} {{\Box_2}^4} {{\Box_3}^3}      \nonumber\\&&\ \ \ \ \mbox{}
  +384 \Box_1 {{\Box_2}^5} {{\Box_3}^3}
  -62 {{\Box_2}^6} {{\Box_3}^3}
  -70 {{\Box_1}^5} {{\Box_3}^4}
  +138 {{\Box_1}^4} \Box_2 {{\Box_3}^4}            \nonumber\\&&\ \ \ \ \mbox{}
  -56 {{\Box_1}^3} {{\Box_2}^2} {{\Box_3}^4}
  +1112 {{\Box_1}^2} {{\Box_2}^3} {{\Box_3}^4}
  -210 \Box_1 {{\Box_2}^4} {{\Box_3}^4}
  +110 {{\Box_2}^5} {{\Box_3}^4}                 \nonumber\\&&\ \ \ \ \mbox{}
  +70 {{\Box_1}^4} {{\Box_3}^5}
  -16 {{\Box_1}^3} \Box_2 {{\Box_3}^5}
  -280 {{\Box_1}^2} {{\Box_2}^2} {{\Box_3}^5}
  -48 \Box_1 {{\Box_2}^3} {{\Box_3}^5}             \nonumber\\&&\ \ \ \ \mbox{}
  -110 {{\Box_2}^4} {{\Box_3}^5}
  -42 {{\Box_1}^3} {{\Box_3}^6}
  -14 {{\Box_1}^2} \Box_2 {{\Box_3}^6}
  +58 \Box_1 {{\Box_2}^2} {{\Box_3}^6}             \nonumber\\&&\ \ \ \ \mbox{}
  +62 {{\Box_2}^3} {{\Box_3}^6}
  +14 {{\Box_1}^2} {{\Box_3}^7}
  -18 {{\Box_2}^2} {{\Box_3}^7}
  -2 \Box_1 {{\Box_3}^8}
  +2 \Box_2 {{\Box_3}^8}
\Big) \nonumber\\&&\ \ \ \ \mbox{}
+\frac{\ln(\Box_2/\Box_3)}{45 {D^4} \Box_1 \Box_3}\Big(
  -2 {{\Box_1}^8}
  +14 {{\Box_1}^7} \Box_2
  -42 {{\Box_1}^6} {{\Box_2}^2}                  \nonumber\\&&\ \ \ \ \mbox{}
  +70 {{\Box_1}^5} {{\Box_2}^3}
  -70 {{\Box_1}^4} {{\Box_2}^4}
  +42 {{\Box_1}^3} {{\Box_2}^5}
  -14 {{\Box_1}^2} {{\Box_2}^6}                  \nonumber\\&&\ \ \ \ \mbox{}
  +2 \Box_1 {{\Box_2}^7}
  +10 {{\Box_1}^6} \Box_2 \Box_3
  -112 {{\Box_1}^5} {{\Box_2}^2} \Box_3
  +282 {{\Box_1}^4} {{\Box_2}^3} \Box_3            \nonumber\\&&\ \ \ \ \mbox{}
  -320 {{\Box_1}^3} {{\Box_2}^4} \Box_3
  +182 {{\Box_1}^2} {{\Box_2}^5} \Box_3
  -48 \Box_1 {{\Box_2}^6} \Box_3
  +4 {{\Box_2}^7} \Box_3                         \nonumber\\&&\ \ \ \ \mbox{}
  +688 {{\Box_1}^4} {{\Box_2}^2} {{\Box_3}^2}
  -1054 {{\Box_1}^3} {{\Box_2}^3} {{\Box_3}^2}
  +142 {{\Box_1}^2} {{\Box_2}^4} {{\Box_3}^2}      \nonumber\\&&\ \ \ \ \mbox{}
  +284 \Box_1 {{\Box_2}^5} {{\Box_3}^2}
  -36 {{\Box_2}^6} {{\Box_3}^2}
  +1354 {{\Box_1}^2} {{\Box_2}^3} {{\Box_3}^3}     \nonumber\\&&\ \ \ \ \mbox{}
  -432 \Box_1 {{\Box_2}^4} {{\Box_3}^3}
  +124 {{\Box_2}^5} {{\Box_3}^3}
  -220 {{\Box_2}^4} {{\Box_3}^4}
\Big) \nonumber\\&&\ \ \ \ \mbox{}
+\frac{\ln(\Box_1/\Box_2)}{(\Box_1-\Box_2)}\Big({{-2}\over {15 \Box_3}}\Big) \nonumber\\&&\ \ \ \ \mbox{}
+\frac1{270 {D^3} \Box_1 \Box_2 \Box_3}\Big(
   {{\Box_1}^7}
  -14 {{\Box_1}^6} \Box_2
  +42 {{\Box_1}^5} {{\Box_2}^2}                  \nonumber\\&&\ \ \ \ \mbox{}
  -70 {{\Box_1}^4} {{\Box_2}^3}
  +70 {{\Box_1}^3} {{\Box_2}^4}
  -42 {{\Box_1}^2} {{\Box_2}^5}
  +14 \Box_1 {{\Box_2}^6}                        \nonumber\\&&\ \ \ \ \mbox{}
  -2 {{\Box_2}^7}
  +102 {{\Box_1}^5} \Box_2 \Box_3
  -414 {{\Box_1}^4} {{\Box_2}^2} \Box_3
  +112 {{\Box_1}^3} {{\Box_2}^3} \Box_3            \nonumber\\&&\ \ \ \ \mbox{}
  +318 {{\Box_1}^2} {{\Box_2}^4} \Box_3
  -252 \Box_1 {{\Box_2}^5} \Box_3
  +46 {{\Box_2}^6} \Box_3
  +1890 {{\Box_1}^3} {{\Box_2}^2} {{\Box_3}^2}     \nonumber\\&&\ \ \ \ \mbox{}
  -2436 {{\Box_1}^2} {{\Box_2}^3} {{\Box_3}^2}
  -846 \Box_1 {{\Box_2}^4} {{\Box_3}^2}
  -126 {{\Box_2}^5} {{\Box_3}^2}                 \nonumber\\&&\ \ \ \ \mbox{}
  +1084 \Box_1 {{\Box_2}^3} {{\Box_3}^3}
  +82 {{\Box_2}^4} {{\Box_3}^3}
\Big)
,\end{fleqnarray}

\begin{fleqnarray}&&\Gamma_{26}(-\Box_1,-\Box_2,-\Box_3) =
\Gamma(-\Box_1,-\Box_2,-\Box_3)\frac1{{D^4}}\Big(
   8 {{\Box_1}^5} \Box_2                         \nonumber\\&&\ \ \ \ \mbox{}
  -32 {{\Box_1}^4} {{\Box_2}^2}
  +24 {{\Box_1}^3} {{\Box_2}^3}
  +48 {{\Box_1}^4} \Box_2 \Box_3
  -48 {{\Box_1}^3} {{\Box_2}^2} \Box_3             \nonumber\\&&\ \ \ \ \mbox{}
  -72 {{\Box_1}^3} \Box_2 {{\Box_3}^2}
  +96 {{\Box_1}^2} {{\Box_2}^2} {{\Box_3}^2}
  -32 {{\Box_1}^2} \Box_2 {{\Box_3}^3}
  +24 \Box_1 \Box_2 {{\Box_3}^4}
\Big) \nonumber\\&&\ \ \ \ \mbox{}
+\frac{\ln(\Box_1/\Box_2)}{9 {D^4} \Box_2}\Big(
  -2 {{\Box_1}^6}
  +22 {{\Box_1}^5} \Box_2
  +118 {{\Box_1}^4} {{\Box_2}^2}                 \nonumber\\&&\ \ \ \ \mbox{}
  -450 {{\Box_1}^3} {{\Box_2}^3}
  +10 {{\Box_1}^5} \Box_3
  -32 {{\Box_1}^4} \Box_2 \Box_3
  +474 {{\Box_1}^3} {{\Box_2}^2} \Box_3            \nonumber\\&&\ \ \ \ \mbox{}
  -20 {{\Box_1}^4} {{\Box_3}^2}
  -36 {{\Box_1}^3} \Box_2 {{\Box_3}^2}
  -624 {{\Box_1}^2} {{\Box_2}^2} {{\Box_3}^2}
  +20 {{\Box_1}^3} {{\Box_3}^3}                  \nonumber\\&&\ \ \ \ \mbox{}
  +80 {{\Box_1}^2} \Box_2 {{\Box_3}^3}
  -10 {{\Box_1}^2} {{\Box_3}^4}
  -34 \Box_1 \Box_2 {{\Box_3}^4}
  +2 \Box_1 {{\Box_3}^5}
\Big) \nonumber\\&&\ \ \ \ \mbox{}
+\frac{\ln(\Box_1/\Box_3)}{9 {D^4} \Box_1 \Box_2}\Big(
  -4 {{\Box_1}^7}
  +44 {{\Box_1}^6} \Box_2
  -22 {{\Box_1}^5} {{\Box_2}^2}
  -174 {{\Box_1}^4} {{\Box_2}^3}                 \nonumber\\&&\ \ \ \ \mbox{}
  +276 {{\Box_1}^3} {{\Box_2}^4}
  -140 {{\Box_1}^2} {{\Box_2}^5}
  +22 \Box_1 {{\Box_2}^6}
  -2 {{\Box_2}^7}                              \nonumber\\&&\ \ \ \ \mbox{}
  +20 {{\Box_1}^6} \Box_3
  -64 {{\Box_1}^5} \Box_2 \Box_3
  +726 {{\Box_1}^4} {{\Box_2}^2} \Box_3
  -912 {{\Box_1}^3} {{\Box_2}^3} \Box_3            \nonumber\\&&\ \ \ \ \mbox{}
  +252 {{\Box_1}^2} {{\Box_2}^4} \Box_3
  -32 \Box_1 {{\Box_2}^5} \Box_3
  +10 {{\Box_2}^6} \Box_3
  -40 {{\Box_1}^5} {{\Box_3}^2}                  \nonumber\\&&\ \ \ \ \mbox{}
  -72 {{\Box_1}^4} \Box_2 {{\Box_3}^2}
  -120 {{\Box_1}^3} {{\Box_2}^2} {{\Box_3}^2}
  +504 {{\Box_1}^2} {{\Box_2}^3} {{\Box_3}^2}
  -36 \Box_1 {{\Box_2}^4} {{\Box_3}^2}             \nonumber\\&&\ \ \ \ \mbox{}
  -20 {{\Box_2}^5} {{\Box_3}^2}
  +40 {{\Box_1}^4} {{\Box_3}^3}
  +160 {{\Box_1}^3} \Box_2 {{\Box_3}^3}
  -552 {{\Box_1}^2} {{\Box_2}^2} {{\Box_3}^3}      \nonumber\\&&\ \ \ \ \mbox{}
  +80 \Box_1 {{\Box_2}^3} {{\Box_3}^3}
  +20 {{\Box_2}^4} {{\Box_3}^3}
  -20 {{\Box_1}^3} {{\Box_3}^4}
  -68 {{\Box_1}^2} \Box_2 {{\Box_3}^4}             \nonumber\\&&\ \ \ \ \mbox{}
  -34 \Box_1 {{\Box_2}^2} {{\Box_3}^4}
  -10 {{\Box_2}^3} {{\Box_3}^4}
  +4 {{\Box_1}^2} {{\Box_3}^5}
  +2 {{\Box_2}^2} {{\Box_3}^5}
\Big) \nonumber\\&&\ \ \ \ \mbox{}
+\frac1{6 {D^3} \Box_1 \Box_2}\Big(
   6 {{\Box_1}^5}
  -70 {{\Box_1}^4} \Box_2
  +64 {{\Box_1}^3} {{\Box_2}^2}
  -26 {{\Box_1}^4} \Box_3                        \nonumber\\&&\ \ \ \ \mbox{}
  +100 {{\Box_1}^3} \Box_2 \Box_3
  -146 {{\Box_1}^2} {{\Box_2}^2} \Box_3
  +44 {{\Box_1}^3} {{\Box_3}^2}
  +24 {{\Box_1}^2} \Box_2 {{\Box_3}^2}             \nonumber\\&&\ \ \ \ \mbox{}
  -36 {{\Box_1}^2} {{\Box_3}^3}
  -34 \Box_1 \Box_2 {{\Box_3}^3}
  +14 \Box_1 {{\Box_3}^4}
  -{{\Box_3}^5}
\Big)
,\end{fleqnarray}

\begin{fleqnarray}&&\Gamma_{27}(-\Box_1,-\Box_2,-\Box_3) =
\Gamma(-\Box_1,-\Box_2,-\Box_3)\frac1{3 {D^6}}\Big(
  -4 {{\Box_1}^9} \Box_2                         \nonumber\\&&\ \ \ \ \mbox{}
  +8 {{\Box_1}^8} {{\Box_2}^2}
  +32 {{\Box_1}^7} {{\Box_2}^3}
  -136 {{\Box_1}^6} {{\Box_2}^4}
  +100 {{\Box_1}^5} {{\Box_2}^5}                 \nonumber\\&&\ \ \ \ \mbox{}
  -32 {{\Box_1}^8} \Box_2 \Box_3
  -200 {{\Box_1}^7} {{\Box_2}^2} \Box_3
  +792 {{\Box_1}^6} {{\Box_2}^3} \Box_3
  -560 {{\Box_1}^5} {{\Box_2}^4} \Box_3            \nonumber\\&&\ \ \ \ \mbox{}
  +60 {{\Box_1}^7} \Box_2 {{\Box_3}^2}
  -352 {{\Box_1}^6} {{\Box_2}^2} {{\Box_3}^2}
  -1292 {{\Box_1}^5} {{\Box_2}^3} {{\Box_3}^2}     \nonumber\\&&\ \ \ \ \mbox{}
  +1584 {{\Box_1}^4} {{\Box_2}^4} {{\Box_3}^2}
  +32 {{\Box_1}^6} \Box_2 {{\Box_3}^3}
  +960 {{\Box_1}^5} {{\Box_2}^2} {{\Box_3}^3}      \nonumber\\&&\ \ \ \ \mbox{}
  -1232 {{\Box_1}^4} {{\Box_2}^3} {{\Box_3}^3}
  +4 {{\Box_1}^5} \Box_2 {{\Box_3}^4}
  -296 {{\Box_1}^4} {{\Box_2}^2} {{\Box_3}^4}
  +556 {{\Box_1}^3} {{\Box_2}^3} {{\Box_3}^4}      \nonumber\\&&\ \ \ \ \mbox{}
  -192 {{\Box_1}^4} \Box_2 {{\Box_3}^5}
  +136 {{\Box_1}^3} {{\Box_2}^2} {{\Box_3}^5}
  +116 {{\Box_1}^3} \Box_2 {{\Box_3}^6}
  -160 {{\Box_1}^2} {{\Box_2}^2} {{\Box_3}^6}      \nonumber\\&&\ \ \ \ \mbox{}
  +64 {{\Box_1}^2} \Box_2 {{\Box_3}^7}
  -24 \Box_1 \Box_2 {{\Box_3}^8}
\Big) \nonumber\\&&\ \ \ \ \mbox{}
+\frac{\ln(\Box_1/\Box_2)}{135 {D^6} \Box_2}\Big(
  -{{\Box_1}^{10}}
  +9 {{\Box_1}^9} \Box_2
  -681 {{\Box_1}^8} {{\Box_2}^2}                 \nonumber\\&&\ \ \ \ \mbox{}
  +1073 {{\Box_1}^7} {{\Box_2}^3}
  +3954 {{\Box_1}^6} {{\Box_2}^4}
  -12530 {{\Box_1}^5} {{\Box_2}^5}               \nonumber\\&&\ \ \ \ \mbox{}
  +9 {{\Box_1}^9} \Box_3
  -84 {{\Box_1}^8} \Box_2 \Box_3
  -2205 {{\Box_1}^7} {{\Box_2}^2} \Box_3           \nonumber\\&&\ \ \ \ \mbox{}
  -16728 {{\Box_1}^6} {{\Box_2}^3} \Box_3
  +40362 {{\Box_1}^5} {{\Box_2}^4} \Box_3
  -30 {{\Box_1}^8} {{\Box_3}^2}                  \nonumber\\&&\ \ \ \ \mbox{}
  +114 {{\Box_1}^7} \Box_2 {{\Box_3}^2}
  +5112 {{\Box_1}^6} {{\Box_2}^2} {{\Box_3}^2}
  -6432 {{\Box_1}^5} {{\Box_2}^3} {{\Box_3}^2}     \nonumber\\&&\ \ \ \ \mbox{}
  -73320 {{\Box_1}^4} {{\Box_2}^4} {{\Box_3}^2}
  +42 {{\Box_1}^7} {{\Box_3}^3}
  +252 {{\Box_1}^6} \Box_2 {{\Box_3}^3}            \nonumber\\&&\ \ \ \ \mbox{}
  +704 {{\Box_1}^5} {{\Box_2}^2} {{\Box_3}^3}
  +28524 {{\Box_1}^4} {{\Box_2}^3} {{\Box_3}^3}
  -660 {{\Box_1}^5} \Box_2 {{\Box_3}^4}            \nonumber\\&&\ \ \ \ \mbox{}
  -180 {{\Box_1}^4} {{\Box_2}^2} {{\Box_3}^4}
  -2160 {{\Box_1}^3} {{\Box_2}^3} {{\Box_3}^4}
  -84 {{\Box_1}^5} {{\Box_3}^5}                  \nonumber\\&&\ \ \ \ \mbox{}
  +276 {{\Box_1}^4} \Box_2 {{\Box_3}^5}
  -7320 {{\Box_1}^3} {{\Box_2}^2} {{\Box_3}^5}
  +126 {{\Box_1}^4} {{\Box_3}^6}                 \nonumber\\&&\ \ \ \ \mbox{}
  +414 {{\Box_1}^3} \Box_2 {{\Box_3}^6}
  +4688 {{\Box_1}^2} {{\Box_2}^2} {{\Box_3}^6}
  -90 {{\Box_1}^3} {{\Box_3}^7}                  \nonumber\\&&\ \ \ \ \mbox{}
  -444 {{\Box_1}^2} \Box_2 {{\Box_3}^7}
  +33 {{\Box_1}^2} {{\Box_3}^8}
  +123 \Box_1 \Box_2 {{\Box_3}^8}
  -5 \Box_1 {{\Box_3}^9}
\Big) \nonumber\\&&\ \ \ \ \mbox{}
+\frac{\ln(\Box_1/\Box_3)}{135 {D^6} \Box_1 \Box_2}\Big(
  -2 {{\Box_1}^{11}}
  +18 {{\Box_1}^{10}} \Box_2
  -663 {{\Box_1}^9} {{\Box_2}^2}                 \nonumber\\&&\ \ \ \ \mbox{}
  +1999 {{\Box_1}^8} {{\Box_2}^3}
  -588 {{\Box_1}^7} {{\Box_2}^4}
  -4852 {{\Box_1}^6} {{\Box_2}^5}                \nonumber\\&&\ \ \ \ \mbox{}
  +7678 {{\Box_1}^5} {{\Box_2}^6}
  -4542 {{\Box_1}^4} {{\Box_2}^7}
  +926 {{\Box_1}^3} {{\Box_2}^8}                 \nonumber\\&&\ \ \ \ \mbox{}
  +18 {{\Box_1}^2} {{\Box_2}^9}
  +9 \Box_1 {{\Box_2}^{10}}
  -{{\Box_2}^{11}}
  +18 {{\Box_1}^{10}} \Box_3                     \nonumber\\&&\ \ \ \ \mbox{}
  -168 {{\Box_1}^9} \Box_2 \Box_3
  -3447 {{\Box_1}^8} {{\Box_2}^2} \Box_3
  -8652 {{\Box_1}^7} {{\Box_2}^3} \Box_3           \nonumber\\&&\ \ \ \ \mbox{}
  +44112 {{\Box_1}^6} {{\Box_2}^4} \Box_3
  -42372 {{\Box_1}^5} {{\Box_2}^5} \Box_3
  +3750 {{\Box_1}^4} {{\Box_2}^6} \Box_3           \nonumber\\&&\ \ \ \ \mbox{}
  +8076 {{\Box_1}^3} {{\Box_2}^7} \Box_3
  -1242 {{\Box_1}^2} {{\Box_2}^8} \Box_3
  -84 \Box_1 {{\Box_2}^9} \Box_3                   \nonumber\\&&\ \ \ \ \mbox{}
  +9 {{\Box_2}^{10}} \Box_3
  -60 {{\Box_1}^9} {{\Box_3}^2}
  +228 {{\Box_1}^8} \Box_2 {{\Box_3}^2}
  +5166 {{\Box_1}^7} {{\Box_2}^2} {{\Box_3}^2}     \nonumber\\&&\ \ \ \ \mbox{}
  -35922 {{\Box_1}^6} {{\Box_2}^3} {{\Box_3}^2}
  -6690 {{\Box_1}^5} {{\Box_2}^4} {{\Box_3}^2}
  +66630 {{\Box_1}^4} {{\Box_2}^5} {{\Box_3}^2}    \nonumber\\&&\ \ \ \ \mbox{}
  -29490 {{\Box_1}^3} {{\Box_2}^6} {{\Box_3}^2}
  +54 {{\Box_1}^2} {{\Box_2}^7} {{\Box_3}^2}
  +114 \Box_1 {{\Box_2}^8} {{\Box_3}^2}            \nonumber\\&&\ \ \ \ \mbox{}
  -30 {{\Box_2}^9} {{\Box_3}^2}
  +84 {{\Box_1}^8} {{\Box_3}^3}
  +504 {{\Box_1}^7} \Box_2 {{\Box_3}^3}
  +2614 {{\Box_1}^6} {{\Box_2}^2} {{\Box_3}^3}     \nonumber\\&&\ \ \ \ \mbox{}
  +39324 {{\Box_1}^5} {{\Box_2}^3} {{\Box_3}^3}
  -66330 {{\Box_1}^4} {{\Box_2}^4} {{\Box_3}^3}
  +10800 {{\Box_1}^3} {{\Box_2}^5} {{\Box_3}^3}    \nonumber\\&&\ \ \ \ \mbox{}
  +1910 {{\Box_1}^2} {{\Box_2}^6} {{\Box_3}^3}
  +252 \Box_1 {{\Box_2}^7} {{\Box_3}^3}
  +42 {{\Box_2}^8} {{\Box_3}^3}                  \nonumber\\&&\ \ \ \ \mbox{}
  -1320 {{\Box_1}^6} \Box_2 {{\Box_3}^4}
  +1980 {{\Box_1}^5} {{\Box_2}^2} {{\Box_3}^4}
  +2880 {{\Box_1}^4} {{\Box_2}^3} {{\Box_3}^4}     \nonumber\\&&\ \ \ \ \mbox{}
  +5040 {{\Box_1}^3} {{\Box_2}^4} {{\Box_3}^4}
  +2160 {{\Box_1}^2} {{\Box_2}^5} {{\Box_3}^4}
  -660 \Box_1 {{\Box_2}^6} {{\Box_3}^4}            \nonumber\\&&\ \ \ \ \mbox{}
  -168 {{\Box_1}^6} {{\Box_3}^5}
  +552 {{\Box_1}^5} \Box_2 {{\Box_3}^5}
  -7764 {{\Box_1}^4} {{\Box_2}^2} {{\Box_3}^5}     \nonumber\\&&\ \ \ \ \mbox{}
  +6732 {{\Box_1}^3} {{\Box_2}^3} {{\Box_3}^5}
  -444 {{\Box_1}^2} {{\Box_2}^4} {{\Box_3}^5}
  +276 \Box_1 {{\Box_2}^5} {{\Box_3}^5}            \nonumber\\&&\ \ \ \ \mbox{}
  -84 {{\Box_2}^6} {{\Box_3}^5}
  +252 {{\Box_1}^5} {{\Box_3}^6}
  +828 {{\Box_1}^4} \Box_2 {{\Box_3}^6}            \nonumber\\&&\ \ \ \ \mbox{}
  -1262 {{\Box_1}^3} {{\Box_2}^2} {{\Box_3}^6}
  -5950 {{\Box_1}^2} {{\Box_2}^3} {{\Box_3}^6}
  +414 \Box_1 {{\Box_2}^4} {{\Box_3}^6}            \nonumber\\&&\ \ \ \ \mbox{}
  +126 {{\Box_2}^5} {{\Box_3}^6}
  -180 {{\Box_1}^4} {{\Box_3}^7}
  -888 {{\Box_1}^3} \Box_2 {{\Box_3}^7}            \nonumber\\&&\ \ \ \ \mbox{}
  +3258 {{\Box_1}^2} {{\Box_2}^2} {{\Box_3}^7}
  -444 \Box_1 {{\Box_2}^3} {{\Box_3}^7}
  -90 {{\Box_2}^4} {{\Box_3}^7}                  \nonumber\\&&\ \ \ \ \mbox{}
  +66 {{\Box_1}^3} {{\Box_3}^8}
  +246 {{\Box_1}^2} \Box_2 {{\Box_3}^8}
  +123 \Box_1 {{\Box_2}^2} {{\Box_3}^8}
  +33 {{\Box_2}^3} {{\Box_3}^8}                  \nonumber\\&&\ \ \ \ \mbox{}
  -10 {{\Box_1}^2} {{\Box_3}^9}
  -5 {{\Box_2}^2} {{\Box_3}^9}
\Big) \nonumber\\&&\ \ \ \ \mbox{}
+\frac1{540 {D^5} \Box_1 \Box_2 \Box_3}\Big(
  -2 {{\Box_1}^{10}}
  +20 {{\Box_1}^9} \Box_2
  -90 {{\Box_1}^8} {{\Box_2}^2}                  \nonumber\\&&\ \ \ \ \mbox{}
  +240 {{\Box_1}^7} {{\Box_2}^3}
  -420 {{\Box_1}^6} {{\Box_2}^4}
  +252 {{\Box_1}^5} {{\Box_2}^5}
  +44 {{\Box_1}^9} \Box_3                        \nonumber\\&&\ \ \ \ \mbox{}
  -260 {{\Box_1}^8} \Box_2 \Box_3
  +6904 {{\Box_1}^7} {{\Box_2}^2} \Box_3
  -19592 {{\Box_1}^6} {{\Box_2}^3} \Box_3          \nonumber\\&&\ \ \ \ \mbox{}
  +12904 {{\Box_1}^5} {{\Box_2}^4} \Box_3
  -318 {{\Box_1}^8} {{\Box_3}^2}
  +2104 {{\Box_1}^7} \Box_2 {{\Box_3}^2}           \nonumber\\&&\ \ \ \ \mbox{}
  +10056 {{\Box_1}^6} {{\Box_2}^2} {{\Box_3}^2}
  +51336 {{\Box_1}^5} {{\Box_2}^3} {{\Box_3}^2}
  -63178 {{\Box_1}^4} {{\Box_2}^4} {{\Box_3}^2}    \nonumber\\&&\ \ \ \ \mbox{}
  +1092 {{\Box_1}^7} {{\Box_3}^3}
  -3956 {{\Box_1}^6} \Box_2 {{\Box_3}^3}
  -23604 {{\Box_1}^5} {{\Box_2}^2} {{\Box_3}^3}    \nonumber\\&&\ \ \ \ \mbox{}
  +48068 {{\Box_1}^4} {{\Box_2}^3} {{\Box_3}^3}
  -2088 {{\Box_1}^6} {{\Box_3}^4}
  -488 {{\Box_1}^5} \Box_2 {{\Box_3}^4}            \nonumber\\&&\ \ \ \ \mbox{}
  +4048 {{\Box_1}^4} {{\Box_2}^2} {{\Box_3}^4}
  -16832 {{\Box_1}^3} {{\Box_2}^3} {{\Box_3}^4}
  +2364 {{\Box_1}^5} {{\Box_3}^5}                \nonumber\\&&\ \ \ \ \mbox{}
  +7420 {{\Box_1}^4} \Box_2 {{\Box_3}^5}
  -10296 {{\Box_1}^3} {{\Box_2}^2} {{\Box_3}^5}
  -1584 {{\Box_1}^4} {{\Box_3}^6}                \nonumber\\&&\ \ \ \ \mbox{}
  -6440 {{\Box_1}^3} \Box_2 {{\Box_3}^6}
  +5844 {{\Box_1}^2} {{\Box_2}^2} {{\Box_3}^6}
  +588 {{\Box_1}^3} {{\Box_3}^7}                 \nonumber\\&&\ \ \ \ \mbox{}
  +1396 {{\Box_1}^2} \Box_2 {{\Box_3}^7}
  -102 {{\Box_1}^2} {{\Box_3}^8}
  +98 \Box_1 \Box_2 {{\Box_3}^8}                   \nonumber\\&&\ \ \ \ \mbox{}
  +8 \Box_1 {{\Box_3}^9}
  -{{\Box_3}^{10}}
\Big)
,\end{fleqnarray}

\begin{fleqnarray}&&\Gamma_{28}(-\Box_1,-\Box_2,-\Box_3) =
\Gamma(-\Box_1,-\Box_2,-\Box_3)\frac1{{D^5}}\Big(
  -16 {{\Box_1}^6} \Box_2 \Box_3                   \nonumber\\&&\ \ \ \ \mbox{}
  +48 {{\Box_1}^5} {{\Box_2}^2} \Box_3
  -32 {{\Box_1}^4} {{\Box_2}^3} \Box_3
  -16 {{\Box_1}^5} \Box_2 {{\Box_3}^2}
  -128 {{\Box_1}^4} {{\Box_2}^2} {{\Box_3}^2}      \nonumber\\&&\ \ \ \ \mbox{}
  +144 {{\Box_1}^3} {{\Box_2}^3} {{\Box_3}^2}
  +112 {{\Box_1}^4} \Box_2 {{\Box_3}^3}
  -144 {{\Box_1}^3} {{\Box_2}^2} {{\Box_3}^3}
  -80 {{\Box_1}^3} \Box_2 {{\Box_3}^4}             \nonumber\\&&\ \ \ \ \mbox{}
  +128 {{\Box_1}^2} {{\Box_2}^2} {{\Box_3}^4}
  -32 {{\Box_1}^2} \Box_2 {{\Box_3}^5}
  +16 \Box_1 \Box_2 {{\Box_3}^6}
\Big) \nonumber\\&&\ \ \ \ \mbox{}
+\frac{\ln(\Box_1/\Box_2)}{45 {D^5} \Box_2}\Big(
  -32 {{\Box_1}^6} {{\Box_2}^2}
  +160 {{\Box_1}^5} {{\Box_2}^3}
  -320 {{\Box_1}^4} {{\Box_2}^4}                 \nonumber\\&&\ \ \ \ \mbox{}
  +4 {{\Box_1}^7} \Box_3
  -72 {{\Box_1}^6} \Box_2 \Box_3
  -1304 {{\Box_1}^5} {{\Box_2}^2} \Box_3           \nonumber\\&&\ \ \ \ \mbox{}
  +2808 {{\Box_1}^4} {{\Box_2}^3} \Box_3
  -24 {{\Box_1}^6} {{\Box_3}^2}
  +200 {{\Box_1}^5} \Box_2 {{\Box_3}^2}            \nonumber\\&&\ \ \ \ \mbox{}
  -88 {{\Box_1}^4} {{\Box_2}^2} {{\Box_3}^2}
  -7800 {{\Box_1}^3} {{\Box_2}^3} {{\Box_3}^2}
  +60 {{\Box_1}^5} {{\Box_3}^3}                  \nonumber\\&&\ \ \ \ \mbox{}
  -80 {{\Box_1}^4} \Box_2 {{\Box_3}^3}
  +4380 {{\Box_1}^3} {{\Box_2}^2} {{\Box_3}^3}
  -80 {{\Box_1}^4} {{\Box_3}^4}                  \nonumber\\&&\ \ \ \ \mbox{}
  -240 {{\Box_1}^3} \Box_2 {{\Box_3}^4}
  -3040 {{\Box_1}^2} {{\Box_2}^2} {{\Box_3}^4}
  +60 {{\Box_1}^3} {{\Box_3}^5}                  \nonumber\\&&\ \ \ \ \mbox{}
  +280 {{\Box_1}^2} \Box_2 {{\Box_3}^5}
  -24 {{\Box_1}^2} {{\Box_3}^6}
  -88 \Box_1 \Box_2 {{\Box_3}^6}
  +4 \Box_1 {{\Box_3}^7}
\Big) \nonumber\\&&\ \ \ \ \mbox{}
+\frac{\ln(\Box_1/\Box_3)}{45 {D^5} \Box_1 \Box_2}\Big(
  -16 {{\Box_1}^7} {{\Box_2}^2}
  +80 {{\Box_1}^6} {{\Box_2}^3}
  -160 {{\Box_1}^5} {{\Box_2}^4}                 \nonumber\\&&\ \ \ \ \mbox{}
  +160 {{\Box_1}^4} {{\Box_2}^5}
  -80 {{\Box_1}^3} {{\Box_2}^6}
  +16 {{\Box_1}^2} {{\Box_2}^7}
  +8 {{\Box_1}^8} \Box_3                         \nonumber\\&&\ \ \ \ \mbox{}
  -144 {{\Box_1}^7} \Box_2 \Box_3
  -844 {{\Box_1}^6} {{\Box_2}^2} \Box_3
  +3048 {{\Box_1}^5} {{\Box_2}^3} \Box_3           \nonumber\\&&\ \ \ \ \mbox{}
  -2700 {{\Box_1}^4} {{\Box_2}^4} \Box_3
  +240 {{\Box_1}^3} {{\Box_2}^5} \Box_3
  +460 {{\Box_1}^2} {{\Box_2}^6} \Box_3            \nonumber\\&&\ \ \ \ \mbox{}
  -72 \Box_1 {{\Box_2}^7} \Box_3
  +4 {{\Box_2}^8} \Box_3
  -48 {{\Box_1}^7} {{\Box_3}^2}
  +400 {{\Box_1}^6} \Box_2 {{\Box_3}^2}            \nonumber\\&&\ \ \ \ \mbox{}
  -2888 {{\Box_1}^5} {{\Box_2}^2} {{\Box_3}^2}
  -1320 {{\Box_1}^4} {{\Box_2}^3} {{\Box_3}^2}
  +6480 {{\Box_1}^3} {{\Box_2}^4} {{\Box_3}^2}     \nonumber\\&&\ \ \ \ \mbox{}
  -2800 {{\Box_1}^2} {{\Box_2}^5} {{\Box_3}^2}
  +200 \Box_1 {{\Box_2}^6} {{\Box_3}^2}
  -24 {{\Box_2}^7} {{\Box_3}^2}                  \nonumber\\&&\ \ \ \ \mbox{}
  +120 {{\Box_1}^6} {{\Box_3}^3}
  -160 {{\Box_1}^5} \Box_2 {{\Box_3}^3}
  +6060 {{\Box_1}^4} {{\Box_2}^2} {{\Box_3}^3}     \nonumber\\&&\ \ \ \ \mbox{}
  -9120 {{\Box_1}^3} {{\Box_2}^3} {{\Box_3}^3}
  +1680 {{\Box_1}^2} {{\Box_2}^4} {{\Box_3}^3}
  -80 \Box_1 {{\Box_2}^5} {{\Box_3}^3}             \nonumber\\&&\ \ \ \ \mbox{}
  +60 {{\Box_2}^6} {{\Box_3}^3}
  -160 {{\Box_1}^5} {{\Box_3}^4}
  -480 {{\Box_1}^4} \Box_2 {{\Box_3}^4}
  -80 {{\Box_1}^3} {{\Box_2}^2} {{\Box_3}^4}       \nonumber\\&&\ \ \ \ \mbox{}
  +2960 {{\Box_1}^2} {{\Box_2}^3} {{\Box_3}^4}
  -240 \Box_1 {{\Box_2}^4} {{\Box_3}^4}
  -80 {{\Box_2}^5} {{\Box_3}^4}                  \nonumber\\&&\ \ \ \ \mbox{}
  +120 {{\Box_1}^4} {{\Box_3}^5}
  +560 {{\Box_1}^3} \Box_2 {{\Box_3}^5}
  -2148 {{\Box_1}^2} {{\Box_2}^2} {{\Box_3}^5}     \nonumber\\&&\ \ \ \ \mbox{}
  +280 \Box_1 {{\Box_2}^3} {{\Box_3}^5}
  +60 {{\Box_2}^4} {{\Box_3}^5}
  -48 {{\Box_1}^3} {{\Box_3}^6}
  -176 {{\Box_1}^2} \Box_2 {{\Box_3}^6}            \nonumber\\&&\ \ \ \ \mbox{}
  -88 \Box_1 {{\Box_2}^2} {{\Box_3}^6}
  -24 {{\Box_2}^3} {{\Box_3}^6}
  +8 {{\Box_1}^2} {{\Box_3}^7}
  +4 {{\Box_2}^2} {{\Box_3}^7}
\Big) \nonumber\\&&\ \ \ \ \mbox{}
+\frac1{135 {D^4} \Box_1 \Box_2 \Box_3}\Big(
   2 {{\Box_1}^8}
  -16 {{\Box_1}^7} \Box_2
  +56 {{\Box_1}^6} {{\Box_2}^2}                  \nonumber\\&&\ \ \ \ \mbox{}
  -112 {{\Box_1}^5} {{\Box_2}^3}
  +70 {{\Box_1}^4} {{\Box_2}^4}
  -28 {{\Box_1}^7} \Box_3
  +212 {{\Box_1}^6} \Box_2 \Box_3                  \nonumber\\&&\ \ \ \ \mbox{}
  -468 {{\Box_1}^5} {{\Box_2}^2} \Box_3
  +284 {{\Box_1}^4} {{\Box_2}^3} \Box_3
  +86 {{\Box_1}^6} {{\Box_3}^2}
  +360 {{\Box_1}^5} \Box_2 {{\Box_3}^2}            \nonumber\\&&\ \ \ \ \mbox{}
  +3690 {{\Box_1}^4} {{\Box_2}^2} {{\Box_3}^2}
  -4136 {{\Box_1}^3} {{\Box_2}^3} {{\Box_3}^2}
  -76 {{\Box_1}^5} {{\Box_3}^3}
  -1900 {{\Box_1}^4} \Box_2 {{\Box_3}^3}           \nonumber\\&&\ \ \ \ \mbox{}
  +4136 {{\Box_1}^3} {{\Box_2}^2} {{\Box_3}^3}
  -70 {{\Box_1}^4} {{\Box_3}^4}
  +1616 {{\Box_1}^3} \Box_2 {{\Box_3}^4}           \nonumber\\&&\ \ \ \ \mbox{}
  -3690 {{\Box_1}^2} {{\Box_2}^2} {{\Box_3}^4}
  +188 {{\Box_1}^3} {{\Box_3}^5}
  +108 {{\Box_1}^2} \Box_2 {{\Box_3}^5}            \nonumber\\&&\ \ \ \ \mbox{}
  -142 {{\Box_1}^2} {{\Box_3}^6}
  -212 \Box_1 \Box_2 {{\Box_3}^6}
  +44 \Box_1 {{\Box_3}^7}
  -2 {{\Box_3}^8}
\Big)
,\end{fleqnarray}

\begin{fleqnarray}&&\Gamma_{29}(-\Box_1,-\Box_2,-\Box_3) =
\Gamma(-\Box_1,-\Box_2,-\Box_3)\frac1{3 {D^6}}\Big(
   48 {{\Box_1}^7} \Box_2 \Box_3                   \nonumber\\&&\ \ \ \ \mbox{}
  -64 {{\Box_1}^6} {{\Box_2}^2} \Box_3
  -176 {{\Box_1}^5} {{\Box_2}^3} \Box_3
  +384 {{\Box_1}^4} {{\Box_2}^4} \Box_3
  -176 {{\Box_1}^3} {{\Box_2}^5} \Box_3            \nonumber\\&&\ \ \ \ \mbox{}
  -64 {{\Box_1}^2} {{\Box_2}^6} \Box_3
  +48 \Box_1 {{\Box_2}^7} \Box_3
  -64 {{\Box_1}^6} \Box_2 {{\Box_3}^2}
  +640 {{\Box_1}^5} {{\Box_2}^2} {{\Box_3}^2}      \nonumber\\&&\ \ \ \ \mbox{}
  -576 {{\Box_1}^4} {{\Box_2}^3} {{\Box_3}^2}
  -576 {{\Box_1}^3} {{\Box_2}^4} {{\Box_3}^2}
  +640 {{\Box_1}^2} {{\Box_2}^5} {{\Box_3}^2}
  -64 \Box_1 {{\Box_2}^6} {{\Box_3}^2}             \nonumber\\&&\ \ \ \ \mbox{}
  -176 {{\Box_1}^5} \Box_2 {{\Box_3}^3}
  -576 {{\Box_1}^4} {{\Box_2}^2} {{\Box_3}^3}
  +1728 {{\Box_1}^3} {{\Box_2}^3} {{\Box_3}^3}     \nonumber\\&&\ \ \ \ \mbox{}
  -576 {{\Box_1}^2} {{\Box_2}^4} {{\Box_3}^3}
  -176 \Box_1 {{\Box_2}^5} {{\Box_3}^3}
  +384 {{\Box_1}^4} \Box_2 {{\Box_3}^4}
  -576 {{\Box_1}^3} {{\Box_2}^2} {{\Box_3}^4}      \nonumber\\&&\ \ \ \ \mbox{}
  -576 {{\Box_1}^2} {{\Box_2}^3} {{\Box_3}^4}
  +384 \Box_1 {{\Box_2}^4} {{\Box_3}^4}
  -176 {{\Box_1}^3} \Box_2 {{\Box_3}^5}
  +640 {{\Box_1}^2} {{\Box_2}^2} {{\Box_3}^5}      \nonumber\\&&\ \ \ \ \mbox{}
  -176 \Box_1 {{\Box_2}^3} {{\Box_3}^5}
  -64 {{\Box_1}^2} \Box_2 {{\Box_3}^6}
  -64 \Box_1 {{\Box_2}^2} {{\Box_3}^6}
  +48 \Box_1 \Box_2 {{\Box_3}^7}
\Big) \nonumber\\&&\ \ \ \ \mbox{}
+\frac{\ln(\Box_1/\Box_2)}{45 {D^6} \Box_1 \Box_2 \Box_3}\Big(
  -8 {{\Box_1}^9} {{\Box_2}^2}
  +56 {{\Box_1}^8} {{\Box_2}^3}
  -168 {{\Box_1}^7} {{\Box_2}^4}                 \nonumber\\&&\ \ \ \ \mbox{}
  +280 {{\Box_1}^6} {{\Box_2}^5}
  -280 {{\Box_1}^5} {{\Box_2}^6}
  +168 {{\Box_1}^4} {{\Box_2}^7}
  -56 {{\Box_1}^3} {{\Box_2}^8}                  \nonumber\\&&\ \ \ \ \mbox{}
  +8 {{\Box_1}^2} {{\Box_2}^9}
  +216 {{\Box_1}^8} {{\Box_2}^2} \Box_3
  -864 {{\Box_1}^7} {{\Box_2}^3} \Box_3
  +1080 {{\Box_1}^6} {{\Box_2}^4} \Box_3           \nonumber\\&&\ \ \ \ \mbox{}
  -1080 {{\Box_1}^4} {{\Box_2}^6} \Box_3
  +864 {{\Box_1}^3} {{\Box_2}^7} \Box_3
  -216 {{\Box_1}^2} {{\Box_2}^8} \Box_3
  -4 {{\Box_1}^9} {{\Box_3}^2}                   \nonumber\\&&\ \ \ \ \mbox{}
  +108 {{\Box_1}^8} \Box_2 {{\Box_3}^2}
  +3600 {{\Box_1}^7} {{\Box_2}^2} {{\Box_3}^2}
  -624 {{\Box_1}^6} {{\Box_2}^3} {{\Box_3}^2}      \nonumber\\&&\ \ \ \ \mbox{}
  -16848 {{\Box_1}^5} {{\Box_2}^4} {{\Box_3}^2}
  +16848 {{\Box_1}^4} {{\Box_2}^5} {{\Box_3}^2}
  +624 {{\Box_1}^3} {{\Box_2}^6} {{\Box_3}^2}      \nonumber\\&&\ \ \ \ \mbox{}
  -3600 {{\Box_1}^2} {{\Box_2}^7} {{\Box_3}^2}
  -108 \Box_1 {{\Box_2}^8} {{\Box_3}^2}
  +4 {{\Box_2}^9} {{\Box_3}^2}
  +28 {{\Box_1}^8} {{\Box_3}^3}                  \nonumber\\&&\ \ \ \ \mbox{}
  -432 {{\Box_1}^7} \Box_2 {{\Box_3}^3}
  -6624 {{\Box_1}^6} {{\Box_2}^2} {{\Box_3}^3}
  +29136 {{\Box_1}^5} {{\Box_2}^3} {{\Box_3}^3}    \nonumber\\&&\ \ \ \ \mbox{}
  -29136 {{\Box_1}^3} {{\Box_2}^5} {{\Box_3}^3}
  +6624 {{\Box_1}^2} {{\Box_2}^6} {{\Box_3}^3}
  +432 \Box_1 {{\Box_2}^7} {{\Box_3}^3}            \nonumber\\&&\ \ \ \ \mbox{}
  -28 {{\Box_2}^8} {{\Box_3}^3}
  -84 {{\Box_1}^7} {{\Box_3}^4}
  +540 {{\Box_1}^6} \Box_2 {{\Box_3}^4}            \nonumber\\&&\ \ \ \ \mbox{}
  -4068 {{\Box_1}^5} {{\Box_2}^2} {{\Box_3}^4}
  -33300 {{\Box_1}^4} {{\Box_2}^3} {{\Box_3}^4}
  +33300 {{\Box_1}^3} {{\Box_2}^4} {{\Box_3}^4}    \nonumber\\&&\ \ \ \ \mbox{}
  +4068 {{\Box_1}^2} {{\Box_2}^5} {{\Box_3}^4}
  -540 \Box_1 {{\Box_2}^6} {{\Box_3}^4}
  +84 {{\Box_2}^7} {{\Box_3}^4}
  +140 {{\Box_1}^6} {{\Box_3}^5}                 \nonumber\\&&\ \ \ \ \mbox{}
  +12780 {{\Box_1}^4} {{\Box_2}^2} {{\Box_3}^5}
  -12780 {{\Box_1}^2} {{\Box_2}^4} {{\Box_3}^5}
  -140 {{\Box_2}^6} {{\Box_3}^5}                 \nonumber\\&&\ \ \ \ \mbox{}
  -140 {{\Box_1}^5} {{\Box_3}^6}
  -540 {{\Box_1}^4} \Box_2 {{\Box_3}^6}
  -6000 {{\Box_1}^3} {{\Box_2}^2} {{\Box_3}^6}     \nonumber\\&&\ \ \ \ \mbox{}
  +6000 {{\Box_1}^2} {{\Box_2}^3} {{\Box_3}^6}
  +540 \Box_1 {{\Box_2}^4} {{\Box_3}^6}
  +140 {{\Box_2}^5} {{\Box_3}^6}                 \nonumber\\&&\ \ \ \ \mbox{}
  +84 {{\Box_1}^4} {{\Box_3}^7}
  +432 {{\Box_1}^3} \Box_2 {{\Box_3}^7}
  -432 \Box_1 {{\Box_2}^3} {{\Box_3}^7}
  -84 {{\Box_2}^4} {{\Box_3}^7}                  \nonumber\\&&\ \ \ \ \mbox{}
  -28 {{\Box_1}^3} {{\Box_3}^8}
  -108 {{\Box_1}^2} \Box_2 {{\Box_3}^8}
  +108 \Box_1 {{\Box_2}^2} {{\Box_3}^8}
  +28 {{\Box_2}^3} {{\Box_3}^8}                  \nonumber\\&&\ \ \ \ \mbox{}
  +4 {{\Box_1}^2} {{\Box_3}^9}
  -4 {{\Box_2}^2} {{\Box_3}^9}
\Big) \nonumber\\&&\ \ \ \ \mbox{}
+\frac1{135 {D^5} \Box_1 \Box_2 \Box_3}\Big(
  -{{\Box_1}^9}
  +13 {{\Box_1}^8} \Box_2
  -50 {{\Box_1}^7} {{\Box_2}^2}                  \nonumber\\&&\ \ \ \ \mbox{}
  +82 {{\Box_1}^6} {{\Box_2}^3}
  -44 {{\Box_1}^5} {{\Box_2}^4}
  -44 {{\Box_1}^4} {{\Box_2}^5}
  +82 {{\Box_1}^3} {{\Box_2}^6}                  \nonumber\\&&\ \ \ \ \mbox{}
  -50 {{\Box_1}^2} {{\Box_2}^7}
  +13 \Box_1 {{\Box_2}^8}
  -{{\Box_2}^9}
  +13 {{\Box_1}^8} \Box_3                        \nonumber\\&&\ \ \ \ \mbox{}
  -154 {{\Box_1}^7} \Box_2 \Box_3
  +64 {{\Box_1}^6} {{\Box_2}^2} \Box_3
  +922 {{\Box_1}^5} {{\Box_2}^3} \Box_3
  -1690 {{\Box_1}^4} {{\Box_2}^4} \Box_3           \nonumber\\&&\ \ \ \ \mbox{}
  +922 {{\Box_1}^3} {{\Box_2}^5} \Box_3
  +64 {{\Box_1}^2} {{\Box_2}^6} \Box_3
  -154 \Box_1 {{\Box_2}^7} \Box_3
  +13 {{\Box_2}^8} \Box_3                        \nonumber\\&&\ \ \ \ \mbox{}
  -50 {{\Box_1}^7} {{\Box_3}^2}
  +64 {{\Box_1}^6} \Box_2 {{\Box_3}^2}
  -5088 {{\Box_1}^5} {{\Box_2}^2} {{\Box_3}^2}     \nonumber\\&&\ \ \ \ \mbox{}
  +5074 {{\Box_1}^4} {{\Box_2}^3} {{\Box_3}^2}
  +5074 {{\Box_1}^3} {{\Box_2}^4} {{\Box_3}^2}
  -5088 {{\Box_1}^2} {{\Box_2}^5} {{\Box_3}^2}     \nonumber\\&&\ \ \ \ \mbox{}
  +64 \Box_1 {{\Box_2}^6} {{\Box_3}^2}
  -50 {{\Box_2}^7} {{\Box_3}^2}
  +82 {{\Box_1}^6} {{\Box_3}^3}
  +922 {{\Box_1}^5} \Box_2 {{\Box_3}^3}            \nonumber\\&&\ \ \ \ \mbox{}
  +5074 {{\Box_1}^4} {{\Box_2}^2} {{\Box_3}^3}
  -17196 {{\Box_1}^3} {{\Box_2}^3} {{\Box_3}^3}
  +5074 {{\Box_1}^2} {{\Box_2}^4} {{\Box_3}^3}     \nonumber\\&&\ \ \ \ \mbox{}
  +922 \Box_1 {{\Box_2}^5} {{\Box_3}^3}
  +82 {{\Box_2}^6} {{\Box_3}^3}
  -44 {{\Box_1}^5} {{\Box_3}^4}
  -1690 {{\Box_1}^4} \Box_2 {{\Box_3}^4}           \nonumber\\&&\ \ \ \ \mbox{}
  +5074 {{\Box_1}^3} {{\Box_2}^2} {{\Box_3}^4}
  +5074 {{\Box_1}^2} {{\Box_2}^3} {{\Box_3}^4}
  -1690 \Box_1 {{\Box_2}^4} {{\Box_3}^4}           \nonumber\\&&\ \ \ \ \mbox{}
  -44 {{\Box_2}^5} {{\Box_3}^4}
  -44 {{\Box_1}^4} {{\Box_3}^5}
  +922 {{\Box_1}^3} \Box_2 {{\Box_3}^5}
  -5088 {{\Box_1}^2} {{\Box_2}^2} {{\Box_3}^5}     \nonumber\\&&\ \ \ \ \mbox{}
  +922 \Box_1 {{\Box_2}^3} {{\Box_3}^5}
  -44 {{\Box_2}^4} {{\Box_3}^5}
  +82 {{\Box_1}^3} {{\Box_3}^6}
  +64 {{\Box_1}^2} \Box_2 {{\Box_3}^6}             \nonumber\\&&\ \ \ \ \mbox{}
  +64 \Box_1 {{\Box_2}^2} {{\Box_3}^6}
  +82 {{\Box_2}^3} {{\Box_3}^6}
  -50 {{\Box_1}^2} {{\Box_3}^7}
  -154 \Box_1 \Box_2 {{\Box_3}^7}                  \nonumber\\&&\ \ \ \ \mbox{}
  -50 {{\Box_2}^2} {{\Box_3}^7}
  +13 \Box_1 {{\Box_3}^8}
  +13 \Box_2 {{\Box_3}^8}
  -{{\Box_3}^9}
\Big)
.\end{fleqnarray}
\arraycolsep=5pt

It should be noted that, in four dimensions, the
basis of twenty nine cubic structures (2.15)--(2.43) is
overcomplete (see Appendix). There exists a constraint
between the purely gravitational structures
(eq. (A.35) of Appendix) which reduces the dimension of the basis
by one. The results above are given in the reduced
basis obtained by elimination of the completely
symmetric part of the structure $\Re_1\Re_2\Re_3({28})$.
This is seen from the fact that the form factor
$\Gamma_{28}$ in eq. (6.40) possesses the property
\begin{eqnarray}
&&\Gamma_{28}(\Box_1,\Box_2,\Box_3)
+\Gamma_{28}(\Box_3,\Box_1,\Box_2)
+\Gamma_{28}(\Box_2,\Box_3,\Box_1) \nonumber\\&&
+\Gamma_{28}(\Box_2,\Box_1,\Box_3)
+\Gamma_{28}(\Box_3,\Box_2,\Box_1)
+\Gamma_{28}(\Box_1,\Box_3,\Box_2) = 0.
\end{eqnarray}

Other properties of the form factors, including their
asymptotic behaviours at large and small arguments, are
studied below. The differential
equations for the basic form factor (6.8), and comments
on the expressions above will be found in sect. 18.

\section{
The $\alpha$-representation of the third-order
form factors in the effective action
}
\setcounter{equation}{0}

\hspace{\parindent}
Below we consider separately the third-order
form factors (6.7) in the effective action.
When written down explicitly as above, they are
very cumbersome. Most compact is their integral
$\alpha$-representation which is in fact the one
obtained initially (see sects. 17--19) and
from which all the other representations are
derived including the one given above.

In the $\alpha$-representation, the functions (6.7)
are given in terms of the integrals
\begin{equation}
\left<\frac{P(\alpha,\Box)}{-\Omega}\right>_3 =
\int_{\alpha \geq 0}
d\alpha_1\,
d\alpha_2\,
d\alpha_3\,
\delta(1
-\alpha_1
-\alpha_2
-\alpha_3)\frac{P(\alpha,\Box)}{-\Omega}
\end{equation}
where
\[ \Omega =
\alpha_2\alpha_3\Box_1
+\alpha_1\alpha_3\Box_2
+\alpha_1\alpha_2\Box_3,
\]
and $P(\alpha,\Box)$ is a polynomial in $\alpha$'s,
boxes and inverse boxes. There are also explicit
contributions of two types: purely tree terms and
terms proportional to (6.9b) with tree coefficients.

The expressions for the twenty nine form factors
(6.7) in the $\alpha$-form are as follows:

\arraycolsep=0pt
\begin{fleqnarray}&&\Gamma_{1}(-\Box_1,-\Box_2,-\Box_3) =
\left<\frac1{-\Omega}\Big(
   {1\over 3}
\Big)\right>_3
,\end{fleqnarray}

\begin{fleqnarray}&&\Gamma_{2}(-\Box_1,-\Box_2,-\Box_3) =
\left<\frac1{-\Omega}\Big(
   {4\over 3}\alpha_1 \alpha_2 \alpha_3
\Big)\right>_3
+\frac13\frac{\ln(\Box_1/\Box_2)}{(\Box_1-\Box_2)}
,\end{fleqnarray}

\begin{fleqnarray}&&\Gamma_{3}(-\Box_1,-\Box_2,-\Box_3) =
\left<\frac1{-\Omega} (
   2\alpha_1 \alpha_2)\right>_3
,\end{fleqnarray}

\begin{fleqnarray}&&\Gamma_{4}(-\Box_1,-\Box_2,-\Box_3) =
\left<\frac1{-\Omega}\Big(
  -{1\over 9}{{\alpha_1}^2}
  +{7\over 9}{{\alpha_1}^2} \alpha_2
  -{{\alpha_1}^2} {{\alpha_2}^2}
\right.\nonumber\\&&\ \ \ \ \mbox{} \left.
  +{1\over {36}}\alpha_3
  -{4\over 9}\alpha_1 \alpha_3
  +{{\alpha_1}^2} \alpha_3
  +{8\over 9}\alpha_1 \alpha_2 \alpha_3
  -2{{\alpha_1}^2} \alpha_2 \alpha_3
\Big)\right>_3
,\end{fleqnarray}

\begin{fleqnarray}&&\Gamma_{5}(-\Box_1,-\Box_2,-\Box_3) =
\left<\frac1{-\Omega}\left(
   {1\over 9}\alpha_1
  +{1\over 9}\alpha_1 \alpha_2
  -{2\over 9}\alpha_1 \alpha_3
\right.\right.\nonumber\\&&\ \ \ \ \mbox{} \left.\left.
+{{\Box_1}\over {\Box_2}}\Big(
   {1\over 9}\alpha_2
  +{1\over 9}{{\alpha_2}^2}
\Big)
\right)\right>_3 \nonumber\\&&\ \ \ \ \mbox{}
+ {1\over {4 \Box_2}} - {{\Box_3}\over {24 \Box_1 \Box_2}}
,\end{fleqnarray}

\begin{fleqnarray}&&\Gamma_{6}(-\Box_1,-\Box_2,-\Box_3) =
\left<\frac1{-\Omega}\Big(
  -{1\over 6}
  +\alpha_1
  -{{\alpha_1}^2}
\Big)\right>_3
,\end{fleqnarray}

\begin{fleqnarray}&&\Gamma_{7}(-\Box_1,-\Box_2,-\Box_3) =
\left<\frac1{-\Omega}\left(
  -{1\over {12}}\alpha_2
  +{5\over 6}\alpha_2 \alpha_3
\right.\right.\nonumber\\&&\ \ \ \ \mbox{}
  -{1\over 4}\alpha_1 \alpha_2 \alpha_3
  -2\alpha_1 {{\alpha_2}^2} \alpha_3
  -2{{\alpha_2}^3} \alpha_3
\nonumber\\&&\ \ \ \ \mbox{} \left.\left.
+{{\Box_2}\over {\Box_1}}\Big(
  -{1\over {12}}\alpha_1
\Big)
\right)\right>_3 \nonumber\\&&\ \ \ \ \mbox{}
+\frac{\ln(\Box_2/\Box_3)}{(\Box_2-\Box_3)}\Big({{-\Box_2}\over {12 \Box_1}}\Big)
,\end{fleqnarray}

\begin{fleqnarray}&&\Gamma_{8}(-\Box_1,-\Box_2,-\Box_3) =
\left<\frac1{-\Omega}\left(
   {1\over 6}\alpha_2
  +{5\over 3}\alpha_1 \alpha_2
\right.\right.\nonumber\\&&\ \ \ \ \mbox{}
  -2{{\alpha_1}^2} \alpha_2
  -4\alpha_1 \alpha_2 \alpha_3
  +8\alpha_1 {{\alpha_2}^2} \alpha_3 \nonumber\\&&\ \ \ \ \mbox{}
+{{\Box_2}\over {\Box_1}}\Big(
  -{1\over 6}\alpha_1
  +{1\over 3}\alpha_1 \alpha_2
  +2{{\alpha_1}^2} \alpha_2  \nonumber\\&&\ \ \ \ \mbox{}\left.\left.\vphantom{\frac12}
  +\alpha_1 \alpha_3
  +8{{\alpha_1}^2} \alpha_2 \alpha_3
\Big)
\right)\right>_3
,\end{fleqnarray}

\begin{fleqnarray}&&\Gamma_{9}(-\Box_1,-\Box_2,-\Box_3) =
\left<\frac1{-\Omega}\left(
  -{1\over {648}}
  +{1\over {216}}\alpha_1
  -{5\over {108}}{{\alpha_1}^2}
\right.\right.\nonumber\\&&\ \ \ \ \mbox{}
  +{2\over {135}}\alpha_2
  +{1\over {36}}\alpha_1 \alpha_2
  +{{35}\over {216}}{{\alpha_1}^2} \alpha_2
  -{{13}\over {1080}}{{\alpha_2}^2} \nonumber\\&&\ \ \ \ \mbox{}
  +{1\over {180}}\alpha_1 {{\alpha_2}^2}
  -{{14}\over {45}}{{\alpha_1}^2} {{\alpha_2}^2}
  +{{11}\over {1080}}{{\alpha_2}^3}
  +{{41}\over {270}}\alpha_1 {{\alpha_2}^3} \nonumber\\&&\ \ \ \ \mbox{}
  -{4\over 9}{{\alpha_1}^2} {{\alpha_2}^3}
  +{{13}\over {360}}\alpha_2 \alpha_3
  +{1\over {72}}\alpha_1 \alpha_2 \alpha_3
  -{{11}\over {135}}{{\alpha_1}^2} \alpha_2 \alpha_3 \nonumber\\&&\ \ \ \ \mbox{}
  -{{19}\over {360}}{{\alpha_2}^2} \alpha_3
  -{1\over 6}\alpha_1 {{\alpha_2}^2} \alpha_3
  -{{113}\over {54}}{{\alpha_1}^2} {{\alpha_2}^2} \alpha_3
  +{{47}\over {540}}{{\alpha_2}^3} \alpha_3          \nonumber\\&&\ \ \ \ \mbox{}
  +{7\over {108}}\alpha_1 {{\alpha_2}^3} \alpha_3
  +{{\alpha_1}^2} {{\alpha_2}^3} \alpha_3
  -{1\over {10}}{{\alpha_2}^2} {{\alpha_3}^2}
  -{1\over {12}}\alpha_1 {{\alpha_2}^2} {{\alpha_3}^2} \nonumber\\&&\ \ \ \ \mbox{}
  +{2\over 3}{{\alpha_1}^2} {{\alpha_2}^2} {{\alpha_3}^2}
  +{1\over 9}{{\alpha_2}^3} {{\alpha_3}^2}           \nonumber\\&&\ \ \ \ \mbox{}
+{{\Box_2}\over {\Box_1}}\Big(
   {2\over {135}}\alpha_1
  -{{11}\over {540}}{{\alpha_1}^2}
  -{{37}\over {540}}\alpha_1 \alpha_2          \nonumber\\&&\ \ \ \ \mbox{}
  -{1\over {36}}{{\alpha_1}^2} \alpha_2
  +{7\over {90}}\alpha_1 {{\alpha_2}^2}
  -{1\over {10}}{{\alpha_1}^2} {{\alpha_2}^2}
  -{1\over {15}}\alpha_1 {{\alpha_2}^3}       \nonumber\\&&\ \ \ \ \mbox{}
  +{1\over 2}{{\alpha_1}^2} {{\alpha_2}^3}
  -{{11}\over {135}}\alpha_1 \alpha_3
  -{1\over {180}}{{\alpha_1}^2} \alpha_3
  -{{43}\over {180}}\alpha_1 \alpha_2 \alpha_3 \nonumber\\&&\ \ \ \ \mbox{}
  +{{14}\over {45}}{{\alpha_1}^2} \alpha_2 \alpha_3
  +{{19}\over {36}}\alpha_1 {{\alpha_2}^2} \alpha_3
  +{{17}\over {36}}{{\alpha_1}^2} {{\alpha_2}^2} \alpha_3
  -{5\over {36}}\alpha_1 {{\alpha_2}^3} \alpha_3  \nonumber\\&&\ \ \ \ \mbox{}
  -{7\over {60}}\alpha_1 {{\alpha_3}^2}
  +{{17}\over {45}}{{\alpha_1}^2} {{\alpha_3}^2}
  -{{31}\over {60}}\alpha_1 \alpha_2 {{\alpha_3}^2}
  +{1\over {36}}{{\alpha_1}^2} \alpha_2 {{\alpha_3}^2}  \nonumber\\&&\ \ \ \ \mbox{}
  +{1\over 2}\alpha_1 {{\alpha_2}^2} {{\alpha_3}^2}
  +{{\alpha_1}^2} {{\alpha_2}^2} {{\alpha_3}^2}
  +{{11}\over {45}}\alpha_1 {{\alpha_3}^3}
  -{{11}\over {18}}{{\alpha_1}^2} {{\alpha_3}^3}   \nonumber\\&&\ \ \ \ \mbox{}\left.\left.
  +{{29}\over {36}}\alpha_1 \alpha_2 {{\alpha_3}^3}
  +{{\alpha_1}^2} \alpha_2 {{\alpha_3}^3}
  +{{\alpha_1}^2} {{\alpha_3}^4}
\Big)
\right)\right>_3 \nonumber\\&&\ \ \ \ \mbox{}
+\frac{\ln(\Box_1/\Box_2)}{(\Box_1-\Box_2)}\Big(\frac1{720}-{{\Box_1}\over {120 \Box_3}}\Big)
- {{\Box_1}\over {2160 \Box_2 \Box_3}}
,\end{fleqnarray}

\begin{fleqnarray}&&\Gamma_{10}(-\Box_1,-\Box_2,-\Box_3) =
\left<\frac1{-\Omega}\Big(
   {1\over 3}\alpha_1 \alpha_2 \alpha_3
\Big)\right>_3 \nonumber\\&&\ \ \ \ \mbox{}
+ {1\over {270 \Box_3}} - {{\Box_1}\over {540 \Box_2 \Box_3}}
,\end{fleqnarray}

\begin{fleqnarray}&&\Gamma_{11}(-\Box_1,-\Box_2,-\Box_3) =
\left<\frac1{-\Omega}\left(
  -{7\over {360}}\alpha_1
  +{1\over {180}}{{\alpha_1}^2}
  +{1\over {360}}{{\alpha_1}^3}
\right.\right.\nonumber\\&&\ \ \ \ \mbox{}
  -{7\over {135}}\alpha_1 \alpha_2
  +{7\over {45}}{{\alpha_1}^2} \alpha_2
  -{{49}\over {1080}}{{\alpha_1}^3} \alpha_2
  -{2\over {15}}{{\alpha_1}^2} {{\alpha_2}^2} \nonumber\\&&\ \ \ \ \mbox{}
  +{{19}\over {108}}{{\alpha_1}^3} {{\alpha_2}^2}
  -{1\over {216}}\alpha_3
  +{1\over {45}}\alpha_1 \alpha_3
  +{1\over {60}}{{\alpha_1}^2} \alpha_3        \nonumber\\&&\ \ \ \ \mbox{}
  -{1\over {270}}{{\alpha_1}^3} \alpha_3
  -{2\over {45}}\alpha_1 \alpha_2 \alpha_3
  +{{22}\over {45}}{{\alpha_1}^2} \alpha_2 \alpha_3
  -{{25}\over {216}}{{\alpha_1}^3} \alpha_2 \alpha_3 \nonumber\\&&\ \ \ \ \mbox{}
  -{{13}\over {36}}{{\alpha_1}^2} {{\alpha_2}^2} \alpha_3
  +{{41}\over {1080}}\alpha_1 {{\alpha_3}^2}
  -{{67}\over {270}}{{\alpha_1}^2} {{\alpha_3}^2}
  +{2\over 9}{{\alpha_1}^3} {{\alpha_3}^2}       \nonumber\\&&\ \ \ \ \mbox{}
  +{1\over 6}\alpha_1 \alpha_2 {{\alpha_3}^2}
  -{{145}\over {216}}{{\alpha_1}^2} \alpha_2 {{\alpha_3}^2} \nonumber\\&&\ \ \ \ \mbox{}
+{{\Box_3}\over {\Box_2}}\Big(
   {1\over {72}}\alpha_2
  -{1\over {12}}\alpha_1 \alpha_2
  +{1\over {45}}{{\alpha_1}^2} \alpha_2   \nonumber\\&&\ \ \ \ \mbox{}
  +{{19}\over {540}}{{\alpha_2}^2}
  +{7\over {45}}\alpha_1 {{\alpha_2}^2}
  -{1\over 3}{{\alpha_1}^2} {{\alpha_2}^2}
  +{1\over 3}{{\alpha_1}^3} {{\alpha_2}^2}   \nonumber\\&&\ \ \ \ \mbox{}
  -{4\over {135}}\alpha_2 \alpha_3
  +{1\over {45}}\alpha_1 \alpha_2 \alpha_3
  -{1\over {45}}{{\alpha_1}^2} \alpha_2 \alpha_3
  +{1\over {45}}{{\alpha_2}^2} \alpha_3         \nonumber\\&&\ \ \ \ \mbox{}
  -{8\over {45}}\alpha_1 {{\alpha_2}^2} \alpha_3
  +{1\over {90}}\alpha_2 {{\alpha_3}^2}
  +{4\over {45}}\alpha_1 \alpha_2 {{\alpha_3}^2}
  -{2\over 9}{{\alpha_1}^2} \alpha_2 {{\alpha_3}^2} \nonumber\\&&\ \ \ \ \mbox{}
  -{4\over {45}}{{\alpha_2}^2} {{\alpha_3}^2}
  +{2\over 9}\alpha_1 {{\alpha_2}^2} {{\alpha_3}^2}
\Big)                                              \nonumber\\&&\ \ \ \ \mbox{}
+{{\Box_1}\over {\Box_2}}\Big(
  -{7\over {540}}\alpha_2
  +{1\over {540}}\alpha_1 \alpha_2
  +{2\over {135}}{{\alpha_2}^2}                       \nonumber\\&&\ \ \ \ \mbox{}
  -{7\over {180}}\alpha_1 {{\alpha_2}^2}
  -{1\over {60}}\alpha_2 \alpha_3
  +{7\over {90}}\alpha_1 \alpha_2 \alpha_3
  -{1\over 9}{{\alpha_1}^2} \alpha_2 \alpha_3    \nonumber\\&&\ \ \ \ \mbox{}
  +{2\over {45}}{{\alpha_2}^2} \alpha_3
  -{{22}\over {45}}\alpha_1 {{\alpha_2}^2} \alpha_3
  +{1\over 3}{{\alpha_1}^2} {{\alpha_2}^2} \alpha_3
  +{2\over {45}}\alpha_2 {{\alpha_3}^2}          \nonumber\\&&\ \ \ \ \mbox{} \left.\left.
  -{4\over 9}\alpha_1 \alpha_2 {{\alpha_3}^2}
  +{2\over 9}{{\alpha_1}^2} \alpha_2 {{\alpha_3}^2}
  -{2\over 9}{{\alpha_2}^3} {{\alpha_3}^2}
\Big)
\right)\right>_3 \nonumber\\&&\ \ \ \ \mbox{}
+\frac1{120}\frac{\ln(\Box_1/\Box_2)}{(\Box_1-\Box_2)}
-{{\Box_3}\over {540 \Box_1 \Box_2}}
,\end{fleqnarray}

\begin{fleqnarray}&&\Gamma_{12}(-\Box_1,-\Box_2,-\Box_3) =
\frac1{\Box_1}\left[
\left<\frac1{-\Omega}\left(
   {8\over 3}{{\alpha_1}^2}
  -4{{\alpha_1}^3}
\right. \right.\right.\nonumber\\&&\ \ \ \ \mbox{}\left.\left.
+{\Box_1 \over {\Box_2}}\Big(
   {1\over 3}\alpha_2
\Big)
+{\Box_1 \over {\Box_3}}\Big(
   {1\over 3}\alpha_3
\Big)
\right)\right>_3 \nonumber\\&&\ \ \ \ \mbox{}\left.
+\frac{\ln(\Box_1/\Box_2)}{(\Box_1-\Box_2)}\Big({\Box_1 \over {3 \Box_3}}\Big)
+\frac{\ln(\Box_1/\Box_3)}{(\Box_1-\Box_3)}\Big({\Box_1 \over {3 \Box_2}}\Big)\right]
,\end{fleqnarray}

\begin{fleqnarray}&&\Gamma_{13}(-\Box_1,-\Box_2,-\Box_3) =
\frac1{\Box_1}\left[
\left<\frac1{-\Omega}\Big(
   2\alpha_1
\Big)\right>_3
+2\frac{\ln(\Box_2/\Box_3)}{(\Box_2-\Box_3)}\right]
,\end{fleqnarray}

\begin{fleqnarray}&&\Gamma_{14}(-\Box_1,-\Box_2,-\Box_3) =
\frac1{\Box_3}
\left<\frac1{-\Omega}\Big(
   2\alpha_3
  -4{{\alpha_3}^2}
\Big)\right>_3
,\end{fleqnarray}

\begin{fleqnarray}&&\Gamma_{15}(-\Box_1,-\Box_2,-\Box_3) =
\frac1{\Box_1}
\left<\frac1{-\Omega}\Big(
   {2\over 3}{{\alpha_1}^2}
  -4{{\alpha_1}^3}
  +4{{\alpha_1}^4}
\Big)\right>_3
,\end{fleqnarray}

\begin{fleqnarray}&&\Gamma_{16}(-\Box_1,-\Box_2,-\Box_3) =
\frac1{\Box_1}
\left<\frac1{-\Omega}\Big(
   {4\over 9}\alpha_1
  +{4\over 3}\alpha_2 {{\alpha_1}^2}
  -{4\over 9}\alpha_1 \alpha_3
  -{4\over 3}{{\alpha_1}^2} \alpha_3
\Big)\right>_3 \nonumber\\&&\ \ \ \ \mbox{}
+{1\over {6\Box_1\Box_2}}
,\end{fleqnarray}

\begin{fleqnarray}&&\Gamma_{17}(-\Box_1,-\Box_2,-\Box_3) =
\frac1{\Box_1}\left[
\left<\frac1{-\Omega}\Big(
   2{{\alpha_1}^2}
\Big)\right>_3
+\frac{\ln(\Box_2/\Box_3)}{(\Box_2-\Box_3)}\right]
,\end{fleqnarray}

\begin{fleqnarray}&&\Gamma_{18}(-\Box_1,-\Box_2,-\Box_3) =
\frac1{\Box_1}
\left<\frac1{-\Omega}\Big(
   2{{\alpha_1}^2}
  -8{{\alpha_1}^2} \alpha_2
  +8{{\alpha_1}^2} \alpha_2 \alpha_3
\Big)\right>_3
,\end{fleqnarray}

\begin{fleqnarray}&&\Gamma_{19}(-\Box_1,-\Box_2,-\Box_3) =
\frac1{\Box_1}\left[
\left<\frac1{-\Omega}\Big(
  -4{{\alpha_1}^2} \alpha_2 \alpha_3
\Big)\right>_3 \right.\nonumber\\&&\ \ \ \ \mbox{} \left.
+\frac16\frac{\ln(\Box_2/\Box_3)}{(\Box_2-\Box_3)}\right]
,\end{fleqnarray}

\begin{fleqnarray}&&\Gamma_{20}(-\Box_1,-\Box_2,-\Box_3) =
\frac1{\Box_1}\left[
\left<\frac1{-\Omega}\Big(
  -{2\over 3}{{\alpha_1}^2}
  +{{10}\over 3}{{\alpha_1}^2} \alpha_2 \right.\right.\nonumber\\&&\ \ \ \ \mbox{}\left.
  -{2\over 3}\alpha_1 {{\alpha_2}^2}
  -{{11}\over 3}{{\alpha_1}^2} \alpha_2 \alpha_3
  +{2\over 3}\alpha_1 {{\alpha_2}^2} \alpha_3
\Big)\right>_3 \nonumber\\&&\ \ \ \ \mbox{}\left.
-\frac16\frac{\ln(\Box_2/\Box_3)}{(\Box_2-\Box_3)}\right]
,\end{fleqnarray}

\begin{fleqnarray}&&\Gamma_{21}(-\Box_1,-\Box_2,-\Box_3) =
\frac1{\Box_1}\left[
\left<\frac1{-\Omega}\Big(
   8{{\alpha_1}^2} \alpha_3
  -16{{\alpha_1}^2} {{\alpha_3}^2}
\Big)\right>_3 \right.\nonumber\\&&\ \ \ \ \mbox{}\left.
-\frac23\frac{\ln(\Box_2/\Box_3)}{(\Box_2-\Box_3)}\right]
,\end{fleqnarray}

\begin{fleqnarray}&&\Gamma_{22}(-\Box_1,-\Box_2,-\Box_3) =
\frac1{\Box_1}\left[
\left<\frac1{-\Omega}\Big(
   {1\over {90}}\alpha_1
  +{1\over 5}{{\alpha_1}^2}
  -{{23}\over {54}}\alpha_1 \alpha_3    \right.\right.\nonumber\\&&\ \ \ \ \mbox{}
  -{{101}\over {90}}{{\alpha_1}^2} \alpha_3
  +{{29}\over {90}}\alpha_1 \alpha_2 \alpha_3
  +{{68}\over {45}}{{\alpha_1}^2} \alpha_2 \alpha_3
  +{1\over 6}\alpha_1 {{\alpha_3}^2}
  +{4\over 3}{{\alpha_1}^2} {{\alpha_3}^2} \nonumber\\&&\ \ \ \ \mbox{} \left.
  +{4\over {45}}\alpha_1 \alpha_2 {{\alpha_3}^2}
  +{2\over 9}\alpha_1 {{\alpha_2}^2} {{\alpha_3}^2}
  +{4\over 9}\alpha_1 {{\alpha_3}^3}
  -{2\over 9}\alpha_1 {{\alpha_3}^4}
\Big)\right>_3 \nonumber\\&&\ \ \ \ \mbox{}\left.
+\frac1{90}\frac{\ln(\Box_2/\Box_3)}{(\Box_2-\Box_3)}\right] \nonumber\\&&\ \ \ \ \mbox{}
+\frac1{\Box_2}\left[
\left<\frac1{-\Omega}\Big(
   {2\over {45}}\alpha_2
  +{2\over 9}{{\alpha_1}^2} \alpha_2
  +{2\over {45}}{{\alpha_2}^2}
  -{2\over {45}}\alpha_1 {{\alpha_2}^2}     \right.\right.  \nonumber\\&&\ \ \ \ \mbox{}
  -{{28}\over {45}}{{\alpha_1}^2} {{\alpha_2}^2}
  -{4\over 9}{{\alpha_1}^2} {{\alpha_2}^3}
  -{7\over {54}}\alpha_2 \alpha_3
  +{7\over {10}}\alpha_1 \alpha_2 \alpha_3       \nonumber\\&&\ \ \ \ \mbox{}
  -{{46}\over {45}}{{\alpha_1}^2} \alpha_2 \alpha_3
  +{1\over {18}}{{\alpha_2}^2} \alpha_3
  +{4\over {45}}\alpha_1 {{\alpha_2}^2} \alpha_3
  +{{38}\over 9}{{\alpha_1}^2} {{\alpha_2}^2} \alpha_3 \nonumber\\&&\ \ \ \ \mbox{}
  -4{{\alpha_1}^2} {{\alpha_2}^3} \alpha_3
  -{7\over {10}}\alpha_2 {{\alpha_3}^2}
  -{{17}\over {15}}\alpha_1 \alpha_2 {{\alpha_3}^2}
  +{{19}\over 9}{{\alpha_1}^2} \alpha_2 {{\alpha_3}^2} \nonumber\\&&\ \ \ \ \mbox{}
  +{8\over {15}}{{\alpha_2}^2} {{\alpha_3}^2}
  +{{11}\over 9}\alpha_1 {{\alpha_2}^2} {{\alpha_3}^2}
  -4{{\alpha_1}^2} {{\alpha_2}^2} {{\alpha_3}^2}
  +{{10}\over 9}\alpha_2 {{\alpha_3}^3}               \nonumber\\&&\ \ \ \ \mbox{}\left.
  +{{11}\over 9}\alpha_1 \alpha_2 {{\alpha_3}^3}
  +{2\over 9}{{\alpha_2}^2} {{\alpha_3}^3}
  -{2\over 9}\alpha_2 {{\alpha_3}^4}
\Big)\right>_3 \nonumber\\&&\ \ \ \ \mbox{}\left.
+\frac1{30}\frac{\ln(\Box_1/\Box_3)}{(\Box_1-\Box_3)}\right] \nonumber\\&&\ \ \ \ \mbox{}
-{{1}\over {270 \Box_1 \Box_2}}
,\end{fleqnarray}

\begin{fleqnarray}&&\Gamma_{23}(-\Box_1,-\Box_2,-\Box_3) =
\frac1{\Box_1}\left[
\left<\frac1{-\Omega}\left(
  -{4\over {45}}\alpha_1
  +{2\over {135}}{{\alpha_1}^2}
  -{{16}\over {135}}\alpha_1 \alpha_2
\right.\right.\right.\nonumber\\&&\ \ \ \ \mbox{}
  +{2\over 9}{{\alpha_1}^2} \alpha_2
  +{4\over {15}}{{\alpha_1}^3} \alpha_2
  -{{34}\over {45}}\alpha_1 {{\alpha_2}^2}
  +{{76}\over {45}}{{\alpha_1}^2} {{\alpha_2}^2} \nonumber\\&&\ \ \ \ \mbox{}
  -{8\over 9}{{\alpha_1}^3} {{\alpha_2}^2}
  +{8\over 9}\alpha_1 {{\alpha_2}^4}
  +{7\over {45}}\alpha_1 \alpha_3
  -{1\over 5}{{\alpha_1}^2} \alpha_3           \nonumber\\&&\ \ \ \ \mbox{}
  -{{19}\over {45}}\alpha_1 \alpha_2 \alpha_3
  -{{28}\over {45}}{{\alpha_1}^2} \alpha_2 \alpha_3
  +{{20}\over 9}\alpha_1 {{\alpha_2}^2} \alpha_3
  -{4\over 3}{{\alpha_1}^2} {{\alpha_2}^2} \alpha_3 \nonumber\\&&\ \ \ \ \mbox{}
  +{1\over {15}}\alpha_1 {{\alpha_3}^2}
  +{8\over {45}}{{\alpha_1}^2} {{\alpha_3}^2}
  +{{32}\over {45}}\alpha_1 \alpha_2 {{\alpha_3}^2}
  -{4\over 9}{{\alpha_1}^2} \alpha_2 {{\alpha_3}^2}  \nonumber\\&&\ \ \ \ \mbox{}
  -{4\over 3}\alpha_1 {{\alpha_2}^2} {{\alpha_3}^2}
  -{8\over {45}}\alpha_1 {{\alpha_3}^3}
  -{4\over 9}{{\alpha_1}^2} {{\alpha_3}^3}
  -{4\over 9}\alpha_1 \alpha_2 {{\alpha_3}^3} \nonumber\\&&\ \ \ \ \mbox{}
+{{\Box_1}\over {\Box_3}}\Big(
  -{2\over {45}}\alpha_3
  -{4\over {45}}\alpha_1 \alpha_3
  -{{14}\over {45}}{{\alpha_1}^2} \alpha_3
  +{4\over 9}{{\alpha_1}^4} \alpha_3 \nonumber\\&&\ \ \ \ \mbox{}
  -{4\over {45}}\alpha_2 \alpha_3
  +{4\over {45}}\alpha_1 \alpha_2 \alpha_3
  +{{52}\over {45}}{{\alpha_1}^2} \alpha_2 \alpha_3
  -{{14}\over {45}}{{\alpha_2}^2} \alpha_3  \nonumber\\&&\ \ \ \ \mbox{}
  +{{52}\over {45}}\alpha_1 {{\alpha_2}^2} \alpha_3
  -{8\over 9}{{\alpha_1}^2} {{\alpha_2}^2} \alpha_3
  +{4\over 9}{{\alpha_2}^4} \alpha_3
  +{{14}\over {15}}{{\alpha_1}^2} {{\alpha_3}^2} \nonumber\\&&\ \ \ \ \mbox{}
  -{4\over {45}}\alpha_1 \alpha_2 {{\alpha_3}^2}
  -{4\over 9}{{\alpha_1}^2} \alpha_2 {{\alpha_3}^2}
  +{{14}\over {15}}{{\alpha_2}^2} {{\alpha_3}^2}
  -{4\over 9}\alpha_1 {{\alpha_2}^2} {{\alpha_3}^2}
\nonumber\\&&\ \ \ \ \mbox{}  \left.\left.
  -{2\over 9}{{\alpha_1}^2} {{\alpha_3}^3}
  -{2\over 9}{{\alpha_2}^2} {{\alpha_3}^3}
\Big)
\right)\right>_3 \nonumber\\&&\ \ \ \ \mbox{}\left.
+\frac{\ln(\Box_1/\Box_2)}{(\Box_1-\Box_2)}\Big({{-\Box_1}\over {30 \Box_3}}\Big)\right] \nonumber\\&&\ \ \ \ \mbox{}
+{1\over {135\Box_1\Box_2}}
,\end{fleqnarray}

\begin{fleqnarray}&&\Gamma_{24}(-\Box_1,-\Box_2,-\Box_3) =
\frac1{\Box_2}\left[
\left<\frac1{-\Omega}\left(
  -{5\over {54}}\alpha_2
  -{{23}\over {270}}\alpha_1 \alpha_2
\right.\right.\right.\nonumber\\&&\ \ \ \ \mbox{}
  +{2\over 5}{{\alpha_1}^2} \alpha_2
  -{4\over {15}}{{\alpha_1}^3} \alpha_2
  +{1\over {270}}{{\alpha_2}^2} \nonumber\\&&\ \ \ \ \mbox{}
  +{{13}\over {270}}\alpha_2 \alpha_3
  -{1\over 5}\alpha_1 \alpha_2 \alpha_3
  +{4\over {15}}\alpha_1 \alpha_2 {{\alpha_3}^2} \nonumber\\&&\ \ \ \ \mbox{}\left.\left.
+{{\Box_2} \over {\Box_1}}\Big(
  -{2\over {45}}\alpha_1
  +{1\over {45}}\alpha_1 \alpha_2
  +{1\over {45}}\alpha_1 \alpha_3
\Big)
\right)\right>_3 \nonumber\\&&\ \ \ \ \mbox{}\left.
+\frac{\ln(\Box_2/\Box_3)}{(\Box_2-\Box_3)}\Big({{-\Box_2}\over {30 \Box_1}}\Big)\right]\nonumber\\&&\ \ \ \ \mbox{}
+{1\over {540\Box_2\Box_3}}
,\end{fleqnarray}

\begin{fleqnarray}&&\Gamma_{25}(-\Box_1,-\Box_2,-\Box_3) =
\frac1{\Box_1}\left[
\left<\frac1{-\Omega}\left(
  -{{13}\over {135}}\alpha_1
  -{{56}\over {135}}\alpha_1 \alpha_2
\right.\right.\right.\nonumber\\&&\ \ \ \ \mbox{}
  +{{28}\over {45}}\alpha_1 {{\alpha_2}^2}
  +{{32}\over {45}}{{\alpha_1}^2} {{\alpha_2}^2}
  +{{16}\over {15}}\alpha_1 {{\alpha_2}^2} \alpha_3  \nonumber\\&&\ \ \ \ \mbox{}
+{{\Box_1} \over {\Box_3}}\Big(
  -{8\over {45}}\alpha_3
  -{{37}\over {135}}\alpha_1 \alpha_3
  +{{16}\over {45}}{{\alpha_1}^3} \alpha_3
  +{{11}\over {135}}\alpha_2 \alpha_3    \nonumber\\&&\ \ \ \ \mbox{}
  +{{28}\over {45}}\alpha_1 \alpha_2 \alpha_3
  -{4\over {45}}{{\alpha_2}^2} \alpha_3
  -{{16}\over {45}}\alpha_1 {{\alpha_2}^2} \alpha_3
  +{1\over {135}}{{\alpha_3}^2}
\nonumber\\&&\ \ \ \ \mbox{} \left.\left.
  +{{32}\over {45}}\alpha_1 {{\alpha_3}^2}
  -{{16}\over {45}}{{\alpha_1}^2} {{\alpha_3}^2}
  -{{32}\over {45}}\alpha_1 \alpha_2 {{\alpha_3}^2}
  +{{16}\over {45}}{{\alpha_2}^2} {{\alpha_3}^2}
\Big)
\right)\right>_3 \nonumber\\&&\ \ \ \ \mbox{}\left.
+\frac{\ln(\Box_1/\Box_2)}{(\Box_1-\Box_2)}\Big({{-2\Box_1}\over {15 \Box_3}}\Big)\right] \nonumber\\&&\ \ \ \ \mbox{}
-{{1}\over {135\Box_1\Box_3}} + {{1}\over {270 \Box_2 \Box_3}}
,\end{fleqnarray}

\begin{fleqnarray}&&\Gamma_{26}(-\Box_1,-\Box_2,-\Box_3) =
\frac1{\Box_1\Box_2}
\left<\frac1{-\Omega}\Big(
   4{{\alpha_1}^2} {{\alpha_2}^2}
\Big)\right>_3
,\end{fleqnarray}

\begin{fleqnarray}&&\Gamma_{27}(-\Box_1,-\Box_2,-\Box_3) =
\frac1{\Box_1\Box_2}
\left<\frac1{-\Omega}\Big(
  -{4\over 3}{{\alpha_1}^4} {{\alpha_2}^2}
  +{8\over 3}{{\alpha_1}^3} {{\alpha_2}^3} \right.\nonumber\\&&\ \ \ \ \mbox{}\left.
  -{4\over 3}{{\alpha_1}^2} {{\alpha_2}^2} {{\alpha_3}^2}
\Big)\right>_3 \nonumber\\&&\ \ \ \ \mbox{}
-{{1}\over {540\Box_1\Box_2\Box_3}}
,\end{fleqnarray}

\begin{fleqnarray}&&\Gamma_{28}(-\Box_1,-\Box_2,-\Box_3) =
\frac1{\Box_1\Box_2}
\left<\frac1{-\Omega}\Big(
   {8\over 3}{{\alpha_1}^2} {{\alpha_2}^2} \alpha_3
\Big)\right>_3  \nonumber\\&&\ \ \ \ \mbox{}
+{1\over {135\Box_1\Box_2\Box_3}}
,\end{fleqnarray}

\begin{fleqnarray}&&\Gamma_{29}(-\Box_1,-\Box_2,-\Box_3) =
\frac1{\Box_1\Box_2\Box_3}
\left<\frac1{-\Omega}\Big(
 {8\over 3}{{\alpha_1}^2} {{\alpha_2}^2} {{\alpha_3}^2}
\Big)\right>_3
.\end{fleqnarray}
\arraycolsep=3pt

The $\alpha$-representation exists also for the
form factors in the heat kernel (sect. 15), and in both
cases it suffers from one and the same shortcoming:
it is not unique in a sense that the vanishing of an
integral like (7.1) does not imply the vanishing
of the polynomial $P(\alpha,\Box)$. The minor cause
of this nonuniqueness is the presence in (7.1) of the
delta-function which confines $P(\alpha,\Box)$ to
$\sum\alpha=1$, and the major one is the fact that
$P(\alpha,\Box)$ depends not only on $\alpha$ but
also on $\Box$. In expressions (7.2)--(7.30),
this dependence manifests itself in the presence
of the factors $\Box_n$ and $1/\Box_m$ ($n,m=1,2,3$)
in the coefficients of the $\alpha$-polynomials.
Because the arguments of the functions $\Gamma_i$
enter not only the kernel $1/(-\Omega)$, (7.1) is not
a proper integral representation. In consequence
of this fact, there exists a hierarchy of
nontrivial identities between the averages of the
form (7.1). Examples of such identities will
be found in sect. 18.

The form factors of the curvature structures with
and without derivatives have different dimension.
In the form factors $\Gamma_1$ to $\Gamma_{11}$
corresponding to the structures without derivatives,
the $\Box$-coefficients of the $\alpha$-polynomials
are always of the form $\Box_n/\Box_m$. As the
analysis in sects. 19, 20 shows, it is these
coefficients that determine the asymptotic
behaviours of the form factors. Each factor
$\Box_n/\Box_m$ causes the logarithmic growth at
large $\Box_n$ and the power growth at small $\Box_m$.
This correlation can be traced by a comparison of the
exact expressions (7.2)--(7.30) with the tables
of asymptotic behaviours in sects. 10, 11. In
calculations with the exact
form factors, it is the coefficients $\Box_n/\Box_m$
that cause the problem of nonuniqueness.

Another problem with the $\alpha$-representation
is the presence of nonanalytic terms that are
not in the $\alpha$-form. There are two types of
such terms:
\begin{equation}
{\rm a)\ }\frac{\ln(\Box_n/\Box_k)}{\Box_n-\Box_k},\hspace{7mm}
{\rm b)\ }\frac{\ln(\Box_n/\Box_k)}{\Box_n-\Box_k}
\frac{\Box_n}{\Box_m}.
\end{equation}
The terms (7.31a) are really independent but the
terms (7.31b) mix up with the $\alpha$-averages
in the limits $\Box_n\rightarrow-\infty$ and
$\Box_m\rightarrow-0$. This can also be traced by
a comparison of
expressions (7.2)--(7.30) with the tables of
asymptotic behaviours below.

In eqs. (7.27)--(7.30), the form factor $\Gamma_{29}$
of the structure with six derivatives and the form
factors $\Gamma_{26}$ to $\Gamma_{28}$ of the
structures with four derivatives contain the overall
factors $1/\Box_1\Box_2\Box_3$ and $1/\Box_1\Box_2$ respectively
which can be attached to the basis structures
themselves to give these structures the standard
dimension. The respectively redefined form factors
are then in no way different from the form factors
of the structures without derivatives. A similar
redefinition in the case of the form factors
$\Gamma_{12}$ to $\Gamma_{25}$ corresponding to the
structures with two derivatives encounters a
difficulty since it is not immediately clear
which of the three inverse boxes
$1/\Box_1, 1/\Box_2, 1/\Box_3$ should be attached to
the curvature structure. It turns out that at
least a partial answer to this question can be given
on the basis of the asymptotic behaviours of the
form factors. With one exception, there exists a
choice (and sometimes more than one) such that
the redefined form factor does not acquire
{\it a growth} at large arguments. In expressions
(7.13)--(7.26) above, the $1/\Box$ satisfying this
criterion is written down as an overall factor.
The exception is $\Gamma_{22}$. This form factor
can only be written as a sum of two each of which
satisfying the above criterion. This is fixed in the
form of expression (7.23). Strictly speaking,
$\Gamma_{25}$ in eq. (7.26) is, in this sense, also
a sum of two form factors but one of the summands
is a pure tree.

Inspection of expressions (7.2)--(7.30) shows
that they obey the following general rule.
Each $1/\Box$ multiplier in the $\alpha$-polynomial
appears only in a product with the like $\alpha$,
e.g. $\alpha_1/\Box_1$, $\alpha_1\alpha_2/\Box_1\Box_2$,
$\alpha_1\alpha_2\alpha_3/\Box_1\Box_2\Box_3$, etc.
This "rule of the like $\alpha$" plays an important
role both in calculations with the exact form
factors and in the forms of their asymptotic
behaviours. The work of this rule is discussed
in detail in sects. 19, 20.

The nonuniqueness of the $\alpha$-representation
makes it unfit for carrying out checks like
the check of the trace anomaly.
The explicit representation in sect. 6 possesses
the advantage of being unique but is cumbersome
and unfit for applications to the expectation-value
problems because the nonlocal operators are not
expressed through the Green function [1,4]. This
compels looking for unique integral representations
of the form factors. Two such representations are
given below.

It should be emphasized that only the symmetrized
form factors $\Gamma_i^{\rm sym}$ make sense
(eq. (6.12)). In a not symmetrized form, various
expressions for $\Gamma_i$ may differ by
terms vanishing after the symmetrization.

\section{The Laplace representation of the
third-order form factors in the effective action}
\setcounter{equation}{0}

\hspace{\parindent}
The Laplace form of the form factors arises naturally
when the loop diagrams are calculated in terms of the
heat kernels. Therefore, this representation
is readily obtained in the present technique.

The Laplace representation {\it proper}
\begin{equation}
f(-\Box_1,-\Box_2,-\Box_3) = \int^\infty_0\!\! d^3 u\,\rho(u_1,u_2,u_3)
{\rm e}^{{\scriptscriptstyle \sum} u\Box},
\end{equation}
\[
\sum u\Box \equiv u_1\Box_1+u_2\Box_2+u_3\Box_3,
\hspace{7mm}
\Box_1,\Box_2,\Box_3 < 0
\]
exists only for functions $f$ decreasing at large
values of each of the arguments (and is insensitive
to a power growth at small values). This representation
is, therefore, useful for studying the large-$\Box$
limit. In the process of giving the form factors the
Laplace guise (sect. 19), their nondecreasing terms
get detached and take the form of Laplace integrals
multiplied by powers of $\Box$'s. It is important that
the Laplace originals in these terms depend on only
two of the three arguments. In consequence of this fact,
the nondecreasing terms factorize into functions of
one or two variables which are, moreover, elementary.

Another property of the form factors, which is a
direct consequence of the "rule of the like
$\alpha$" (see sect. 7), is that all their Laplace
originals are rational.

The expressions for the form factors (6.7)
in the Laplace form are as follows:

\arraycolsep=0pt
\begin{fleqnarray}&&\Gamma_{1}(-\Box_1,-\Box_2,-\Box_3)=\int^\infty_0\!\! d^3 u\,\rho_{1}(u_1,u_2,u_3){\rm e}^{{\scriptscriptstyle \sum} u\Box},\end{fleqnarray}

\begin{fleqnarray}&&\Gamma_{2}(-\Box_1,-\Box_2,-\Box_3)=\int^\infty_0\!\! d^3 u\,\rho_{2}(u_1,u_2,u_3){\rm e}^{{\scriptscriptstyle \sum} u\Box}\nonumber\\&&\ \ \ \ \mbox{}
+\frac{\Box_1}{3}\int^\infty_0\!\! d^3 u\,{{ ( u_2 + u_3 ) }^{-1}}{\rm e}^{{\scriptscriptstyle \sum} u\Box},\end{fleqnarray}

\begin{fleqnarray}&&\Gamma_{3}(-\Box_1,-\Box_2,-\Box_3)=\int^\infty_0\!\! d^3 u\,\rho_{3}(u_1,u_2,u_3){\rm e}^{{\scriptscriptstyle \sum} u\Box},\end{fleqnarray}

\begin{fleqnarray}&&\Gamma_{4}(-\Box_1,-\Box_2,-\Box_3)=\int^\infty_0\!\! d^3 u\,\rho_{4}(u_1,u_2,u_3){\rm e}^{{\scriptscriptstyle \sum} u\Box},\end{fleqnarray}

\begin{fleqnarray}&&\Gamma_{5}(-\Box_1,-\Box_2,-\Box_3)=\int^\infty_0\!\! d^3 u\,\rho_{5}(u_1,u_2,u_3){\rm e}^{{\scriptscriptstyle \sum} u\Box}\nonumber\\&&\ \ \ \ \mbox{}
-\frac{\Box_1}{6}\int^\infty_0\!\! d^3 u\,{{ ( u_1 + u_3 )^{-1} }}{\rm e}^{{\scriptscriptstyle \sum} u\Box}\nonumber\\&&\ \ \ \ \mbox{}
-{{\Box_1\Box_3}\over 4}\int^\infty_0\!\! d^3 u\,{\rm e}^{{\scriptscriptstyle \sum} u\Box}\nonumber\\&&\ \ \ \ \mbox{}
+ {{{{\Box_3}^2}}\over {24}}\int^\infty_0\!\! d^3 u\,{\rm e}^{{\scriptscriptstyle \sum} u\Box},\end{fleqnarray}

\begin{fleqnarray}&&\Gamma_{6}(-\Box_1,-\Box_2,-\Box_3)=\int^\infty_0\!\! d^3 u\,\rho_{6}(u_1,u_2,u_3){\rm e}^{{\scriptscriptstyle \sum} u\Box},\end{fleqnarray}

\begin{fleqnarray}&&\Gamma_{7}(-\Box_1,-\Box_2,-\Box_3)=\int^\infty_0\!\! d^3 u\,\rho_{7}(u_1,u_2,u_3){\rm e}^{{\scriptscriptstyle \sum} u\Box},\end{fleqnarray}

\begin{fleqnarray}&&\Gamma_{8}(-\Box_1,-\Box_2,-\Box_3)=\int^\infty_0\!\! d^3 u\,\rho_{8}(u_1,u_2,u_3){\rm e}^{{\scriptscriptstyle \sum} u\Box}\nonumber\\&&\ \ \ \ \mbox{}
+\frac{\Box_2}{3}\int^\infty_0\!\! d^3 u\, { {{( u_2 + u_3 ) }^{-3}}}
{( -{{u_2}^2} - 4 u_2 u_3 - {{u_3}^2} ) }
{\rm e}^{{\scriptscriptstyle \sum} u\Box},\end{fleqnarray}

\begin{fleqnarray}&&\Gamma_{9}(-\Box_1,-\Box_2,-\Box_3)=\int^\infty_0\!\! d^3 u\,\rho_{9}(u_1,u_2,u_3){\rm e}^{{\scriptscriptstyle \sum} u\Box}\nonumber\\&&\ \ \ \ \mbox{}
+ \frac{\Box_1}{720}\int^\infty_0\!\! d^3 u\, \left[ {{( u_2 + u_3 ) }^{-1}}
\right.\nonumber\\&&\ \ \ \ \mbox{}\left.\hspace{23mm}
+	   {4 {{( u_1 + u_2 ) }^{-4}}}
          ({-5 {{u_1}^3} + 20 {{u_1}^2} u_2 - 2 u_1 {{u_2}^2}})
               \right]
{\rm e}^{{\scriptscriptstyle \sum} u\Box}\nonumber\\&&\ \ \ \ \mbox{}
+ {{{{\Box_1}^2}}\over {2160}}\int^\infty_0\!\! d^3 u\,{\rm e}^{{\scriptscriptstyle \sum} u\Box},\end{fleqnarray}

\begin{fleqnarray}&&\Gamma_{10}(-\Box_1,-\Box_2,-\Box_3)=\int^\infty_0\!\! d^3 u\,\rho_{10}(u_1,u_2,u_3){\rm e}^{{\scriptscriptstyle \sum} u\Box}\nonumber\\&&\ \ \ \ \mbox{}
- {{\Box_1 \Box_2}\over {270}}\int^\infty_0\!\! d^3 u\,{\rm e}^{{\scriptscriptstyle \sum} u\Box}\nonumber\\&&\ \ \ \ \mbox{}
+ {{{{\Box_1}^2}}\over {540}}\int^\infty_0\!\! d^3 u\,{\rm e}^{{\scriptscriptstyle \sum} u\Box},\end{fleqnarray}

\begin{fleqnarray}&&\Gamma_{11}(-\Box_1,-\Box_2,-\Box_3)=\int^\infty_0\!\! d^3 u\,\rho_{11}(u_1,u_2,u_3){\rm e}^{{\scriptscriptstyle \sum} u\Box}\nonumber\\&&\ \ \ \ \mbox{}
+ \frac{\Box_3}{1080}\int^\infty_0\!\! d^3 u\,
\left[
{{9 ( u_1 + u_2 )^{-1} }}
\right.\nonumber\\&&\ \ \ \ \mbox{}\hspace{28mm}
+ {{{( u_1 + u_3 ) }^{-5}}}
(-18 {{u_1}^4} - 103 {{u_1}^3} u_3
- 57 {{u_1}^2} {{u_3}^2}
\nonumber\\&&\ \ \ \ \mbox{}\left.\hspace{70mm}
                      - 33 u_1 {{u_3}^3} - 13 {{u_3}^4})
\right]{\rm e}^{{\scriptscriptstyle \sum} u\Box}
\nonumber\\&&\ \ \ \ \mbox{}
+ \frac{\Box_1}{1080}\int^\infty_0\!\! d^3 u\,
  {{{( u_1 + u_3 ) }^{-5}}}
(-{{u_1}^4} + 153 {{u_1}^3} u_3 + 165 {{u_1}^2} {{u_3}^2}
\nonumber\\&&\ \ \ \ \mbox{}
\hspace{70mm}
+ 71 u_1 {{u_3}^3} + 12 {{u_3}^4})
{\rm e}^{{\scriptscriptstyle \sum} u\Box}\nonumber\\&&\ \ \ \ \mbox{}
+ {{{{\Box_3}^2}}\over {540}}\int^\infty_0\!\! d^3 u\,{\rm e}^{{\scriptscriptstyle \sum} u\Box},\end{fleqnarray}

\begin{fleqnarray}&&\Gamma_{12}(-\Box_1,-\Box_2,-\Box_3)=\int^\infty_0\!\! d^3 u\,\rho_{12}(u_1,u_2,u_3){\rm e}^{{\scriptscriptstyle \sum} u\Box},\end{fleqnarray}

\begin{fleqnarray}&&\Gamma_{13}(-\Box_1,-\Box_2,-\Box_3)=\int^\infty_0\!\! d^3 u\,\rho_{13}(u_1,u_2,u_3){\rm e}^{{\scriptscriptstyle \sum} u\Box},\end{fleqnarray}

\begin{fleqnarray}&&\Gamma_{14}(-\Box_1,-\Box_2,-\Box_3)=\int^\infty_0\!\! d^3 u\,\rho_{14}(u_1,u_2,u_3){\rm e}^{{\scriptscriptstyle \sum} u\Box},\end{fleqnarray}

\begin{fleqnarray}&&\Gamma_{15}(-\Box_1,-\Box_2,-\Box_3)=\int^\infty_0\!\! d^3 u\,\rho_{15}(u_1,u_2,u_3){\rm e}^{{\scriptscriptstyle \sum} u\Box},\end{fleqnarray}

\begin{fleqnarray}&&\Gamma_{16}(-\Box_1,-\Box_2,-\Box_3)=\int^\infty_0\!\! d^3 u\,\rho_{16}(u_1,u_2,u_3){\rm e}^{{\scriptscriptstyle \sum} u\Box}\nonumber\\&&\ \ \ \ \mbox{}
-{{\Box_3}\over 6}\int^\infty_0\!\! d^3 u\,{\rm e}^{{\scriptscriptstyle \sum} u\Box},\end{fleqnarray}

\begin{fleqnarray}&&\Gamma_{17}(-\Box_1,-\Box_2,-\Box_3)=\int^\infty_0\!\! d^3 u\,\rho_{17}(u_1,u_2,u_3){\rm e}^{{\scriptscriptstyle \sum} u\Box},\end{fleqnarray}

\begin{fleqnarray}&&\Gamma_{18}(-\Box_1,-\Box_2,-\Box_3)=\int^\infty_0\!\! d^3 u\,\rho_{18}(u_1,u_2,u_3){\rm e}^{{\scriptscriptstyle \sum} u\Box},\end{fleqnarray}

\begin{fleqnarray}&&\Gamma_{19}(-\Box_1,-\Box_2,-\Box_3)=\int^\infty_0\!\! d^3 u\,\rho_{19}(u_1,u_2,u_3){\rm e}^{{\scriptscriptstyle \sum} u\Box},\end{fleqnarray}

\begin{fleqnarray}&&\Gamma_{20}(-\Box_1,-\Box_2,-\Box_3)=\int^\infty_0\!\! d^3 u\,\rho_{20}(u_1,u_2,u_3){\rm e}^{{\scriptscriptstyle \sum} u\Box},\end{fleqnarray}

\begin{fleqnarray}&&\Gamma_{21}(-\Box_1,-\Box_2,-\Box_3)=\int^\infty_0\!\! d^3 u\,\rho_{21}(u_1,u_2,u_3){\rm e}^{{\scriptscriptstyle \sum} u\Box},\end{fleqnarray}

\begin{fleqnarray}&&\Gamma_{22}(-\Box_1,-\Box_2,-\Box_3)=\int^\infty_0\!\! d^3 u\,\rho_{22}(u_1,u_2,u_3){\rm e}^{{\scriptscriptstyle \sum} u\Box}\nonumber\\&&\ \ \ \ \mbox{}
+{{\Box_2}\over {270}}\int^\infty_0\!\! d^3 u\,{\rm e}^{{\scriptscriptstyle \sum} u\Box},\end{fleqnarray}

\begin{fleqnarray}&&\Gamma_{23}(-\Box_1,-\Box_2,-\Box_3)=\int^\infty_0\!\! d^3 u\,\rho_{23}(u_1,u_2,u_3){\rm e}^{{\scriptscriptstyle \sum} u\Box}\nonumber\\&&\ \ \ \ \mbox{}
-{{\Box_3}\over {135}}\int^\infty_0\!\! d^3 u\,{\rm e}^{{\scriptscriptstyle \sum} u\Box},\end{fleqnarray}

\begin{fleqnarray}&&\Gamma_{24}(-\Box_1,-\Box_2,-\Box_3)=\int^\infty_0\!\! d^3 u\,\rho_{24}(u_1,u_2,u_3){\rm e}^{{\scriptscriptstyle \sum} u\Box}\nonumber\\&&\ \ \ \ \mbox{}
-{{\Box_1}\over {540}}\int^\infty_0\!\! d^3 u\,{\rm e}^{{\scriptscriptstyle \sum} u\Box},\end{fleqnarray}

\begin{fleqnarray}&&\Gamma_{25}(-\Box_1,-\Box_2,-\Box_3)=\int^\infty_0\!\! d^3 u\,\rho_{25}(u_1,u_2,u_3){\rm e}^{{\scriptscriptstyle \sum} u\Box}\nonumber\\&&\ \ \ \ \mbox{}
-{{\Box_1}\over {270}}\int^\infty_0\!\! d^3 u\,{\rm e}^{{\scriptscriptstyle \sum} u\Box}\nonumber\\&&\ \ \ \ \mbox{}
+ {{\Box_2}\over {135}}\int^\infty_0\!\! d^3 u\,{\rm e}^{{\scriptscriptstyle \sum} u\Box},\end{fleqnarray}

\begin{fleqnarray}&&\Gamma_{26}(-\Box_1,-\Box_2,-\Box_3)=\int^\infty_0\!\! d^3 u\,\rho_{26}(u_1,u_2,u_3){\rm e}^{{\scriptscriptstyle \sum} u\Box},\end{fleqnarray}

\begin{fleqnarray}&&\Gamma_{27}(-\Box_1,-\Box_2,-\Box_3)=\int^\infty_0\!\! d^3 u\,\rho_{27}(u_1,u_2,u_3){\rm e}^{{\scriptscriptstyle \sum} u\Box},\end{fleqnarray}

\begin{fleqnarray}&&\Gamma_{28}(-\Box_1,-\Box_2,-\Box_3)=\int^\infty_0\!\! d^3 u\,\rho_{28}(u_1,u_2,u_3){\rm e}^{{\scriptscriptstyle \sum} u\Box},\end{fleqnarray}

\begin{fleqnarray}&&\Gamma_{29}(-\Box_1,-\Box_2,-\Box_3)=\int^\infty_0\!\! d^3 u\,\rho_{29}(u_1,u_2,u_3){\rm e}^{{\scriptscriptstyle \sum} u\Box}\end{fleqnarray}

where

\begin{fleqnarray}&&\rho_{1} =
\frac1{3}\Big[u_1 u_2 + u_1 u_3 + u_2 u_3\Big]^{-1}
,\end{fleqnarray}

\begin{fleqnarray}&&\rho_{2} =
\frac1{3}\Big[{{( u_1 u_2 + u_1 u_3 + u_2 u_3 ) }^4}
\Big]^{-1}\Big(
   4 {{u_1}^2} {{u_2}^2} {{u_3}^2}
\Big),\end{fleqnarray}

\begin{fleqnarray}&&\rho_{3} =
\Big[{{( u_1 u_2 + u_1 u_3 + u_2 u_3 ) }^3}
\Big]^{-1}\Big(
   2 u_1 u_2 {{u_3}^2}
\Big),\end{fleqnarray}

\begin{fleqnarray}&&\rho_{4} =
\frac1{36}\Big[{{( u_1 u_2 + u_1 u_3 + u_2 u_3 ) }^5}
\Big]^{-1}\Big(
   {{u_1}^4} {{u_2}^4}
  -10 {{u_1}^4} {{u_2}^3} u_3 \nonumber\\&&\ \ \ \ \mbox{}
  +6 {{u_1}^4} {{u_2}^2} {{u_3}^2}
  +6 {{u_1}^3} {{u_2}^3} {{u_3}^2}
  +14 {{u_1}^4} u_2 {{u_3}^3}
  +6 {{u_1}^3} {{u_2}^2} {{u_3}^3} \nonumber\\&&\ \ \ \ \mbox{}
  -4 {{u_1}^4} {{u_3}^4}
  +20 {{u_1}^3} u_2 {{u_3}^4}
  -12 {{u_1}^2} {{u_2}^2} {{u_3}^4}
\Big),\end{fleqnarray}

\begin{fleqnarray}&&\rho_{5} =
\frac1{6}\Big[{{( u_1 + u_3 ) }^2}
     {{( u_2 + u_3 ) }^2}
     {{( u_1 u_2 + u_1 u_3 + u_2 u_3 ) }^3}\Big]^{-1}
\nonumber\\&&\ \ \ \ \mbox{}\times
\Big(
   3 {{u_1}^3} {{u_2}^3} {{u_3}^2}
  +10 {{u_1}^3} {{u_2}^2} {{u_3}^3}
  +5 {{u_1}^3} u_2 {{u_3}^4}
  +6 {{u_1}^2} {{u_2}^2} {{u_3}^4}  \nonumber\\&&\ \ \ \ \mbox{}
  +{{u_1}^3} {{u_3}^5}
  +3 {{u_1}^2} u_2 {{u_3}^5}
\Big),\end{fleqnarray}

\begin{fleqnarray}&&\rho_{6} =
\frac1{6}\Big[{{( u_1 u_2 + u_1 u_3 + u_2 u_3 ) }^3}
\Big]^{-1}\Big(
  -{{u_1}^2} {{u_2}^2}
  +2 {{u_1}^2} u_2 u_3  \nonumber\\&&\ \ \ \ \mbox{}
  -2 {{u_1}^2} {{u_3}^2}
  +4 u_1 u_2 {{u_3}^2}
\Big),\end{fleqnarray}

\begin{fleqnarray}&&\rho_{7} =
\frac1{12}\Big[{{( u_2 + u_3 ) }^2}
     {{( u_1 u_2 + u_1 u_3 + u_2 u_3 ) }^5}\Big]^{-1}
\Big(
  -8 {{u_1}^4} {{u_2}^5} u_3 \nonumber\\&&\ \ \ \ \mbox{}
  +16 {{u_1}^4} {{u_2}^4} {{u_3}^2}
  -4 {{u_1}^3} {{u_2}^5} {{u_3}^2}
  +24 {{u_1}^4} {{u_2}^3} {{u_3}^3}
  -12 {{u_1}^3} {{u_2}^4} {{u_3}^3} \nonumber\\&&\ \ \ \ \mbox{}
  -4 {{u_1}^2} {{u_2}^5} {{u_3}^3}
  -4 {{u_1}^2} {{u_2}^4} {{u_3}^4}
  -10 u_1 {{u_2}^5} {{u_3}^4}
  -{{u_2}^5} {{u_3}^5}
\Big),\end{fleqnarray}

\begin{fleqnarray}&&\rho_{8} =
\Big[{{( u_2 + u_3 ) }^4}
     {{( u_1 u_2 + u_1 u_3 + u_2 u_3 ) }^5}\Big]^{-1}
\Big(
  -4 {{u_1}^3} {{u_2}^7} {{u_3}^2} \nonumber\\&&\ \ \ \ \mbox{}
  +20 {{u_1}^3} {{u_2}^6} {{u_3}^3}
  -4 {{u_1}^2} {{u_2}^7} {{u_3}^3}
  +80 {{u_1}^3} {{u_2}^5} {{u_3}^4}
  +24 {{u_1}^2} {{u_2}^6} {{u_3}^4} \nonumber\\&&\ \ \ \ \mbox{}
  +28 {{u_1}^2} {{u_2}^5} {{u_3}^5}
  +20 u_1 {{u_2}^6} {{u_3}^5}
  +2 {{u_2}^6} {{u_3}^6}
\Big),\end{fleqnarray}

\begin{fleqnarray}&&\rho_{9} =
\frac1{540}\Big[{{( u_1 + u_2 ) }^4}
     {{( u_1 + u_3 ) }^4} {{( u_2 + u_3 ) }^4}
     {{( u_1 u_2 + u_1 u_3 + u_2 u_3 ) }^7}\Big]^{-1}
\nonumber\\&&\ \ \ \ \mbox{}\times
\Big(
   10 {{u_1}^{14}} {{u_2}^{10}}
  +40 {{u_1}^{13}} {{u_2}^{11}}
  +3 {{u_1}^{12}} {{u_2}^{12}}
  -50 {{u_1}^{14}} {{u_2}^9} u_3  \nonumber\\&&\ \ \ \ \mbox{}
  -250 {{u_1}^{13}} {{u_2}^{10}} u_3
  -1094 {{u_1}^{12}} {{u_2}^{11}} u_3
  -30 {{u_1}^{14}} {{u_2}^8} {{u_3}^2} \nonumber\\&&\ \ \ \ \mbox{}
  +250 {{u_1}^{13}} {{u_2}^9} {{u_3}^2}
  -1700 {{u_1}^{12}} {{u_2}^{10}} {{u_3}^2}
  -2088 {{u_1}^{11}} {{u_2}^{11}} {{u_3}^2} \nonumber\\&&\ \ \ \ \mbox{}
  +240 {{u_1}^{14}} {{u_2}^7} {{u_3}^3}
  +4020 {{u_1}^{13}} {{u_2}^8} {{u_3}^3}
  +7830 {{u_1}^{12}} {{u_2}^9} {{u_3}^3} \nonumber\\&&\ \ \ \ \mbox{}
  +2910 {{u_1}^{11}} {{u_2}^{10}} {{u_3}^3}
  +180 {{u_1}^{14}} {{u_2}^6} {{u_3}^4}
  +6900 {{u_1}^{13}} {{u_2}^7} {{u_3}^4}  \nonumber\\&&\ \ \ \ \mbox{}
  +20070 {{u_1}^{12}} {{u_2}^8} {{u_3}^4}
  +11340 {{u_1}^{11}} {{u_2}^9} {{u_3}^4}
  -2172 {{u_1}^{10}} {{u_2}^{10}} {{u_3}^4} \nonumber\\&&\ \ \ \ \mbox{}
  -30 {{u_1}^{14}} {{u_2}^5} {{u_3}^5}
  +5040 {{u_1}^{13}} {{u_2}^6} {{u_3}^5}
  +10176 {{u_1}^{12}} {{u_2}^7} {{u_3}^5} \nonumber\\&&\ \ \ \ \mbox{}
  -47400 {{u_1}^{11}} {{u_2}^8} {{u_3}^5}
  -160176 {{u_1}^{10}} {{u_2}^9} {{u_3}^5}
  -1464 {{u_1}^{12}} {{u_2}^6} {{u_3}^6} \nonumber\\&&\ \ \ \ \mbox{}
  -153500 {{u_1}^{11}} {{u_2}^7} {{u_3}^6}
  -523172 {{u_1}^{10}} {{u_2}^8} {{u_3}^6} \nonumber\\&&\ \ \ \ \mbox{}
  -377568 {{u_1}^9} {{u_2}^9} {{u_3}^6}
  -371690 {{u_1}^{10}} {{u_2}^7} {{u_3}^7} \nonumber\\&&\ \ \ \ \mbox{}
  -1452596 {{u_1}^9} {{u_2}^8} {{u_3}^7}
  -330957 {{u_1}^8} {{u_2}^8} {{u_3}^8}
\Big),\end{fleqnarray}

\begin{fleqnarray}&&\rho_{10} =
\frac1{3}\Big[{{( u_1 u_2 + u_1 u_3 + u_2 u_3 ) }^4}
  \Big]^{-1}\Big(
   {{u_1}^2} {{u_2}^2} {{u_3}^2}
\Big),\end{fleqnarray}

\begin{fleqnarray}&&\rho_{11} =
\frac1{90}\Big[{{( u_1 + u_3 ) }^3}
     {{( u_2 + u_3 ) }^3}
     {{( u_1 u_2 + u_1 u_3 + u_2 u_3 ) }^5}\Big]^{-1}
\nonumber\\&&\ \ \ \ \mbox{}\times
\Big(
  -30 {{u_1}^6} {{u_2}^6} {{u_3}^2}
  -190 {{u_1}^6} {{u_2}^5} {{u_3}^3}
  -215 {{u_1}^6} {{u_2}^4} {{u_3}^4}
  -270 {{u_1}^5} {{u_2}^5} {{u_3}^4} \nonumber\\&&\ \ \ \ \mbox{}
  -106 {{u_1}^6} {{u_2}^3} {{u_3}^5}
  -505 {{u_1}^5} {{u_2}^4} {{u_3}^5}
  -28 {{u_1}^6} {{u_2}^2} {{u_3}^6}
  -143 {{u_1}^5} {{u_2}^3} {{u_3}^6}  \nonumber\\&&\ \ \ \ \mbox{}
  -170 {{u_1}^4} {{u_2}^4} {{u_3}^6}
  -8 {{u_1}^6} u_2 {{u_3}^7}
  +21 {{u_1}^5} {{u_2}^2} {{u_3}^7}
  -{{u_1}^6} {{u_3}^8}                \nonumber\\&&\ \ \ \ \mbox{}
  +11 {{u_1}^5} u_2 {{u_3}^8}
  +60 {{u_1}^4} {{u_2}^2} {{u_3}^8}
  +30 {{u_1}^3} {{u_2}^3} {{u_3}^8}
  +2 {{u_1}^5} {{u_3}^9}              \nonumber\\&&\ \ \ \ \mbox{}
  +10 {{u_1}^4} u_2 {{u_3}^9}
  +20 {{u_1}^3} {{u_2}^2} {{u_3}^9}
\Big),\end{fleqnarray}

\begin{fleqnarray}&&\rho_{12} =
\frac1{3}\Big[( u_1 + u_2 )  ( u_1 + u_3 )
     {{( u_1 u_2 + u_1 u_3 + u_2 u_3 ) }^3}\Big]^{-1}
\nonumber\\&&\ \ \ \ \mbox{}\times
\Big(
   {{u_1}^4} {{u_2}^3}
  +3 {{u_1}^4} {{u_2}^2} u_3
  +4 {{u_1}^3} {{u_2}^3} u_3
  +3 {{u_1}^4} u_2 {{u_3}^2} \nonumber\\&&\ \ \ \ \mbox{}
  +12 {{u_1}^3} {{u_2}^2} {{u_3}^2}
  +9 {{u_1}^2} {{u_2}^3} {{u_3}^2}
  +{{u_1}^4} {{u_3}^3}
  +4 {{u_1}^3} u_2 {{u_3}^3}  \nonumber\\&&\ \ \ \ \mbox{}
  +9 {{u_1}^2} {{u_2}^2} {{u_3}^3}
  +6 u_1 {{u_2}^3} {{u_3}^3}
\Big),\end{fleqnarray}

\begin{fleqnarray}&&\rho_{13} =
\Big[( u_2 + u_3 )
     ( u_1 u_2 + u_1 u_3 + u_2 u_3 ) \Big]^{-1}\Big(
   2 u_2 u_3
\Big),\end{fleqnarray}

\begin{fleqnarray}&&\rho_{14} =
\Big[{{( u_1 u_2 + u_1 u_3 + u_2 u_3 ) }^2}
  \Big]^{-1}\Big(
   2 u_1 u_2 u_3
\Big),\end{fleqnarray}

\begin{fleqnarray}&&\rho_{15} =
\frac1{3}\Big[{{( u_1 u_2 + u_1 u_3 + u_2 u_3 ) }^4}
  \Big]^{-1}\Big(
   {{u_1}^2} {{u_2}^3} {{u_3}^2}   \nonumber\\&&\ \ \ \ \mbox{}
  +{{u_1}^2} {{u_2}^2} {{u_3}^3}
  -2 u_1 {{u_2}^3} {{u_3}^3}
\Big),\end{fleqnarray}

\begin{fleqnarray}&&\rho_{16} =
\frac1{3}\Big[{{( u_1 u_2 + u_1 u_3 + u_2 u_3 ) }^3}
  \Big]^{-1}\Big(
  -2 {{u_1}^3} u_2 u_3           \nonumber\\&&\ \ \ \ \mbox{}
  -2 {{u_1}^3} {{u_3}^2}
  -6 {{u_1}^2} u_2 {{u_3}^2}
\Big),\end{fleqnarray}

\begin{fleqnarray}&&\rho_{17} =
\Big[( u_2 + u_3 )
     {{( u_1 u_2 + u_1 u_3 + u_2 u_3 ) }^2}\Big]^{-1}\Big(
   {{u_2}^2} {{u_3}^2}
\Big),\end{fleqnarray}

\begin{fleqnarray}&&\rho_{18} =
\frac1{3}\Big[{{( u_1 u_2 + u_1 u_3 + u_2 u_3 ) }^4}
  \Big]^{-1}\Big(
  -2 {{u_1}^4} {{u_2}^3}
  -10 {{u_1}^4} {{u_2}^2} u_3 \nonumber\\&&\ \ \ \ \mbox{}
  -8 {{u_1}^3} {{u_2}^3} u_3
  -16 {{u_1}^3} {{u_2}^2} {{u_3}^2}
  -18 {{u_1}^2} {{u_2}^3} {{u_3}^2}
  -6 u_1 {{u_2}^3} {{u_3}^3}
\Big),\end{fleqnarray}

\begin{fleqnarray}&&\rho_{19} =
\frac1{6}\Big[( u_2 + u_3 )
     {{( u_1 u_2 + u_1 u_3 + u_2 u_3 ) }^4}\Big]^{-1}
\Big(
   2 {{u_1}^4} {{u_2}^4} \nonumber\\&&\ \ \ \ \mbox{}
  +12 {{u_1}^4} {{u_2}^3} u_3
  +8 {{u_1}^3} {{u_2}^4} u_3
  +10 {{u_1}^4} {{u_2}^2} {{u_3}^2}
  +40 {{u_1}^3} {{u_2}^3} {{u_3}^2} \nonumber\\&&\ \ \ \ \mbox{}
  +12 {{u_1}^2} {{u_2}^4} {{u_3}^2}
  +12 {{u_1}^2} {{u_2}^3} {{u_3}^3}
  +8 u_1 {{u_2}^4} {{u_3}^3}
  +{{u_2}^4} {{u_3}^4}
\Big),\end{fleqnarray}

\begin{fleqnarray}&&\rho_{20} =
\frac1{6}\Big[( u_2 + u_3 )
     {{( u_1 u_2 + u_1 u_3 + u_2 u_3 ) }^4}\Big]^{-1}
\nonumber\\&&\ \ \ \ \mbox{}\times
\Big(
  -4 {{u_1}^3} {{u_2}^4} u_3
  -2 {{u_1}^2} {{u_2}^4} {{u_3}^2}
  -2 {{u_1}^2} {{u_2}^3} {{u_3}^3}
  -{{u_2}^4} {{u_3}^4}
\Big),\end{fleqnarray}

\begin{fleqnarray}&&\rho_{21} =
\frac1{3}\Big[( u_2 + u_3 )
     {{( u_1 u_2 + u_1 u_3 + u_2 u_3 ) }^4}\Big]^{-1}
\nonumber\\&&\ \ \ \ \mbox{}\times
\Big(
  -2 {{u_1}^4} {{u_2}^4}
  -12 {{u_1}^4} {{u_2}^3} u_3
  -8 {{u_1}^3} {{u_2}^4} u_3
  -20 {{u_1}^4} {{u_2}^2} {{u_3}^2} \nonumber\\&&\ \ \ \ \mbox{}
  -40 {{u_1}^3} {{u_2}^3} {{u_3}^2}
  -24 {{u_1}^2} {{u_2}^4} {{u_3}^2}
  -12 {{u_1}^4} u_2 {{u_3}^3}
  -40 {{u_1}^3} {{u_2}^2} {{u_3}^3}  \nonumber\\&&\ \ \ \ \mbox{}
  -36 {{u_1}^2} {{u_2}^3} {{u_3}^3}
  -8 u_1 {{u_2}^4} {{u_3}^3}
  -2 {{u_1}^4} {{u_3}^4}
  -8 {{u_1}^3} u_2 {{u_3}^4}         \nonumber\\&&\ \ \ \ \mbox{}
  -12 {{u_1}^2} {{u_2}^2} {{u_3}^4}
  -8 u_1 {{u_2}^3} {{u_3}^4}
  -2 {{u_2}^4} {{u_3}^4}
\Big),\end{fleqnarray}

\begin{fleqnarray}&&\rho_{22} =
\frac1{90}\Big[( u_1 + u_2 )  ( u_1 + u_3 )
     ( u_2 + u_3 )
     {{( u_1 u_2 + u_1 u_3 + u_2 u_3 ) }^6}\Big]^{-1}
\nonumber\\&&\ \ \ \ \mbox{}\times
\Big(
  -3 {{u_1}^8} {{u_2}^6}
  -3 {{u_1}^7} {{u_2}^7}
  -3 {{u_1}^6} {{u_2}^8}
  -18 {{u_1}^8} {{u_2}^5} u_3 \nonumber\\&&\ \ \ \ \mbox{}
  -39 {{u_1}^7} {{u_2}^6} u_3
  -39 {{u_1}^6} {{u_2}^7} u_3
  -18 {{u_1}^5} {{u_2}^8} u_3
  -45 {{u_1}^8} {{u_2}^4} {{u_3}^2} \nonumber\\&&\ \ \ \ \mbox{}
  -153 {{u_1}^7} {{u_2}^5} {{u_3}^2}
  -218 {{u_1}^6} {{u_2}^6} {{u_3}^2}
  -164 {{u_1}^5} {{u_2}^7} {{u_3}^2}
  -51 {{u_1}^4} {{u_2}^8} {{u_3}^2}  \nonumber\\&&\ \ \ \ \mbox{}
  -30 {{u_1}^8} {{u_2}^3} {{u_3}^3}
  -285 {{u_1}^7} {{u_2}^4} {{u_3}^3}
  -617 {{u_1}^6} {{u_2}^5} {{u_3}^3}
  -644 {{u_1}^5} {{u_2}^6} {{u_3}^3}  \nonumber\\&&\ \ \ \ \mbox{}
  -330 {{u_1}^4} {{u_2}^7} {{u_3}^3}
  -78 {{u_1}^3} {{u_2}^8} {{u_3}^3}
  -435 {{u_1}^6} {{u_2}^4} {{u_3}^4}
  -1286 {{u_1}^5} {{u_2}^5} {{u_3}^4}  \nonumber\\&&\ \ \ \ \mbox{}
  -996 {{u_1}^4} {{u_2}^6} {{u_3}^4}
  -403 {{u_1}^3} {{u_2}^7} {{u_3}^4}
  -63 {{u_1}^2} {{u_2}^8} {{u_3}^4}
  -717 {{u_1}^4} {{u_2}^5} {{u_3}^5}   \nonumber\\&&\ \ \ \ \mbox{}
  -845 {{u_1}^3} {{u_2}^6} {{u_3}^5}
  -239 {{u_1}^2} {{u_2}^7} {{u_3}^5}
  -24 u_1 {{u_2}^8} {{u_3}^5}
  -175 {{u_1}^2} {{u_2}^6} {{u_3}^6}   \nonumber\\&&\ \ \ \ \mbox{}
  -42 u_1 {{u_2}^7} {{u_3}^6}
  -3 {{u_2}^8} {{u_3}^6}
  -2 {{u_2}^7} {{u_3}^7}
\Big),\end{fleqnarray}

\begin{fleqnarray}&&\rho_{23} =
\frac1{90}\Big[( u_1 + u_2 )
     {{( u_1 u_2 + u_1 u_3 + u_2 u_3 ) }^5}\Big]^{-1}
\nonumber\\&&\ \ \ \ \mbox{}\times
\Big(
   6 {{u_1}^6} {{u_2}^4}
  +3 {{u_1}^5} {{u_2}^5}
  +28 {{u_1}^6} {{u_2}^3} u_3
  +58 {{u_1}^5} {{u_2}^4} u_3 \nonumber\\&&\ \ \ \ \mbox{}
  +42 {{u_1}^6} {{u_2}^2} {{u_3}^2}
  +152 {{u_1}^5} {{u_2}^3} {{u_3}^2}
  +110 {{u_1}^4} {{u_2}^4} {{u_3}^2}
  +24 {{u_1}^6} u_2 {{u_3}^3}    \nonumber\\&&\ \ \ \ \mbox{}
  +124 {{u_1}^5} {{u_2}^2} {{u_3}^3}
  +300 {{u_1}^4} {{u_2}^3} {{u_3}^3}
  +4 {{u_1}^6} {{u_3}^4}
  +24 {{u_1}^5} u_2 {{u_3}^4}    \nonumber\\&&\ \ \ \ \mbox{}
  +60 {{u_1}^4} {{u_2}^2} {{u_3}^4}
  +40 {{u_1}^3} {{u_2}^3} {{u_3}^4}
\Big),\end{fleqnarray}

\begin{fleqnarray}&&\rho_{24} =
\frac1{30}\Big[( u_2 + u_3 )
     {{( u_1 u_2 + u_1 u_3 + u_2 u_3 ) }^4}\Big]^{-1}
\nonumber\\&&\ \ \ \ \mbox{}\times
\Big(
   2 {{u_1}^3} {{u_2}^5}
  +10 {{u_1}^3} {{u_2}^4} u_3
  +8 {{u_1}^2} {{u_2}^5} u_3
  +20 {{u_1}^3} {{u_2}^3} {{u_3}^2} \nonumber\\&&\ \ \ \ \mbox{}
  +32 {{u_1}^2} {{u_2}^4} {{u_3}^2}
  +8 u_1 {{u_2}^5} {{u_3}^2}
  +24 {{u_1}^2} {{u_2}^3} {{u_3}^3}
  +16 u_1 {{u_2}^4} {{u_3}^3}       \nonumber\\&&\ \ \ \ \mbox{}
  +2 {{u_2}^5} {{u_3}^3}
  +{{u_2}^4} {{u_3}^4}
\Big),\end{fleqnarray}

\begin{fleqnarray}&&\rho_{25} =
\frac1{15}\Big[( u_1 + u_2 )  ( u_1 + u_3 )
     {{( u_1 u_2 + u_1 u_3 + u_2 u_3 ) }^4}\Big]^{-1}
\nonumber\\&&\ \ \ \ \mbox{}\times
\Big(
   2 {{u_1}^6} {{u_2}^3}
  +10 {{u_1}^6} {{u_2}^2} u_3
  +12 {{u_1}^5} {{u_2}^3} u_3
  +20 {{u_1}^5} {{u_2}^2} {{u_3}^2} \nonumber\\&&\ \ \ \ \mbox{}
  +30 {{u_1}^4} {{u_2}^3} {{u_3}^2}
  +10 {{u_1}^3} {{u_2}^3} {{u_3}^3}
\Big),\end{fleqnarray}

\begin{fleqnarray}&&\rho_{26} =
\frac1{6}\Big[{{( u_1 u_2 + u_1 u_3 + u_2 u_3 ) }^3}
  \Big]^{-1}\Big(
   {{u_1}^3} {{u_2}^3}
  +6 {{u_1}^3} {{u_2}^2} u_3 \nonumber\\&&\ \ \ \ \mbox{}
  +4 {{u_1}^3} u_2 {{u_3}^2}
  +6 {{u_1}^2} {{u_2}^2} {{u_3}^2}
\Big),\end{fleqnarray}

\begin{fleqnarray}&&\rho_{27} =
\frac1{540}\Big[{{( u_1 u_2 + u_1 u_3 + u_2 u_3 ) }^5}
  \Big]^{-1}\Big(
   {{u_1}^5} {{u_2}^5}
  +10 {{u_1}^5} {{u_2}^4} u_3   \nonumber\\&&\ \ \ \ \mbox{}
  +50 {{u_1}^5} {{u_2}^3} {{u_3}^2}
  +20 {{u_1}^4} {{u_2}^4} {{u_3}^2}
  +62 {{u_1}^5} {{u_2}^2} {{u_3}^3}
  +210 {{u_1}^4} {{u_2}^3} {{u_3}^3} \nonumber\\&&\ \ \ \ \mbox{}
  +22 {{u_1}^5} u_2 {{u_3}^4}
  +100 {{u_1}^4} {{u_2}^2} {{u_3}^4}
  +150 {{u_1}^3} {{u_2}^3} {{u_3}^4}
  +2 {{u_1}^5} {{u_3}^5}             \nonumber\\&&\ \ \ \ \mbox{}
  +10 {{u_1}^4} u_2 {{u_3}^5}
  +20 {{u_1}^3} {{u_2}^2} {{u_3}^5}
\Big),\end{fleqnarray}

\begin{fleqnarray}&&\rho_{28} =
\frac1{135}\Big[{{( u_1 u_2 + u_1 u_3 + u_2 u_3 ) }^4}
  \Big]^{-1}\Big(
   2 {{u_1}^4} {{u_2}^4}
  +16 {{u_1}^4} {{u_2}^3} u_3       \nonumber\\&&\ \ \ \ \mbox{}
  +6 {{u_1}^4} {{u_2}^2} {{u_3}^2}
  +24 {{u_1}^3} {{u_2}^3} {{u_3}^2}
  -8 {{u_1}^4} u_2 {{u_3}^3}
  -24 {{u_1}^3} {{u_2}^2} {{u_3}^3}  \nonumber\\&&\ \ \ \ \mbox{}
  -2 {{u_1}^4} {{u_3}^4}
  -8 {{u_1}^3} u_2 {{u_3}^4}
  -6 {{u_1}^2} {{u_2}^2} {{u_3}^4}
\Big),\end{fleqnarray}

\begin{fleqnarray}&&\rho_{29} =
\frac1{45}\Big[{{( u_1 u_2 + u_1 u_3 + u_2 u_3 ) }^4}
  \Big]^{-1}\Big(
  -{{u_1}^4} {{u_2}^4} u_3
  -4 {{u_1}^4} {{u_2}^3} {{u_3}^2}  \nonumber\\&&\ \ \ \ \mbox{}
  -4 {{u_1}^4} {{u_2}^2} {{u_3}^3}
  -4 {{u_1}^3} {{u_2}^3} {{u_3}^3}
\Big).\end{fleqnarray}
\arraycolsep=5pt

Although the nondecreasing form factors are not completely
presentable in the Laplace form, the explicit
dependence on the $\Box$'s outside the
kernel $\exp(\sum u \Box)$ is purely local.

Apart from the nondecreasing terms which are
elementary, the Laplace representation is unique.
The calculations with the form factors can, therefore,
be carried out in terms of the Laplace originals but if,
in these calculations, the form factors are multiplied
by new powers of $\Box$'s, the whole procedure of
absorbing these multipliers in the Laplace originals
and detaching the nondecreasing terms should be repeated.
This situation is encountered in the calculation of the
trace anomaly (sect. 12), and it makes the Laplace
representation inconvenient for this calculation.
In addition, this representation is well adapted only
to the euclidean signature of the metric. When going
over to the lorentzian signature, setting the retarded
boundary conditions [1,4] for the kernel $\exp(\sum u \Box)$
is embarrassing. Only in the spectral representation are all
the drawbacks removed.

\section{The generalized spectral representation
of the third-order form factors in the
effective action}
\setcounter{equation}{0}

\hspace{\parindent}
Spectral representation of the form factors
is of special importance for applications
because it allows going over to the lorentzian
signature in the expectation-value equations [1,4].
The generalized spectral representation retains this
quality and, in addition, makes it possible to
overcome difficulties connected with the
discontinuous nature of the spectral weight in
the third-order form factors [3] (see sect. 20).

In the generalized spectral representation,
there is one
extra integration over a parameter entering
the spectral weight. Namely, for each of the
arguments $\Box_1,\Box_2,\Box_3$ in the triple form
factors we introduce the following spectral
integral
\begin{equation}
{\cal S}(y,\Box)=
\frac y2\int^\infty_0dm\,
\frac{J_1(ym)}{m^2-\Box}
\end{equation}
depending on a parameter $y$, where $J_1$ is the
order-1 Bessel function. Next, we introduce the
operator
\begin{equation}
C=y^2\frac{\partial}{\partial y^2},
\end{equation}
and denote, for short,
\begin{equation}
{\cal S}_n={\cal S}(y_n,\Box_n),
\hspace{7mm}
C_n={y_n}^2\frac{\partial}{\partial {y_n}^2},
\hspace{7mm}
n=1,2,3.
\end{equation}
The third-order form factors $\Gamma_i$ are then
expressed through integrals of the form
\begin{equation}
2\int^\infty_0dy^2\,\left(\frac4{y^2}\right)^3
P(C_1,C_2,C_3){\cal S}_1{\cal S}_2{\cal S}_3
\end{equation}
where  $P(C_1,C_2,C_3)$ is a polynomial, and it is
understood that $C_1, C_2, C_3$ act on
${\cal S}_1, {\cal S}_2, {\cal S}_3$ respectively with
subsequently setting
$y_1=y_2=y_3=y$.

The spectral representation (both generalized
and ordinary) is sensitive to the behaviour of a
function at small arguments. Only the functions that
behave in each argument like
${\cal O}/\Box,\ {\cal O}\rightarrow0$ at
$\Box\rightarrow-0$ admit this representation.
\footnote{\normalsize
The exception to this rule pointed out in
sect. 20 does not concern the triple and
double spectral forms used here.}
In fact, as pointed out in sect. 7, the form factors
$\Gamma_1$ to $\Gamma_{11}$ contain the
coefficients $\Box_n/\Box_m$ which cause the
$1/\Box_m$ behaviour at small $\Box_m$
(and the $\ln(-\Box_n)$ behaviour at large $\Box_n$).
In the spectral technique, the terms with these
coefficients get detached and take the form
\begin{equation}
\frac{\Box_n}{\Box_m}\int^\infty_0dy^2\,\left(\frac4{y^2}\right)^2
P(C_n,C_k){\cal S}_n{\cal S}_k,\hspace{7mm}
k\neq m,
\hspace{7mm}
k\neq n.
\end{equation}
The double-spectral integral in (9.5) gives
the coefficient of the $1/\Box_m$ asymptotic
behaviour as a function of $\Box_n$ and $\Box_k$.
The total form factor is a sum of the
triple-spectral contributions (9.4), double-spectral
contributions (9.5), and tree terms.

The form factors $\Gamma_{12}$ to $\Gamma_{29}$
are given below in their redefined versions, with the overall
$1/\Box$ factors (see sect. 7). Since, in this version,
$\Gamma_{22}$ is a sum of two form factors, it has
two contributions of the form (9.4) with the
polynomials $P_{22-1}$ and $P_{22-2}$.

The expressions for the form factors (6.7)
in the generalized spectral form are as
follows:
\arraycolsep=0pt
\mathindent=2pt
\begin{fleqnarray}&&\Gamma_{1}(-\Box_1,-\Box_2,-\Box_3) =
2\int^\infty_0dy^2\,\left(\frac4{y^2}\right)^3\,P_{1}(C_1,C_2,C_3){\cal S}_1{\cal S}_2{\cal S}_3
,\end{fleqnarray}

\begin{fleqnarray}&&\Gamma_{2}(-\Box_1,-\Box_2,-\Box_3) =
2\int^\infty_0dy^2\,\left(\frac4{y^2}\right)^3\,P_{2}(C_1,C_2,C_3){\cal S}_1{\cal S}_2{\cal S}_3 \nonumber\\&&\ \ \ \ \mbox{}
+ \int^\infty_0dy^2\,\left(\frac4{y^2}\right)^2\,\left({-\frac{1}3} C_1C_2\right){\cal S}_1{\cal S}_2
,\end{fleqnarray}

\begin{fleqnarray}&&\Gamma_{3}(-\Box_1,-\Box_2,-\Box_3) =
2\int^\infty_0dy^2\,\left(\frac4{y^2}\right)^3\,P_{3}(C_1,C_2,C_3){\cal S}_1{\cal S}_2{\cal S}_3
,\end{fleqnarray}

\begin{fleqnarray}&&\Gamma_{4}(-\Box_1,-\Box_2,-\Box_3) =
2\int^\infty_0dy^2\,\left(\frac4{y^2}\right)^3\,P_{4}(C_1,C_2,C_3){\cal S}_1{\cal S}_2{\cal S}_3
,\end{fleqnarray}

\begin{fleqnarray}&&\Gamma_{5}(-\Box_1,-\Box_2,-\Box_3) =
2\int^\infty_0dy^2\,\left(\frac4{y^2}\right)^3\,P_{5}(C_1,C_2,C_3){\cal S}_1{\cal S}_2{\cal S}_3 \nonumber\\&&\ \ \ \ \mbox{}
+ \frac{\Box_1}{\Box_2}\int^\infty_0dy^2\,\left(\frac4{y^2}\right)^2\,\left(
   {1\over 24}C_1 C_3
\right){\cal S}_1{\cal S}_3 \nonumber\\&&\ \ \ \ \mbox{}
+ {1\over {4 \Box_2}} - {{\Box_3}\over {24 \Box_1 \Box_2}}
,\end{fleqnarray}

\begin{fleqnarray}&&\Gamma_{6}(-\Box_1,-\Box_2,-\Box_3) =
2\int^\infty_0dy^2\,\left(\frac4{y^2}\right)^3\,P_{6}(C_1,C_2,C_3){\cal S}_1{\cal S}_2{\cal S}_3
,\end{fleqnarray}

\begin{fleqnarray}&&\Gamma_{7}(-\Box_1,-\Box_2,-\Box_3) =
2\int^\infty_0dy^2\,\left(\frac4{y^2}\right)^3\,P_{7}(C_1,C_2,C_3){\cal S}_1{\cal S}_2{\cal S}_3
,\end{fleqnarray}

\begin{fleqnarray}&&\Gamma_{8}(-\Box_1,-\Box_2,-\Box_3) =
2\int^\infty_0dy^2\,\left(\frac4{y^2}\right)^3\,P_{8}(C_1,C_2,C_3){\cal S}_1{\cal S}_2{\cal S}_3 \nonumber\\&&\ \ \ \ \mbox{}
+ \frac{\Box_2}{\Box_1}\int^\infty_0dy^2\,\left(\frac4{y^2}\right)^2\,\left(
   {1\over 3}C_2 C_3
  -{7\over {24}}{{C_2}^2} C_3
\right.\nonumber\\&&\ \ \ \ \mbox{}\left.
  +{1\over {24}}{{C_2}^3} C_3
  -{7\over {24}}C_2 {{C_3}^2}
  +{1\over 6}{{C_2}^2} {{C_3}^2}
  +{1\over {24}}C_2 {{C_3}^3}
\right){\cal S}_2{\cal S}_3
,\end{fleqnarray}

\begin{fleqnarray}&&\Gamma_{9}(-\Box_1,-\Box_2,-\Box_3) =
2\int^\infty_0dy^2\,\left(\frac4{y^2}\right)^3\,P_{9}(C_1,C_2,C_3){\cal S}_1{\cal S}_2{\cal S}_3 \nonumber\\&&\ \ \ \ \mbox{}
+ \frac{\Box_1}{\Box_2}\int^\infty_0dy^2\,\left(\frac4{y^2}\right)^2\,\left(
  -{1\over {720}}C_1 C_3
  +{{17}\over {2160}}{{C_1}^2} C_3
\right.\nonumber\\&&\ \ \ \ \mbox{}
  +{1\over {2160}}{{C_1}^3} C_3
  +{1\over {4320}}C_1 {{C_3}^2}
  -{1\over {80}}{{C_1}^2} {{C_3}^2}
  -{1\over {2160}}{{C_1}^3} {{C_3}^2}
\nonumber\\&&\ \ \ \ \mbox{}\left.
  +{1\over {432}}C_1 {{C_3}^3}
  +{1\over {216}}{{C_1}^2} {{C_3}^3}
  -{1\over {864}}C_1 {{C_3}^4}
\right){\cal S}_1{\cal S}_3 \nonumber\\&&\ \ \ \ \mbox{}
+ \int^\infty_0dy^2\,\left(\frac4{y^2}\right)^2\,\left({-\frac1{720}} C_1C_2\right){\cal S}_1{\cal S}_2 \nonumber\\&&\ \ \ \ \mbox{}
- {{\Box_1}\over {2160 \Box_2 \Box_3}}
,\end{fleqnarray}

\begin{fleqnarray}&&\Gamma_{10}(-\Box_1,-\Box_2,-\Box_3) =
2\int^\infty_0dy^2\,\left(\frac4{y^2}\right)^3\,P_{10}(C_1,C_2,C_3){\cal S}_1{\cal S}_2{\cal S}_3 \nonumber\\&&\ \ \ \ \mbox{}
+ {1\over {270 \Box_3}} - {{\Box_1}\over {540 \Box_2 \Box_3}}
,\end{fleqnarray}

\begin{fleqnarray}&&\Gamma_{11}(-\Box_1,-\Box_2,-\Box_3) =
2\int^\infty_0dy^2\,\left(\frac4{y^2}\right)^3\,P_{11}(C_1,C_2,C_3){\cal S}_1{\cal S}_2{\cal S}_3 \nonumber\\&&\ \ \ \ \mbox{}
+ \frac{\Box_1}{\Box_2}\int^\infty_0dy^2\,\left(\frac4{y^2}\right)^2\,\left(
  -{7\over {320}}C_1 C_3
  +{{3289}\over {103680}}{{C_1}^2} C_3
\right.\nonumber\\&&\ \ \ \ \mbox{}
  -{{49}\over {4320}}{{C_1}^3} C_3
  +{{191}\over {103680}}{{C_1}^4} C_3
  -{1\over {8640}}{{C_1}^5} C_3
  +{{3049}\over {103680}}C_1 {{C_3}^2}    \nonumber\\&&\ \ \ \ \mbox{}
  -{{3949}\over {103680}}{{C_1}^2} {{C_3}^2}
  +{{307}\over {34560}}{{C_1}^3} {{C_3}^2}
  -{{71}\over {103680}}{{C_1}^4} {{C_3}^2} \nonumber\\&&\ \ \ \ \mbox{}
  -{{1213}\over {103680}}C_1 {{C_3}^3}
  +{{157}\over {11520}}{{C_1}^2} {{C_3}^3}
  -{{11}\over {6912}}{{C_1}^3} {{C_3}^3}
\nonumber\\&&\ \ \ \ \mbox{}\left.
  +{{143}\over {103680}}C_1 {{C_3}^4}
  -{{17}\over {11520}}{{C_1}^2} {{C_3}^4}
  +{1\over {103680}}C_1 {{C_3}^5}
\right){\cal S}_1{\cal S}_3 \nonumber\\&&\ \ \ \ \mbox{}
+ \frac{\Box_3}{\Box_1}\int^\infty_0dy^2\,\left(\frac4{y^2}\right)^2\,\left(
   {{149}\over {8640}}C_2 C_3
  -{{1973}\over {103680}}{{C_2}^2} C_3
\right.\nonumber\\&&\ \ \ \ \mbox{}
  +{{767}\over {103680}}{{C_2}^3} C_3
  -{{163}\over {103680}}{{C_2}^4} C_3
  +{{13}\over {103680}}{{C_2}^5} C_3   \nonumber\\&&\ \ \ \ \mbox{}
  -{{2573}\over {103680}}C_2 {{C_3}^2}
  +{{2009}\over {103680}}{{C_2}^2} {{C_3}^2}
  -{{41}\over {11520}}{{C_2}^3} {{C_3}^2} \nonumber\\&&\ \ \ \ \mbox{}
  +{{11}\over {34560}}{{C_2}^4} {{C_3}^2}
  +{{227}\over {17280}}C_2 {{C_3}^3}
  -{{263}\over {34560}}{{C_2}^2} {{C_3}^3}  \nonumber\\&&\ \ \ \ \mbox{}
  +{{19}\over {34560}}{{C_2}^3} {{C_3}^3}
  -{{283}\over {103680}}C_2 {{C_3}^4}
\nonumber\\&&\ \ \ \ \mbox{}\left.
  +{{103}\over {103680}}{{C_2}^2} {{C_3}^4}
  +{1\over {5760}}C_2 {{C_3}^5}
\right){\cal S}_2{\cal S}_3 \nonumber\\&&\ \ \ \ \mbox{}
+ \int^\infty_0dy^2\,\left(\frac4{y^2}\right)^2\,\left({-\frac1{120}} C_1C_2\right){\cal S}_1{\cal S}_2 \nonumber\\&&\ \ \ \ \mbox{}
-{{\Box_3}\over {540 \Box_1 \Box_2}}
,\end{fleqnarray}

\begin{fleqnarray}&&\Gamma_{12}(-\Box_1,-\Box_2,-\Box_3) =
\frac1{\Box_1}2\int^\infty_0dy^2\,\left(\frac4{y^2}\right)^3\,P_{12}(C_1,C_2,C_3){\cal S}_1{\cal S}_2{\cal S}_3
,\end{fleqnarray}

\begin{fleqnarray}&&\Gamma_{13}(-\Box_1,-\Box_2,-\Box_3) =
\frac1{\Box_1}\left[
2\int^\infty_0dy^2\,\left(\frac4{y^2}\right)^3\,P_{13}(C_1,C_2,C_3){\cal S}_1{\cal S}_2{\cal S}_3
\right. \nonumber\\&&\ \ \ \ \mbox{}\left.
+ \int^\infty_0dy^2\,\left(\frac4{y^2}\right)^2\,\left({-2} C_2C_3\right){\cal S}_2{\cal S}_3 \right]
,\end{fleqnarray}

\begin{fleqnarray}&&\Gamma_{14}(-\Box_1,-\Box_2,-\Box_3) =
\frac1{\Box_3} 2\int^\infty_0dy^2\,\left(\frac4{y^2}\right)^3\,P_{14}(C_1,C_2,C_3){\cal S}_1{\cal S}_2{\cal S}_3
,\end{fleqnarray}

\begin{fleqnarray}&&\Gamma_{15}(-\Box_1,-\Box_2,-\Box_3) =
\frac1{\Box_1} 2\int^\infty_0dy^2\,\left(\frac4{y^2}\right)^3\,P_{15}(C_1,C_2,C_3){\cal S}_1{\cal S}_2{\cal S}_3
,\end{fleqnarray}

\begin{fleqnarray}&&\Gamma_{16}(-\Box_1,-\Box_2,-\Box_3) =
\frac1{\Box_1} 2\int^\infty_0dy^2\,\left(\frac4{y^2}\right)^3\,P_{16}(C_1,C_2,C_3){\cal S}_1{\cal S}_2{\cal S}_3 \nonumber\\&&\ \ \ \ \mbox{}
+{1\over {6\Box_1\Box_2}}
,\end{fleqnarray}

\begin{fleqnarray}&&\Gamma_{17}(-\Box_1,-\Box_2,-\Box_3) =
\frac1{\Box_1}\left[
2\int^\infty_0dy^2\,\left(\frac4{y^2}\right)^3\,P_{17}(C_1,C_2,C_3){\cal S}_1{\cal S}_2{\cal S}_3
\right. \nonumber\\&&\ \ \ \ \mbox{}\left.
+ \int^\infty_0dy^2\,\left(\frac4{y^2}\right)^2\,\left({-} C_2C_3\right){\cal S}_2{\cal S}_3 \right]
,\end{fleqnarray}

\begin{fleqnarray}&&\Gamma_{18}(-\Box_1,-\Box_2,-\Box_3) =
\frac1{\Box_1} 2\int^\infty_0dy^2\,\left(\frac4{y^2}\right)^3\,P_{18}(C_1,C_2,C_3){\cal S}_1{\cal S}_2{\cal S}_3
,\end{fleqnarray}

\begin{fleqnarray}&&\Gamma_{19}(-\Box_1,-\Box_2,-\Box_3) =
\frac1{\Box_1}\left[
2\int^\infty_0dy^2\,\left(\frac4{y^2}\right)^3\,P_{19}(C_1,C_2,C_3){\cal S}_1{\cal S}_2{\cal S}_3
\right. \nonumber\\&&\ \ \ \ \ \ \ \ \mbox{}\left.
+ \int^\infty_0dy^2\,\left(\frac4{y^2}\right)^2\,\left({-\frac16} C_2C_3\right){\cal S}_2{\cal S}_3 \right]
,\end{fleqnarray}

\begin{fleqnarray}&&\Gamma_{20}(-\Box_1,-\Box_2,-\Box_3) =
\frac1{\Box_1}\left[
2\int^\infty_0dy^2\,\left(\frac4{y^2}\right)^3\,P_{20}(C_1,C_2,C_3){\cal S}_1{\cal S}_2{\cal S}_3
\right.\nonumber\\&&\ \ \ \ \ \ \ \ \mbox{}\left.
+ \int^\infty_0dy^2\,\left(\frac4{y^2}\right)^2\,\left({\frac16} C_2C_3\right){\cal S}_2{\cal S}_3 \right]
,\end{fleqnarray}

\begin{fleqnarray}&&\Gamma_{21}(-\Box_1,-\Box_2,-\Box_3) =
\frac1{\Box_1}\left[
2\int^\infty_0dy^2\,\left(\frac4{y^2}\right)^3\,P_{21}(C_1,C_2,C_3){\cal S}_1{\cal S}_2{\cal S}_3
 \right.\nonumber\\&&\ \ \ \ \ \ \ \ \mbox{}\left.
+ \int^\infty_0dy^2\,\left(\frac4{y^2}\right)^2\,\left({\frac23} C_2C_3\right){\cal S}_2{\cal S}_3 \right]
,\end{fleqnarray}

\begin{fleqnarray}&&\Gamma_{22}(-\Box_1,-\Box_2,-\Box_3) =
\frac1{\Box_1}\left.[
2\int^\infty_0dy^2\,\left(\frac4{y^2}\right)^3\,P_{22-1}(C_1,C_2,C_3){\cal S}_1{\cal S}_2{\cal S}_3
\right.\nonumber\\&&\ \ \ \ \ \ \ \ \mbox{}\left.
+ \int^\infty_0dy^2\,\left(\frac4{y^2}\right)^2\,\left({-\frac1{90}} C_2C_3\right){\cal S}_2{\cal S}_3 \right] \nonumber\\&&\ \ \ \ \mbox{}
+\frac1{\Box_2}\left[
2\int^\infty_0dy^2\,\left(\frac4{y^2}\right)^3\,P_{22-2}(C_1,C_2,C_3){\cal S}_1{\cal S}_2{\cal S}_3
\right.\nonumber\\&&\ \ \ \ \ \ \ \ \mbox{}\left.
+ \int^\infty_0dy^2\,\left(\frac4{y^2}\right)^2\,\left({-\frac1{30}} C_1C_2\right){\cal S}_1{\cal S}_2 \right] \nonumber\\&&\ \ \ \ \mbox{}
-{{1}\over {270 \Box_1 \Box_3}}
,\end{fleqnarray}

\begin{fleqnarray}&&\Gamma_{23}(-\Box_1,-\Box_2,-\Box_3) =
\frac1{\Box_1}
2\int^\infty_0dy^2\,\left(\frac4{y^2}\right)^3\,P_{23}(C_1,C_2,C_3){\cal S}_1{\cal S}_2{\cal S}_3
 \nonumber\\&&\ \ \ \ \mbox{}
+{1\over {135\Box_1\Box_2}}
,\end{fleqnarray}

\begin{fleqnarray}&&\Gamma_{24}(-\Box_1,-\Box_2,-\Box_3) =
\frac1{\Box_2}
2\int^\infty_0dy^2\,\left(\frac4{y^2}\right)^3\,P_{24}(C_1,C_2,C_3){\cal S}_1{\cal S}_2{\cal S}_3
 \nonumber\\&&\ \ \ \ \mbox{}
+{1\over {540\Box_2\Box_3}}
,\end{fleqnarray}

\begin{fleqnarray}&&\Gamma_{25}(-\Box_1,-\Box_2,-\Box_3) =
\frac1{\Box_1}
2\int^\infty_0dy^2\,\left(\frac4{y^2}\right)^3\,P_{25}(C_1,C_2,C_3){\cal S}_1{\cal S}_2{\cal S}_3
\nonumber\\&&\ \ \ \ \mbox{}
-{{1}\over {135\Box_1\Box_3}} + {{1}\over {270 \Box_2 \Box_3}}
,\end{fleqnarray}

\begin{fleqnarray}&&\Gamma_{26}(-\Box_1,-\Box_2,-\Box_3) =
\frac1{\Box_1\Box_2}
2\int^\infty_0dy^2\,\left(\frac4{y^2}\right)^3\,P_{26}(C_1,C_2,C_3){\cal S}_1{\cal S}_2{\cal S}_3
,\end{fleqnarray}

\begin{fleqnarray}&&\Gamma_{27}(-\Box_1,-\Box_2,-\Box_3) =
\frac1{\Box_1\Box_2}
2\int^\infty_0dy^2\,\left(\frac4{y^2}\right)^3\,P_{27}(C_1,C_2,C_3){\cal S}_1{\cal S}_2{\cal S}_3
 \nonumber\\&&\ \ \ \ \mbox{}
-{{1}\over {540\Box_1\Box_2\Box_3}}
,\end{fleqnarray}

\begin{fleqnarray}&&\Gamma_{28}(-\Box_1,-\Box_2,-\Box_3) =
\frac1{\Box_1\Box_2}
2\int^\infty_0dy^2\,\left(\frac4{y^2}\right)^3\,P_{28}(C_1,C_2,C_3){\cal S}_1{\cal S}_2{\cal S}_3
\nonumber\\&&\ \ \ \ \mbox{}
+{1\over {135\Box_1\Box_2\Box_3}}
,\end{fleqnarray}

\[\Gamma_{29}(-\Box_1,-\Box_2,-\Box_3) =
\frac1{\Box_1\Box_2\Box_3}
2\int^\infty_0dy^2\,\left(\frac4{y^2}\right)^3\,P_{29}(C_1,C_2,C_3){\cal S}_1{\cal S}_2{\cal S}_3
\]
\begin{fleqnarray}&& {\rm where}
\end{fleqnarray}
\arraycolsep=5pt
\mathindent=\parindent
\begin{fleqnarray}&& P_{1} =
   {1\over 3}C_1 C_2 C_3
,\end{fleqnarray}

\begin{fleqnarray}&& P_{2} =
   {2\over 9}C_1 C_2 C_3
  -{2\over 3}{{C_1}^2} C_2 C_3
  +{2\over 3}{{C_1}^2} {{C_2}^2} C_3
  -{2\over 9}{{C_1}^2} {{C_2}^2} {{C_3}^2}
,\end{fleqnarray}

\begin{fleqnarray}&& P_{3} =
   C_1 C_2 C_3
  -{{C_1}^2} C_2 C_3
  -C_1 {{C_2}^2} C_3
  +{{C_1}^2} {{C_2}^2} C_3
,\end{fleqnarray}

\begin{fleqnarray}&& P_{4} =
   {{11}\over {108}}C_1 C_2 C_3
  -{5\over {36}}{{C_1}^2} C_2 C_3
  -{1\over {108}}{{C_1}^3} C_2 C_3 \nonumber\\&&\ \ \ \ \mbox{}
  -{{19}\over {216}}{{C_1}^2} {{C_2}^2} C_3
  +{{11}\over {54}}{{C_1}^3} {{C_2}^2} C_3
  -{1\over {24}}{{C_1}^3} {{C_2}^3} C_3 \nonumber\\&&\ \ \ \ \mbox{}
  -{{13}\over {108}}C_1 C_2 {{C_3}^2}
  +{{17}\over {108}}{{C_1}^2} C_2 {{C_3}^2}
  -{1\over {12}}{{C_1}^3} C_2 {{C_3}^2}   \nonumber\\&&\ \ \ \ \mbox{}
  +{{11}\over {108}}{{C_1}^2} {{C_2}^2} {{C_3}^2}
  -{1\over {12}}{{C_1}^3} {{C_2}^2} {{C_3}^2}
,\end{fleqnarray}

\begin{fleqnarray}&& P_{5} =
  -{1\over 6}C_1 C_3
  +{1\over 3}{{C_1}^2} C_3
  -{1\over 6}{{C_1}^3} C_3 \nonumber\\&&\ \ \ \ \mbox{}
  +{1\over 9}C_1 C_2 C_3
  -{2\over 9}{{C_1}^2} C_2 C_3
  +{1\over {18}}{{C_1}^3} C_2 C_3
  +{1\over {18}}{{C_1}^2} {{C_2}^2} C_3 \nonumber\\&&\ \ \ \ \mbox{}
  +{1\over 9}C_1 C_2 {{C_3}^2}
  -{1\over 9}{{C_1}^2} C_2 {{C_3}^2}
,\end{fleqnarray}

\begin{fleqnarray}&& P_{6} =
  -{1\over 6}C_1 C_2 C_3
  +{1\over 2}{{C_1}^2} C_2 C_3
  -{1\over 2}{{C_1}^3} C_2 C_3
,\end{fleqnarray}

\begin{fleqnarray}&& P_{7} =
   {1\over {12}}C_2 C_3
  -{3\over 8}C_1 C_2 C_3
  +{5\over {24}}{{C_1}^2} C_2 C_3
  -{1\over 6}{{C_2}^2} C_3       \nonumber\\&&\ \ \ \ \mbox{}
  +{7\over 6}C_1 {{C_2}^2} C_3
  -{1\over 2}{{C_1}^2} {{C_2}^2} C_3
  +{1\over {12}}{{C_2}^3} C_3
  -{7\over {12}}C_1 {{C_2}^3} C_3   \nonumber\\&&\ \ \ \ \mbox{}
  +{1\over {12}}{{C_1}^2} {{C_2}^3} C_3
  +{1\over {12}}C_1 {{C_2}^4} C_3
  -{{19}\over {24}}C_1 {{C_2}^2} {{C_3}^2}
  +{7\over {24}}{{C_1}^2} {{C_2}^2} {{C_3}^2} \nonumber\\&&\ \ \ \ \mbox{}
  +{7\over {12}}C_1 {{C_2}^3} {{C_3}^2}
  -{1\over {12}}{{C_1}^2} {{C_2}^3} {{C_3}^2}
  -{1\over {12}}C_1 {{C_2}^4} {{C_3}^2}
,\end{fleqnarray}

\begin{fleqnarray}&& P_{8} =
  -2C_2 C_3
  +{5\over 3}C_1 C_2 C_3
  +{1\over 6}{{C_1}^2} C_2 C_3
  -{1\over 3}{{C_1}^3} C_2 C_3   \nonumber\\&&\ \ \ \ \mbox{}
  +6{{C_2}^2} C_3
  -{{14}\over 3}C_1 {{C_2}^2} C_3
  +{1\over 6}{{C_1}^2} {{C_2}^2} C_3
  +{1\over 3}{{C_1}^3} {{C_2}^2} C_3  \nonumber\\&&\ \ \ \ \mbox{}
  -{{11}\over 3}{{C_2}^3} C_3
  +3C_1 {{C_2}^3} C_3
  -{1\over 3}{{C_1}^2} {{C_2}^3} C_3
  +{5\over 6}{{C_2}^4} C_3            \nonumber\\&&\ \ \ \ \mbox{}
  -{2\over 3}C_1 {{C_2}^4} C_3
  -{8\over 3}{{C_2}^2} {{C_3}^2}
  +2C_1 {{C_2}^2} {{C_3}^2}
  -{1\over 3}{{C_1}^2} {{C_2}^2} {{C_3}^2}  \nonumber\\&&\ \ \ \ \mbox{}
  +{{11}\over 6}{{C_2}^3} {{C_3}^2}
  -{5\over 3}C_1 {{C_2}^3} {{C_3}^2}
  +{1\over 3}{{C_1}^2} {{C_2}^3} {{C_3}^2}
  -{1\over 3}{{C_2}^4} {{C_3}^2}         \nonumber\\&&\ \ \ \ \mbox{}
  +{1\over 3}C_1 {{C_2}^4} {{C_3}^2}
,\end{fleqnarray}

\begin{fleqnarray}&& P_{9} =
  -{{13}\over {72}}C_1 C_2
  +{{1801}\over {2160}}{{C_1}^2} C_2
  -{{3209}\over {4320}}{{C_1}^3} C_2    \nonumber\\&&\ \ \ \ \mbox{}
  +{{265}\over {864}}{{C_1}^4} C_2
  -{{43}\over {864}}{{C_1}^5} C_2
  +{1\over {4320}}{{C_1}^6} C_2
  -{{1117}\over {1440}}{{C_1}^2} {{C_2}^2} \nonumber\\&&\ \ \ \ \mbox{}
  +{{4639}\over {4320}}{{C_1}^3} {{C_2}^2}
  -{{31}\over {96}}{{C_1}^4} {{C_2}^2}
  +{{47}\over {1440}}{{C_1}^5} {{C_2}^2}   \nonumber\\&&\ \ \ \ \mbox{}
  +{1\over {864}}{{C_1}^6} {{C_2}^2}
  -{{67}\over {240}}{{C_1}^3} {{C_2}^3}
  +{{127}\over {1080}}{{C_1}^4} {{C_2}^3}   \nonumber\\&&\ \ \ \ \mbox{}
  -{1\over {144}}{{C_1}^5} {{C_2}^3}
  -{7\over {864}}{{C_1}^4} {{C_2}^4}
  +{{77}\over {720}}C_1 C_2 C_3          \nonumber\\&&\ \ \ \ \mbox{}
  -{{4859}\over {12960}}{{C_1}^2} C_2 C_3
  +{{5783}\over {12960}}{{C_1}^3} C_2 C_3
  -{{631}\over {2592}}{{C_1}^4} C_2 C_3   \nonumber\\&&\ \ \ \ \mbox{}
  +{{43}\over {864}}{{C_1}^5} C_2 C_3
  -{1\over {240}}{{C_1}^6} C_2 C_3
  +{7\over {108}}{{C_1}^2} {{C_2}^2} C_3  \nonumber\\&&\ \ \ \ \mbox{}
  -{{59}\over {240}}{{C_1}^3} {{C_2}^2} C_3
  +{{1553}\over {12960}}{{C_1}^4} {{C_2}^2} C_3
  -{{47}\over {4320}}{{C_1}^5} {{C_2}^2} C_3   \nonumber\\&&\ \ \ \ \mbox{}
  +{{139}\over {2592}}{{C_1}^3} {{C_2}^3} C_3
  -{1\over {32}}{{C_1}^4} {{C_2}^3} C_3
  +{1\over {360}}{{C_1}^5} {{C_2}^3} C_3    \nonumber\\&&\ \ \ \ \mbox{}
  +{1\over {720}}{{C_1}^4} {{C_2}^4} C_3
  +{{373}\over {3240}}{{C_1}^2} {{C_2}^2} {{C_3}^2}
  -{{41}\over {810}}{{C_1}^3} {{C_2}^2} {{C_3}^2} \nonumber\\&&\ \ \ \ \mbox{}
  -{{61}\over {12960}}{{C_1}^4} {{C_2}^2} {{C_3}^2}
  +{{19}\over {12960}}{{C_1}^3} {{C_2}^3} {{C_3}^2}
  +{1\over {720}}{{C_1}^4} {{C_2}^3} {{C_3}^2}  \nonumber\\&&\ \ \ \ \mbox{}
  +{1\over {1080}}{{C_1}^3} {{C_2}^3} {{C_3}^3}
,\end{fleqnarray}

\begin{fleqnarray}&& P_{10} =
   {1\over {18}}C_1 C_2 C_3
  -{1\over 6}{{C_1}^2} C_2 C_3
  +{1\over 6}{{C_1}^2} {{C_2}^2} C_3
  -{1\over {18}}{{C_1}^2} {{C_2}^2} {{C_3}^2}
,\end{fleqnarray}

\begin{fleqnarray}&& P_{11} =
   {{59}\over {540}}C_1 C_3
  -{{53}\over {216}}{{C_1}^2} C_3
  +{{47}\over {270}}{{C_1}^3} C_3   \nonumber\\&&\ \ \ \ \mbox{}
  -{{37}\over {1080}}{{C_1}^4} C_3
  -{1\over {540}}{{C_1}^5} C_3
  -{{29}\over {3240}}C_1 C_2 C_3    \nonumber\\&&\ \ \ \ \mbox{}
  +{{163}\over {6480}}{{C_1}^2} C_2 C_3
  -{{449}\over {12960}}{{C_1}^3} C_2 C_3
  +{7\over {12960}}{{C_1}^4} C_2 C_3    \nonumber\\&&\ \ \ \ \mbox{}
  +{1\over {360}}{{C_1}^5} C_2 C_3
  +{{119}\over {6480}}{{C_1}^2} {{C_2}^2} C_3
  -{7\over {1296}}{{C_1}^3} {{C_2}^2} C_3  \nonumber\\&&\ \ \ \ \mbox{}
  +{1\over {648}}{{C_1}^4} {{C_2}^2} C_3
  +{1\over {4320}}{{C_1}^3} {{C_2}^3} C_3
  -{{19}\over {12960}}{{C_1}^4} {{C_2}^3} C_3 \nonumber\\&&\ \ \ \ \mbox{}
  -{{11}\over {72}}C_1 {{C_3}^2}
  +{{319}\over {1080}}{{C_1}^2} {{C_3}^2}
  -{{109}\over {540}}{{C_1}^3} {{C_3}^2}  \nonumber\\&&\ \ \ \ \mbox{}
  +{{17}\over {540}}{{C_1}^4} {{C_3}^2}
  +{1\over {270}}{{C_1}^5} {{C_3}^2}
  +{{31}\over {2160}}C_1 C_2 {{C_3}^2} \nonumber\\&&\ \ \ \ \mbox{}
  -{{373}\over {6480}}{{C_1}^2} C_2 {{C_3}^2}
  +{{83}\over {1728}}{{C_1}^3} C_2 {{C_3}^2}
  +{{47}\over {5184}}{{C_1}^4} C_2 {{C_3}^2} \nonumber\\&&\ \ \ \ \mbox{}
  -{1\over {360}}{{C_1}^5} C_2 {{C_3}^2}
  +{{319}\over {12960}}{{C_1}^2} {{C_2}^2} {{C_3}^2}
  -{{223}\over {8640}}{{C_1}^3} {{C_2}^2} {{C_3}^2}  \nonumber\\&&\ \ \ \ \mbox{}
  +{5\over {5184}}{{C_1}^4} {{C_2}^2} {{C_3}^2}
  +{{13}\over {4320}}{{C_1}^3} {{C_2}^3} {{C_3}^2}
  +{{83}\over {1080}}C_1 {{C_3}^3}               \nonumber\\&&\ \ \ \ \mbox{}
  -{{103}\over {1080}}{{C_1}^2} {{C_3}^3}
  +{{73}\over {1080}}{{C_1}^3} {{C_3}^3}
  -{1\over {360}}{{C_1}^4} {{C_3}^3}      \nonumber\\&&\ \ \ \ \mbox{}
  -{1\over {540}}{{C_1}^5} {{C_3}^3}
  -{{349}\over {12960}}C_1 C_2 {{C_3}^3}
  +{{127}\over {8640}}{{C_1}^2} C_2 {{C_3}^3} \nonumber\\&&\ \ \ \ \mbox{}
  -{{197}\over {25920}}{{C_1}^3} C_2 {{C_3}^3}
  -{1\over {216}}{{C_1}^4} C_2 {{C_3}^3}
  -{{13}\over {960}}{{C_1}^2} {{C_2}^2} {{C_3}^3} \nonumber\\&&\ \ \ \ \mbox{}
  +{{193}\over {25920}}{{C_1}^3} {{C_2}^2} {{C_3}^3}
  -{1\over {60}}C_1 {{C_3}^4}
  +{7\over {1080}}{{C_1}^2} {{C_3}^4}       \nonumber\\&&\ \ \ \ \mbox{}
  -{1\over {72}}{{C_1}^3} {{C_3}^4}
  +{1\over {60}}C_1 C_2 {{C_3}^4}
  +{1\over {180}}{{C_1}^2} C_2 {{C_3}^4}  \nonumber\\&&\ \ \ \ \mbox{}
  +{1\over {540}}{{C_1}^3} {{C_3}^5}
  -{1\over {540}}C_1 C_2 {{C_3}^5}
  -{1\over {540}}{{C_1}^2} C_2 {{C_3}^5}
,\end{fleqnarray}

\begin{fleqnarray}&& P_{12} =
  -{1\over 3}C_1 C_2
  +{2\over 3}{{C_1}^2} C_2
  -{1\over 3}{{C_1}^3} C_2    \nonumber\\&&\ \ \ \ \mbox{}
  -{1\over 3}C_1 C_3
  +{2\over 3}{{C_1}^2} C_3
  -{1\over 3}{{C_1}^3} C_3
  -{4\over 3}C_1 C_2 C_3      \nonumber\\&&\ \ \ \ \mbox{}
  +{{10}\over 3}{{C_1}^2} C_2 C_3
  -{8\over 3}{{C_1}^3} C_2 C_3
  +{2\over 3}{{C_1}^4} C_2 C_3
,\end{fleqnarray}

\begin{fleqnarray}&& P_{13} =
   2C_1 C_2 C_3
  -2{{C_1}^2} C_2 C_3
,\end{fleqnarray}

\begin{fleqnarray}&& P_{14} =
  -2C_1 C_2 C_3
  +4C_1 C_2 {{C_3}^2}
  -2C_1 C_2 {{C_3}^3}
,\end{fleqnarray}

\begin{fleqnarray}&& P_{15} =
   {2\over 3}C_1 C_2 C_3
  -2{{C_1}^2} C_2 C_3
  +{{13}\over 6}{{C_1}^3} C_2 C_3  \nonumber\\&&\ \ \ \ \mbox{}
  -{{C_1}^4} C_2 C_3
  +{1\over 6}{{C_1}^5} C_2 C_3
,\end{fleqnarray}

\begin{fleqnarray}&& P_{16} =
   {2\over 9}C_1 C_2 C_3
  -{2\over 9}{{C_1}^2} C_2 C_3
  -{4\over 9}C_1 {{C_2}^2} C_3  \nonumber\\&&\ \ \ \ \mbox{}
  +{2\over 3}{{C_1}^2} {{C_2}^2} C_3
  -{2\over 9}{{C_1}^3} {{C_2}^2} C_3
  +{2\over 3}C_1 C_2 {{C_3}^2}      \nonumber\\&&\ \ \ \ \mbox{}
  -{8\over 9}{{C_1}^2} C_2 {{C_3}^2}
  +{2\over 9}{{C_1}^3} C_2 {{C_3}^2}
,\end{fleqnarray}

\begin{fleqnarray}&& P_{17} =
   2C_1 C_2 C_3
  -3{{C_1}^2} C_2 C_3
  +{{C_1}^3} C_2 C_3
,\end{fleqnarray}

\begin{fleqnarray}&& P_{18} =
   {4\over 3}C_1 {{C_2}^2} C_3
  -2{{C_1}^2} {{C_2}^2} C_3
  +{2\over 3}{{C_1}^3} {{C_2}^2} C_3
  +{2\over 3}C_1 {{C_2}^2} {{C_3}^2} \nonumber\\&&\ \ \ \ \mbox{}
  -{{C_1}^2} {{C_2}^2} {{C_3}^2}
  +{1\over 3}{{C_1}^3} {{C_2}^2} {{C_3}^2}
,\end{fleqnarray}

\begin{fleqnarray}&& P_{19} =
  -{1\over 3}C_1 C_2 C_3
  +{1\over 2}{{C_1}^2} C_2 C_3
  -{1\over 6}{{C_1}^3} C_2 C_3  \nonumber\\&&\ \ \ \ \mbox{}
  +{2\over 3}C_1 {{C_2}^2} C_3
  -{{C_1}^2} {{C_2}^2} C_3
  +{1\over 3}{{C_1}^3} {{C_2}^2} C_3  \nonumber\\&&\ \ \ \ \mbox{}
  -{1\over 3}C_1 {{C_2}^2} {{C_3}^2}
  +{1\over 2}{{C_1}^2} {{C_2}^2} {{C_3}^2}
  -{1\over 6}{{C_1}^3} {{C_2}^2} {{C_3}^2}
,\end{fleqnarray}

\begin{fleqnarray}&& P_{20} =
  -{1\over {36}}C_1 C_2 C_3
  -{1\over {24}}{{C_1}^2} C_2 C_3
  +{5\over {72}}{{C_1}^3} C_2 C_3
  -{{11}\over {36}}C_1 {{C_2}^2} C_3  \nonumber\\&&\ \ \ \ \mbox{}
  +{5\over 9}{{C_1}^2} {{C_2}^2} C_3
  -{1\over 4}{{C_1}^3} {{C_2}^2} C_3
  -{1\over {12}}C_1 {{C_2}^3} C_3
  +{1\over {12}}{{C_1}^2} {{C_2}^3} C_3  \nonumber\\&&\ \ \ \ \mbox{}
  -{2\over 9}C_1 {{C_2}^2} {{C_3}^2}
  +{3\over 8}{{C_1}^2} {{C_2}^2} {{C_3}^2}
  -{{11}\over {72}}{{C_1}^3} {{C_2}^2} {{C_3}^2}
  -{1\over {36}}C_1 {{C_2}^3} {{C_3}^2} \nonumber\\&&\ \ \ \ \mbox{}
  +{1\over {36}}{{C_1}^2} {{C_2}^3} {{C_3}^2}
,\end{fleqnarray}

\begin{fleqnarray}&& P_{21} =
   {4\over 3}C_1 C_2 {{C_3}^2}
  -2{{C_1}^2} C_2 {{C_3}^2}
  +{2\over 3}{{C_1}^3} C_2 {{C_3}^2}  \nonumber\\&&\ \ \ \ \mbox{}
  -{4\over 3}C_1 C_2 {{C_3}^3}
  +2{{C_1}^2} C_2 {{C_3}^3}
  -{2\over 3}{{C_1}^3} C_2 {{C_3}^3}
,\end{fleqnarray}

\begin{fleqnarray}&& P_{22-1} =
   {{22}\over {135}}C_1 C_2 C_3
  -{1\over 4}{{C_1}^2} C_2 C_3
  +{{47}\over {540}}{{C_1}^3} C_2 C_3   \nonumber\\&&\ \ \ \ \mbox{}
  -{{46}\over {135}}C_1 {{C_2}^2} C_3
  +{{241}\over {540}}{{C_1}^2} {{C_2}^2} C_3
  -{{19}\over {180}}{{C_1}^3} {{C_2}^2} C_3 \nonumber\\&&\ \ \ \ \mbox{}
  +{{53}\over {270}}C_1 {{C_2}^3} C_3
  -{{34}\over {135}}{{C_1}^2} {{C_2}^3} C_3
  +{1\over {18}}{{C_1}^3} {{C_2}^3} C_3     \nonumber\\&&\ \ \ \ \mbox{}
  -{1\over {540}}C_1 {{C_2}^5} C_3
  +{1\over {540}}{{C_1}^2} {{C_2}^5} C_3
  +{{28}\over {135}}C_1 {{C_2}^2} {{C_3}^2} \nonumber\\&&\ \ \ \ \mbox{}
  -{{73}\over {270}}{{C_1}^2} {{C_2}^2} {{C_3}^2}
  +{{17}\over {270}}{{C_1}^3} {{C_2}^2} {{C_3}^2}
  -{2\over {135}}C_1 {{C_2}^3} {{C_3}^2}       \nonumber\\&&\ \ \ \ \mbox{}
  +{2\over {135}}{{C_1}^2} {{C_2}^3} {{C_3}^2}
  +{1\over {540}}C_1 {{C_2}^3} {{C_3}^3}
  -{1\over {540}}{{C_1}^2} {{C_2}^3} {{C_3}^3}
,\end{fleqnarray}

\begin{fleqnarray}&& P_{22-2} =
   {{19}\over {90}}C_1 C_2 C_3
  -{1\over {36}}{{C_1}^2} C_2 C_3
  -{4\over {135}}{{C_1}^3} C_2 C_3       \nonumber\\&&\ \ \ \ \mbox{}
  -{{17}\over {45}}C_1 {{C_2}^2} C_3
  +{{71}\over {360}}{{C_1}^2} {{C_2}^2} C_3
  -{1\over {72}}{{C_1}^3} {{C_2}^2} C_3     \nonumber\\&&\ \ \ \ \mbox{}
  +{{227}\over {1080}}C_1 {{C_2}^3} C_3
  -{{29}\over {540}}{{C_1}^2} {{C_2}^3} C_3
  +{7\over {1080}}{{C_1}^3} {{C_2}^3} C_3    \nonumber\\&&\ \ \ \ \mbox{}
  -{1\over {24}}C_1 {{C_2}^4} C_3
  +{{11}\over {1080}}{{C_1}^2} {{C_2}^4} C_3
  -{1\over {540}}C_1 {{C_2}^5} C_3          \nonumber\\&&\ \ \ \ \mbox{}
  -{{59}\over {270}}C_1 C_2 {{C_3}^2}
  -{1\over {10}}{{C_1}^2} C_2 {{C_3}^2}
  +{7\over {90}}{{C_1}^3} C_2 {{C_3}^2}    \nonumber\\&&\ \ \ \ \mbox{}
  +{{529}\over {1080}}C_1 {{C_2}^2} {{C_3}^2}
  -{{11}\over {54}}{{C_1}^2} {{C_2}^2} {{C_3}^2}
  +{1\over {216}}{{C_1}^3} {{C_2}^2} {{C_3}^2} \nonumber\\&&\ \ \ \ \mbox{}
  -{{131}\over {540}}C_1 {{C_2}^3} {{C_3}^2}
  +{{17}\over {360}}{{C_1}^2} {{C_2}^3} {{C_3}^2}
  -{1\over {1080}}{{C_1}^3} {{C_2}^3} {{C_3}^2} \nonumber\\&&\ \ \ \ \mbox{}
  +{{47}\over {1080}}C_1 {{C_2}^4} {{C_3}^2}
  -{{11}\over {1080}}{{C_1}^2} {{C_2}^4} {{C_3}^2}
  +{1\over {540}}C_1 {{C_2}^5} {{C_3}^2}         \nonumber\\&&\ \ \ \ \mbox{}
  -{1\over {90}}C_1 C_2 {{C_3}^3}
  +{7\over {45}}{{C_1}^2} C_2 {{C_3}^3}
  -{{31}\over {540}}{{C_1}^3} C_2 {{C_3}^3}    \nonumber\\&&\ \ \ \ \mbox{}
  -{{109}\over {1080}}C_1 {{C_2}^2} {{C_3}^3}
  -{{11}\over {1080}}{{C_1}^2} {{C_2}^2} {{C_3}^3}
  +{2\over {135}}{{C_1}^3} {{C_2}^2} {{C_3}^3}  \nonumber\\&&\ \ \ \ \mbox{}
  +{7\over {216}}C_1 {{C_2}^3} {{C_3}^3}
  +{7\over {1080}}{{C_1}^2} {{C_2}^3} {{C_3}^3}
  -{1\over {180}}{{C_1}^3} {{C_2}^3} {{C_3}^3}  \nonumber\\&&\ \ \ \ \mbox{}
  -{1\over {540}}C_1 {{C_2}^4} {{C_3}^3}
  +{1\over {54}}C_1 C_2 {{C_3}^4}
  -{1\over {36}}{{C_1}^2} C_2 {{C_3}^4}       \nonumber\\&&\ \ \ \ \mbox{}
  +{1\over {108}}{{C_1}^3} C_2 {{C_3}^4}
  -{1\over {90}}C_1 {{C_2}^2} {{C_3}^4}
  +{1\over {60}}{{C_1}^2} {{C_2}^2} {{C_3}^4}  \nonumber\\&&\ \ \ \ \mbox{}
  -{1\over {180}}{{C_1}^3} {{C_2}^2} {{C_3}^4}
,\end{fleqnarray}

\begin{fleqnarray}&& P_{23} =
  -{3\over 5}C_1 C_2
  +{{1021}\over {540}}{{C_1}^2} C_2
  -{{1219}\over {540}}{{C_1}^3} C_2   \nonumber\\&&\ \ \ \ \mbox{}
  +{{233}\over {180}}{{C_1}^4} C_2
  -{{23}\over {60}}{{C_1}^5} C_2
  +{8\over {135}}{{C_1}^6} C_2      \nonumber\\&&\ \ \ \ \mbox{}
  -{1\over {270}}{{C_1}^7} C_2
  +{{277}\over {540}}C_1 {{C_2}^2}
  -{{173}\over {135}}{{C_1}^2} {{C_2}^2}  \nonumber\\&&\ \ \ \ \mbox{}
  +{{113}\over {108}}{{C_1}^3} {{C_2}^2}
  -{3\over {10}}{{C_1}^4} {{C_2}^2}
  +{1\over {45}}{{C_1}^5} {{C_2}^2}     \nonumber\\&&\ \ \ \ \mbox{}
  -{{17}\over {108}}C_1 {{C_2}^3}
  +{{29}\over {90}}{{C_1}^2} {{C_2}^3}
  -{{89}\over {540}}{{C_1}^3} {{C_2}^3}  \nonumber\\&&\ \ \ \ \mbox{}
  -{1\over {135}}{{C_1}^4} {{C_2}^3}
  +{1\over {135}}{{C_1}^5} {{C_2}^3}
  +{1\over {27}}C_1 {{C_2}^4}          \nonumber\\&&\ \ \ \ \mbox{}
  -{2\over {27}}{{C_1}^2} {{C_2}^4}
  +{1\over {27}}{{C_1}^3} {{C_2}^4}
  -{1\over {270}}C_1 {{C_2}^5}         \nonumber\\&&\ \ \ \ \mbox{}
  +{1\over {135}}{{C_1}^2} {{C_2}^5}
  -{1\over {270}}{{C_1}^3} {{C_2}^5}
  +{{43}\over {135}}C_1 C_2 C_3       \nonumber\\&&\ \ \ \ \mbox{}
  -{{101}\over {135}}{{C_1}^2} C_2 C_3
  +{{16}\over {27}}{{C_1}^3} C_2 C_3
  -{{26}\over {135}}{{C_1}^4} C_2 C_3  \nonumber\\&&\ \ \ \ \mbox{}
  +{4\over {135}}{{C_1}^5} C_2 C_3
  -{{41}\over {90}}C_1 {{C_2}^2} C_3
  +{{161}\over {270}}{{C_1}^2} {{C_2}^2} C_3  \nonumber\\&&\ \ \ \ \mbox{}
  -{5\over {54}}{{C_1}^3} {{C_2}^2} C_3
  -{7\over {135}}{{C_1}^4} {{C_2}^2} C_3
  +{1\over {270}}{{C_1}^5} {{C_2}^2} C_3   \nonumber\\&&\ \ \ \ \mbox{}
  +{{41}\over {135}}C_1 {{C_2}^3} C_3
  -{{47}\over {135}}{{C_1}^2} {{C_2}^3} C_3
  +{1\over {30}}{{C_1}^3} {{C_2}^3} C_3       \nonumber\\&&\ \ \ \ \mbox{}
  +{1\over {90}}{{C_1}^4} {{C_2}^3} C_3
  -{2\over {27}}C_1 {{C_2}^4} C_3
  +{2\over {27}}{{C_1}^2} {{C_2}^4} C_3      \nonumber\\&&\ \ \ \ \mbox{}
  +{1\over {135}}C_1 {{C_2}^5} C_3
  -{1\over {135}}{{C_1}^2} {{C_2}^5} C_3
  +{1\over 9}C_1 C_2 {{C_3}^2}            \nonumber\\&&\ \ \ \ \mbox{}
  -{{133}\over {540}}{{C_1}^2} C_2 {{C_3}^2}
  +{{79}\over {540}}{{C_1}^3} C_2 {{C_3}^2}
  -{7\over {540}}{{C_1}^4} C_2 {{C_3}^2}   \nonumber\\&&\ \ \ \ \mbox{}
  +{1\over {540}}{{C_1}^5} C_2 {{C_3}^2}
  +{1\over {108}}C_1 {{C_2}^2} {{C_3}^2}
  +{1\over {15}}{{C_1}^2} {{C_2}^2} {{C_3}^2} \nonumber\\&&\ \ \ \ \mbox{}
  -{{41}\over {540}}{{C_1}^3} {{C_2}^2} {{C_3}^2}
  -{{19}\over {540}}C_1 {{C_2}^3} {{C_3}^2}
  +{1\over {45}}{{C_1}^2} {{C_2}^3} {{C_3}^2}  \nonumber\\&&\ \ \ \ \mbox{}
  +{7\over {540}}{{C_1}^3} {{C_2}^3} {{C_3}^2}
  -{{23}\over {270}}C_1 C_2 {{C_3}^3}
  +{{14}\over {135}}{{C_1}^2} C_2 {{C_3}^3}   \nonumber\\&&\ \ \ \ \mbox{}
  -{1\over {54}}{{C_1}^3} C_2 {{C_3}^3}
  +{1\over {30}}C_1 {{C_2}^2} {{C_3}^3}
  -{1\over {27}}{{C_1}^2} {{C_2}^2} {{C_3}^3}  \nonumber\\&&\ \ \ \ \mbox{}
  +{1\over {270}}{{C_1}^3} {{C_2}^2} {{C_3}^3}
  -{1\over {90}}C_1 {{C_2}^3} {{C_3}^3}
  +{1\over {90}}{{C_1}^2} {{C_2}^3} {{C_3}^3} \nonumber\\&&\ \ \ \ \mbox{}
  +{1\over {54}}C_1 C_2 {{C_3}^4}
  -{1\over {45}}{{C_1}^2} C_2 {{C_3}^4}
  +{1\over {270}}{{C_1}^3} C_2 {{C_3}^4}    \nonumber\\&&\ \ \ \ \mbox{}
  -{1\over {270}}C_1 {{C_2}^2} {{C_3}^4}
  +{1\over {270}}{{C_1}^2} {{C_2}^2} {{C_3}^4}
,\end{fleqnarray}

\begin{fleqnarray}&& P_{24} =
   {1\over {90}}C_2 C_3
  -{7\over {135}}C_1 C_2 C_3
  -{{13}\over {540}}{{C_1}^2} C_2 C_3   \nonumber\\&&\ \ \ \ \mbox{}
  +{1\over {90}}{{C_1}^4} C_2 C_3
  -{1\over {90}}{{C_2}^2} C_3
  +{1\over {20}}C_1 {{C_2}^2} C_3
  +{{13}\over {540}}{{C_1}^2} {{C_2}^2} C_3 \nonumber\\&&\ \ \ \ \mbox{}
  -{1\over {90}}{{C_1}^4} {{C_2}^2} C_3
  -{1\over {90}}{{C_2}^3} C_3
  +{1\over {540}}C_1 {{C_2}^3} C_3
  +{1\over {90}}{{C_2}^4} C_3            \nonumber\\&&\ \ \ \ \mbox{}
  +{1\over {90}}C_2 {{C_3}^2}
  -{{13}\over {540}}C_1 C_2 {{C_3}^2}
  -{1\over {45}}{{C_2}^2} {{C_3}^2}
  +{{13}\over {540}}C_1 {{C_2}^2} {{C_3}^2} \nonumber\\&&\ \ \ \ \mbox{}
  +{1\over {90}}{{C_2}^3} {{C_3}^2}
  +{1\over {90}}C_1 C_2 {{C_3}^3}
  -{1\over {90}}{{C_1}^2} C_2 {{C_3}^3}
  -{1\over {90}}C_1 {{C_2}^2} {{C_3}^3}  \nonumber\\&&\ \ \ \ \mbox{}
  +{1\over {90}}{{C_1}^2} {{C_2}^2} {{C_3}^3}
,\end{fleqnarray}

\begin{fleqnarray}&& P_{25} =
  -{5\over {27}}C_1 C_2
  +{{191}\over {270}}{{C_1}^2} C_2
  -{{44}\over {45}}{{C_1}^3} C_2
  +{{53}\over {90}}{{C_1}^4} C_2    \nonumber\\&&\ \ \ \ \mbox{}
  -{4\over {27}}{{C_1}^5} C_2
  +{2\over {135}}{{C_1}^6} C_2
  +{1\over {10}}C_1 {{C_2}^2}
  -{{31}\over {135}}{{C_1}^2} {{C_2}^2}  \nonumber\\&&\ \ \ \ \mbox{}
  +{{43}\over {270}}{{C_1}^3} {{C_2}^2}
  -{4\over {135}}{{C_1}^4} {{C_2}^2}
  +{4\over {135}}C_1 {{C_2}^3}           \nonumber\\&&\ \ \ \ \mbox{}
  -{2\over {27}}{{C_1}^2} {{C_2}^3}
  +{8\over {135}}{{C_1}^3} {{C_2}^3}
  -{2\over {135}}{{C_1}^4} {{C_2}^3}     \nonumber\\&&\ \ \ \ \mbox{}
  +{7\over {30}}C_1 C_2 C_3
  -{{58}\over {135}}{{C_1}^2} C_2 C_3
  +{{53}\over {270}}{{C_1}^3} C_2 C_3   \nonumber\\&&\ \ \ \ \mbox{}
  +{2\over {135}}{{C_1}^4} C_2 C_3
  -{2\over {135}}{{C_1}^5} C_2 C_3
  -{{22}\over {45}}C_1 {{C_2}^2} C_3   \nonumber\\&&\ \ \ \ \mbox{}
  +{8\over {15}}{{C_1}^2} {{C_2}^2} C_3
  -{2\over {135}}{{C_1}^3} {{C_2}^2} C_3
  -{4\over {135}}{{C_1}^4} {{C_2}^2} C_3  \nonumber\\&&\ \ \ \ \mbox{}
  +{2\over 9}C_1 {{C_2}^3} C_3
  -{4\over {15}}{{C_1}^2} {{C_2}^3} C_3
  +{2\over {45}}{{C_1}^3} {{C_2}^3} C_3    \nonumber\\&&\ \ \ \ \mbox{}
  +{2\over {15}}C_1 {{C_2}^2} {{C_3}^2}
  -{2\over {15}}{{C_1}^2} {{C_2}^2} {{C_3}^2}
  -{2\over {45}}C_1 {{C_2}^3} {{C_3}^2}    \nonumber\\&&\ \ \ \ \mbox{}
  +{2\over {45}}{{C_1}^2} {{C_2}^3} {{C_3}^2}
,\end{fleqnarray}

\begin{fleqnarray}&& P_{26} =
   {2\over 3}C_1 C_2 C_3
  -2{{C_1}^2} C_2 C_3
  +{2\over 3}{{C_1}^3} C_2 C_3        \nonumber\\&&\ \ \ \ \mbox{}
  +{3\over 2}{{C_1}^2} {{C_2}^2} C_3
  -{{C_1}^3} {{C_2}^2} C_3
  +{1\over 6}{{C_1}^3} {{C_2}^3} C_3
,\end{fleqnarray}

\begin{fleqnarray}&& P_{27} =
   {4\over {135}}C_1 C_2 C_3
  -{{17}\over {135}}{{C_1}^2} C_2 C_3
  +{7\over {90}}{{C_1}^3} C_2 C_3    \nonumber\\&&\ \ \ \ \mbox{}
  -{1\over {135}}{{C_1}^4} C_2 C_3
  -{1\over {270}}{{C_1}^5} C_2 C_3
  +{{37}\over {270}}{{C_1}^2} {{C_2}^2} C_3 \nonumber\\&&\ \ \ \ \mbox{}
  -{{97}\over {540}}{{C_1}^3} {{C_2}^2} C_3
  +{7\over {270}}{{C_1}^4} {{C_2}^2} C_3
  +{1\over {180}}{{C_1}^5} {{C_2}^2} C_3    \nonumber\\&&\ \ \ \ \mbox{}
  +{7\over {108}}{{C_1}^3} {{C_2}^3} C_3
  -{7\over {270}}{{C_1}^4} {{C_2}^3} C_3
  -{1\over {540}}{{C_1}^5} {{C_2}^3} C_3  \nonumber\\&&\ \ \ \ \mbox{}
  +{1\over {270}}{{C_1}^4} {{C_2}^4} C_3
  +{1\over {45}}C_1 C_2 {{C_3}^2}
  -{1\over {15}}{{C_1}^2} C_2 {{C_3}^2}   \nonumber\\&&\ \ \ \ \mbox{}
  +{1\over {45}}{{C_1}^3} C_2 {{C_3}^2}
  +{1\over {20}}{{C_1}^2} {{C_2}^2} {{C_3}^2}
  -{1\over {30}}{{C_1}^3} {{C_2}^2} {{C_3}^2} \nonumber\\&&\ \ \ \ \mbox{}
  +{1\over {180}}{{C_1}^3} {{C_2}^3} {{C_3}^2}
  -{1\over {135}}C_1 C_2 {{C_3}^3}
  +{1\over {45}}{{C_1}^2} C_2 {{C_3}^3}    \nonumber\\&&\ \ \ \ \mbox{}
  -{1\over {135}}{{C_1}^3} C_2 {{C_3}^3}
  -{1\over {60}}{{C_1}^2} {{C_2}^2} {{C_3}^3}
  +{1\over {90}}{{C_1}^3} {{C_2}^2} {{C_3}^3} \nonumber\\&&\ \ \ \ \mbox{}
  -{1\over {540}}{{C_1}^3} {{C_2}^3} {{C_3}^3}
,\end{fleqnarray}

\begin{fleqnarray}&& P_{28} =
   {4\over {45}}C_1 C_2 C_3
  -{4\over {15}}{{C_1}^2} C_2 C_3
  +{4\over {45}}{{C_1}^3} C_2 C_3       \nonumber\\&&\ \ \ \ \mbox{}
  +{1\over 5}{{C_1}^2} {{C_2}^2} C_3
  -{2\over {15}}{{C_1}^3} {{C_2}^2} C_3
  +{1\over {45}}{{C_1}^3} {{C_2}^3} C_3   \nonumber\\&&\ \ \ \ \mbox{}
  -{4\over {45}}C_1 C_2 {{C_3}^2}
  +{4\over {15}}{{C_1}^2} C_2 {{C_3}^2}
  -{4\over {45}}{{C_1}^3} C_2 {{C_3}^2}  \nonumber\\&&\ \ \ \ \mbox{}
  -{1\over 5}{{C_1}^2} {{C_2}^2} {{C_3}^2}
  +{2\over {15}}{{C_1}^3} {{C_2}^2} {{C_3}^2}
  -{1\over {45}}{{C_1}^3} {{C_2}^3} {{C_3}^2}
,\end{fleqnarray}

\begin{fleqnarray}&& P_{29} =
   {4\over {135}}C_1 C_2 C_3
  -{2\over {15}}{{C_1}^2} C_2 C_3
  +{2\over {45}}{{C_1}^3} C_2 C_3     \nonumber\\&&\ \ \ \ \mbox{}
  +{1\over 5}{{C_1}^2} {{C_2}^2} C_3
  -{1\over {15}}{{C_1}^3} {{C_2}^2} C_3
  -{1\over {15}}{{C_1}^2} {{C_2}^3} C_3  \nonumber\\&&\ \ \ \ \mbox{}
  +{1\over {45}}{{C_1}^3} {{C_2}^3} C_3
  -{1\over {10}}{{C_1}^2} {{C_2}^2} {{C_3}^2}
  +{1\over {10}}{{C_1}^3} {{C_2}^2} {{C_3}^2} \nonumber\\&&\ \ \ \ \mbox{}
  -{1\over {30}}{{C_1}^3} {{C_2}^3} {{C_3}^2}
  +{1\over {270}}{{C_1}^3} {{C_2}^3} {{C_3}^3}
.\end{fleqnarray}

For calculations within the generalized spectral
technique, it is important to know the
asymptotic behaviours of the integral (9.1).
They are as follows:
\begin{equation}
{\cal S}=-\frac1{2\Box}
+{\rm O}\left({\rm e}^{-y\sqrt{-\Box}}\right),\hspace{7mm}
y\rightarrow\infty,
\end{equation}
and
\begin{eqnarray}
{\cal S}&=&{\cal S}^M
+{\rm O}\left(y^{2(M+1)}\ln y^2\right),\hspace{7mm}
y\rightarrow0, \\[\baselineskip]
{\cal S}^M&=&
\sum^M_{m=1}
\frac{(y^2)^m(-\Box)^{m-1}}{4^m(m!)^2}
\left[\frac m2
\ln\Big(-\frac14y^2\Box\Big)
-m\psi(m+1)+\frac12
\right],
\hspace{7mm}
M\geq1 \nonumber\\
\end{eqnarray}
where $\psi(x)$ is the Euler $\psi$-function.

The generalized spectral representation can be
extended to products of the form factors with
any positive powers of $\Box$'s. This is achieved by
a repeated use of the relation
\begin{equation}
\Box{\cal S}=-\frac12-\frac4{y^2}(C-1)C{\cal S}
\end{equation}
along with the commutation rule
\begin{equation}
P(C)\left(\frac1{y^2}\right)^N =
\left(\frac1{y^2}\right)^N P(C-N)
\end{equation}
valid for any polynomial $P(C)$. The result
for a product of
$\Box_1^{M_1}\Box_2^{M_2}\Box_3^{M_3}$
with the integrals in (9.4) or (9.5) is then again
a sum of integrals of the form
\begin{equation}
\int^\infty_0dy^2\,\left(\frac4{y^2}\right)^N
P(C,\dots C)\underbrace{{\cal S}\dots{\cal S}}_{K}.
\end{equation}

Owing to this property, the generalized spectral
representation solves the notorious problem
of nonuniqueness. An example showing how it makes
manifest the hidden identities between the form
factors will be considered in sect. 12. The only
arbitrariness that remains in this representation
corresponds to the possibility of integration by
parts over $y^2$ and is expressed by the identity
\begin{equation}
C_1+C_2+C_3=2
\end{equation}
in the integral (9.4),
\begin{equation}
C_n+C_k=1
\end{equation}
in the integral (9.5), and
\begin{equation}
\sum C=N-1
\end{equation}
in the integral (9.70). This arbitrariness is
removed by excluding everywhere a
$C$ with any chosen (but one and the same) index.

The formal use of the identity (9.73) is safe if,
in eq. (9.70), $N\leq K$, as is the case in the
expressions for the form factors above. However,
in the case of products of the form factors with
the positive powers of $\Box$'s, the integral (9.70)
may appear with $N>K$. The polynomial
$P(C,\dots C)$ will then contain the factors
\begin{equation}
\prod^M_{p=1}(C-p)^2
\end{equation}
improving the asymptotic behaviour of the
respective ${\cal S}$ so that the integral (9.70)
converge at the lower limit. The ${\cal S}$ on which
the operator (9.74) acts can be replaced by
$({\cal S}-{\cal S}^M)$ with ${\cal S}^M$ in (9.67). Only after this
replacement has been made, can the identity (9.73)
be applied to exclude $C$ in the operator (9.74).

The last limitation is connected with the fact
that ${\cal S}$ does not decrease at $y\rightarrow\infty$
but the action of at least one $C$ makes it
decreasing exponentially. One must, therefore,
ensure that at least one ${\cal S}$ in the integral (9.70)
be accompanied by at least one like $C$.
In the expressions for the form factors above, this
condition is always fulfilled.

For the details and derivations see sect. 20.

\section{ The large-$\Box$  asymptotic behaviours
of the third-order form factors}
\setcounter{equation}{0}

\hspace{\parindent}
After varying of one of the curvatures in the effective
action (6.1), the respective $\Box$ argument of the
form factor will become distinguished for it will refer
to the observation point of the current. Obviously, each
of the $\Box$ arguments will, in its turn, find itself
in this role while the other arguments will refer to the
points of internal integrations. Therefore, of
interest are the asymptotic behaviours of the form
factors $\Gamma_i$ in one (each) of the three
arguments $\Box_1,\Box_2,\Box_3$ with the two others fixed.
Here we present such asymptotic behaviours of
$\Gamma_i$ at $\Box_m \rightarrow-\infty$, and in the
next section at $\Box_m \rightarrow-0$ ($m=1,2,3$).

Although the total dimension of the form factors
$\Gamma_1$ to $\Gamma_{11}$ is $\Box^{-1}$, in individual
arguments they grow, generally, like ${\Box_m}^{+1}$ and
$\ln(-\Box_m)$, $\Box_m \rightarrow-\infty$. The growth is,
however, present only in the gravitational form factors;
the nongravitational ones tend to a constant at
$\Box_m \rightarrow-\infty$. The constant in the
asymptotic behaviour, as a function of two other
$\Box$'s, is either a tree or the function (7.31a).

In the asymptotic expressions below, O denotes
{\it decreasing} terms. The behaviour of the form factors
$\Gamma_1$ to $\Gamma_{11}$ is obtained with this
accuracy. The behaviour of the form factors
$\Gamma_{12}$ to $\Gamma_{29}$ is obtained with a
higher accuracy so that for the  redefined form
factors of the standard dimension the accuracy be O.
Such a redefinition is discussed in sect. 7, and the
form in which the results for
$\Gamma_{12}$ to $\Gamma_{29}$ are presented below
corresponds to the form of the exact expressions in
sect. 7.

The asymptotic expressions for the form factors (6.7)
at large $(-\Box_m)$ are as follows:

\begin{fleqnarray}\Gamma^{\rm sym}_{1}&=&{\rm O},\hspace{7mm}
\Box_1\rightarrow-\infty\ \ {\rm or}\ \ \Box_2\rightarrow-\infty\ \ {\rm or}\ \ \Box_3\rightarrow-\infty\end{fleqnarray}

\begin{fleqnarray}\Gamma^{\rm sym}_{2} &=&
\left\{\begin{tabular}{l}$
     {{\ln (\Box_2/\Box_3)}\over {9 ( \Box_2 - \Box_3 ) }}
+{\rm O},\hspace{7mm} \Box_1\rightarrow-\infty $\\[5mm]$
     {{\ln (\Box_1/\Box_3)}\over {9 ( \Box_1 - \Box_3 ) }}
+{\rm O},\hspace{7mm}  \Box_2\rightarrow-\infty $\\[5mm]$
     {{\ln (\Box_1/\Box_2)}\over {9 ( \Box_1 - \Box_2 ) }}
+{\rm O},\hspace{7mm}  \Box_3\rightarrow-\infty
$\end{tabular}\right.\end{fleqnarray}

\begin{fleqnarray}\Gamma^{\rm sym}_{3}&=&{\rm O},\hspace{7mm} \Box_1\rightarrow-\infty\ \ {\rm or}\ \
\Box_2\rightarrow-\infty\ \ {\rm or}\ \ \Box_3\rightarrow-\infty\end{fleqnarray}

\begin{fleqnarray}\Gamma^{\rm sym}_{4}&=&{\rm O},\hspace{7mm} \Box_1\rightarrow-\infty\ \ {\rm or}\ \
\Box_2\rightarrow-\infty\ \ {\rm or}\ \ \Box_3\rightarrow-\infty\end{fleqnarray}

\begin{fleqnarray}\Gamma^{\rm sym}_{5} &=&
\left\{\begin{tabular}{l}$
     - {{\ln (\Box_1/\Box_3)}\over {12 \Box_2}} + {1\over {8 \Box_2}}
+{\rm O},\hspace{7mm} \Box_1\rightarrow-\infty $\\[5mm]$
     - {{\ln (\Box_2/\Box_3)}\over {12 \Box_1}} + {1\over {8 \Box_1}}
+{\rm O},\hspace{7mm} \Box_2\rightarrow-\infty $\\[5mm]$
     {{-\Box_3}\over {24 \Box_1 \Box_2}} +
{1\over {8 \Box_1}} + {1\over {8 \Box_2}}
+{\rm O},\hspace{7mm} \Box_3\rightarrow-\infty
$\end{tabular}\right.\end{fleqnarray}

\begin{fleqnarray}\Gamma^{\rm sym}_{6}&=&{\rm O},\hspace{7mm} \Box_1\rightarrow-\infty\ \ {\rm or}\ \ \Box_2\rightarrow-\infty\ \
{\rm or}\ \ \Box_3\rightarrow-\infty\end{fleqnarray}

\begin{fleqnarray}\Gamma^{\rm sym}_{7}&=&{\rm O},\hspace{7mm} \Box_1\rightarrow-\infty\ \ {\rm or}\ \ \Box_2\rightarrow-\infty\ \
{\rm or}\ \ \Box_3\rightarrow-\infty\end{fleqnarray}

\begin{fleqnarray}\Gamma^{\rm sym}_{8}&=&
\left\{\begin{tabular}{l}$
{\rm O},\hspace{7mm} \Box_1\rightarrow-\infty $\\[5mm]$
     - {{\ln (\Box_2/\Box_3)}\over {6 \Box_1}} - {{1}\over {6 \Box_1}}
+{\rm O},\hspace{7mm}  \Box_2\rightarrow-\infty $\\[5mm]$
     - {{\ln (\Box_3/\Box_2)}\over {6 \Box_1}} - {{1}\over {6 \Box_1}}
+{\rm O},\hspace{7mm}  \Box_3\rightarrow-\infty
$\end{tabular}\right.\end{fleqnarray}

\begin{fleqnarray}\Gamma^{\rm sym}_{9} &=&
\left\{\begin{tabular}{l}$
     {{-\Box_1}\over {6480 \Box_2 \Box_3}} +
      {{\ln (\Box_2/\Box_3)}\over {2160 ( \Box_2 - \Box_3 ) }} $\\[3mm]$
      +{1 \over {1080 \Box_2}}
      +{1 \over {1080 \Box_3}}
+{\rm O},\hspace{7mm} \Box_1\rightarrow-\infty $\\[5mm]$
     {{-\Box_2}\over {6480 \Box_1 \Box_3}} +
      {{\ln (\Box_1/\Box_3)}\over {2160 ( \Box_1 - \Box_3 ) }} $\\[3mm]$
      +{1 \over {1080 \Box_1}}
      +{1 \over {1080 \Box_3}}
+{\rm O},\hspace{7mm} \Box_2\rightarrow-\infty $\\[5mm]$
     {{-\Box_3}\over {6480 \Box_1 \Box_2}} +
      {{\ln (\Box_1/\Box_2)}\over {2160 ( \Box_1 - \Box_2 ) }} $\\[3mm]$
      +{1 \over {1080 \Box_1}}
      +{1 \over {1080 \Box_2}}
+{\rm O},\hspace{7mm} \Box_3\rightarrow-\infty
$\end{tabular}\right.\end{fleqnarray}

\begin{fleqnarray}\Gamma^{\rm sym}_{10} &=&
\left\{\begin{tabular}{l}$
     {{-\Box_1}\over {1620 \Box_2 \Box_3}} +
      {1\over{810\Box_2}} + {1\over{810\Box_3}}
+{\rm O},\hspace{7mm} \Box_1\rightarrow-\infty $\\[5mm]$
     {{-\Box_2}\over {1620 \Box_1 \Box_3}} +
      {1\over{810\Box_1}} + {1\over{810\Box_3}}
+{\rm O},\hspace{7mm} \Box_2\rightarrow-\infty $\\[5mm]$
     {{-\Box_3}\over {1620 \Box_1 \Box_2}} +
      {1\over{810\Box_1}} + {1\over{810\Box_2}}
+{\rm O},\hspace{7mm} \Box_3\rightarrow-\infty
$\end{tabular}\right.\end{fleqnarray}

\begin{fleqnarray}\Gamma^{\rm sym}_{11} &=&
\left\{\begin{tabular}{l}$
       {{\ln (\Box_1/\Box_3)}\over {180 \Box_2}}
       + {{19}\over {2160 \Box_2}}
+{\rm O},\hspace{7mm} \Box_1\rightarrow-\infty $\\[5mm]$
       {{\ln (\Box_2/\Box_3)}\over {180 \Box_1}}
      + {{19}\over {2160 \Box_1}}
+{\rm O},\hspace{7mm} \Box_2\rightarrow-\infty $\\[5mm]$
      {{-\Box_3}\over {540 \Box_1 \Box_2}}
      - {{\ln (\Box_3/\Box_1)}\over {120 \Box_2}}
      - {{\ln (\Box_3/\Box_2)}\over {120 \Box_1}} $\\[3mm]$
      + {{\ln (\Box_1/\Box_2)}\over {120 ( \Box_1 - \Box_2 ) }}
      + {1\over {2160 \Box_1}} + {1\over {2160 \Box_2}}
+{\rm O}, $\\[3mm]$ \Box_3\rightarrow-\infty
$\end{tabular}\right.\end{fleqnarray}

\begin{fleqnarray}\Gamma^{\rm sym}_{12}&=&\frac1{\Box_1}{\rm O},\hspace{7mm}
 \Box_1\rightarrow-\infty\ \ {\rm or}\ \ \Box_2\rightarrow-\infty\ \ {\rm or}\ \ \Box_3\rightarrow-\infty\end{fleqnarray}

\begin{fleqnarray}\Gamma^{\rm sym}_{13}&=&\frac1{\Box_1}\left(2\frac{\ln(\Box_2/\Box_3)}{(\Box_2-\Box_3)}+{\rm O}\right),\nonumber\\&&\ \ \ \ \ \ \ \ \mbox{}
 \Box_1\rightarrow-\infty\ \ {\rm or}\ \ \Box_2\rightarrow-\infty\ \ {\rm or}\ \ \Box_3\rightarrow-\infty\end{fleqnarray}

\begin{fleqnarray}\Gamma^{\rm sym}_{14}&=&\frac1{\Box_3}{\rm O},\hspace{7mm}
 \Box_1\rightarrow-\infty\ \ {\rm or}\ \ \Box_2\rightarrow-\infty\ \ {\rm or}\ \ \Box_3\rightarrow-\infty\end{fleqnarray}

\begin{fleqnarray}\Gamma^{\rm sym}_{15}&=&\frac1{\Box_1}{\rm O},\hspace{7mm}
 \Box_1\rightarrow-\infty\ \ {\rm or}\ \ \Box_2\rightarrow-\infty\ \ {\rm or}\ \ \Box_3\rightarrow-\infty\end{fleqnarray}

\begin{fleqnarray}\Gamma^{\rm sym}_{16}&=&\frac1{\Box_1}\left(\frac1{12\Box_2}+{\rm O}\right)
+(1\leftrightarrow2),\nonumber\\&&\ \ \ \ \ \ \ \ \mbox{}
 \Box_1\rightarrow-\infty\ \ {\rm or}\ \ \Box_2\rightarrow-\infty\ \ {\rm or}\ \ \Box_3\rightarrow-\infty\end{fleqnarray}

\begin{fleqnarray}\Gamma^{\rm sym}_{17}&=&\frac1{\Box_1}\left(\frac{\ln(\Box_2/\Box_3)}{(\Box_2-\Box_3)}+{\rm O}\right),\nonumber\\&&\ \ \ \ \ \ \ \ \mbox{}
\Box_1\rightarrow-\infty\ \ {\rm or}\ \ \Box_2\rightarrow-\infty\ \ {\rm or}\ \ \Box_3\rightarrow-\infty\end{fleqnarray}

\begin{fleqnarray}\Gamma^{\rm sym}_{18}&=&\frac1{\Box_1}{\rm O},\hspace{7mm}
 \Box_1\rightarrow-\infty\ \ {\rm or}\ \ \Box_2\rightarrow-\infty\ \ {\rm or}\ \ \Box_3\rightarrow-\infty\end{fleqnarray}

\begin{fleqnarray}\Gamma^{\rm sym}_{19}&=&\frac1{\Box_1}\left(\frac16\frac{\ln(\Box_2/\Box_3)}{(\Box_2-\Box_3)}+{\rm O}\right),
\nonumber\\&&\ \ \ \ \ \ \ \ \mbox{}
 \Box_1\rightarrow-\infty\ \ {\rm or}\ \ \Box_2\rightarrow-\infty\ \ {\rm or}\ \ \Box_3\rightarrow-\infty\end{fleqnarray}

\begin{fleqnarray}\Gamma^{\rm sym}_{20}&=&\frac1{\Box_1}\left(-\frac16\frac{\ln(\Box_2/\Box_3)}{(\Box_2-\Box_3)}+{\rm O}\right),
\nonumber\\&&\ \ \ \ \ \ \ \ \mbox{}
 \Box_1\rightarrow-\infty\ \ {\rm or}\ \ \Box_2\rightarrow-\infty\ \ {\rm or}\ \ \Box_3\rightarrow-\infty\end{fleqnarray}

\begin{fleqnarray}\Gamma^{\rm sym}_{21}&=&\frac1{\Box_1}\left(-\frac23\frac{\ln(\Box_2/\Box_3)}{(\Box_2-\Box_3)}+{\rm O}\right),
\nonumber\\&&\ \ \ \ \ \ \ \ \mbox{}
 \Box_1\rightarrow-\infty\ \ {\rm or}\ \ \Box_2\rightarrow-\infty\ \ {\rm or}\ \ \Box_3\rightarrow-\infty\end{fleqnarray}

\begin{fleqnarray}\Gamma^{\rm sym}_{22}&=&
  \frac1{\Box_1}\left(\frac1{90}\frac{\ln(\Box_2/\Box_3)}{(\Box_2-\Box_3)}+{\rm O}\right)  \nonumber\\&&\mbox{}
  +\left[\frac1{\Box_2}\left(\frac1{60}\frac{\ln(\Box_1/\Box_3)}{(\Box_1-\Box_3)}
  -\frac1{540\Box_1}
  +{\rm O}\right)+(2\leftrightarrow3)\right]
 ,\nonumber\\&&\ \ \ \ \ \ \ \ \mbox{}
 \Box_1\rightarrow-\infty\ \ {\rm or}\ \ \Box_2\rightarrow-\infty\ \ {\rm or}\ \ \Box_3\rightarrow-\infty\end{fleqnarray}

\begin{fleqnarray}\Gamma^{\rm sym}_{23}&=&\frac1{\Box_1}
\left(\frac1{270\Box_2}+{\rm O}\right)+(1\leftrightarrow2),\nonumber\\&&\ \ \ \ \ \ \ \ \mbox{}
\Box_1\rightarrow-\infty\ \ {\rm or}\ \ \Box_2\rightarrow-\infty\ \ {\rm or}\ \ \Box_3\rightarrow-\infty\end{fleqnarray}

\begin{fleqnarray}\Gamma^{\rm sym}_{24}&=&\frac1{\Box_2}
\left(\frac1{1080\Box_3}+{\rm O}\right)+(2\leftrightarrow3),\nonumber\\&&\ \ \ \ \ \ \ \ \mbox{}
 \Box_1\rightarrow-\infty\ \ {\rm or}\ \ \Box_2\rightarrow-\infty\ \ {\rm or}\ \ \Box_3\rightarrow-\infty\end{fleqnarray}

\begin{fleqnarray}\Gamma^{\rm sym}_{25}&=&
 \frac1{270\Box_2\Box_3}
+\frac1{\Box_1}\left(
 -\frac1{270\Box_2}
 -\frac1{270\Box_3}
 +{\rm O}\right),\nonumber\\&&\ \ \ \ \ \ \ \ \mbox{}
 \Box_1\rightarrow-\infty\ \ {\rm or}\ \ \Box_2\rightarrow-\infty\ \ {\rm or}\ \ \Box_3\rightarrow-\infty\end{fleqnarray}

\begin{fleqnarray}\Gamma^{\rm sym}_{26}&=&\frac1{\Box_1\Box_2}{\rm O},\hspace{7mm}
 \Box_1\rightarrow-\infty\ \ {\rm or}\ \ \Box_2\rightarrow-\infty\ \ {\rm or}\ \ \Box_3\rightarrow-\infty\end{fleqnarray}

\begin{fleqnarray}\Gamma^{\rm sym}_{27}&=&\frac1{\Box_1\Box_2}\left(
-\frac1{540\Box_3}+
{\rm O}\right),\nonumber\\&&\ \ \ \ \ \ \ \ \mbox{}
 \Box_1\rightarrow-\infty\ \ {\rm or}\ \ \Box_2\rightarrow-\infty\ \ {\rm or}\ \ \Box_3\rightarrow-\infty\end{fleqnarray}

\begin{fleqnarray}\Gamma^{\rm sym}_{28}&=&\frac1{\Box_1\Box_2}\left(
\frac1{135\Box_3}+{\rm O}\right),\nonumber\\&&\ \ \ \ \ \ \ \ \mbox{}
 \Box_1\rightarrow-\infty\ \ {\rm or}\ \ \Box_2\rightarrow-\infty\ \ {\rm or}\ \ \Box_3\rightarrow-\infty\end{fleqnarray}

\begin{fleqnarray}\Gamma^{\rm sym}_{29}&=&\frac1{\Box_1\Box_2\Box_3}{\rm O},\nonumber\\&&\ \ \ \ \ \ \ \ \mbox{}
 \Box_1\rightarrow-\infty\ \ {\rm or}\ \ \Box_2\rightarrow-\infty\ \ {\rm or}\ \ \Box_3\rightarrow-\infty.\end{fleqnarray}

For the derivation of these results see sect. 19.
The nondecreasing terms in the asymptotic expressions
above are in complete agreement with the ones
appearing in the Laplace representation of the
form factors (sect. 8).

\section{The small-$\Box$ asymptotic behaviours
of the third-order form factors}
\setcounter{equation}{0}

\hspace{\parindent}
The small-$\Box$ asymptotic behaviours
of the form factors will be the first among
the present results used in the study of the
gravitational expectation-value equations [4].
In an asymptotically flat space-time, they should
determine the behaviour of the vacuum current
at spatial and null infinities. The rate of the
energy radiation by the gravitational collapse,
as a nonlocal functional of the curvature [17],
should, in the first place, follow from these results.
Furthermore, to lowest order in  the curvature this
calculation has been carried out for the spherically
symmetric state, and it was shown that the Hawking
stable component of the radiation is contained in
the third-order form factors (see [5]).

These applications determine also the accuracy with
which the asymptotic behaviours of the form
factors should be calculated. Terms ${\rm O}\left(1\right)$
at $\Box\rightarrow -0$ give already vanishing
contributions at the asymptotically flat infinity.
In fact, the gravitational form factors behave,
generally, like $1/\Box_m$ and
$\ln(-\Box_m)$ at $\Box_m\rightarrow-0$.
The coefficients of these behaviours are functions
of two other $\Box$'s, for which we introduce
the following notations. In the form factor
$\Gamma_i$, the coefficient of the
$1/\Box_m$ behaviour will be denoted $A_i(m)$,
and  the coefficient of the $\ln(-\Box_m)$
behaviour $B_i(m)$. These functions
are of the following general form ($m=1,2,3$):
\begin{eqnarray} A_i(1)&=&a_i(1)\left.\frac{\ln(j_2\Box_2/j_3\Box_3)}{(j_2\Box_2-j_3\Box_3)}\right|_{j=1},\end{eqnarray}
\begin{eqnarray} A_i(2)&=&a_i(2)\left.\frac{\ln(j_1\Box_1/j_3\Box_3)}{(j_1\Box_1-j_3\Box_3)}\right|_{j=1},\end{eqnarray}
\begin{eqnarray} A_i(3)&=&a_i(3)\left.\frac{\ln(j_1\Box_1/j_2\Box_2)}{(j_1\Box_1-j_2\Box_2)}\right|_{j=1},\end{eqnarray}
\begin{eqnarray} B_i(1)&=&b_i(1)\left.\frac{\ln(j_2\Box_2/j_3\Box_3)}{(j_2\Box_2-j_3\Box_3)}\right|_{j=1},\end{eqnarray}
\begin{eqnarray} B_i(2)&=&b_i(2)\left.\frac{\ln(j_1\Box_1/j_3\Box_3)}{(j_1\Box_1-j_3\Box_3)}\right|_{j=1},\end{eqnarray}
\begin{eqnarray} B_i(3)&=&b_i(3)\left.\frac{\ln(j_1\Box_1/j_2\Box_2)}{(j_1\Box_1-j_2\Box_2)}\right|_{j=1}\end{eqnarray}
where $a_i(m)$ and $b_i(m)$ are
polynomials in $\partial/\partial j$.
The asymptotic expressions below are supplied
with the forms of these polynomials. Some of
the functions $A_i(m)$ or $B_i(m)$ vanish,
and in addition to these functions, the coefficients
of the leading asymptotic behaviours have tree
contributions.

With ${\rm O}$ denoting terms ${\rm O}\left(1\right)$, the
asymptotic expressions for the form factors
(6.7) at small $\Box_m$ are as follows:

\begin{fleqnarray}&&\Gamma^{\rm sym}_{1} =
\left\{\begin{tabular}{l}$
\ln(-\Box_1)
B_{1}(1)
 +
{\rm O},\hspace{7mm} \Box_1\rightarrow-0 $\\[5mm]$
\ln(-\Box_2)
B_{1}(2)
 +
{\rm O},\hspace{7mm} \Box_2\rightarrow-0 $\\[5mm]$
\ln(-\Box_3)
B_{1}(3)
 +
{\rm O},\hspace{7mm} \Box_3\rightarrow-0
$\end{tabular}\right.\end{fleqnarray}

\begin{fleqnarray}&& b_{1}(1) =
   {1\over 3}
,\end{fleqnarray}

\begin{fleqnarray}&& b_{1}(2) =
   {1\over 3}
,\end{fleqnarray}

\begin{fleqnarray}&& b_{1}(3) =
   {1\over 3}
,\end{fleqnarray}

\begin{fleqnarray}&&\Gamma^{\rm sym}_{2} =
\left\{\begin{tabular}{l}$
\ln(-\Box_1)\Big(
-{{1}\over {9 \Box_2}} - {1\over {9 \Box_3}}
\Big) +
{\rm O},\hspace{7mm} \Box_1\rightarrow-0 $\\[5mm]$
\ln(-\Box_2)\Big(
-{{1}\over {9 \Box_1}} - {1\over {9 \Box_3}}
\Big) +
{\rm O},\hspace{7mm} \Box_2\rightarrow-0 $\\[5mm]$
\ln(-\Box_3)\Big(
-{{1}\over {9 \Box_1}} - {1\over {9 \Box_2}}
\Big) +
{\rm O},\hspace{7mm} \Box_3\rightarrow-0
$\end{tabular}\right.\end{fleqnarray}

\begin{fleqnarray}&&\Gamma^{\rm sym}_{3} =
\left\{\begin{tabular}{l}$
{\rm O},\hspace{7mm} \Box_1\rightarrow-0\ \ {\rm or}\ \ \Box_2\rightarrow-0 $\\[5mm]$
\ln(-\Box_3)
B_{3}(3)
 +
{\rm O},\hspace{7mm} \Box_3\rightarrow-0
$\end{tabular}\right.\end{fleqnarray}

\begin{fleqnarray}&& b_{3}(3) =
    \frac{\partial}{\partial j_1}\frac{\partial}{\partial j_2}
,\end{fleqnarray}

\begin{fleqnarray}&&\Gamma^{\rm sym}_{4} =
\left\{\begin{tabular}{l}$
\ln(-\Box_1)
B_{4}(1)
 +
{\rm O},\hspace{7mm} \Box_1\rightarrow-0 $\\[5mm]$
\ln(-\Box_2)
B_{4}(2)
 +
{\rm O},\hspace{7mm} \Box_2\rightarrow-0 $\\[5mm]$
\ln(-\Box_3)
B_{4}(3)
 +
{\rm O},\hspace{7mm} \Box_3\rightarrow-0
$\end{tabular}\right.\end{fleqnarray}

\begin{fleqnarray}&& b_{4}(1) =
  -{1\over {36}}
    \frac{\partial^2}{\partial j_2^2}
  -{1\over {36}}
    \frac{\partial}{\partial j_3}
  -{1\over 9}
    \frac{\partial}{\partial j_2}\frac{\partial}{\partial j_3}
  -{1\over {12}}
    \frac{\partial^2}{\partial j_2^2}\frac{\partial}{\partial j_3}
,\end{fleqnarray}

\begin{fleqnarray}&& b_{4}(2) =
  -{1\over {36}}
    \frac{\partial^2}{\partial j_1^2}
  -{1\over {36}}
    \frac{\partial}{\partial j_3}
  -{1\over 9}
    \frac{\partial}{\partial j_1}\frac{\partial}{\partial j_3}
  -{1\over {12}}
    \frac{\partial^2}{\partial j_1^2}\frac{\partial}{\partial j_3}
,\end{fleqnarray}

\begin{fleqnarray}&& b_{4}(3) =
  -{1\over {36}}
    \frac{\partial^2}{\partial j_1^2}
  -{7\over {108}}
    \frac{\partial^2}{\partial j_1^2}\frac{\partial}{\partial j_2}
  -{1\over {36}}
    \frac{\partial^2}{\partial j_2^2}       \nonumber\\&&\ \ \ \ \ \ \ \ \mbox{}
  -{7\over {108}}
    \frac{\partial}{\partial j_1}\frac{\partial^2}{\partial j_2^2}
  -{1\over {24}}
    \frac{\partial^2}{\partial j_1^2}\frac{\partial^2}{\partial j_2^2}
,\end{fleqnarray}

\begin{fleqnarray}&&\Gamma^{\rm sym}_{5} =
\left\{\begin{tabular}{l}$
\frac1{\Box_1}\Big(
{1\over 8} - {{\Box_3}\over {24 \Box_2}} +
A_{5}(1)
\Big) +
{\rm O},\hspace{7mm} \Box_1\rightarrow-0 $\\[5mm]$
\frac1{\Box_2}\Big(
{1\over 8} - {{\Box_3}\over {24 \Box_1}} +
A_{5}(2)
\Big) +
{\rm O},\hspace{7mm} \Box_2\rightarrow-0 $\\[5mm]$
\ln(-\Box_3)\Big(
{1\over {12 \Box_1}} + {1\over {12 \Box_2}}
\Big) +
{\rm O},\hspace{7mm} \Box_3\rightarrow-0
$\end{tabular}\right.\end{fleqnarray}

\begin{fleqnarray}&& a_{5}(1) =
-\frac1{12}\Box_2
,\end{fleqnarray}

\begin{fleqnarray}&& a_{5}(2) =
-\frac1{12}\Box_1
,\end{fleqnarray}

\begin{fleqnarray}&&\Gamma^{\rm sym}_{6} =
\left\{\begin{tabular}{l}$
\ln(-\Box_1)
B_{6}(1)
 +
{\rm O},\hspace{7mm} \Box_1\rightarrow-0 $\\[5mm]$
\ln(-\Box_2)
B_{6}(2)
 +
{\rm O},\hspace{7mm} \Box_2\rightarrow-0 $\\[5mm]$
\ln(-\Box_3)
B_{6}(3)
 +
{\rm O},\hspace{7mm} \Box_3\rightarrow-0
$\end{tabular}\right.\end{fleqnarray}

\begin{fleqnarray}&& b_{6}(1) =
  -{1\over 6}
  -{1\over 2}
    \frac{\partial}{\partial j_2}
  -{1\over 4}
    \frac{\partial^2}{\partial j_2^2}
,\end{fleqnarray}

\begin{fleqnarray}&& b_{6}(2) =
  -{1\over 6}
  -{1\over 2}
    \frac{\partial}{\partial j_1}
  -{1\over 4}
    \frac{\partial^2}{\partial j_1^2}
,\end{fleqnarray}

\begin{fleqnarray}&& b_{6}(3) =
  -{1\over 6}
  -{1\over 2}
    \frac{\partial}{\partial j_1}
  -{1\over 4}
    \frac{\partial^2}{\partial j_1^2}
  -{1\over 2}
    \frac{\partial}{\partial j_2}
  -{1\over 4}
    \frac{\partial^2}{\partial j_2^2}
,\end{fleqnarray}

\begin{fleqnarray}&&\Gamma^{\rm sym}_{7} =
\left\{\begin{tabular}{l}$
\ln(-\Box_1)
B_{7}(1)
 +
{\rm O},\hspace{7mm} \Box_1\rightarrow-0 $\\[5mm]$
{\rm O},\hspace{7mm} \Box_2\rightarrow-0\ \ {\rm or}\ \ \Box_3\rightarrow-0
$\end{tabular}\right.\end{fleqnarray}

\begin{fleqnarray}&& b_{7}(1) =
   {1\over {24}}
    \frac{\partial}{\partial j_2}
  +{1\over {96}}
    \frac{\partial^2}{\partial j_2^2}
  +{1\over {24}}
    \frac{\partial}{\partial j_3}
  +{{19}\over {48}}
    \frac{\partial}{\partial j_2}\frac{\partial}{\partial j_3}  \nonumber\\&&\ \ \ \ \ \ \ \ \mbox{}
  -{1\over {24}}
    \frac{\partial^3}{\partial j_2^3}\frac{\partial}{\partial j_3}
  +{1\over {96}}
    \frac{\partial^2}{\partial j_3^2}
  -{1\over {24}}
    \frac{\partial}{\partial j_2}\frac{\partial^3}{\partial j_3^3}  \nonumber\\&&\ \ \ \ \mbox{}
+{{\Box_3}\over {\Box_2}}\left(
   {1\over {96}}
    \frac{\partial^2}{\partial j_2^2}
  -{1\over {96}}
    \frac{\partial}{\partial j_2}\frac{\partial}{\partial j_3}
\right)              \nonumber\\&&\ \ \ \ \mbox{}
+{{\Box_2}\over {\Box_3}}\left(
  -{1\over {96}}
    \frac{\partial}{\partial j_2}\frac{\partial}{\partial j_3}
  +{1\over {96}}
    \frac{\partial^2}{\partial j_3^2}
\right)
,\end{fleqnarray}

\begin{fleqnarray}&&\Gamma^{\rm sym}_{8} =
\left\{\begin{tabular}{l}$
\frac1{\Box_1}
A_{8}(1)
 +
{\rm O},\hspace{7mm} \Box_1\rightarrow-0 $\\[5mm]$
\ln(-\Box_2)\Big(
{1\over {6 \Box_1}}
\Big) +
{\rm O},\hspace{7mm} \Box_2\rightarrow-0 $\\[5mm]$
\ln(-\Box_3)\Big(
{1\over {6 \Box_1}}
\Big) +
{\rm O},\hspace{7mm} \Box_3\rightarrow-0
$\end{tabular}\right.\end{fleqnarray}

\begin{fleqnarray}&& a_{8}(1) =
\Box_3\left(
   {1\over {12}}
  +{1\over 4}
    \frac{\partial}{\partial j_2}
  +{1\over 4}
    \frac{\partial}{\partial j_3}
  -{1\over 6}
    \frac{\partial}{\partial j_2}\frac{\partial}{\partial j_3}
\right)        \nonumber\\&&\ \ \ \ \ \ \ \ \mbox{}
+\Box_2\left(
   {1\over {12}}
  +{1\over 4}
    \frac{\partial}{\partial j_2}
  +{1\over 4}
    \frac{\partial}{\partial j_3}
  -{1\over 6}
    \frac{\partial}{\partial j_2}\frac{\partial}{\partial j_3}
\right)
,\end{fleqnarray}

\begin{fleqnarray}&&\Gamma^{\rm sym}_{9} =
\left\{\begin{tabular}{l}$
\frac1{\Box_1}\Big(
-{{\Box_2}\over {6480 \Box_3}} - {{\Box_3}\over {6480 \Box_2}} +
A_{9}(1)
\Big) +
\ln(-\Box_1)\Big(
{1\over {1080 \Box_2}}$\\[3mm]$\ \ \ \  + {1\over {1080 \Box_3}} +
B_{9}(1)
\Big) +
{\rm O},\hspace{7mm} \Box_1\rightarrow-0 $\\[5mm]$
\frac1{\Box_2}\Big(
-{{\Box_1}\over {6480 \Box_3}} - {{\Box_3}\over {6480 \Box_1}} +
A_{9}(2)
\Big) +
\ln(-\Box_2)\Big(
{1\over {1080 \Box_1}}$\\[3mm]$\ \ \ \  + {1\over {1080 \Box_3}} +
B_{9}(2)
\Big) +
{\rm O},\hspace{7mm} \Box_2\rightarrow-0 $\\[5mm]$
\frac1{\Box_3}\Big(
-{{\Box_1}\over {6480 \Box_2}} - {{\Box_2}\over {6480 \Box_1}} +
A_{9}(3)
\Big) +
\ln(-\Box_3)\Big(
{1\over {1080 \Box_1}}$\\[3mm]$\ \ \ \  + {1\over {1080 \Box_2}} +
B_{9}(3)
\Big) +
{\rm O},\hspace{7mm} \Box_3\rightarrow-0
$\end{tabular}\right.\end{fleqnarray}

\begin{fleqnarray}&& a_{9}(1) =
\Box_3\left(
   {1\over {1296}}
    \frac{\partial^3}{\partial j_2^3}
  -{1\over {324}}
    \frac{\partial^2}{\partial j_2^2}\frac{\partial}{\partial j_3}
  +{1\over {3240}}
    \frac{\partial}{\partial j_2}\frac{\partial^2}{\partial j_3^2}
\right)         \nonumber\\&&\ \ \ \ \ \ \ \ \mbox{}
+\Box_2\left(
   {1\over {3240}}
    \frac{\partial^2}{\partial j_2^2}\frac{\partial}{\partial j_3}
  -{1\over {324}}
    \frac{\partial}{\partial j_2}\frac{\partial^2}{\partial j_3^2}
  +{1\over {1296}}
    \frac{\partial^3}{\partial j_3^3}
\right)
,\end{fleqnarray}

\begin{fleqnarray}&& b_{9}(1) =
  -{{49}\over {4665600}}
    \frac{\partial^6}{\partial j_2^6}
  +{1\over {25920}}
    \frac{\partial^5}{\partial j_2^5}\frac{\partial}{\partial j_3}
  +{1\over {20736}}
    \frac{\partial^4}{\partial j_2^4}\frac{\partial^2}{\partial j_3^2}  \nonumber\\&&\ \ \ \ \ \ \ \ \mbox{}
  -{{17}\over {72900}}
    \frac{\partial^3}{\partial j_2^3}\frac{\partial^3}{\partial j_3^3}
  +{1\over {20736}}
    \frac{\partial^2}{\partial j_2^2}\frac{\partial^4}{\partial j_3^4}
  +{1\over {25920}}
    \frac{\partial}{\partial j_2}\frac{\partial^5}{\partial j_3^5}      \nonumber\\&&\ \ \ \ \ \ \ \ \mbox{}
  -{{49}\over {4665600}}
    \frac{\partial^6}{\partial j_3^6}             \nonumber\\&&\ \ \ \ \mbox{}
+{{\Box_3}\over {\Box_2}}\left(
  -{{13}\over {1555200}}
    \frac{\partial^6}{\partial j_2^6}
  +{{29}\over {518400}}
    \frac{\partial^5}{\partial j_2^5}\frac{\partial}{\partial j_3}
  -{{17}\over {388800}}
    \frac{\partial^4}{\partial j_2^4}\frac{\partial^2}{\partial j_3^2}
\right.\nonumber\\&&\ \ \ \ \ \ \ \ \mbox{}\left.
  -{1\over {8100}}
    \frac{\partial^3}{\partial j_2^3}\frac{\partial^3}{\partial j_3^3}
  +{{133}\over {1555200}}
    \frac{\partial^2}{\partial j_2^2}\frac{\partial^4}{\partial j_3^4}
  -{{23}\over {1555200}}
    \frac{\partial}{\partial j_2}\frac{\partial^5}{\partial j_3^5}
\right)                    \nonumber\\&&\ \ \ \ \mbox{}
+{{\Box_2}\over {\Box_3}}\left(
  -{{23}\over {1555200}}
    \frac{\partial^5}{\partial j_2^5}\frac{\partial}{\partial j_3}
  +{{133}\over {1555200}}
    \frac{\partial^4}{\partial j_2^4}\frac{\partial^2}{\partial j_3^2}
  -{1\over {8100}}
    \frac{\partial^3}{\partial j_2^3}\frac{\partial^3}{\partial j_3^3}
\right. \nonumber\\&&\ \ \ \ \ \ \ \ \mbox{}\left.
  -{{17}\over {388800}}
    \frac{\partial^2}{\partial j_2^2}\frac{\partial^4}{\partial j_3^4}
  +{{29}\over {518400}}
    \frac{\partial}{\partial j_2}\frac{\partial^5}{\partial j_3^5}
  -{{13}\over {1555200}}
    \frac{\partial^6}{\partial j_3^6}
\right)
,\end{fleqnarray}

\begin{fleqnarray}&& a_{9}(2) =
\Box_3\left(
   {1\over {1296}}
    \frac{\partial^3}{\partial j_1^3}
  -{1\over {324}}
    \frac{\partial^2}{\partial j_1^2}\frac{\partial}{\partial j_3}
  +{1\over {3240}}
    \frac{\partial}{\partial j_1}\frac{\partial^2}{\partial j_3^2}
\right)         \nonumber\\&&\ \ \ \ \ \ \ \ \mbox{}
+\Box_1\left(
   {1\over {3240}}
    \frac{\partial^2}{\partial j_1^2}\frac{\partial}{\partial j_3}
  -{1\over {324}}
    \frac{\partial}{\partial j_1}\frac{\partial^2}{\partial j_3^2}
  +{1\over {1296}}
    \frac{\partial^3}{\partial j_3^3}
\right)
,\end{fleqnarray}

\begin{fleqnarray}&& b_{9}(2) =
  -{{49}\over {4665600}}
    \frac{\partial^6}{\partial j_1^6}
  +{1\over {25920}}
    \frac{\partial^5}{\partial j_1^5}\frac{\partial}{\partial j_3}
  +{1\over {20736}}
    \frac{\partial^4}{\partial j_1^4}\frac{\partial^2}{\partial j_3^2}      \nonumber\\&&\ \ \ \ \ \ \ \ \mbox{}
  -{{17}\over {72900}}
    \frac{\partial^3}{\partial j_1^3}\frac{\partial^3}{\partial j_3^3}
  +{1\over {20736}}
    \frac{\partial^2}{\partial j_1^2}\frac{\partial^4}{\partial j_3^4}
  +{1\over {25920}}
    \frac{\partial}{\partial j_1}\frac{\partial^5}{\partial j_3^5}        \nonumber\\&&\ \ \ \ \ \ \ \ \mbox{}
  -{{49}\over {4665600}}
    \frac{\partial^6}{\partial j_3^6}               \nonumber\\&&\ \ \ \ \mbox{}
+{{\Box_3}\over {\Box_1}}\left(
  -{{13}\over {1555200}}
    \frac{\partial^6}{\partial j_1^6}
  +{{29}\over {518400}}
    \frac{\partial^5}{\partial j_1^5}\frac{\partial}{\partial j_3}
  -{{17}\over {388800}}
    \frac{\partial^4}{\partial j_1^4}\frac{\partial^2}{\partial j_3^2}
\right.\nonumber\\&&\ \ \ \ \ \ \ \ \mbox{}\left.
  -{1\over {8100}}
    \frac{\partial^3}{\partial j_1^3}\frac{\partial^3}{\partial j_3^3}
  +{{133}\over {1555200}}
    \frac{\partial^2}{\partial j_1^2}\frac{\partial^4}{\partial j_3^4}
  -{{23}\over {1555200}}
    \frac{\partial}{\partial j_1}\frac{\partial^5}{\partial j_3^5}
\right)  \nonumber\\&&\ \ \ \ \mbox{}
+{{\Box_1}\over {\Box_3}}\left(
  -{{23}\over {1555200}}
    \frac{\partial^5}{\partial j_1^5}\frac{\partial}{\partial j_3}
  +{{133}\over {1555200}}
    \frac{\partial^4}{\partial j_1^4}\frac{\partial^2}{\partial j_3^2}
  -{1\over {8100}}
    \frac{\partial^3}{\partial j_1^3}\frac{\partial^3}{\partial j_3^3}
\right.
\nonumber\\&&\ \ \ \ \ \ \ \ \mbox{}
\left.
  -{{17}\over {388800}}
    \frac{\partial^2}{\partial j_1^2}\frac{\partial^4}{\partial j_3^4}
  +{{29}\over {518400}}
    \frac{\partial}{\partial j_1}\frac{\partial^5}{\partial j_3^5}
  -{{13}\over {1555200}}
    \frac{\partial^6}{\partial j_3^6}
\right)
,\end{fleqnarray}

\begin{fleqnarray}&& a_{9}(3) =
\Box_2\left(
   {1\over {1296}}
    \frac{\partial^3}{\partial j_1^3}
  -{1\over {324}}
    \frac{\partial^2}{\partial j_1^2}\frac{\partial}{\partial j_2}
  +{1\over {3240}}
    \frac{\partial}{\partial j_1}\frac{\partial^2}{\partial j_2^2}
\right)
\nonumber\\&&\ \ \ \ \ \ \ \ \mbox{}
+\Box_1\left(
   {1\over {3240}}
    \frac{\partial^2}{\partial j_1^2}\frac{\partial}{\partial j_2}
  -{1\over {324}}
    \frac{\partial}{\partial j_1}\frac{\partial^2}{\partial j_2^2}
  +{1\over {1296}}
    \frac{\partial^3}{\partial j_2^3}
\right)
,\end{fleqnarray}

\begin{fleqnarray}&& b_{9}(3) =
  -{{49}\over {4665600}}
    \frac{\partial^6}{\partial j_1^6}
  +{1\over {25920}}
    \frac{\partial^5}{\partial j_1^5}\frac{\partial}{\partial j_2}
  +{1\over {20736}}
    \frac{\partial^4}{\partial j_1^4}\frac{\partial^2}{\partial j_2^2}   \nonumber\\&&\ \ \ \ \ \ \ \ \mbox{}
  -{{17}\over {72900}}
    \frac{\partial^3}{\partial j_1^3}\frac{\partial^3}{\partial j_2^3}
  +{1\over {20736}}
    \frac{\partial^2}{\partial j_1^2}\frac{\partial^4}{\partial j_2^4}
  +{1\over {25920}}
    \frac{\partial}{\partial j_1}\frac{\partial^5}{\partial j_2^5}       \nonumber\\&&\ \ \ \ \ \ \ \ \mbox{}
  -{{49}\over {4665600}}
    \frac{\partial^6}{\partial j_2^6} \nonumber\\&&\ \ \ \ \mbox{}
+{{\Box_2}\over {\Box_1}}\left(
  -{{13}\over {1555200}}
    \frac{\partial^6}{\partial j_1^6}
  +{{29}\over {518400}}
    \frac{\partial^5}{\partial j_1^5}\frac{\partial}{\partial j_2}
  -{{17}\over {388800}}
    \frac{\partial^4}{\partial j_1^4}\frac{\partial^2}{\partial j_2^2}
\right.\nonumber\\&&\ \ \ \ \ \ \ \ \mbox{}
\left.
  -{1\over {8100}}
    \frac{\partial^3}{\partial j_1^3}\frac{\partial^3}{\partial j_2^3}
  +{{133}\over {1555200}}
    \frac{\partial^2}{\partial j_1^2}\frac{\partial^4}{\partial j_2^4}
  -{{23}\over {1555200}}
    \frac{\partial}{\partial j_1}\frac{\partial^5}{\partial j_2^5}
\right)        \nonumber\\&&\ \ \ \ \mbox{}
+{{\Box_1}\over {\Box_2}}\left(
  -{{23}\over {1555200}}
    \frac{\partial^5}{\partial j_1^5}\frac{\partial}{\partial j_2}
  +{{133}\over {1555200}}
    \frac{\partial^4}{\partial j_1^4}\frac{\partial^2}{\partial j_2^2}
  -{1\over {8100}}
    \frac{\partial^3}{\partial j_1^3}\frac{\partial^3}{\partial j_2^3}
\right.\nonumber\\&&\ \ \ \ \ \ \ \ \mbox{}\left.
  -{{17}\over {388800}}
    \frac{\partial^2}{\partial j_1^2}\frac{\partial^4}{\partial j_2^4}
  +{{29}\over {518400}}
    \frac{\partial}{\partial j_1}\frac{\partial^5}{\partial j_2^5}
  -{{13}\over {1555200}}
    \frac{\partial^6}{\partial j_2^6}
\right)
,\end{fleqnarray}

\begin{fleqnarray}&&\Gamma^{\rm sym}_{10} =
\left\{\begin{tabular}{l}$
\frac1{\Box_1}\Big(
{1\over {810}} - {{\Box_2}\over {1620 \Box_3}}
- {{\Box_3}\over {1620 \Box_2}}
\Big) +
{\rm O},\hspace{7mm} \Box_1\rightarrow-0 $\\[5mm]$
\frac1{\Box_2}\Big(
{1\over {810}} - {{\Box_1}\over {1620 \Box_3}}
- {{\Box_3}\over {1620 \Box_1}}
\Big) +
{\rm O},\hspace{7mm} \Box_2\rightarrow-0 $\\[5mm]$
\frac1{\Box_3}\Big(
{1\over {810}} - {{\Box_1}\over {1620 \Box_2}}
- {{\Box_2}\over {1620 \Box_1}}
\Big) +
{\rm O},\hspace{7mm} \Box_3\rightarrow-0
$\end{tabular}\right.\end{fleqnarray}

\begin{fleqnarray}&&\Gamma^{\rm sym}_{11} =
\left\{\begin{tabular}{l}$
\frac1{\Box_1}\Big(
{{-\Box_3}\over {540 \Box_2}} +
A_{11}(1)
\Big) +
{\rm O},\hspace{7mm} \Box_1\rightarrow-0 $\\[5mm]$
\frac1{\Box_2}\Big(
{{-\Box_3}\over {540 \Box_1}} +
A_{11}(2)
\Big) +
{\rm O},\hspace{7mm} \Box_2\rightarrow-0 $\\[5mm]$
\ln(-\Box_3)\Big(
-{{1}\over {180 \Box_1}} - {1\over {180 \Box_2}}
\Big) +
{\rm O},\hspace{7mm} \Box_3\rightarrow-0
$\end{tabular}\right.\end{fleqnarray}

\begin{fleqnarray}&& a_{11}(1) =
\Box_3\left(
  -{{17}\over {1080}}
  -{{17}\over {2160}}
    \frac{\partial}{\partial j_2}
  +{{11}\over {2160}}
    \frac{\partial^2}{\partial j_2^2}
  +{1\over {720}}
    \frac{\partial^3}{\partial j_2^3}
\right.\nonumber\\&&\ \ \ \ \ \ \ \ \mbox{}
  -{1\over {180}}
    \frac{\partial}{\partial j_3}
  +{1\over {540}}
    \frac{\partial}{\partial j_2}\frac{\partial}{\partial j_3}
  -{1\over {2160}}
    \frac{\partial^2}{\partial j_2^2}\frac{\partial}{\partial j_3}
  +{1\over {1080}}
    \frac{\partial^2}{\partial j_3^2}
\nonumber\\&&\ \ \ \ \ \ \ \ \mbox{}
\left.
  +{1\over {360}}
    \frac{\partial}{\partial j_2}\frac{\partial^2}{\partial j_3^2}
  +{1\over {1080}}
    \frac{\partial^2}{\partial j_2^2}\frac{\partial^2}{\partial j_3^2}
\right)                 \nonumber\\&&\ \ \ \ \mbox{}
+\Box_2\left(
   {1\over {360}}
  -{1\over {360}}
    \frac{\partial}{\partial j_2}
  -{1\over {2160}}
    \frac{\partial}{\partial j_3}
  +{1\over {270}}
    \frac{\partial}{\partial j_2}\frac{\partial}{\partial j_3}
  -{1\over {1080}}
    \frac{\partial^2}{\partial j_2^2}\frac{\partial}{\partial j_3}
\right.\nonumber\\&&\ \ \ \ \ \ \ \ \mbox{}\left.
  -{1\over {540}}
    \frac{\partial^2}{\partial j_3^2}
  -{1\over {108}}
    \frac{\partial}{\partial j_2}\frac{\partial^2}{\partial j_3^2}
  -{1\over {1080}}
    \frac{\partial^2}{\partial j_2^2}\frac{\partial^2}{\partial j_3^2}
\right)
,\end{fleqnarray}

\begin{fleqnarray}&& a_{11}(2) =
\Box_3\left(
  -{{17}\over {1080}}
  -{{17}\over {2160}}
    \frac{\partial}{\partial j_1}
  +{{11}\over {2160}}
    \frac{\partial^2}{\partial j_1^2}
  +{1\over {720}}
    \frac{\partial^3}{\partial j_1^3}
\right.\nonumber\\&&\ \ \ \ \ \ \ \ \mbox{}
  -{1\over {180}}
    \frac{\partial}{\partial j_3}
  +{1\over {540}}
    \frac{\partial}{\partial j_1}\frac{\partial}{\partial j_3}
  -{1\over {2160}}
    \frac{\partial^2}{\partial j_1^2}\frac{\partial}{\partial j_3}
  +{1\over {1080}}
    \frac{\partial^2}{\partial j_3^2}
\nonumber\\&&\ \ \ \ \ \ \ \ \mbox{}
\left.
  +{1\over {360}}
    \frac{\partial}{\partial j_1}\frac{\partial^2}{\partial j_3^2}
  +{1\over {1080}}
    \frac{\partial^2}{\partial j_1^2}\frac{\partial^2}{\partial j_3^2}
\right)                 \nonumber\\&&\ \ \ \ \mbox{}
+\Box_1\left(
   {1\over {360}}
  -{1\over {360}}
    \frac{\partial}{\partial j_1}
  -{1\over {2160}}
    \frac{\partial}{\partial j_3}
  +{1\over {270}}
    \frac{\partial}{\partial j_1}\frac{\partial}{\partial j_3}
  -{1\over {1080}}
    \frac{\partial^2}{\partial j_1^2}\frac{\partial}{\partial j_3}
\right.\nonumber\\&&\ \ \ \ \ \ \ \ \mbox{}
\left.
  -{1\over {540}}
    \frac{\partial^2}{\partial j_3^2}
  -{1\over {108}}
    \frac{\partial}{\partial j_1}\frac{\partial^2}{\partial j_3^2}
  -{1\over {1080}}
    \frac{\partial^2}{\partial j_1^2}\frac{\partial^2}{\partial j_3^2}
\right)
,\end{fleqnarray}

\begin{fleqnarray}&&\Gamma^{\rm sym}_{12} =
\left\{\begin{tabular}{l}$
\ln(-\Box_1)\Big(
{{-2}\over {3 \Box_2 \Box_3}} +
B_{12}(1)
\Big) +
{\rm O},\hspace{7mm} \Box_1\rightarrow-0 $\\[5mm]$
\ln(-\Box_2)\Big(
{{-1}\over {3 \Box_1 \Box_3}} +
B_{12}(2)
\Big) +
{\rm O},\hspace{7mm} \Box_2\rightarrow-0 $\\[5mm]$
\ln(-\Box_3)\Big(
{{-1}\over {3 \Box_1 \Box_2}} +
B_{12}(3)
\Big) +
{\rm O},\hspace{7mm} \Box_3\rightarrow-0
$\end{tabular}\right.\end{fleqnarray}

\begin{fleqnarray}&& b_{12}(1) =
{1\over {\Box_2}}\left(
  -{1\over 3}
    \frac{\partial}{\partial j_2}
\right)
+{1\over {\Box_3}}\left(
  -{1\over 3}
    \frac{\partial}{\partial j_3}
\right)
,\end{fleqnarray}

\begin{fleqnarray}&& b_{12}(2) =
{1\over {\Box_1}}\left(
   {5\over 4}
    \frac{\partial^2}{\partial j_1^2}
  +{2\over 3}
    \frac{\partial^3}{\partial j_1^3}
  +{1\over {12}}
    \frac{\partial}{\partial j_1}\frac{\partial}{\partial j_3}
\right)  \nonumber\\&&\ \ \ \ \ \ \ \ \mbox{}
+{1\over {\Box_3}}\left(
  -{1\over 3}
    \frac{\partial}{\partial j_3}
  +{1\over {12}}
    \frac{\partial}{\partial j_1}\frac{\partial}{\partial j_3}
  -{1\over {12}}
    \frac{\partial^2}{\partial j_3^2}
\right)
,\end{fleqnarray}

\begin{fleqnarray}&& b_{12}(3) =
{1\over {\Box_1}}\left(
   {5\over 4}
    \frac{\partial^2}{\partial j_1^2}
  +{2\over 3}
    \frac{\partial^3}{\partial j_1^3}
  +{1\over {12}}
    \frac{\partial}{\partial j_1}\frac{\partial}{\partial j_2}
\right)  \nonumber\\&&\ \ \ \ \ \ \ \ \mbox{}
+{1\over {\Box_2}}\left(
  -{1\over 3}
    \frac{\partial}{\partial j_2}
  +{1\over {12}}
    \frac{\partial}{\partial j_1}\frac{\partial}{\partial j_2}
  -{1\over {12}}
    \frac{\partial^2}{\partial j_2^2}
\right)
,\end{fleqnarray}

\begin{fleqnarray}&&\Gamma^{\rm sym}_{13} =
\left\{\begin{tabular}{l}$
\ln(-\Box_1)
B_{13}(1)
 +
{\rm O},\hspace{7mm} \Box_1\rightarrow-0 $\\[5mm]$
\ln(-\Box_2)\Big(
{{-2}\over {\Box_1 \Box_3}} +
B_{13}(2)
\Big) +
{\rm O},\hspace{7mm} \Box_2\rightarrow-0 $\\[5mm]$
\ln(-\Box_3)\Big(
{{-2}\over {\Box_1 \Box_2}} +
B_{13}(3)
\Big) +
{\rm O},\hspace{7mm} \Box_3\rightarrow-0
$\end{tabular}\right.\end{fleqnarray}

\begin{fleqnarray}&& b_{13}(1) =
{1\over {\Box_2}}\left(
  -{1\over 2}
    \frac{\partial^2}{\partial j_2^2}
  +{1\over 2}
    \frac{\partial}{\partial j_2}\frac{\partial}{\partial j_3}
\right)
+{1\over {\Box_3}}\!\left(
   {1\over 2}
    \frac{\partial}{\partial j_2}\frac{\partial}{\partial j_3}
  -{1\over 2}
    \frac{\partial^2}{\partial j_3^2}
\right)\!
,\end{fleqnarray}

\begin{fleqnarray}&& b_{13}(2) =
{1\over {\Box_1}}\left(
  -2
    \frac{\partial}{\partial j_1}
\right)
,\end{fleqnarray}

\begin{fleqnarray}&& b_{13}(3) =
{1\over {\Box_1}}\left(
  -2
    \frac{\partial}{\partial j_1}
\right)
,\end{fleqnarray}

\begin{fleqnarray}&&\Gamma^{\rm sym}_{14} =
\left\{\begin{tabular}{l}$
\ln(-\Box_1)
B_{14}(1)
 +
{\rm O},\hspace{7mm} \Box_1\rightarrow-0 $\\[5mm]$
\ln(-\Box_2)
B_{14}(2)
 +
{\rm O},\hspace{7mm} \Box_2\rightarrow-0 $\\[5mm]$
\ln(-\Box_3)
B_{14}(3)
 +
{\rm O},\hspace{7mm} \Box_3\rightarrow-0
$\end{tabular}\right.\end{fleqnarray}

\begin{fleqnarray}&& b_{14}(1) =
{1\over {\Box_3}}\left(
  -2
    \frac{\partial}{\partial j_3}
  -2
    \frac{\partial^2}{\partial j_3^2}
\right)
,\end{fleqnarray}

\begin{fleqnarray}&& b_{14}(2) =
{1\over {\Box_3}}\left(
  -2
    \frac{\partial}{\partial j_3}
  -2
    \frac{\partial^2}{\partial j_3^2}
\right)
,\end{fleqnarray}

\begin{fleqnarray}&& b_{14}(3) =
{1\over {\Box_1}}\!\left(
  -{1\over 2}
    \frac{\partial^2}{\partial j_1^2}
  +{1\over 2}
    \frac{\partial}{\partial j_1}\frac{\partial}{\partial j_2}
\right)
+{1\over {\Box_2}}\!\left(
   {1\over 2}
    \frac{\partial}{\partial j_1}\frac{\partial}{\partial j_2}
  -{1\over 2}
    \frac{\partial^2}{\partial j_2^2}
\right)\!
,\end{fleqnarray}

\begin{fleqnarray}&&\Gamma^{\rm sym}_{15} =
\left\{\begin{tabular}{l}$
{\rm O},\hspace{7mm} \Box_1\rightarrow-0 $\\[5mm]$
\ln(-\Box_2)
B_{15}(2)
 +
{\rm O},\hspace{7mm} \Box_2\rightarrow-0 $\\[5mm]$
\ln(-\Box_3)
B_{15}(3)
 +
{\rm O},\hspace{7mm} \Box_3\rightarrow-0
$\end{tabular}\right.\end{fleqnarray}

\begin{fleqnarray}&& b_{15}(2) =
{1\over {\Box_1}}\left(
   {1\over 3}
    \frac{\partial^2}{\partial j_1^2}
  +{2\over 3}
    \frac{\partial^3}{\partial j_1^3}
  +{1\over 6}
    \frac{\partial^4}{\partial j_1^4}
\right)
,\end{fleqnarray}

\begin{fleqnarray}&& b_{15}(3) =
{1\over {\Box_1}}\left(
   {1\over 3}
    \frac{\partial^2}{\partial j_1^2}
  +{2\over 3}
    \frac{\partial^3}{\partial j_1^3}
  +{1\over 6}
    \frac{\partial^4}{\partial j_1^4}
\right)
,\end{fleqnarray}

\begin{fleqnarray}&&\Gamma^{\rm sym}_{16} =
\left\{\begin{tabular}{l}$
\frac1{\Box_1}\Big(
{1\over {6 \Box_2}} +
A_{16}(1)
\Big) +
{\rm O},\hspace{7mm} \Box_1\rightarrow-0 $\\[5mm]$
\frac1{\Box_2}\Big(
{1\over {6 \Box_1}} +
A_{16}(2)
\Big) +
{\rm O},\hspace{7mm} \Box_2\rightarrow-0 $\\[5mm]$
\ln(-\Box_3)\Big(
{1\over {3 \Box_1 \Box_2}}
\Big) +
{\rm O},\hspace{7mm} \Box_3\rightarrow-0
$\end{tabular}\right.\end{fleqnarray}

\begin{fleqnarray}&& a_{16}(1) =
   {1\over 3}
    \frac{\partial}{\partial j_2}
,\end{fleqnarray}

\begin{fleqnarray}&& a_{16}(2) =
   {1\over 3}
    \frac{\partial}{\partial j_1}
,\end{fleqnarray}

\begin{fleqnarray}&&\Gamma^{\rm sym}_{17} =
\left\{\begin{tabular}{l}$
{\rm O},\hspace{7mm} \Box_1\rightarrow-0 $\\[5mm]$
\ln(-\Box_2)\Big(
-{1\over {\Box_1 \Box_3}} +
B_{17}(2)
\Big) +
{\rm O},\hspace{7mm} \Box_2\rightarrow-0 $\\[5mm]$
\ln(-\Box_3)\Big(
-{1\over {\Box_1 \Box_2}} +
B_{17}(3)
\Big) +
{\rm O},\hspace{7mm} \Box_3\rightarrow-0
$\end{tabular}\right.\end{fleqnarray}

\begin{fleqnarray}&& b_{17}(2) =
{1\over {\Box_1}}\left(
    \frac{\partial^2}{\partial j_1^2}
\right)
,\end{fleqnarray}

\begin{fleqnarray}&& b_{17}(3) =
{1\over {\Box_1}}\left(
    \frac{\partial^2}{\partial j_1^2}
\right)
,\end{fleqnarray}

\begin{fleqnarray}&&\Gamma^{\rm sym}_{18} =
\left\{\begin{tabular}{l}$
\frac1{\Box_1}
A_{18}(1)
 +
{\rm O},\hspace{7mm} \Box_1\rightarrow-0 $\\[5mm]$
\ln(-\Box_2)
B_{18}(2)
 +
{\rm O},\hspace{7mm} \Box_2\rightarrow-0 $\\[5mm]$
\ln(-\Box_3)
B_{18}(3)
 +
{\rm O},\hspace{7mm} \Box_3\rightarrow-0
$\end{tabular}\right.\end{fleqnarray}

\begin{fleqnarray}&& a_{18}(1) =
  -1
  -{2\over 3}
    \frac{\partial}{\partial j_2}
  -{2\over 3}
    \frac{\partial}{\partial j_3}
  -{1\over 3}
    \frac{\partial}{\partial j_2}\frac{\partial}{\partial j_3}
,\end{fleqnarray}

\begin{fleqnarray}&& b_{18}(2) =
{1\over {\Box_1}}\left(
    \frac{\partial^2}{\partial j_1^2}
  +{2\over 3}
    \frac{\partial^2}{\partial j_1^2}\frac{\partial}{\partial j_3}
\right)
,\end{fleqnarray}

\begin{fleqnarray}&& b_{18}(3) =
{1\over {\Box_1}}\left(
    \frac{\partial^2}{\partial j_1^2}
  +{2\over 3}
    \frac{\partial^2}{\partial j_1^2}\frac{\partial}{\partial j_2}
\right)
,\end{fleqnarray}

\begin{fleqnarray}&&\Gamma^{\rm sym}_{19} =
\left\{\begin{tabular}{l}$
\frac1{\Box_1}
A_{19}(1)
 +
{\rm O},\hspace{7mm} \Box_1\rightarrow-0 $\\[5mm]$
\ln(-\Box_2)\Big(
{{-1}\over {6 \Box_1 \Box_3}}
\Big) +
{\rm O},\hspace{7mm} \Box_2\rightarrow-0 $\\[5mm]$
\ln(-\Box_3)\Big(
{{-1}\over {6 \Box_1 \Box_2}}
\Big) +
{\rm O},\hspace{7mm} \Box_3\rightarrow-0
$\end{tabular}\right.\end{fleqnarray}

\begin{fleqnarray}&& a_{19}(1) =
   {1\over 6}
  +{1\over 6}
    \frac{\partial}{\partial j_2}\frac{\partial}{\partial j_3}
,\end{fleqnarray}

\begin{fleqnarray}&&\Gamma^{\rm sym}_{20} =
\left\{\begin{tabular}{l}$
\ln(-\Box_1)
B_{20}(1)
 +
{\rm O},\hspace{7mm} \Box_1\rightarrow-0 $\\[5mm]$
\ln(-\Box_2)\Big(
{1\over {6 \Box_1 \Box_3}} +
B_{20}(2)
\Big) +
{\rm O},\hspace{7mm} \Box_2\rightarrow-0 $\\[5mm]$
\ln(-\Box_3)\Big(
{1\over {6 \Box_1 \Box_2}} +
B_{20}(3)
\Big) +
{\rm O},\hspace{7mm} \Box_3\rightarrow-0
$\end{tabular}\right.\end{fleqnarray}

\begin{fleqnarray}&& b_{20}(1) =
{1\over {\Box_2}}\left(
   {1\over {144}}
    \frac{\partial^4}{\partial j_2^4}
  -{5\over {144}}
    \frac{\partial^3}{\partial j_2^3}\frac{\partial}{\partial j_3}
  +{5\over {144}}
    \frac{\partial^2}{\partial j_2^2}\frac{\partial^2}{\partial j_3^2}
  -{1\over {144}}
    \frac{\partial}{\partial j_2}\frac{\partial^3}{\partial j_3^3}
\right)  \nonumber\\&&\ \ \ \ \ \ \ \ \mbox{}
+{1\over {\Box_3}}\left(
  -{1\over {144}}
    \frac{\partial^3}{\partial j_2^3}\frac{\partial}{\partial j_3}
  +{5\over {144}}
    \frac{\partial^2}{\partial j_2^2}\frac{\partial^2}{\partial j_3^2}
\right.\nonumber\\&&\ \ \ \ \ \ \ \ \mbox{}\ \ \ \ \left.
  -{5\over {144}}
    \frac{\partial}{\partial j_2}\frac{\partial^3}{\partial j_3^3}
  +{1\over {144}}
    \frac{\partial^4}{\partial j_3^4}
\right)
,\end{fleqnarray}

\begin{fleqnarray}&& b_{20}(2) =
{1\over {\Box_1}}\left(
  -{1\over 3}
    \frac{\partial^2}{\partial j_1^2}
  -{5\over {18}}
    \frac{\partial^2}{\partial j_1^2}\frac{\partial}{\partial j_3}
  +{1\over {18}}
    \frac{\partial}{\partial j_1}\frac{\partial^2}{\partial j_3^2}
\right)
,\end{fleqnarray}

\begin{fleqnarray}&& b_{20}(3) =
{1\over {\Box_1}}\left(
  -{1\over 3}
    \frac{\partial^2}{\partial j_1^2}
  -{5\over {18}}
    \frac{\partial^2}{\partial j_1^2}\frac{\partial}{\partial j_2}
  +{1\over {18}}
    \frac{\partial}{\partial j_1}\frac{\partial^2}{\partial j_2^2}
\right)
,\end{fleqnarray}

\begin{fleqnarray}&&\Gamma^{\rm sym}_{21} =
\left\{\begin{tabular}{l}$
\frac1{\Box_1}
A_{21}(1)
 +
{\rm O},\hspace{7mm} \Box_1\rightarrow-0 $\\[5mm]$
\ln(-\Box_2)\Big(
{2\over {3 \Box_1 \Box_3}} +
B_{21}(2)
\Big) +
{\rm O},\hspace{7mm} \Box_2\rightarrow-0 $\\[5mm]$
\ln(-\Box_3)\Big(
{2\over {3 \Box_1 \Box_2}}
\Big) +
{\rm O},\hspace{7mm} \Box_3\rightarrow-0
$\end{tabular}\right.\end{fleqnarray}

\begin{fleqnarray}&& a_{21}(1) =
  -{2\over 3}
  +{4\over 3}
    \frac{\partial}{\partial j_3}
  +{2\over 3}
    \frac{\partial^2}{\partial j_3^2}
,\end{fleqnarray}

\begin{fleqnarray}&& b_{21}(2) =
{1\over {\Box_1}}\left(
  -{4\over 3}
    \frac{\partial^2}{\partial j_1^2}\frac{\partial}{\partial j_3}
  -{2\over 3}
    \frac{\partial^2}{\partial j_1^2}\frac{\partial^2}{\partial j_3^2}
\right)
,\end{fleqnarray}

\begin{fleqnarray}&&\Gamma^{\rm sym}_{22} =
\left\{\begin{tabular}{l}$
\frac1{\Box_1}\Big(
-{{1}\over {540 \Box_2}} - {1\over {540 \Box_3}} +
A_{22}(1)
\Big) +
{\rm O},\hspace{7mm} \Box_1\rightarrow-0 $\\[5mm]$
\frac1{\Box_2}\Big(
{{-1}\over {540 \Box_1}} +
A_{22}(2)
\Big)$\\[3mm]$ +
\ln(-\Box_2)\Big(
{{-1}\over {36 \Box_1 \Box_3}} +
B_{22}(2)
\Big) +
{\rm O},\hspace{7mm} \Box_2\rightarrow-0 $\\[5mm]$
\frac1{\Box_3}\Big(
{{-1}\over {540 \Box_1}} +
A_{22}(3)
\Big)$\\[3mm]$ +
\ln(-\Box_3)\Big(
{{-1}\over {36 \Box_1 \Box_2}} +
B_{22}(3)
\Big) +
{\rm O},\hspace{7mm} \Box_3\rightarrow-0
$\end{tabular}\right.\end{fleqnarray}

\begin{fleqnarray}&& a_{22}(1) =
  -{1\over {60}}
,\end{fleqnarray}

\begin{fleqnarray}&& a_{22}(2) =
   {1\over {360}}
    \frac{\partial^3}{\partial j_1^3}
  +{1\over {72}}
    \frac{\partial^2}{\partial j_1^2}\frac{\partial}{\partial j_3}
  +{1\over {120}}
    \frac{\partial}{\partial j_1}\frac{\partial^2}{\partial j_3^2}
  +{1\over {360}}
    \frac{\partial^3}{\partial j_3^3}
,\end{fleqnarray}

\begin{fleqnarray}&& b_{22}(2) =
{1\over {\Box_1}}\left(
   {{13}\over {64800}}
    \frac{\partial^6}{\partial j_1^6}
  +{{89}\over {194400}}
    \frac{\partial^5}{\partial j_1^5}\frac{\partial}{\partial j_3}
  -{{47}\over {38880}}
    \frac{\partial^4}{\partial j_1^4}\frac{\partial^2}{\partial j_3^2}
\right.\nonumber\\&&\ \ \ \ \ \ \ \ \mbox{}
\left.
  +{1\over {21600}}
    \frac{\partial^3}{\partial j_1^3}\frac{\partial^3}{\partial j_3^3}
  +{{29}\over {194400}}
    \frac{\partial^2}{\partial j_1^2}\frac{\partial^4}{\partial j_3^4}
  +{1\over {38880}}
    \frac{\partial}{\partial j_1}\frac{\partial^5}{\partial j_3^5}
\right)               \nonumber\\&&\ \ \ \ \mbox{}
+{1\over {\Box_3}}\left(
   {1\over {3600}}
    \frac{\partial^5}{\partial j_1^5}\frac{\partial}{\partial j_3}
  -{{31}\over {194400}}
    \frac{\partial^4}{\partial j_1^4}\frac{\partial^2}{\partial j_3^2}
  +{7\over {38880}}
    \frac{\partial^3}{\partial j_1^3}\frac{\partial^3}{\partial j_3^3}
\right.\nonumber\\&&\ \ \ \ \ \ \ \ \mbox{}
\left.
  -{{37}\over {64800}}
    \frac{\partial^2}{\partial j_1^2}\frac{\partial^4}{\partial j_3^4}
  +{{11}\over {48600}}
    \frac{\partial}{\partial j_1}\frac{\partial^5}{\partial j_3^5}
  +{1\over {38880}}
    \frac{\partial^6}{\partial j_3^6}
\right)
,\end{fleqnarray}

\begin{fleqnarray}&& a_{22}(3) =
   {1\over {360}}
    \frac{\partial^3}{\partial j_1^3}
  +{1\over {72}}
    \frac{\partial^2}{\partial j_1^2}\frac{\partial}{\partial j_2}
  +{1\over {120}}
    \frac{\partial}{\partial j_1}\frac{\partial^2}{\partial j_2^2}
  +{1\over {360}}
    \frac{\partial^3}{\partial j_2^3}
,\end{fleqnarray}

\begin{fleqnarray}&& b_{22}(3) =
{1\over {\Box_1}}\left(
   {{13}\over {64800}}
    \frac{\partial^6}{\partial j_1^6}
  +{{89}\over {194400}}
    \frac{\partial^5}{\partial j_1^5}\frac{\partial}{\partial j_2}
  -{{47}\over {38880}}
    \frac{\partial^4}{\partial j_1^4}\frac{\partial^2}{\partial j_2^2}
\right.\nonumber\\&&\ \ \ \ \ \ \ \ \mbox{}\left.
  +{1\over {21600}}
    \frac{\partial^3}{\partial j_1^3}\frac{\partial^3}{\partial j_2^3}
  +{{29}\over {194400}}
    \frac{\partial^2}{\partial j_1^2}\frac{\partial^4}{\partial j_2^4}
  +{1\over {38880}}
    \frac{\partial}{\partial j_1}\frac{\partial^5}{\partial j_2^5}
\right)                   \nonumber\\&&\ \ \ \ \mbox{}
+{1\over {\Box_2}}\left(
   {1\over {3600}}
    \frac{\partial^5}{\partial j_1^5}\frac{\partial}{\partial j_2}
  -{{31}\over {194400}}
    \frac{\partial^4}{\partial j_1^4}\frac{\partial^2}{\partial j_2^2}
  +{7\over {38880}}
    \frac{\partial^3}{\partial j_1^3}\frac{\partial^3}{\partial j_2^3}
\right.\nonumber\\&&\ \ \ \ \ \ \ \ \mbox{}\left.
  -{{37}\over {64800}}
    \frac{\partial^2}{\partial j_1^2}\frac{\partial^4}{\partial j_2^4}
  +{{11}\over {48600}}
    \frac{\partial}{\partial j_1}\frac{\partial^5}{\partial j_2^5}
  +{1\over {38880}}
    \frac{\partial^6}{\partial j_2^6}
\right)
,\end{fleqnarray}

\begin{fleqnarray}&&\Gamma^{\rm sym}_{23} =
\left\{\begin{tabular}{l}$
\frac1{\Box_1}\Big(
{1\over {135 \Box_2}} +
A_{23}(1)
\Big) +
{\rm O},\hspace{7mm} \Box_1\rightarrow-0 $\\[5mm]$
\frac1{\Box_2}\Big(
{1\over {135 \Box_1}} +
A_{23}(2)
\Big) +
{\rm O},\hspace{7mm} \Box_2\rightarrow-0 $\\[5mm]$
\ln(-\Box_3)\Big(
{{-1}\over {45 \Box_1 \Box_2}}
\Big) +
{\rm O},\hspace{7mm} \Box_3\rightarrow-0
$\end{tabular}\right.\end{fleqnarray}

\begin{fleqnarray}&& a_{23}(1) =
   {1\over {90}}
    \frac{\partial^2}{\partial j_2^2}
  +{2\over {45}}
    \frac{\partial}{\partial j_2}\frac{\partial}{\partial j_3}
  +{1\over {60}}
    \frac{\partial^2}{\partial j_3^2}
,\end{fleqnarray}

\begin{fleqnarray}&& a_{23}(2) =
   {1\over {90}}
    \frac{\partial^2}{\partial j_1^2}
  +{2\over {45}}
    \frac{\partial}{\partial j_1}\frac{\partial}{\partial j_3}
  +{1\over {60}}
    \frac{\partial^2}{\partial j_3^2}
,\end{fleqnarray}

\begin{fleqnarray}&&\Gamma^{\rm sym}_{24} =
\left\{\begin{tabular}{l}$
\ln(-\Box_1)\Big(
{{-1}\over {30 \Box_2 \Box_3}}
\Big) +
{\rm O},\hspace{7mm} \Box_1\rightarrow-0 $\\[5mm]$
\frac1{\Box_2}\Big(
{1\over {540 \Box_3}} +
A_{24}(2)
\Big) +
{\rm O},\hspace{7mm} \Box_2\rightarrow-0 $\\[5mm]$
\frac1{\Box_3}\Big(
{1\over {540 \Box_2}} +
A_{24}(3)
\Big) +
{\rm O},\hspace{7mm} \Box_3\rightarrow-0
$\end{tabular}\right.\end{fleqnarray}

\begin{fleqnarray}&& a_{24}(2) =
   {1\over {60}}
    \frac{\partial^2}{\partial j_1^2}
  +{1\over {20}}
    \frac{\partial}{\partial j_1}\frac{\partial}{\partial j_3}
  +{1\over {60}}
    \frac{\partial^2}{\partial j_3^2}
,\end{fleqnarray}

\begin{fleqnarray}&& a_{24}(3) =
   {1\over {60}}
    \frac{\partial^2}{\partial j_1^2}
  +{1\over {20}}
    \frac{\partial}{\partial j_1}\frac{\partial}{\partial j_2}
  +{1\over {60}}
    \frac{\partial^2}{\partial j_2^2}
,\end{fleqnarray}

\begin{fleqnarray}&&\Gamma^{\rm sym}_{25} =
\left\{\begin{tabular}{l}$
\frac1{\Box_1}\Big(
-{{1}\over {270 \Box_2}} - {1\over {270 \Box_3}} +
A_{25}(1)
\Big)$\\[3mm]$ +
\ln(-\Box_1)\Big(
{1\over {15 \Box_2 \Box_3}}
\Big) +
{\rm O},\hspace{7mm} \Box_1\rightarrow-0 $\\[5mm]$
\frac1{\Box_2}\Big(
-{{1}\over {270 \Box_1}} + {1\over {270 \Box_3}}
\Big) +
{\rm O},\hspace{7mm} \Box_2\rightarrow-0 $\\[5mm]$
\frac1{\Box_3}\Big(
-{{1}\over {270 \Box_1}} + {1\over {270 \Box_2}}
\Big) +
{\rm O},\hspace{7mm} \Box_3\rightarrow-0
$\end{tabular}\right.\end{fleqnarray}

\begin{fleqnarray}&& a_{25}(1) =
   {1\over {30}}
    \frac{\partial^2}{\partial j_2^2}
  +{2\over {15}}
    \frac{\partial}{\partial j_2}\frac{\partial}{\partial j_3}
  +{1\over {30}}
    \frac{\partial^2}{\partial j_3^2}
,\end{fleqnarray}

\begin{fleqnarray}&&\Gamma^{\rm sym}_{26} =
\left\{\begin{tabular}{l}$
\frac1{\Box_1}
A_{26}(1)
 +
{\rm O},\hspace{7mm} \Box_1\rightarrow-0 $\\[5mm]$
\frac1{\Box_2}
A_{26}(2)
 +
{\rm O},\hspace{7mm} \Box_2\rightarrow-0 $\\[5mm]$
\ln(-\Box_3)
B_{26}(3)
 +
{\rm O},\hspace{7mm} \Box_3\rightarrow-0
$\end{tabular}\right.\end{fleqnarray}

\begin{fleqnarray}&& a_{26}(1) =
{1\over {\Box_2}}\left(
  -{1\over 6}
    \frac{\partial^2}{\partial j_2^2}
\right)
,\end{fleqnarray}

\begin{fleqnarray}&& a_{26}(2) =
{1\over {\Box_1}}\left(
  -{1\over 6}
    \frac{\partial^2}{\partial j_1^2}
\right)
,\end{fleqnarray}

\begin{fleqnarray}&& b_{26}(3) =
{1\over {\Box_1 \Box_2}}\left(
   {1\over 6}
    \frac{\partial^2}{\partial j_1^2}\frac{\partial^2}{\partial j_2^2}
\right)
,\end{fleqnarray}

\begin{fleqnarray}&&\Gamma^{\rm sym}_{27} =
\left\{\begin{tabular}{l}$
\frac1{\Box_1}\Big(
{{-1}\over {540 \Box_2 \Box_3}} +
A_{27}(1)
\Big) +
{\rm O},\hspace{7mm} \Box_1\rightarrow-0 $\\[5mm]$
\frac1{\Box_2}\Big(
{{-1}\over {540 \Box_1 \Box_3}} +
A_{27}(2)
\Big) +
{\rm O},\hspace{7mm} \Box_2\rightarrow-0 $\\[5mm]$
\frac1{\Box_3}\Big(
{{-1}\over {540 \Box_1 \Box_2}}
\Big) +
\ln(-\Box_3)
B_{27}(3)
 +
{\rm O},\hspace{7mm} \Box_3\rightarrow-0
$\end{tabular}\right.\end{fleqnarray}

\begin{fleqnarray}&& a_{27}(1) =
{1\over {\Box_2}}\left(
   {1\over {180}}
    \frac{\partial^2}{\partial j_2^2}
  +{1\over {135}}
    \frac{\partial^3}{\partial j_2^3}
  +{1\over {1080}}
    \frac{\partial^4}{\partial j_2^4}
  +{1\over {540}}
    \frac{\partial^2}{\partial j_2^2}\frac{\partial^2}{\partial j_3^2}
\right)
,\end{fleqnarray}

\begin{fleqnarray}&& a_{27}(2) =
{1\over {\Box_1}}\left(
   {1\over {180}}
    \frac{\partial^2}{\partial j_1^2}
  +{1\over {135}}
    \frac{\partial^3}{\partial j_1^3}
  +{1\over {1080}}
    \frac{\partial^4}{\partial j_1^4}
  +{1\over {540}}
    \frac{\partial^2}{\partial j_1^2}\frac{\partial^2}{\partial j_3^2}
\right)
,\end{fleqnarray}

\begin{fleqnarray}&& b_{27}(3)\! = \!
{1\over {\Box_1 \Box_2}}\!\left(\!
  -{1\over {1080}}
    \frac{\partial^4}{\partial j_1^4}\frac{\partial^2}{\partial j_2^2}
  +\!{1\over {270}}
    \frac{\partial^3}{\partial j_1^3}\frac{\partial^3}{\partial j_2^3}
  -\!{1\over {1080}}
    \frac{\partial^2}{\partial j_1^2}\frac{\partial^4}{\partial j_2^4}
\right)\!
,\end{fleqnarray}

\begin{fleqnarray}&&\Gamma^{\rm sym}_{28} =
\left\{\begin{tabular}{l}$
\frac1{\Box_1}\Big(
{1\over {135 \Box_2 \Box_3}} +
A_{28}(1)
\Big) +
{\rm O},\hspace{7mm} \Box_1\rightarrow-0 $\\[5mm]$
\frac1{\Box_2}\Big(
{1\over {135 \Box_1 \Box_3}} +
A_{28}(2)
\Big) +
{\rm O},\hspace{7mm} \Box_2\rightarrow-0 $\\[5mm]$
\frac1{\Box_3}\Big(
{1\over {135 \Box_1 \Box_2}}
\Big) +
{\rm O},\hspace{7mm} \Box_3\rightarrow-0
$\end{tabular}\right.\end{fleqnarray}

\begin{fleqnarray}&& a_{28}(1) =
{1\over {\Box_2}}\left(
   {1\over {45}}
    \frac{\partial^2}{\partial j_2^2}\frac{\partial}{\partial j_3}
\right)
,\end{fleqnarray}

\begin{fleqnarray}&& a_{28}(2) =
{1\over {\Box_1}}\left(
   {1\over {45}}
    \frac{\partial^2}{\partial j_1^2}\frac{\partial}{\partial j_3}
\right)
,\end{fleqnarray}

\begin{fleqnarray}&&\Gamma^{\rm sym}_{29} =
\left\{\begin{tabular}{l}$
\frac1{\Box_1}
A_{29}(1)
 +
{\rm O},\hspace{7mm} \Box_1\rightarrow-0 $\\[5mm]$
\frac1{\Box_2}
A_{29}(2)
 +
{\rm O},\hspace{7mm} \Box_2\rightarrow-0 $\\[5mm]$
\frac1{\Box_3}
A_{29}(3)
 +
{\rm O},\hspace{7mm} \Box_3\rightarrow-0
$\end{tabular}\right.\end{fleqnarray}

\begin{fleqnarray}&& a_{29}(1) =
{1\over {\Box_2 \Box_3}}\left(
  -{1\over {270}}
    \frac{\partial^2}{\partial j_2^2}\frac{\partial^2}{\partial j_3^2}
\right)
,\end{fleqnarray}

\begin{fleqnarray}&& a_{29}(2) =
{1\over {\Box_1 \Box_3}}\left(
  -{1\over {270}}
    \frac{\partial^2}{\partial j_1^2}\frac{\partial^2}{\partial j_3^2}
\right)
,\end{fleqnarray}

\begin{fleqnarray}&& a_{29}(3) =
{1\over {\Box_1 \Box_2}}\left(
  -{1\over {270}}
    \frac{\partial^2}{\partial j_1^2}\frac{\partial^2}{\partial j_2^2}
\right)
.\end{fleqnarray}

For the derivation of these results see sect. 20
where also the spectral form of the
functions (11.1)--(11.6) is given.

\section{The trace anomaly in four dimensions}
\setcounter{equation}{0}

\hspace{\parindent}
A crucial check of the results above is a derivation
of the trace anomaly for a conformal invariant
quantum field in four dimensions. To have as
many curvature structures as possible involved in
the check, we choose the following quantum field
model ($\omega=2$):

\begin{equation}
S [ \varphi ] = \frac12\int\! dx\, g^{1/2}\, \,
\left(\nabla_{\mu}\varphi^{T}
\nabla^{\mu}\varphi+\frac{R}{6}\varphi^{T}\varphi
+\frac{\lambda^2}{4!}
(\varphi^{T}\varphi)^2 \right),
\end{equation}

\begin{equation}
 \varphi=\left(\begin{array}{c}\varphi_1 \\ \varphi_2
\end{array} \right), \end{equation}
\begin{equation} \nabla_{\mu}\varphi
= \partial_{\mu}\varphi + {A}_{\mu}\hat{G}\varphi,\hspace{7mm}
\nabla_{\mu}\varphi^{T} =
 \partial_{\mu}\varphi^{T}+A_\mu
\varphi^{T}\hat{G}^T,
\end{equation}
\begin{equation} \hat{G}=\left(\begin{array}{cc} 0&1 \\ -1& 0
\end{array} \right) \end{equation}
where (12.1) is the euclidean action of the complex scalar
quantum field
\begin{equation} \varphi=\varphi_1+i\varphi_2, \end{equation}
rewritten in terms of the real components. The
electromagnetic and gravitational fields in (12.1)
are classical.

The action (12.1) is invariant under the local
conformal transformations
\begin{equation}
\delta_\sigma g^{\mu\nu}(x)=\sigma(x)g^{\mu\nu}(x),\hspace{7mm}
\delta_\sigma\varphi(x)=\frac12\sigma(x)\varphi(x),\hspace{7mm}
\delta_\sigma A_\mu(x)=0
\end{equation}
with the parameter $\sigma(x)$. The hessian of
the action (12.1) has the form (1.2) (times a local
matrix) in which the potential is
\begin{equation}
\hat{P}=-\frac{2\lambda^2}{4!}
\left(\begin{array}{cc} 3{\varphi_1}^2
+{\varphi_2}^2&2{\varphi_1}{\varphi_2} \\
2{\varphi_1}{\varphi_2} & 3{\varphi_2}^2
+{\varphi_1}^2\end{array}\right),
\end{equation}
and the commutator curvature is
\begin{equation}
\hat{\cal R}_{\mu\nu}=\hat{G}(
\partial_\mu A_\nu-
\partial_\nu A_\mu) \end{equation}
with $\hat{G}$ in (12.4).

>From (12.6)--(12.8) we find the conformal
transformation laws for the curvatures and
$\Box$-operators:
\begin{fleqnarray}&&
\delta_\sigma \hat{P}=\sigma\hat{P},\hspace{7mm} \delta_\sigma\hat{\cal R}_{\mu\nu}=0, \\[\baselineskip]&&
\delta_\sigma R_{\mu\nu}=
(\nabla_\mu\nabla_\nu+\frac12g_{\mu\nu}\Box)\sigma, \\[\baselineskip]&&
\delta_\sigma R=(3\Box+R)\sigma, \\[\baselineskip]&&
(\delta_\sigma\Box)\hat{P}=\sigma\Box\hat{P}
 -\nabla_\alpha\sigma\nabla^\alpha\hat{P}, \\[\baselineskip]&&
(\delta_\sigma\Box)\hat{\cal R}_{\mu\nu} =
\sigma\Box\hat{\cal R}_{\mu\nu}
+\hat{\cal R}_{\mu\nu}\Box\sigma \nonumber\\&&\mbox{}\hspace{24mm}
+\nabla_\mu\sigma\nabla^\alpha\hat{\cal R}_{\alpha\nu}
-\nabla_\nu\sigma\nabla^\alpha\hat{\cal R}_{\alpha\mu},  \\[\baselineskip]&&
(\delta_\sigma\Box)R_{\mu\nu} =
\sigma\Box R_{\mu\nu}
+R_{\mu\nu}\Box\sigma
+\nabla_\alpha\sigma\nabla^\alpha R_{\mu\nu} \nonumber\\&&\mbox{}\hspace{24mm}
+\nabla_{(\mu}\sigma\nabla_{\nu)} R
-2\nabla^\alpha\sigma\nabla_{(\mu}R_{\nu)\alpha},  \\[\baselineskip]&&
(\delta_\sigma\Box)R=\sigma\Box R-\nabla_\alpha\sigma\nabla^\alpha R.
\end{fleqnarray}
Having got these laws, one may already forget the
particular content of the model, and merely consider
the transformation (12.9)--(12.15) in the
effective action. For the dimensionally regularized
\footnote{\normalsize
The dimensional regularization was used in paper II
for the derivation of the quadratic terms in $W$.
In fact, important is only belonging of the
regularization to one of the two alternative
classes discussed in [22].}
one-loop effective action (6.1), the result should
be exactly
\begin{equation}
-\delta_\sigma W=-\frac1{2(4\pi)^2}\int\! dx\, g^{1/2}\,
\sigma(x)\,{\rm tr}\,\hat{a}_2(x,x)
\end{equation}
where $\hat{a}_2(x,x)$ is the second DeWitt
coefficient
\footnote{\normalsize
Since the function $\sigma(x)$ is arbitrary in
any compact domain, the anomaly (12.16)
provides a check of ${\rm tr}\,\hat{a}_2(x,x)$ itself
whereas in sect. 4 we dealt only with the
integral
\[\int\! dx\, g^{1/2}\, {\rm tr}\,\hat{a}_2(x,x). \]
Hence the differences between the
expressions (4.47) and (12.17).}
at coincident points [6,7]:
\begin{equation}
\hat{a}_2(x,x) = \frac16\Box\hat{P}
+\frac1{180}\Box R\hat{1}
+\frac1{180}(R^2_{\alpha\beta\mu\nu}
-R^2_{\alpha\beta})\hat{1}
+\frac1{12}\hat{\cal R}^2_{\mu\nu}
+\frac12\hat{P}^2.
\end{equation}
Expression (12.16) is the general form of the
conformal anomaly in four dimensions [18--22].
For the model above,
\begin{equation}
\delta_\sigma=\int dx\,\left(
\sigma(x)g^{\mu\nu}
\frac{\delta}{\delta g^{\mu\nu}}
+\frac12\sigma(x)\varphi
\frac{\delta}{\delta\varphi}
\right),
\end{equation}
and
\begin{equation}
g^{-1/2}\left(
g^{\mu\nu}
\frac{\delta W}{\delta g^{\mu\nu}}
+\frac12\varphi
\frac{\delta W}{\delta\varphi}
\right) =
\frac1{2(4\pi)^2}{\rm tr}\,\hat{a}_2(x,x).
\end{equation}

In the present technique, eq. (12.16) can be
obtained only with a given accuracy ${\rm O}[\Re^n]$
and with the Riemann tensor expressed through the
Ricci tensor. To lowest order, one may use the
expression for $R^2_{\alpha\beta\mu\nu}$
given in Appendix A of paper II. After elimination
of the Riemann tensor from (12.17), eq. (12.16)
takes the form
\begin{eqnarray}
-\delta_\sigma W &=&
{\frac1{2(4\pi)^2}\int\! dx\, g^{1/2}\, \,{\rm tr}\,}
\left\{
-\frac16(\Box\hat{P})\sigma
\right.\nonumber\\&&\mbox{}
-\frac1{180}(\Box R)\sigma\hat{1}
-\frac1{12}\hat{\cal R}^2_{\mu\nu}\sigma
-\frac12\hat{P}^2\sigma\nonumber\\&&\mbox{}
-\frac1{180}
\left(1+2\frac{\Box_1}{\Box_2}
-4\frac{\Box_3}{\Box_1}+\frac{{\Box_3}^2}{\Box_1\Box_2}\right)
{R_1^{\mu\nu}R_{2\mu\nu}\sigma_3\hat{1}}
\nonumber\\&&\mbox{}
-\frac1{45}\left(\frac2{\Box_1}-\frac{\Box_3}{\Box_1\Box_2}\right)
{\nabla^\mu R^{\nu\lambda}_1
         \nabla_\nu R_{2\mu\lambda}\sigma_3\hat{1}}
\nonumber\\&&\mbox{}\left.
-\frac1{45}\frac1{\Box_1\Box_2}
{\nabla_\alpha\nabla_\beta R_{1\mu\nu}
         \nabla^\mu\nabla^\nu R_2^{\alpha\beta}\sigma_3\hat{1}}
\right\}
+{\rm O}[\Re^3]
\end{eqnarray}
where the notation in the nonlocal terms is the
same as before with $\sigma$ playing the role of
the third curvature. It is the latter equation
that will be checked below by a direct calculation
with $W$ in (6.1).

We begin this check with calculating the
result of the transformation (12.9)--(12.15)
in the quadratic terms of $W$. For the quadratic
terms of (6.1) we have
\begin{eqnarray}&&
\delta_\sigma\int\! dx\, g^{1/2}\, {\rm tr}\left\{
\sum^5_{i=1}\gamma_i(-\Box_2)\Re_1\Re_2({i})\right\}
=\int\! dx\, g^{1/2}\, {\rm tr}\left\{
-\frac3{18}\hat{P}\Box\sigma \right.\nonumber\\&&\mbox{}\hspace{19mm}
+\frac1{30}R^{\mu\nu}\left[\gamma(-\Box)
+\frac{16}{15}\right]\Big(\nabla_\mu\nabla_\nu\sigma
+\frac12g_{\mu\nu}\Box\sigma\Big)\hat{1} \nonumber\\&&\mbox{}\hspace{19mm}
-\frac3{90}R\left[\gamma(-\Box)+\frac{37}{30}\right]
\Box\sigma\hat{1} \nonumber\\&&\mbox{}\hspace{19mm}
+\frac1{12}
\hat{\cal R}_{\mu\nu}
\Big(\delta_\sigma\gamma(-\Box)\Big)
\hat{\cal R}^{\mu\nu}
+\frac1{2}
\hat{P}
\Big(\delta_\sigma\gamma(-\Box)\Big)
\hat{P} \nonumber\\&&\mbox{}\hspace{19mm}
\left.
+\frac1{60}
R_{\mu\nu}
\Big(\delta_\sigma\gamma(-\Box)\Big)
R^{\mu\nu}\hat{1}
-\frac1{180}
R
\Big(\delta_\sigma\gamma(-\Box)\Big)
R \hat{1}\right\}
\end{eqnarray}
where
\begin{equation}
\gamma(-\Box)=-\ln\left(-\frac{\Box}{\mu^2}\right).
\end{equation}
In the term linear in $R^{\mu\nu}$, for being able
to use the Bianchi identity, one must commute
$\gamma(-\Box)$ with $\nabla_\mu\nabla_\nu$, and the
commutator cannot be neglected. As a result of
this commutation, the linear nonlocal terms cancel,
and we obtain
\begin{eqnarray}&&
\delta_\sigma\int\! dx\, g^{1/2}\, {\rm tr}\left\{
\sum^5_{i=1}\gamma_i(-\Box_2)\Re_1\Re_2({i})\right\}
\nonumber\\&&\ \ \ \ \ \ \ \ \mbox{}
=\int\! dx\, g^{1/2}\, {\rm tr}\left\{\right.
-\frac1{6}(\Box\hat{P})\sigma
-\frac1{180}(\Box R)\sigma\hat{1} \nonumber\\&&\ \ \ \ \ \ \ \ \mbox{}\ \ \ \
+\frac1{30}R_{\mu\nu}\Big[\gamma(-\Box),
\nabla^\mu\nabla^\nu\Big]\sigma\hat{1} \nonumber\\&&\ \ \ \ \ \ \ \ \mbox{}\ \ \ \
+\frac1{12}
\hat{\cal R}_{\mu\nu}
\Big(\delta_\sigma\gamma(-\Box)\Big)
\hat{\cal R}^{\mu\nu}
+\frac1{2}
\hat{P}
\Big(\delta_\sigma\gamma(-\Box)\Big)
\hat{P} \nonumber\\&&\ \ \ \ \ \ \ \ \mbox{}\ \ \ \
\left.
+\frac1{60}
R_{\mu\nu}
\Big(\delta_\sigma\gamma(-\Box)\Big)
R^{\mu\nu}\hat{1}
-\frac1{180}
R
\Big(\delta_\sigma\gamma(-\Box)\Big)
R \hat{1}\right\}
\end{eqnarray}
where the first two terms correctly reproduce the
linear contributions to the anomaly, and the
remaining terms are already quadratic in the
curvature.

For the calculation of the quadratic terms in
(12.23) we use the spectral representation
\begin{equation}
\gamma(-\Box) =
\int^\infty_0dm^2\,
\Big(\frac1{m^2-\Box}-\frac1{m^2+\mu^2}\Big)
\end{equation}
and the commutation (variation) rule for the
inverse operator
\begin{equation}
\Big[
\frac1{m^2-\Box}
,\nabla^\mu\nabla^\nu\Big] = -
\frac1{m^2-\Box}
\Big[-\Box,\nabla^\mu\nabla^\nu\Big]
\frac1{m^2-\Box},
\end{equation}
\begin{equation}
\delta_\sigma
\frac1{m^2-\Box} = -
\frac1{m^2-\Box}
\delta_\sigma(-\Box)
\frac1{m^2-\Box}
\end{equation}
where, within the required accuracy, the
factors on the right-hand sides can already be
commuted freely. Doing the spectral-mass integral
then gives
\begin{eqnarray}&&
\int\! dx\, g^{1/2}\,
R_{\mu\nu}[\gamma(-\Box),
\nabla^\mu\nabla^\nu]\sigma \nonumber\\&&\ \ \ \ \mbox{}
=-\int\! dx\, g^{1/2}\, \frac{\ln(\Box_1/\Box_3)}{(\Box_1-\Box_3)}
[\Box_3,\nabla^\mu_3\nabla^\nu_3]
R_{1\mu\nu}\sigma_3
+{\rm O}[R^3..],
\end{eqnarray}
and, similarly,
\begin{eqnarray}&&
\int\! dx\, g^{1/2}\, {\rm tr}\,
\Re_1\Big(\delta_\sigma\gamma(-\Box_2)\Big)\Re_2 \nonumber\\&&\ \ \ \ \ \ \ \ \mbox{}
=-\int\! dx\, g^{1/2}\, {\rm tr}\,\frac{\ln(\Box_1/\Box_2)}{(\Box_1-\Box_2)}
(\delta_\sigma\Box_2)\Re_1\Re_2
+{\rm O}[\Re^3].
\end{eqnarray}
There remain to be used in (12.28) the
transformation laws (12.12)--(12.15), and
in (12.27) the expression for the commutator
\begin{eqnarray}
[\Box,\nabla_\mu\nabla_\nu]\sigma &=&
2\nabla_{(\mu}R_{\nu)\alpha}\nabla^\alpha\sigma
+2R_{\alpha(\mu}\nabla_{\nu)}\nabla^\alpha\sigma \nonumber\\&&\mbox{}
-\nabla_\alpha R_{\mu\nu}\nabla^\alpha\sigma
-2R_{\alpha\nu\beta\mu}
\nabla^\alpha\nabla^\beta\sigma
\end{eqnarray}
in which the Riemann tensor should be expressed
through the Ricci tensor.

The final result for (12.23) is
\begin{eqnarray}&&
\delta_\sigma\int\! dx\, g^{1/2}\, {\rm tr}\left\{
\sum^5_{i=1}\gamma_i(-\Box_2)\Re_1\Re_2({i})\right\}  \nonumber\\&&\ \ \ \ \ \ \ \ \mbox{}
=\int\! dx\, g^{1/2}\, {\rm tr}\left\{
-\frac1{6}(\Box\hat{P})\sigma
-\frac1{180}(\Box R)\sigma\hat{1}
\right.
\nonumber\\&&\ \ \ \ \ \ \ \ \mbox{}\ \ \ \ \ \ \ \
\left.
+\sum^{10}_{i=1}
M_i(\Box_1,\Box_2,\Box_3)\Re_1\Re_2\sigma_3({i})\right\}
+{\rm O}[\Re^3]
\end{eqnarray}
where $\Re_1\Re_2\sigma_3({i})$ are the following ten tensor
structures:
\begin{eqnarray}
\Re_1\Re_2\sigma_3({1}) &=&
{R_1R_2\sigma_3\hat{1}}
, \\[2mm]
\Re_1\Re_2\sigma_3({2}) &=&
{R_1^{\mu\nu}R_{2\mu\nu}\sigma_3\hat{1}}
, \\[2mm]
\Re_1\Re_2\sigma_3({3}) &=&
{R_1^{\mu\nu}\nabla_\nu\nabla_\mu R_2\sigma_3\hat{1}}
, \\[2mm]
\Re_1\Re_2\sigma_3({4}) &=&
{\nabla^\mu R^{\nu\lambda}_1
         \nabla_\nu R_{2\mu\lambda}\sigma_3\hat{1}}
, \\[2mm]
\Re_1\Re_2\sigma_3({5}) &=&
{\nabla_\alpha\nabla_\beta R_{1\mu\nu}
      \nabla^\mu\nabla^\nu R_2^{\alpha\beta}\sigma_3\hat{1}}
, \\[2mm]
\Re_1\Re_2\sigma_3({6}) &=&
{\hat{P}_1 R_2\sigma_3}
, \\[2mm]
\Re_1\Re_2\sigma_3({7}) &=&
{\nabla_\alpha\nabla_\beta\hat{P}_1 R_2^{\alpha\beta}\sigma_3}
, \\[2mm]
\Re_1\Re_2\sigma_3({8}) &=&
{\hat{P}_1\hat{P}_2\sigma_3}
, \\[2mm]
\Re_1\Re_2\sigma_3({9}) &=&
{\hat{\cal R}_1^{\mu\nu}\hat{\cal R}_{2\mu\nu}\sigma_3}
, \\[2mm]
\Re_1\Re_2\sigma_3({10}) &=&
{\nabla_\mu\hat{\cal R}_1^{\mu\alpha}\nabla^\nu\hat{\cal R}_{2\nu\alpha}\sigma_3}
,
\end{eqnarray}
and for the form factors
$M_i(\Box_1,\Box_2,\Box_3)$ we obtain
\begin{fleqnarray}&&
M_1 =
\frac{(-2\Box_1+5\Box_3)}{720}\frac{\ln(\Box_1/\Box_2)}{(\Box_1-\Box_2)}
+\frac{(\Box_1+\Box_2-\Box_3)}{120}\frac{\ln(\Box_1/\Box_3)}{(\Box_1-\Box_3)}
,\end{fleqnarray}
\begin{fleqnarray}&&
M_2 =
\frac{(-2\Box_1-\Box_3)}{120}\frac{\ln(\Box_1/\Box_2)}{(\Box_1-\Box_2)} \nonumber\\&&\ \ \ \ \ \ \ \ \mbox{}\ \ \ \
+\frac{(\Box_1-\Box_2-\Box_3)(\Box_1-\Box_3)}{60\Box_2}\frac{\ln(\Box_1/\Box_3)}{(\Box_1-\Box_3)}
,\end{fleqnarray}
\begin{fleqnarray}&&
M_3 = -\frac1{30}\frac{\ln(\Box_1/\Box_3)}{(\Box_1-\Box_3)}-\frac1{30}\frac{\ln(\Box_2/\Box_3)}{(\Box_2-\Box_3)}
,\end{fleqnarray}
\begin{fleqnarray}&&
M_4 = -\frac1{30}\frac{\ln(\Box_1/\Box_2)}{(\Box_1-\Box_2)}
+\frac{(\Box_1-\Box_2-\Box_3)}{15\Box_2}\frac{\ln(\Box_1/\Box_3)}{(\Box_1-\Box_3)}
,\end{fleqnarray}
\begin{fleqnarray}&&
M_5 = \frac1{15\Box_2}\frac{\ln(\Box_1/\Box_3)}{(\Box_1-\Box_3)}
,\end{fleqnarray}
\begin{fleqnarray}&&
M_6 = 0
,\end{fleqnarray}
\begin{fleqnarray}&&
M_7 = 0
,\end{fleqnarray}
\begin{fleqnarray}&&
M_8 = \frac{(-2\Box_2-\Box_3)}4\frac{\ln(\Box_1/\Box_2)}{(\Box_1-\Box_2)}
,\end{fleqnarray}
\begin{fleqnarray}&&
M_9 = -\frac{\Box_3}{12}\frac{\ln(\Box_1/\Box_2)}{(\Box_1-\Box_2)}
,\end{fleqnarray}
\begin{fleqnarray}&&
M_{10} = \frac16\frac{\ln(\Box_1/\Box_2)}{(\Box_1-\Box_2)}
.\end{fleqnarray}

The conformal transformation in the cubic terms
of the effective action (6.1) is easier to carry
out because, within the required accuracy, only the
curvatures in $\Re_1\Re_2\Re_3$ need be varied.
The result is again a sum of contributions of the
ten tensor structures (12.31)--(12.40):
\begin{eqnarray}&&
\delta_\sigma\int\! dx\, g^{1/2}\, {\rm tr}\left\{
\sum^{29}_{i=1}\Gamma_{i}(-\Box_1,-\Box_2,-\Box_3)\Re_1\Re_2\Re_3({i})\right\}  \nonumber\\&&\ \ \ \ \mbox{}
=\int\! dx\, g^{1/2}\, {\rm tr}\left\{
\sum^{10}_{i=1}
N_i(\Box_1,\Box_2,\Box_3)\Re_1\Re_2\sigma_3({i})\right\}
+{\rm O}[\Re^3]
\end{eqnarray}
where the form factors $N_i(\Box_1,\Box_2,\Box_3)$ are the following
combinations of the form factors
$\Gamma_{i}(-\Box_1,-\Box_2,-\Box_3)$:

\begin{fleqnarray} N_{1} &=& (-\Box_1 + 2\Box_3){\Gamma^{\rm sym}_{11}}{\big|}_{\Box_1\leftrightarrow\Box_3}\nonumber\\&&\mbox{} +
  \frac14\Big((\Box_1 - \Box_2 - \Box_3)(-\Box_1 + \Box_2 - \Box_3)
  \nonumber\\&&\mbox{}\ \ \ \ + \Box_3(-\Box_1 - \Box_2 + \Box_3)\Big){\Gamma^{\rm sym}_{22}}{\big|}_{\Box_1\leftrightarrow\Box_3}\nonumber\\&&\mbox{} +
  \left(-\frac14(\Box_1 - \Box_2 - \Box_3)^2 - \frac12\Box_2\Box_3
\right.
   \nonumber\\&&\mbox{}\ \ \ \
+ \frac14(\Box_1 - \Box_2 - \Box_3)
(-\Box_1 - \Box_2 + \Box_3)\nonumber\\&&\mbox{}\ \ \ \ \left. -
  \frac14\Box_3(-\Box_1 - \Box_2 + \Box_3)\right){\Gamma^{\rm sym}_{23}}{\big|}_{\Box_1\leftrightarrow\Box_3}\nonumber\\&&\mbox{} +
 \frac1{16}(-\Box_1 - \Box_2 + \Box_3)^2{\Gamma^{\rm sym}_{25}}{\big|}_{\Box_1\leftrightarrow\Box_3}\nonumber\\&&\mbox{} +
 \frac12(-\Box_1 + \Box_3)\left(\frac12(\Box_1 - \Box_2 - \Box_3)^2
  + \Box_2\Box_3\right){\Gamma^{\rm sym}_{27}}{\big|}_{\Box_1\leftrightarrow\Box_3}\nonumber\\&&\mbox{} +
 \frac1{16}(-\Box_1 + \Box_2 - \Box_3)(-\Box_1 - \Box_2 + \Box_3)^2
{\Gamma^{\rm sym}_{28}}{\big|}_{\Box_1\leftrightarrow\Box_3}\nonumber\\&&\mbox{}
 + 9\Box_3\Gamma^{\rm sym}_{9} +
 \frac38(-\Box_1 - \Box_2 + \Box_3)\Gamma^{\rm sym}_{10}\nonumber\\&&\mbox{}
 -\frac32\Box_3(-\Box_1 - \Box_2 + \Box_3)\Gamma^{\rm sym}_{22}\nonumber\\&&\mbox{}
 + \frac14(\Box_1 - 2\Box_3)(-\Box_1 - \Box_2 + \Box_3)\Gamma^{\rm sym}_{24}\nonumber\\&&\mbox{} +
 \frac18(\Box_1 - \Box_2 - \Box_3)(-\Box_1 - \Box_2 + \Box_3)
\Gamma^{\rm sym}_{25} \nonumber\\&&\mbox{} +
 \frac1{32}(-\Box_1 - \Box_2 + \Box_3)
  \Big((\Box_1 - \Box_2 - \Box_3)(-\Box_1 + \Box_2 - \Box_3)
  \nonumber\\&&\mbox{}\ \ \ \  + \Box_3(-\Box_1 - \Box_2 + \Box_3)\Big) \Gamma^{\rm sym}_{28}
,\end{fleqnarray}

\begin{fleqnarray} N_{2} &=& \frac14\Big((\Box_1 - \Box_2 - \Box_3)
(-\Box_1 + \Box_2 - \Box_3)
  \nonumber\\&&\mbox{}\ \ \ \  + \Box_3(-\Box_1 - \Box_2 + \Box_3)\Big)
  {\Gamma^{\rm sym}_{24}}{\big|}_{\Box_1\leftrightarrow\Box_3}\nonumber\\&&\mbox{} + \frac32\Box_3\Gamma^{\rm sym}_{10} + 3\Box_3\Gamma^{\rm sym}_{11}
,\end{fleqnarray}

\begin{fleqnarray} N_{3} &=& -\frac12(-\Box_1 - \Box_2 + \Box_3)
{\Gamma^{\rm sym}_{24}}{\big|}_{\Box_1\leftrightarrow\Box_2}\nonumber\\&&\mbox{} +
   \frac12 (-\Box_1 + \Box_2 - \Box_3){\Gamma^{\rm sym}_{25}}{\big|}_{\Box_1\leftrightarrow\Box_2} \nonumber\\&&\mbox{}
     + \frac12(-\Box_1 - \Box_2 + \Box_3){\Gamma^{\rm sym}_{25}}{\big|}_{\Box_1\leftrightarrow\Box_3}\nonumber\\&&\mbox{} +
   \frac14(-\Box_1 + \Box_2 - \Box_3)(-\Box_1 - \Box_2 + \Box_3)
{\Gamma^{\rm sym}_{28}}{\big|}_{\Box_1\leftrightarrow\Box_3}\nonumber\\&&\mbox{}
     + 2{\Gamma^{\rm sym}_{11}}{\big|}_{\Box_2\leftrightarrow\Box_3} +
    (-\Box_1 + \Box_2 - \Box_3){\Gamma^{\rm sym}_{23}}{\big|}_{\Box_2\leftrightarrow\Box_3}\nonumber\\&&\mbox{} +
    \left(\frac12(-\Box_1 + \Box_2 - \Box_3)^2 + \Box_1\Box_3\right)
{\Gamma^{\rm sym}_{27}}{\big|}_{\Box_2\leftrightarrow\Box_3}\nonumber\\&&\mbox{} +
\frac14\Big((\Box_1 - \Box_2 - \Box_3)(-\Box_1 - \Box_2 + \Box_3)\nonumber\\&&\mbox{}\ \ \ \
            - \Box_3(-\Box_1 - \Box_2 + \Box_3)\Big){\Gamma^{\rm sym}_{28}}{\big|}_{\Box_2\leftrightarrow\Box_3}\nonumber\\&&\mbox{} +
    3\Gamma^{\rm sym}_{10} - 6\Box_3\Gamma^{\rm sym}_{22} + (\Box_1 - 2\Box_3)\Gamma^{\rm sym}_{24}\nonumber\\&&\mbox{} +
\frac12(\Box_1 - \Box_2 - 2\Box_3)\Gamma^{\rm sym}_{25}\nonumber\\&&\mbox{} +
\frac14\Big((\Box_1 - \Box_2 - \Box_3)(-\Box_1 + \Box_2 - \Box_3)\nonumber\\&&\mbox{}\ \ \ \
      + \Box_3(-\Box_1 - \Box_2 + \Box_3)\Big)\Gamma^{\rm sym}_{28}\nonumber\\&&\mbox{} +
    \frac38(-\Box_1 + \Box_3)\Big((-\Box_1 + \Box_2 - \Box_3)^2
+ 2\Box_1\Box_3\Big)\Gamma^{\rm sym}_{29}
,\end{fleqnarray}

\begin{fleqnarray} N_{4} &=& \frac12\Box_3{\Gamma^{\rm sym}_{25}}{\big|}_{\Box_1\leftrightarrow\Box_3}
+ 3\Gamma^{\rm sym}_{10} + 3\Box_3\Gamma^{\rm sym}_{23} +
    (\Box_1 - \Box_2 - 2\Box_3)\Gamma^{\rm sym}_{25}\nonumber\\&&\mbox{}
     + \frac14\Big((\Box_1 - \Box_2 - \Box_3)(-\Box_1 + \Box_2
- \Box_3)\nonumber\\&&\mbox{}\ \ \ \  +
         \Box_3(-\Box_1 - \Box_2 + \Box_3)\Big)\Gamma^{\rm sym}_{28}
,\end{fleqnarray}

\begin{fleqnarray} N_{5} &=& {\Gamma^{\rm sym}_{25}}{\big|}_{\Box_1\leftrightarrow\Box_3} - (\Box_1 - \Box_2 + 2\Box_3)
{\Gamma^{\rm sym}_{28}}{\big|}_{\Box_1\leftrightarrow\Box_3}\nonumber\\&&\mbox{}
    - 2\Gamma^{\rm sym}_{24} +
    3\Box_3\Gamma^{\rm sym}_{27}\nonumber\\&&\mbox{}
   + \frac34\Big((-\Box_1 + \Box_2 - \Box_3)^2 + 2\Box_1\Box_3\Big)\Gamma^{\rm sym}_{29}
,\end{fleqnarray}

\begin{fleqnarray} N_{6} &=& -\frac34\Box_3(-\Box_1 - \Box_2 + \Box_3)
{\Gamma^{\rm sym}_{15}}\big|_{\Box_1\rightarrow\Box_2,
\Box_2\rightarrow\Box_3,\Box_3\rightarrow\Box_1}
  \nonumber\\&&\mbox{} + 6\Box_3{\Gamma^{\rm sym}_{4}}{\big|}_{\Box_1\leftrightarrow\Box_3} +
    (-\Box_1 + 2\Box_3){\Gamma^{\rm sym}_{5}}{\big|}_{\Box_1\leftrightarrow\Box_3}\nonumber\\&&\mbox{}
  + \frac14\Big((\Box_1 - \Box_2 - \Box_3)(-\Box_1 + \Box_2 - \Box_3)\nonumber\\&&\mbox{}\ \ \ \  +
         \Box_3(-\Box_1 - \Box_2 + \Box_3)\Big){\Gamma^{\rm sym}_{15}}{\big|}_{\Box_1\leftrightarrow\Box_3} \nonumber\\&&\mbox{} +
    \left(\frac14(\Box_1 - \Box_2 - \Box_3)\Box_3 \right.\nonumber\\&&\mbox{}\ \ \ \ \left.
     + \frac12(\Box_1 - \Box_2 - \Box_3)(-\Box_1 + \Box_3)\right)
{\Gamma^{\rm sym}_{16}}{\big|}_{\Box_1\leftrightarrow\Box_3}\nonumber\\&&\mbox{} \!+
    \frac12(-\Box_1 + \Box_3)\!\left(\frac12(\Box_1 - \Box_2 - \Box_3)^2
      + \Box_2\Box_3\right)\!{\Gamma^{\rm sym}_{26}}{\big|}_{\Box_1\leftrightarrow\Box_3}
\!,\end{fleqnarray}

\begin{fleqnarray} N_{7} &=& -3\Box_3{\Gamma^{\rm sym}_{15}}\big|_{\Box_1\rightarrow\Box_2,
\Box_2\rightarrow\Box_3,\Box_3\rightarrow\Box_1}
+ 2{\Gamma^{\rm sym}_{5}}{\big|}_{\Box_1\leftrightarrow\Box_3}\nonumber\\&&\mbox{} +
    (\Box_1 - \Box_2 - \Box_3){\Gamma^{\rm sym}_{16}}{\big|}_{\Box_1\leftrightarrow\Box_3}\nonumber\\&&\mbox{}
 +\left(\frac12(\Box_1 - \Box_2 - \Box_3)^2
+ \Box_2\Box_3\right){\Gamma^{\rm sym}_{26}}{\big|}_{\Box_1\leftrightarrow\Box_3}
,\end{fleqnarray}

\begin{fleqnarray} N_{8} &=& \left(\frac14(\Box_1 - \Box_2 - \Box_3)^2
  + \frac12\Box_2\Box_3\right){\Gamma^{\rm sym}_{17}}{\big|}_{\Box_1\leftrightarrow\Box_3}
  + 3\Box_3\Gamma^{\rm sym}_{6}
,\end{fleqnarray}

\begin{fleqnarray} N_{9} &=& 3\Box_3{\Gamma^{\rm sym}_{7}}{\big|}_{\Box_1\leftrightarrow\Box_3}
   + \frac14(\Box_1 + \Box_2 + 3\Box_3){\Gamma^{\rm sym}_{8}}{\big|}_{\Box_1\leftrightarrow\Box_3}\nonumber\\&&\mbox{} -
    \frac12\Box_1\Box_2{\Gamma^{\rm sym}_{18}}{\big|}_{\Box_1\leftrightarrow\Box_3} \nonumber\\&&\mbox{}
     + \frac14\Big((\Box_1 - \Box_2 - \Box_3)(-\Box_1
+ \Box_2 - \Box_3)\nonumber\\&&\mbox{}\ \ \ \  +
         \Box_3(-\Box_1 - \Box_2 + \Box_3)\Big){\Gamma^{\rm sym}_{19}}{\big|}_{\Box_1\leftrightarrow\Box_3}\nonumber\\&&\mbox{} +
    \frac14\Big(\Box_2(\Box_1 - \Box_2 - \Box_3) - \Box_2\Box_3\Big)
{\Gamma^{\rm sym}_{21}}{\big|}_{\Box_1\leftrightarrow\Box_3}
,\end{fleqnarray}

\begin{fleqnarray} N_{10} &=&
{\Gamma^{\rm sym}_{8}}{\big|}_{\Box_1\leftrightarrow\Box_3} + \frac12(-\Box_1 - \Box_2 + 2\Box_3)
{\Gamma^{\rm sym}_{18}}{\big|}_{\Box_1\leftrightarrow\Box_3}\nonumber\\&&\mbox{} +
3\Box_3{\Gamma^{\rm sym}_{20}}{\big|}_{\Box_1\leftrightarrow\Box_3} + \frac12(\Box_1 - \Box_2 - \Box_3)
{\Gamma^{\rm sym}_{21}}{\big|}_{\Box_1\leftrightarrow\Box_3}
.\end{fleqnarray}

The total result of (12.30) and (12.51) is the
following conformal variation of the effective
action (6.1):
\begin{eqnarray}
-\delta_\sigma W &=&
{\frac1{2(4\pi)^2}\int\! dx\, g^{1/2}\, \,{\rm tr}\,}
\left\{
-\frac16(\Box\hat{P})\sigma
-\frac1{180}(\Box R)\sigma\hat{1} \right.\nonumber\\&&\mbox{}\left.
+\sum^{10}_{i=1}(M_i+N_i)\Re_1\Re_2\sigma_3({i})\right\}
+{\rm O}[\Re^3]
\end{eqnarray}
where the quadratic terms are determined by the sum
$M_i+N_i$. There remain to be calculated the linear
combinations of the third-order form factors
$\Gamma_{i}$ in (12.52)--(12.61). This is the most important
part of the calculation because it checks both:
our results for the third-order form factors and our
capability of working with them. The simplest is to use
the explicit forms of $\Gamma_{i}$ in (6.13)--(6.41).
This way of calculation is most straightforward and
least instructive. It is, nevertheless, gratifying
to observe that all terms with the basic third-order
form factor $\Gamma(-\Box_1,-\Box_2,-\Box_3)$ cancel in the
combinations $N_i$, all terms with the second-order
form factors $\ln(\Box_n/\Box_m)$ cancel in the
combinations $N_i+M_i$, and there remain only trees:
\mathindent=0pt
\arraycolsep=0pt
\begin{fleqnarray} \frac12\Big(M_{1}+N_{1}+
{M_{1}}{\big|}_{\Box_1\leftrightarrow\Box_2}+{N_{1}}{\big|}_{\Box_1\leftrightarrow\Box_2}\Big) &=& 0 ,\end{fleqnarray}
\begin{fleqnarray} \frac12\Big(M_{2}+N_{2}+
{M_{2}}{\big|}_{\Box_1\leftrightarrow\Box_2}+{N_{2}}{\big|}_{\Box_1\leftrightarrow\Box_2}\Big)&&=
   -\frac1{180}-\frac{\Box_1}{180\Box_2}-\frac{\Box_2}{180\Box_1}\nonumber\\&&\mbox{}
   +\frac{\Box_3}{90\Box_1}+\frac{\Box_3}{90\Box_2}
   -\frac{{\Box_3}^2}{180\Box_1\Box_2}\, ,\end{fleqnarray}
\begin{fleqnarray}  M_{3}+N_{3} &=& 0 ,\end{fleqnarray}
\begin{fleqnarray} \frac12\Big(M_{4}+N_{4}+{M_{4}}{\big|}_{\Box_1\leftrightarrow\Box_2}+{N_{4}}{\big|}_{\Box_1\leftrightarrow\Box_2}\Big)
&=&
  -\frac1{45\Box_1}-\frac1{45\Box_2}+\frac{\Box_3}{45\Box_1\Box_2} ,\end{fleqnarray}
\begin{fleqnarray} \frac12\Big(M_{5}+N_{5}+{M_{5}}{\big|}_{\Box_1\leftrightarrow\Box_2}+{N_{5}}{\big|}_{\Box_1\leftrightarrow\Box_2}\Big)
 &=&
   -\frac1{45\Box_1\Box_2} ,\end{fleqnarray}
\begin{fleqnarray}  M_{6}+N_{6} &=& 0 ,\end{fleqnarray}
\begin{fleqnarray}  M_{7}+N_{7} &=& 0 ,\end{fleqnarray}
\begin{fleqnarray} \frac12\Big(M_{8}+N_{8}+{M_{8}}{\big|}_{\Box_1\leftrightarrow\Box_2}
+{N_{8}}{\big|}_{\Box_1\leftrightarrow\Box_2}\Big) &=& -\frac12,\end{fleqnarray}
\begin{fleqnarray} \frac12\Big(M_{9}+N_{9}+{M_{9}{\big|}_{\Box_1\leftrightarrow\Box_2}}
+{N_{9}}{\big|}_{\Box_1\leftrightarrow\Box_2}\Big) &=& -\frac1{12},\end{fleqnarray}
\begin{fleqnarray} \frac12\Big(M_{10}+N_{10}+{M_{10}}{\big|}_{\Box_1\leftrightarrow\Box_2}
+{N_{10}}{\big|}_{\Box_1\leftrightarrow\Box_2}\Big) &=& 0 .\end{fleqnarray}
\arraycolsep=5pt
\mathindent=\parindent
The symmetrizations on the left-hand sides of
these equations correspond to the symmetries of the
tensor structures (12.31)--(12.40). With the form
factors (12.63)--(12.72), eq. (12.62) takes the
final form
\begin{eqnarray}
-\delta_\sigma W &=&
{\frac1{2(4\pi)^2}\int\! dx\, g^{1/2}\, \,{\rm tr}\,}
\left\{
-\frac16(\Box\hat{P})\sigma
-\frac1{180}(\Box R)\sigma\hat{1} \right.\nonumber\\&&\mbox{}
-\frac1{180}
 \left(1+\frac{2\Box_1}{\Box_2}-\frac{4\Box_3}{\Box_1}+\frac{{\Box_3}^2}{\Box_1\Box_2}\right)
\Re_1\Re_2\sigma_3({2}) \nonumber\\&&\mbox{}
-\frac1{45}\left(\frac2{\Box_1}-\frac{\Box_3}{\Box_1\Box_2}\right)\Re_1\Re_2\sigma_3({4}) \nonumber\\&&\mbox{}
-\frac1{45}\frac1{\Box_1\Box_2}\Re_1\Re_2\sigma_3({5})
-\frac12\Re_1\Re_2\sigma_3({8}) \nonumber\\&&\mbox{}\left.
-\frac1{12}\Re_1\Re_2\sigma_3({9})
\right\}
+{\rm O}[\Re^3]
\end{eqnarray}
which is precisely the trace anomaly (12.20).

The derivation of the anomaly is not, however,
an end in itself. It is also important that this
and similar derivations be feasible within the
working technique used in applications. The
generalization of the spectral representation
elaborated in sect. 20 and summarized in sect. 9
serves this purpose. The task is again to calculate
the linear combinations (12.52)--(12.61) of the form
factors $\Gamma_{i}$ but in terms of integral originals.
As noted repeatedly in the present paper, the
difficulty that such a calculation encounters is
connected with the fact that the coefficients
of the linear combinations of the form factors contain
$\Box$'s. Because these  $\Box$'s appear outside the
kernel of the integral representation, there
exist nontrivial linear identities between the
form factors not expressible in terms of
the originals. Examples of such identities are
considered in sect. 18. In the calculation of
expressions (12.52)--(12.61), this difficulty
is present in full measure. The result should be
an almost complete cancellation leading to
eqs. (12.63)--(12.72), and this cancellation
is based entirely on the hidden identities of the
said type. Below we illustrate the derivation
of the anomaly within the generalized spectral
technique by deriving one of these identities.

The example we shall consider is given in eq. (18.11)
of sect. 18. In terms of the $\alpha$-integrals,
the following linear combination of the third-order
form factors:
\begin{fleqnarray}
L&=&(\Box_1-\Box_2)
\left<\frac{\alpha_3}{-\Omega}\right>_3
+\Box_3
\left<\frac{(\alpha_1-\alpha_2)}{-\Omega}\right>_3
\end{fleqnarray}
should equal the second-order form factor
\begin{fleqnarray}
L&=&\ln ({\Box_2}/{\Box_1}).
\end{fleqnarray}
In terms of the generalized spectral integrals,
expression (12.74) is of the form (see sect. 9)
\begin{fleqnarray}
L &=& -(\Box_1-\Box_2)2\int^\infty_0dy^2\,\left(\frac4{y^2}\right)^3\,
(C_3-1)C_1C_2C_3{{\cal S}_1}{{\cal S}_2}{{\cal S}_3}   \nonumber\\&&\ \ \ \ \ \ \ \ \mbox{}
-\Box_32\int^\infty_0dy^2\,\left(\frac4{y^2}\right)^3\,
(C_1-C_2)C_1C_2C_3{{\cal S}_1}{{\cal S}_2}{{\cal S}_3}.
\end{fleqnarray}
In order to bring it to the final form (12.75),
one  must first absorb the $\Box$'s by using
relations (9.68) and (9.69). The result is
\begin{fleqnarray}
L &=&2\int^\infty_0dy^2\,\left(\frac4{y^2}\right)^4\,\left[
(C_3-1)(C_1-1)^2
-(C_3-1)(C_2-1)^2
\right.
\nonumber\\&&\mbox{}\hspace{30mm}
\left.
+(C_1-C_2)(C_3-1)^2\right]
C_1C_2C_3{{\cal S}_1}{{\cal S}_2}{{\cal S}_3}
\end{fleqnarray}
where the convergence at the lower limit holds
owing to the presence of the operator
$(C-1)^2$ in each term. Next, one must use in
(12.77) the relation (9.73)
\begin{equation}
C_1+C_2+C_3=3
\end{equation}
to remove the operators $C$ from one of the
${\cal S}\,$'s, say ${{\cal S}_1}$, but then, in the term
with $(C_1-1)^2$, the bare ${{\cal S}_1}$ will
appear with a subtraction:
\begin{fleqnarray}
L&=&2\int^\infty_0dy^2\,\left(\frac4{y^2}\right)^4\,\left\{
(C_3-1)(C_2+C_3-2)^2
C_1C_2C_3({{\cal S}_1}-{{{\cal S}_1}}^{\bf 1}){{\cal S}_2}{{\cal S}_3}
\right.
\nonumber\\&&\mbox{}\hspace{30mm}
-\Big[(C_3-1)(C_2-1)^2
+(2C_2+C_3-3)(C_3-1)^2\Big] \nonumber\\&&\mbox{}\hspace{35mm}
\times\left.
C_1C_2C_3{{\cal S}_1}{{\cal S}_2}{{\cal S}_3}
\right\}
\end{fleqnarray}
where ${\cal S}^{\bf 1}$ is ${\cal S}^M$ in (9.67) with
$M=1$. At this stage, the triple-spectral contribution
cancels, and we obtain
\begin{fleqnarray}
L &=& -2\int^\infty_0dy^2\,\left(\frac4{y^2}\right)^4\,\Big[
(C_3-1)(C_2-1)^2
+2(C_2-1)(C_3-1)^2
\nonumber\\&&\mbox{}\hspace{35mm}
+(C_3-1)^3
\Big]C_1C_2C_3{{{\cal S}_1}}^{\bf 1}{{\cal S}_2}{{\cal S}_3}.
\end{fleqnarray}
Next, one must again use the relation (12.78),
to remove the operators $C$ from ${{\cal S}_2}$, but, in
the term with $(C_2-1)^2$, the bare ${{\cal S}_2}$ will
appear with a subtraction. At this stage, the
double-spectral contribution vanishes owing
to the fact that
\begin{equation}
(C_1-1)^2{{\cal S}_1}^{\bf 1}=0,
\end{equation}
and we obtain
\begin{fleqnarray}
L &=& 2\int^\infty_0dy^2\,\left(\frac4{y^2}\right)^4\,
(C_1-C_2)(C_3-1)^2
C_1C_2C_3{{\cal S}_1}^{\bf 1}
{{\cal S}_2}^{\bf 1}{{\cal S}_3}.
\end{fleqnarray}
With ${\cal S}^{\bf 1}$ in (9.67), this integral
takes the form
\begin{fleqnarray}
L &=& \frac12
\ln ({\Box_2}/{\Box_1})
\int^\infty_0dy^2\,\left(\frac4{y^2}\right)^2\, (C_3-1)^2C_3{{\cal S}_3}
\end{fleqnarray}
where the integrand is a total derivative. Hence
\begin{fleqnarray}
L &=& -8
\ln({\Box_2}/{\Box_1})\left.\left[
\frac1{y^2}(C_3-1)C_3{{\cal S}_3}\right]
\right|_{y=0} =
\ln({\Box_2}/{\Box_1}).
\end{fleqnarray}
These are the typical cancellations occurring in
expressions (12.52)--(12.61). The amount of
calculations with the spectral forms is much smaller
than with the explicit forms of sect. 6, and the
result is again eqs. (12.63)--(12.72). In this way
the anomaly is derived in the generalized spectral
technique.

Finally, the third way of deriving the anomaly
that we present is making the conformal
transformation in the trace of the heat kernel.
The expansion of ${\rm Tr} K(s)$ in powers of the
curvatures is given in eq. (2.1). To enable
a comparison with the effective action (6.1), one
must subtract from the heat kernel the terms
of zeroth and first order in the curvature (see a
remark to eq. (6.1)). These are the first two
terms of expression (2.1). For ${\rm Tr} K(s)$
with these terms subtracted we introduce
the notation ($\omega=2$)
\begin{equation}
{\rm Tr} K'(s)={\rm Tr} K(s)-
\frac1{(4\pi s)^2}
\int\! dx\, g^{1/2}\, {\rm tr}\,(\hat{1}+s\hat{P}).
\end{equation}
The second-order terms in (2.1) transform like in
eq. (12.23) but, instead of the form factor
$\gamma(-\Box)$, one has to deal with the
form factors
\begin{equation}
f(-s\Box),\hspace{7mm}
\frac{f(-s\Box)-1}{s\Box},\hspace{7mm}
\frac{f(-s\Box)-1-\frac16s\Box}{(s\Box)^2}
\end{equation}
where $f(-s\Box)$ is given in (2.9). The
counterparts of eqs. (12.27) and (12.28)
are, in this case
\footnote{\normalsize
For the derivation see sect. 14, eq. (14.8).
},
\begin{fleqnarray}&&
\int\! dx\, g^{1/2}\,
R_{\mu\nu}[f(-s\Box),
\nabla^\mu\nabla^\nu]\sigma \nonumber\\&&\mbox{}
=\int\! dx\, g^{1/2}\,
\frac{f(-s\Box_1)-f(-s\Box_3)}{\Box_1-\Box_3}
[\Box_3,\nabla^\mu_3\nabla^\nu_3]
R_{1\mu\nu}\sigma_3
+{\rm O}[R^3..],
\end{fleqnarray}
\begin{fleqnarray}&&
\int\! dx\, g^{1/2}\, {\rm tr}\,
\Re_1\Big(\delta_\sigma f(-s\Box_2)\Big)\Re_2 \nonumber\\&&\ \ \ \ \mbox{}
=\int\! dx\, g^{1/2}\, {\rm tr}\,
\frac{f(-s\Box_1)-f(-s\Box_2)}{\Box_1-\Box_2}
(\delta_\sigma\Box_2)\Re_1\Re_2
+{\rm O}[\Re^3].
\end{fleqnarray}
The third-order terms in (2.1) transform in a
way completely similar to the above. An important
distinction from the previous case is that the linear
nonlocal terms do not cancel.

The total result for ${\rm Tr} K'(s)$ divided by s is
of the form
\begin{eqnarray}
\frac1s\delta_\sigma{\rm Tr} K'(s) &=&
\frac1{(4\pi)^2}\int\! dx\, g^{1/2}\, {\rm tr}\,
\Big\{
\sigma\Box t_1(s,\Box)\hat{P}
+\sigma\Box t_2(s,\Box)R\hat{1}
\nonumber\\&&\mbox{}
+\sum^{10}_{i=1}
T_i(s,\Box_1,\Box_2,\Box_3)\Re_1\Re_2\sigma_3({i})\Big\}
+{\rm O}[\Re^3]
\end{eqnarray}
where $\Re_1\Re_2\sigma_3({i})$ are the tensor structures
(12.31)--(12.40), and the functions $t_1, t_2, T_i$
are obtained
as certain combinations
of the form factors in the heat
kernel. The differential equations for
these form factors can next be used the same way as
in sect. 5
\footnote{\normalsize
One makes the substitutions (5.7)--(5.9).}
to bring the functions $t_1, t_2, T_i$
to the form of total derivatives:
\begin{equation}
t_1=\frac{d}{ds}\widetilde{t}_1,\hspace{7mm}
t_2=\frac{d}{ds}\widetilde{t}_2,\hspace{7mm}
T_i=\frac{d}{ds}\widetilde{T}_i.
\end{equation}
The final result for
$\widetilde{t}_1,\widetilde{t}_2,\widetilde{T}_i$
is worth presenting in full. In terms of the basic
form factors in the heat kernel:
\begin{equation}
f(-s\Box),\hspace{7mm} F(-s\Box_1,-s\Box_2,-s\Box_3)
\end{equation}
(eqs. (2.9) and (2.75)), and the determinant
$D$ in eq. (6.11), we obtain
\begin{fleqnarray}&&
\widetilde{t}_1 = \frac{f(-s\Box)-1}{s\Box},
\end{fleqnarray}
\begin{fleqnarray}&&
\widetilde{t}_2 = \frac1{12}\frac{f(-s\Box)-1}{s\Box}
-\frac12 \frac{f(-s\Box)-1-\frac16s\Box}{(s\Box)^2},
\end{fleqnarray}
\begin{fleqnarray}&& \widetilde{T}_{1}=
F(-s\Box_1,-s\Box_2,-s\Box_3)
\frac{1}{36{D^4}}
({\Box_3}^8-4 {\Box_1} {\Box_3}^7-16 {\Box_1}^2 {\Box_3}^6
\nonumber\\&&\ \ \ \ \mbox{}
+68 {\Box_1}^3 {\Box_3}^5
-100 {\Box_1}^4 {\Box_3}^4
+68 {\Box_1}^5 {\Box_3}^3-16 {\Box_1}^6 {\Box_3}^2
-4 {\Box_1}^7 {\Box_3}
\nonumber\\&&\ \ \ \ \mbox{}
+2 {\Box_1}^8
+16 {\Box_1} {\Box_2}{\Box_3}^6
-60 {\Box_1}^2 {\Box_2} {\Box_3}^5
+8 {\Box_1}^3 {\Box_2} {\Box_3}^4+68 {\Box_1}^4 {\Box_2} {\Box_3}^3
\nonumber\\&&\ \ \ \ \mbox{}
-48 {\Box_1}^5 {\Box_2} {\Box_3}^2
-4 {\Box_1}^6 {\Box_2} {\Box_3}
+8 {\Box_1}^7 {\Box_2}+
96 {\Box_1}^2 {\Box_2}^2 {\Box_3}^4
\nonumber\\&&\ \ \ \ \mbox{}
-136 {\Box_1}^3{\Box_2}^2 {\Box_3}^3
+36 {\Box_1}^5 {\Box_2}^2 {\Box_3}
-16 {\Box_1}^6 {\Box_2}^2
+64 {\Box_1}^3 {\Box_2}^3 {\Box_3}^2
\nonumber\\&&\ \ \ \ \mbox{}
-28 {\Box_1}^4{\Box_2}^3 {\Box_3}
-40 {\Box_1}^5 {\Box_2}^3
+46 {\Box_1}^4 {\Box_2}^4)
\nonumber\\&&\ \ \ \ \mbox{}
-\frac1s\left(F(-s\Box_1,-s\Box_2,-s\Box_3)-\frac12\right)
\frac{1}{3{D^3}}
(3 {\Box_3}^5-16 {\Box_1} {\Box_3}^4
\nonumber\\&&\ \ \ \ \mbox{}
+4 {\Box_3}^3 {\Box_1}^2+24 {\Box_3}^2 {\Box_1}^3-26 {\Box_3} {\Box_1}^4
+8 {\Box_1}^5+28 {\Box_3}^3 {\Box_1} {\Box_2}
\nonumber\\&&\ \ \ \ \mbox{}
-52 {\Box_3}^2 {\Box_2} {\Box_1}^2+12 {\Box_2} {\Box_1}^4+26
{\Box_3} {\Box_1}^2 {\Box_2}^2-20 {\Box_2}^2 {\Box_1}^3)
\nonumber\\&&\ \ \ \ \mbox{}
+\frac1{s^2}\left(F(-s\Box_1,-s\Box_2,-s\Box_3)-\frac12-\frac1{24}s(\Box_1+\Box_2+\Box_3)\right)
\frac{2}{{D^2}}
\nonumber\\&&\ \ \ \ \mbox{}
\times
({\Box_3}^2-4 {\Box_1} {\Box_3}+2 {\Box_1}^2+4 {\Box_1} {\Box_2})
\nonumber\\&&\ \ \ \ \mbox{}
+\left(\frac{f(-s\Box_1)-1}{s\Box_1}\right)
\frac{{\Box_1}}{6{D^4}}
(-{\Box_3}^2 {\Box_1}^5+{\Box_1} {\Box_2}^6-{\Box_2} {\Box_1}^6
+{\Box_2}^7-{\Box_1}^7
\nonumber\\&&\ \ \ \ \mbox{}
+{\Box_3}^7-37 {\Box_3}^2 {\Box_2}^
4 {\Box_1}+9 {\Box_2}^5 {\Box_3}^2+3 {\Box_3} {\Box_1}^6
+5 {\Box_3}^4 {\Box_1}^3-5 {\Box_1}^4 {\Box_3}^3
\nonumber\\&&\ \ \ \ \mbox{}
-3 {\Box_3}^6 {\Box_1}
-5 {\Box_1}^5 {\Box_2}^2-27 {\Box_1}^3 {\Box_2}^4
+27 {\Box_1}^4 {\Box_2}^3+5 {\Box_1}^2 {\Box_2}^5
-5 {\Box_3}^6 {\Box_2}
\nonumber\\&&\ \ \ \ \mbox{}
-3 {\Box_3} {\Box_1}^4 {\Box_2}^2
+42 {\Box_3}^3 {\Box_1}^2 {\Box_2}^2
-14 {\Box_1}^2 {\Box_2}^3 {\Box_3}^2-12 {\Box_3}^3 {\Box_1}^3 {\Box_2}
\nonumber\\&&\ \ \ \ \mbox{}
-23 {\Box_3}^4 {\Box_1}^2 {\Box_2}
+20{\Box_3} {\Box_1}^3 {\Box_2}^3-57 {\Box_1} {\Box_2}^2 {\Box_3}^4
+22 {\Box_3}^5 {\Box_1} {\Box_2}
\nonumber\\&&\ \ \ \ \mbox{}
-11 {\Box_3} {\Box_1}^2 {\Box_2}^4
+6 {\Box_3} {\Box_1} {\Box_2}^5
+68 {\Box_3}^3 {\Box_1} {\Box_2}^3+29 {\Box_3}^2 {\Box_1}^4 {\Box_2}
\nonumber\\&&\ \ \ \ \mbox{}
+14 {\Box_1}^3 {\Box_2}^2 {\Box_3}^2
-10 {\Box_3} {\Box_1}^5 {\Box_2}+{\Box_3}^5{\Box_1}^2
-5 {\Box_2}^4 {\Box_3}^3+9 {\Box_2}^2 {\Box_3}^5
\nonumber\\&&\ \ \ \ \mbox{}
-5 {\Box_3}^4 {\Box_2}^3
-5 {\Box_2}^6 {\Box_3})
\nonumber\\&&\ \ \ \ \mbox{}
-\left(\frac{f(-s\Box_3)-1}{s\Box_3}\right)
\frac{{\Box_3}}{24{D^4}}
({\Box_3}^7+2 {\Box_3}^6 {\Box_1}
-38 {\Box_3}^5 {\Box_1}^2+90 {\Box_3}^4 {\Box_1}^3
\nonumber\\&&\ \ \ \ \mbox{}
-90 {\Box_1}^4 {\Box_3}^3+38 {\Box_3}^2 {\Box_1}^5
-2 {\Box_3} {\Box_1}^6-2 {\Box_1}^7+10 {\Box_3}^5 {\Box_1} {\Box_2}
\nonumber\\&&\ \ \ \ \mbox{}
-50 {\Box_3}^4 {\Box_1}^2 {\Box_2}+24 {\Box_3}^3{\Box_1}^3 {\Box_2}
-2 {\Box_3}^2 {\Box_1}^4 {\Box_2}+20 {\Box_3} {\Box_1}^5 {\Box_2}
\nonumber\\&&\ \ \ \ \mbox{}
-14 {\Box_2} {\Box_1}^6+66 {\Box_3}^3 {\Box_1}^2 {\Box_2}^2
-36 {\Box_1}^3 {\Box_2}^2 {\Box_3}^2-62 {\Box_3} {\Box_1}^4 {\Box_2}^2
\nonumber\\&&\ \ \ \ \mbox{}
+54 {\Box_1}^5 {\Box_2}^2
+44 {\Box_3} {\Box_1}^3 {\Box_2}^3
-38 {\Box_1}^4 {\Box_2}^3)
\nonumber\\&&\ \ \ \ \mbox{}
-\left(\frac{f(-s\Box_1)-1-\frac16s\Box_1}{(s\Box_1)^2}\right)
\frac{2{\Box_1}^2}{{D^3}}
({\Box_3}^4-4 {\Box_2} {\Box_3}^3
-6 {\Box_1}^2 {\Box_3}^2
\nonumber\\&&\ \ \ \ \mbox{}
+6 {\Box_2}^2 {\Box_3}^2
+8 {\Box_1} {\Box_2} {\Box_3}^2-
16 {\Box_1} {\Box_2}^2 {\Box_3}
-4 {\Box_2}^3 {\Box_3}+8 {\Box_1}^3 {\Box_3}
\nonumber\\&&\ \ \ \ \mbox{}
-2{\Box_1}^2 {\Box_2}^2
-4 {\Box_1}^3 {\Box_2}
-3 {\Box_1}^4+8 {\Box_1} {\Box_2}^3
+{\Box_2}^4)
\nonumber\\&&\ \ \ \ \mbox{}
+\left(\frac{f(-s\Box_3)-1-\frac16s\Box_3}{(s\Box_3)^2}\right)
\frac{{\Box_3}}{4{D^3}}
(7 {\Box_3}^5-22 {\Box_1} {\Box_3}^4
-20 {\Box_3}^3 {\Box_1}^2
\nonumber\\&&\ \ \ \ \mbox{}
+52 {\Box_3}^2 {\Box_1}^3-26 {\Box_3} {\Box_1}^4+
2 {\Box_1}^5
+36 {\Box_3}^3 {\Box_1} {\Box_2}-52 {\Box_3}^2 {\Box_2} {\Box_1}^2
\nonumber\\&&\ \ \ \ \mbox{}
+8 {\Box_3} {\Box_1}^3 {\Box_2}-6 {\Box_2} {\Box_1}^4
+18 {\Box_3} {\Box_1}^2
{\Box_2}^2+4 {\Box_2}^2 {\Box_1}^3)
,\end{fleqnarray}
\begin{fleqnarray}&& \widetilde{T}_{2}=
\frac1{s^2}\left(F(-s\Box_1,-s\Box_2,-s\Box_3)-\frac12-\frac1{24}s(\Box_1+\Box_2+\Box_3)\right)
\frac{2}{{\Box_1}{\Box_2}}
\nonumber\\&&\ \ \ \ \mbox{}
-\left(\frac{f(-s\Box_3)-1}{s\Box_3}\right)
 \frac{{\Box_3} (-{\Box_3}+2 {\Box_1})}{8{\Box_2}{\Box_1}}
\nonumber\\&&\ \ \ \ \mbox{}
+\left(\frac{f(-s\Box_3)-1-\frac16s\Box_3}{(s\Box_3)^2}\right)
 \frac{{\Box_3} (-5 {\Box_3}+2 {\Box_1})}{4{\Box_2}{\Box_1}}
,\end{fleqnarray}
\begin{fleqnarray}&& \widetilde{T}_{3}=
-F(-s\Box_1,-s\Box_2,-s\Box_3)
\frac{2{\Box_1}}{3{D^4}}
({\Box_1}^6-9 {\Box_1}^4 {\Box_2}^2-9 {\Box_3}^2 {\Box_1}^4
\nonumber\\&&\ \ \ \ \mbox{}
-2 {\Box_1}^4 {\Box_2} {\Box_3}+16 {\Box_1}^3 {\Box_2}^3
+8 {\Box_1}^3 {\Box_2}^2 {\Box_3}+8 {\Box_1}^3 {\Box_2} {\Box_3}^2
+16 {\Box_3}^3 {\Box_1}^3
\nonumber\\&&\ \ \ \ \mbox{}
+10 {\Box_3}^2 {\Box_1}^2 {\Box_2}^2
-9 {\Box_1}^2 {\Box_2}^4-9 {\Box_3}^4 {\Box_1}^2
-12 {\Box_1}^2 {\Box_2} {\Box_3}^3-12 {\Box_1}^2 {\Box_2}^3 {\Box_3}
\nonumber\\&&\ \ \ \ \mbox{}
+8 {\Box_1} {\Box_2}^4 {\Box_3}-8 {\Box_1} {\Box_2}^2{\Box_3}^3
-8 {\Box_1} {\Box_2}^3 {\Box_3}^2+8 {\Box_1} {\Box_2} {\Box_3}^4
-2 {\Box_2}^5 {\Box_3}
\nonumber\\&&\ \ \ \ \mbox{}
-{\Box_2}^2 {\Box_3}^4-2 {\Box_3}^5 {\Box_2}+{\Box_3}^6+4
{\Box_2}^3 {\Box_3}^3-{\Box_3}^2 {\Box_2}^4+{\Box_2}^6)
\nonumber\\&&\ \ \ \ \mbox{}
+\frac1s\left(F(-s\Box_1,-s\Box_2,-s\Box_3)-\frac12\right)
\frac{4}{3{D^3}}
(40 {\Box_1}^2 {\Box_2} {\Box_3}+19 {\Box_1}^4
\nonumber\\&&\ \ \ \ \mbox{}
-22 {\Box_1}^3 {\Box_2}-22 {\Box_1}^3 {\Box_3}
-12 {\Box_1}^2 {\Box_3}^2-12 {\Box_1}^2 {\Box_2}^2
-14 {\Box_1} {\Box_2}^2 {\Box_3}
\nonumber\\&&\ \ \ \ \mbox{}
-14 {\Box_1} {\Box_3}^2 {\Box_2}
+14 {\Box_1} {\Box_3}^3+14 {\Box_1} {\Box_2}^3
+{\Box_2}^4+6 {\Box_2}^2{\Box_3}^2
-4 {\Box_2}^3 {\Box_3}
\nonumber\\&&\ \ \ \ \mbox{}
-4 {\Box_2} {\Box_3}^3+{\Box_3}^4)
\nonumber\\&&\ \ \ \ \mbox{}
-\frac1{s^2}\left(F(-s\Box_1,-s\Box_2,-s\Box_3)-\frac12-\frac1{24}s(\Box_1+\Box_2+\Box_3)\right)   \frac{48 {\Box_1}}{{D^2}}
\nonumber\\&&\ \ \ \ \mbox{}
+\left(\frac{f(-s\Box_1)-1}{s\Box_1}\right)
\frac{4{\Box_1}^2}{3{D^4}}
({\Box_1}^5+{\Box_1}^4 {\Box_3}+{\Box_1}^4 {\Box_2}
-8 {\Box_1}^3 {\Box_3}^2
\nonumber\\&&\ \ \ \ \mbox{}
-8 {\Box_1}^3 {\Box_2}^2
+8 {\Box_1}^2 {\Box_2}^3+8 {\Box_1}^2 {\Box_3}^3
-{\Box_1} {\Box_2}^4-4 {\Box_1} {\Box_2}^3 {\Box_3}
\nonumber\\&&\ \ \ \ \mbox{}
-{\Box_1} {\Box_3}^4
-4 {\Box_2} {\Box_3}^3 {\Box_1}+10 {\Box_1} {\Box_2}^2{\Box_3}^2
+3 {\Box_2}^4 {\Box_3}-{\Box_2}^5-{\Box_3}^5
\nonumber\\&&\ \ \ \ \mbox{}
-2 {\Box_2}^2 {\Box_3}^3-2 {\Box_2}^3 {\Box_3}^2+3 {\Box_2} {\Box_3}^4)
\nonumber\\&&\ \ \ \ \mbox{}
-\left(\frac{f(-s\Box_2)-1}{s\Box_2}\right)
\frac{1}{6{\Box_1}{D^4}}
({\Box_3}^8+{\Box_1}^8+28 {\Box_1}^6 {\Box_3}^2
-56 {\Box_1}^5 {\Box_3}^3
\nonumber\\&&\ \ \ \ \mbox{}
-56 {\Box_1}^3 {\Box_3}^5
+70 {\Box_1}^4 {\Box_3}^4+28 {\Box_1}^2 {\Box_3}^6
-8 {\Box_1}^7 {\Box_3}-8 {\Box_1} {\Box_3}^7
\nonumber\\&&\ \ \ \ \mbox{}
-61 {\Box_1}^3 {\Box_2} {\Box_3}^4-7 {\Box_2} {\Box_3}^7
+35 {\Box_2}^4 {\Box_3}^4-35 {\Box_2}^3 {\Box_3}^5
+21 {\Box_2}^2 {\Box_3}^6
\nonumber\\&&\ \ \ \ \mbox{}
+99 {\Box_1}^4 {\Box_2}^4
-99 {\Box_1}^5 {\Box_2}^3-29 {\Box_1}^3{\Box_2}^5
-{\Box_1}^2 {\Box_2}^6+29 {\Box_1}^6 {\Box_2}^2
\nonumber\\&&\ \ \ \ \mbox{}
+{\Box_1}^7 {\Box_2}-{\Box_1} {\Box_2}^7
-{\Box_2}^7 {\Box_3}-21 {\Box_2}^5 {\Box_3}^3
+7 {\Box_2}^6 {\Box_3}^2+23 {\Box_1}^6 {\Box_2} {\Box_3}
\nonumber\\&&\ \ \ \ \mbox{}
-27 {\Box_1}^2 {\Box_2} {\Box_3}^5-5 {\Box_1}^4 {\Box_2}^2 {\Box_3}^2
+139 {\Box_1}^4 {\Box_2} {\Box_3}^3-99 {\Box_1}^5 {\Box_2} {\Box_3}^2
\nonumber\\&&\ \ \ \ \mbox{}
+31 {\Box_1} {\Box_2} {\Box_3}^6+100 {\Box_1}^3 {\Box_2}^2 {\Box_3}^3
+50 {\Box_1}^3 {\Box_2}^3 {\Box_3}^2 +81 {\Box_1}^4 {\Box_2}^3 {\Box_3}
\nonumber\\&&\ \ \ \ \mbox{}
-45 {\Box_1}^2 {\Box_2}^2 {\Box_3}^4+66 {\Box_1}^2 {\Box_2}^4 {\Box_3}^2
-39 {\Box_1}^2 {\Box_2}^5 {\Box_3}-47 {\Box_1} {\Box_2}^5 {\Box_3}^2
\nonumber\\&&\ \ \ \ \mbox{}
+14 {\Box_1} {\Box_2}^6 {\Box_3}+60 {\Box_1} {\Box_2}^4 {\Box_3}^3
-4 {\Box_1}^3 {\Box_2}^4 {\Box_3}-66 {\Box_1}^5 {\Box_2}^2 {\Box_3}
\nonumber\\&&\ \ \ \ \mbox{}
-15 {\Box_1} {\Box_2}^3 {\Box_3}^4-34 {\Box_1} {\Box_2}^2 {\Box_3}^5
+18 {\Box_1}^2 {\Box_2}^3 {\Box_3}^3)
\nonumber\\&&\ \ \ \ \mbox{}
-\left(\frac{f(-s\Box_3)-1}{s\Box_3}\right)
\frac{{\Box_3}}{6{\Box_1}{D^4}}
({\Box_1}^6 {\Box_3}+29 {\Box_1}^4 {\Box_3}^3
-{\Box_3}^7+9 {\Box_1}^7+{\Box_2}^7
\nonumber\\&&\ \ \ \ \mbox{}
-7 {\Box_2}^6 {\Box_3}-35 {\Box_2}^4{\Box_3}^3
-21 {\Box_2}^2 {\Box_3}^5+7 {\Box_2} {\Box_3}^6
+35 {\Box_2}^3 {\Box_3}^4
\nonumber\\&&\ \ \ \ \mbox{}
+7 {\Box_3}^6 {\Box_1}
-43 {\Box_1}^5 {\Box_3}^2-27 {\Box_1}^5 {\Box_2}^2
+27 {\Box_1}^3 {\Box_3}^4-29 {\Box_1}^2 {\Box_3}^5
\nonumber\\&&\ \ \ \ \mbox{}
-17 {\Box_1}^6 {\Box_2}-9 {\Box_1} {\Box_2}^6
+45 {\Box_1}^2 {\Box_2}^5-101 {\Box_1}^3 {\Box_2}^4
+99 {\Box_1}^4 {\Box_2}^3
\nonumber\\&&\ \ \ \ \mbox{}
+21 {\Box_2}^5 {\Box_3}^2
+6 {\Box_1}^5 {\Box_2} {\Box_3}-26 {\Box_2} {\Box_3}^5 {\Box_1}
+30{\Box_1}^2 {\Box_2}^2 {\Box_3}^3
\nonumber\\&&\ \ \ \ \mbox{}
+25 {\Box_1} {\Box_2}^2 {\Box_3}^4
-81 {\Box_1}^2 {\Box_2}^4 {\Box_3}+84 {\Box_1}^3 {\Box_2}^3 {\Box_3}
+20 {\Box_1}{\Box_2}^3 {\Box_3}^3
\nonumber\\&&\ \ \ \ \mbox{}
+34 {\Box_1}^3 {\Box_2}^2 {\Box_3}^2
-55 {\Box_1} {\Box_2}^4 {\Box_3}^2+2 {\Box_1}^2 {\Box_2}^3 {\Box_3}^2
-41 {\Box_1}^4 {\Box_2}^2 {\Box_3}
\nonumber\\&&\ \ \ \ \mbox{}
+41 {\Box_1}^4 {\Box_2} {\Box_3}^2
-44 {\Box_1}^3 {\Box_2} {\Box_3}^3+33 {\Box_1}^2 {\Box_2} {\Box_3}^4
+38 {\Box_1} {\Box_2}^5 {\Box_3})
\nonumber\\&&\ \ \ \ \mbox{}
-\left(\frac{f(-s\Box_1)-1-\frac16s\Box_1}{(s\Box_1)^2}\right)
\frac{16 {\Box_1}^3}{{D^3}}
(-2 {\Box_3}^2+4 {\Box_2} {\Box_3}
+3 {\Box_1}^2
\nonumber\\&&\ \ \ \ \mbox{}
-{\Box_1} {\Box_2}-2 {\Box_2}^2-{\Box_3} {\Box_1})
\nonumber\\&&\ \ \ \ \mbox{}
+\left(\frac{f(-s\Box_2)-1-\frac16s\Box_2}{(s\Box_2)^2}\right)
\frac{1}{{\Box_1}{D^3}}
({\Box_1}^6+{\Box_3}^6
+58 {\Box_1}^4 {\Box_2}^2
\nonumber\\&&\ \ \ \ \mbox{}
+15 {\Box_3}^2 {\Box_1}^4
-18 {\Box_1}^3 {\Box_2}^3-20 {\Box_3}^3 {\Box_1}^3
-27 {\Box_1}^2 {\Box_2}^4+15 {\Box_3}^4 {\Box_1}^2
\nonumber\\&&\ \ \ \ \mbox{}
-{\Box_2}^5 {\Box_3}+10 {\Box_2}^2 {\Box_3}^4
-5 {\Box_3}^5 {\Box_2}-10 {\Box_2}^3{\Box_3}^3
+5 {\Box_3}^2 {\Box_2}^4
\nonumber\\&&\ \ \ \ \mbox{}
+31 {\Box_1}^4 {\Box_2} {\Box_3}
-34 {\Box_1}^3 {\Box_2} {\Box_3}^2-88 {\Box_1}^3 {\Box_2}^2 {\Box_3}
+6 {\Box_1}^2{\Box_2} {\Box_3}^3
\nonumber\\&&\ \ \ \ \mbox{}
+12 {\Box_3}^2 {\Box_1}^2 {\Box_2}^2
+22 {\Box_1} {\Box_2}^4 {\Box_3}+8 {\Box_1} {\Box_2}^2 {\Box_3}^3
-6 {\Box_1}^2 {\Box_2}^3 {\Box_3}
\nonumber\\&&\ \ \ \ \mbox{}
-30 {\Box_1} {\Box_2}^3 {\Box_3}^2
+11 {\Box_1} {\Box_2} {\Box_3}^4-6 {\Box_1}^5 {\Box_3}
-9 {\Box_1}^5 {\Box_2}
\nonumber\\&&\ \ \ \ \mbox{}
-5 {\Box_1} {\Box_2}^5-6 {\Box_1} {\Box_3}^5)
\nonumber\\&&\ \ \ \ \mbox{}
-\left(\frac{f(-s\Box_3)-1-\frac16s\Box_3}{(s\Box_3)^2}\right)
\frac{{\Box_3}}{{\Box_1}{D^3}}
(-{\Box_2}^5-10 {\Box_2}^3 {\Box_3}^2
-5 {\Box_2} {\Box_3}^4
\nonumber\\&&\ \ \ \ \mbox{}
+10 {\Box_2}^2 {\Box_3}^3
+{\Box_3}^5-{\Box_1} {\Box_3}^4+3{\Box_1}^5
-6 {\Box_1}^2 {\Box_2} {\Box_3}^2-4 {\Box_2} {\Box_3}^3 {\Box_1}
\nonumber\\&&\ \ \ \ \mbox{}
-18 {\Box_1}^2 {\Box_2}^3+22 {\Box_1}^3 {\Box_2}^2
+18 {\Box_1} {\Box_2}^2{\Box_3}^2+76 {\Box_1}^3 {\Box_2} {\Box_3}
\nonumber\\&&\ \ \ \ \mbox{}
-20 {\Box_1} {\Box_2}^3 {\Box_3}-18 {\Box_1}^2 {\Box_2}^2 {\Box_3}
+5 {\Box_2}^4 {\Box_3}-13 {\Box_1}^4 {\Box_2}
\nonumber\\&&\ \ \ \ \mbox{}
-43 {\Box_1}^4 {\Box_3}-2 {\Box_1}^3 {\Box_3}^2
+42 {\Box_1}^2 {\Box_3}^3+7 {\Box_1} {\Box_2}^4)
\nonumber\\&&\ \ \ \ \mbox{}
-\frac1{{\Box_2}-{\Box_3}}\left(\frac{f(-s\Box_2)-1}{s\Box_2}-\frac{f(-s\Box_3)-1}{s\Box_3}\right)
\frac{{\Box_3}}{6{\Box_1}}
\nonumber\\&&\mbox{}\ \
+\frac1{{\Box_2}-{\Box_3}}\Big(\frac{f(-s\Box_2)-1-\frac16s\Box_2}{(s\Box_2)^2}  -\frac{f(-s\Box_3)-1-\frac16s\Box_3}{(s\Box_3)^2}\Big)
\frac{{\Box_3}}{{\Box_1}}
,\end{fleqnarray}
\begin{fleqnarray}&& \widetilde{T}_{4}=
-\frac1s\left(F(-s\Box_1,-s\Box_2,-s\Box_3)-\frac12\right)
\frac{8}{{D^2}}
(-{\Box_3}^2+2 {\Box_1}^2-2 {\Box_2} {\Box_1})
\nonumber\\&&\ \ \ \ \mbox{}
+\frac1{s^2}\left(F(-s\Box_1,-s\Box_2,-s\Box_3)-\frac12-\frac1{24}s(\Box_1+\Box_2+\Box_3)\right) \nonumber\\&&\ \ \ \ \ \ \ \ \mbox{}\times
\frac{8 (-{\Box_3}+2 {\Box_1})}{{D}{\Box_1}{\Box_2}}
\nonumber\\&&\ \ \ \ \mbox{}
-\left(\frac{f(-s\Box_3)-1}{s\Box_3}\right)
\frac{{\Box_3}}{2{\Box_2}{\Box_1}}
\nonumber\\&&\ \ \ \ \mbox{}
+\left(\frac{f(-s\Box_1)-1-\frac16s\Box_1}{(s\Box_1)^2}\right)
\frac{32 {\Box_1}^2 ({\Box_3}+{\Box_1}-{\Box_2})}{{D^2}}
\nonumber\\&&\ \ \ \ \mbox{}
+\left(\frac{f(-s\Box_3)-1-\frac16s\Box_3}{(s\Box_3)^2}\right)   \frac{{\Box_3}}{{D^2}{\Box_2}{\Box_1}}
(5 {\Box_3}^4-32 {\Box_3}^3 {\Box_1}
+36 {\Box_3}^2 {\Box_1}^2
\nonumber\\&&\ \ \ \ \mbox{}
-16 {\Box_3} {\Box_1}^3
+2 {\Box_1}^4-20 {\Box_3}^2 {\Box_1} {\Box_2}
+16 {\Box_2} {\Box_3} {\Box_1}^2
\nonumber\\&&\ \ \ \ \mbox{}
-8 {\Box_2} {\Box_1}^3
+6 {\Box_1}^2 {\Box_2}^2)
,\end{fleqnarray}
\begin{fleqnarray}&& \widetilde{T}_{5}=
F(-s\Box_1,-s\Box_2,-s\Box_3)
\frac{4 {\Box_1} {\Box_2}}{D^4}
({\Box_3}^4-4 {\Box_1}^2 {\Box_3}^2
+2 {\Box_1}^4
\nonumber\\&&\ \ \ \ \mbox{}
+4 {\Box_1} {\Box_2} {\Box_3}^2
-8 {\Box_2} {\Box_1}^3+6 {\Box_1}^2 {\Box_2}^2)
\nonumber\\&&\ \ \ \ \mbox{}
-\frac1s\left(F(-s\Box_1,-s\Box_2,-s\Box_3)-\frac12\right)
\frac{16}{D^3}
({\Box_3}^3-{\Box_1} {\Box_3}^2
-4 {\Box_1}^2 {\Box_3}
\nonumber\\&&\ \ \ \ \mbox{}
+3 {\Box_1}^3
+4 {\Box_1} {\Box_2} {\Box_3}-3 {\Box_1}^2 {\Box_2})
\nonumber\\&&\ \ \ \ \mbox{}
+\frac1{s^2}\left(F(-s\Box_1,-s\Box_2,-s\Box_3)-\frac12-\frac1{24}s(\Box_1+\Box_2+\Box_3)\right)
\frac{8}{D^2{\Box_1}{\Box_2}}
\nonumber\\&&\ \ \ \ \mbox{}
\times
({\Box_3}^2-4 {\Box_1} {\Box_3}+2 {\Box_1}^2+4 {\Box_1} {\Box_2})
\nonumber\\&&\ \ \ \ \mbox{}
-\left(\frac{f(-s\Box_1)-1}{s\Box_1}\right)
\frac{16 {\Box_1}^2 {\Box_2}}{D^4}
({\Box_1}^3+{\Box_1}^2 {\Box_3}
-3 {\Box_1}^2 {\Box_2}+3 {\Box_1} {\Box_2}^2
\nonumber\\&&\ \ \ \ \mbox{}
-{\Box_1} {\Box_3}^2-2 {\Box_1} {\Box_2} {\Box_3}
-{\Box_2}^3+{\Box_2}^2 {\Box_3}+{\Box_2} {\Box_3}^2-{\Box_3}^3)
\nonumber\\&&\ \ \ \ \mbox{}
-\left(\frac{f(-s\Box_3)-1}{s\Box_3}\right)
\frac{{\Box_3}}{2D^4{\Box_1}{\Box_2}}
(-{\Box_3}^7+14 {\Box_1} {\Box_3}^6
-42 {\Box_1}^2 {\Box_3}^5
\nonumber\\&&\ \ \ \ \mbox{}
+70 {\Box_1}^3 {\Box_3}^4
-70 {\Box_1}^4 {\Box_3}^3+42 {\Box_1}^5 {\Box_3}^2
-14 {\Box_3} {\Box_1}^6+2 {\Box_1}^7
\nonumber\\&&\ \ \ \ \mbox{}
-26 {\Box_1} {\Box_2} {\Box_3}^5+50 {\Box_1}^2 {\Box_2} {\Box_3}^4
+40 {\Box_1}^3 {\Box_2} {\Box_3}^3-110 {\Box_1}^4 {\Box_2} {\Box_3}^2
\nonumber\\&&\ \ \ \ \mbox{}
+76 {\Box_1}^5 {\Box_2} {\Box_3}-18 {\Box_1}^6 {\Box_2}
+30 {\Box_1}^2 {\Box_2}^2 {\Box_3}^3+68 {\Box_1}^3 {\Box_2}^2 {\Box_3}^2
\nonumber\\&&\ \ \ \ \mbox{}
-178 {\Box_1}^4 {\Box_2}^2 {\Box_3}
+42 {\Box_1}^5 {\Box_2}^2+116 {\Box_1}^3 {\Box_2}^3 {\Box_3}
-26 {\Box_1}^4 {\Box_2}^3)
\nonumber\\&&\ \ \ \ \mbox{}
+\left(\frac{f(-s\Box_1)-1-\frac16s\Box_1}{(s\Box_1)^2}\right)
 \frac{32 {\Box_1}^2}{D^3}
(3 {\Box_1}^2-{\Box_1} {\Box_3}
-{\Box_1} {\Box_2}
\nonumber\\&&\ \ \ \ \mbox{}
+4 {\Box_2} {\Box_3}-2 {\Box_3}^2-2 {\Box_2}^2)
\nonumber\\&&\ \ \ \ \mbox{}
+\left(\frac{f(-s\Box_3)-1-\frac16s\Box_3}{(s\Box_3)^2}\right)
\frac{{\Box_3}}{{\Box_2}{\Box_1} D^3}
(-5 {\Box_3}^5+42 {\Box_1} {\Box_3}^4
-68 {\Box_1}^2 {\Box_3}^3
\nonumber\\&&\ \ \ \ \mbox{}
+52 {\Box_1}^3 {\Box_3}^2
-18 {\Box_1}^4 {\Box_3}+2{\Box_1}^5
+20 {\Box_1} {\Box_2} {\Box_3}^3
-52 {\Box_1}^2 {\Box_2} {\Box_3}^2
\nonumber\\&&\ \ \ \ \mbox{}
-24 {\Box_1}^3 {\Box_2} {\Box_3}-6 {\Box_1}^4 {\Box_2}
+42 {\Box_1}^2 {\Box_2}^2 {\Box_3}+4 {\Box_1}^3 {\Box_2}^2)
,\end{fleqnarray}
\begin{fleqnarray}&& \widetilde{T}_{6}=
-F(-s\Box_1,-s\Box_2,-s\Box_3)
\frac{1}{3 D^2}
({\Box_2}^4-4 {\Box_2}^3 {\Box_3}
+2 {\Box_1} {\Box_2}^3
+6 {\Box_2}^2 {\Box_3}^2
\nonumber\\&&\ \ \ \ \mbox{}
-6 {\Box_1}^2 {\Box_2}^2-2 {\Box_1} {\Box_2}^2{\Box_3}
+4 {\Box_1}^2 {\Box_2} {\Box_3}+2 {\Box_1}^3 {\Box_2}
-4 {\Box_2} {\Box_3}^3
\nonumber\\&&\ \ \ \ \mbox{}
-2 {\Box_1} {\Box_2} {\Box_3}^2
+{\Box_3}^4-6 {\Box_3}^2 {\Box_1}^2+{\Box_1}^4+
2 {\Box_1} {\Box_3}^3+2 {\Box_3} {\Box_1}^3)
\nonumber\\&&\ \ \ \ \mbox{}
+\frac1s\left(F(-s\Box_1,-s\Box_2,-s\Box_3)-\frac12\right)
\frac{4 {\Box_1}}{D}
\nonumber\\&&\ \ \ \ \mbox{}
+\left(\frac{f(-s\Box_1)-1}{s\Box_1}\right)
\frac{4 {\Box_1}^2}{D^2}
({\Box_1} {\Box_2}+{\Box_1} {\Box_3}
-{\Box_2}^2-{\Box_3}^2+2 {\Box_2} {\Box_3})
\nonumber\\&&\ \ \ \ \mbox{}
-\left(\frac{f(-s\Box_2)-1}{s\Box_2}\right)
\frac{{\Box_2}}{D^2}
({\Box_1}^3+{\Box_2} {\Box_1}^2
-{\Box_1}^2 {\Box_3}-{\Box_2}^2 {\Box_1}
\nonumber\\&&\ \ \ \ \mbox{}
-{\Box_1} {\Box_3}^2+2 {\Box_2} {\Box_1} {\Box_3}
-{\Box_2}^3-3 {\Box_2}{\Box_3}^2+3 {\Box_3} {\Box_2}^2+{\Box_3}^3)
\nonumber\\&&\ \ \ \ \mbox{}
-\left(\frac{f(-s\Box_3)-1}{s\Box_3}\right)
\frac{{\Box_3}}{D^2}
({\Box_1}^3-{\Box_2} {\Box_1}^2
+{\Box_1}^2 {\Box_3}-{\Box_2}^2 {\Box_1}
\nonumber\\&&\ \ \ \ \mbox{}
-{\Box_1} {\Box_3}^2+2 {\Box_2} {\Box_1} {\Box_3}
+{\Box_2}^3+3 {\Box_2}{\Box_3}^2-3 {\Box_3} {\Box_2}^2-{\Box_3}^3)
,\end{fleqnarray}
\begin{fleqnarray}&& \widetilde{T}_{7}=
F(-s\Box_1,-s\Box_2,-s\Box_3)
\frac{4 {\Box_2}}{D^2}
({\Box_3}^2+2 {\Box_1} {\Box_3}
-2 {\Box_2} {\Box_3}+{\Box_1}^2
\nonumber\\&&\ \ \ \ \mbox{}
-2 {\Box_1} {\Box_2}+{\Box_2}^2)
\nonumber\\&&\ \ \ \ \mbox{}
-\frac1s\left(F(-s\Box_1,-s\Box_2,-s\Box_3)-\frac12\right)
\frac{8}{D}
\nonumber\\&&\ \ \ \ \mbox{}
+\left(\frac{f(-s\Box_1)-1}{s\Box_1}\right)
\frac{2}{{\Box_2} D^2}
({\Box_1}^3 {\Box_2}+{\Box_1}^3 {\Box_3}
-7 {\Box_1}^2 {\Box_2}^2-3 {\Box_1}^2 {\Box_3}^2
\nonumber\\&&\ \ \ \ \mbox{}
-6 {\Box_1}^2 {\Box_2} {\Box_3}-11 {\Box_1}{\Box_2}^2 {\Box_3}
+3 {\Box_1} {\Box_3}^3+7 {\Box_1} {\Box_2}^3
+{\Box_1} {\Box_2} {\Box_3}^2
\nonumber\\&&\ \ \ \ \mbox{}
+4 {\Box_2}^3 {\Box_3}
-6 {\Box_2}^2 {\Box_3}^2+4 {\Box_2} {\Box_3}^3-
{\Box_2}^4-{\Box_3}^4)
\nonumber\\&&\ \ \ \ \mbox{}
+\left(\frac{f(-s\Box_2)-1}{s\Box_2}\right)
\frac{8 {\Box_2}^2 ({\Box_3}+{\Box_1}-{\Box_2})}{{D^2}}
\nonumber\\&&\ \ \ \ \mbox{}
-\left(\frac{f(-s\Box_3)-1}{s\Box_3}\right)
\frac{2 {\Box_3}}{{\Box_2}{D^2}}
({\Box_1}^3-3 {\Box_3} {\Box_1}^2
-5 {\Box_2} {\Box_1}^2+2 {\Box_3} {\Box_1} {\Box_2}
\nonumber\\&&\ \ \ \ \mbox{}
+7 {\Box_1} {\Box_2}^2+3 {\Box_1} {\Box_3}^2
+3 {\Box_2}{\Box_3}^2+{\Box_2}^2 {\Box_3}-{\Box_3}^3-3 {\Box_2}^3)
\nonumber\\&&\ \ \ \ \mbox{}
-\frac1{{\Box_1}-{\Box_3}}\left(\frac{f(-s\Box_1)-1}{s\Box_1}-\frac{f(-s\Box_3)-1}{s\Box_3}\right)
 \frac{2 {\Box_3}}{{\Box_2}}
,\end{fleqnarray}
\begin{fleqnarray}&& \widetilde{T}_{8}=
F(-s\Box_1,-s\Box_2,-s\Box_3)
,\end{fleqnarray}
\begin{fleqnarray}&& \widetilde{T}_{9}=
-F(-s\Box_1,-s\Box_2,-s\Box_3)
\frac{2 {\Box_1} {\Box_2}}{D^2}
(-{\Box_3}^2+2 {\Box_1}^2-2 {\Box_1} {\Box_2})
\nonumber\\&&\ \ \ \ \mbox{}
+\frac1s\left(F(-s\Box_1,-s\Box_2,-s\Box_3)-\frac12\right)
\frac{2 (-{\Box_3}+2 {\Box_1})}{D}
\nonumber\\&&\ \ \ \ \mbox{}
+\left(\frac{f(-s\Box_1)-1}{s\Box_1}\right)
 \frac{8 {\Box_2} {\Box_1}^2 ({\Box_3}-{\Box_2}+{\Box_1})}{D^2}
\nonumber\\&&\ \ \ \ \mbox{}
-\left(\frac{f(-s\Box_3)-1}{s\Box_3}\right)
\frac{{\Box_3}}{D^2}
(-{\Box_3}^3+6 {\Box_3}^2 {\Box_1}
-6 {\Box_3} {\Box_1}^2+2 {\Box_1}^3
\nonumber\\&&\ \ \ \ \mbox{}
+6 {\Box_3} {\Box_2} {\Box_1}-2 {\Box_2} {\Box_1}^2)
,\end{fleqnarray}
\begin{fleqnarray}&& \widetilde{T}_{10}=
-F(-s\Box_1,-s\Box_2,-s\Box_3)
\frac{2}{D^2}
({\Box_3}^3-2 {\Box_1} {\Box_3}^2
-2 {\Box_1}^2 {\Box_3}+2 {\Box_1}^3
\nonumber\\&&\ \ \ \ \mbox{}
+2 {\Box_1} {\Box_2} {\Box_3}-2 {\Box_2} {\Box_1}^2)
\nonumber\\&&\ \ \ \ \mbox{}
+\frac1s\left(F(-s\Box_1,-s\Box_2,-s\Box_3)-\frac12\right)
\frac{8}{D}
\nonumber\\&&\ \ \ \ \mbox{}
-\left(\frac{f(-s\Box_1)-1}{s\Box_1}\right)
\frac{8 {\Box_1}}{D^2}
({\Box_3}^2-2 {\Box_2} {\Box_3}-{\Box_1}^2+{\Box_2}^2)
\nonumber\\&&\ \ \ \ \mbox{}
+\left(\frac{f(-s\Box_3)-1}{s\Box_3}\right)
\frac{4 {\Box_3} ({\Box_3}^2-2 {\Box_1}^2+2 {\Box_1} {\Box_2})}{D^2}
.\end{fleqnarray}

The conformal variation of the effective action
(6.1) can now be obtained as
\begin{equation}
-\delta_\sigma W = \frac12
\int^\infty_0\frac{ds}s\delta_\sigma{\rm Tr} K'(s).
\end{equation}
>From (12.89) and (12.90) we find
\begin{eqnarray}
-\delta_\sigma W &=& -
{\frac1{2(4\pi)^2}\int\! dx\, g^{1/2}\, \,{\rm tr}\,}
\Big\{
\sigma\Box\widetilde{t}_1(0,\Box)\hat{P}
+\sigma\Box\widetilde{t}_2(0,\Box)R\hat{1}  \nonumber\\&&\mbox{}
+\sum^{10}_{i=1}
\widetilde{T}_i(0,\Box_1,\Box_2,\Box_3)
\Re_1\Re_2\sigma_3({i})\Big\}
+{\rm O}[\Re^3]
\end{eqnarray}
where use is made of the fact that the functions
$\widetilde{t}_1,\widetilde{t}_2,\widetilde{T}_i$
as given in eqs. (12.92)--(12.103) vanish at
$s\rightarrow\infty$. The behaviours of these
functions at $s=0$ follow from the results of sect. 4 :
\arraycolsep=0pt
\begin{fleqnarray} \widetilde{t}_1(0,\Box) &=& \frac16, \end{fleqnarray}
\begin{fleqnarray} \widetilde{t}_2(0,\Box) &=& \frac1{180}, \end{fleqnarray}
\begin{fleqnarray} \frac12\Big(\widetilde{T}_{1}+
{\widetilde{T}_{1}}{\big|}_{\Box_1\leftrightarrow\Box_2}\Big) &=& 0 ,\hspace{7mm} s=0\end{fleqnarray}
\begin{fleqnarray} \frac12\Big(\widetilde{T}_{2}+
{\widetilde{T}_{2}}{\big|}_{\Box_1\leftrightarrow\Box_2}\Big) &=&
   \frac1{180}+\frac{\Box_1}{180\Box_2}+\frac{\Box_2}{180\Box_1}\nonumber\\&&\mbox{}
   -\frac{\Box_3}{90\Box_1}-\frac{\Box_3}{90\Box_2}
   +\frac{{\Box_3}^2}{180\Box_1\Box_2} ,\hspace{7mm} s=0\end{fleqnarray}
\begin{fleqnarray}  \widetilde{T}_{3} &=& 0 ,\hspace{7mm} s=0\end{fleqnarray}
\begin{fleqnarray} \frac12(\widetilde{T}_{4}+{\widetilde{T}_{4}}{\big|}_{\Box_1\leftrightarrow\Box_2}) &=&
  \frac1{45\Box_1}+\frac1{45\Box_2}
-\frac{\Box_3}{45\Box_1\Box_2} ,\hspace{7mm} s=0\end{fleqnarray}
\begin{fleqnarray} \frac12\Big(\widetilde{T}_{5}+{\widetilde{T}_{5}}{\big|}_{\Box_1\leftrightarrow\Box_2}\Big) &=&
   \frac1{45\Box_1\Box_2} ,\hspace{7mm} s=0\end{fleqnarray}
\begin{fleqnarray} \widetilde{T}_{6} &=& 0 ,\hspace{7mm} s=0\end{fleqnarray}
\begin{fleqnarray} \widetilde{T}_{7} &=& 0 ,\hspace{7mm} s=0\end{fleqnarray}
\begin{fleqnarray} \frac12\Big(\widetilde{T}_{8}
+{\widetilde{T}_{8}}{\big|}_{\Box_1\leftrightarrow\Box_2}\Big) &=& \frac12,\hspace{7mm} s=0\end{fleqnarray}
\begin{fleqnarray} \frac12\Big(\widetilde{T}_{9}
+{\widetilde{T}_{9}}{\big|}_{\Box_1\leftrightarrow\Box_2}\Big) &=& \frac1{12},\hspace{7mm} s=0\end{fleqnarray}
\begin{fleqnarray} \frac12\Big(\widetilde{T}_{10}
+{\widetilde{T}_{10}}{\big|}_{\Box_1\leftrightarrow\Box_2}\Big) &=& 0,\hspace{7mm} s=0.\end{fleqnarray}
\arraycolsep=5pt
With these expressions inserted in (12.105), one
arrives at eq. (12.73) which is the correct trace
anomaly.

Because the conformal transformation is inhomogeneous
in the curvature, the expansion in powers of the
curvature does not preserve the exact conformal
properties of the effective action. These
properties can only be recovered order by
order. One can try to remove this shortcoming of
covariant perturbation theory by using the ideas
of ref. [22] but such an improvement is
already beyond the scope of the present paper.

This concludes the presentation of the results
concerning third order in the curvature, and the
remaining part of the paper is devoted to their
derivations. We start in the next section with
perturbation theory for the heat kernel, and
subsequently outline in succession the techniques
used for obtaining the results in sects. 2 and 6--11.

\section{Third order of perturbation theory for the
trace of the heat kernel}
\setcounter{equation}{0}

\hspace{\parindent}In covariant perturbation theory [1,2],
the heat kernel is first expanded in powers of
perturbations:
\begin{equation}
K(s) = \sum^\infty_{n=0} K_n(s)
\end{equation}
where $K_n(s)$ is a term of $n$-th power in the perturbations
of metric, connection and potential
\begin{equation}
h^{\mu\nu},\hspace{7mm} \widehat{\Gamma}_\mu,\hspace{7mm}
\hat{P}-\frac16R\hat{1}.
\end{equation}
When calculated by the algorithm of paper II, the trace
of $K_n(s)$ is obtained in the form
\begin{eqnarray}
{\rm Tr} K_n(s) &=& \frac1{(4\pi s)^\omega}\frac1n
\int dx\,\widetilde{g}^{1/2}(x)\,
\int_{\alpha_i\geq0}d^n\alpha\,
  \delta(1-\sum^n_1\alpha_i)   \nonumber\\&&\mbox{}
\times{\rm tr}\Big\{\exp\Big[s\Omega_n
  (\alpha_1,\dots\alpha_n|{\widetilde{\nabla}}^i)\Big]\nonumber\\&&\mbox{}\ \ \ \
\times\sum^n_{l=0}s^l\widehat{B}^l_n
(\alpha_1,\dots\alpha_n|x_i)\Big\}\Big|_{x_i=x}
\end{eqnarray}
(eq. (5.46) of paper II), or, with the notation
\begin{equation}
\int_{\alpha_i\geq0}d^n\alpha\,
  \delta(1-\sum^n_1\alpha_i)
  f(\alpha_1,\dots\alpha_n|x_i)\Big|_{x_i=x}
= \big<f\big>_n,
\end{equation}
\begin{equation}
{\rm Tr} K_n(s) = \frac1{(4\pi s)^\omega}\frac1n
\int dx\,\widetilde{g}^{1/2}\,
\sum^n_{l=0}s^l{\rm tr}\big<
{\rm e}^{s\Omega_n}
\widehat{B}^l_n\big>_n.
\end{equation}
Here ${\widetilde{g}}$ and ${\widetilde{\nabla}}$ are auxiliary, flat, metric and
covariant derivative,\break
$\Omega_n(\alpha_1,\dots\alpha_n|{\widetilde{\nabla}}^i)$ is an
operator of second order in ${\widetilde{\nabla}}^i$, and
${\widetilde{\nabla}}^i$ acts on the perturbation number $i$
contained in ${\hat{B}}^l_n$. Each term in
$\widehat{B}^l_n(\alpha_1,\dots\alpha_n|x_i)$
where $i$ ranges from 1 to $n$ is a product
of $n$ perturbations (13.2) at the points
$x_1,\dots x_n$ respectively, and the label
$i$ on a perturbation means that the perturbation
is at the point $x_i$. For example,
\[
h_1^{\mu\nu}{\hat{\Gamma}}_{2\alpha}\hat{P}_3
= h^{\mu\nu}(x_1)
{\hat{\Gamma}}_\alpha(x_2)\hat{P}(x_3).
\]
After the action of ${\widetilde{\nabla}}^i$, all points $x_i$
are made coincident with the integration point $x$
in (13.3) or (13.5). The possibility of integration
over $x$ by parts is then expressed by the identity
\begin{equation}
\sum^n_{i=1}{\widetilde{\nabla}}^i = 0
\end{equation}
which is used to put the third-order form factors
in the form given below.

Since each perturbation $h^{\mu\nu}$ or
${\hat{\Gamma}}_\mu$ is an infinite series in the curvature, the
expansion to a given order in the curvature involves
all lower orders in perturbations. Therefore, we begin
with quoting the results of [2] for
$n=1$ and
$n=2$ in (13.3). For $n=1$,
\begin{eqnarray}
\Omega_1&=&0,\\[\baselineskip]
{\hat{B}}^0_1 &=& -\frac12h\hat{1},\\[\baselineskip]
{\hat{B}}^1_1 &=& \frac13{\widetilde{\nabla}}_\mu{\widetilde{\nabla}}_\nu h^{\mu\nu}\hat{1}
-\frac1{12}{\widetilde{\vphantom{I}\Box}} h\hat{1}-{\widetilde{\nabla}}_\mu{\hat{\Gamma}}^\mu
+\hat{P}-\frac16 R\hat{1}.
\end{eqnarray}
Here and below
\begin{equation}
h=h^{\mu\nu}
\widetilde{g}_{\mu\nu},\hspace{7mm}
{\widetilde{\vphantom{I}\Box}}={\widetilde{g}}^{\mu\nu}{\widetilde{\nabla}}_\mu{\widetilde{\nabla}}_\nu,
\end{equation}
and the indices of ${\widetilde{\nabla}}_\mu$ and the perturbations
are raised and lowered with the flat metric
${\widetilde{g}}_{\mu\nu}$, except for
\begin{equation}
{\hat{\Gamma}}^\mu\equiv({\widetilde{g}}^{\mu\nu}+h^{\mu\nu}){\hat{\Gamma}}_\nu.
\end{equation}

For $n=2$,
\begin{eqnarray}&&
\Omega_2(\alpha_1,\alpha_2|{\widetilde{\nabla}}^i)
  = \alpha_1\alpha_2{\widetilde{\vphantom{I}\Box}}_2,
\\[\baselineskip]&&
{\hat{B}}^0_2(\alpha_1,\alpha_2|x_i) =
  \Big(\frac14h_1h_2+\frac12h_{1\mu\nu}h_2^{\mu\nu}\Big)\hat{1},
\\[\baselineskip]&&
{\hat{B}}^1_2(\alpha_1,\alpha_2|x_i) =
  -\alpha^2_1({\widetilde{\nabla}}_\mu{\widetilde{\nabla}}_\nu h^{\mu\nu}_1)h_2\hat{1} \nonumber\\&&\ \ \ \ \mbox{}
  -2\alpha_1\alpha_2({\widetilde{\nabla}}_\nu h^{\nu\mu}_1)
   {\widetilde{g}}_{\mu\alpha}({\widetilde{\nabla}}_\beta h_2^{\beta\alpha})\hat{1}
  -2{\hat{\Gamma}}_1^\mu{\widetilde{g}}_{\mu\nu}{\hat{\Gamma}}_2^\nu
  +2\alpha_2 h_1 ({\widetilde{\nabla}}_\mu{\hat{\Gamma}}_2^\mu) \nonumber\\&&\ \ \ \ \mbox{}
  +4\alpha_1({\widetilde{\nabla}}_\mu h_1^{\mu\nu}){\widetilde{g}}_{\nu\alpha}
   {\hat{\Gamma}}_2^\alpha
  -h_1\Big(\hat{P}_2-\frac16R_2\hat{1}\Big),
\\[\baselineskip]&&
{\hat{B}}^2_2(\alpha_1,\alpha_2|x_i) =
  \hat{1}(\alpha_1\alpha_2)^2
    ({\widetilde{\nabla}}_\mu{\widetilde{\nabla}}_\nu h_1^{\mu\nu})
    ({\widetilde{\nabla}}_\alpha{\widetilde{\nabla}}_\beta h_2^{\alpha\beta})\nonumber\\&&\ \ \ \ \mbox{}
  +4\alpha_1\alpha_2
    ({\widetilde{\nabla}}_\mu{\hat{\Gamma}}_1^\mu)
    ({\widetilde{\nabla}}_\nu{\hat{\Gamma}}_2^\nu)
  -4\alpha_1^2\alpha_2
    ({\widetilde{\nabla}}_\mu{\widetilde{\nabla}}_\nu h_1^{\mu\nu})
    ({\widetilde{\nabla}}_\alpha{\hat{\Gamma}}_2^\alpha) \nonumber\\&&\ \ \ \ \mbox{}
  +2\alpha_1^2
    ({\widetilde{\nabla}}_\mu{\widetilde{\nabla}}_\nu h_1^{\mu\nu})\Big(\hat{P}_2-\frac16R_2\hat{1}\Big) \nonumber\\&&\ \ \ \ \mbox{}
  -4\alpha_2\Big(\hat{P}_1-\frac16R_1\hat{1}\Big)
    ({\widetilde{\nabla}}_\mu{\hat{\Gamma}}_2^\mu)
  +\Big(\hat{P}_1-\frac16R_1\hat{1}\Big)\Big(\hat{P}_2-\frac16R_2\hat{1}\Big).
\end{eqnarray}

A routine calculation by the algorithm of
paper II gives for $n=3$
\begin{eqnarray}&&
\Omega_3(\alpha_1,\alpha_2,\alpha_3|{\widetilde{\nabla}}^i) =
  \alpha_2\alpha_3{\widetilde{\vphantom{I}\Box}}_1
  +\alpha_1\alpha_3{\widetilde{\vphantom{I}\Box}}_2
  +\alpha_1\alpha_2{\widetilde{\vphantom{I}\Box}}_3,
\\[\baselineskip]&&
{\hat{B}}^0_3(\alpha_1,\alpha_2,\alpha_3|x_i) =
  -\hat{1}\Big(\frac18h_1h_2h_3
  +\frac34h_1 h_2^{\mu\nu}h_{3\mu\nu}
  +h_{1\mu}^\alpha h_{2\nu}^\mu h_{3\alpha}^\nu\Big),
\\[\baselineskip]&&
{\hat{B}}^1_3(\alpha_1,\alpha_2,\alpha_3|x_i) = \nonumber\\&&\ \ \ \ \mbox{}
  \hat{1}\left[3({\widetilde{g}}_{\alpha\beta}{\widetilde{g}}_{\mu\nu}
    +2{\widetilde{g}}_{\mu\alpha}{\widetilde{g}}_{\nu\beta})
     \Big(D^1_\lambda D^2_\sigma
      +\frac14 D^2_\lambda D^2_\sigma\Big)\right]
    h_1^{\alpha\beta}h_2^{\mu\nu}h_3^{\lambda\sigma} \nonumber\\&&\ \ \ \ \mbox{}
  -3\Big({\widetilde{g}}^{(2)}_{\alpha\nu\lambda\sigma} D^1_\mu
     +{\widetilde{g}}^{(2)}_{\mu\nu\alpha\sigma} D^2_\lambda
     +\frac12{\widetilde{g}}^{(2)}_{\mu\nu\lambda\sigma} D^3_\alpha\Big)
      {\hat{\Gamma}}_1^\alpha h_2^{\mu\nu} h_3^{\lambda\sigma}   \nonumber\\&&\ \ \ \ \mbox{}
  +3({\widetilde{g}}_{\alpha\beta}{\hat{\Gamma}}_1^\alpha{\hat{\Gamma}}_2^\beta h_3
      +2{\hat{\Gamma}}_1^\alpha{\hat{\Gamma}}_2^\beta h_{3\alpha\beta})
  +\frac34{\widetilde{g}}^{(2)}_{\mu\nu\alpha\beta}\Big(\hat{P}_1-\frac16R_1\hat{1}\Big)
     h_2^{\mu\nu}h_3^{\alpha\beta},
\\[\baselineskip]&&
{\hat{B}}^2_3(\alpha_1,\alpha_2,\alpha_3|x_i) =  \nonumber\\&&\ \ \ \ \mbox{}
  -\hat{1}\Big(\frac32
      {\widetilde{g}}_{\alpha\beta}
      D^1_\mu D^1_\nu D^2_\lambda D^2_\sigma
    +6 {\widetilde{g}}_{\mu\alpha}
      D^3_\beta D^1_\nu D^2_\lambda D^2_\sigma\Big)
    h_1^{\alpha\beta}h_2^{\mu\nu}h_3^{\lambda\sigma}  \nonumber\\&&\ \ \ \ \mbox{}
  +12{\widetilde{g}}_{\alpha\beta}D^2_\mu
     {\hat{\Gamma}}_1^\alpha{\hat{\Gamma}}_2^\beta{\hat{\Gamma}}_3^\mu \nonumber\\&&\ \ \ \ \mbox{}
  +(3{\widetilde{g}}_{\mu\nu} D^3_\alpha D^2_\lambda D^2_\sigma
    +3{\widetilde{g}}_{\lambda\sigma} D^3_\alpha D^1_\mu D^1_\nu
    +6{\widetilde{g}}_{\alpha\mu} D^1_\nu D^2_\lambda D^2_\sigma   \nonumber\\&&\ \ \ \ \ \ \ \ \mbox{}
    +6{\widetilde{g}}_{\alpha\lambda} D^1_\mu D^1_\nu D^2_\sigma
    +12{\widetilde{g}}_{\mu\lambda} D^3_\alpha D^1_\nu D^2_\sigma )
    {\hat{\Gamma}}_1^\alpha h_2^{\mu\nu}h_3^{\lambda\sigma}  \nonumber\\&&\ \ \ \ \mbox{}
  -( 6{\widetilde{g}}_{\mu\nu} D^3_\alpha D^1_\beta
    +12{\widetilde{g}}_{\mu\beta} D^3_\alpha D^2_\nu  \nonumber\\&&\ \ \ \ \ \ \ \ \mbox{}
    +6{\widetilde{g}}_{\alpha\beta} D^2_\mu D^2_\nu
    +12{\widetilde{g}}_{\alpha\mu} D^1_\beta D^2_\nu )
    {\hat{\Gamma}}_1^\alpha{\hat{\Gamma}}_2^\beta h_3^{\mu\nu}    \nonumber\\&&\ \ \ \ \mbox{}
  -\frac32\Big(\hat{P}_1-\frac16R_1\hat{1}\Big)\Big(\hat{P}_2-\frac16R_2\hat{1}\Big) h_3   \nonumber\\&&\ \ \ \ \mbox{}
  +( 3{\widetilde{g}}_{\mu\nu} D^1_\alpha
    +6{\widetilde{g}}_{\mu\alpha} D^2_\nu
    -3{\widetilde{g}}_{\mu\nu} D^3_\alpha
    -6{\widetilde{g}}_{\mu\alpha} D^1_\nu )
    \Big(\hat{P}_1-\frac16R_1\hat{1}\Big){\hat{\Gamma}}_2^\alpha h_3^{\mu\nu}  \nonumber\\&&\ \ \ \ \mbox{}
  -6{\widetilde{g}}_{\alpha\beta}\Big(\hat{P}_1-\frac16R_1\hat{1}\Big){\hat{\Gamma}}_2^\alpha{\hat{\Gamma}}_3^\beta \nonumber\\&&\ \ \ \ \mbox{}
  -\frac32(
      {\widetilde{g}}_{\mu\nu} D^2_\lambda D^2_\sigma
     +{\widetilde{g}}_{\lambda\sigma} D^1_\mu D^1_\nu
     +4{\widetilde{g}}_{\mu\lambda} D^1_\nu D^2_\sigma )
     \Big(\hat{P}_1-\frac16R_1\hat{1}\Big) h_2^{\mu\nu} h_3^{\lambda\sigma} ,
\\[\baselineskip]&&
{\hat{B}}^3_3(\alpha_1,\alpha_2,\alpha_3|x_i) =
  D^3_\alpha D^3_\beta D^1_\mu
    D^1_\nu D^2_\lambda D^2_\sigma
    h_1^{\alpha\beta} h_2^{\mu\nu} h_3^{\lambda\sigma}\hat{1} \nonumber\\&&\ \ \ \ \mbox{}
  -8 D^3_\alpha D^1_\beta D^2_\mu
     {\hat{\Gamma}}_1^\alpha {\hat{\Gamma}}_2^\beta {\hat{\Gamma}}_3^\mu
  -6 D^3_\alpha D^1_\mu D^1_\nu D^2_\lambda D^2_\sigma
     {\hat{\Gamma}}_1^\alpha h_2^{\mu\nu} h_3^{\lambda\sigma}  \nonumber\\&&\ \ \ \ \mbox{}
  +12 D^3_\alpha D^1_\beta D^2_\mu D^2_\nu
     {\hat{\Gamma}}_1^\alpha{\hat{\Gamma}}_2^\beta h_3^{\mu\nu}
  +\Big(\hat{P}_1-\frac16R_1\hat{1}\Big)\Big(\hat{P}_2-\frac16R_2\hat{1}\Big)\Big(\hat{P}_3-\frac16R_3\hat{1}\Big)  \nonumber\\&&\ \ \ \ \mbox{}
  +3 D^2_\alpha D^2_\beta\Big(\hat{P}_1-\frac16R_1\hat{1}\Big)\Big(\hat{P}_2-\frac16R_2\hat{1}\Big) h_3^{\alpha\beta} \nonumber\\&&\ \ \ \ \mbox{}
  -6 D^2_\mu\Big(\hat{P}_1-\frac16R_1\hat{1}\Big)\Big(\hat{P}_2-\frac16R_2\hat{1}\Big){\hat{\Gamma}}^\mu_3  \nonumber\\&&\ \ \ \ \mbox{}
  +6( D^3_\alpha D^1_\mu D^1_\nu
     -D^1_\alpha D^2_\mu D^2_\nu )
    \Big(\hat{P}_1-\frac16R_1\hat{1}\Big){\hat{\Gamma}}^\alpha_2 h_3^{\mu\nu}  \nonumber\\&&\ \ \ \ \mbox{}
  +12 D^1_\alpha D^2_\beta \Big(\hat{P}_1-\frac16R_1\hat{1}\Big){\hat{\Gamma}}_2^\alpha{\hat{\Gamma}}_3^\beta \nonumber\\&&\ \ \ \ \mbox{}
  +3 D^1_\mu D^1_\nu D^2_\lambda D^2_\sigma
    \Big(\hat{P}_1-\frac16R_1\hat{1}\Big) h_2^{\mu\nu} h_3^{\lambda\sigma}
\end{eqnarray}
where
\begin{eqnarray}
D^1_\mu &=& \alpha_3{\widetilde{\nabla}}^2_\mu-\alpha_2{\widetilde{\nabla}}^3_\mu,\\[\baselineskip]
D^2_\mu &=& \alpha_1{\widetilde{\nabla}}^3_\mu-\alpha_3{\widetilde{\nabla}}^1_\mu,\\[\baselineskip]
D^3_\mu &=& \alpha_2{\widetilde{\nabla}}^1_\mu-\alpha_1{\widetilde{\nabla}}^2_\mu,
\end{eqnarray}
and
\begin{equation}
{\widetilde{g}}^{(2)}_{\mu\nu\alpha\beta} =
 {\widetilde{g}}_{\mu\alpha}{\widetilde{g}}_{\nu\beta}
+{\widetilde{g}}_{\mu\beta}{\widetilde{g}}_{\nu\alpha}
+{\widetilde{g}}_{\mu\nu}{\widetilde{g}}_{\alpha\beta}.
\end{equation}

\section{Expansion of ${\rm Tr} K(s)$ in powers of
the curvatures to third order}
\setcounter{equation}{0}

\hspace{\parindent}
Next step is replacing the perturbations $h^{\mu\nu}$
and ${\hat{\Gamma}}_\mu$ by their expressions through the
curvatures, and eliminating the auxiliary
quantities ${\widetilde{g}}_{\alpha\beta}$ ant ${\widetilde{\nabla}}_\alpha$.
The iterational solutions for $h^{\mu\nu}$ and
${\hat{\Gamma}}_\mu$ are needed now to third order in the curvature
whereas in paper II they were obtained to second order
(eqs. (4.28), (4.29) of II). It is, of course, possible
to work out these solutions to third order but we shall
avoid this calculation by using the following trick.
Let us rewrite eqs. (13.1), (13.5) as follows:
\begin{eqnarray}
{\rm Tr} K(s) &=& \frac1{(4\pi s)^\omega}
  \int d x\, {\widetilde{g}}^{1/2}\,{\rm tr}\,\left[\hat{1} +{\hat{B}}^0_1
  +\frac12\big<{\hat{B}}^0_2\big>_2
  +\frac13\big<{\hat{B}}^0_3\big>_3\right] \nonumber\\
 &+&\frac{s}{(4\pi s)^\omega}
    \int d x\, {\widetilde{g}}^{1/2}\,{\rm tr}\,\left[{\hat{B}}^1_1
  +\frac12\big<\Omega_2{\hat{B}}^0_2\big>_2
\right.
\nonumber\\&&\mbox{}\hspace{40mm}
\left.
  +\frac12\big<{\hat{B}}^1_2\big>_2
  +\frac13\big<\Omega_3{\hat{B}}^0_3\big>_3
  +\frac13\big<{\hat{B}}^1_3\big>_3  \right] \nonumber\\
 &+&\frac{s^2}{2(4\pi s)^\omega}
    \int d x\, {\widetilde{g}}^{1/2}\,{\rm tr}\,\left[
    \Big<\frac{{\rm e}^{s\Omega_2}-1-s\Omega_2}{s^2}
          {\hat{B}}^0_2\Big>_2
\right.
\nonumber\\&&\mbox{}\hspace{40mm}
\left.
  +\Big<\frac{{\rm e}^{s\Omega_2}-1}{s}{\hat{B}}^1_2\Big>_2
  +\big<{\rm e}^{s\Omega_2}{\hat{B}}^2_2\big>_2\right] \nonumber\\
 &+&\frac{s^3}{3(4\pi s)^\omega}
    \int d x\, {\widetilde{g}}^{1/2}\,{\rm tr}\,\left[
    \Big<\frac{{\rm e}^{s\Omega_3}-1-s\Omega_3}{s^3}
          {\hat{B}}^0_3\Big>_3
\right.
\nonumber\\&&\mbox{}\hspace{40mm}
  +\Big<\frac{{\rm e}^{s\Omega_3}-1}{s^2}{\hat{B}}^1_3\Big>_3
  +\Big<\frac{{\rm e}^{s\Omega_3}}{s}{\hat{B}}^2_3\Big>_3
\nonumber\\&&\mbox{}\hspace{40mm}
\left.
  +\big<{\rm e}^{s\Omega_3}{\hat{B}}^3_3\big>_3\right]
+{\rm O}[\Re^4].
\end{eqnarray}
The purpose of this decomposition is to single out
from the outset the terms of zeroth and first order in the
curvature which can only be local and coincident with
the coefficients $\hat{a}_0(x,x)$ and
$\hat{a}_1(x,x)$ of the Schwinger-DeWitt expansion.
The connection between  covariant perturbation theory
and the Schwinger-DeWitt technique was discussed in
paper II where an expression was obtained for the DeWitt
coefficients in terms of perturbation theory
(eq. (8.8)  of II). In particular,
\begin{fleqnarray}&&
\int\! dx\, g^{1/2}\, {\rm tr}\,\hat{a}_0(x,x) = \int dx\,{\widetilde{g}}^{1/2}\,{\rm tr}\,\Big[
  \hat{1} + {\hat{B}}^0_1
\nonumber\\&&\mbox{}\hspace{58mm}
+\frac12\big<{\hat{B}}^0_2\big>_2
+\frac13\big<{\hat{B}}^0_3\big>_3
\Big]+{\rm O}[\Re^4],
\end{fleqnarray}
\begin{fleqnarray}&&
\int\! dx\, g^{1/2}\, {\rm tr}\,\hat{a}_1(x,x) = \int dx\,{\widetilde{g}}^{1/2}\,{\rm tr}\,\left[
  {\hat{B}}^1_1
+\frac12\big<\Omega_2{\hat{B}}^0_2\big>_2
+\frac12\big<{\hat{B}}^1_2\big>_2
\right.
\nonumber\\&&\mbox{}\hspace{55mm}
\left.
+\frac13\big<\Omega_3{\hat{B}}^0_3\big>_3
+\frac13\big<{\hat{B}}^1_3\big>_3
\right]+{\rm O}[\Re^4],
\end{fleqnarray}
and, therefore, for the first two integrals in (14.1)
we can use the known exact results
\begin{eqnarray}
\int\! dx\, g^{1/2}\, {\rm tr}\,\hat{a}_0(x,x) &=& \int\! dx\, g^{1/2}\, {\rm tr}\,\hat{1}, \\[\baselineskip]
\int\! dx\, g^{1/2}\, {\rm tr}\,\hat{a}_1(x,x) &=& \int\! dx\, g^{1/2}\, {\rm tr}\,\hat{P}.
\end{eqnarray}
To second order in the curvature, it was checked in paper II
that these results really follow from eqs. (14.2) and (14.3).
To third order in the curvature such a check would require
a knowledge of the iterational solutions for $h^{\mu\nu}$
and ${\hat{\Gamma}}_\mu$ to third order. Instead of deriving these
solutions, we shall take eqs. (14.4) and (14.5) for
granted. Then, after elimination of the first two
integrals in expression (14.1), the remaining terms
in this expression are already of second and third order
in perturbations, and, therefore, the knowledge of
$h^{\mu\nu}$ and ${\hat{\Gamma}}_\mu$ to second order in the
curvature is sufficient for their calculation.

Consider now the third integral in (14.1) which
involves terms of second order in perturbations:
${\hat{B}}^0_2, {\hat{B}}^1_2, {\hat{B}}^2_2$. It is straightforward to
substitute in (13.13)--(13.15) the equations of
paper II expressing $h^{\mu\nu}$ and ${\hat{\Gamma}}_\mu$
through $R_{\mu\nu}$ and $\hat{\cal R}_{\mu\nu}$ (use is made of
eqs. (4.28)--(4.34), (7.3)--(7.5) of II) but, in the
second-order form factor itself
\begin{equation}
{\rm e}^{s\Omega_2} = {\rm e}^{s\alpha_1\alpha_2{\widetilde{\vphantom{I}\Box}}},
\end{equation}
when expressing ${\widetilde{\vphantom{I}\Box}}$ through $\Box$, the terms linear
in the curvature should also be retained. The expression
of ${\widetilde{\vphantom{I}\Box}}$ through $\Box$ is generally of the form
\begin{equation}
{\widetilde{\vphantom{I}\Box}} = \Box + {\cal O}(\Re,\nabla) + {\rm O}[\Re^2]
\end{equation}
where ${\cal O}(\Re,\nabla)$ is an operator containing
the curvature linearly. Its explicit forms are given
in paper II for all cases where $\Box$ acts on a
scalar, tensor, matrix, etc (eqs. (4.30)--(4.34) of II).
The expansion of the form factor ${\rm e}^{s\Omega_2}$
is then accomplished as follows:
\begin{eqnarray}&&
\int\! dx\, g^{1/2}\, \Re_1({\rm e}^{a{\widetilde{\vphantom{I}\Box}}}-{\rm e}^{a\Box})\Re_2 \nonumber\\&&\ \ \ \ \mbox{}
=\int\! dx\, g^{1/2}\, \Re_1\int^a_0dt\,{\rm e}^{(a-t)\Box}
{\cal O}(\Re,\nabla){\rm e}^{t\Box}\Re_2
+ {\rm O}[\Re^4] \nonumber\\&&\ \ \ \ \mbox{}
=\int\! dx\, g^{1/2}\, \int^a_0dt\,{\rm e}^{(a-t)\Box_1+t\Box_2}
{\cal O}(\Re_3,\nabla_2)\Re_1\Re_2
+ {\rm O}[\Re^4] \nonumber\\&&\ \ \ \ \mbox{}
=\int\! dx\, g^{1/2}\, \frac{{\rm e}^{a\Box_2}-{\rm e}^{a\Box_1}}{\Box_2-\Box_1}
{\cal O}(\Re_3,\nabla_2)\Re_1\Re_2
+ {\rm O}[\Re^4]
\end{eqnarray}
with $a=s\alpha_1\alpha_2$. Here the numbers on the arguments
of ${\cal O}(\Re_3,\nabla_2)$ mean that ${\cal O}$ as an
operator acts  on $\Re_2$, and the curvature that it contains
acquires the number 3. Expression (14.8) represents a
third-order contribution with the form factor of a new
type.

In this way, for the terms with the second-order form factor
in (14.1), we obtain the expansions which extend to third
order eqs. (7.7), (7.8) and (7.9) of paper II:
\mathindent=0pt
\begin{fleqnarray}&&
\int d x\, {\widetilde{g}}^{1/2}\,{\rm tr}\,
    \Big<\frac{{\rm e}^{s\Omega_2}-1-s\Omega_2}{s^2}
          {\hat{B}}^0_2\Big>_2
\nonumber\\&&\ \ \ \ \mbox{}
=\int\! dx\, g^{1/2}\, {\rm tr}\Big<{\cal C}_2(R_1R_2
  +2R_1^{\mu\nu}R_{2\mu\nu})\hat{1}\Big>_2
\nonumber\\&&\ \ \ \ \mbox{}
+\int\! dx\, g^{1/2}\, {\rm tr}\left<\left(-\frac1{\Box_2}{\cal C}_1+2\frac{\Box_1}{\Box_2\Box_3}{\cal C}_3
-s{\cal W}_{12}+\frac{\Box_1}{\Box_3}s{\cal W}_{12}\right)R_1 R_2 R_3\hat{1}
\right.
\nonumber\\&&\ \ \ \ \ \ \ \ \mbox{}
+4\frac{\Box_1}{\Box_2\Box_3}{\cal C}_1R_{1\,\alpha}^\mu R_{2\,\beta}^{\alpha} R_{3\,\mu}^\beta\hat{1} \nonumber\\&&\ \ \ \ \ \ \ \ \mbox{}
+\left(3\frac{\Box_3}{\Box_1\Box_2}{\cal C}_3-2\frac1{\Box_1}{\cal C}_3+2\frac{\Box_1}{\Box_2\Box_3}{\cal C}_1
  -2\frac1{\Box_2}{\cal C}_1-s{\cal W}_{12}\right)
R_1^{\mu\nu}R_{2\,\mu\nu}R_3\hat{1} \nonumber\\&&\ \ \ \ \ \ \ \ \mbox{}
+\left(-4\frac1{\Box_2\Box_3}{\cal C}_1-4\frac1{\Box_3}s{\cal W}_{12}
  +2\frac1{\Box_1}s{\cal W}_{23}\right)
R_1^{\alpha\beta}\nabla_\alpha R_2 \nabla_\beta R_3\hat{1} \nonumber\\&&\ \ \ \ \ \ \ \ \mbox{}
+\left(8\frac1{\Box_1\Box_3}{\cal C}_2-4\frac1{\Box_1\Box_2}{\cal C}_3+4\frac1{\Box_3}s{\cal W}_{12}\right)
\nabla^\mu R_1^{\nu\alpha}\nabla_\nu R_{2\,\mu\alpha}R_3\hat{1} \nonumber\\&&\ \ \ \ \ \ \ \ \mbox{}
+\left(8\frac1{\Box_1\Box_2}{\cal C}_3+4\frac1{\Box_2\Box_3}{\cal C}_1+4\frac1{\Box_1}s{\cal W}_{23}\right)
R_1^{\mu\nu}\nabla_\mu R_2^{\alpha\beta}\nabla_\nu R_{3\,\alpha\beta}\hat{1}
\nonumber\\&&\ \ \ \ \ \ \ \ \mbox{}
\left.
+\left(16\frac1{\Box_1\Box_3}{\cal C}_2+8\frac1{\Box_2\Box_3}{\cal C}_1+16\frac1{\Box_3}s{\cal W}_{12}\right)
R_1^{\mu\nu}\nabla_\alpha R_{2\,\beta\mu}\nabla^\beta R_{3\,\nu}^\alpha\hat{1}
\right>_2
\nonumber\\&&\ \ \ \ \ \ \ \ \mbox{}
+ {\rm O}[\Re^4],
\end{fleqnarray}
\begin{fleqnarray}&&
\int d x\, {\widetilde{g}}^{1/2}\,{\rm tr}\,
    \Big<\frac{{\rm e}^{s\Omega_2}-1}{s}
          {\hat{B}}^1_2\Big>_2
\nonumber\\&&\ \ \ \ \mbox{}
=\int\! dx\, g^{1/2}\, {\rm tr}\Big<{\cal B}_2\Big[\hat{\cal R}_1^{\mu\nu}\hat{\cal R}_{2\mu\nu}
\nonumber\\&&\ \ \ \ \ \ \ \ \mbox{}\ \ \ \
 +\Big(2\alpha_1\alpha_2-2\alpha_1^2+\frac13\Big)
R_1R_2\hat{1}
 -2R_1\hat{P}_2\Big]\Big>_2 \nonumber\\&&\ \ \ \ \mbox{}
+\int\! dx\, g^{1/2}\, {\rm tr}\left<
  \left(-2\frac1{\Box_2}{\cal B}_1+2\frac1{\Box_1}{\cal B}_3-\frac{\Box_3}{\Box_1\Box_2}{\cal B}_3
\right.\right.
\nonumber\\&&\ \ \ \ \ \ \ \ \mbox{}\ \ \ \
\left.
 +s{\cal V}_{23}+\frac{\Box_2}{\Box_1}s{\cal V}_{23}-\frac{\Box_3}{\Box_1}s{\cal V}_{23}\right)R_1 R_2 \hat{P}_3 \nonumber\\&&\ \ \ \ \ \ \ \ \mbox{}
+\left(2\frac1{\Box_1}{\cal B}_3-3\frac{\Box_3}{\Box_1\Box_2}{\cal B}_3\right)R_1^{\mu\nu}R_{2\,\mu\nu}\hat{P}_3 \nonumber\\&&\ \ \ \ \ \ \ \ \mbox{}
+\left(\frac1{\Box_1}{\cal B}_2-\frac12s{\cal V}_{23}\right)R_1\hat{\cal R}^{\mu\nu}_2\hat{\cal R}_{3\,\mu\nu} \nonumber\\&&\ \ \ \ \ \ \ \ \mbox{}
-4\frac1{\Box_1}{\cal B}_3R_1^{\alpha\beta}\hat{\cal R}_{2\,\alpha}^{\ \ \ \,\mu}\hat{\cal R}_{3\,\beta\mu} \nonumber\\&&\ \ \ \ \ \ \ \ \mbox{}
+\left[\Big(2\alpha_1\alpha_2-2\alpha_1^2+\frac16\Big)
   \frac{\Box_1}{\Box_2\Box_3}{\cal B}_1
   +2(\alpha^2_1-\alpha_1\alpha_2)\frac1{\Box_2}{\cal B}_1
\right.
\nonumber\\&&\ \ \ \ \ \ \ \ \mbox{}\ \ \ \
\left.
 -\Big(\alpha_1\alpha_2-\alpha_1^2+\frac16\Big)
   s{\cal V}_{23}\right]R_1 R_2 R_3\hat{1} \nonumber\\&&\ \ \ \ \ \ \ \ \mbox{}
+\left[\Big(2\alpha_1\alpha_2-4\alpha_1^2+\frac12\Big)\frac{\Box_3}{\Box_1\Box_2}{\cal B}_3
\right.
\nonumber\\&&\ \ \ \ \ \ \ \ \mbox{}\ \ \ \
\left.
  +\Big(4\alpha_1^2-4\alpha_1\alpha_2-\frac13\Big)\frac1{\Box_1}{\cal B}_3\right]
   R_1^{\mu\nu}R_{2\,\mu\nu}R_3\hat{1} \nonumber\\&&\ \ \ \ \ \ \ \ \mbox{}
+\left(-4\frac1{\Box_1\Box_2}{\cal B}_3-4\frac1{\Box_1}s{\cal V}_{23}\right)R_1^{\mu\nu}\nabla_\mu R_2\nabla_\nu \hat{P}_3  \nonumber\\&&\ \ \ \ \ \ \ \ \mbox{}
-4\frac1{\Box_1\Box_2}{\cal B}_3\nabla^\mu R_1^{\nu\alpha}\nabla_\nu R_{2\,\mu\alpha}\hat{P}_3 \nonumber\\&&\ \ \ \ \ \ \ \ \mbox{}
-8\frac1{\Box_1\Box_3}{\cal B}_2 R_{1\,\alpha\beta}\nabla_\mu\hat{\cal R}_2^{\mu\alpha}\nabla_\nu\hat{\cal R}_3^{\nu\beta} \nonumber\\&&\ \ \ \ \ \ \ \ \mbox{}
+2\frac1{\Box_1}s{\cal V}_{23} R_1^{\alpha\beta}\nabla_\alpha\hat{\cal R}_2^{\mu\nu}\nabla_\beta\hat{\cal R}_{3\,\mu\nu}  \nonumber\\&&\ \ \ \ \ \ \ \ \mbox{}
+\left[-\Big(4\alpha^2_1-\frac23\Big)\frac1{\Box_1\Box_2}{\cal B}_3
\right.
\nonumber\\&&\ \ \ \ \ \ \ \ \mbox{}\ \ \ \
\left.
   +\Big(4\alpha_1\alpha_2
   -4\alpha_1^2+\frac23\Big)\frac1{\Box_1}s{\cal V}_{23}\right]R_1^{\alpha\beta}\nabla_\alpha R_2 \nabla_\beta R_3\hat{1}  \nonumber\\&&\ \ \ \ \ \ \ \ \mbox{}
+\Big(8\alpha_1^2-8\alpha_1\alpha_2-\frac23\Big)
  \frac1{\Box_1\Box_2}{\cal B}_3 \nabla^\mu R_1^{\nu\alpha}\nabla_\nu R_{2\,\mu\alpha}R_3\hat{1} \nonumber\\&&\ \ \ \ \ \ \ \ \mbox{}
-4\frac1{\Box_1}s{\cal V}_{23}
  \hat{\cal R}^{\alpha\beta}_1
  \nabla_\alpha\hat{\cal R}_2^{\mu\nu}
  \nabla_\beta\hat{\cal R}_{3\mu\nu} \nonumber\\&&\ \ \ \ \ \ \ \ \mbox{}
+4\frac1{\Box_1\Box_2\Box_3}({\cal B}_1+{\cal B}_2)
  \nabla_\mu\nabla_\lambda\hat{\cal R}_1^{\lambda\nu}
  \nabla^\alpha\hat{\cal R}_{2\alpha\nu}
  \nabla_\sigma\hat{\cal R}_3^{\sigma\mu}  \nonumber\\&&\ \ \ \ \ \ \ \ \mbox{}
+8\alpha_1\frac1{\Box_2\Box_3}{\cal B}_1
  \nabla^\alpha\hat{\cal R}_{1\alpha\beta}
  R_2^{\beta\mu}\nabla_\mu R_3
\nonumber\\&&\ \ \ \ \ \ \ \ \mbox{}
\left.
+4\frac1{\Box_1}s{\cal V}_{23}(
  \nabla_\alpha R_{1\beta\mu}-
  \nabla_\mu R_{1\beta\alpha})
  \nabla^\beta\hat{\cal R}_2^{\alpha\nu}
  \hat{\cal R}_{3\,\cdot\,\nu}^{\ \mu}
\right>_2
+{\rm O}[\Re^4],
\end{fleqnarray}
\begin{fleqnarray}&&
\int d x\, {\widetilde{g}}^{1/2}\,{\rm tr}\,\Big<{\rm e}^{s\Omega_2}{\hat{B}}^2_2\Big>_2
\nonumber\\&&\ \ \ \ \mbox{}
=\int\! dx\, g^{1/2}\, {\rm tr}\left<{\cal A}_2\left[\hat{P}_1\hat{P}_2 +
   \Big(2\alpha^2_1-\frac13\Big)R_1\hat{P}_2
\right.\right.
\nonumber\\&&\ \ \ \ \ \ \ \ \mbox{}
\left.\left.
  +\Big(\alpha^2_1\alpha^2_2-\frac13\alpha^2_1+\frac1{36}\Big)
  R_1R_2\hat{1}\right]\right>_2 \nonumber\\&&\ \ \ \ \mbox{}
+\int\! dx\, g^{1/2}\, {\rm tr}\left<
\left[\alpha^2_1\frac{\Box_3}{\Box_1\Box_2}{\cal A}_3
 +2\alpha_1^2\frac1{\Box_2}{\cal A}_1-2\alpha_1^2\frac1{\Box_2}{\cal A}_3
\right.\right.
 \nonumber\\&&\ \ \ \ \ \ \ \ \mbox{}\ \ \ \ \left.
 -\frac13\frac1{\Box_2}{\cal A}_1
 +\Big(\alpha_1^2-\frac16\Big)\left(\frac{\Box_3}{\Box_1}-\frac{\Box_2}{\Box_1}-1\right)
s\,{\cal U}_{23}\right]R_1 R_2 \hat{P}_3 \nonumber\\&&\ \ \ \ \ \ \ \ \mbox{}
+\alpha^2_1\left(\frac{\Box_3}{\Box_1\Box_2}{\cal A}_3-2\frac1{\Box_1}{\cal A}_3\right)R_1^{\mu\nu}R_{2\,\mu\nu}\hat{P}_3
+\frac1{\Box_3}{\cal A}_1\hat{P}_1\hat{P}_2 R_3 \nonumber\\&&\ \ \ \ \ \ \ \ \mbox{}
+\left[\Big(\alpha_1^2\alpha_2^2-\frac16\alpha_1^2\Big)\frac{\Box_1}{\Box_2\Box_3}{\cal A}_1
  -\Big(\alpha_1^2\alpha_1^2-\frac1{36}\Big)\frac1{\Box_2}{\cal A}_1
\right.
\nonumber\\&&\ \ \ \ \ \ \ \ \mbox{}\ \ \ \ \left.
-\frac12\Big(\alpha_1^2\alpha_2^2-\frac13\alpha_1^2+\frac1{36}\Big)
  s\,{\cal U}_{23}\right]R_1 R_2 R_3\hat{1} \nonumber\\&&\ \ \ \ \ \ \ \ \mbox{}
+\Big(\alpha_1^2\alpha_2^2-\frac16\alpha_1^2\Big)
  \left(\frac{\Box_3}{\Box_1\Box_2}{\cal A}_3-2\frac1{\Box_1}{\cal A}_3\right)R_1^{\mu\nu}R_{2\,\mu\nu}R_3\hat{1} \nonumber\\&&\ \ \ \ \ \ \ \ \mbox{}
+4\alpha_2\frac1{\Box_1\Box_2}{\cal A}_3\nabla_\mu \hat{\cal R}_1^{\mu\alpha}\nabla^\nu \hat{\cal R}_{2\,\nu\alpha}\hat{P}_3 \nonumber\\&&\ \ \ \ \ \ \ \ \mbox{}
+4\Big(\alpha_1^2-\frac16\Big)\frac1{\Box_1}s\,{\cal U}_{23}R_1^{\mu\nu}\nabla_\mu R_2\nabla_\nu \hat{P}_3 \nonumber\\&&\ \ \ \ \ \ \ \ \mbox{}
-4\alpha_1^2\frac1{\Box_1\Box_2}{\cal A}_3\nabla^\mu R_1^{\nu\alpha}\nabla_\nu R_{2\,\mu\alpha}\hat{P}_3 \nonumber\\&&\ \ \ \ \ \ \ \ \mbox{}
-2\frac1{\Box_1}s\,{\cal U}_{23}R_1^{\mu\nu}\nabla_\mu\nabla_\nu\hat{P}_2\hat{P}_3 \nonumber\\&&\ \ \ \ \ \ \ \ \mbox{}
+4\alpha_2\Big(\alpha_1^2-\frac16\Big)\frac1{\Box_2\Box_3}{\cal A}_1 R_1\nabla_\alpha\hat{\cal R}_2^{\alpha\mu}\nabla^\beta\hat{\cal R}_{3\,\beta\mu} \nonumber\\&&\ \ \ \ \ \ \ \ \mbox{}
+2\Big(\alpha_1^2\alpha_2^2-\frac13\alpha_1^2+\frac1{36}\Big)
  \frac1{\Box_1}s\,{\cal U}_{23} R_1^{\alpha\beta}\nabla_\alpha R_2 \nabla_\beta R_3\hat{1} \nonumber\\&&\ \ \ \ \ \ \ \ \mbox{}
-4\Big(\alpha_1^2\alpha_2^2-\frac16\alpha_1^2\Big)
  \frac1{\Box_1\Box_2}{\cal A}_3 \nabla^\mu R_1^{\nu\alpha}\nabla_\nu R_{2\,\mu\alpha}R_3\hat{1} \nonumber\\&&\ \ \ \ \ \ \ \ \mbox{}
-4\alpha_2\Big(\alpha_1^2-\frac16\Big)\frac1{\Box_1\Box_2}{\cal A}_3
  \nabla^\nu\hat{\cal R}_{1\nu\mu}\nabla^\mu R_2 R_3 \nonumber\\&&\ \ \ \ \ \ \ \ \mbox{}
-4\frac1{\Box_1}s\,{\cal U}_{23}\hat{\cal R}_1^{\mu\nu} \nabla_\mu\hat{P}_2\nabla_\nu\hat{P}_3
\nonumber\\&&\ \ \ \ \ \ \ \ \mbox{}
\left.
-4\alpha_2\frac1{\Box_1\Box_2}{\cal A}_3
  \nabla^\nu\hat{\cal R}_{1\nu\mu}\nabla^\mu R_2\hat{P}_3\right>_2
+{\rm O}[\Re^4]
\end{fleqnarray}
\mathindent=\parindent
where ($m,n=1,2,3; \ m\neq n$)
\begin{eqnarray}
{\cal A}_{m} &=& {\rm e}^{\alpha_1\alpha_2\Box_{m}}, \\[\baselineskip]
{\cal B}_{m} &=& \frac{{\rm e}^{\alpha_1\alpha_2\Box_{m}}-1}{s\Box_m}, \\[\baselineskip]
{\cal C}_{m} &=& \frac{{\rm e}^{\alpha_1\alpha_2\Box_{m}}-1-s\alpha_1\alpha_2\Box_m}
	     {(s\Box_m)^2},
\end{eqnarray}
\begin{eqnarray}
{\cal U}_{mn} &=&
\frac{{\rm e}^{\alpha_1\alpha_2\Box_{m}}-{\rm e}^{\alpha_1\alpha_2\Box_{n}}}{s(\Box_m-\Box_n)},
\\[\baselineskip]
{\cal V}_{mn} &=& \frac1{s(\Box_m-\Box_n)}
\Big(\frac{{\rm e}^{\alpha_1\alpha_2\Box_{m}}-1}{s\Box_m}-\frac{{\rm e}^{\alpha_1\alpha_2\Box_{n}}-1}{s\Box_n}
\Big),
\\[\baselineskip]
{\cal W}_{mn} &=& \frac1{s(\Box_m-\Box_n)}
\Big(\frac{{\rm e}^{\alpha_1\alpha_2\Box_{m}}-1-s\alpha_1\alpha_2\Box_m}
	     {(s\Box_m)^2}
\nonumber\\&& \hspace{20mm}\mbox{}-\frac{{\rm e}^{\alpha_1\alpha_2\Box_{n}}-1-s\alpha_1\alpha_2\Box_n}
	     {(s\Box_n)^2}\Big)
\end{eqnarray}
(the numbers on the form factors refer to the numbers
on the boxes appearing in them), and the averaging
$\big<\ \big>_2$ is defined in (13.4).

Finally, the last integral in (14.1), with
${\hat{B}}^0_3,{\hat{B}}^1_3,{\hat{B}}^2_3,{\hat{B}}^3_3$, is already of
third order in perturbations. Therefore, it is sufficient
to substitute in eqs.~(13.16)--(13.20) the lowest-order
expressions
\begin{eqnarray}
h^{\mu\nu} &=& 2\frac1{\Box} R^{\mu\nu} + {\rm O}[\Re^2],\\
{\hat{\Gamma}}_\mu &=& \nabla^\nu\frac1{\Box}\hat{\cal R}_{\nu\mu}
  + {\rm O}[\Re^2], \\
{\widetilde{\nabla}}_\mu &=& \nabla_\mu + {\rm O}[\Re],\hspace{7mm}
  {\widetilde{\vphantom{I}\Box}} = \Box + {\rm O}[\Re].
\end{eqnarray}
The final result is presented below. As distinct
from the form factors (14.12)--(14.17) coming from the
second order, the form factor appearing in the last
terms of (14.1):
\begin{equation}
{\rm e}^{s\Omega_3} = \exp[s(
\alpha_2\alpha_3\Box_1
+\alpha_1\alpha_3\Box_2
+\alpha_1\alpha_2\Box_3)]+{\rm O}[\Re]
\end{equation}
is an irreducible nonlocal function of all
the three boxes.

On the whole, there appear initially forty different cubic
structures with derivatives that do not contract in the
$\Box$ operators. However, by using eq. (13.6) and
its consequences
\begin{equation}
\nabla_1+\nabla_2+\nabla_3 = 0,
\end{equation}
\begin{equation}
2\nabla_1\nabla_2 = \Box_3-\Box_1-\Box_2,
\end{equation}
the Jacobi and Bianchi identities
\begin{equation}
\nabla_\lambda \hat{\cal R}_{\beta\mu}
+\nabla_\mu \hat{\cal R}_{\lambda\beta}
+\nabla_\beta \hat{\cal R}_{\mu\lambda} = 0,
\end{equation}
\begin{equation}
\nabla^\mu R_{\mu\alpha} = \frac12 \nabla_\alpha R,
\end{equation}
and the possibility of discarding terms ${\rm O}[\Re]$
when commuting the covariant derivatives (since their
contribution is already ${\rm O}[\Re^4]$),
we reduce the number of independent cubic structures
to thirty three
\footnote{\normalsize
Below, in sect. 16, it will be shown that the
contributions of four of these structures vanish.
In this way the final basis consisting of twenty
nine structures (2.15)--(2.43) is obtained. For a
particular space-time dimension,
the dimension of the basis
of nonlocal invariants can be smaller
(see Appendix).
}.
Here are some examples of this reduction.

The identity
\begin{eqnarray}&&
{\rm tr}\,
  \hat{\cal R}_1^{\alpha\beta}
  \nabla_\alpha\hat{\cal R}_2^{\mu\nu}
  \nabla_\beta\hat{\cal R}_{3\mu\nu} =  {\rm tr}\Big(
\Box_1
\hat{\cal R}^\mu_{1\alpha}\hat{\cal R}^\alpha_{2\beta}
\hat{\cal R}^\beta_{3\mu}
\nonumber\\&&\mbox{}\hspace{40mm}
+\hat{\cal R}_{3\alpha\beta}
 \nabla_\mu\hat{\cal R}_1^{\mu\alpha}\nabla_\nu\hat{\cal R}_2^{\nu\beta}
+\hat{\cal R}_{2\alpha\beta}
 \nabla_\mu\hat{\cal R}_3^{\mu\alpha}\nabla_\nu\hat{\cal R}_1^{\nu\beta}\Big)
\nonumber\\&&\mbox{}\hspace{40mm}
+{\rm O}[\Re^4] +\
{\rm a\ total\ derivative}
\end{eqnarray}
transforms the left-hand side to the structures
2 and 12 of the table (2.15)--(2.43). The identity
\begin{eqnarray}&&
{\rm tr}\nabla_\mu\nabla_\lambda\hat{\cal R}_1^{\lambda\nu}
\nabla^\alpha\hat{\cal R}_{2\alpha\nu}
\nabla_\sigma\hat{\cal R}_3^{\sigma\mu} = -\frac12{\rm tr}\Big(
\Box_1\hat{\cal R}_{1\alpha\beta}
\nabla_\mu\hat{\cal R}_2^{\mu\alpha}
\nabla_\nu\hat{\cal R}_3^{\nu\beta}
\nonumber\\&&\mbox{}\hspace{40mm}
+\Box_2\hat{\cal R}_{2\alpha\beta}
\nabla_\mu\hat{\cal R}_3^{\mu\alpha}
\nabla_\nu\hat{\cal R}_1^{\nu\beta}
-\Box_3\hat{\cal R}_{3\alpha\beta}
\nabla_\mu\hat{\cal R}_1^{\mu\alpha}
\nabla_\nu\hat{\cal R}_2^{\nu\beta} \Big)
\nonumber\\&&\mbox{}\hspace{40mm}
+{\rm O}[\Re^4] +\
{\rm a\ total\ derivative}
\end{eqnarray}
transforms the left-hand side to the structure 12
of the table. The identities
\begin{fleqnarray}&&
\nabla_\alpha R_1^{\beta\mu}
\nabla_\beta\hat{\cal R}_2^{\alpha\nu}
\hat{\cal R}_{3\mu\nu} =
-\frac12 R_1^{\alpha\beta}
\nabla_\alpha\hat{\cal R}_2^{\mu\nu}
\nabla_\beta\hat{\cal R}_{3\mu\nu}
\nonumber\\&&\mbox{}\hspace{20mm}
+R_1^{\mu\nu}
\nabla_\mu\nabla_\lambda\hat{\cal R}_2^{\lambda\alpha}
\hat{\cal R}_{3\alpha\nu}
+{\rm O}[\Re^4] +\
{\rm a\ total\ derivative},
\end{fleqnarray}
\begin{fleqnarray}&&
{\rm tr}\nabla^\mu R_{1\beta\alpha}
\nabla^\beta\hat{\cal R}_2^{\alpha\nu}
\hat{\cal R}_{3\mu\nu}
\nonumber\\&&\mbox{}\hspace{16mm}
={\rm tr}\Big[-\frac18(\Box_1+\Box_2+\Box_3)
R_1\hat{\cal R}_2^{\mu\nu}\hat{\cal R}_{3\mu\nu}
+\frac12 R_1^{\alpha\beta}
\nabla_\alpha\hat{\cal R}_2^{\mu\nu}
\nabla_\beta\hat{\cal R}_{3\mu\nu}
\nonumber\\&&\mbox{}\hspace{26mm}
-\frac12R_1\nabla_\alpha\hat{\cal R}_2^{\alpha\mu}
 \nabla^\beta\hat{\cal R}_{3\beta\mu}
-R_1^{\mu\nu}\nabla_\mu\nabla_\lambda
  \hat{\cal R}_3^{\lambda\alpha}\hat{\cal R}_{2\alpha\nu}\Big]
\nonumber\\&&\mbox{}\hspace{26mm}
+{\rm O}[\Re^4] +\
{\rm a\ total\ derivative}
\end{fleqnarray}
transform their left-hand sides to the structures
7, 19, 20 and 21 of the table. The identity
\begin{eqnarray}&&
{\rm tr} R_1^{\mu\nu}
\nabla_\mu\nabla_\lambda\hat{\cal R}_2^{\lambda\alpha}
\nabla_\nu\nabla^\sigma\hat{\cal R}_{3\sigma\alpha}
 \nonumber\\&&\ \ \ \ \ \ \ \ \mbox{}
={\rm tr}\Big[-\Box_2\Box_3 R_1^{\mu\nu}
  \hat{\cal R}_{2\mu}{}^\alpha\hat{\cal R}_{3\nu\alpha}
+\frac12(\Box_1-\Box_2-\Box_3)R_1^{\mu\nu}
  \nabla^\alpha\hat{\cal R}_{2\alpha\mu}
  \nabla^\beta\hat{\cal R}_{3\beta\nu} \nonumber\\&&\ \ \ \ \ \ \ \ \mbox{}\ \ \ \ \ \
-\Box_3R_1^{\mu\nu}
 \nabla_\mu\nabla_\lambda\hat{\cal R}_2^{\lambda\alpha}
 \hat{\cal R}_{3\alpha\nu}
-\Box_2R_1^{\mu\nu}
 \nabla_\mu\nabla_\lambda\hat{\cal R}_3^{\lambda\alpha}
 \hat{\cal R}_{2\alpha\nu}\Big] \nonumber\\&&\ \ \ \ \ \ \ \ \mbox{}\ \ \ \ \ \
+{\rm O}[\Re^4] +\
{\rm a\ total\ derivative}
\end{eqnarray}
transforms its left-hand side to the structures
8, 18 and 21 of the table.

The identity (14.26) is obtained by using (14.22) for
both $\nabla$'s on the left-hand side to write
\[
2\nabla^2_{[\alpha}\nabla^3_{\beta]} =
\nabla^1_{[\alpha}\nabla^2_{\beta]}
-\nabla^1_{[\alpha}\nabla^3_{\beta]} + {\rm O}[\Re],
\]
and next applying in succession: the Jacobi
identity to $\hat{\cal R}_2$ and $\hat{\cal R}_3$, (14.22) to $\nabla$
other than $\nabla_\alpha$ or $\nabla_\beta$, the Jacobi
identity to $\hat{\cal R}_1$, and again (14.22) to $\nabla$
other than $\nabla_\alpha$ or $\nabla_\beta$. The
identity (14.27) is obtained by writing
\[
\nabla^{\mu}\hat{\cal R}_1^{\lambda\nu} =
\nabla^{[\mu}\hat{\cal R}_1^{\lambda\nu]}
+\nabla^{(\mu}\hat{\cal R}_1^{\lambda\nu)},
\]
applying (14.24) to the first term, (14.22) to
the second term, next the Jacobi identity to
$\hat{\cal R}_2$ and $\hat{\cal R}_3$, and noting that what is left over
forms a total derivative up to ${\rm O}[\Re^4]$. The
identity (14.28) is obtained by first removing
the derivative acting on $R_1$, and next using the
Jacobi identity for $\hat{\cal R}_3$. The identity
(14.29) is obtained by replacing the derivative
acting on $R_1$, so as to use (14.25), and
next removing the derivative from the Ricci scalar,
applying the Jacobi identity to $\hat{\cal R}_2$ and
$\hat{\cal R}_3$ in the unwanted terms, and using (14.23). The
identity (14.30) is obtained by applying (14.24) to both
$\hat{\cal R}$'s, using (14.23), and again applying (14.24).

The contribution of one of the structures:
\begin{eqnarray}&&
\nabla_\lambda\hat{\cal R}_1^{\lambda\sigma}
\nabla_\sigma\nabla_\alpha\nabla_\beta R_2^{\mu\nu}
\nabla_\mu\nabla_\nu R_3^{\alpha\beta} =
-\nabla_\lambda\hat{\cal R}_1^{\lambda\sigma}
\nabla_\alpha\nabla_\beta R_2^{\mu\nu}
\nabla_\sigma\nabla_\mu\nabla_\nu R_3^{\alpha\beta}
\nonumber\\&&\mbox{}\hspace{55mm}
+{\rm O}[\Re^4] +\
{\rm a\ total\ derivative}
\end{eqnarray}
vanishes because its form factor turns out to be
symmetric under a permutation of the labels 2 and 3
whereas the structure itself is antisymmetric
under this permutation.

The final result of the calculations above
is as follows.

\section{The $\alpha$~-representation of the form factors
in the trace of the heat kernel}
\setcounter{equation}{0}

\hspace{\parindent}We obtain
\begin{eqnarray} {\rm Tr} K(s) &=& \frac1{(4\pi s)^\omega}\int\! dx\, g^{1/2}\, {\rm tr}\Big\{\hat{1}+s\hat{P}
+s^2\sum^5_{i=1}f_{i}(-s\Box_2)\Re_1\Re_2({i})\nonumber\\&&\mbox{}
+\left(s^3\sum^{11}_{i=1}+s^4\sum^{25}_{i=12}
+s^5\sum^{28}_{i=26}\right) F_{i}(-s\Box_1,-s\Box_2,-s\Box_3)\Re_1\Re_2\Re_3({i})\nonumber\\&&\mbox{}
+s^6 F_{29}(-s\Box_1,-s\Box_2,-s\Box_3)\Re_1\Re_2\Re_3({29})\nonumber\\&&\mbox{}
+\Big[s^4 \sum^{32}_{i=30} F_{i}(-s\Box_1,-s\Box_2,-s\Box_3)\Re_1\Re_2\Re_3({i})\nonumber\\&&\mbox{}
+s^5 F_{33}(-s\Box_1,-s\Box_2,-s\Box_3)\Re_1\Re_2\Re_3({33})\Big]
+{\rm O}[\Re^4]
\Big\}
\end{eqnarray}
where $\Re_1\Re_2({i})$ with $i=1$ to $5$ are quadratic structures of the
table (2.2)--(2.6), $\Re_1\Re_2\Re_3({i})$ with $i=1$ to $29$ are cubic structures
of the table (2.15)--(2.43), and there are four additional
cubic structures linear in $\hat{\cal R}_{\mu\nu}$:
\begin{eqnarray}
\Re_1\Re_2\Re_3({30})&=&\nabla_\beta \hat{\cal R}_1^{\beta\alpha}\nabla_\alpha R_2 R_3,\\[\baselineskip]
\Re_1\Re_2\Re_3({31})&=&\nabla_\mu \hat{\cal R}_1^{\mu\alpha}R_{2\,\alpha\beta}\nabla^\beta R_3,\\[\baselineskip]
\Re_1\Re_2\Re_3({32})&=&\hat{P}_1 \nabla_\beta \hat{\cal R}_2^{\beta\alpha}\nabla_\alpha R_3,\\[\baselineskip]
\Re_1\Re_2\Re_3({33})&=&\nabla_\alpha \hat{\cal R}_1^{\alpha\beta}\nabla_\beta R_2^{\mu\nu}\nabla_\mu \nabla_\nu R_3.
\end{eqnarray}

The form factors
\begin{equation}
f_i (-s\Box), \hspace{7mm} F_i (-s\Box_1,-s\Box_2,-s\Box_3)
\end{equation}
are obtained as integrals over the parameters
\begin{fleqnarray}&&
\big< (\ldots) \big>_2 =
   \int_{\alpha\geq 0} d\alpha_1\, d\alpha_2 \,\delta(1-\alpha_1-\alpha_2)(\ldots),
\end{fleqnarray}
\begin{fleqnarray}&&
\big< (\ldots) \big>_3 =
   \int_{\alpha\geq 0} d\alpha_1 \,d\alpha_2 \,d\alpha_3\,
\delta(1-\alpha_1-\alpha_2-\alpha_3)(\ldots),
\end{fleqnarray}
and, in this form, are represented by two nonlocal kernels:
\begin{equation}
\exp (s\alpha_1\alpha_2\Box)
\end{equation}
and
\begin{equation}
\exp (s\Omega),
\hspace{7mm}
\Omega=\alpha_2 \alpha_3 \Box_1 + \alpha_1 \alpha_3 \Box_2 + \alpha_1 \alpha_2 \Box_3 .
\end{equation}
The function (15.9) appears in the combinations
\begin{equation}
{\cal A,\ \ \ \  B,\ \ \ \  C,\ \ \ \  U,\ \ \ \  V,\ \ \ \  W}
\end{equation}
introduced in (14.12)--(14.17), and the function (15.10) appears
in the combinations (cf.(14.1))
\begin{equation}
{\rm e}^{s\Omega},\hspace{7mm} {\rm e}^{s\Omega}-1,\hspace{7mm} {\rm e}^{s\Omega}-1-s\Omega
\end{equation}
which figure explicitly in the expressions below. The
coefficients of these functions are polynomials in $\alpha$~'s,
boxes, and inverse boxes.

In this representation, the second-order form factors are of
the form
\begin{eqnarray}
f_1 &=& \left\langle{\cal C}\right\rangle_{2},\\[\baselineskip]
f_2 &=& \left\langle \frac12 \left(\alpha_1^2\alpha_2^2 -\frac13 \alpha_1^2
+\frac1{36}\right){\cal A}\right.\nonumber\\&&\mbox{}
+\left.\left(\alpha_1\alpha_2-\alpha_1^2+\frac16\right){\cal B}+\frac12
{\cal C}\right\rangle_{2},\\[\baselineskip]
f_3&=&\left\langle\left(\alpha_1^2-\frac16\right){\cal A}-{\cal B}\right\rangle_{2},\\[\baselineskip]
f_4&=&\left\langle\frac12{\cal A}\right\rangle_{2},\\[\baselineskip]
f_5&=&\left\langle\frac12{\cal B}\right\rangle_{2},
\end{eqnarray}
and the results for the third-order  form factors are as follows:
\arraycolsep=0pt
\begin{fleqnarray}&& F_{1}=\left\langle{1\over 3}{\rm e}^{s\Omega}\right\rangle_{3} ,\end{fleqnarray}
\begin{fleqnarray}&&F_{2}=
\left\langle
{{4\over 3} \alpha_1 \alpha_2 \alpha_3}{\rm e}^{s\Omega}\right\rangle_{3}
+\left\langle-2{\cal V}_{12}\right\rangle_{2},
\end{fleqnarray}
\begin{fleqnarray}&&F_{3}=\left\langle
2 \alpha_1 \alpha_2
{\rm e}^{s\Omega}\right\rangle_{3},
\end{fleqnarray}
\begin{fleqnarray}&&F_{4}=
\left\langle{1\over {\Box_1 \Box_2}}\frac{({\rm e}^{s\Omega}-1)}{s^2}+
\left[{1\over {\Box_1}}\left({1\over 3}
- {{\alpha_1}^2} - {{\alpha_2}^2} + 3 \alpha_1 \alpha_3 + \alpha_2 \alpha_3 -
       {{\alpha_3}^2}\right)\right.\right.\nonumber\\&&\ \ \ \ \mbox{}+\left.
{{\Box_3}\over {\Box_1 \Box_2}} (-{3 \alpha_2 \alpha_3})\right]\frac{{\rm e}^{s\Omega}}{s}
+\left[\left(
{1\over {36}}+{1\over 2}{{{\alpha_1}^4}} -{1\over 6}{\alpha_1 \alpha_2} -
 {1\over 3}{{{\alpha_2}^2}}\right.\right.\nonumber\\&&\ \ \ \ \mbox{}
- {1\over 2}{{{\alpha_1}^2} {{\alpha_2}^2}}-\left.
 {1\over 6}{\alpha_2 \alpha_3} + {{\alpha_2}^3} \alpha_3 - {1\over 6}{{{\alpha_3}^2}} +
\alpha_1 \alpha_2{{\alpha_3}^2}+{3\over 2}{{{\alpha_2}^2}{{\alpha_3}^2}}\right)\nonumber\\&&\ \ \ \ \mbox{}
+{{\Box_2}\over {\Box_1}}
  \left( -{1\over 2}{{{\alpha_1}^4}} - {1\over 6}{\alpha_1 \alpha_2} \right.-
        {1\over 2}{{{\alpha_1}^3} \alpha_2} - {1\over 2}{{{\alpha_1}^2}
{{\alpha_2}^2}} +
  {1\over 2}{\alpha_1 {{\alpha_2}^3}}\nonumber\\&&\ \ \ \ \mbox{}+{1\over 6}{\alpha_1 \alpha_3} -
{{\alpha_1}^3} \alpha_3 -
{1\over 2}{{{\alpha_1}^2} \alpha_2 \alpha_3} +\left. {1\over 2}{\alpha_1
{{\alpha_2}^2} \alpha_3} +
    {1\over 2}{{{\alpha_1}^2} {{\alpha_3}^2}} + \alpha_1 \alpha_2 {{\alpha_3}^2}
\right)\nonumber\\&&\ \ \ \ \mbox{}
+ {{\Box_3}\over {\Box_1}}\left( {1\over 2}{{{\alpha_1}^4}}
+ {1\over 6}{\alpha_1 \alpha_2} +
    {{\alpha_1}^3} \alpha_2 \right.+ {3\over 2}{ {{\alpha_1}^2} {{\alpha_2}^2}} -
        {1\over 6}{\alpha_1 \alpha_3} \nonumber\\&&\ \ \ \ \mbox{}+ {{\alpha_1}^3} \alpha_3 +
2 {{\alpha_1}^2} \alpha_2 \alpha_3 +
        \alpha_1 {{\alpha_2}^2} \alpha_3 - {1\over 2}{{{\alpha_1}^2} {{\alpha_3}^2}} -
        \left.\alpha_1 \alpha_2 {{\alpha_3}^2} \right)\nonumber\\&&\ \ \ \ \mbox{}+\left.\left.
  {{ {{\Box_3}^2}}\over {\Box_1 \Box_2}}\left(-{1\over 2}{ {{\alpha_1}^3}
\alpha_2  } -
        {1\over 2}{{{\alpha_1}^2} {{\alpha_2}^2}} -
{3\over 2}{ {{\alpha_1}^2} \alpha_2 \alpha_3}
\right)\right]{\rm e}^{s\Omega}\right\rangle_{3}\nonumber\\&&\ \ \ \ \mbox{}
+\left\langle{1\over {\Box_1}}\left(
{-{1\over 6} + {{\alpha_1}^2}}\right)\frac{{\cal A}_2}{s}
+\left[{1\over {\Box_1}}(-{{\alpha_1}^2})+
 {{\Box_3}\over {\Box_1 \Box_2}}\left(\frac12{{\alpha_1}^2}\right)\right]
\frac{{\cal A}_3}{s}\right.\nonumber\\&&\ \ \ \ \mbox{}
-{1\over {\Box_1}}\frac{{\cal B}_2}{s}
+\left({1\over {\Box_1}} -\frac12 {{\Box_3}\over {\Box_1 \Box_2}} \right)
\frac{{\cal B}_3}{s}\nonumber\\&&\ \ \ \ \mbox{}
+\left[\frac{\Box_3}{\Box_1}\left(\frac12\alpha_1^2 -\frac{1}{12}\right)+
\frac{\Box_2}{\Box_1}\left(-\frac12\alpha_1^2 +\frac{1}{12}\right)\right.
\nonumber\\&&\ \ \ \ \mbox{}\left.
+\left(-\frac12 \alpha_1^2 +\frac{1}{12}\right)\right]
{\cal U}_{23}\left.
+\left(\frac12+\frac12\frac{\Box_2}{\Box_1}
-\frac12\frac{\Box_3}{\Box_1}\right){\cal V}_{23}
\right\rangle_{2},
\end{fleqnarray}
\begin{fleqnarray}&&F_{5}=
\left\langle 2{1\over {\Box_1 \Box_2}}\frac{({\rm e}^{s\Omega}-1)}{s^2}\right\rangle_{3}
+\left\langle\left[{1\over {\Box_1}}(-{{{\alpha_1}^2}}) +
{{\Box_3}\over {\Box_1 \Box_2}}(\frac12{{{\alpha_1}^2}})\right]\right.
\frac{{\cal A}_3}{s}\nonumber\\&&\ \ \ \ \mbox{}
+\left.\left({1\over {\Box_1}} -\frac32{{ \Box_3}\over { \Box_1 \Box_2}}
\right)
\frac{{\cal B}_3}{s}\right\rangle_{2}
,\end{fleqnarray}
\begin{fleqnarray}&&F_{6}=\left\langle-{1\over {\Box_3}}\frac{{\rm e}^{s\Omega}}{s}+\left[
\left(-{1\over 6} + {{\alpha_1}^2} + \alpha_1 \alpha_3\right) +
  {{\Box_1}\over {\Box_3}}(\alpha_1 \alpha_3 - \alpha_2 \alpha_3) \right]{\rm e}^{s\Omega}
\right\rangle_{3}\nonumber\\&&\ \ \ \ \mbox{}
+\left\langle\frac12{{1}\over {\Box_3}}\frac{{\cal A}_1}{s}\right\rangle_{2}
,\end{fleqnarray}
\begin{fleqnarray}&&F_{7}=\left\langle
{1\over {\Box_1}}\left(
{2 {{\alpha_2}^2} -2 \alpha_1 \alpha_3 -4 \alpha_2 \alpha_3}\right)\frac{{\rm e}^{s\Omega}}{s}\right.\nonumber\\&&\ \ \ \ \mbox{}
+\left[\left(-{1\over 3}{ \alpha_2 \alpha_3 } + {{\alpha_1}^2} \alpha_2 \alpha_3 +
  2 \alpha_1 {{\alpha_2}^2} \alpha_3 + 2 {{\alpha_2}^3} \alpha_3\right)\right.\nonumber\\&&\ \ \ \ \mbox{}
+\left.\left.
  {{\Box_3}\over {\Box_1}}
\left( -2 \alpha_1 {{\alpha_2}^2} \alpha_3 + 2 \alpha_1 \alpha_2 {{\alpha_3}^2} \right)
\right]{\rm e}^{s\Omega}\right\rangle_{3}\nonumber\\&&\ \ \ \ \mbox{}
+\left\langle\frac12{1\over {\Box_1}}
\frac{{\cal B}_2}{s}
+\frac12 \frac{\Box_2}{\Box_1}{\cal V}_{23}\right\rangle_{2}
,\end{fleqnarray}
\begin{fleqnarray}&&F_{8}=\left\langle
{1\over {\Box_1}}(-{4 {{\alpha_1}^2} + 16 \alpha_1 \alpha_2})\frac{{\rm e}^{s\Omega}}{s}
+\left[(-4 {{\alpha_1}^2} \alpha_2 \alpha_3) +
    {{\Box_3}\over {\Box_1}}(8 {{\alpha_1}^2} \alpha_2 \alpha_3)  \right]{\rm e}^{s\Omega}
\right\rangle_{3}\nonumber\\&&\ \ \ \ \mbox{}
+\left\langle-2 {1\over {\Box_1}}\frac{{\cal B}_3}{s}\right\rangle_{2}
,\end{fleqnarray}
\begin{fleqnarray}&&F_{9}=
\left\langle-\frac13{1\over {\Box_1 \Box_2 \Box_3}}\frac{({\rm e}^{s\Omega}-1-s\Omega)}{s^3}\right.
+ {1\over {\Box_1 \Box_2}}
\left(-{1\over 6} + {{\alpha_1}^2} - 2 \alpha_1 \alpha_2 \right.\nonumber\\&&\ \ \ \ \mbox{}\left.
- 2 \alpha_2 \alpha_3 + {3\over 2}{ {{\alpha_3}^2}}\right)\frac{({\rm e}^{s\Omega}-1)}{s^2}
+\left[{1\over {\Box_1}}
\left(-{1\over {36}} +{1\over 6}{{{\alpha_1}^2}}
-{1\over 2}{{{\alpha_1}^4}}\right.\right.\nonumber\\&&\ \ \ \ \mbox{}
   + {{\alpha_1}^3} \alpha_2 + {1\over 3}{{{\alpha_2}^2}} + 2 \alpha_1 {{\alpha_2}^3} -
       {1\over 2}{{{\alpha_2}^4}}-{1\over 2}{\alpha_1 \alpha_3} - {1\over 6}
{\alpha_2 \alpha_3}
       \nonumber\\&&\ \ \ \ \mbox{}-{{\alpha_1}^2} \alpha_2 \alpha_3 + 4 \alpha_1 {{\alpha_2}^2} \alpha_3 +
{{\alpha_2}^3} \alpha_3 -
 \left.{5\over 2}{{{\alpha_1}^2} {{\alpha_3}^2}} + {3\over 2}{{{\alpha_2}^2}
{{\alpha_3}^2}}
       \right)\nonumber\\&&\ \ \ \ \mbox{}
+ {{\Box_1}\over {\Box_2 \Box_3}}\left( -2 {{\alpha_1}^3} \alpha_2 +
        {3\over 2}{{{\alpha_1}^2} {{\alpha_2}^2}} + {1\over 2}{{{\alpha_2}^4}}
\right.+
        {1\over 2}{\alpha_1 \alpha_3} \nonumber\\&&\ \ \ \ \mbox{}- {1\over 2}{{{\alpha_1}^2}
\alpha_2 \alpha_3} -\left.
   \left.\alpha_1 {{\alpha_2}^2} \alpha_3 - {1\over 2}{{{\alpha_2}^2} {{\alpha_3}^2}} -
        \alpha_1 {{\alpha_3}^3} \right)\right]\frac{{\rm e}^{s\Omega}}{s}\nonumber\\&&\ \ \ \ \mbox{}
+\left[\left(-{1\over {648}}\right.
+{1\over {24}}{{{\alpha_1}^2}} - {1\over {12}}{{{\alpha_1}^4}} -\right.
  {1\over 6}{{{\alpha_1}^3} \alpha_2} + {1\over {36}}{\alpha_1 \alpha_3} \nonumber\\&&\ \ \ \ \mbox{}-
  {1\over 6}{{{\alpha_1}^2} \alpha_2 \alpha_3} - {{\alpha_1}^3} {{\alpha_2}^2} \alpha_3 -
  {1\over 6}{{{\alpha_1}^2} {{\alpha_3}^2}} -\left.
  {1\over 3}{{{\alpha_1}^2} {{\alpha_2}^2} {{\alpha_3}^2}}\right)\nonumber\\&&\ \ \ \ \mbox{}
+{{\Box_1}\over {\Box_2}}
  \left({1\over {36}}{\alpha_1 \alpha_2} - {1\over {12}}{{{\alpha_1}^3} \alpha_2}
\right.+
        {1\over 6}{{{\alpha_1}^2} {{\alpha_2}^2}} - {1\over 2}{{{\alpha_1}^4}
{{\alpha_2}^2}} -
        {1\over {12}}{\alpha_1 {{\alpha_2}^3}} \nonumber\\&&\ \ \ \ \mbox{}- {1\over 2}{{{\alpha_1}^2}
{{\alpha_2}^4}}
 - {1\over {36}}{\alpha_2 \alpha_3} + {1\over {12}}{{{\alpha_1}^2} \alpha_2 \alpha_3}
	- {1\over 2}{{{\alpha_1}^4} \alpha_2 \alpha_3} - {1\over 4}
{\alpha_1 {{\alpha_2}^2} \alpha_3}
\nonumber\\&&\ \ \ \ \mbox{} + {1\over 2}{\alpha_1 {{\alpha_2}^4} \alpha_3} - {1\over 3}{\alpha_1 \alpha_2
	{{\alpha_3}^2}} + {1\over 2}{{{\alpha_1}^3} \alpha_2 {{\alpha_3}^2}} -
	{1\over 3}{{{\alpha_2}^2} {{\alpha_3}^2}} +
        2 {{\alpha_1}^2} {{\alpha_2}^2} {{\alpha_3}^2} \nonumber\\&&\ \ \ \ \mbox{}+
 {5\over 2}{ \alpha_1 {{\alpha_2}^3} {{\alpha_3}^2}} + {{\alpha_2}^4} {{\alpha_3}^2} +
        {{\alpha_1}^2} \alpha_2 {{\alpha_3}^3} + 3 \alpha_1 {{\alpha_2}^2} {{\alpha_3}^3} +
 {{\alpha_2}^3} {{\alpha_3}^3}\nonumber\\&&\ \ \ \ \mbox{} \left.+\alpha_1 \alpha_2 {{\alpha_3}^4}
+ {{\alpha_2}^2} {{\alpha_3}^4}
        \right)  +{{{{\Box_1}^2}}\over {\Box_2 \Box_3}}
  \left({1\over 4}{\alpha_1 \alpha_2 {{\alpha_3}^2}} \right.- {{\alpha_1}^3} \alpha_2
{{\alpha_3}^2}\nonumber\\&&\ \ \ \ \mbox{} +
        {1\over {12}}{{{\alpha_2}^2} {{\alpha_3}^2}} -
        {{\alpha_1}^2} {{\alpha_2}^2} {{\alpha_3}^2} - 2 \alpha_1 {{\alpha_2}^3}
{{\alpha_3}^2} +
        {1\over {12}}{\alpha_2 {{\alpha_3}^3}} \nonumber\\&&\ \ \ \ \mbox{}-
        {1\over 2}{{{\alpha_1}^2} \alpha_2 {{\alpha_3}^3}} -\left.\left.
        {1\over 2}{{{\alpha_2}^3} {{\alpha_3}^3}} - {1\over 2}{\alpha_1
\alpha_2 {{\alpha_3}^4}} -
        \left.{1\over 2}{{{\alpha_2}^2} {{\alpha_3}^4}} \right)
  \right]{\rm e}^{s\Omega}\right\rangle_{3}\nonumber\\&&\ \ \ \ \mbox{}
+\left\langle{{\Box_1}\over {\Box_2 \Box_3}}
\left(-{1\over {12}}{{{\alpha_1}^2}} + {1\over 2}{{{\alpha_1}^2}
{{\alpha_2}^2}}
	\right)\frac{{\cal A}_1}{s}\right.
+{1\over {\Box_1}}\left({{1\over{72}} - {1\over 2}{{{\alpha_1}^2}
{{\alpha_2}^2}}}
\right)\frac{{\cal A}_2}{s}\nonumber\\&&\ \ \ \ \mbox{}
+{{\Box_1}\over {\Box_2 \Box_3}}\left( {1\over {12}}
 - {{\alpha_1}^2} + \alpha_1 \alpha_2 \right)\frac{{\cal B}_1}{s}
+{1\over {\Box_1}}
({{{\alpha_1}^2} - \alpha_1 \alpha_2})\frac{{\cal B}_2}{s}\nonumber\\&&\ \ \ \ \mbox{}
+{{\Box_1}\over {\Box_2 \Box_3}}\frac{{\cal C}_1}{s}
-\frac12{1\over {\Box_1}}\frac{{\cal C}_2}{s}
+\left(\frac{1}{12}\alpha_1^2 -{1\over 4}\alpha_1^2 \alpha_2^2 -
{1\over {144}}\right){\cal U}_{12}\nonumber\\&&\ \ \ \ \mbox{}
+\left({1\over{2}}\alpha_1^2 - {1\over{2}}\alpha_1 \alpha_2
- {1\over{12}}\right){\cal V}_{12}\left.
+\left({1\over{2}}\frac{\Box_1}{\Box_3}-{1\over{2}}\right)
{\cal W}_{12}\right\rangle_{2}
,\end{fleqnarray}
\begin{fleqnarray}&&F_{10}=
\left\langle-\frac83{1\over {\Box_1 \Box_2 \Box_3}}\frac{({\rm e}^{s\Omega}-1-s\Omega)}{s^3}\right\rangle_{3}
+\left\langle2{{\Box_1}\over {\Box_2 \Box_3}}\frac{{\cal C}_1}{s}\right\rangle_{2}
,\end{fleqnarray}
\begin{fleqnarray}&&F_{11}=\left\langle-2{1\over {\Box_1 \Box_2 \Box_3}}\frac{({\rm e}^{s\Omega}-1-s\Omega)}{s^3}\right.
+\left[{1\over {\Box_1 \Box_2}}\left(
-{1\over 3} + 2 {{\alpha_1}^2} + 2 \alpha_1 \alpha_3 + {{\alpha_3}^2}\right)
\right.\nonumber\\&&\ \ \ \ \mbox{}+\left.\left.
  {1\over {\Box_1 \Box_3}}({-2 \alpha_1 \alpha_3 + 2 \alpha_2 \alpha_3})\right]
\frac{({\rm e}^{s\Omega}-1)}{s^2}\right\rangle_{3}\nonumber\\&&\ \ \ \ \mbox{}
+\left\langle\left[{1\over {\Box_1}}
\left({\frac16{{{\alpha_1}^2}} - {{\alpha_1}^2} {{\alpha_2}^2}}\right)
+{{\Box_3}\over {\Box_1 \Box_2}}
\left(-{1\over {12}}{{{\alpha_1}^2}} + {1\over 2}{{{\alpha_1}^2}
{{\alpha_2}^2}} \right)
  \right]\frac{{\cal A}_3}{s}\right.\nonumber\\&&\ \ \ \ \mbox{}
+\left[
 {1\over {\Box_1}}\left({-{1\over 6} + 2 {{\alpha_1}^2} - 2 \alpha_1 \alpha_2}
\right)
  +{{\Box_3}\over{\Box_1 \Box_2}}\left( {1\over 4} - 2 {{\alpha_1}^2} +
\alpha_1 \alpha_2
\right) \right]\frac{{\cal B}_3}{s}\nonumber\\&&\ \ \ \ \mbox{}
+\left(-{1\over {\Box_2}} +{{\Box_1}\over {\Box_2 \Box_3}}\right)
\frac{{\cal C}_1}{s}
+\left(-{1\over {\Box_1}}+\frac32{{\Box_3}\over{\Box_1\Box_2}}\right)
\frac{{\cal C}_3}{s}
- \left.\frac{1}{2}{\cal W}_{12}\right\rangle_{2}
,\end{fleqnarray}
\begin{fleqnarray}&&F_{12}=
\left\langle
{1\over {\Box_2 \Box_3}}({-2 \alpha_1 + 2 \alpha_2 + 2 \alpha_3})\frac{{\rm e}^{s\Omega}}{s^2}\right.
+\left[{1\over {\Box_2}}(2 \alpha_1 \alpha_2 \alpha_3)\right.\nonumber\\&&\ \ \ \ \mbox{}
\left.\left.+{1\over {\Box_3}}(2 \alpha_1 \alpha_2 \alpha_3) +
    {{\Box_1}\over {\Box_2 \Box_3}}(-2 \alpha_1 \alpha_2 \alpha_3) \right]
\frac{{\rm e}^{s\Omega}}{s}\right\rangle_{3}\nonumber\\&&\ \ \ \ \mbox{}
+\left\langle-2{1\over {\Box_2 \Box_3}}\frac{{\cal B}_1}{s^2}
-2\frac{1}{\Box_3}\frac{{\cal V}_{12}}{s}
-2\frac{1}{\Box_2}\frac{{\cal V}_{13}}{s}\right\rangle_{2}
,\end{fleqnarray}
\begin{fleqnarray}&&F_{13}=\left\langle{1\over {\Box_1}}({2 \alpha_1})\frac{{\rm e}^{s\Omega}}{s}\right\rangle_{3}
+\left\langle -2\frac{1}{\Box_1}\frac{{\cal U}_{23}}{s}\right\rangle_{2}
,\end{fleqnarray}
\begin{fleqnarray}&&F_{14}=
\left\langle -2{{1}\over {\Box_1 \Box_2}}\frac{{\rm e}^{s\Omega}}{s^2}+
\left[ {1\over {\Box_1}}(2 \alpha_1 \alpha_2) + {1\over {\Box_2}}(2 \alpha_1
\alpha_2)\right.\right.\nonumber\\&&\ \ \ \ \mbox{} \left.\left.
+{{\Box_3}\over {\Box_1 \Box_2}}(-2 \alpha_1 \alpha_2) \right] \frac{{\rm e}^{s\Omega}}{s}
\right\rangle_{3}
+\left\langle{1\over {\Box_1 \Box_2}}({2 \alpha_2})\frac{{\cal A}_3}{s^2}\right\rangle_{2}
,\end{fleqnarray}
\begin{fleqnarray}&&F_{15}=
\left\langle{1\over {\Box_1 \Box_2}}({-4 \alpha_1 + 12 {{\alpha_1}^2}})\frac{{\rm e}^{s\Omega}}{s^2}
+\left[
{1\over {\Box_1}}\left( {2\over 3}{{{\alpha_1}^2}} \right.- 2{{\alpha_1}^4}-
2{{\alpha_1}^3}\alpha_2 \right.\right.\nonumber\\&&\ \ \ \ \mbox{} \left.
-2{{\alpha_1}^2} \alpha_2 \alpha_3 - 2{{\alpha_1}^2} {{\alpha_3}^2} \right) +
  {1\over {\Box_2}}{\left( -2{{\alpha_1}^3} \alpha_2   + 2{{\alpha_1}^2}
{{\alpha_2}^2} +
        2{{\alpha_1}^2} \alpha_2 \alpha_3 \right) } \nonumber\\&&\ \ \ \ \mbox{}+\left.
  {{\Box_3}\over {\Box_1 \Box_2}}
\left(2{{\alpha_1}^3}\alpha_2-2{{\alpha_1}^2}{{\alpha_2}^2} -
\left. 2{{\alpha_1}^2} \alpha_2 \alpha_3 \right)\right]\frac{{\rm e}^{s\Omega}}{s}\right\rangle_{3}\nonumber\\&&\ \ \ \ \mbox{}
+\left\langle -2{1\over {\Box_1 \Box_2}}\frac{{\cal B}_3}{s^2}
+\frac{1}{\Box_1}\left(2\alpha_1^2 -\frac13 \right)\frac{{\cal U}_{23}}{s}
-2\frac{1}{\Box_1}\frac{{\cal V}_{23}}{s}\right\rangle_{2}
,\end{fleqnarray}
\begin{fleqnarray}&&F_{16}=\left\langle
{1\over {\Box_1 \Box_2}}({8 \alpha_1 \alpha_2})\frac{{\rm e}^{s\Omega}}{s^2}\right\rangle_{3}
+\left\langle{1\over {\Box_1 \Box_2}}
(-{2 {{\alpha_1}^2}})\frac{{\cal A}_3}{s^2}\right.\nonumber\\&&\ \ \ \ \mbox{}\left.
+2{1\over {\Box_1 \Box_2}}\frac{{\cal B}_3}{s^2}\right\rangle_{2},
\end{fleqnarray}
\begin{fleqnarray}&&F_{17}=\left\langle
{1\over {\Box_1}}({2 {{\alpha_1}^2}})\frac{{\rm e}^{s\Omega}}{s}\right\rangle_{3}
+\left\langle-\frac{1}{\Box_1}\frac{{\cal U}_{23}}{s}\right\rangle_{2}
,\end{fleqnarray}
\begin{fleqnarray}&&F_{18}=
\left\langle 4{1\over {\Box_1 \Box_2 \Box_3}}\frac{({\rm e}^{s\Omega}-1)}{s^3}\right.
+\left[{1\over {\Box_2 \Box_3}}({-8 \alpha_1\alpha_2} +2\alpha_1^2)\right.\nonumber\\&&\ \ \ \ \mbox{}+
\left.{1\over{\Box_1 \Box_2}}( -4\alpha_1^2+{8 \alpha_1\alpha_2}+{8 \alpha_1\alpha_3})
\right]\frac{{\rm e}^{s\Omega}}{s^2}
+\left[\frac{1}{\Box_1}({4 {{\alpha_1}^2} \alpha_2 \alpha_3})\right.\nonumber\\&&\ \ \ \ \mbox{}\left.
+\left.\frac{1}{\Box_2}({-8 {{\alpha_1}^2} \alpha_2 \alpha_3})+
\frac{\Box_1}{\Box_2\Box_3}({2 {{\alpha_1}^2} \alpha_2 \alpha_3})+
\frac{\Box_3}{\Box_1\Box_2}({4 {{\alpha_1}^2} \alpha_2 \alpha_3})\right]
\frac{{\rm e}^{s\Omega}}{s}\right\rangle_{3}\nonumber\\&&\ \ \ \ \mbox{}+
\left\langle-4{1\over {\Box_1 \Box_3}}\frac{{\cal B}_2}{s^2}\right\rangle_{2}
,\end{fleqnarray}
\begin{fleqnarray}&&F_{19}=
\left\langle
{1\over {\Box_1}}(-{4 {{\alpha_1}^2} \alpha_2 \alpha_3})\frac{{\rm e}^{s\Omega}}{s}\right\rangle_{3}
+\left\langle-\frac{1}{\Box_1}\frac{{\cal V}_{23}}{s}\right\rangle_{2}
,\end{fleqnarray}
\begin{fleqnarray}&&F_{20}=
\left\langle 2{1\over {\Box_1 \Box_2 \Box_3}}\frac{({\rm e}^{s\Omega}-1)}{s^3}\right.
+\left[{1\over {\Box_1 \Box_2}}
\left(2 {{\alpha_2}^2} - 4 \alpha_1 \alpha_3 \right.- 8 \alpha_2 \alpha_3
+ 2 {{\alpha_3}^2}\right)\nonumber\\&&\ \ \ \ \mbox{}
+{1\over {\Box_2 \Box_3}}\left.
\left({{1\over 3} - {{\alpha_1}^2} - 4 {{\alpha_2}^2} + 4 \alpha_2 \alpha_3}\right)
\right]\frac{{\rm e}^{s\Omega}}{s^2}\nonumber\\&&\ \ \ \ \mbox{}
+\left[{1\over {\Box_2}}\left(-{2\over 3}{\alpha_2 \alpha_3} +
 2 {{\alpha_1}^2} \alpha_2 \alpha_3 +  4 \alpha_1 {{\alpha_2}^2} \alpha_3
+ 2 {{\alpha_2}^3} \alpha_3 + 2 \alpha_2 {{\alpha_3}^3}\right)\right.\nonumber\\&&\ \ \ \ \mbox{}
+{{\Box_1}\over{\Box_2 \Box_3}}
\left( {1\over 3}{\alpha_2 \alpha_3} - {{\alpha_1}^2} \alpha_2 \alpha_3 -
        2 \alpha_1 {{\alpha_2}^2} \alpha_3 - 2 {{\alpha_2}^3} \alpha_3 \right)
 \nonumber\\&&\ \ \ \ \mbox{}\left.\left.+{{\Box_3}\over {\Box_1 \Box_2}}
 \left( -2 \alpha_1 {{\alpha_2}^2} \alpha_3 +
        2 \alpha_1 \alpha_2 {{\alpha_3}^2} \right)\right]\frac{{\rm e}^{s\Omega}}{s}\right\rangle_{3}\nonumber\\&&\ \ \ \ \mbox{}+
\left\langle{1\over {\Box_2 \Box_3}}\left(
{-{1\over 3}{\alpha_2} + 2 {{\alpha_1}^2} \alpha_2}\right)\frac{{\cal A}_1}{s^2}
+\frac{1}{\Box_1}\frac{{\cal V}_{23}}{s}\right\rangle_{2}
,\end{fleqnarray}
\begin{fleqnarray}&&F_{21}=
\left\langle
{1\over {\Box_1 \Box_2}}({-8 {{\alpha_1}^2} + 16 \alpha_1 \alpha_3})\frac{{\rm e}^{s\Omega}}{s^2}
+\left[{1\over {\Box_1}}({8 {{\alpha_1}^2} \alpha_2 \alpha_3})\right.\right.\nonumber\\&&\ \ \ \ \mbox{}
+\left.\left.{1\over {\Box_2}}(-{8 {{\alpha_1}^2} \alpha_2 \alpha_3})
+{{\Box_3}\over {\Box_1 \Box_2}}({8 {{\alpha_1}^2} \alpha_2 \alpha_3})\right]
\frac{{\rm e}^{s\Omega}}{s}\right\rangle_{3}\nonumber\\&&\ \ \ \ \mbox{}
+\left\langle 4\frac{1}{\Box_1}\frac{{\cal V}_{23}}{s}\right\rangle_{2}
,\end{fleqnarray}
\begin{fleqnarray}&&F_{22}=
\left\langle{1\over {\Box_1 \Box_2 \Box_3}}\left({-10 {{\alpha_1}^2} + 24 \alpha_1 \alpha_3
+ 4 \alpha_2 \alpha_3}\right)\frac{({\rm e}^{s\Omega}-1)}{s^3}\right.\nonumber\\&&\ \ \ \ \mbox{}
+\left[{1\over {\Box_1 \Box_2}}\left(2{{\alpha_1}^4}
-{4\over 3}{{{\alpha_1}^2}} \right.\right.
+ {2\over 3}{\alpha_1 \alpha_2} -
      12 {{\alpha_1}^3} \alpha_2 + 18 {{\alpha_1}^2} {{\alpha_2}^2} +
    {2\over 3}{\alpha_1 \alpha_3}\nonumber\\&&\ \ \ \ \mbox{} - 4 {{\alpha_1}^3} \alpha_3 + 12 {{\alpha_1}^2}
\alpha_2 \alpha_3 +
      4 \alpha_1 {{\alpha_2}^2} \alpha_3 - 6 {{\alpha_1}^2} {{\alpha_3}^2} -\left.
      4 \alpha_1 \alpha_2 {{\alpha_3}^2}\right)\nonumber\\&&\ \ \ \ \mbox{} +
  {1\over {\Box_2 \Box_3}}\left(8 {{\alpha_1}^3} \alpha_2 - 12 {{\alpha_1}^2}
{{\alpha_2}^2}
- 12 {{\alpha_1}^2} \alpha_2 \alpha_3 -
  \left.4 \alpha_1 \alpha_2 {{\alpha_3}^2}\right)\right]\frac{{\rm e}^{s\Omega}}{s^2}\nonumber\\&&\ \ \ \ \mbox{}
+\left[
{1\over {\Box_1}}\left(-{1\over {18}}{{{\alpha_1}^2}} + {1\over 3}
{{{\alpha_1}^4}}
+\right.{1\over 3}{{{\alpha_1}^3} \alpha_2} + {1\over 3}{{{\alpha_1}^2}
\alpha_2 \alpha_3}
-\right. 2 {{\alpha_1}^4} \alpha_2 \alpha_3 \nonumber\\&&\ \ \ \ \mbox{}+ 2 {{\alpha_1}^3} {{\alpha_2}^2}
\alpha_3 +
       {1\over 3}{{{\alpha_1}^2} {{\alpha_3}^2}} - 2 {{\alpha_1}^4}
{{\alpha_3}^2} \left.-
       2 {{\alpha_1}^3} {{\alpha_3}^3}\right) \nonumber\\&&\ \ \ \ \mbox{}+{1\over {\Box_2}}
  \left({1\over 3}{{{\alpha_1}^3} \alpha_2} \right.- {1\over 3}
{{{\alpha_1}^2} {{\alpha_2}^2}} +
      2 {{\alpha_1}^4} {{\alpha_2}^2} - 2 {{\alpha_1}^3} {{\alpha_2}^3} -
{1\over 3}{{{\alpha_1}^2} \alpha_2 \alpha_3} \nonumber\\&&\ \ \ \ \mbox{}+ 2 {{\alpha_1}^4} \alpha_2 \alpha_3 -
      4 {{\alpha_1}^3} {{\alpha_2}^2} \alpha_3
\left.- 2 {{\alpha_1}^3} \alpha_2 {{\alpha_3}^2}\right)
    +{{\Box_1}\over {\Box_2 \Box_3}}\left(4 {{\alpha_1}^3} \alpha_2
{{\alpha_3}^2}\right)  \nonumber\\&&\ \ \ \ \mbox{}
  +{{\Box_2}\over {\Box_1 \Box_3}}\left( -{1\over 3}{{{\alpha_1}^3}
\alpha_3  } +\right.
        {1\over 3}{{{\alpha_1}^2} \alpha_2 \alpha_3} - 2 {{\alpha_1}^4} \alpha_2 \alpha_3 -
        2 {{\alpha_1}^3} {{\alpha_2}^2} \alpha_3 \nonumber\\&&\ \ \ \ \mbox{}
+ {1\over 3}{{{\alpha_1}^2} {{\alpha_3}^2}} -
 \left.\left.\left.2 {{\alpha_1}^4} {{\alpha_3}^2} + 2 {{\alpha_1}^3}
{{\alpha_3}^3} \right)
\right]\frac{{\rm e}^{s\Omega}}{s}\right\rangle_{3}
+\left\langle{1\over {\Box_1 \Box_2}}\left({{1\over 3} - 2
{{\alpha_1}^2}}\right)\frac{{\cal B}_3}{s^2}
\right.\nonumber\\&&\ \ \ \ \mbox{}
-2{1\over {\Box_2 \Box_3}}\frac{{\cal C}_1}{s^2}
+\frac{1}{\Box_1}\left(\alpha_1^2 \alpha_2^2
-{1\over{3}}\alpha_1^2+{1\over{36}}\right)
\frac{{\cal U}_{23}}{s}\nonumber\\&&\ \ \ \ \mbox{}
+\frac{1}{\Box_1}\left(2 \alpha_1 \alpha_2 -2 \alpha_1^2+{1\over{3}}\right)
\frac{{\cal V}_{23}}{s}
+\frac{1}{\Box_1}\frac{{\cal W}_{23}}{s}
-2\frac{1}{\Box_2}\left.\frac{{\cal W}_{13}}{s}\right\rangle_{2}
,\end{fleqnarray}
\begin{fleqnarray}&&F_{23}=
\left\langle{1\over {\Box_1 \Box_2 \Box_3}}
\left({8 {{\alpha_1}^2} - 16 \alpha_1 \alpha_2 - 8 \alpha_2 \alpha_3}\right)
\frac{({\rm e}^{s\Omega}-1)}{s^3}\right.\nonumber\\&&\ \ \ \ \mbox{}
+\left[{1\over{\Box_1 \Box_2}}
\left(-{4\over 3}{\alpha_1 \alpha_2}
+ 8 {{\alpha_1}^3} \alpha_2 + 4 \alpha_1 \alpha_2 \alpha_3\right)\right.\nonumber\\&&\ \ \ \ \mbox{}
\left.\left.
+ {1\over {\Box_2 \Box_3}}\left( 8{{\alpha_1}^2} \alpha_2 \alpha_3
- 8\alpha_1 {{\alpha_2}^2} \alpha_3 \right)\right]\frac{{\rm e}^{s\Omega}}{s^2}\right\rangle_{3}\nonumber\\&&\ \ \ \ \mbox{}
+\left\langle{1\over {\Box_1 \Box_2}}\left(
{{1\over 3}{{{\alpha_1}^2}} -2 {{\alpha_1}^2} {{\alpha_2}^2}}\right)
\frac{{\cal A}_3}{s^2}\right.
+{1\over {\Box_1 \Box_2}}
\left({-{1\over 3} + 4 {{\alpha_1}^2} - 4 \alpha_1 \alpha_2}\right)
\frac{{\cal B}_3}{s^2}\nonumber\\&&\ \ \ \ \mbox{}
+4{1\over {\Box_1 \Box_3}}\frac{{\cal C}_2}{s^2}
-2{{1}\over {\Box_1 \Box_2}}\frac{{\cal C}_3}{s^2}\left.
+2\frac{1}{\Box_3}\frac{{\cal W}_{12}}{s}\right\rangle_{2}
,\end{fleqnarray}
\begin{fleqnarray}&&F_{24}=\left\langle{1\over {\Box_1 \Box_2 \Box_3}}
(-{4 {{\alpha_1}^2}})\frac{({\rm e}^{s\Omega}-1)}{s^3}\right\rangle_{3}\nonumber\\&&\ \ \ \ \mbox{}
+\left\langle 2{1\over {\Box_2 \Box_3}}\frac{{\cal C}_1}{s^2}
+4{1\over {\Box_1 \Box_2}}\frac{{\cal C}_3}{s^2}
+2\frac{1}{\Box_1}\frac{{\cal W}_{23}}{s}\right\rangle_{2}
,\end{fleqnarray}
\begin{fleqnarray}&&F_{25}=\left\langle
{1\over {\Box_1 \Box_2 \Box_3}}({-16 \alpha_2 \alpha_3})\frac{({\rm e}^{s\Omega}-1)}{s^3}\right\rangle_{3}\nonumber\\&&\ \ \ \ \mbox{}
+\left\langle 4{1\over {\Box_2 \Box_3}}\frac{{\cal C}_1}{s^2}
+8{1\over {\Box_1 \Box_3}}\frac{{\cal C}_2}{s^2}
+8\frac{1}{\Box_3}\frac{{\cal W}_{12}}{s}\right\rangle_{2}
,\end{fleqnarray}
\begin{fleqnarray}&& F_{26}=\left\langle
{1\over {\Box_1 \Box_2}}({4 {{\alpha_1}^2} {{\alpha_2}^2}})\frac{{\rm e}^{s\Omega}}{s^2}\right\rangle_{3}
,\end{fleqnarray}
\begin{fleqnarray}&&F_{27}=\left\langle
{1\over {\Box_1 \Box_2 \Box_3}}
\left({8 {{\alpha_1}^3} \alpha_2 - 12 {{\alpha_1}^2} {{\alpha_2}^2}
- 8 {{\alpha_1}^2} \alpha_2 \alpha_3}\right)\frac{{\rm e}^{s\Omega}}{s^3}\right.\nonumber\\&&\ \ \ \ \mbox{}
+\left[{1\over {\Box_1 \Box_2}}\left(
-{2\over 3}{{{\alpha_1}^2} {{\alpha_2}^2}} + 4 {{\alpha_1}^4} {{\alpha_2}^2}
+\right.
     4 {{\alpha_1}^3} {{\alpha_2}^2} \alpha_3\right)\nonumber\\&&\ \ \ \ \mbox{}\left.
+\left.{1\over {\Box_1 \Box_3}}\left(-4 {{\alpha_1}^3} {{\alpha_2}^2} \alpha_3
+ 4 {{\alpha_1}^2} {{\alpha_2}^3} \alpha_3\right)\right]\frac{{\rm e}^{s\Omega}}{s^2}\right\rangle_{3}
,\end{fleqnarray}
\begin{fleqnarray}&& F_{28}=
\left\langle{1\over {\Box_1 \Box_2 \Box_3}}
(-{16 \alpha_1 \alpha_2 {{\alpha_3}^2}})\frac{{\rm e}^{s\Omega}}{s^3}\right\rangle_{3}
,\end{fleqnarray}
\begin{fleqnarray}&& F_{29}=
\left\langle{1\over { \Box_1 \Box_2 \Box_3}}
\left(\frac83{ {{\alpha_1}^2} {{\alpha_2}^2} {{\alpha_3}^2}}\right)\frac{{\rm e}^{s\Omega}}{s^3}
\right\rangle_{3}
,\end{fleqnarray}
\begin{fleqnarray}&& F_{30}=\left\langle
\Big[\frac{1}{\Box_2\Box_3}(2\alpha_3^2-4\alpha_1\alpha_3^2-4\alpha_1^2\alpha_3)\right.
\nonumber\\&&\ \ \ \ \mbox{}+
{1\over {\Box_1 \Box_2}} \left( -{1\over 3} - 2{{\alpha_1}^3}
- 4 {{\alpha_1}^2} \alpha_2\right. + 2{{\alpha_2}^2} \nonumber\\&&\ \ \ \ \mbox{}
+ 2{{\alpha_2}^3} + 2{{\alpha_1}^2} \alpha_3 \left.\left.
+ 4 \alpha_1 {{\alpha_3}^2} - 2\alpha_2 {{\alpha_3}^2} \right)\Big]\frac{{\rm e}^{s\Omega}}{s^2}
\right\rangle_{3}\nonumber\\&&\ \ \ \ \mbox{}
+\left[
{\Box_3\over {\Box_1 \Box_2}}\left(- {1\over 3}{\alpha_1 \alpha_2} + 2\alpha_1
{{\alpha_2}^3} +
       2{{\alpha_1}^2} \alpha_2 \alpha_3 \right) \right.\nonumber\\&&\ \ \ \ \mbox{}+
{\Box_1\over {\Box_2 \Box_3}}(-2\alpha_1\alpha_2\alpha_3^2)+
{1\over {\Box_1}}\left(2\alpha_1^2\alpha_3^2-2\alpha_1\alpha_2\alpha_3^2-\frac13
\alpha_1\alpha_3\right)\nonumber\\&&\ \ \ \ \mbox{}
\left.\left.+{1\over {\Box_2}}\left(\frac13\alpha_1\alpha_2-2\alpha_1\alpha_2\alpha_3
-2\alpha_1\alpha_2^3+4\alpha_1\alpha_2\alpha_3^2\right)\right]\frac{{\rm e}^{s\Omega}}{s}\right\rangle_{3}\nonumber\\&&\ \ \ \ \mbox{}
+\left\langle{1\over {\Box_1 \Box_2}}\left({{1\over 3}{\alpha_2} - 2 {{\alpha_1}^2}
\alpha_2}\right)
\frac{{\cal A}_3}{s^2}\right\rangle_{2}
,\end{fleqnarray}
\begin{fleqnarray}&& F_{31}=
\left\langle-4{1\over {\Box_1 \Box_2 \Box_3}}\frac{({\rm e}^{s\Omega}-1)}{s^3}\right.
+\left[
{1\over{\Box_1 \Box_2}}{( -4\alpha_1 \alpha_2+ 4{{\alpha_2}^2} ) }\right.\nonumber\\&&\ \ \ \ \mbox{}
 +\left.\left.{1\over {\Box_1 \Box_3}}({4 \alpha_2 \alpha_3}) +
  {1\over{\Box_2 \Box_3}}(-{4 \alpha_2 \alpha_3})\right]\frac{{\rm e}^{s\Omega}}{s^2}\right\rangle_{3}\nonumber\\&&\ \ \ \ \mbox{}
+\left\langle{1\over {\Box_2 \Box_3}}({4 \alpha_1})\frac{{\cal B}_1}{s^2}\right\rangle_{2}
,\end{fleqnarray}
\begin{fleqnarray}&& F_{32}=\left\langle
2{1\over {\Box_2 \Box_3}}\frac{{\rm e}^{s\Omega}}{s^2}\right.
+\left[{1\over {\Box_2}}(2\alpha_1 \alpha_2 -2{{\alpha_2}^2}) +\right.
{1\over {\Box_3}}({-2 \alpha_2 \alpha_3})\nonumber\\&&\ \ \ \ \mbox{}\left.\left.
+ {\Box_1\over {\Box_2 \Box_3}}({2\alpha_2\alpha_3})\right]\frac{{\rm e}^{s\Omega}}{s}\right\rangle_{3}
+\left\langle-{1\over {\Box_2 \Box_3}}\frac{{\cal A}_1}{s^2}\right\rangle_{2}
,\end{fleqnarray}
\begin{fleqnarray}&& F_{33}=\left\langle
{1\over {\Box_1 \Box_2 \Box_3}}{( 8\alpha_1 \alpha_2 -4{{\alpha_2}^2})}\frac{{\rm e}^{s\Omega}}{s^3}
+\left[{1\over {\Box_1 \Box_3}}
(-{4 \alpha_1 {{\alpha_2}^2} \alpha_3})\right.\right.\nonumber\\&&\ \ \ \ \mbox{}+\left.\left.
  {1\over {\Box_2 \Box_3}}({4 \alpha_1 {{\alpha_2}^2} \alpha_3})+
{1\over{\Box_1 \Box_2}}{(4{{\alpha_1}^2} {{\alpha_2}^2} -4\alpha_1 {{\alpha_2}^3} )}
\right]\frac{{\rm e}^{s\Omega}}{s^2}\right\rangle_{3}\!.
\end{fleqnarray}
\arraycolsep=5pt

The $\alpha$~-representations  is the starting point for all the
further derivations. Therefore, we present here several reference
formulae concerning the \hbox{$\alpha$-integrals.} One has
\begin{fleqnarray}&&
\left\langle \alpha_1^n \alpha_2^m \right\rangle_{2} = \frac{n!m!}{(n+m+1)!},
\end{fleqnarray}
\begin{fleqnarray}&&
\left\langle \alpha_1^n \alpha_2^m \ln(\alpha_1\alpha_2)\right\rangle_{2}=
 \frac{n!m!}{(n+m+1)!}
\nonumber\\&&\mbox{}\hspace{50mm}
\times\left(\sum^{n}_{k=1}\frac1k+\sum^{m}_{k=1}\frac1k
-2\sum^{n+m+1}_{k=1}\frac1k\right),
\end{fleqnarray}
\begin{fleqnarray}&&
\left\langle \alpha_1^n \alpha_2^m \alpha_3^k \right\rangle_{3} = \frac{n!m!k!}{(n+m+k+2)!}.
\end{fleqnarray}
These equations are, in particular, useful for obtaining
the early-time asymptotic behaviours of the form factors.
The late-time behaviour was studied in paper II.
The relevant results are
\begin{fleqnarray}&&
\left\langle P(\alpha_1,\alpha_2)\exp(s\alpha_1\alpha_2\Box)\right\rangle_{2}=
-\frac1s \frac{P(1,0)+P(0,1)}{\Box}+{\rm O}\left(\frac{1}{s^2}\right),
\nonumber\\&&\mbox{}\hspace{80mm}
s\rightarrow\infty
\end{fleqnarray}
\begin{fleqnarray}&&
\left\langle P(\alpha_1,\alpha_2,\alpha_3)\exp(s\Omega)\right\rangle_{3}=
\frac1{s^2}
\left[{{P(1,0,0)}\over{\Box_2\Box_3}}
+{{P(0,1,0)}\over{\Box_1\Box_3}}+{{P(0,0,1)}\over{\Box_1\Box_2}}\right]
\nonumber\\&&\mbox{}\hspace{65mm}
+{\rm O}\left(\frac{1}{s^3}\right),\hspace{7mm}
s\rightarrow\infty
\end{fleqnarray}
where $P$~'s are polynomials in $\alpha$. It follows,
in particular, that the leading asymptotic terms are absent
if a monomial in $\alpha$ contains at least two unlike $\alpha$~'s.
This fact is useful when checking the infrared finiteness
of the effective action in two dimensions.

\section{Reduction of the form factors in ${\rm Tr} K(s)$
to the basic form factors}
\setcounter{equation}{0}

\hspace{\parindent}
The problem with the $\alpha$-representation is that the
$s\Box$-arguments of the form factors appear not only
in the kernels (15.9) and (15.10).
As seen from the expressions above,
they enter also the coefficients
of the polynomials in $\alpha$ onto which
the form factors are mapped.
For this reason, the $\alpha$-representation  is not unique
in a sense that, even with the delta-function in (15.8)
taken into account, the vanishing of an integral like
\begin{equation}
\left\langle P(\alpha,\Box) {\rm e}^{s\Omega}\right\rangle_{3}
\end{equation}
does not imply the vanishing of the polynomial
$P(\alpha,\Box)$. This nonuniqueness obscures the
properties of the form factors and makes difficult various checks
like the check of the trace anomaly.
In particular, the fact that the contributions
of the structures (15.2)--(15.5) vanish (see below)
is not seen from (15.47)--(15.50).
The origin of the $\Box$'s in the coefficients is the
tree formulae which express the perturbations through the
curvatures.
The problem of nonuniqueness persists in all representations
of the form factors in the heat kernel and effective action.
Most of the further work with the form factors is aimed
at removing this defect.

One way of obtaining a unique representation for the form factors
in the heat kernel is elimination of all polynomials in $\alpha$.
All form factors will then be explicitly expressed through
the basic ones
\begin{equation}
f(\xi)=\left\langle{\rm e}^{-\alpha_1\alpha_2\xi}\right\rangle_{2}, \hspace{7mm} \xi=-s\Box,
\end{equation}
\begin{equation}
F(\xi_1,\xi_2,\xi_3)=\left\langle{\rm e}^{s\Omega}\right\rangle_{3},\hspace{7mm} \xi_{m} =-s\Box_{m} .
\end{equation}
The technique of eliminating the polynomials
in $\alpha$ is as follows.

After a use of the delta-function in (15.7) and (15.8), there remain
to be considered the contributions of the monomials:
\begin{fleqnarray}&&
\left\langle\alpha_1^n {\rm e}^{-\alpha_1\alpha_2\xi} \right\rangle_{2} =
\int^1_0 \! d\alpha\,\alpha^n \exp\Big[-\alpha(1-\alpha)\xi\Big],
\end{fleqnarray}
\begin{fleqnarray}&&
\left\langle\alpha_1^n \alpha_2^m {\rm e}^{s\Omega} \right\rangle_{3}=
\int^1_0 \! d\alpha_2\int^{1-\alpha_2}_0 \! d\alpha_1 \,\alpha_1^n \alpha_2^m \nonumber\\&&\ \ \ \ \mbox{} \times
\exp\Big[-\alpha_2(1-\alpha_1-\alpha_2)\xi_1-\alpha_1(1-\alpha_1-\alpha_2)\xi_2-\alpha_1\alpha_2\xi_3\Big].
\end{fleqnarray}
For the case (16.4), eq. (7.12) of paper II:
\begin{equation}
\int^1_0\!
d\alpha\,
\frac{d}{d\alpha}\alpha^n \exp\Big[-\alpha(1-\alpha)\xi\Big]=
\left\{
   \begin{array}{l}
          0,\hspace{7mm} n=0\\
          1,\hspace{7mm} n>0
       \end{array}
\right.
\end{equation}
yields the following recurrence relations
\begin{fleqnarray}&&
\left\langle\alpha_1 {\rm e}^{-\alpha_1\alpha_2\xi} \right\rangle_{2} =
\frac12 \left\langle {\rm e}^{-\alpha_1\alpha_2\xi} \right\rangle_{2},
\end{fleqnarray}
\begin{fleqnarray}&&
\left\langle\alpha_1^n {\rm e}^{-\alpha_1\alpha_2\xi} \right\rangle_{2} =
\frac12 \left\langle \alpha_1^{n-1}{\rm e}^{-\alpha_1\alpha_2\xi}\right\rangle_{2}\nonumber\\&&\ \ \ \ \ \ \ \ \mbox{}
-\frac12 (n-1)\left\langle\alpha_1^{n-2}\left(\frac{{\rm e}^{-\alpha_1\alpha_2\xi}-1}{\xi}
\right)\right\rangle_{2},\hspace{7mm} n\geq 2
\end{fleqnarray}
which make it possible to express all integrals (16.4) through
the basic form factor (16.2). Note that this procedure
automatically leads to the appearance
of the form factors with subtractions (14.13)--(14.14).
The recurrence relations for them are similar:
\begin{fleqnarray}&&
\left\langle\alpha_1\left(\frac{{\rm e}^{-\alpha_1\alpha_2\xi}-1}{\xi}\right)\right\rangle_{2}=
\frac12 \left\langle\left(\frac{{\rm e}^{-\alpha_1\alpha_2\xi}-1}{\xi}\right)\right\rangle_{2},
\end{fleqnarray}
\begin{fleqnarray}&&
\left\langle\alpha_1^n \left(\frac{{\rm e}^{-\alpha_1\alpha_2\xi}-1}{\xi}\right)\right\rangle_{2}=
\frac12 \left\langle\alpha_1^{n-1}
\left(\frac{{\rm e}^{-\alpha_1\alpha_2\xi}-1}{\xi}\right)\right\rangle_{2}\nonumber\\&&\ \ \ \ \ \ \ \ \mbox{}
-\frac12 (n-1)
\left\langle\alpha_1^{n-2}\left(\frac{{\rm e}^{-\alpha_1\alpha_2\xi}-1+\alpha_1\alpha_2\xi}
{\xi^2}\right)\right\rangle_{2},\hspace{7mm} n\geq 2
\end{fleqnarray}
as one can check with the aid of eq. (15.51).
The appearance of the subtractions is explained by
analyticity of the integral (16.4) in $\xi$ at $\xi=0$.
Since the recurrence relations imply a division by $\xi$,
the appearing subtractions maintain the analyticity.
For the form factors with subtractions one has
\begin{fleqnarray}&&
\left\langle \frac{{\rm e}^{-\alpha_1\alpha_2\xi}-1}{\xi}\right\rangle_{2}=
\frac{f(\xi)-1}{\xi},
\end{fleqnarray}
\begin{fleqnarray}&&
\left\langle\frac{{\rm e}^{-\alpha_1\alpha_2\xi}-1+\alpha_1\alpha_2\xi}{\xi^2}\right\rangle_{2}=
\frac{f(\xi)-1+\frac16 \xi}{\xi^2}
\end{fleqnarray}
in terms of (16.2).

Elimination of the polynomials in $\alpha$ from  the third-order
form factors is based on integration by parts in (16.5):
\begin{eqnarray}&&
\int^1_0 \! d\alpha_2 \int^{1-\alpha_2}_0 \! d\alpha_1
\frac{d}{d\alpha_1}\alpha_1^n \alpha_2^m \exp\left(\left.s\Omega
\right|_{\alpha_3=1-\alpha_1-\alpha_2}\right)
\nonumber\\&&\ \ \ \ \ \ \ \ \mbox{}=
\left\{
   \begin{array}{ll}
\left\langle\alpha_2^m\left({\rm e}^{-\alpha_1\alpha_2\xi_3}-{\rm e}^{-\alpha_1\alpha_2\xi_1}\right)\right\rangle_{2},
&\hspace{7mm}
n=0
\\[2mm]
\left\langle\alpha_1^n\alpha_2^m {\rm e}^{-\alpha_1\alpha_2\xi_3}\right\rangle_{2},&\hspace{7mm} n>0,
       \end{array}
\right.
\end{eqnarray}
\begin{eqnarray}&&
\int^1_0 \! d\alpha_2 \int^{1-\alpha_2}_0 \! d\alpha_1
\frac{d}{d\alpha_2}\alpha_1^n \alpha_2^m \exp\left(\left.s\Omega
\right|_{\alpha_3=1-\alpha_1-\alpha_2}\right)\nonumber\\&&\ \ \ \ \ \ \ \ \mbox{}=
\left\{
   \begin{array}{ll}
\left\langle\alpha_1^n\left({\rm e}^{-\alpha_1\alpha_2\xi_3}-{\rm e}^{-\alpha_1\alpha_2\xi_2}\right)\right\rangle_{2},
&
\hspace{7mm}
m=0
\\[2mm]
\left\langle\alpha_1^n\alpha_2^m {\rm e}^{-\alpha_1\alpha_2\xi_3}\right\rangle_{2}  ,&
\hspace{7mm}
m>0
       \end{array}
\right.
\end{eqnarray}
where the second-order form factors appearing on the right-hand
sides are subject to the recurrence relations above.
By performing the differentiations on the left-hand sides of
(16.13),(16.14), one obtains two linear algebraic equations for
the quantities
\[
\left\langle\alpha_1^{n+1}\alpha_2^{m} {\rm e}^{s\Omega} \right\rangle_{3},
\hspace{7mm}
\left\langle\alpha_1^{n}\alpha_2^{m+1} {\rm e}^{s\Omega} \right\rangle_{3}
\]
containing the highest-order monomials. The discriminant of this
linear system is
\begin{equation}
\Delta=\xi_1^2+\xi_2^2+\xi_3^2-2\xi_1\xi_2-2\xi_1\xi_3-2\xi_2\xi_3,
\end{equation}
and the recurrence relations obtained in this way
are of the form
\begin{fleqnarray}
\left\langle\alpha_1^{n+1}\alpha_2^{m} {\rm e}^{s\Omega} \right\rangle_{3}&=&
-\frac{\xi_1(\xi_3+\xi_2-\xi_1)}{\Delta}
\left\langle\alpha_1^{n}\alpha_2^{m} {\rm e}^{s\Omega} \right\rangle_{3}\nonumber\\&&\mbox{}
+2n\frac{\xi_1}{\Delta}
\left\langle\alpha_1^{n-1}\alpha_2^{m} {\rm e}^{s\Omega} \right\rangle_{3}\nonumber\\&&\mbox{}
+m\frac{(\xi_3-\xi_2-\xi_1)}{\Delta}
\left\langle\alpha_1^{n}\alpha_2^{m-1} {\rm e}^{s\Omega} \right\rangle_{3}\nonumber\\&&\mbox{}
-\frac{(\xi_3+\xi_1-\xi_2)}{\Delta}
\left\langle\alpha_1^{n}\alpha_2^{m} {\rm e}^{-\alpha_1\alpha_2\xi_3}\right\rangle_{2}
+\beta(n,m),
\end{fleqnarray}
\begin{fleqnarray}&&
\beta(n,m)=0,\hspace{7mm} n>0,\hspace{7mm} m>0
\end{fleqnarray}
\begin{fleqnarray}&&
\beta(n,0)=
\frac{(\xi_3-\xi_2-\xi_1)}{\Delta}
\left\langle\alpha_1^{n} {\rm e}^{-\alpha_1\alpha_2\xi_2}\right\rangle_{2},
\hspace{7mm} n>0
\end{fleqnarray}
\begin{fleqnarray}&&
\beta(0,m)=
2\frac{\xi_1}{\Delta}
\left\langle\alpha_1^{m} {\rm e}^{-\alpha_1\alpha_2\xi_1}\right\rangle_{2},
\hspace{7mm}
m>0
\end{fleqnarray}
\begin{fleqnarray}&&
\beta(0,0)=
2\frac{\xi_1}{\Delta}
\left\langle {\rm e}^{-\alpha_1\alpha_2\xi_1}\right\rangle_{2}
+\frac{(\xi_3-\xi_2-\xi_1)}{\Delta}
\left\langle {\rm e}^{-\alpha_1\alpha_2\xi_2}\right\rangle_{2},
\end{fleqnarray}
\begin{fleqnarray}
\left\langle\alpha_1^{n}\alpha_2^{m+1} {\rm e}^{s\Omega} \right\rangle_{3}&=&
-\frac{\xi_2(\xi_3+\xi_1-\xi_2)}{\Delta}
\left\langle\alpha_1^{n}\alpha_2^{m} {\rm e}^{s\Omega} \right\rangle_{3}\nonumber\\&&\mbox{}
+2m\frac{\xi_2}{\Delta}
\left\langle\alpha_1^{n}\alpha_2^{m-1} {\rm e}^{s\Omega} \right\rangle_{3}\nonumber\\&&\mbox{}
+n\frac{(\xi_3-\xi_2-\xi_1)}{\Delta}
\left\langle\alpha_1^{n-1}\alpha_2^{m} {\rm e}^{s\Omega} \right\rangle_{3}\nonumber\\&&\mbox{}
-\frac{(\xi_3+\xi_2-\xi_1)}{\Delta}
\left\langle\alpha_1^{n}\alpha_2^{m} {\rm e}^{-\alpha_1\alpha_2\xi_3}\right\rangle_{2}
+\delta(n,m),
\end{fleqnarray}
\begin{fleqnarray}&&
\delta(n,m)=0,\hspace{7mm} n>0,\hspace{7mm} m>0
\end{fleqnarray}
\begin{fleqnarray}&&
\delta(n,0)=
2\frac{\xi_2}{\Delta}
\left\langle\alpha_1^{n} {\rm e}^{-\alpha_1\alpha_2\xi_2}\right\rangle_{2},
\hspace{7mm}
n>0
\end{fleqnarray}
\begin{fleqnarray}&&
\delta(0,m)=
\frac{(\xi_3-\xi_2-\xi_1)}{\Delta}
\left\langle\alpha_1^{m} {\rm e}^{-\alpha_1\alpha_2\xi_1}\right\rangle_{2},
\hspace{7mm}
m>0
\end{fleqnarray}
\begin{fleqnarray}&&
\delta(0,0)=
2\frac{\xi_2}{\Delta}
\left\langle {\rm e}^{-\alpha_1\alpha_2\xi_2}\right\rangle_{2}
+\frac{(\xi_3-\xi_2-\xi_1)}{\Delta}
\left\langle {\rm e}^{-\alpha_1\alpha_2\xi_1}\right\rangle_{2}.
\end{fleqnarray}
Together with (16.7)--(16.8) these relations make it possible
to express all integrals (16.5) through the basic form factors
(16.2)
and (16.3). Again one can show that the $\alpha$-polynomials do not
destroy the combinations  (15.11) and (15.12) in which the form
factors
appear. The recurrence relations with subtractions obtained
with the aid of eq. (15.53) are of the form
\begin{fleqnarray}
\left\langle\alpha_1^{n+1}\alpha_2^{m} \Big({\rm e}^{s\Omega}-1\Big) \right\rangle_{3}&=&
-\frac{\xi_1(\xi_3+\xi_2-\xi_1)}{\Delta}
\left\langle\alpha_1^{n}\alpha_2^{m}\Big({\rm e}^{s\Omega}-1\Big) \right\rangle_{3}
\nonumber\\&& \hspace{-20pt}\mbox{}
+2n\frac{\xi_1}{\Delta}
\left\langle\alpha_1^{n-1}\alpha_2^{m} \Big({\rm e}^{s\Omega}-1-s\Omega\Big) \right\rangle_{3}
\nonumber\\&& \hspace{-20pt}\mbox{}
+m\frac{(\xi_3-\xi_2-\xi_1)}{\Delta}
\left\langle\alpha_1^{n}\alpha_2^{m-1} \Big({\rm e}^{s\Omega}-1-s\Omega\Big)\right\rangle_{3}
\nonumber\\&& \hspace{-20pt}\mbox{}
-\frac{(\xi_3+\xi_1-\xi_2)}{\Delta}
\left\langle\alpha_1^{n}\alpha_2^{m} \Big({\rm e}^{-\alpha_1\alpha_2\xi_3}-1+\alpha_1\alpha_2\xi_3
\Big)\right\rangle_{2}
\nonumber\\&& \hspace{-20pt}\mbox{}
+\gamma(n,m),
\end{fleqnarray}
\begin{fleqnarray}&&
\gamma(n,m)=0,\hspace{7mm} n>0,\hspace{7mm} m>0
\end{fleqnarray}
\begin{fleqnarray}&&
\gamma(n,0)=
\frac{(\xi_3-\xi_2-\xi_1)}{\Delta}
\left\langle\alpha_1^{n} \Big({\rm e}^{-\alpha_1\alpha_2\xi_2}-1+\alpha_1\alpha_2\xi_2\Big)\right\rangle_{2},
\nonumber\\&& \hspace{80mm}\mbox{}
n>0
\end{fleqnarray}
\begin{fleqnarray}&&
\gamma(0,m)=
2\frac{\xi_1}{\Delta}
\left\langle\alpha_1^{m} \Big({\rm e}^{-\alpha_1\alpha_2\xi_1}-1+\alpha_1\alpha_2\xi_1\Big)\right\rangle_{2},
\hspace{7mm}  m>0
\end{fleqnarray}
\begin{fleqnarray}&&
\gamma(0,0)=
2\frac{\xi_1}{\Delta}
\left\langle \Big({\rm e}^{-\alpha_1\alpha_2\xi_1}-1+\alpha_1\alpha_2\xi_1\Big) \right\rangle_{2}
\nonumber\\&& \hspace{27mm}\mbox{}
+\frac{(\xi_3-\xi_2-\xi_1)}{\Delta}
\left\langle\Big( {\rm e}^{-\alpha_1\alpha_2\xi_2}-1+\alpha_1\alpha_2\xi_2 \Big)\right\rangle_{2},
\end{fleqnarray}
\begin{fleqnarray}
\left\langle\alpha_1^{n}\alpha_2^{m+1} \Big({\rm e}^{s\Omega}-1\Big) \right\rangle_{3}&=&
-\frac{\xi_2(\xi_3+\xi_1-\xi_2)}{\Delta}
\left\langle\alpha_1^{n}\alpha_2^{m}\Big({\rm e}^{s\Omega}-1\Big) \right\rangle_{3}
\nonumber\\&& \hspace{-20pt}\mbox{}
+2m\frac{\xi_2}{\Delta}
\left\langle\alpha_1^{n}\alpha_2^{m-1} \Big({\rm e}^{s\Omega}-1-s\Omega\Big) \right\rangle_{3}
\nonumber\\&& \hspace{-20pt}\mbox{}
+n\frac{(\xi_3-\xi_2-\xi_1)}{\Delta}
\left\langle\alpha_1^{n-1}\alpha_2^{m} \Big({\rm e}^{s\Omega}-1-s\Omega\Big)\right\rangle_{3}
\nonumber\\&& \hspace{-20pt}\mbox{}
-\frac{(\xi_3+\xi_2-\xi_1)}{\Delta}
\left\langle\alpha_1^{n}\alpha_2^{m} \Big({\rm e}^{-\alpha_1\alpha_2\xi_3}-1+\alpha_1\alpha_2\xi_3
\Big)\right\rangle_{2}
\nonumber\\&& \hspace{-20pt}\mbox{}
+\sigma(n,m),
\end{fleqnarray}
\begin{fleqnarray}&&
\sigma(n,m)=0,\hspace{7mm} n>0,\hspace{7mm} m>0
\end{fleqnarray}
\begin{fleqnarray}&&
\sigma(n,0)=
2\frac{\xi_2}{\Delta}
\left\langle\alpha_1^{n} \Big({\rm e}^{-\alpha_1\alpha_2\xi_2}-1+\alpha_1\alpha_2\xi_2\Big)\right\rangle_{2},
\hspace{7mm} n>0
\end{fleqnarray}
\begin{fleqnarray}&&
\sigma(0,m)=
\frac{(\xi_3-\xi_2-\xi_1)}{\Delta}
\left\langle\alpha_1^{m} \Big({\rm e}^{-\alpha_1\alpha_2\xi_1}-1+\alpha_1\alpha_2\xi_1\Big)\right\rangle_{2},
\nonumber\\&& \hspace{80mm}\mbox{}
 m>0
\end{fleqnarray}
\begin{fleqnarray}&&
\sigma(0,0)=
2\frac{\xi_2}{\Delta}
\left\langle\Big( {\rm e}^{-\alpha_1\alpha_2\xi_2}-1+\alpha_1\alpha_2\xi_2\Big)\right\rangle_{2}
\nonumber\\&& \hspace{27mm}\mbox{}
+\frac{(\xi_3-\xi_2-\xi_1)}{\Delta}
\left\langle\Big( {\rm e}^{-\alpha_1\alpha_2\xi_1}-1+\alpha_1\alpha_2\xi_1\Big)\right\rangle_{2},
\end{fleqnarray}
and, for the combinations (15.12) themselves, one has
\begin{fleqnarray}&&
\left\langle\Big({\rm e}^{s\Omega}-1\Big) \right\rangle_{3}=F(\xi_1,\xi_2,\xi_3) -\frac12,
\end{fleqnarray}
\begin{fleqnarray}&&
\left\langle\Big({\rm e}^{s\Omega}-1-s\Omega\Big) \right\rangle_{3}=F(\xi_1,\xi_2,\xi_3) -\frac12+\frac{1}{24}
(\xi_1+\xi_2+\xi_3)
\end{fleqnarray}
in terms of (16.3). However, the analyticity in $\xi_1,\xi_2,\xi_3$
is now maintained by a more general mechanism.
The analyticity holds only in the sum of the form factors on the
right-hand side of (16.16) (and, similarly, (16.21), (16.26),
(16.31)), and it is a nontrivial fact that, when these form factors
are expanded in power series in $\xi$, the denominator $\Delta$
gets always cancelled. The mechanism of maintaining
analyticity is based on the existence of linear differential
equations which the functions (16.2) and (16.3) satisfy.

The differential equations for the basic form factors can be derived
with the aid of the recurrence relations above. From (16.2),
one has
\begin{fleqnarray}
-\frac{d}{d\xi}f(\xi)&=&
\left\langle\alpha_1 \alpha_2 {\rm e}^{-\alpha_1\alpha_2\xi}\right\rangle_{2}
\end{fleqnarray}
which, by means of (16.8), leads to the following equation
for the function~\hbox{$f(\xi)$:}
\begin{fleqnarray}
-\frac{d}{d\xi}f(\xi)&=&\frac14 f(\xi)+\frac12
\frac{f(\xi)-1}{\xi}.
\end{fleqnarray}
The form factor (16.12) with two subtractions is expressed through
the second derivative of $f(\xi)$:
\begin{fleqnarray}
\frac{d^2}{d\xi^2}f(\xi)&=&\frac1{16}f(\xi)
+\frac14\frac{f(\xi)-1}{\xi}
+\frac34\frac{f(\xi)-1+\frac16\xi}{\xi^2}.
\end{fleqnarray}
Similarly, one obtains the equation for the form factor (16.3):
\begin{fleqnarray}
-\frac{\partial}{\partial\xi_1}F(\xi_1,\xi_2,\xi_3)&=&
\frac1{{\Delta}^2}\Big[(\xi_1-\xi_2-\xi_3)\Delta\nonumber\\&&\mbox{}
+\xi_2\xi_3(2\xi_2\xi_3-\xi_2^2-\xi_3^2+\xi_1^2)\Big]F(\xi_1,\xi_2,\xi_3)\nonumber\\&&\mbox{}
+\frac12\frac{8\xi_1\xi_2\xi_3+(\xi_2+\xi_3-\xi_1)\Delta}{\Delta^2}f(\xi_1)\nonumber\\&&\mbox{}
+2\frac{\xi_2\xi_3(\xi_3-\xi_2-\xi_1)}{\Delta^2}f(\xi_2)\nonumber\\&&\mbox{}
+2\frac{\xi_2\xi_3(\xi_2-\xi_3-\xi_1)}{\Delta^2}f(\xi_3).
\end{fleqnarray}
The function $F(\xi_1,\xi_2,\xi_3)$ is completely symmetric  in $\xi_1,\xi_2,\xi_3$ and,
therefore, satisfies two other equations, with $\partial/\partial\xi_2$
and $\partial/\partial\xi_3$, derivable from  (16.41) by symmetry.
Finally, as a consequence of these equations, one can derive an
equation
for the form factor (16.3) as a function of $s$:
\begin{fleqnarray}
-s\frac{\partial}{\partial s}F(-s\Box_1,-s\Box_2,-s\Box_3) &=&
\left(s\frac{\Box_1\Box_2\Box_3}{D}+1\right)F(-s\Box_1,-s\Box_2,-s\Box_3)\nonumber\\&&\mbox{}
+\frac{\Box_1(\Box_3+\Box_2-\Box_1)}{2D}f(-s\Box_1)\nonumber\\&&\mbox{}
+\frac{\Box_2(\Box_3+\Box_1-\Box_2)}{2D}f(-s\Box_2)\nonumber\\&&\mbox{}
+\frac{\Box_3(\Box_1+\Box_2-\Box_3)}{2D}f(-s\Box_3),
\end{fleqnarray}
\begin{equation}
D={\Box_1}^2+{\Box_2}^2+{\Box_3}^2-2\Box_1\Box_2-2\Box_1\Box_3-2\Box_2\Box_3.
\end{equation}

By applying the reduction technique above to expressions
(15.13)--(15.50),
the form factors $f_i$ with $i=1$ to $5$ and $F_i$ with $i=1$
to $29$
are brought to their final forms presented in the tables
(2.10)--(2.14) and (2.77)--(2.105). For the form factors $F_i$
with $i=30$ to $33$ the following results are obtained:
\mathindent=0pt
\arraycolsep=0pt
\begin{fleqnarray}&&
F_{30}=
 F(\xi_1,\xi_2,\xi_3)\Big[
\frac{1}{3\Delta^4}
(-{\xi_3}-{\xi_2}+{\xi_1})({\xi_1}^6-5{\xi_1}^5{\xi_2}
+8{\xi_1}^3{\xi_2}^3
\nonumber\\&&\ \ \ \ \mbox{}
-13{\xi_1}^2{\xi_2}^4+5{\xi_3}^5{\xi_1}
-6{\xi_3}^5{\xi_2}+24{\xi_3}^4{\xi_2}^2-36{\xi_2}^3{\xi_3}^3
+24{\xi_2}^4{\xi_3}^2
\nonumber\\&&\ \ \ \ \mbox{}
+8{\xi_1}^3{\xi_3}^3+4{\xi_1}^4{\xi_3}^2
-13{\xi_1}^2{\xi_3}^4-5{\xi_1}^5{\xi_3}+4{\xi_1}^4{\xi_2}^2-12{\xi_1}^3{\xi_3}^2{\xi_2}
\nonumber\\&&\ \ \ \ \mbox{}
+32{\xi_1}^2{\xi_3}^3{\xi_2}-38{\xi_1}^2{\xi_3}^2{\xi_2}^2
-6{\xi_1}^4{\xi_3}{\xi_2}-12{\xi_1}^3{\xi_3}{\xi_2}^2+32{\xi_1}^2{\xi_2}^3{\xi_3}
\nonumber\\&&\ \ \ \ \mbox{}
-3{\xi_3}^4{\xi_1}{\xi_2}-2{\xi_1}{\xi_3}^3{\xi_2}^2
-2{\xi_1}{\xi_2}^3{\xi_3}^2-3{\xi_1}{\xi_2}^4{\xi_3}-6{\xi_2}^5{\xi_3}+5{\xi_1}{\xi_2}^5)
\nonumber\\&&\ \ \ \ \mbox{}
-\frac{2}{3{\xi_1}\Delta^3}
(84{\xi_3}^2{\xi_1}{\xi_2}^2+17{\xi_1}^2{\xi_3}{\xi_2}^2
-68{\xi_2}^3{\xi_3}{\xi_1}-63{\xi_1}^3{\xi_2}^2
+7{\xi_1}^2{\xi_2}^3
\nonumber\\&&\ \ \ \ \mbox{}
+26{\xi_1}{\xi_2}^4-68{\xi_3}^3{\xi_1}{\xi_2}
+17{\xi_3}^2{\xi_1}^2{\xi_2}+26{\xi_1}^3{\xi_2}{\xi_3}+
31{\xi_1}^4{\xi_2}-12{\xi_3}^3{\xi_2}^2
\nonumber\\&&\ \ \ \ \mbox{}
-12{\xi_3}^2{\xi_2}^3+18{\xi_2}^4{\xi_3}
-6{\xi_2}^5+18{\xi_3}^4{\xi_2}+7{\xi_3}^3{\xi_1}
^2-63{\xi_3}^2{\xi_1}^3+31{\xi_1}^4{\xi_3}
\nonumber\\&&\ \ \ \ \mbox{}
+5{\xi_1}^5+26{\xi_3}^4{\xi_1}-6{\xi_3}^5)
\Big]\nonumber\\&&\ \ \ \ \mbox{}
+\left(F(\xi_1,\xi_2,\xi_3)-\frac12\right)\frac{4}{{\xi_1}{\xi_2}{\xi_3}\Delta}({\xi_1}^2+2{\xi_2}{\xi_3})\nonumber\\&&\ \ \ \ \mbox{}
+f({\xi_1})\frac{2}{3\Delta^4}
({\xi_1}^6-5{\xi_1}^5{\xi_2}+8{\xi_1}^3{\xi_2}^3-13{\xi_1}^2{\xi_2}^4
+5{\xi_3}^5{\xi_1}-6{\xi_3}^5{\xi_2}\nonumber\\&&\ \ \ \ \mbox{}
+24{\xi_3}^4{\xi_2}^2-36{\xi_2}^3{\xi_3}^3+24{\xi_2}^4{\xi_3}^2
+8{\xi_1}^3{\xi_3}^3+4{\xi_1}^4{\xi_3}^2
-13{\xi_1}^2{\xi_3}^4\nonumber\\&&\ \ \ \ \mbox{}-5{\xi_1}^5{\xi_3}+4{\xi_1}^4{\xi_2}^2
-12{\xi_1}^3{\xi_3}^2{\xi_2}
+32{\xi_1}^2{\xi_3}^3{\xi_2}-38{\xi_1}^2{\xi_3}^2{\xi_2}^2\nonumber\\&&\ \ \ \ \mbox{}
-6{\xi_1}^4{\xi_3}{\xi_2}-12{\xi_1}^3{\xi_3}{\xi_2}^2
+32{\xi_1}^2{\xi_2}^3{\xi_3}-3{\xi_3}^4{\xi_1}{\xi_2}
-2{\xi_1}{\xi_3}^3{\xi_2}^2\nonumber\\&&\ \ \ \ \mbox{}
-2{\xi_1}{\xi_2}^3{\xi_3}^2-3{\xi_1}{\xi_2}^4{\xi_3}
-6{\xi_2}^5{\xi_3}+5{\xi_1}{\xi_2}^5)\nonumber\\&&\ \ \ \ \mbox{}
-f({\xi_2})\frac{1}{12{\xi_1}\Delta^4}
(20{\xi_2}^3{\xi_3}^3{\xi_1}-66{\xi_2}^3{\xi_3}^2{\xi_1}^2
+121{\xi_2}^4{\xi_3}^2{\xi_1}+148
{\xi_2}^3{\xi_1}^3{\xi_3}\nonumber\\&&\ \ \ \ \mbox{}+81{\xi_2}^4{\xi_1}^2{\xi_3}
-74{\xi_2}^5{\xi_3}{\xi_1}-151{\xi_2}^2{\xi_3}^4{\xi_1}
+194{\xi_2}^2{\xi_3}^3{\xi_1}^2
+34{\xi_2}^2{\xi_3}^2{\xi_1}^3\nonumber\\&&\ \ \ \ \mbox{}-151{\xi_2}^2{\xi_1}^4{\xi_3}
-35{\xi_2}^3{\xi_3}^4+35{\xi_2}^4{\xi_3}^3
-21{\xi_2}^5{\xi_3}^2+29{\xi_2}^3{\xi_1}^4
-69{\xi_2}^4{\xi_1}^3\nonumber\\&&\ \ \ \ \mbox{}+19{\xi_2}^5{\xi_1}^2+7{\xi_2}^6{\xi_3}
+21{\xi_2}^2{\xi_3}^5
+53{\xi_2}^2{\xi_1}^5
-47{\xi_1}^6{\xi_2}+91{\xi_1}^3{\xi_3}^4\nonumber\\&&\ \ \ \ \mbox{}-189{\xi_1}^4{\xi_3}^3+165
{\xi_1}^5{\xi_3}^2
-9{\xi_1}{\xi_3}^6-65{\xi_1}^6{\xi_3}
-225{\xi_1}^2{\xi_2}{\xi_3}^4\nonumber\\&&\ \ \ \ \mbox{}+180{\xi_1}^3{\xi_2}{\xi_3}^3
-41{\xi_1}^4{\xi_2}{\xi_3}^2+54{\xi_1}^5{\xi_2}{\xi_3}+86{\xi_1}{\xi_2}{\xi_3}^5\nonumber\\&&\ \ \ \ \mbox{}
-3{\xi_1}^2{\xi_3}^5-7{\xi_3}^6{\xi_2}
+{\xi_3}^7+9{\xi_1}^7+7{\xi_2}^6{\xi_1}-{\xi_2}^7)\nonumber\\&&\ \ \ \ \mbox{}
-f({\xi_3})\frac{1}{12{\xi_1}\Delta^4}
(20{\xi_2}^3{\xi_3}^3{\xi_1}+194{\xi_2}^3{\xi_3}^2{\xi_1}^2
-151{\xi_2}^4{\xi_3}^2{\xi_1}+180
{\xi_2}^3{\xi_1}^3{\xi_3}\nonumber\\&&\ \ \ \ \mbox{}
-225{\xi_2}^4{\xi_1}^2{\xi_3}
+86{\xi_2}^5{\xi_3}{\xi_1}+121{\xi_2}^2{\xi_3}^4{\xi_1}
-66{\xi_2}^2{\xi_3}^3{\xi_1}^2
+34{\xi_2}^2{\xi_3}^2{\xi_1}^3\nonumber\\&&\ \ \ \ \mbox{}
-41{\xi_2}^2{\xi_1}^4{\xi_3}+35{\xi_2}^3{\xi_3}^4-35{\xi_2}^4{\xi_3}^3+
21{\xi_2}^5{\xi_3}^2
-189{\xi_2}^3{\xi_1}^4\nonumber\\&&\ \ \ \ \mbox{}
+91{\xi_2}^4{\xi_1}^3-3{\xi_2}^5{\xi_1}^2-7{\xi_2}^6{\xi_3}
-21{\xi_2}^2{\xi_3}^5+165{\xi_2}^2{\xi_1}^5-65{\xi_1}^6{\xi_2}\nonumber\\&&\ \ \ \ \mbox{}
-69{\xi_1}^3{\xi_3}^4+29{\xi_1}^4{\xi_3}^3+53{\xi_1}^5{\xi_3}^2
+7{\xi_1}{\xi_3}^6-47{\xi_1}^6{\xi_3}+81{\xi_1}^2{\xi_2}{\xi_3}^4\nonumber\\&&\ \ \ \ \mbox{}
+148{\xi_1}^3{\xi_2}{\xi_3}^3-151{\xi_1}^4{\xi_2}{\xi_3}^2
+54{\xi_1}^5{\xi_2}{\xi_3}-74{\xi_1}{\xi_2}{\xi_3}^5\nonumber\\&&\ \ \ \ \mbox{}
+19{\xi_1}^2{\xi_3}^5+7{\xi_3}^6{\xi_2}
-{\xi_3}^7+9{\xi_1}^7-9{\xi_2}^6{\xi_1}+{\xi_2}^7)\nonumber\\&&\ \ \ \ \mbox{}
+\left(\frac{f(\xi_1)-1}{\xi_1}\right) \frac{1}{{\xi_2}{\xi_3}\Delta^3}
({\xi_1}^5{\xi_2}+6{\xi_1}^3{\xi_2}^3
-4{\xi_1}^2{\xi_2}^4+{\xi_3}^5{\xi_1}-6{\xi_3}^5{\xi_2}\nonumber\\&&\ \ \ \ \mbox{}+24{\xi_3}^4{\xi_2}^2
-36{\xi_2}^3{\xi_3}^3+24{\xi_2}^4{\xi_3}^2
+6{\xi_1}^3{\xi_3}^3-4{\xi_1}^4{\xi_3}^2-4{\xi_1}^2{\xi_3}^4
+{\xi_1}^5{\xi_3}\nonumber\\&&\ \ \ \ \mbox{}
-4{\xi_1}^4{\xi_2}^2-46{\xi_1}^3{\xi_3}^2{\xi_2}
+60{\xi_1}^2{\xi_3}^3{\xi_2}-80{\xi_1}^2{\xi_3}^2{\xi_2}^2-
14{\xi_1}^4{\xi_3}{\xi_2}\nonumber\\&&\ \ \ \ \mbox{}
-46{\xi_1}^3{\xi_3}{\xi_2}^2
+60{\xi_1}^2{\xi_2}^3{\xi_3}+5{\xi_3}^4{\xi_1}{\xi_2}
-6{\xi_1}{\xi_3}^3{\xi_2}^2
-6{\xi_1}{\xi_2}^3{\xi_3}^2\nonumber\\&&\ \ \ \ \mbox{}+5{\xi_1}{\xi_2}^4{\xi_3}-6{\xi_2}^5{\xi_3}+
{\xi_1}{\xi_2}^5)\nonumber\\&&\ \ \ \ \mbox{}
+\left(\frac{f(\xi_2)-1}{\xi_2}\right)\frac{1}{2{\xi_1}{\xi_3}\Delta^3}
(-{\xi_3}^6+2{\xi_1}^5{\xi_2}+12{\xi_1}^3{\xi_2}^3
-8{\xi_1}^2{\xi_2}^4+3{\xi_3}^5{\xi_1}\nonumber\\&&\ \ \ \ \mbox{}-7{\xi_3}^
5{\xi_2}+6{\xi_3}^4{\xi_2}^2+34{\xi_2}^3{\xi_3}^3
-53{\xi_2}^4{\xi_3}^2-2{\xi_1}^3{\xi_3}^3+3{\xi_1}^4{\xi_3}^2
-2{\xi_1}^2{\xi_3}^4\nonumber\\&&\ \ \ \ \mbox{}-{\xi_1}^5{\xi_3}
-8{\xi_1}^4{\xi_2}^2+32{\xi_1}^3{\xi_3}^2{\xi_2}
-106{\xi_1}^2{\xi_3}^3{\xi_2}
+66{\xi_1}^2{\xi_3}^2{\xi_2}^2+9{\xi_1}^4{\xi_3}{\xi_2}\nonumber\\&&\ \ \ \ \mbox{}
+70{\xi_1}^3{\xi_3}{\xi_2}^2
-14{\xi_1}^2{\xi_2}^3{\xi_3}+70{\xi_3}^4{\xi_1}{\xi_2}-
102{\xi_1}{\xi_3}^3{\xi_2}^2
+112{\xi_1}{\xi_2}^3{\xi_3}^2\nonumber\\&&\ \ \ \ \mbox{}-85{\xi_1}{\xi_2}^4{\xi_3}
+21{\xi_2}^5{\xi_3}+2{\xi_1}{\xi_2}^5)\nonumber\\&&\ \ \ \ \mbox{}
-\left(\frac{f(\xi_3)-1}{\xi_3}\right)\frac{1}{2{\xi_1}{\xi_2}\Delta^3}
({\xi_1}^5{\xi_2}+2{\xi_1}^3{\xi_2}^3
+2{\xi_1}^2{\xi_2}^4-2{\xi_3}^5{\xi_1}\nonumber\\&&\ \ \ \ \mbox{}-21{\xi_3}^5{\xi_2}+53{\xi_3}
^4{\xi_2}^2-34{\xi_2}^3{\xi_3}^3-6{\xi_2}^4{\xi_3}^2
-12{\xi_1}^3{\xi_3}^3+8{\xi_1}^4{\xi_3}^2\nonumber\\&&\ \ \ \ \mbox{}
+8{\xi_1}^2{\xi_3}
^4-2{\xi_1}^5{\xi_3}-3{\xi_1}^4{\xi_2}^2
-70{\xi_1}^3{\xi_3}^2{\xi_2}+14{\xi_1}^2{\xi_3}^3{\xi_2}
-66{\xi_1}^2{\xi_3}^2{\xi_2}^2\nonumber\\&&\ \ \ \ \mbox{}
-9{\xi_1}^4{\xi_3}{\xi_2}-32{\xi_1}^3{\xi_3}{\xi_2}^2
+106{\xi_1}^2{\xi_2}^3{\xi_3}+85{\xi_3}^4{\xi_1}{\xi_2}
-112{\xi_1}{\xi_3}^3{\xi_2}^2
\nonumber\\&&\ \ \ \ \mbox{}+102{\xi_1}{\xi_2}^3{\xi_3}^2
-70{\xi_1}{\xi_2}^4{\xi_3}+7{\xi_2}^5{\xi_3}-3{\xi_1}{\xi_2}^5
+{\xi_2}^6)
\nonumber\\&&\ \ \ \ \mbox{}
+ \left(\frac{f(\xi_1)-1+\frac16\xi_1}{\xi_1^2}\right)\frac{4{\xi_1}^2}{{\xi_2}{\xi_3}\Delta^2}
(-{\xi_3}^2+2{\xi_2}{\xi_3}-2{\xi_1}{\xi_3}-{\xi_2}^2
-2{\xi_1}{\xi_2}+3{\xi_1}^2)
\nonumber\\&&\ \ \ \ \mbox{}
-\left(\frac{f(\xi_2)-1+\frac16\xi_2}{\xi_2^2}\right)\frac{4{\xi_2}}{{\xi_3}\Delta^2}
({\xi_3}^2-2{\xi_1}{\xi_3}-3{\xi_2}^2+{\xi_1}^2
+2{\xi_1}{\xi_2}+2{\xi_2}{\xi_3})
\nonumber\\&&\ \ \ \ \mbox{}
-\left(\frac{f(\xi_3)-1+\frac16\xi_3}{\xi_3^2}\right)\frac{4{\xi_3}}{{\xi_2}\Delta^2}
(-2{\xi_1}{\xi_2}+2{\xi_1}{\xi_3}+2{\xi_2}{\xi_3}+{\xi_1}^2
+{\xi_2}^2-3{\xi_3}^2),
\nonumber\\&&\mbox{}
\end{fleqnarray}
\begin{fleqnarray}&&
F_{31}(\xi_1,\xi_2,\xi_3)=0,
\end{fleqnarray}
\begin{fleqnarray}&&
F_{32}(\xi_1,\xi_2,\xi_3)=0,
\end{fleqnarray}
\begin{fleqnarray}&&
F_{33}(\xi_1,\xi_2,\xi_3)=0.
\end{fleqnarray}
\mathindent=\parindent
\arraycolsep=5pt
The nonvanishing form factor $F_{30}$ is, however, symmetric
under a permutation of the labels 2 and 3:
\begin{equation}
F_{30}(\xi_1,\xi_2,\xi_3)=
F_{30}({\xi_1},{\xi_3},{\xi_2})
\end{equation}
as one can check by a direct inspection of expression  (16.44).
On the other hand, the structure 30 in eq. (15.2) is
antisymmetric under
this permutation:
\begin{eqnarray}&&
\Re_1\Re_2\Re_3({30})=\nabla_\beta \hat{\cal R}_1^{\beta\alpha}\nabla_\alpha R_2 R_3\nonumber\\&&\ \ \ \ \ \ \ \ \mbox{}=
-\nabla_\beta \hat{\cal R}_1^{\beta\alpha}
		\nabla_\alpha R_3
		 R_2 +{\rm O}[\Re^4]+
{\rm a\ total\ derivative},
\end{eqnarray}
and, therefore, the contribution of this structure  vanishes
\begin{equation}
\int\! dx\, g^{1/2}\, {\rm tr} F_{30}(-s\Box_1,-s\Box_2,-s\Box_3) \Re_1\Re_2\Re_3({30})={\rm O}[\Re^4]
\end{equation}
   by the same mechanism as the contribution of the structure
(14.31).
The difference is only that neither of the properties (16.45)
--(16.48)
is seen before the form factors are brought to a unique
representation by eliminating the $\alpha$-polynomials.
Thus the contributions of four extra structures:
$\Re_1\Re_2\Re_3({i})$ with $i=30$ to $33$ in eq. (15.1) vanish,
and there remain only the contributions of the twenty nine cubic
structures presented in the final table (2.15)--(2.43).

Note that all the structures (14.31) and (15.2)--(15.5) whose
contributions vanish are linear in $\hat{\cal R}_{\mu\nu}$. In the final
result for the trace of the heat kernel, there  remains only
one cubic
structure linear in $\hat{\cal R}_{\mu\nu}$:
\begin{equation}
\Re_1\Re_2\Re_3({13})=\hat{\cal R}_1^{\mu\nu} \nabla_\mu\hat{P}_2\nabla_\nu\hat{P}_3
\end{equation}
(eq. (2.27) of the final table).
Its form factor is symmetric under a
permutation of the labels 2 and 3:
\begin{equation}
F_{13}(\xi_1,\xi_2,\xi_3)=F_{13}({\xi_1},{\xi_3},{\xi_2})
\end{equation}
(eq. (2.89)) but, because all the three curvatures in (16.51)
 are matrices, the structure (16.51) possesses no antisymmetry
under this
permutation. Its contribution can be written down as
\begin{eqnarray}
&&\int\! dx\, g^{1/2}\, {\rm tr} F_{13}(-s\Box_1,-s\Box_2,-s\Box_3) \Re_1\Re_2\Re_3({13})
\nonumber\\&&\ \ \ \ \ \ \ \ \mbox{}=
\frac12\int\! dx\, g^{1/2}\,  {\rm tr} F \hat{\cal R}^{\mu\nu}\Big[\nabla_\mu \hat{P},\nabla_\nu \hat{P}\Big],
\end{eqnarray}
and it does not vanish in the general case, as one can
convince oneself
by considering simple examples.

\section{Third-order form factors in the
effective action ($\omega=2$).
Finiteness}
\setcounter{equation}{0}

\hspace{\parindent}
By integrating the form factors
\[
F_i(-s\Box_1,-s\Box_2,-s\Box_3)
\]
over $s$ with the appropriate weights (following
from eqs. (1.9) and (15.1)) we obtain the
third-order form factors in the effective action:
\begin{equation}
G_i(-\Box_1,-\Box_2,-\Box_3)=
(4\pi)^2\int^\infty_0\frac{d s}{s}
\frac{s^{p_i}}{(4\pi s)^\omega}
F_i(-s\Box_1,-s\Box_2,-s\Box_3),
\end{equation}
\[
p_i=\left\{
\begin{tabular}{l}
3,\ $i=1$\ {\rm to}\ 11,\\
4,\ $i=12$\ {\rm to}\ 25,\\
5,\ $i=26$\ {\rm to}\ 28,\\
6,\ $i=29$.
\end{tabular}
\right.
\]
The normalization $(4\pi)^2$ is intended for the case
$\omega=2$ which is the only case we shall be interested in.

Inspection of eqs. (15.18)--(15.46) for
the form factors
$F_i$ in the $\alpha$-representation shows that all
$G_i$ in (17.1) have one and the same structure and
can generically be presented in the form
\begin{eqnarray}&&
G(-\Box_1,-\Box_2,-\Box_3) =
(4\pi)^{2-\omega}\int^\infty_0 d s\,\left\{
\left<
a\frac{{\rm e}^{s\Omega}}{s^{\omega-2}}
+b\frac{{\rm e}^{s\Omega}}{s^{\omega-1}}
\right.\right.
\nonumber\\&& \hspace{50mm}\mbox{}
\left.
+\overline{b}\frac{{\rm e}^{s\Omega}-1}{s^{\omega}}
+\overline{\overline{b}}\frac{{\rm e}^{s\Omega}-1-s\Omega}{s^{\omega+1}}
\right\rangle_{3}  \nonumber\\&& \hspace{35mm}\mbox{}
+\sum^3_{n=1}\left\langle
g_n\frac{{\rm e}^{s\alpha_1\alpha_2\Box_n}}{s^{\omega-1}}
+{\overline{g}}_n\frac{{\rm e}^{s\alpha_1\alpha_2\Box_n}-1}{s^{\omega}\Box_n}
\right.
\nonumber\\&& \hspace{50mm}\mbox{}\left.
+{\overline{\overline{g}}}_n\frac{{\rm e}^{s\alpha_1\alpha_2\Box_n}-1-s\alpha_1\alpha_2\Box_n}
{s^{\omega+1}{\Box_n}^2}\right\rangle_{2} \nonumber\\&&\ \ \ \ \mbox{}
+\sum_{1\leq n<m\leq 3}\frac1{\Box_n-\Box_m}\left\langle
h_{nm}\frac1{s^{\omega-1}}(
{\rm e}^{s\alpha_1\alpha_2\Box_n}
-{\rm e}^{s\alpha_1\alpha_2\Box_m})
\right.
\nonumber\\&& \hspace{30mm}\mbox{}
+{\overline{h}}_{nm}\frac1{s^{\omega}}\Big(
\frac{{\rm e}^{s\alpha_1\alpha_2\Box_n}-1}{\Box_n}
-\frac{{\rm e}^{s\alpha_1\alpha_2\Box_m}-1}{\Box_m}\Big)
\nonumber\\&&\ \ \ \ \mbox{}
\left.\left.
+{\overline{\overline{h}}}_{nm}\frac1{s^{\omega+1}}\Big(
\frac{{\rm e}^{s\alpha_1\alpha_2\Box_n}
-1-s\alpha_1\alpha_2\Box_n}{{\Box_n}^2}
-\frac{{\rm e}^{s\alpha_1\alpha_2\Box_m}
-1-s\alpha_1\alpha_2\Box_m}{{\Box_m}^2}\Big)\right\rangle_{2}\right\}
\nonumber\\&&\ \ \ \ \mbox{}
\end{eqnarray}
where
\begin{equation}
a,\ \ b,\ \ \overline{b},\ \ \overline{\overline{b}},\ \
g_n,\ \ {\overline{g}}_n,\ \ {\overline{\overline{g}}}_n,\ \
h_{nm},\ \ {\overline{h}}_{nm},\ \ {\overline{\overline{h}}}_{nm}
\end{equation}
are functions only of $\alpha$'s and $\Box$'s.
If, instead of the $\alpha$-representation for the
form factors $F_i$, one starts with their explicit
forms in eqs. (2.77)--(2.105), the result for $G_i$
will anyway have the form (17.2) with the only
difference that the coefficients (17.3) will be
functions only
of $\Box$'s, and then the $\alpha$-averages
in (17.2) will be identified directly with the basic form
factors $f(\xi)$ and $F(\xi_1,\xi_2,\xi_3)$ (cf. eqs. (16.2),
(16.3), (16.11), (16.12), (16.36), (16.37)).
However, in this case, the functions $f(\xi)$
and $F(\xi_1,\xi_2,\xi_3)$ themselves should be put back in the
$\alpha$-form because the next step in the
calculation is commuting the
$\alpha$- and $s$-integrations.

At $\omega=2$, the $a$-term and all $h$-terms in (17.2)
are finite whereas all $b$-terms and all $g$-terms
are ultraviolet divergent. On the other hand,
it is well known
that, in four dimensions, the ultraviolet divergences
of the one-loop effective action are limited to terms
of zeroth, first and second orders in the curvature
[6]; terms of third order should be finite already.
{\em They are finite indeed}. There are two mechanisms
by which the ultraviolet divergences appearing in the
third-order form factors (17.2) cancel. In the form
factors $G_i$ of all structures except purely
gravitational the divergences appear only because of the
division of the heat kernel into $b$-terms and
$g$-terms (and subdivisions into
$b, \overline{b}, \overline{\overline{b}}$ and $g, \overline{g}, \overline{\overline{g}}$). In the
sum of these terms the divergences cancel, and the
form factors $G_i$ themselves are finite. The situation
with the purely gravitational structures is
different. Their form factors
$G_i$ are actually divergent
\footnote{\normalsize
The reason why these divergences appear at all is the
presence of the Riemann tensor in the DeWitt
coefficient $a_2(x,x)$ which governs the ultraviolet
divergences in four dimensions (see sect. 4).
Since covariant perturbation theory expands the Riemann
tensor in an infinite series in powers of the Ricci
tensor (see paper II and Appendix below), the divergent
term with the Riemann tensor brings divergent
contributions to the third and all higher orders in the
curvature. The problem vanishes, however, if one
takes into account the Gauss-Bonnet identity which,
in four dimensions, eliminates the Riemann tensor
from the (integrated) $a_2(x,x)$. Automatically
eliminated then are also all divergent contributions
of higher orders in the curvature. At each order,
there exists a nonlocal constraint which ensures
this elimination. The hierarchy of these constraints
is generated by the expansion of the Gauss-Bonnet
invariant (see eq. (A.38) of Appendix).
}
but the divergences cancel in the sum
\[
\sum_i G_i\Re_1\Re_2\Re_3({i})
\]
owing to a nonlocal constraint which holds between
the purely gravitational structures in four dimensions
(see Appendix). Below, these conclusions are confirmed by
a direct calculation.

Although the final quantity of interest is finite,
for dealing with intermediate divergent quantities,
it is convenient to use the method of dimensional
regularization. By applying the rules
of dimensional regularization
to the integrals
in (17.2) we obtain (see e.g. [7])
\begin{fleqnarray}&&
(4\pi)^{2-\omega}\int^\infty_0\!ds\,
\frac{{\rm e}^{sE}}{s^{\omega-2}} =
\int^\infty_0\!ds\,{\rm e}^{sE}
+{\rm O}(2-\omega),
\end{fleqnarray}
\begin{fleqnarray}&&
(4\pi)^{2-\omega}\int^\infty_0\!ds\,
\frac{{\rm e}^{sE}}{s^{\omega-1}} =
\Big(\frac1{2-\omega}+\ln4\pi\Big)\nonumber\\&& \hspace{50mm}\mbox{}
-E\int^\infty_0\!ds\,\ln s\,{\rm e}^{sE}
+{\rm O}(2-\omega),
\end{fleqnarray}
\begin{fleqnarray}&&
(4\pi)^{2-\omega}\int^\infty_0\!ds\,
\frac{{\rm e}^{sE}-1}{s^{\omega}} =
\Big(\frac1{2-\omega}+\ln4\pi\Big)E+E\nonumber\\&& \hspace{50mm}\mbox{}
-E^2\int^\infty_0\!ds\,\ln s\,{\rm e}^{sE}
+{\rm O}(2-\omega),
\end{fleqnarray}
\begin{fleqnarray}&&
(4\pi)^{2-\omega}\int^\infty_0\!ds\,
\frac{{\rm e}^{sE}-1-sE}{s^{\omega+1}} =
\Big(\frac1{2-\omega}+\ln4\pi\Big)\frac12E^2+\frac34E^2\nonumber\\&& \hspace{50mm}\mbox{}
-\frac12E^3\int^\infty_0\!ds\,\ln s\,{\rm e}^{sE}
+{\rm O}(2-\omega),
\end{fleqnarray}
and
\begin{fleqnarray}&&
\int^\infty_0\!ds\,{\rm e}^{sE} = -\frac1E,
\end{fleqnarray}
\begin{fleqnarray}&&
\int^\infty_0\!ds\,\ln s\,{\rm e}^{sE} =
\frac1E\Big({\rm C}+\ln(-E)\Big)
\end{fleqnarray}
where
\begin{equation}
E=\left\{
\begin{tabular}{l}
$\Omega$ \\ $\alpha_1\alpha_2\Box$
\end{tabular}
\right.,\hspace{7mm} E<0,
\end{equation}
and ${\rm C}$ is the Euler constant. As a result, the
 form factor (17.2) takes the form
\begin{eqnarray}&&
G(-\Box_1,-\Box_2,-\Box_3) = \Big(\frac1{2-\omega}+\ln4\pi-{\rm C}\Big) \nonumber\\&&\ \ \ \ \ \ \ \ \mbox{}
\times\left[
\Big<b+\overline{b}\Omega+\frac12\overline{\overline{b}}\Omega^2\Big>_3
+\sum^3_{n=1}\Big<
g_n+\alpha_1\alpha_2{\overline{g}}_n+\frac12(\alpha_1\alpha_2)^2{\overline{\overline{g}}}_n\Big>_2
\right] \nonumber\\&&\ \ \ \ \mbox{}
-\left[
\Big<\ln(-\Omega)\Big(b+\overline{b}\Omega+\frac12\overline{\overline{b}}\Omega^2\Big)\Big>_3
\right.
\nonumber\\&&\ \ \ \ \ \ \ \ \mbox{}
+\sum^3_{n=1}\ln(-\Box_n)\Big<
g_n+\alpha_1\alpha_2{\overline{g}}_n+\frac12(\alpha_1\alpha_2)^2{\overline{\overline{g}}}_n\Big>_2
\Big] \nonumber\\&&\ \ \ \ \mbox{}
+\Big<a\frac1{-\Omega}\Big>_3
-\sum_{1\leq n<m\leq 3}\frac{\ln(\Box_n/\Box_m)}
{\Box_n-\Box_m}
\Big<h_{nm}
+\alpha_1\alpha_2{\overline{h}}_{nm}
+\frac12(\alpha_1\alpha_2)^2{\overline{\overline{h}}}_{nm}\Big>_2 \nonumber\\&&\ \ \ \ \mbox{}
+r(\Box_1,\Box_2,\Box_3)
\end{eqnarray}
where
\begin{eqnarray}&&
r(\Box_1,\Box_2,\Box_3) =
\Big<\overline{b}\Omega+\frac34\overline{\overline{b}}\Omega^2\Big>_3
+\sum^3_{n=1}\left\langle
{\overline{g}}_n\alpha_1\alpha_2+\frac34{\overline{\overline{g}}}_n(\alpha_1\alpha_2)^2\right\rangle_2  \nonumber\\&&\ \ \ \ \ \ \ \ \mbox{}
-\sum^3_{n=1}\left\langle (\ln\alpha_1\alpha_2)\Big(
g_n+\alpha_1\alpha_2{\overline{g}}_n+\frac12(\alpha_1\alpha_2)^2{\overline{\overline{g}}}_n\Big)\right\rangle_2
\end{eqnarray}
is a rational (tree) function of $\Box$'s which can be
calculated explicitly with the aid of (15.51)--(15.53).

The first group of terms in (17.11), with the pole in
$\omega$, represents the logarithmic divergences. The second
group contains accompanying them by dimension
$\ln \Box$ terms. Since the divergences must cancel,
the log's with an arbitrary scaling must cancel as well;
only $\log$'s of ratios, like $\ln(\Box_n/\Box_m)$,
may and do survive. To single out the log's with
an arbitrary scaling, we denote for short
\begin{fleqnarray}&&
b+\overline{b}\Omega+\frac12\overline{\overline{b}}\Omega^2 = {\widetilde{b}},
\end{fleqnarray}
\begin{fleqnarray}&&
g_n+\alpha_1\alpha_2{\overline{g}}_n+\frac12(\alpha_1\alpha_2)^2{\overline{\overline{g}}}_n
={\widetilde{g}}_n,
\end{fleqnarray}
\begin{fleqnarray}&&
h_{nm}
+\alpha_1\alpha_2{\overline{h}}_{nm}
+\frac12(\alpha_1\alpha_2)^2{\overline{\overline{h}}}_{nm}
=\widetilde{h}_{nm}
\end{fleqnarray}
and write
\mathindent=0pt
\begin{fleqnarray}&&
\Big<\ln(-\Omega){\widetilde{b}}\Big>_3 =
\frac13\left(\sum^3_{n=1}\ln(-\Box_n)\right)
\Big<{\widetilde{b}}\Big>_3
+\Big<\ln\Big(
\frac{-\Omega}{(-\Box_1\Box_2\Box_3)^{1/3}}\Big)
{\widetilde{b}}\Big>_3,
\end{fleqnarray}
\begin{fleqnarray}&&
\sum^3_{n=1}\ln(-\Box_n)\Big<{\widetilde{g}}_n\Big>_2 =
\frac13\left(\sum^3_{n=1}\ln(-\Box_n)\right)
\sum^3_{m=1}\Big<{\widetilde{g}}_m\Big>_2 \nonumber\\&& \hspace{40mm}\mbox{}
+\frac13\sum_{1\leq n<m\leq 3}
\Big(\ln(\Box_n/\Box_m)\Big)
\Big<{\widetilde{g}}_n-{\widetilde{g}}_m \Big>_2 .
\end{fleqnarray}

Then we have
\begin{fleqnarray}&&
G(-\Box_1,-\Box_2,-\Box_3) = G^{\rm div}+G^{\rm fin},
\end{fleqnarray}
\begin{fleqnarray}&&
G^{\rm div} =
\left(\frac1{2-\omega}+\ln4\pi-{\rm C}
-\frac13\sum^3_{n=1}\ln(-\Box_n)\right)
\Big[\Big<{\widetilde{b}}\Big>_3
+\sum^3_{m=1}\Big<{\widetilde{g}}_m \Big>_2\Big],
\end{fleqnarray}
\begin{fleqnarray}&&
G^{\rm fin} = \left\langle a\frac1{-\Omega}\right\rangle_3
-\left\langle{\widetilde{b}}\Big(\ln(-\Omega)
-\frac13\sum^3_{n=1}\ln(-\Box_n)\Big)\right\rangle_3   \nonumber\\&&\ \ \ \ \mbox{}
-\frac13\sum_{1\leq n<m\leq 3}
\Big<{\widetilde{g}}_n-{\widetilde{g}}_m \Big>_2
\ln(\Box_n/\Box_m)
-\sum_{1\leq n<m\leq 3}
\Big<\widetilde{h}_{nm}\Big>_2
\frac{\ln(\Box_n/\Box_m)}{\Box_n-\Box_m} \nonumber\\&&\ \ \ \ \mbox{}
+r(\Box_1,\Box_2,\Box_3).
\end{fleqnarray}
\mathindent=\parindent

The averages in (17.19) are $\alpha$-integrals
of pure polynomials and are easily calculated
for each $G_i$ with the aid of (15.51), (15.53).
The result of this calculation can be presented in
the form
\begin{eqnarray}&&
\int\! dx\, g^{1/2}\, {\rm tr}\,\sum^{29}_{i=1} G^{\rm div}_i \Re_1\Re_2\Re_3({i}) \nonumber\\&&\ \ \ \ \mbox{} =
\int\! dx\, g^{1/2}\, {\rm tr}\,
\left(\frac1{2-\omega}+\ln4\pi-{\rm C}
-\frac13\sum^3_{n=1}\ln(-\Box_n)
\right)
\nonumber\\&&\ \ \ \ \mbox{}\ \
\times\left\{
\frac1{360}\frac{\Box_1}{\Box_2\Box_3}\Re_1\Re_2\Re_3({9})
\right.
\nonumber\\&&\ \ \ \ \ \ \ \ \mbox{}
+\frac1{45}\Big(\frac12\frac{\Box_1}{\Box_2\Box_3}-\frac1{\Box_1}\Big)\Re_1\Re_2\Re_3({10}) \nonumber\\&&\ \ \ \ \ \ \ \ \mbox{}
+\frac1{90}\Big(-\frac12\frac{\Box_3}{\Box_1\Box_2}+\frac1{\Box_1}\Big)\Re_1\Re_2\Re_3({11}) \nonumber\\&&\ \ \ \ \ \ \ \ \mbox{}
+\frac1{30}\Big(-\frac12\frac1{\Box_2\Box_3}-\frac13\frac1{\Box_1\Box_2}\Big)\Re_1\Re_2\Re_3({22}) \nonumber\\&&\ \ \ \ \ \ \ \ \mbox{}
+\frac1{45}\frac1{\Box_1\Box_2}\Re_1\Re_2\Re_3({23})
+\frac1{45}\frac1{\Box_2\Box_3}\Re_1\Re_2\Re_3({24}) \nonumber\\&&\ \ \ \ \ \ \ \ \mbox{}
+\frac1{45}\Big(-\frac1{\Box_2\Box_3}+2\frac1{\Box_1\Box_2}\Big)\Re_1\Re_2\Re_3({25})\nonumber\\&&\ \ \ \ \ \ \ \ \mbox{}
+\frac1{45}\Big(-\frac1{\Box_1\Box_2\Box_3}\Big)\Re_1\Re_2\Re_3({27}) \nonumber\\&&\ \ \ \ \ \ \ \ \mbox{}
+\frac2{45}\Big(-\frac1{\Box_1\Box_2\Box_3}\Big)\Re_1\Re_2\Re_3({28}) \Big\}
\end{eqnarray}
whence it is seen that $G^{\rm div}$ (properly
symmetrized) is nonvanishing only for purely
gravitational structures, and, among the latter,
the structure with the maximum number of derivatives:
$\Re_1\Re_2\Re_3({29})$ has $G^{\rm div}=0$. The fact that
$G^{\rm div}_{29}=0$ is explainable from the
viewpoint of the analysis carried out in
Appendix. If in the identity (A.35) of Appendix
one puts
\begin{equation}
{\cal F}^{\rm sym}(\Box_1,\Box_2,\Box_3) =
-\frac2{45}\frac1{\Box_1\Box_2\Box_3}
\left(\frac1{2-\omega}+\ln4\pi-{\rm C}
-\frac13\sum^3_{n=1}\ln(-\Box_n)\right)
\end{equation}
the result will be precisely the right-hand side of (17.21).
Hence
\begin{equation}
\int\! dx\, g^{1/2}\, {\rm tr}\,\sum^{29}_{i=1} G^{\rm div}_i \Re_1\Re_2\Re_3({i})  = 0.
\end{equation}

As discussed in Appendix, the identity (A.35) does in
fact mean that, in four dimensions, the basis of nonlocal
gravitational invariants can be reduced by one structure.
{\em It is in the overcomplete basis that the gravitational
form factors contain divergences}. Our final results in
sects. 6--9 are given in the reduced basis obtained by
eliminating the completely symmetric part of the structure
28. In the heat kernel, this reduction of the basis
amounts to replacing the form factors $F_i$ by the
following modified ones:
\begin{eqnarray}
F^{\rm mod}_{i} &=& F_{i},\hspace{7mm}
i\neq 9,10,11,22,23,24,25,27,28, \\[\baselineskip]
F^{\rm mod}_{9} &=& F_{9}
-\frac13\frac{({\Box_1}^2+{\Box_2}^2+{\Box_3}^2)}{{\Box_1}{\Box_2}{\Box_3}}
\left\langle\frac13\alpha_1\alpha_2\alpha_3\frac{{\rm e}^{s\Omega}}s\right\rangle_3,\\[\baselineskip]
F^{\rm mod}_{10} &=& F_{10}
-\frac43\frac1{\Box_1\Box_2\Box_3}({\Box_1}^2+{\Box_2}^2+{\Box_3}^2 \nonumber\\&&\ \ \ \ \mbox{}
-2{\Box_1}{\Box_2}-2{\Box_1}{\Box_3}-2{\Box_2}{\Box_3})
\left\langle\frac13\alpha_1\alpha_2\alpha_3\frac{{\rm e}^{s\Omega}}s\right\rangle_3,\\[\baselineskip]
F^{\rm mod}_{11} &=& F_{11}
-2\frac{({\Box_1}+{\Box_2}-{\Box_3})}{{\Box_1}{\Box_2}}
\left\langle\frac13\alpha_1\alpha_2\alpha_3\frac{{\rm e}^{s\Omega}}s\right\rangle_3,\\[\baselineskip]
F^{\rm mod}_{22} &=& F_{22}
+2\frac{(3{\Box_1}+{\Box_2}+{\Box_3})}{{\Box_1}{\Box_2}{\Box_3}}
\left\langle\frac13\alpha_1\alpha_2\alpha_3\frac{{\rm e}^{s\Omega}}{s^2}\right\rangle_3,\\[\baselineskip]
F^{\rm mod}_{23} &=& F_{23}
-8\frac1{\Box_1\Box_2}
\left\langle\frac13\alpha_1\alpha_2\alpha_3\frac{{\rm e}^{s\Omega}}{s^2}\right\rangle_3,\\[\baselineskip]
F^{\rm mod}_{24} &=& F_{24}
-8\frac1{\Box_2\Box_3}
\left\langle\frac13\alpha_1\alpha_2\alpha_3\frac{{\rm e}^{s\Omega}}{s^2}\right\rangle_3,\\[\baselineskip]
F^{\rm mod}_{25} &=& F_{25}
-8\frac{({\Box_2}+{\Box_3}-{\Box_1})}{{\Box_1}{\Box_2}{\Box_3}}
\left\langle\frac13\alpha_1\alpha_2\alpha_3\frac{{\rm e}^{s\Omega}}{s^2}\right\rangle_3,\\[\baselineskip]
F^{\rm mod}_{27} &=& F_{27}
+8\frac1{\Box_1\Box_2\Box_3}
\left\langle\frac13\alpha_1\alpha_2\alpha_3\frac{{\rm e}^{s\Omega}}{s^3}\right\rangle_3,\\[\baselineskip]
F^{\rm mod}_{28} &=& F_{28}
+16\frac1{\Box_1\Box_2\Box_3}
\left\langle\frac13\alpha_1\alpha_2\alpha_3\frac{{\rm e}^{s\Omega}}{s^3}\right\rangle_3.
\end{eqnarray}
For the respectively modified form factors in
the effective action we introduce the notation
\begin{equation}
\Gamma_{i}(-\Box_1,-\Box_2,-\Box_3) =
(4\pi)^2\int^\infty_0\frac{d s}{s}
\frac{s^{p_i}}{(4\pi s)^\omega}\,
F_i^{\rm mod}(-s\Box_1,-s{\Box_2},-s{\Box_3})
\end{equation}
where the exponents $p_i$ are the same as in (17.1).

It is easy to make sure that
\begin{equation}
\int\! dx\, g^{1/2}\, {\rm tr}\sum^{29}_{i=1}s^{p_i}
F^{\rm mod}_i\Re_1\Re_2\Re_3({i})
=
\int\! dx\, g^{1/2}\, {\rm tr}\sum^{29}_{i=1}s^{p_i}
F_i\Re_1\Re_2\Re_3({i}) ,
\end{equation}
and, respectively,
\begin{equation}
\int\! dx\, g^{1/2}\, {\rm tr}\sum^{29}_{i=1}
\Gamma_i\Re_1\Re_2\Re_3({i})
=
\int\! dx\, g^{1/2}\, {\rm tr}\sum^{29}_{i=1}
G_i\Re_1\Re_2\Re_3({i})
\end{equation}
since (17.35) is a special case of the identity
(A.35) corresponding to
\begin{equation}
{\cal F}^{\rm sym}(\Box_1,\Box_2,\Box_3) =
\frac{16}{\Box_1\Box_2\Box_3}
\left\langle\frac13\alpha_1\alpha_2\alpha_3\frac{{\rm e}^{s\Omega}}{s^3}\right\rangle_3.
\end{equation}
Thus, the above modification of the form factors
has no effect on the trace of the heat kernel and
the effective action {\em in four dimensions}.
On the other hand, with the expression (15.45)
for $F_{28}$, the completely symmetric part of
$F_{28}^{\rm mod}$ in (17.33) is
\begin{equation}
-\frac{16}{\Box_1\Box_2\Box_3}
\Big<\frac13(
\alpha_1\alpha_2\alpha_3^2
+\alpha_1\alpha_3\alpha_2^2
+\alpha_2\alpha_3\alpha_1^2
)\frac{{\rm e}^{s\Omega}}{s^3}\Big>_3
+\frac{16}{\Box_1\Box_2\Box_3}
\Big<\frac13\alpha_1\alpha_2\alpha_3\frac{{\rm e}^{s\Omega}}{s^3}\Big>_3
=0
\end{equation}
by virtue of the delta-function
$\delta(\sum\alpha-1)$ contained in $\Big<\,\Big>_3$.
Since removal of  the symmetric part of $G_{28}$
removes $G^{\rm div}_{28}$, it automatically, via
the identity (A.35), removes all $G^{\rm div}_i$.
As a result, in the reduced basis, the form
factors $\Gamma_i$ {\it themselves} are finite.

\section{Reduction of the form factors in
\hbox{$W$ ($\omega=2$)} to the basic form factors}
\setcounter{equation}{0}

\hspace{\parindent}
In  terms of the representation (17.2), the
transition from $G_i$ to $\Gamma_i$ changes only the
coefficient $b$. Since this change turns into zero
the coefficients of all $G^{\rm div}_i$ in (17.19),
the form factors $\Gamma_i$ have the same form as
$G^{\rm fin}$ in (17.20) with the modified $b$.
Each
\begin{equation}
\Gamma_{i}(-\Box_1,-\Box_2,-\Box_3),\hspace{7mm} i=1\ {\rm\ to}\ 29
\end{equation}
is, therefore, a sum of contributions of the following
five types:
\begin{equation}
\left\langle P(\alpha,\Box)\frac1{-\Omega}\right\rangle_3,
\end{equation}
\begin{equation}
\left\langle P(\alpha,\Box)
\Big(\ln(-\Omega)
-\frac13\sum^3_{n=1}\ln(-\Box_n)
\Big)
\right\rangle_3,
\end{equation}
\begin{equation}
\sum_{1\leq m<n\leq 3}
\Big<P_{nm}(\alpha,\Box)\Big>_2
\ln(\Box_n/\Box_m),
\end{equation}
\begin{equation}
\sum_{1\leq m<n\leq 3}
\Big<P_{nm}(\alpha,\Box)\Big>_2
\frac{\ln(\Box_n/\Box_m)}{\Box_n-\Box_m},
\end{equation}
\begin{equation}
\Big<P(\alpha,\Box)\Big>_3,\hspace{7mm}
\Big<P(\alpha,\Box)\Big>_2
\end{equation}
where $P(\alpha,\Box)$ are polynomials in
$\alpha$'s, boxes and inverse boxes.  In (18.4)--(18.6),
the $\alpha$-averages can be calculated explicitly.
Below we summarize a technique by which i) the
contributions (18.3), (18.4) and (18.6) can be put in
the form (18.2), and ii) the contributions of the form
(18.2) can be expressed in either an algebraic or a
differential way through elementary functions and
the basic third-order form factor
\begin{equation}
\Gamma(-\Box_1,-\Box_2,-\Box_3) =
\int^\infty_0ds\,F
(-s\Box_1,-s\Box_2,-s\Box_3)=
\left\langle\frac1{-\Omega}\right\rangle_3.
\end{equation}
The contributions of the type (18.5) are of a
special origin (see sect. 14) and remain unaffected
by these transformations.

The reduction is mainly based on the formulae derived
in paper III. Eqs. (2.17), (2.18), (2.19) and (4.11)
of paper III, after a minor modification and
adaptation to the present notation, read
\begin{fleqnarray}&&
\left\langle\frac{\alpha_1^{n_1}\alpha_2^{n_2}\alpha_3^{n_3}}{-\Omega}\right\rangle_3
=
\frac1{(n_1+n_2+n_3)!}\nonumber\\&&\ \ \ \ \mbox{}\times
(-\Box_1)^{n_1}\frac{\partial^{n_1}}
  {\partial\Box_1^{n_1}}
(-\Box_2)^{n_2}\frac{\partial^{n_2}}
  {\partial\Box_2^{n_2}}
(-\Box_3)^{n_3}\frac{\partial^{n_3}}
  {\partial\Box_3^{n_3}}\Gamma(-\Box_1,-\Box_2,-\Box_3),
\nonumber\\&&\mbox{}
\end{fleqnarray}
\begin{fleqnarray}&&
\left<\alpha_1^{n_1}\alpha_2^{n_2}\alpha_3^{n_3}
\left(\ln(-\Omega)
-\frac13\sum^3_{m=1}\ln(-\Box_m)
\right)\right>_3
=
\frac{n_1!n_2!n_3!}{(n_1+n_2+n_3+2)!}\nonumber\\&& \hspace{20mm}\mbox{}\times
\left(
\frac43\sum^{n_1}_{l=1}\frac1l
+\frac43\sum^{n_2}_{l=1}\frac1l
+\frac43\sum^{n_3}_{l=1}\frac1l
-2\sum^{n_1+n_2+n_3+2}_{l=3}\frac1l\right) \nonumber\\&&\ \ \ \ \mbox{}
-\frac2{(n_1+n_2+n_3+2)!}
(-\Box_1)^{n_1+1}\frac{\partial}
  {\partial{\Box_1}^{n_1}}
(-\Box_2)^{n_2+1}\frac{\partial}
  {\partial{\Box_2}^{n_2}}
\nonumber\\&&\ \ \ \ \mbox{}\ \ \times
(-\Box_3)^{n_3+1}\frac{\partial}
  {\partial{\Box_3}^{n_3}}
\frac1{\Box_1\Box_2\Box_3}
\left<\ln(-\Omega)
-\frac13\sum^3_{m=1}\ln(-\Box_m)
\right>_3,
\end{fleqnarray}
\begin{fleqnarray}&&
\left<\ln(-\Omega)
-\frac13\sum^3_{m=1}\ln(-\Box_m)
\right>_3
\nonumber\\&&\ \ \ \ \ \ \ \ \mbox{}
=-\frac32-\frac16
\sum^3_{m=1}\Box_m\Big(1+\Box_m
\frac{\partial}{\partial\Box_m}\Big)
\Gamma(-\Box_1,-\Box_2,-\Box_3),
\end{fleqnarray}
\begin{fleqnarray}&&
\ln(\Box_1/\Box_2) \nonumber\\&& \hspace{3mm}\mbox{}=
\left[
\Box_1\Box_3\Big(
\frac{\partial}{\partial\Box_1}
+\frac{\partial}{\partial\Box_3}\Big)
-\Box_2\Box_3\Big(
\frac{\partial}{\partial\Box_2}
+\frac{\partial}{\partial\Box_3}\Big)
\right]\Gamma(-\Box_1,-\Box_2,-\Box_3),
\nonumber\\&&\mbox{}
\end{fleqnarray}
and the identity
\begin{equation}
\frac12=\left\langle\frac{-\Omega}{-\Omega}\right\rangle_3
\end{equation}
adds here one more relation:
\begin{equation}
-1=\Box_1\Box_2\Box_3\Big(
\frac{\partial}{\partial\Box_1}
\frac{\partial}{\partial\Box_2}
+
\frac{\partial}{\partial\Box_2}
\frac{\partial}{\partial\Box_3}
+
\frac{\partial}{\partial\Box_1}
\frac{\partial}{\partial\Box_3}
\Big)\Gamma(-\Box_1,-\Box_2,-\Box_3).
\end{equation}

These relations make it possible to
express all contributions (18.2), (18.3), (18.4),
(18.6) through the derivatives of the basic form
factor $\Gamma$, and, on the other hand, relation (18.8)
establishes a one-to-one correspondence
between derivatives of $\Gamma$ and averages of the
form (18.2). Therefore, all form factors (18.1)
can be put in either of the two equivalent forms
\begin{fleqnarray}&&
\Gamma_{i}(-\Box_1,-\Box_2,-\Box_3)=
\left<\frac{P_i(\alpha,\Box)}{-\Omega}\right>_3
+{\rm terms}\ (18.5),
\end{fleqnarray}
\begin{fleqnarray}&&
\Gamma_{i}(-\Box_1,-\Box_2,-\Box_3)
\nonumber\\&&\ \ \ \ \ \ \ \ \mbox{}
=P'_i
\Big(\Box,\frac{\partial}
{\partial\Box}\Big)\Gamma(-\Box_1,-\Box_2,-\Box_3)
+{\rm terms}\ (18.5).
\end{fleqnarray}
In this way the results in sect. 7 are obtained.
The final expressions for $\Gamma_{i}$ in the
$\alpha$-representation given in sect. 7 differ
from (18.14) only in that some tree terms of the type
(18.6), those responsible for the power growth
of the form factors, are written down explicitly.
Singling out of such terms from the general expression
(18.14) is discussed in the next section in connection
with the asymptotic behaviour of the form factors.

The representations (18.14) and (18.15) are not unique.
Differentiation of eq. (18.11) and similar equations with
permuted indices gives rise to a hierarchy of identities
between the derivatives of $\Gamma$, or the averages of
the form (18.2). For example, by differentiating (18.11)
once with respect to $\Box_3$, one obtains the identity
\begin{eqnarray}&&
\left[
\Box_1\frac{\partial}{\partial\Box_1}
-\Box_2\frac{\partial}{\partial\Box_2}
+(\Box_1-\Box_2)\frac{\partial}{\partial\Box_3}
+\Box_1\Box_3\Big(
\frac{\partial^2}{\partial\Box_1\partial\Box_3}
+\frac{\partial^2}{\partial\Box_3^2}\Big)
\right.
\nonumber\\&&\ \ \ \ \ \ \ \ \mbox{}
\left.
-\Box_2\Box_3\Big(
\frac{\partial^2}{\partial\Box_2\partial\Box_3}
+\frac{\partial^2}{\partial\Box_3^2}\Big)
\right]\Gamma(-\Box_1,-\Box_2,-\Box_3)=0,
\end{eqnarray}
or
\begin{eqnarray}&&\hspace{-5mm}
\left\langle\frac1{-\Omega}\Big((\Box_1-\Box_2)(2\alpha_3^2-\alpha_3)
+\Box_3(\alpha_1-\alpha_2)(2\alpha_3-1)\Big)\right\rangle_3=0.
\end{eqnarray}
This arbitrariness, and the constraint $\sum\alpha=1$
have been used to bring the form factors to their
final forms (7.2)--(7.30) and, in particular, to
secure the fulfilment of the "rule of the like $\alpha$"
observed in sect. 7. The fact that expressions (18.14)
{\em can} be transformed so that this rule hold is
a nontrivial property of the form factors which
has important implications.

The cause of nonuniqueness of the representations
(18.14) and (18.15) is the existence of linear
differential equations satisfied by the function (18.7).
For the derivation of these equations one may use the
equations for the basic form factor in the heat kernel
\[ F=F(-s\Box_1,-s\Box_2,-s\Box_3). \]
By combining eqs. (16.41) and (16.42), one has
\begin{eqnarray}
\frac{\partial}{\partial\Box_1}F &=&
\frac{(\Box_2-\Box_3)^2-\Box_1^2}{D\Box_1}
\frac{\partial}{\partial s}(sF)
+\frac{\Box_2+\Box_3-\Box_1}{D}F \nonumber\\&&\mbox{}
+\frac1{2D}\Big(\Box_1f(-s\Box_1)-\Box_2f(-s\Box_2)\Big)\nonumber\\&&\mbox{}
+\frac1{2D}\Big(\Box_1f(-s\Box_1)-\Box_3f(-s\Box_3)\Big)\nonumber\\&&\mbox{}
+\frac{(\Box_2-\Box_3)}{2D\Box_1}
\Big(\Box_3f(-s\Box_3)-\Box_2f(-s\Box_2)\Big).
\end{eqnarray}
When this equation is integrated over $s$ from 0 to
$\infty$, the total derivative term with $F$
vanishes by virtue of the asymptotic behaviours (3.2)
and (4.2). Moreover, since the leading asymptotic
behaviour (3.1) of the function $f$ cancels in the
appearing differences, the integrals of these
differences converge and give
\begin{equation}
\int^\infty_0\!ds\,
\Big(\Box_1f(-s\Box_1)-\Box_2f(-s\Box_2)\Big)
=-2\ln({\Box_1}/{\Box_2}).
\end{equation}
The result is the following equation for the function (18.7):
\begin{eqnarray}&&
\frac{\partial}{\partial\Box_1}\Gamma =
\frac{\Box_2+\Box_3-\Box_1}D\Gamma \nonumber\\&&\mbox{}
+\frac1D\ln({\Box_2}/{\Box_1})
+\frac1D\ln({\Box_3}/{\Box_1})
+\frac{\Box_2-\Box_3}{D\Box_1}\ln({\Box_2}/{\Box_3})
\end{eqnarray}
and two other equations, with
$\partial/\partial\Box_2$ and
$\partial/\partial\Box_3$,
obtained from (18.20) by symmetry.

After the use of these equations in (18.15), all third-order
form factors become expressed in an algebraic way
through elementary functions and the function $\Gamma$,
and this representation is unique already. In this way
the explicit expressions (6.13)--(6.41) are obtained.

\section{Derivation of the large-$\Box$ asymptotic
be\-havi\-ours and Laplace originals of the form
factors in $W$ ($\omega=2$)}
\setcounter{equation}{0}

\hspace{\parindent}
The Laplace representation for the basic form factor
(18.7) is obtained by writing
\begin{equation}
\Gamma(-\Box_1,-\Box_2,-\Box_3)=
\int^\infty_0\!ds\,
\int_{\alpha_i\geq0}\!d^3\alpha\,
\delta\Big(1-\sum^3_1\alpha_i\Big){\rm e}^{s\Omega}
\end{equation}
and making the replacement of variables
\begin{equation}
s,\ \alpha_1,\ \alpha_2,\ \alpha_3\ \rightarrow\ u_1,\ u_2,\ u_3 :
\end{equation}
\begin{eqnarray}
\alpha_1&=&\frac{u_2u_3}{u_1u_2+u_2u_3+u_1u_3}, \nonumber\\[\baselineskip]
\alpha_2&=&\frac{u_1u_3}{u_1u_2+u_2u_3+u_1u_3}, \nonumber\\[\baselineskip]
\alpha_3&=&\frac{u_1u_2}{u_1u_2+u_2u_3+u_1u_3}, \nonumber\\[\baselineskip]
s&=&\frac{(u_1u_2+u_2u_3+u_1u_3)^2}{u_1u_2u_3}, \nonumber
\end{eqnarray}
\[
\Big(0\leq s<\infty,\
\alpha_i\geq0,\
\sum^3_1\alpha_i=1
\Big)\ \rightarrow\
(0\leq u_i<\infty),\nonumber
\]
\[
ds\,d^3\alpha\,
\delta\Big(1-\sum^3_1\alpha_i\Big) =
\frac{d^3u}{u_1u_2+u_2u_3+u_1u_3},
\]
\[
s\Omega=u_1{\Box_1}+u_2{\Box_2}+u_3{\Box_3}\equiv \sum u\Box.
\]
The result is
\begin{equation}
\Gamma(-\Box_1,-\Box_2,-\Box_3)=\int^\infty_0\!\! d^3 u\,\frac1{u_1u_2+u_2u_3+u_1u_3}{\rm e}^{{\scriptscriptstyle \sum} u\Box}.
\end{equation}
The Laplace originals of all form factors (18.14)
can then be obtained by introducing the generating
function
\begin{eqnarray}
Z(j_1,j_2,j_3)& = &
\Gamma(-j_1{\Box_1},-j_2{\Box_2},-j_3{\Box_3})\nonumber\\&&\mbox{}
=\int^\infty_0\!\! d^3 u\,\frac{{\rm e}^{{\scriptscriptstyle \sum} u\Box}}{u_1u_2j_3+u_2u_3j_1+u_1u_3j_2}
\end{eqnarray}
and noting that
\begin{equation}
\left.
\left<\frac{\alpha_1^{n_1}\alpha_2^{n_2}\alpha_3^{n_3}}{-\Omega}\right>_3
=\frac{(-1)^{n_1+n_2+n_3}}
{(n_1+n_2+n_3)!}
\left(\frac{\partial}{\partial j_1}\right)^{n_1}
\left(\frac{\partial}{\partial j_2}\right)^{n_2}
\left(\frac{\partial}{\partial j_3}\right)^{n_3}
Z\right|_{j=1}
\end{equation}
as is obvious from the replacement
(19.2) or eq. (18.8). To complete representing
the functions (18.14) in the form of Laplace
integrals, eq. (19.5) should be supplemented
with the inverse Laplace transformation for the
terms of the type (18.5). For the further use
we present this transformation in the form of
 the generating equation
\begin{equation}
\frac1{j_1{\Box_1}}
\frac1{(j_2{\Box_2}-j_3{\Box_3})}
\ln\left(\frac{j_2{\Box_2}}{j_3{\Box_3}}\right)
=\frac1{j_1}\int^\infty_0\!\! d^3 u\,
\frac{{\rm e}^{{\scriptscriptstyle \sum} u\Box}}{j_2u_3+j_3u_2}
\end{equation}
which can next  be differentiated with respect to $j$.

The problem is, however, that again the $\Box$
arguments of the form factors enter not only
the kernel ${\rm e}^{{\scriptscriptstyle \sum} u\Box}$ but also the coefficients of the
$\alpha$-polynomials in (18.14). Originally
\footnote{\normalsize
In the final expressions (7.2)--(7.30), the
$\alpha$-integrals with the coefficients (19.7b)
and (19.8b) are transformed into tree terms by
using eqs. (19.21), (19.22) below.
},
in the form factors of the structures without
derivatives, these coefficients
are of the form
\begin{equation}
{\rm a})\ \frac{\Box_k}{\Box_n},\hspace{7mm}
{\rm b})\ \frac{{\Box_k}^2}{\Box_n\Box_m};\hspace{7mm}
m,n,k=1,2,3
\end{equation}
for the structures with two derivatives
they are of the form
\begin{equation}
{\rm a})\ \frac1{\Box_n},\hspace{7mm}
{\rm b})\ \frac{\Box_k}{\Box_n\Box_m},\hspace{7mm}
m,n,k=1,2,3,
\end{equation}
for the structures with four derivatives
of the form
\begin{equation}
\frac{\Box_m}{\Box_n},\hspace{7mm}
m,n=1,2,3,
\end{equation}
and, for the structure 29 with six derivatives,
the $\Box$ coefficient in the form factor is
\begin{equation}
\frac1{\Box_1\Box_2\Box_3}.
\end{equation}
The task is now to try to absorb these $\Box$
and $1/\Box$ multipliers in the Laplace originals;
otherwise the representation will not be unique
and all advantages of  dealing with the
integral originals will be lost. Indeed, with the
aid of (19.5), the identities like (18.17)
can immediately be translated in the language of
Laplace integrals to give relations of the form
\[ \int^\infty_0\!\! d^3 u\,\rho(u,\Box){\rm e}^{{\scriptscriptstyle \sum} u\Box}=0 \]
with nonvanishing $\rho(u,\Box)$.

As will be seen below, for the Laplace representation,
the $1/\Box$ multipliers in (19.7)--(19.10) present no
problem; the problem arises only with the positive
powers of $\Box_k$ in (19.7) and (19.8) because they
enhance the behaviour of the form factors at large
$(-\Box_k)$. The point is that the Laplace representation
exist only for functions {\it decreasing} at large
values of their arguments whereas, at small values, any
power growth is admissible.

To study the behaviour of the form factors at large
negative $\Box_k$, it suffices to consider the
generating function (19.4). In the form
\begin{eqnarray}
Z &=& \int^\infty_0du_3\,Y(u_3){\rm e}^{u_3{\Box_3}}, \\
Y(u_3) &=& \int^\infty_0\!du_1\,du_2\,
\frac{{\rm e}^{u_1{\Box_1}+u_2{\Box_2}}}
{u_1u_2j_3+u_2u_3j_1+u_1u_3j_2},
\end{eqnarray}
the behaviour of $Z$ at large $(-{\Box_3})$ is determined by the
behaviour of $Y(u_3)$ at small $u_3$. For the asymptotic
behaviour of $Y(x)$ at small $x$ one obtains
\begin{eqnarray} Y(x)&&= \frac1{j_3}
\ln\Big(-x\frac{j_1{\Box_1}}{j_3}\Big)
\ln\Big(-x\frac{j_2{\Box_2}}{j_3}\Big) \nonumber\\&&\mbox{}
-\frac{\Gamma'(1)}{j_3}
\ln\Big(x^2\frac{j_1j_2{\Box_1}{\Box_2}}{j_3^2}\Big)
+\frac{\Gamma''(1)}{j_3}+{\rm O}\left(x\right),
\hspace{5mm}
 x\rightarrow0
\end{eqnarray}
where $\Gamma'(1), \Gamma''(1)$ are derivatives of
the Euler $\Gamma$-function at the point 1. Hence the
asymptotic behaviour of $Z$ at large negative
${\Box_3}$ is
\begin{eqnarray}&&
Z = -\frac1{j_3{\Box_3}}\ln^2(-{\Box_3})
+\frac1{j_3{\Box_3}}
\ln (-{\Box_3})
\ln\Big(\frac{j_1j_2{\Box_1}{\Box_2}}{j_3^2}\Big) \nonumber\\&&\mbox{}
-\frac1{j_3{\Box_3}}
\left[
\ln\Big(-\frac{j_1{\Box_1}}{j_3}\Big)
\ln\Big(-\frac{j_2{\Box_2}}{j_3}\Big)
+2\zeta(2)\right]  +{\rm O}\left(\frac1{{\Box_3}^2}\right),\hspace{7mm}
-{\Box_3}\rightarrow\infty
\nonumber\\&&\mbox{}
\end{eqnarray}
where $\zeta(2)$ is the Riemann $\zeta$-function at the
point 2. By using (19.5), (19.14) and the expressions
(7.2)--(7.30) for $\Gamma_i$ in the $\alpha$-representation,
one can obtain  the asymptotic behaviours of all form
factors at large values of each of the three arguments.
The  final table of asymptotic behaviours is given in sect.
10.

Thus, apart from the multipliers (19.7)--(19.10), the
behaviour of the form factors is generally
$\ln^2(-\Box)/\Box,\ \Box\rightarrow-\infty$ in each of the
arguments. The presence of the multipliers (19.7)--(19.10)
changes in this behaviour not only the power of $\Box$
but also the power of $\ln(-\Box)$. The cause is the
"rule of the like $\alpha$" mentioned in sect 7.
By this rule, each $1/\Box$ multiplier
appears only in a product with the like $\alpha$,
e.g. $\alpha_1/{\Box_1},\ \alpha_1\alpha_2/{\Box_1}{\Box_2}$, etc. Since
each $\alpha$ is equivalent to a derivative with respect
to $j$, the negative powers of $\Box$ in (19.7)--(19.10)
appear only in the combinations
\begin{fleqnarray}&&
\frac1{\Box_1}
\frac{\partial}{\partial j_1}
Z =
\frac1{j_1j_3{\Box_1}{\Box_3}}
\ln\Big(\frac{j_3{\Box_3}}{j_2{\Box_2}}\Big)
+{\rm O}\left(\frac1{{\Box_3}^2}\right),\hspace{7mm}
-{\Box_3}\rightarrow\infty
\end{fleqnarray}
\begin{fleqnarray}&&
\frac1{\Box_1\Box_2}
\frac{\partial}{\partial j_1}
\frac{\partial}{\partial j_2}
Z =
-\frac1{\Box_1\Box_2\Box_3}
\frac1{j_1j_2j_3}
+{\rm O}\left(\frac1{{\Box_3}^2}\right),\hspace{4mm}
-{\Box_3}\rightarrow\infty
\end{fleqnarray}
and similar combinations with permuted indices.
As seen from (19.7)--(19.10), the terms leading at large
$(-{\Box_3})$ are always proportional to $1/{\Box_1}{\Box_2}$
and, therefore, by (19.16), have a purely power
asymptotic behaviour. We conclude that the
large-$\Box$ behaviour of the form factors
in individual $\Box$ arguments is generally
\begin{equation}
\Gamma_{1-11}\propto\Box_k,\hspace{7mm}
-\Box_k\rightarrow\infty
\end{equation}
for the structures without derivatives,
\begin{equation}
\Gamma_{12-25}\propto{\rm const},\hspace{7mm}
-\Box_k\rightarrow\infty
\end{equation}
for the structures with two derivatives,
\begin{equation}
\Gamma_{26-28}\propto\frac1{\Box_k},\hspace{7mm}
-\Box_k\rightarrow\infty
\end{equation}
for the structures with four derivatives, and
\begin{equation}
\Gamma_{29}\propto\frac1{{\Box_k}^2},\hspace{7mm}
-\Box_k\rightarrow\infty
\end{equation}
for the structure with six derivatives
\footnote{\normalsize
Notwithstanding that the total dimension of the
form factors is
$\Box^{-1}$ for $\Gamma_{1-11}$,
$\Box^{-2}$ for $\Gamma_{12-25}$,
$\Box^{-3}$ for $\Gamma_{26-28}$, and
$\Box^{-4}$ for $\Gamma_{29}$.
}.

Since the form factors (19.17) and (19.18) do not
decrease at large values of their arguments, they
do not admit a Laplace representation. The best one
can do is to single out the nondecreasing terms
and treat them separately. The first step is to get rid
of the coefficients (19.7b) and (19.8b) which
cause the strongest growth. This can be done
as follows:
\begin{fleqnarray}&&
\Gamma_{1-11}+{\rm O}\left(\ln(-{\Box_3})\right) =
\left\langle\frac{{\Box_3}^2\alpha_1\alpha_2}{{\Box_1}{\Box_2}}
\frac{P(\alpha)}{-\Omega}\right\rangle_3  \nonumber\\&&\ \ \ \ \ \ \ \ \mbox{}
=\left\langle\frac{\Box_3}{\Box_1\Box_2}(\Omega-{\Box_1}\alpha_2\alpha_3-{\Box_2}\alpha_1\alpha_3)
\frac{P(\alpha)}{-\Omega}\right\rangle_3  \nonumber\\&&\ \ \ \ \ \ \ \ \mbox{}
=-\frac{\Box_3}{\Box_1\Box_2}\Big<P(\alpha)\Big>_3
-\left\langle\Big(
\frac{{\Box_3}\alpha_2\alpha_3}{{\Box_2}}
+\frac{{\Box_3}\alpha_1\alpha_3}{\Box_1}
\Big)
\frac{P(\alpha)}{-\Omega}\right\rangle_3  ,
\end{fleqnarray}
and, similarly,
\begin{fleqnarray}&&
\Gamma_{12-25}+{\rm O}\left(\frac{\ln(-{\Box_3})}{{\Box_3}}\right) =
\left\langle\frac{{\Box_3}\alpha_1\alpha_2}{{\Box_1}{\Box_2}}
\frac{P(\alpha)}{-\Omega}\right\rangle_3  \nonumber\\&&\ \ \ \ \ \ \ \ \mbox{}
=-\frac1{\Box_1\Box_2}\Big<P(\alpha)\Big>_3
-\left\langle\Big(
\frac{\alpha_2}{{\Box_2}}
+\frac{\alpha_1}{\Box_1}
\Big)
\frac{\alpha_3P(\alpha)}{-\Omega}\right\rangle_3
\end{fleqnarray}
where the "rule of the like $\alpha$" works again.
Here the purely tree terms on the right-hand sides
are the leading asymptotic terms (19.17)
and (19.18) respectively. Their contributions in the
Laplace representation can, at best, be written
down as
\begin{fleqnarray}&&
\Gamma_{1-11}+{\rm O}\left(\ln(-{\Box_3})\right) \propto
{\Box_3}^2\int^\infty_0\!\! d^3 u\,{\rm e}^{{\scriptscriptstyle \sum} u\Box},
\end{fleqnarray}
\begin{fleqnarray}&&
\Gamma_{12-25}+{\rm O}\left(\frac{\ln(-{\Box_3})}{\Box_3}\right) \propto
{\Box_3}\int^\infty_0\!\! d^3 u\,{\rm e}^{{\scriptscriptstyle \sum} u\Box}.
\end{fleqnarray}

In the remaining terms, the $1/\Box$ multipliers
in (19.7)--(19.10) can easily be absorbed in the
Laplace originals by integration by parts.
Generally, this leads to the appearance of
logarithmic originals but, owing to the
"rule of the like $\alpha$", we need only
\arraycolsep=0pt
\begin{fleqnarray}&&
\frac1{\Box_1}
\frac{\partial}{\partial j_1}
Z = \int^\infty_0\!\! d^3 u\,
\frac{u_1}{j_1}
\frac{{\rm e}^{{\scriptscriptstyle \sum} u\Box}}{u_1u_2j_3+u_2u_3j_1+u_1u_3j_2},
\end{fleqnarray}
\begin{fleqnarray}&&
\frac1{\Box_1\Box_2}
\frac{\partial}{\partial j_1}
\frac{\partial}{\partial j_2}
Z = \int^\infty_0\!\! d^3 u\,
\frac{u_1u_2}{j_1j_2}
\frac{{\rm e}^{{\scriptscriptstyle \sum} u\Box}}{u_1u_2j_3+u_2u_3j_1+u_1u_3j_2},
\end{fleqnarray}
\begin{fleqnarray}&&
\frac1{\Box_1\Box_2\Box_3}
\frac{\partial}{\partial j_1}
\frac{\partial}{\partial j_2}
\frac{\partial}{\partial j_3}
Z = \int^\infty_0\!\! d^3 u\,
\frac{u_1u_2u_3}{j_1j_2j_3}
\frac{{\rm e}^{{\scriptscriptstyle \sum} u\Box}}{u_1u_2j_3+u_2u_3j_1+u_1u_3j_2},
\end{fleqnarray}
\arraycolsep=3pt
and the Laplace originals remain rational.

There remain to be considered the terms with the
 coefficient (19.7a). Such terms exist only in
the form factors of the structures without derivatives,
and, as follows from (19.15), they grow like
$(\ln(-\Box_k)+{\rm const}),\ \Box_k\rightarrow-\infty$.
To single out the growing contribution, eq. (19.25)
should be rewritten as
\begin{eqnarray}&&
\frac1{\Box_1}
\frac{\partial}{\partial j_1}Z
\nonumber\\&&\ \ \ \ \mbox{}
= -\int^\infty_0\!\! d^3 u\,{\rm e}^{{\scriptscriptstyle \sum} u\Box} u_2u_3
( {u_1u_2j_3+u_2u_3j_1+u_1u_3j_2})^{-1}
(j_2u_3+j_3u_2)^{-1}
\nonumber\\&&\ \ \ \ \mbox{}
+\frac1{j_1}\int^\infty_0\!\! d^3 u\,\frac{{\rm e}^{{\scriptscriptstyle \sum} u\Box}}{j_2u_3+j_3u_2}.
\end{eqnarray}
Here the second integral contains the asymptotic
term (19.15) whereas the first integral is already
${\rm O}\left(1/{\Box_3}^2\right),\ {\Box_3}\rightarrow-\infty$. Therefore,
upon multiplication
by ${\Box_3}$, the multiplier ${\Box_3}$ can be absorbed in the
first integral of (19.28) by integration by parts.
We obtain
\begin{eqnarray}&&
\frac{\Box_3}{\Box_1}
\frac{\partial}{\partial j_1}Z
\nonumber\\&&\ \ \ \ \mbox{}
= \int^\infty_0\!\! d^3 u\,{\rm e}^{{\scriptscriptstyle \sum} u\Box} u_2[
j_3u_1u_2
( {u_1u_2j_3+u_2u_3j_1+u_1u_3j_2})^{-2}
(j_2u_3+j_3u_2)^{-1}    \nonumber\\&&\ \ \ \ \ \ \ \ \mbox{}\ \ \ \ \ \ \ \ \ \ \ \
-j_2u_3( {u_1u_2j_3+u_2u_3j_1+u_1u_3j_2})^{-1}
(j_2u_3+j_3u_2)^{-2}] \nonumber\\&&\ \ \ \ \mbox{}
+\frac{{\Box_3}}{j_1}\int^\infty_0\!\! d^3 u\,\frac{{\rm e}^{{\scriptscriptstyle \sum} u\Box}}{j_2u_3+j_3u_2}
\end{eqnarray}
which is a sum of a decreasing Laplace
integral and the growing contribution. Note
that the latter contribution is precisely
of the form (19.6) multiplied by ${\Box_3}$.

In some cases, as a result of the action of
the derivatives $\partial/\partial j$, in the leading
asymptotic behaviour $(\ln(-{\Box_3})+{\rm const})$ of
(19.29), the ``$\ln(-{\Box_3})$'' cancels, and
the ``const'' remains:
\begin{equation}
P_{\rm special}\left(
\frac{\partial}{\partial j}\right)\frac{\Box_3}{\Box_1}
\frac{\partial}{\partial j_1}
Z = \frac1{\Box_1}+{\rm O}\left(\frac1{\Box_3}\right),\hspace{7mm}
{\Box_3}\rightarrow-\infty.
\end{equation}
The nondecreasing tree term $1/{\Box_1}$
that appears here
can be singled out explicitly already at the level
of the $\alpha$-representation by using the identities
of sect. 18
\footnote{\normalsize
Such tree terms and the tree terms
obtained in
 (19.21), (19.22) figure in the final
expressions of sect. 7 for the
$\alpha$-representation of the form factors.
}.
In the Laplace representation, it can be put
in the form
\begin{equation}
\frac1{\Box_1}=-{\Box_2}{\Box_3}\int^\infty_0\!\! d^3 u\,{\rm e}^{{\scriptscriptstyle \sum} u\Box}
\end{equation}
similar to (19.23), (19.24).

There are only two types of nondecreasing
contributions: (19.6) multiplied by
either ${\Box_2}$ or ${\Box_3}$, and the tree terms
(19.23), (19.24), (19.31). Both factorize into
elementary functions of one or two variables.
The nonfactorizable  triple form factors are
given by proper Laplace integrals with
rational originals.

In this way the final results in sect. 8 are obtained.

\section{Derivation of the small--$\Box$ asymptotic
behaviors, and the generalized spectral te\-chnique
for the form factors in $W(\omega=2)$}
\setcounter{equation}{0}

\hspace{\parindent}
The spectral representation of the third-order form factors
was studied in paper III. For the basic form factor (18.7)
it is of the form [3, 23]
\begin{equation}
\Gamma(-\Box_1,-\Box_2,-\Box_3)= \int_0^\infty {
{d{m_1}^2\, d{m_2}^2\, d{m_3}^2\, \rho (m_1,m_2,m_3)
}\over{
({m_1}^2-\Box_1)({m_2}^2-\Box_2)({m_3}^2-\Box_3)
}}
\end{equation}
with the discontinuous spectral weight
\begin{equation}
\rho (m_1,m_2,m_3)=\left\{
\begin{array}{ll}
1/4\pi S& {\rm if\ there\ exists\ a\ triangle}\\
&{\rm with\ the\ sides} \ m_1,\ m_2,\ m_3,\\
0&{\rm  otherwise},
\end{array}
\right.
\end{equation}
where $m_1= \sqrt{{m_1}^2},\ m_2= \sqrt{{m_2}^2},\
m_3= \sqrt{{m_3}^2},$ and $ S $ is the area of a triangle with
the sides $m_1, \ m_2, \ m_3 $.

The form factors with $\alpha$-polyno\-mials are then
obtained by noting that
\begin{equation}
(-\Box)^n \frac{\partial^n}{\partial\Box^n} \frac1{{m}^2-\Box }=
\left(\frac{\partial}{\partial m^2}\right)^n \frac{(m^2)^n}{{m}^2-\Box }
\end{equation}
and using eq. (18.8):
\mathindent=0pt
\begin{fleqnarray}&&
\left\langle \frac{\alpha_1^{n_1}\alpha_2^{n_2}\alpha_3^{n_3}}{-\Omega}\right\rangle_{3}=
\frac{1}{(n_1+n_2+n_3)!}
\int_0^\infty
d{m_1}^2 \,d{m_2}^2 \,d{m_3}^2\, \rho(m_1,m_2,m_3)\nonumber\\&&\mbox{}\  \times
\left(\frac{\partial}{\partial {m_1}^2}\right)^{n_1}
\left(\frac{\partial}{\partial {m_2}^2}\right)^{n_2}
\left(\frac{\partial}{\partial {m_3}^2}\right)^{n_3}
 {\frac{({m_1}^2)^{n_1}}{{m_1}^2-\Box_1}}
 {\frac{({m_2}^2)^{n_2}}{{m_2}^2-\Box_2}}
 {\frac{({m_3}^2)^{n_3}}{{m_3}^2-\Box_3}}.
\end{fleqnarray}
\mathindent=\parindent
Although the problem of expressing the form factors through the Green
function $1/({m}^2-\Box)$ is thereby solved, the action of the derivatives
$\partial/\partial m^2$ results in the appearance of powers of
$1/({m}^2-\Box)$. A natural next step would be to integrate in (20.4) by
parts in order that the derivatives  $\partial/\partial m^2$ act
on $\rho$ but, as seen from the Heron formula for the area of a
triangle
\begin{eqnarray}
\frac1{4\pi S}&=& \frac1{\pi}\Big[(m_1+m_2+m_3)
(m_1+m_2-m_3)\nonumber\\&&\mbox{}\times
(m_1+m_3-m_2)(m_3+m_2-m_1)\Big]^{-\frac12},
\end{eqnarray}
this integration by parts will result in a divergence of the integral
at the integration boundary
\begin{equation}
(m_1=m_2+m_3)\bigcup (m_2=m_1+m_3)\bigcup (m_3=m_1+m_2).
\end{equation}
On the other hand, with the derivatives acting on the Green
functions, the representation in eq. (20.4) is different for
different
$n_1,n_2, n_3 $, and , because of this nonuniqueness, unfit for
verification of hidden identities between the form factors.

The success comes with the recognition of the following remarkable
fact (see, e.g., [24]):
\begin{equation}
\int_0^\infty d y^2
J_0(y m_1)
J_0(y m_2)
J_0(y m_3)
= 4\rho (m_1,m_2,m_3)
\end{equation}
where $J_0$ is the order-0 Bessel function, and $\rho$ is given in
(20.2). The use of this relation in (20.1) gives
\footnote{\normalsize
Eq. (20.8) can be proved independently by using the integral
representation for ${\cal K}_0$. }
\begin{equation}
\Gamma(-\Box_1,-\Box_2,-\Box_3)=2\int_0^\infty d y^2
{\cal K}_{0} (y\, \sqrt{-\Box_1}) {\cal K}_{0} (y\, \sqrt{-\Box_2}) {\cal K}_{0} (y\, \sqrt{-\Box_3})
\end{equation}
where ${\cal K}_0$ is the order-0 Macdonald
function for which we  have the
spectral representation
\begin{equation}
{\cal K}_{0} (y\, \sqrt{-\Box}) = \frac12
\int_0^\infty
\frac{d m^2 J_0(ym)}{{m}^2-\Box}.
\end{equation}

Eq. (20.8) is a {\em generalized} spectral representation in which
there is one extra integration over a parameter entering the spectral
weight. This integration will always be considered as the last one.
Two important advantages of this representation are that i) the
boundary (20.6) disappears owing to the properties of the integral
(20.7), and ii) in the integral over the parameter, the triple
form factor factorizes into functions of one variable.

The generalized spectral representation of the form factors with
$\alpha$-poly\-no\-mials is obtained by noting that
\begin{equation}
(-\Box)^n \frac{\partial^n}{\partial\Box^n}
{\cal K}_{0} (y\, \sqrt{-\Box}) =(-y^2)^n
\left(\frac{\partial}{\partial y^2}\right)^n {\cal K}_{0} (y\, \sqrt{-\Box})
\end{equation}
and using eq. (18.8):
\begin{eqnarray}&&
\left\langle \frac{\alpha_1^{n_1}\alpha_2^{n_2}\alpha_3^{n_3}}{-\Omega}\right\rangle_{3}
\nonumber\\&&\ \ \ \ \mbox{}
= \frac{2}{(n_1+n_2+n_3)!}
\int_0^\infty dy^2 (-y^2)^{n_1+n_2+n_3}
\left(\frac{\partial}{\partial y_1^2}\right)^{n_1}
\left(\frac{\partial}{\partial y_2^2}\right)^{n_2}
\left(\frac{\partial}{\partial y_3^2}\right)^{n_3}
\nonumber\\&&\ \ \ \ \ \ \ \ \mbox{}\times\left.
{\cal K}_{0} (y_1\,\sqrt{-\Box_1}) {\cal K}_{0} (y_2\,\sqrt{-\Box_2}) {\cal K}_{0} (y_3\,\sqrt{-\Box_3}) \right|_{y_1=y_2=y_3=y}.
\end{eqnarray}
This is equivalent to introducing again the
generating function (19.4)
for which we have
\begin{equation}
Z(j_1, j_2,j_3)=
2\int_0^\infty\! dy^2
{\cal K}_{0} (y \, j_1^{1/2}\,\sqrt{-\Box_1}) {\cal K}_{0} (y \, j_2^{1/2}\,\sqrt{-\Box_2}) {\cal K}_{0} (y \, j_3^{1/2}\,\sqrt{-\Box_3}).
\end{equation}

In what follows, it will be convenient to express the derivatives in
(20.11) in terms of the conformal operator
\begin{equation}
C \equiv y^2 \frac{d}{dy^2}.
\end{equation}
For the calculations with $C$, it will suffice to use the
commutation rule
\begin{equation}
P(C)\left(\frac1{y^2}\right)^N=
\left(\frac1{y^2}\right)^N
P(C-N)
\end{equation}
where $P(C)$ is an arbitrary polynomial in $C$. For example, it is
easy to see that the operator in (20.10) expands in powers of $C$ as
follows:
\begin{equation}
(-y^2)^n
\left(\frac{\partial}{\partial y^2}\right)^n
=(-1)^n C(C-1)(C-2)\ldots (C-n+1).
\end{equation}

In terms of $C$, the form factor  with the general
$\alpha$-poly\-no\-mial takes the form
\begin{eqnarray}&&
\left\langle \frac{P(\alpha_1,\alpha_2,\alpha_3)}{-\Omega}\right\rangle_{3}=
2\int_0^\infty dy^2 \, Q(C_1,C_2,C_3)\nonumber\\&&\ \ \ \ \ \ \ \ \mbox{}\ \ \ \ \times
{\cal K}_{0} (y_1\,\sqrt{-\Box_1}) {\cal K}_{0} (y_2\,\sqrt{-\Box_2}) {\cal K}_{0} (y_3\,\sqrt{-\Box_3})
\end{eqnarray}
where both $P$ and $Q$ are polynomials, and the operators
$C_1,C_2,C_3$ act on $y_1,y_2,y_3$ respectively with subsequently
setting
$y_1=y_2=y_3=y$. The only arbitrariness of this representation is
the one corresponding to the constraint
\begin{equation}
\alpha_1+\alpha_2+\alpha_3=1.
\end{equation}
Now it becomes the arbitrariness of integration by parts in $y^2$,
and the constraint (20.17) turns into the identity
\begin{equation}
C_1+C_2+C_3=-1.
\end{equation}
The general rule of integration by parts in the expression
\begin{equation}
\int_0^\infty\! dy^2
\left(\frac1{y^2}\right)^N
Q(C_1,\ldots C_n)f(y_1,\ldots y_n) \Big|_{y_1=\ldots=y}
\end{equation}
is
\begin{equation}
C_1+\ldots +C_n=N-1
\end{equation}
{\em provided that the boundary contributions vanish}. In the
integral (20.16), the latter condition is always fulfilled because
the action of the operators $C$ cannot deteriorate the behaviours
\begin{equation}
{\cal K}_{0} (y\, \sqrt{-\Box}) ={\rm O}\left(\ln y\right), \hspace{7mm} y\rightarrow 0,
\end{equation}
\begin{equation}
{\cal K}_{0} (y\, \sqrt{-\Box}) ={\rm O}\left({\rm e}^{-y\sqrt{-\Box}}\right), \hspace{7mm} y\rightarrow \infty.
\end{equation}
In a more general case, the validity of (20.20) may depend on which
of the $C$'s will be expressed through the others. Such case will be
encountered below.

Eq. (20.16) should be supplemented with the spectral form of the
second-order form factor in (18.5). In the ordinary spectral
representation of this form factor
\begin{eqnarray}
\frac{\ln(\Box_1/\Box_2)}{(\Box_1-\Box_2)} &=&
-\int_0^\infty
\frac{dm^2}{({m^2-\Box_1})({m^2-\Box_2})}\nonumber\\
&=&-\int_0^\infty\! d{m_1}^2\, d{m_2}^2\,
\frac{\delta({m_1}^2-{m_2}^2)}{({m_1}^2-\Box_1)({m_2}^2-\Box_2)}
\end{eqnarray}
the double-spectral weight is again discontinuous, while its
generalized spectral representation is similar to (20.8):
\begin{eqnarray}
\frac{\ln(\Box_1/\Box_2)}{(\Box_1-\Box_2)} &=&
-\int_0^\infty\!
dy^2\, {\cal K}_{0} (y\, \sqrt{-\Box_1}) {\cal K}_{0} (y\, \sqrt{-\Box_2}).
\end{eqnarray}
Even a tree can be put in the generalized spectral form:
\begin{equation}
\frac1{\Box}=-\frac12
\int_0^\infty\!
dy^2\, {\cal K}_{0} (y\, \sqrt{-\Box})
\end{equation}
but this is already a signal that the use of this representation
needs a reserve.

The reserve concerns the asymptotic behaviours of generalized
spectral integrals at small $\Box$'s. For the ordinary spectral
representation to exist the function should behave like
${\cal O}/\Box$ where ${\cal O}\rightarrow 0$  at
$\Box\rightarrow -0$. This is also true of the generalized triple and
double spectral forms but not of the single one. Indeed,
the behaviours
of the triple and double forms (20.8) and (20.24) at
$\Box_k\rightarrow -0\ (k=1,2,3)$ follow the behaviour of the
respective ${\cal K}_0$ because, when this ${\cal K}_0$ is
expanded at small $y$, the responsibility for the convergence at the
upper limit rests with another ${\cal K}_0$. With a single
${\cal K}_0$, as in (20.25), this is no longer the case. Hence the
imitation spectral weight of a massless Green function.

 In  the case of the triple form factors, the representation (20.12)
readily gives the small-$\Box$ asymptotic behaviour of the
generating function. By expanding one of the ${\cal K}_0$'s, we
obtain
\begin{eqnarray}
Z&=&-\ln(-j_3\Box_3)\int_0^\infty\! dy^2\, {\cal K}_{0} (y \, j_1^{1/2}\,\sqrt{-\Box_1}) {\cal K}_{0} (y \, j_2^{1/2}\,\sqrt{-\Box_2})
\nonumber\\&& \hspace{50mm}\mbox{}
+{\rm O}\left(1\right), \hspace{7mm}
-\Box_3\rightarrow 0.
\end{eqnarray}
Together with (19.5), this gives the asymptotic behaviours of all
$\alpha$-averages with accuracy ${\rm O}\left(1\right)$. However, because of the
presence of the multipliers $\Box_{n}/\Box_{k}$ in the
$\alpha$-polynomials, to have this accuracy in the total form factor,
the generating function must be known with accuracy ${\rm O}\left(\Box_{k}\right),
\ \Box_{k}\rightarrow-0$.
Owing to the ``rule of the like $\alpha$'', we need
only
\begin{eqnarray}
\frac1{\Box_3}
\frac{\partial}{\partial j_3}Z&=&
-\frac{1}{j_3 \Box_3}
\int_0^\infty\! dy^2\, {\cal K}_{0} (y \, j_1^{1/2}\,\sqrt{-\Box_1}) {\cal K}_{0} (y \, j_2^{1/2}\,\sqrt{-\Box_2})
\nonumber\\&&\mbox{}
+\frac14\ln(-j_3 \Box_3)
\int_0^\infty\! dy^2\, y^2\,{\cal K}_{0} (y \, j_1^{1/2}\,\sqrt{-\Box_1}) {\cal K}_{0} (y \, j_2^{1/2}\,\sqrt{-\Box_2}) \nonumber\\&& \hspace{40mm}\mbox{}
+{\rm O}\left(1\right), \hspace{7mm} -\Box_3\rightarrow 0
\end{eqnarray}
(cf. eq. (19.15)). The coefficients of the asymptotic
terms in (20.26)
and (20.27) can be calculated explicitly: we have eq. (20.24) and
\begin{eqnarray}
\int_0^\infty dy^2\, y^2\,{\cal K}_0(ay) {\cal K}_0(by)& =&
\frac{4}{(a^2-b^2)^2}\left[(a^2+b^2)
\frac{\ln(a^2/b^2)}{a^2-b^2}-2\right],\nonumber\\&&\ \ \ \ \ \ \ \ \mbox{}\ \ \ \ \ \ \ \
a>0,\hspace{7mm} b>0.
\end{eqnarray}
In this way the final results in sect. 11 are obtained.

The large-$\Box$ behaviours of the generalized and ordinary spectral
integrals differ drastically. The ordinary spectral representation
exists only for functions that behave like ${\cal O}\rightarrow 0$
at $\Box \rightarrow -\infty$ whereas the generalized one can stand
any power growth at large arguments. For our purposes, this is an
important advantage because it enables us to absorb in the originals
any positive powers of $\Box$'s. Indeed, from the Bessel equation for
${\cal K}_0$
\begin{equation}
-\frac{4}{z^2}
\left(z^2\frac{d}{dz^2}\right)
\left(z^2\frac{d}{dz^2}\right)
{\cal K}_0+
{\cal K}_0 = 0
\end{equation}
with the argument $z=y\sqrt{-\Box}$, we obtain
\begin{equation}
\Box
{\cal K}_0
= -\frac{4}{y^2}C^2 {\cal K}_0 ,
\end{equation}
\begin{equation}
\Box^2
{\cal K}_0
= \left(-\frac{4}{y^2}\right)^2 (C-1)^2 C^2 {\cal K}_0 ,
\end{equation}
and, generally,
\begin{equation}
\Box^M
{\cal K}_0
= \left(-\frac{4}{y^2}\right)^M \left[\prod^{M-1}_{p=0}
(C-p)^2\right]
{\cal K}_0.
\end{equation}
In combination with (20.16), this gives, for example,
\begin{eqnarray}&&
\left\langle \frac{\Box_1^M P(\alpha_1,\alpha_2,\alpha_3)}{-\Omega}\right\rangle_{3}=
2\int_0^\infty\! dy^2
\left(-\frac{4}{y^2}\right)^M
Q(C_1-M,C_2,C_3)
\nonumber\\&& \hspace{1mm}\mbox{}\times
\left[\prod^{M-1}_{p=0}(C_1-p)^2\right]
{\cal K}_{0} (y_1\,\sqrt{-\Box_1}) {\cal K}_{0} (y_2\,\sqrt{-\Box_2}) {\cal K}_{0} (y_3\,\sqrt{-\Box_3}).
\end{eqnarray}

Thus, the situation is opposite to the one in the case of the Laplace
representation. Now, the positive powers of $\Box$ can be easily
absorbed in the originals, and the negative powers require a special
procedure of detaching the inadmissibly growing terms. The procedure
is as follows. Owing to the ``rule of the like $\alpha$'', a
multiplier $1/\Box$ may appear only in a product with the operator
$C$. By using eq. (20.9), we find
\begin{eqnarray}
\frac{C}{\Box}\,{\cal K}_{0} (y\, \sqrt{-\Box})&=&-\frac12
\int_0^\infty\! dm^2\,
\frac{ym}{2}J_1(ym)\frac1{({m}^2-\Box)\Box}\nonumber\\
&=&-\frac12
\int_0^\infty\! dm^2\,
\frac{y J_1(ym)}{2m}\frac1{{m}^2-\Box}-\frac1{2\Box}
\end{eqnarray}
where $J_1$ is the order-1 Bessel function, and the tree term
$1/2\Box$ that got detached is the leading asymptotic term at
$\Box \rightarrow -0$. The remaining spectral integral with $J_1$
is already ${\cal O}/\Box,\ \Box \rightarrow -0$.

Since the multipliers $1/\Box_{k},\ k=1,2,3$, do appear in our form
factors (and higher powers of $1/\Box_{k}$ never appear), we introduce
\begin{equation}
{\cal S}(y,\Box)=\frac12\int_0^\infty\! dm^2\,
\frac{y J_1(ym)}{2m}\,\frac1{{m}^2-\Box}
\end{equation}
as a basic spectral integral instead of (20.9). Eq. (20.34) expresses
${\cal S}$ through ${\cal K}_0$:
\begin{equation}
{\cal S}=-\frac{C}{\Box}{\cal K}_0-\frac{1}{2\Box}
\end{equation}
but then the Bessel equation (20.30) expresses also
${\cal K}_0$ through ${\cal S}$:
\begin{equation}
{\cal K}_0=\frac{4}{y^2}C{\cal S}.
\end{equation}
The latter relation should be used in all expressions like (20.16),
(20.24), (20.33), etc. to replace everywhere ${\cal K}_0$ by ${\cal S}$.
The substitutions to be made are
\begin{equation}
P(C){\cal K}_0=\frac{4}{y^2}P(C-1)C{\cal S},
\end{equation}
\begin{equation}
\frac1{\Box} P(C) C {\cal K}_0=-P(C){\cal S}-\frac{P(0)}{2\Box},
\end{equation}
\begin{equation}
\Box^M P(C) {\cal K}_0=-
\left(-\frac{4}{y^2}\right)^{M+1}P(C-M-1)
\left[\prod^{M}_{p=1}(C-p)^2\right]C{\cal S}
\end{equation}
with  any polynomial $P(C)$. This gives the final form of the
representation.

An important point is that ${\cal S}$, as distinct from
${\cal K}_0$, does not decrease at $y\rightarrow\infty$. As seen from
(20.36),
\begin{equation}
{\cal S}=-\frac1{2\Box}+{\rm O}\left({\rm e}^{-y\sqrt{-\Box}}\right),\ y\rightarrow\infty
\end{equation}
but the action of at least one $C$ makes ${\cal S}$ decreasing
exponentially. Therefore, the generalized spectral integrals must
necessarily contain at least one differentiated ${\cal S}$, which means
that more than two $1/\Box$ multipliers cannot be absorbed in the
triple form factors. Denoting for short
\begin{equation} {\cal S}(y_1,\Box_1)={\cal S}_1,\hspace{7mm}
{\cal S}(y_2,\Box_2)={\cal S}_2,\hspace{7mm}
{\cal S}(y_3,\Box_3)={\cal S}_3
\end{equation}
and omitting, for simplicity, the $\alpha$-polynomial,  we  have
\begin{fleqnarray}&&
\left\langle \frac{1}{-\Omega}\right\rangle_{3}=
2\int_0^\infty\! dy^2
\left(\frac{4}{y^2}\right)^3
C_1 C_2 C_3  {\cal S}_1 {\cal S}_2 {\cal S}_3,
\end{fleqnarray}
\begin{fleqnarray}&&
\left\langle \frac{1}{-\Omega}\frac{\alpha_1}{\Box_1}\right\rangle_{3}=
2\int_0^\infty\! dy^2
\left(\frac{4}{y^2}\right)^2
C_2 C_3  {\cal S}_1 {\cal S}_2 {\cal S}_3
\nonumber\\&&\ \ \ \ \ \ \ \ \mbox{}\ \ \ \
+\frac1{\Box_1}
\int_0^\infty\! dy^2
\left(\frac{4}{y^2}\right)^2
C_2 C_3   {\cal S}_2 {\cal S}_3,
\end{fleqnarray}
\begin{fleqnarray}&&
\left\langle \frac{1}{-\Omega}\frac{\alpha_1\alpha_2}{\Box_1\Box_2}\right\rangle_{3}=
\int_0^\infty\! dy^2
\left(\frac{4}{y^2}\right)
C_3  {\cal S}_1 {\cal S}_2 {\cal S}_3
\nonumber\\&&\ \ \ \ \ \ \ \ \mbox{}\ \ \ \
+\frac{1}{2\Box_1}
\int_0^\infty\! dy^2
\left(\frac{4}{y^2}\right)
C_3   {\cal S}_2 {\cal S}_3
\nonumber\\&&\ \ \ \ \ \ \ \ \mbox{}\ \ \ \
+\frac{1}{2\Box_2}
\int_0^\infty\! dy^2
\left(\frac{4}{y^2}\right)
C_3   {\cal S}_1 {\cal S}_3
\nonumber\\&&\ \ \ \ \ \ \ \ \mbox{}\ \ \ \
+\frac{1}{4\Box_1\Box_2}
\int_0^\infty\! dy^2
\left(\frac{4}{y^2}\right)
C_3  {\cal S}_3
\end{fleqnarray}
but one more application of \hbox{eq. (20.39)}, to
incorporate the factor
$(\alpha_1\alpha_2\alpha_3)/$
\linebreak
$(\Box_1\Box_2\Box_3)$, would result in the appearance of a
divergent integral with three bare ${\cal S}$.

If the form factors of the structures with derivatives are brought to
the standard dimension by the redefinition discussed in sect. 7,
then one has to deal only with the ratio $\Box_{m}/\Box_{n}$. The
expression for the triple form factor in this case (with the
$\alpha$-polynomial omitted) is
\begin{eqnarray}
\left\langle \frac{1}{-\Omega}\frac{\alpha_1\Box_2}{\Box_1}\right\rangle_{3}&=&
-2\int_0^\infty\! dy^2
\left(\frac{4}{y^2}\right)^3
(C_2-1)^2 C_2 C_3  {\cal S}_1 {\cal S}_2 {\cal S}_3+
\nonumber\\&&\mbox{}
+\frac{\Box_2}{\Box_1}
\int_0^\infty\! dy^2
\left(\frac{4}{y^2}\right)^2
C_2 C_3  {\cal S}_2 {\cal S}_3.
\end{eqnarray}
The integrals with one or two ${\cal S}$, that get detached in
(20.44)--(20.46), reproduce the leading asymptotic terms of the form
factors. In this way the results in sect. 9 are obtained.

The last point to be discussed is integration
by parts over $y$ in the
case like (20.33) where the total power of $\Box_1,\Box_2,\Box_3$ in the
coefficient of the $\alpha$-polynomial is positive. This case is
encountered in the calculation of the trace anomaly (sect. 12).
It will suffice to consider, as an example, eq. (20.33) with the
$\alpha$~-polynomial omitted. In terms of ${\cal S}$,
\begin{equation}
\left\langle \frac{\Box_1^M}{-\Omega}\right\rangle_{3}=
-2\int_0^\infty\! dy^2
\left(-\frac{4}{y^2}\right)^{M+3}
\left[\prod^{M}_{p=1}(C_1-p)^2\right]
C_1 C_2 C_3  {\cal S}_1 {\cal S}_2 {\cal S}_3.
\end{equation}
The convergence of this integral at the lower limit is based on the
following property of the operator $C$:
\begin{equation}
(C-p)^{q+1}
\big(y^2\big)^p\Big(\ln y^2\Big)^q=0.
\end{equation}
Since the asymptotic expansion at small $y$ of both ${\cal K}_0$
and ${\cal S}$ is a series of $\big(y^2\big)^p$ and
$\big(y^2\big)^p \ln y^2$ which starts in the case of
${\cal K}_0$ with $p=0$, and in the case of ${\cal S}$ with $p=1$, we have
at $y \rightarrow 0$
\begin{fleqnarray}&&
{\cal S}={\rm O}\left(y^2 \ln y^2\right),
\end{fleqnarray}
\begin{fleqnarray}&&
(C-1)^2{\cal S}={\rm O}\left(y^4 \ln y^2\right),
\end{fleqnarray}
\begin{fleqnarray}&&
(C-2)^2(C-1)^2{\cal S}={\rm O}\left(y^6 \ln y^2\right),
\end{fleqnarray}
and, generally,
\begin{equation}
\left[\prod^{M}_{p=1}(C-p)^2\right]
{\cal S}=
{\rm O}\left(y^{2M+2} \ln y^2\right).
\end{equation}
The problem is, however, that the integration by parts cannot be
applied to the operators $(C_1-p)^2$ in (20.47)
because the result will
be divergent. On the other hand,
to remove the arbitrariness connected
with the identity (20.20), one may need   to have in (20.47)
the bare ${\cal S}_1$. The procedure of integration  by parts should
then be
modified as follows.
Let us denote ${\cal S}^M$ the first $M$ terms of the
asymptotic series for ${\cal S}$ so that
\begin{equation}
({\cal S}-{\cal S}^M)=
{\rm O}\left(y^{2(M+1)} \ln y^2\right).
\end{equation}
Since
\begin{equation}
\left[\prod^{M}_{p=1}(C-p)^2\right]
{\cal S}^M=0,
\end{equation}
we may replace in (20.47) ${\cal S}_1$ by  $({\cal S}_1-{\cal S}_1^M)$.
After this replacement, the integration by
parts can already be done by
the rule (20.20), and the result is
\begin{eqnarray}
\left\langle \frac{\Box_1^M}{-\Omega}\right\rangle_{3}&=&
2\int_0^\infty\! dy^2
\left(-\frac{4}{y^2}\right)^{M+3}
({\cal S}_1-{\cal S}_1^M)
\left[\prod^{M}_{p=1}(C_2+C_3-p-1)^2\right]\nonumber\\&&\ \ \ \ \ \ \ \ \mbox{}\times
(C_2+ C_3-M-2)C_2 C_3 {\cal S}_2 {\cal S}_3.
\end{eqnarray}
No boundary terms ever appear but the bare ${\cal S}$ appears with
a subtraction. The goal is, nevertheless, reached:
the representation like in (20.55) is unique and can be used for a
verification of hidden identities between the form factors.
An example
of such a verification is given in sect. 12. The only question that
may arise is that the subtraction ${\cal S}^M$ is not in the spectral
form but putting it in the spectral form presents no problem
since the
only nonanalytic function of $\Box$ in ${\cal S}^M$ is $\ln(-\Box)$.

The generalized spectral representation
makes it possible to carry out
calculations like the check of the trace anomaly within the working
technique used in applications.

\section*{Appendix: Identities for nonlocal cubic invariants}
\setcounter{equation}{0}
\renewcommand{\theequation}{A.\arabic{equation}}

\hspace{\parindent}
We begin with the local identities for a tensor possessing
the symmetries of the Weyl tensor:
\begin{eqnarray}
&&C_{\alpha\beta\gamma\delta} =
C_{[\alpha\beta]\gamma\delta} =
C_{\alpha\beta[\gamma\delta]} ,\\[\baselineskip]
&&C_{\alpha\beta\gamma\delta} =
C_{\gamma\delta\alpha\beta} ,\\[\baselineskip]
&&C_{\alpha\beta\gamma\delta} +
C_{\alpha\delta\beta\gamma} +
C_{\alpha\gamma\delta\beta} = 0 ,\\[\baselineskip]
&&C^\alpha{}_{\beta\alpha\delta} = 0.
\end{eqnarray}
Here the Ricci identity (A.3) can be rewritten as
\begin{equation}
C_{\alpha[\beta\gamma]\delta} =
-\frac12C_{\alpha\delta\beta\gamma} \end{equation}
and, with the use of (A.1), as
\begin{equation}
C_{\alpha[\beta\gamma\delta]} = 0 \end{equation}
where the complete antisymmetrization in three
indices is meant. Eq. (A.5) is useful when forming
contractions because it shows that the contractions
of the form
\[ C^{\,\cdot\,\alpha\beta.}
C_{\,\cdot\,\cdot\,\alpha\beta} {\rm\ and\ }
C^{\,\cdot\,\cdot\,\alpha\beta}C_{\,\cdot\,\cdot\,\alpha\beta} \]
express through one another.

In view of applications to the gravitational equations
[25], we first list all possible cubic contractions
having two free indices. The symmetries above allow only
\begin{eqnarray}
J_{1}{}^\nu_\mu&=&C_{\mu\beta\gamma\delta}
       C^{\nu\beta\alpha\sigma}
       C_{\alpha\sigma}{}^{\gamma\delta},\\[\baselineskip]
J_{2}{}^\nu_\mu&=&C_{\mu\beta\gamma\delta}
       C^{\nu\alpha\gamma\sigma}
       C_\alpha{}^\beta{}_\sigma{}^\delta,\\[\baselineskip]
J_{3}{}^\nu_\mu&=&C_{\mu\beta\gamma\delta}
       C^{\nu\alpha\gamma\sigma}
       C_\sigma{}^\beta{}_\alpha{}^\delta,\\[\baselineskip]
J_{4}{}^\nu_\mu&=&C_{\mu\gamma\beta\delta}
       C^{\nu\beta\alpha\sigma}
       C_{\alpha\sigma}{}^{\gamma\delta},\\[\baselineskip]
J_{5}{}^\nu_\mu&=&C^\nu{}_{\sigma\mu\kappa}
       C^{\sigma\alpha\beta\gamma}
       C^\kappa{}_{\alpha\beta\gamma},
\end{eqnarray}
and, furthermore, by (A.5), one has
\begin{equation}
J_{4}{}^\nu_\mu=\frac12J_{1}{}^\nu_\mu,
\end{equation}
and, by applying the Ricci identity to the last $C$
in (A.8), one obtains
\begin{equation}
J_{3}{}^\nu_\mu=J_{2}{}^\nu_\mu
-\frac12J_{4}{}^\nu_\mu=J_{2}{}^\nu_\mu-\frac14J_{1}{}^\nu_\mu.
\end{equation}
Thus, for an arbitrary space-time dimension $2\omega$,
there are three independent contractions:
$J_{1}{}^\nu_\mu, J_{2}{}^\nu_\mu, J_{5}{}^\nu_\mu$.

For a particular space-time dimension, the number of
independent contractions can be smaller because of the
existence of identities obtained by antisymmetrization
of $(2\omega+1)$ indices. Note that such an antisymmetrization
must not involve more than two indices of each $C$ tensor;
otherwise the identity will be satisfied trivially by
virtue of (A.6). Hence, for three $C$ tensors, the number
of indices involved in the antisymmetrization should not
exceed six, and, therefore, the space-time dimension
$2\omega$ for which nontrivial identities exist cannot
exceed five. For $2\omega\leq5$ we have
\begin{equation}
C_{[\alpha\beta}{}^{\gamma\delta}
C_{\gamma\delta}{}^{\kappa\mu}
C_{\kappa\nu]}{}^{\alpha\beta}
\equiv0,\hspace{7mm} 2\omega\leq5
\end{equation}
with the complete antisymmetrization of six lower
indices. When written down explicitly, this
identity takes the form
\begin{equation}
J_{1}{}^\nu_\mu-4J_{3}{}^\nu_\mu-2J_{5}{}^\nu_\mu=0,
\hspace{7mm} 2\omega\leq5\end{equation}
or, by (A.13), the form
\begin{equation}
J_{2}{}^\nu_\mu=\frac12J_{1}{}^\nu_\mu-\frac12J_{5}{}^\nu_\mu,
\hspace{7mm} 2\omega\leq5\end{equation}
and reduces the number of independent contractions down
to two: $J_{1}{}^\nu_\mu$ and $J_{5}{}^\nu_\mu$.

Finally, for $2\omega=4$ (the lowest dimension in which
a nonvanishing Weyl tensor exists), the identity (A.14)
becomes a linear combination of the identities
\begin{equation}
C_{[\alpha\beta}{}^{\gamma\delta}
C_{\gamma\delta}{}^{\kappa\mu}
C_{\kappa]\nu}{}^{\alpha\beta}
\equiv0,\hspace{7mm} 2\omega=4
\end{equation}
with the antisymmetrization over only five indices,
and there is one more identity, quadratic in $C$ :
\begin{equation}
C_{[\alpha\beta}{}^{\gamma\delta}
C_{\gamma\delta}{}^{\alpha\beta}
\delta_{\mu]}^\nu
\equiv0,\hspace{7mm} 2\omega=4.
\end{equation}
Its explicit form is
\begin{equation}
C^{\alpha\beta\gamma\nu}
C_{\alpha\beta\gamma\mu}
=\frac14\delta^\nu_\mu
C_{\alpha\beta\gamma\delta}
C^{\alpha\beta\gamma\delta}
,\hspace{7mm} 2\omega=4.
\end{equation}
When this relation is used in (A.11), the result is
\begin{equation}J_{5}{}^\nu_\mu=0
,\hspace{7mm} 2\omega=4 \end{equation}
by (A.4). Thus, in four dimensions, there is only one
independent contraction: $J_{1}{}^\nu_\mu$.

Similar results hold for invariants except that the
complete contraction of $J_5$ in (A.11) is zero for
any space-time dimension. Therefore, initially one has
four different $C^3$ invariants
\begin{eqnarray}
I_1&=&C_{\mu\beta\gamma\delta}
       C^{\mu\beta\alpha\sigma}
       C_{\alpha\sigma}{}^{\gamma\delta},\\[\baselineskip]
I_2&=&C_{\mu\beta\gamma\delta}
       C^{\mu\alpha\gamma\sigma}
       C_\alpha{}^\beta{}_\sigma{}^\delta,\\[\baselineskip]
I_3&=&C_{\mu\beta\gamma\delta}
       C^{\mu\alpha\gamma\sigma}
       C_\sigma{}^\beta{}_\alpha{}^\delta,\\[\baselineskip]
I_4&=&C_{\mu\gamma\beta\delta}
       C^{\mu\beta\alpha\sigma}
       C_{\alpha\sigma}{}^{\gamma\delta}
\end{eqnarray}
with the relations
\begin{equation} I_3=I_2-\frac14I_1,\
I_4=\frac12I_1,\end{equation}
and, for $2\omega\leq5$, the identity (A.14)
(contracted in $\mu,\nu$) adds one more relation:
\begin{equation} I_2=\frac12I_1,\hspace{7mm} 2\omega\leq5.\end{equation}
When going over from $2\omega=5$ to $2\omega=4$,
the identity (A.19) leads to no further reduction.
Thus, the dimension of the basis of local $C^3$
invariants is 2 for $2\omega>5$, and 1 for both
$2\omega=5$ and $2\omega=4$.

For invariants with the Riemann tensor, the counting
is different because, in this case, the quadratic
identity (A.19) begins working. For $2\omega\leq5$,
the identity (A.26) with the Weyl tensor expressed
through the Riemann tensor reduces the number of
independent cubic invariants by one. For $2\omega=4$,
the identity (A.19) contracted with the Ricci tensor:
\begin{equation}
C^{\alpha\beta\gamma\nu}
C_{\alpha\beta\gamma\mu}
R^\mu_\nu
=\frac14R
C_{\alpha\beta\gamma\delta}
C^{\alpha\beta\gamma\delta}
,\hspace{7mm} 2\omega=4
\end{equation}
reduces this number by one more. These results
agree with the group-theoretic analysis carried
out in [26]. According to [26], the dimension
of the basis of local cubic invariants with the
Riemann tensor (without derivatives) is 8 for
$2\omega>5$, 7 for $2\omega=5$, and 6 for $2\omega=4$.

We shall now concentrate on the space-time
dimension $2\omega=4$ and go over to the consideration
of {\it nonlocal} invariants cubic in the curvature.
Owing to their algebraic nature, the identities above
admit easily a nonlocal generalization. Indeed, the
two cubic identities obtained by antisymmetrizations in four
dimensions: eq. (A.17) contracted in $\mu,\nu$, and
eq. (A.18) contracted with $R^\mu_\nu$ can
in fact be
written down for three {\em different} tensors
and, in particular, for the curvature tensors at
{\em three different points}. One can then multiply
them by arbitrary form factors and next make the
points coincident. It will, in addition, be more convenient to deal now
with the Riemann tensor rather than the Weyl tensor.The nonlocal identities obtained in this way are of the form 
\begin{equation}\widetilde{\cal F}(\Box_1,\Box_2,\Box_3)
R_{1[\alpha\beta}{}^{\gamma\delta}
R_{2\gamma\delta}{}^{\kappa\mu}
R_{3\kappa]\mu}{}^{\alpha\beta}
=0, \ 2\omega=4
\end{equation}
\begin{equation}\widetilde{\widetilde{\cal F}}(\Box_1,\Box_2,\Box_3)
R_{1[\alpha\beta}{}^{\gamma\delta}
R_{2\gamma\delta}{}^{\alpha\beta}
R_{3\mu]}^\mu
=0, \ 2\omega=4
\end{equation}
with arbitrary $\widetilde{\cal F}(\Box_1,\Box_2,\Box_3)$ and $\widetilde{\widetilde{\cal F}}(\Box_1,\Box_2,\Box_3)$.
Since nothing is involved here except inexistence
of five different indices in four dimensions, these
identities are obviously correct in the present
nonlocal form as well.

When written down explicitly, the left-hand sides of eqs. (A.28),
(A.29) take the form (for arbitrary dimension)
\begin{eqnarray}&&
\widetilde{\cal F}(\Box_1,\Box_2,\Box_3)
R_{1[\alpha\beta}{}^{\gamma\delta}
R_{2\gamma\delta}{}^{\kappa\mu}
R_{3\kappa]\mu}{}^{\alpha\beta}\equiv
\frac1{15}\widetilde{\cal F}(\Box_1,\Box_2,\Box_3)
\nonumber\\ &&\ \ \ \ \ \ \ \
\Big[ R_{1\alpha\beta\gamma\delta}
R_2^{\gamma\delta\mu\nu}R_{3\mu\nu}{}^{\alpha\beta}
-2R_{1\alpha\beta\gamma\delta}
  R_2^{\alpha\mu\gamma\nu}R_3{}^\beta{}_\mu{}^\delta{}_\nu
-3R_{1\alpha\beta}R_2^{\alpha\pi\rho\kappa}
  R_3^\beta{}_{\pi\rho\kappa}
-R_{2\alpha\beta}R_1^{\alpha\pi\rho\kappa}
  R_3^\beta{}_{\pi\rho\kappa}
\nonumber\\ &&\ \ \ \ \ \ \ \
-R_{3\alpha\beta}R_1^{\alpha\pi\rho\kappa}
  R_2^\beta{}_{\pi\rho\kappa}
+\frac12
 R_1R_{2\alpha\beta\gamma\delta}
    R_3^{\alpha\beta\gamma\delta}
+2R_{2\alpha\beta\gamma\delta}
  R_1^{\alpha\gamma}R_3^{\beta\delta}
\nonumber\\ &&\ \ \ \ \ \ \ \
+2R_{3\alpha\beta\gamma\delta}
  R_1^{\alpha\gamma}R_2^{\beta\delta}
+2R_{1\alpha\beta}R_2^{\beta\gamma}R_3{}^\alpha_\gamma
- R_1R_{2\alpha\beta}R_3^{\alpha\beta}
\Big],                                                                      
\end{eqnarray}

\begin{eqnarray}&&
\widetilde{\widetilde{\cal F}}(\Box_1,\Box_2,\Box_3)
R_{1[\alpha\beta}{}^{\gamma\delta}
R_{2\gamma\delta}{}^{\alpha\beta}
R_{3\mu]}^\mu\equiv
-\frac2{15}\widetilde{\widetilde{\cal F}}(\Box_1,\Box_2,\Box_3)\nonumber\\&&\ \ \ \ \ \ \ \ \mbox{}\times\Big[
R_{1\alpha\beta\gamma\mu}
R_2^{\alpha\beta\gamma}{}_\nu R_3^{\mu\nu}\nonumber\\&&\ \ \ \ \ \ \ \ \mbox{}
-\frac14R_{1\alpha\beta\gamma\delta}
    R_2^{\alpha\beta\gamma\delta}R_3
-2R_{1\alpha\beta}R_2^{\beta\gamma}R_3{}^\alpha_\gamma\nonumber\\&&\ \ \ \ \ \ \ \ \mbox{}
-R_{1\alpha\beta\gamma\delta}
  R_2^{\alpha\gamma}R_3^{\beta\delta}
-R_{2\alpha\beta\gamma\delta}
  R_1^{\alpha\gamma}R_3^{\beta\delta}\nonumber\\&&\ \ \ \ \ \ \ \ \mbox{}
+R_{1\alpha\beta}R_2^{\alpha\beta}R_3
+\frac12R_1R_{2\alpha\beta}R_3^{\alpha\beta}
+\frac12R_2R_{1\alpha\beta}R_3^{\alpha\beta}\nonumber\\&&\ \ \ \ \ \ \ \ \mbox{}
-\frac14R_1R_2R_3
\Big].
\end{eqnarray}
Since these identities contain arbitrary form
factors, they suggest that, in virtue of (A.28)-(A.29), in four dimensions,
the basis of nonlocal gravitational invariants
may be redundant. However, to convert
relations (A.28), (A.29) into constraints between
the basis structures one must make one more step:
eliminate the Riemann tensor.

As discussed in paper II, the Riemann tensor
expresses through the Ricci tensor in a nonlocal
way once the boundary conditions for the gravitational
field are specified. Obtaining this expression
amounts to solving iteratively a differentiated
Bianchi identity (eq. (A.3) in Appendix A of paper II).
For the present case of a positive-signature
asymptotically euclidean space, the solution to lowest
order in the curvature is given in paper II. This accuracy
is sufficient for obtaining relations between
cubic invariants via the identities (A.30) and (A.31).
However, for the discussion of the Schwinger-DeWitt
coefficients in sect. 4 above, the solution for the
Riemann tensor is needed with accuracy ${\rm O}[R^3_{..}]$.
By making one more iteration, we obtain the needed
expression which extends to second order eq. (A.4)
of paper II:
\mathindent=0pt
\arraycolsep=0pt
\begin{fleqnarray}&& R^{\alpha\beta\mu\nu}=\frac1{\Box}\left\{
       \frac12\Big(
 \nabla^\mu \nabla^\alpha R^{\nu\beta}
+\nabla^\alpha \nabla^\mu R^{\nu\beta}
-\nabla^\nu \nabla^\alpha R^{\mu\beta}
-\nabla^\alpha \nabla^\nu R^{\mu\beta}
\right.
\nonumber\\&& \hspace{20mm}\mbox{}
-\nabla^\mu \nabla^\beta R^{\nu\alpha}
-\nabla^\beta \nabla^\mu R^{\nu\alpha}
+\nabla^\nu \nabla^\beta R^{\mu\alpha}
+\nabla^\beta \nabla^\nu R^{\mu\alpha}
\Big)
\nonumber\\&& \hspace{20mm}\mbox{}
+2R^{[\mu}_{\lambda}\Big(\nabla^{\lambda}
\nabla^{[\alpha}\frac1{\Box} R^{\beta]\nu]}\Big)
+2R^{[\alpha}_{\lambda}\Big(
\nabla^{\lambda}\nabla^{[\mu}\frac1{\Box} R^{\beta]\nu]}\Big)
\nonumber\\&& \hspace{20mm}\mbox{}
-2R^{[\mu}_{\lambda}\Big(
\nabla^{\nu]}\nabla^{[\alpha}\frac1{\Box} R^{\beta]\lambda}\Big)
-2R^{[\alpha}_{\lambda}\Big(
\nabla^{\beta]}\nabla^{[\mu}\frac1{\Box} R^{\nu]\lambda}\Big)
\nonumber\\&& \hspace{10mm}\mbox{}
-8\Big(\nabla^{\lambda}\nabla^{[\alpha}\frac1{\Box} R^{\beta]}_{\sigma}\Big)
\Big(\nabla_{\lambda}\nabla^{[\mu}\frac1{\Box} R^{\nu]\sigma}\Big)
-8\Big(\nabla^{\lambda}\nabla^{[\alpha}\frac1{\Box} R^{[\mu}_{\sigma}\Big)
\Big(\nabla_{\lambda}\nabla^{\beta]}\frac1{\Box} R^{\nu]\sigma}\Big)
\nonumber\\&& \hspace{10mm}\mbox{}
-8\Big(\nabla^{\lambda}\nabla^{[\mu}\frac1{\Box} R^{[\alpha}_{\sigma}\Big)
\Big(\nabla_{\lambda}\nabla^{\nu]}\frac1{\Box} R^{\beta]\sigma}\Big)
+8\Big(\nabla_{\lambda}\nabla^{[\alpha}\frac1{\Box} R^{\beta]}_{\sigma}\Big)
\Big(\nabla^{\sigma}\nabla^{[\mu}\frac1{\Box} R^{\nu]\lambda}\Big)
\nonumber\\&& \hspace{10mm}\mbox{}
-8\Big(\nabla^{\sigma}\nabla^{[\mu}\frac1{\Box} R^{[\alpha}_{\lambda}\Big)
\Big(\nabla^{\lambda}\nabla^{\beta]}\frac1{\Box} R^{\nu]}_{\sigma}\Big)
+8\Big(\nabla^{\lambda}\nabla^{\sigma}\frac1{\Box} R^{[\mu[\alpha}\Big)
\Big(\nabla_{\lambda}\nabla^{\beta]}\frac1{\Box} R^{\nu]}_{\sigma}\Big)
\nonumber\\&& \hspace{10mm}\mbox{}
+8\Big(\nabla^{\lambda}\nabla^{\sigma}\frac1{\Box} R^{[\mu[\alpha\Big)}
\Big(\nabla_{\lambda}\nabla^{\nu]}\frac1{\Box} R^{\beta]}_{\sigma}\Big)
+8\Big(\nabla_{\lambda}\nabla^{[\alpha}\frac1{\Box} R^{[\mu}_{\sigma}\Big)
\Big(\nabla^{\nu]}\nabla^{\beta]}\frac1{\Box} R^{\lambda\sigma}\Big)
\nonumber\\&& \hspace{10mm}\mbox{}
+8\Big(\nabla_{\lambda}\nabla^{[\mu}\frac1{\Box} R^{[\alpha}_{\sigma}\Big)
\Big(\nabla^{\nu]}\nabla^{\beta]}\frac1{\Box} R^{\lambda\sigma}\Big)
-8\Big(\nabla_{\lambda}\nabla_{\sigma}\frac1{\Box} R^{[\mu[\alpha}\Big)
\Big(\nabla^{\nu]}\nabla^{\beta]}\frac1{\Box} R^{\lambda\sigma}\Big)
\nonumber\\&& \hspace{10mm}\mbox{}
\left.
-8\Big(\nabla_{\lambda}\nabla_{\sigma}\frac1{\Box} R^{[\mu[\alpha}\Big)
\Big(\nabla^{\lambda}\nabla^{\sigma}\frac1{\Box} R^{\nu]\beta]}\Big)
-8\Big(\nabla^{[\mu}\nabla^{[\alpha}\frac1{\Box} R_{\lambda\sigma}\Big)
\Big(\nabla^{\nu]}\nabla^{\beta]}\frac1{\Box} R^{\lambda\sigma}\Big)
\right\}
\nonumber\\&& \hspace{10mm}\mbox{}
+{\rm O}[R^3_{..}].
\end{fleqnarray}
Here the antisymmetrizations on the right-hand
side are with respect to $\mu\nu$ and $\alpha\beta$.

Elimination of the Riemann tensor from the
identities (A.30) and (A.31) with the use of
(A.32) (to lowest order) brings these identities
to the following form:
\begin{eqnarray}&&
\widetilde{\cal F}(\Box_1,\Box_2,\Box_3)
R_{1[\alpha\beta}{}^{\gamma\delta}
R_{2\gamma\delta}{}^{\kappa\mu}
R_{3\kappa]\mu}{}^{\alpha\beta}
\equiv
-\frac1{15}\left\{
\frac18
\widetilde{\cal F}'
\left(\frac{\Box_1}{\Box_2\Box_3}+\frac{\Box_2}{\Box_1\Box_3}
+\frac{\Box_3}{\Box_1\Box_2}\right)
   {R_1 R_2 R_3}\right.
\nonumber\\&&\ \
+\widetilde{\cal F}'
\left(
-\frac1{\Box_1}
-\frac1{\Box_2}
-\frac1{\Box_3}
+\frac12\frac{\Box_2}{\Box_1\Box_3}
+\frac12\frac{\Box_1}{\Box_2\Box_3}
+\frac12\frac{\Box_3}{\Box_1\Box_2}
\right)
   {R_{1\,\alpha}^\mu R_{2\,\beta}^{\alpha} R_{3\,\mu}^\beta}
\nonumber\\ &&\ \
+\frac14
\Big(\frac1{\Box_1}+\frac1{\Box_2}
 -\frac{\Box_3}{\Box_1\Box_2}\Big)
\left[\widetilde{\cal F}'
           +\widetilde{\cal F}'{\big|}_{2\leftrightarrow 3}
+\widetilde{\cal F}'
{\big|}_{1\leftrightarrow 3}
\right]   {R_1^{\mu\nu}R_{2\,\mu\nu}R_3}
\nonumber\\&&\ \
+\frac14\Big(\frac1{\Box_1\Box_2}
 +\frac1{\Box_1\Box_3}+\frac3{\Box_2\Box_3}\Big)
\left[
    -\widetilde{\cal F}'
    -\widetilde{\cal F}'
                    {\big|}_{1\leftrightarrow 2}
    -\widetilde{\cal F}'
                    {\big|}_{1\leftrightarrow 3}
\right]  {R_1^{\alpha\beta}
\nabla_\alpha R_2 \nabla_\beta R_3}
\nonumber\\ &&\ \
+\frac1{\Box_1\Box_2}\left[
  \widetilde{\cal F}'
           +\widetilde{\cal F}'{\big|}_{2\leftrightarrow 3}
  +\widetilde{\cal F}'{\big|}_{1\leftrightarrow 3}
\right]  {\nabla^\mu R_1^{\nu\alpha}
\nabla_\nu R_{2\,\mu\alpha}R_3}
\nonumber\\&&\ \
+\frac1{\Box_2\Box_3}\left[
    \widetilde{\cal F}'+\widetilde{\cal F}'
          {\big|}_{1\leftrightarrow 2}
+\widetilde{\cal F}'
                    {\big|}_{1\leftrightarrow 3}
\right]  {R_1^{\mu\nu}\nabla_\mu R_2^{\alpha\beta}\nabla_\nu
         R_{3\,\alpha\beta}}
\nonumber\\&&\ \
+\Big(\frac1{\Box_1\Box_2}+\frac1{\Box_1\Box_3}
      -\frac1{\Box_2\Box_3}\Big)\left[
    \widetilde{\cal F}'
    +\widetilde{\cal F}'{\big|}_{1\leftrightarrow 2}
    +\widetilde{\cal F}'{\big|}_{1\leftrightarrow 3}
\right]  {R_1^{\mu\nu}\nabla_\alpha R_{2\,\beta\mu}\nabla^\beta
       R_{3\,\nu}^\alpha}
\nonumber\\&&\ \
+\frac1{\Box_1\Box_2\Box_3}\left[
-\widetilde{\cal F}'
           -\widetilde{\cal F}'{\big|}_{2\leftrightarrow 3}
	   -\widetilde{\cal F}'
{\big|}_{1\leftrightarrow 3}
\right]  {\nabla_\alpha\nabla_\beta R_1^{\mu\nu}
\nabla_\mu\nabla_\nu
       R_2^{\alpha\beta} R_3}    \nonumber\\&&\ \
\left.+\frac2{\Box_1\Box_2\Box_3}\left[
-\widetilde{\cal F}'
           -\widetilde{\cal F}'{\big|}_{2\leftrightarrow 3}
	   -\widetilde{\cal F}
{\big|}_{1\leftrightarrow 3}
\right]  {\nabla_\mu R_1^{\alpha\lambda}
\nabla_\nu R_{2\,\lambda}^\beta
     \nabla_\alpha\nabla_\beta R_3^{\mu\nu}}
\right\}\nonumber\\
&&\ \ +{\rm a\ total\ derivative}+{\rm O}[R^4_{..}],\ \ \ \ \ \
\widetilde{\cal F}'\equiv(\Box_1-\Box_2-\Box_3)
\widetilde{\cal F},
\end{eqnarray}
\begin{eqnarray}
&&\widetilde{\widetilde{\cal F}}(\Box_1,\Box_2,\Box_3)
R_{1[\alpha\beta}{}^{\gamma\delta}
R_{2\gamma\delta}{}^{\alpha\beta}
R_{3\mu]}^\mu
\equiv
-\frac2{15}\left\{
\frac18\widetilde{\widetilde{\cal F}}
\left(\frac{\Box_1}{\Box_2}+\frac{\Box_2}{\Box_1}
+\frac{{\Box_3}^2}{\Box_1\Box_2}\right)
   {R_1 R_2 R_3} \right.  \nonumber\\
&&\ \
+\widetilde{\widetilde{\cal F}}\left(
-1+\frac12\frac{\Box_1}{\Box_2}+\frac12\frac{\Box_2}{\Box_1}
       -\frac{\Box_3}{\Box_1}-\frac{\Box_3}{\Box_2}
       +\frac12\frac{{\Box_3}^2}{\Box_1\Box_2}\right)
   {R_{1\,\alpha}^\mu R_{2\,\beta}^{\alpha} R_{3\,\mu}^\beta}
\nonumber\\
&&\ \
+\frac14\left[
    \widetilde{\widetilde{\cal F}}\Big(
\frac{\Box_3}{\Box_1}+\frac{\Box_3}{\Box_2}
-\frac{{\Box_3}^2}{\Box_1\Box_2}\Big)
   +{\widetilde{\widetilde{\cal F}}}
{\big|}_{1\leftrightarrow 3}
\Big(1-\frac{\Box_3}{\Box_2}+\frac{\Box_1}{\Box_2}\Big)\right.
\left.
   +{\widetilde{\widetilde{\cal F}}}
{\big|}_{2\leftrightarrow 3}
\Big(1-\frac{\Box_3}{\Box_1}+\frac{\Box_2}{\Box_1}\Big)\right]
   {R_1^{\mu\nu}R_{2\,\mu\nu}R_3}
\nonumber\\&&\ \
+\frac14\left[
    -\Big(\widetilde{\widetilde{\cal F}}
+{\widetilde{\widetilde{\cal F}}}
{\big|}_{1\leftrightarrow 2}\Big)
     \Big(\frac1{\Box_1}+\frac3{\Box_2}
+\frac{\Box_3}{\Box_1\Box_2}\Big)  \right.
\left.
    -{\widetilde{\widetilde{\cal F}}}
{\big|}_{1\leftrightarrow 3}
     \Big(\frac1{\Box_2}+\frac1{\Box_3}
+3\frac{\Box_1}{\Box_2\Box_3}\Big)
   \right]
   {R_1^{\alpha\beta}
\nabla_\alpha R_2 \nabla_\beta R_3}
\nonumber\\
&&\ \
+\left(\widetilde{\widetilde{\cal F}}
\frac{\Box_3}{\Box_1\Box_2}
+{\widetilde{\widetilde{\cal F}}}
{\big|}_{1\leftrightarrow 3}\frac1{\Box_1}
   +{\widetilde{\widetilde{\cal F}}}
{\big|}_{1\leftrightarrow 3}\frac1{\Box_2}\right)
   {\nabla^\mu R_1^{\nu\alpha}
\nabla_\nu R_{2\,\mu\alpha}R_3}
\nonumber\\&&\ \
+\left[
   {\widetilde{\widetilde{\cal F}}}
{\big|}_{1\leftrightarrow 3}
\frac{\Box_1}{\Box_2\Box_3}
   +\Big(\widetilde{\widetilde{\cal F}}
+{\widetilde{\widetilde{\cal F}}}
{\big|}_{1\leftrightarrow 2}\Big)\frac1{\Box_2}
  \right]
  {R_1^{\mu\nu}\nabla_\mu R_2^{\alpha\beta}\nabla_\nu
         R_{3\,\alpha\beta}}
\nonumber\\&&\ \
+\left[
   {\widetilde{\widetilde{\cal F}}}
{\big|}_{1\leftrightarrow 3}
\Big(\frac1{\Box_2}+\frac1{\Box_3}
-\frac{\Box_1}{\Box_2\Box_3}\Big)\right.
\left.
   +\Big(\widetilde{\widetilde{\cal F}}
+{\widetilde{\widetilde{\cal F}}}
{\big|}_{1\leftrightarrow 2}\Big)
    \Big(\frac1{\Box_1}-\frac1{\Box_2}
+\frac{\Box_3}{\Box_1\Box_2}\Big)
 \right]
  {R_1^{\mu\nu}\nabla_\alpha R_{2\,\beta\mu}\nabla^\beta
       R_{3\,\nu}^\alpha}  \nonumber\\&&\ \
+\left(-\widetilde{\widetilde{\cal F}}
\frac1{\Box_1\Box_2}
-{\widetilde{\widetilde{\cal F}}}
{\big|}_{1\leftrightarrow 3}
\frac1{\Box_2\Box_3}-{\widetilde{\widetilde{\cal F}}}
{\big|}_{2\leftrightarrow 3}\frac1{\Box_1\Box_3}\right)
  {\nabla_\alpha\nabla_\beta R_1^{\mu\nu}
\nabla_\mu\nabla_\nu
       R_2^{\alpha\beta} R_3}    \nonumber\\&&\ \
+2\left(-\widetilde{\widetilde{\cal F}}\frac1{\Box_1\Box_2}
-{\widetilde{\widetilde{\cal F}}}
{\big|}_{1\leftrightarrow 3}
\frac1{\Box_2\Box_3}-{\widetilde{\widetilde{\cal F}}}
{\big|}_{2\leftrightarrow 3}\frac1{\Box_1\Box_3}\right)
  {\nabla_\mu R_1^{\alpha\lambda} \nabla_\nu R_{2\,\lambda}^\beta
     \nabla_\alpha\nabla_\beta R_3^{\mu\nu}}
\Big\}\nonumber\\
&&\ \
+{\rm a\ total\ derivative}
+{\rm O}[R^4_{..}],               
\end{eqnarray}
\arraycolsep=3pt
\mathindent=\leftmargini
where
\[
{\widetilde{\cal F}}'
{\big|}_{1\leftrightarrow 2}
\equiv\widetilde{\cal F}'
(\Box_2,\Box_1,\Box_3),\ {\mathrm etc}.
\]
It is now seen that, up to total derivatives and terms ${\rm
O}[R^4_{..}]$, (A.33) is an equivalent form of (A.34)
corresponding to
\[
\widetilde{\widetilde{\cal F}}=
\frac12\frac1{\Box_3}\widetilde{\cal
F}'=\frac12\frac{\Box_1-\Box_2-\Box_3}{\Box_3}
\widetilde{\cal F}.
\]

Thus, assuming integration over the space-time which will
make irrelevant total derivatives, we conclude that,
at third order in the curvature, of the two generally
different identities for nonlocal invariants, existing
in four dimensions, only one is independent: eq. (A.29).
This equation is brought to its final form by putting
\[
\widetilde{\widetilde{\cal F}}(\Box_1,\Box_2,\Box_3)=-\frac16\Box_1\Box_2{\cal F}(\Box_1,\Box_2,\Box_3)
\]
where ${\cal F}(\Box_1,\Box_2,\Box_3)$ is a new arbitrary function,
and taking into account the symmetries of the
tensor structures entering (A.34). The result is
the following constraint between the basis
invariants listed in the table (2.15)-(2.43):
\begin{eqnarray}&&
\int\! dx\, g^{1/2}\, {\rm tr}\,{\cal F}^{\rm sym}(\Box_1,\Box_2,\Box_3)\nonumber\\
&&\times\Big\{
-\frac1{48}({\Box_1}^2+{\Box_2}^2+{\Box_3}^2)\Re_1\Re_2\Re_3({9})\nonumber\\&&\ \ \ \ \mbox{}
-\frac1{12}({\Box_1}^2+{\Box_2}^2+{\Box_3}^2
  -2{\Box_1}{\Box_2}-2{\Box_2}{\Box_3}-2{\Box_1}{\Box_3})\Re_1\Re_2\Re_3({10})\nonumber\\&&\ \ \ \ \mbox{}
-\frac18{\Box_3}({\Box_1}+{\Box_2}-{\Box_3})\Re_1\Re_2\Re_3({11})\nonumber\\&&\ \ \ \ \mbox{}
+\frac18(3{\Box_1}+{\Box_2}+{\Box_3})\Re_1\Re_2\Re_3({22})
-\frac12{\Box_3}\Re_1\Re_2\Re_3({23})\nonumber\\&&\ \ \ \ \mbox{}
-\frac12{\Box_1}\Re_1\Re_2\Re_3({24})
-\frac12({\Box_2}+{\Box_3}-{\Box_1})\Re_1\Re_2\Re_3({25})\nonumber\\&&\ \ \ \ \mbox{}
+\frac12\Re_1\Re_2\Re_3({27})\nonumber\\&&\ \ \ \ \mbox{}
+\Re_1\Re_2\Re_3({28})
\Big\}+{\rm O}[\Re^4]=0,\ 2\omega=4
\end{eqnarray}
where ${\cal F}^{\rm sym}(\Box_1,\Box_2,\Box_3)$ is a {\em
completely symmetric} but otherwise arbitrary
function. This constraint, valid in four dimensions,
reduces the basis of nonlocal gravitational
invariants by one structure. With its aid one can
exclude everywhere either the structure 9 or 10 or the
{\em completely symmetric} (in the labels 1,2,3) part
of anyone of the remaining purely gravitational
structures except $\Re_1\Re_2\Re_3({29})$. The latter structure
which is the only one containing six derivatives
is absent from the constraint (A.35) and is,
therefore, inexcludable. This can be explained by the fact
that its local version is the only independent
contraction of three Weyl tensors: eq. (A.21) (with eq. (A.32) used).

As seen from eq. (A.35), elimination of any structure
except 27 and 28 will result in the appearance of new
nonanalytic terms in the $\alpha$-representation
of the form factors which may complicate obtaining
further representations like the Laplace and
spectral ones. Therefore, in the text, the constraint
(A.35) is used to eliminate the completely symmetric
part of the structure 28.

At least apparently, eqs. (A.28) and (A.29) are not
the most general nonlocal identities that can be written
down by antisymmetrizing five indices. More
generally, one can apply this procedure to three
tensors with
arbitrary indices and arbitrary number of uncontracted
derivatives. Therefore, to make sure that there are no
more constraints between the basis invariants, an independent
check is needed. Since, at third order in the curvature,
the maximum number of derivatives that do not contract
in the box operators is six, we begin this check
with nonlocal structures having three Ricci tensors and
six derivatives. There exist only two such, and only
one of them is independent:
\begin{equation}
\hat{1}\nabla_\alpha\nabla_\beta R_1^{\gamma\delta}
   \nabla_\gamma\nabla_\delta R_2^{\mu\nu}
   \nabla_\mu\nabla_\nu R_3^{\alpha\beta} = \Re_1\Re_2\Re_3({29}),
\end{equation}
\begin{equation}
\hat{1}\nabla_\alpha\nabla_\beta R_1^{\gamma\delta}
   \nabla_\gamma\nabla_\mu R_2^{\alpha\nu}
   \nabla_\delta\nabla_\nu R_3^{\beta\mu} = - \Re_1\Re_2\Re_3({29}) + \dots
\end{equation}
where the ellipses $\dots$ stand for total derivatives and
terms with derivatives contracting in the box operators.
Eq. (A.37) is obtained by three integrations by parts
applied to $\nabla_\alpha, \nabla_\mu$ and $\nabla_\delta$.
Since not more than one index of each Ricci tensor
and not more than one derivative acting on each Ricci tensor
may participate in the antisymmetrization (otherwise
the result will be either trivial or ${\rm O}[R^4_{..}]$), there
are only two possible 5 - antisymmetrizations of (A.36):
\[
\nabla_{[\alpha} \nabla^\beta R_{1\delta}^\gamma
\nabla_\gamma \nabla^\delta R_{2\nu}^\mu
\nabla_{\mu]} \nabla^\nu R_3^{\alpha\beta} = 0,
\]
\[
\nabla_\alpha \nabla_\beta R_{1[\delta}^\gamma
\nabla_\gamma \nabla^\delta R_{2\nu}^\mu
\nabla_{\mu} \nabla^\nu R_{3\beta]}^{\alpha} = 0.
\]
In each of these cases, upon calculation, the terms
(A.36) and (A.37) appear in a sum with equal coefficients
and, therefore, cancel. This proves that the structure
with six derivatives remains unconstrained. Among the
invariants with three Ricci tensors and four derivatives,
only two are independent: the basis structures 27 and 28,
and only the latter admits nontrivial 5-antisymmetrizations.
There is, moreover, only one such:
\[
\nabla_{[\mu} R_{1\lambda}^\alpha
\nabla^\nu R_{2\beta}^\lambda
\nabla_\alpha\nabla^\beta R^\mu_{3\nu]} = 0.
\]
Upon calculation and multiplication by an arbitrary form
factor, the latter identity gives precisely the constraint
(A.35). The invariants with the commutator curvature and
four derivatives are reducible (see sect. 14) and,
therefore, absent from the basis. Finally, invariants
with two derivatives do not admit a nontrivial
5-antisymmetrization since, for that, one needs at
least ten indices: five uncontracted to be involved in
the antisymmetrization, and five more to make a
complete contraction.

Thus, in four dimensions, there is only one constraint
between the basis structures, and the dimension of the basis
of nonlocal cubic invariants which is generally 29
and in the case of the gravitational invariants 10
becomes respectively 28 and 9.

The nonlocal identity obtained above has a direct
relation to the Gauss-Bonnet identity in four
dimensions. Indeed, by calculating the square of the
Riemann tensor with the aid of eq. (A.32), one
finds for arbitrary$2\omega$ :
\begin{eqnarray}&&
\int\! dx\, g^{1/2}\, \,\Big(
R_{\alpha\beta\gamma\delta}
R^{\alpha\beta\gamma\delta}
-4 R_{\mu\nu} R^{\mu\nu} + R^2
\Big)
\nonumber\\&&\ \ \ \ \mbox{}
=\int\! dx\, g^{1/2}\, \, \left[\frac12\frac{\Box_1}{\Box_2\Box_3}
   {R_1 R_2 R_3} \right.  \nonumber\\&& \hspace{15mm}\mbox{}
+2\Big(\frac{\Box_1}{\Box_2\Box_3}-2\frac1{\Box_1}\Big)
   {R_{1\,\alpha}^\mu R_{2\,\beta}^{\alpha}
R_{3\,\mu}^\beta} \nonumber\\&& \hspace{15mm}\mbox{}
+\Big(2\frac1{\Box_1}-\frac{\Box_3}{\Box_1\Box_2}\Big)
   {R_1^{\mu\nu}R_{2\,\mu\nu}R_3}  \nonumber\\&& \hspace{15mm}\mbox{}
+\Big(-2\frac1{\Box_1\Box_3}-3\frac1{\Box_2\Box_3}\Big)
   {R_1^{\alpha\beta}\nabla_\alpha R_2 \nabla_\beta R_3} \nonumber\\&& \hspace{15mm}\mbox{}
+4\frac1{\Box_1\Box_2}
   {\nabla^\mu R_1^{\nu\alpha}\nabla_\nu R_{2\,\mu\alpha}R_3} \nonumber\\&& \hspace{15mm}\mbox{}
+4\frac1{\Box_2\Box_3}
  {R_1^{\mu\nu}\nabla_\mu R_2^{\alpha\beta}\nabla_\nu
         R_{3\,\alpha\beta}}  \nonumber\\&& \hspace{15mm}\mbox{}
+4\Big(2\frac1{\Box_1\Box_2}-\frac1{\Box_2\Box_3}\Big)
  {R_1^{\mu\nu}\nabla_\alpha R_{2\,\beta\mu}\nabla^\beta
       R_{3\,\nu}^\alpha}  \nonumber\\&& \hspace{15mm}\mbox{}
-4\frac1{\Box_1\Box_2\Box_3}
  {\nabla_\alpha\nabla_\beta R_1^{\mu\nu}\nabla_\mu\nabla_\nu
       R_2^{\alpha\beta} R_3}    \nonumber\\&& \hspace{15mm}\mbox{}
-8\frac1{\Box_1\Box_2\Box_3}
  {\nabla_\mu R_1^{\alpha\lambda} \nabla_\nu R_{2\,\lambda}^\beta
     \nabla_\alpha\nabla_\beta R_3^{\mu\nu}}\Big]
+{\rm O}[R^4_{..}].
\end{eqnarray}
In agreement with the result of paper II, a contribution
of second order in the curvature is absent from this expression
for any space-time dimension. The third-order contribution
(A.38) does not generally vanish but vanishes in four
dimensions because it coincides with the left-hand side of
the identity (A.35) if in the latter one puts
\[
{\cal F}^{\rm sym}(\Box_1,\Box_2,\Box_3) = -8({\rm tr}\hat{1})^{-1}\frac1{\Box_1\Box_2\Box_3}.
\]
Comparison of eqs. (A.38) and (A.34) gives
\begin{eqnarray}&&
\int\! dx\, g^{1/2}\, \,\Big(
R_{\alpha\beta\gamma\delta}
R^{\alpha\beta\gamma\delta}
-4 R_{\mu\nu} R^{\mu\nu} + R^2
\Big) \nonumber\\&&\ \ \ \ \mbox{}\ \ \ \
=\int\! dx\, g^{1/2}\, \,
\Big(-10\frac1{\Box_3}\Big)
R_{1[\alpha\beta}{}^{\gamma\delta}
R_{2\gamma\delta}{}^{\alpha\beta}
R_{3\mu]}^\mu +
{\rm O}[R^4_{..}].
\end{eqnarray}
This relation {\em valid for any number of space-time dimensions}
elucidates the mechanism by which the Gauss-Bonnet
identity arises in four dimensions.

\section*{Acknowledgments}

\hspace{\parindent}
The key idea in sect. 20 of factorizing the Heron
spectral weight via eq. (20.7) was suggested to the grateful
authors by Armen Mirzabekian who uses this representation
in the derivation of the Hawking effect from the nonlocal
effective equations (work in progress partially reported
in [4,5]). Special thanks to Gabor Kunstatter and Don
Page whose efforts in bringing the authors together
at the hard times of Russian diaspora made possible
the completion of this work.

\end{document}